# Conceptual Design Report of Super Tau-Charm Facility: The Accelerator

Jiancong Bao[1], Anton Bogomyagkov[12], Zexin Cao[1], Mingxuan Chang[4], Fangzhou Chen[2], Guanghua Chen[2], Qi Chen[1], Qushan Chen[6], Zhi Chen[1], Kuanjun Fan[6], Hailiang Gong[8], Duan Gu[2], Hao Guo[3], Tengjun Guo[1], Chongchao He[9], Tianlong He[1], Kaiwen Hou[7], Hao Hu[6], Tongning Hu[6], Xiaocheng Hu[1], Dazhang Huang[2], Pengwei Huang[7], Ruixuan Huang[1], Zhicheng Huang[1], Hangzhou Li[1], Renkai Li[7], Sangya Li[1], Weiwei Li[1], Xuan Li[2], Xunfeng Li[5], Yu Liang[1], Chao Liu[1], Tao Liu[1], Xiaoyu Liu[1], Xuyang Liu[10], Yuan Liu[6], Huihui Lv[2], Qing Luo[1], Tao Luo[9], Mikhail Skamarokha[12], Shaohang Ma[1], Wenbin Ma[3], Masahito Hosaka[11], Xuece Miao[9], Yihao Mo[1], Kazuhito Ohmi[11], Jian Pang[1], Guoxi Pei[1], Zhijun Qi[9], Fenglei Shang[1], Lei Shang[1], Caitu Shi[1], Kun Sun[8], Li Sun[1], Jingyu Tang[1*], Anxin Wang[1], Chengzhe Wang[1], Hongjin Wang[1], Lei Wang[4], Qian Wang[1], Shengyuan Wang[1], Shikang Wang[1], Ziyu Wang[1], Shaoqing Wei[3], Yelong Wei[1], Jun Wu[2], Sang Wu[2], Chunjie Xie[1], Ziyu Xiong[1], Xin Xu[1], Jun Yang[6], Penghui Yang[1], Tao Yang[3], Lixin Yin[2], Chen Yu[1], Ze Yu[1], Youjin Yuan[4], Yifeng Zeng[6], Ailin Zhang[1], Haiyan Zhang[1], Jialian Zhang[6], Linhao Zhang[1], Ning Zhang[2], Ruiyang Zhang[8], Xiaoyang Zhang[7], Yihao Zhang[1], Yangcheng Zhao[1], Jingxin Zheng[3], Demin Zhou[11], Hao Zhou[1], Yimei Zhou[2], Zeran Zhou[1], Bing Zhu[6], Xinghao Zhu[1], Zi'an Zhu[1], Ye Zou[1]

[1] University of Science and Technology of China, Hefei, Anhui, China
[2] Shanghai Advanced Research Institute, Shanghai, China
[3] Hefei Institute of Physical Sciences, Hefei, Anhui, China
[4] Institute of Modern Physics, Chinese Academy of Sciences, Lanzhou, Gansu, China
[5] Institute of Engineering Thermophysics, Chinese Academy of Sciences, Beijing, China
[6] Huazhong University of Science and Technology, Wuhan, Hubei, China
[7] Tsinghua University, Beijing, China
[8] Anhui University of Science and Technology, Huainan, Anhui, China
[9] Institute of Advanced Light Source Facilities, Shenzhen, Guangdong, China
[10] Flight Technology Research Institute, Beijing, China, China
[11] High Energy Accelerator Research Organization, Tsukuba, Japan
[12] Budker Institute of Nuclear Physics, Novosibirsk, Russia

[*] Corresponding author: jytang@ustc.edu.cn (Jingyu Tang)

**Abstract**: **Abstract**: Electron-positron colliders operating in the GeV region of center-of-mass energies or the Tau-Charm energy region, have been proven to enable competitive frontier research, due to its several unique features. With the progress of high energy physics in the last two decades, a new-generation Tau-Charm factory, Super Tau Charm Facility (STCF) has been actively promoting by the particle physics community in China. STCF holds great potential to address fundamental questions such as the essence of color confinement and the matter-antimatter asymmetry in the universe in the next decades. The main design goals of STCF are with a center-of-mass energy ranging from 2 to 7 GeV and a peak luminosity surpassing $5 \times 10^{34}\ cm^{-2}s^{-1}$ that is optimized at a center-of-mass energy of 4 GeV, which is about 50 times that of the currently operating Tau-Charm factory - BEPCII. The STCF accelerator is composed of two main parts: a double-ring collider with the crab-waist collision scheme and an injector that provides top-up injections for both electron and positron beams. As a typical third-generation electron-positron circular collider, the STCF accelerator faces many challenges in both accelerator physics and technology. In this paper, the conceptual design of the STCF accelerator complex is presented, including the ongoing efforts and plans for technological R&D, as well as the required infrastructure. The STCF project aims to secure support from the Chinese central government for its construction during the 15th Five-Year Plan (2026-2030) in China.

CONTENTS

# 1 General Overview

## 1.1 Scientific Objectives

The high brightness frontier is one of the three major frontiers in contemporary international particle physics research and holds an indispensable position. The current high-brightness cutting-edge experimental facilities all together revolve key scientific problems, and are both competitive and complementary. The operational and on-construction facilities include: LHC/LHCb at CERN, SuperKEKB/ Belle II at KEK, BEPCII/BESIII at IHEP, FAIR/PANDA at GSI, and CEBAF/GlueX at JLab.

LHCb and Belle II primarily focus on B physics, producing large samples of hadrons containing bottom quarks, while also yielding significant numbers of charm hadrons and tau leptons. PANDA and GlueX - based on proton-antiproton annihilation using an antiproton beam on a fixed target and high-energy photoproduction using polarized photons on a proton target, respectively - also cover the tau-charm energy region. The BESIII experiment has concentrated on charm hadron and tau lepton physics; however, the current BEPCII accelerator lacks the upgrade potential to meet the demands of particle physics research in the next 20–30 years. Over the past decade, Chinese scientists have actively advanced the concept of a new-generation tau-charm factory - known as the **Super Tau-Charm Facility (STCF) -** and are striving to launch it as a major national science infrastructure project during the 15th Five-Year Plan period, with the goal of sustaining China's global leadership in tau-charm physics established over the past three decades. In addition to China, Russia is also pursuing a similar initiative - the SCTF project.

STCF aims to collect an enormous and unique dataset in the tau-charm energy region, with an expected integrated luminosity of approximately 1 $ab^{-1}$ per year. The primary physics goals include:

***Studies of CP Violation and New Sources of CPV***

CP violation is a key ingredient in explaining the matter-antimatter asymmetry in the universe. STCF will produce billions of quantum-entangled pairs of neutral D mesons, tau leptons, and hyperons, offering a unique environment for precision CPV studies. In particular, hyperon CP violation searches at STCF are expected to reach world-leading sensitivity (better than $10^{-4}$), directly challenging the Standard Model (SM) predictions.

***Hadron Spectroscopy and Exotic States***

Just as spectroscopy is central to atomic and molecular physics, hadron spectroscopy plays a crucial role in understanding QCD confinement. STCF will yield trillions of (charmonium-like) quark-antiquark states, enabling precise studies of the light hadron spectrum, charmonium-like structures, and the systematics of exotic hadrons.

***Nucleon Structure and Formation***

Probing the internal structure of nucleons is essential for uncovering the properties of matter and QCD confinement. STCF will perform threshold scans of baryonic final states to measure nucleon electromagnetic form factors, baryon decay constants, and strong phases, with at least an order-of-magnitude improvement over existing measurements.

***Precision Measurements of Fundamental Parameters and Searches for New Physics***

STCF will significantly improve the precision of fundamental quantities such as the tau lepton mass, strong phases in neutral D meson decays, and the magnetic dipole moments of baryons and leptons. The sensitivity to potential new physics will be enhanced by up to two orders of magnitude compared to current limits.

## 1.2 Accelerator Design Objectives

The STCF accelerator complex is designed to fulfill the above physics goals through high-luminosity operation in the center-of-mass energy range of 2.0-7.0 GeV. The baseline accelerator design targets include:

- **Beam energy range for both electrons and positrons**: Tunable from 1.0 to 3.5 GeV per beam (2.0 to 7.0 GeV in c.m. energy)
- **Luminosity**: $\geq 5 \times 10^{34}$ $cm^{-2}s^{-1}$ at 2.0 GeV beam energy
- **Operational mode**: Top-up (constant current) injection mode
- **Future upgrade potential**: Design provisions for future luminosity enhancement and other possibilities such as polarized beam and monochrmonization

## 1.3 Accelerator Conceptual Scheme

The core design objective of a new-generation electron-positron collider is to largely increase luminosity. Compared to the currently operating BEPCII, STCF should achieve approximately two orders of magnitude enhancement in luminosity, which demands both new design concepts and advanced technologies. Since around 2010, the international community has converged on a set of key features for the so-called third-generation $e^+e^-$ colliders: a large crossing angle, the use of a crab-waist collision scheme, extremely small vertical beta function ($\beta_y^*$), low beam emittance, and high beam currents. As a third-generation low-energy collider, STCF also adopts this design philosophy while addressing unique challenges specific to the tau-charm energy region.

To meet the physics requirements - center-of-mass energy ranging from 2 to 7 GeV and luminosity $\geq 5 \times 10^{34}$ $cm^{-2}s^{-1}$ at the optimized energy of 4 GeV - STCF will adopt the most advanced designs and techniques for third-generation c+e- colliders. This includes: a double-ring configuration with separate storage rings for electrons and positrons with sufficient circumferences of above 800 m; a large Piwinski angle (total crossing angle of 60 mrad) and a crab-waist collision scheme; high stored beam currents (~2 A); low emittance (horizontal emittance about 5 nm·rad, with transverse coupling below 1%); extremely small $\beta$-functions at the interaction point ($\beta_y^*$< 1 mm), resulting in vertical beam spot sizes of about 100 nm.

In the crab-waist scheme, sextupole magnets near the interaction point (IP) rotate the beam-waist orientation, which, combined with strong focusing optics, significantly reduces the dynamic and momentum acceptances of the ring. This not only complicates beam injection but also leads to an extremely short beam lifetime (<300 s), primarily limited by the Touschek scattering. This beam lifetime is far shorter than in other electron storage rings (colliders and synchrotron radiation light sources), and imposes great challenges on both the collider ring physics design and technical design.

In addition, STCF is designed to operate in a top-up mode, requiring frequent beam injection to maintain quasi-constant current. The very short beam lifetime in the collider rings poses great challenges to the beam injection and the design of the injector. Currently, two injection schemes for the collider rings are under parallel development: a conventional off-axis injection scheme and a state-of-the-art bunch swap-out injection scheme. They require very different bunch charges for the beam injection, about 1 nC for the off-axis injection scheme and about 8 nC for the swap-out injection scheme. Accordingly, the injector should be designed to follow the injection schemes.

STCF adopts an injector scheme with full-energy linac, namely the linacs will accelerate both the electron beam and the positron beam to the injection energy of the collider rings, from 1.0 to 3.5 GeV, according to the collider operation energy. This is to meet the frequent beam injections in the collider rings. While the electron beam pre-accelerated in the linac can be injected into the collider electron ring, the positron beam is first produced by a high-energy and high-current electron beam. As a secondary particle, the positron collection efficiency is very low, and the beam property is very poor. The most challenging aspects for the injector design are with the positron beam. Different injector designs are carried out for the off-axis injection scheme and the swap-out injection scheme in the collider rings:

- **For the Off-Axis Injection Scheme**: In this scheme, each time a single electron/positron bunch of ≤1.5 nC is injected using traditional methods. The electron beam is generated by a photocathode (PC) RF electron gun that provides a good beam quality and is accelerated to the injection energy directly. The positron beam is generated using a high-charge, high-energy electron beam on a positron production target. Here, the electron beam is from a thermionic-cathode (TC) RF electron gun and accelerated to 1.5 GeV. These positrons are collected, accelerated to 1.0 GeV, then damped in a damping ring, and reaccelerated to the required injection energy (see Fig. 1.2-1, top).
- **For the Bunch Swap-out Injection Scheme**: In this scheme, at each injection, a high-charge (8.5 nC) bunch for both electrons and positrons is required. The electron beam for direction injection in the collider electron ring is still from a photocathode gun, but with a higher bunch charge of 8.5 nC. The electron beam for generating positrons is again from a thermionic gun, but with an even higher bunch charge of 11.6 nC and accelerated to 2.5 GeV with a high repetition rate of 90 Hz. Afterwards, the positrons are collected, captured, accelerated to 1 GeV, and accumulated in an accumulation ring with strong damping to achieve low emittance before the acceleration in the main linac and final injection into the collider positron ring (see Fig. 1.2-1, middle).

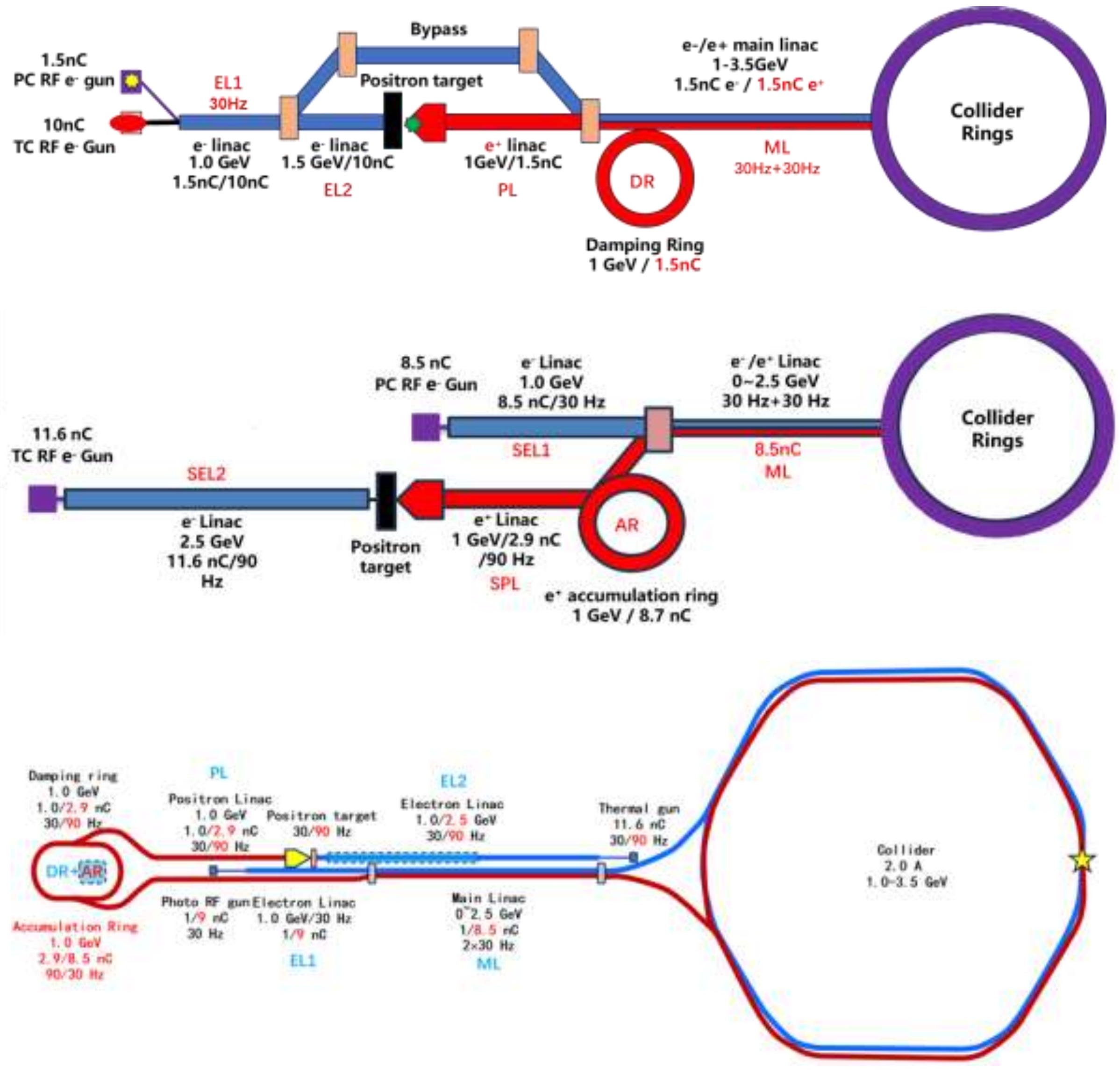

Figure 1.2-1: STCF accelerator general layouts. Top (a): Collider rings with the off-axis injection scheme; Middle (b): Collider rings with the bunch swap-out injection scheme; Bottom (c): Compatible injector scheme for both the off-axis and bunch swap-out injections

No matter whether the off-axis injection or swap-out injection is used, the main linac should accelerate both the electron beam and positron beam alternately at a maximum of 30 Hz each. It will be capable of adjusting the energy of both electron and positron beams across the 1.0-3.5 GeV range, according to the operating energy of the collider rings.

Based on feasibility studies, a **compatible injector scheme** (see Fig. 1.2-1, bottom) has been proposed. With this scheme, the linac tunnel hosts two linacs in two opposite directions. The off-axis injection mode is chosen as the baseline scheme, since it is a mature and cheaper solution. The electron beam energy to drive the positron target is reduced to 1 GeV, which produces positron bunches of 1 nC. The bunch charges for both electrons and positrons for the injection into the collider rings are 1 nC. However, if the future study and operation find that the bunch swap-out injection mode is needed to tackle the great challenge related to the strong coupling between the injected beam and beam-beam effect, the injector can be upgraded to the full bunch swap-out injection mode straightforwardly by increasing the drive electron beam energy to 2.5 GeV and the repetition rate to 90 Hz. The positron bunch charge is accumulated to 8.5 nC by adding an accumulation ring to the damping ring to form a dual-ring damping system in the same tunnel.

## 1.4 Key Physics Design Challenges and Technologies

As a new-generation electron-positron collider, the STCF accelerator complex not only adopts advanced design concepts but also requires the development of new accelerator physics methodologies and enabling novel accelerator technologies. Based on the preliminary studies and international collaborations, several critical areas have been identified where state-of-the-art design techniques and technological innovations must be applied, or existing methods significantly evolved, in order to meet the stringent demands of the STCF construction and operation.

### 1.4.1 Collider Ring Physics Design

The most challenging aspect of the collider ring design lies in the interaction region. On one hand, ultra-strong focusing is needed to achieve extremely small vertical beta functions ($\beta_y^* \leq 1$ mm) at the interaction point. On the other hand, this strong focusing induces large local chromaticities, which, along with sextupoles for chromatic correction, fringe fields from superconducting magnets, and crab sextupoles, significantly degrade the dynamic aperture and momentum aperture of both the electron and positron rings. These effects become particularly critical in the regime of low emittance and high beam current, where the Touschek scattering severely limits beam lifetime.

This design complexity is shared across multiple third-generation $e^+e^-$ collider projects worldwide, each facing varying degrees of difficulty. Even SuperKEKB, the only third-generation collider currently in operation, has struggled to reach its design beam parameters. There is a growing international consensus that new-generation colliders must address a wide set of interdependent physics mechanisms—strong nonlinearities, Touschek effects, collective instabilities, beam-beam interactions, injection dynamics, machine errors, beam collimation, and radiation damping—not in isolation, but through comprehensive and integrated design approaches. Traditionally, they are studied independently. Existing simulation tools must therefore be upgraded to accommodate such complex, multi-physics studies.

### 1.4.2 Injector Physics Design

To accommodate the wide operational energy range (1-3.5 GeV) and short beam lifetime of the collider rings—particularly under the bunch swap-out injection scheme—the injector system must deliver electron and positron beams with the following characteristics: variable energy (1-3.5 GeV), high repetition rate (~30 Hz), low emittance (<30 nm·rad) and high bunch charge (8.5 nC). Meeting these requirements poses significant challenges in the injector design. For instance: how to optimize the linac design to minimize the emittance growth and energy spread when accelerating high-charge electron bunches from a thermionic gun; how to design a positron production and accumulation chain—including target, capture optics, damping, and acceleration—to yield high-quality positron bunches with sufficient charge and low emittance; how to suppress emittance degradation in the beam transport lines for high-intensity electron and positron bunches during injection.

### 1.4.3 Key Technologies

While most of the STCF accelerator technical systems will adopt mature and proven accelerator technologies wherever possible to ensure construction feasibility and adherence to schedule, several key technologies—though having certain development base or already under development—remain immature and must be developed or advanced in big efforts to meet the facility's performance goals and ensure the technological advancement. The project team is actively organizing vigorous R&D efforts in these areas, aiming for critical breakthroughs in the coming years to ensure the implementation of the engineering construction. In parallel, backup solutions to those key technologies are being prepared to guarantee the timely advancement of the construction project and compliance with final performance specifications, even if the R&D of the key technologies faces delays or setbacks.

The key enabling technologies required for the STCF accelerator include:

**Twin-Aperture Superconducting Magnets for the Interaction Region**: These include the final focus quadrupoles as well as integrated corrector coils, higher-order field coils, anti-solenoids, and compensation solenoids.

**Collider Ring RF Systems**: Room-temperature cavities designed to support high circulating currents (~2 A), with deep higher-order mode (HOM) suppression, high-power RF couplers, and stable low-level RF (LLRF) controls.

**Beam Diagnostics and Feedback Systems**: Fast-response, low-noise feedback systems and high-precision 3D bunch profiling instrumentation.

**Other Advanced Accelerator Technologies**: Ultra-fast kicker magnets with pulse bottom widths smaller than 6 ns; Ultra-high vacuum systems with low impedance, capable of operating under very intense synchrotron radiation; High bunch-charge (>8 nC) photocathode electron guns; High-power positron production targets; Klystrons and accelerating structures operating at up to 90 Hz repetition rate; Complex mechanical systems for the MDI (Machine–Detector Interface) region, etc.

# 2 Collider Ring Accelerator Physics

## 2.1 Collision Scheme and Global Parameters

The core objective of designing an electron–positron collider is to maximize luminosity. The luminosity, $L$, can be expressed as follows [1]:

$$L = \frac{\gamma n_b I_b}{2 e r_e \beta_y^*} \xi_y H \tag{1}$$

where:

- $\gamma$ is the relativistic Lorentz factor,
- $r_e$ is the classical electron radius,
- $I_b$ is the bunch current,
- $n_b$ is the number of bunches per ring,
- $\xi_y$ is the vertical beam-beam tune shift parameter,
- $H$ is the hourglass reduction factor.

According to the scientific objectives of STCF [2], which demand a center-of-mass energy range of 2-7 GeV and a design luminosity not lower than $0.5 \times 10^{35}\ \mathrm{cm^{-2}s^{-1}}$ at the optimized energy point of 4 GeV, the STCF accelerator will adopt the most advanced third-generation design philosophy for circular $e^+e^-$ colliders. Specifically, it will follow the double-ring scheme of second-generation colliders and incorporate a large crossing angle with the crab-waist collision scheme [3].

This collision scheme is characterized by a flat beam profile at the interaction point (IP), with extreme vertical compression (on the order of hundreds of nanometers) and also significant horizontal compression (on the order of tens of microns). Due to the large crossing angle, the overlap region of the two bunches at the IP is very small, which permits the use of relatively long bunch lengths without suffering severe luminosity degradation from the hourglass effect.

To mitigate the transverse and transverse-longitudinal coupling instabilities induced by the large crossing angle, crab sextuples are placed at specific phase advance locations on either side of the IP. These magnets couple the horizontal and vertical motions of the bunches such that the vertical focus of particles becomes correlated with their horizontal position, and all vertical foci fall along the optical axis of the opposing bunch—this is the so-called crab-waist collision scheme. Figure 2.1-1 illustrates the principle of the crab-waist collision scheme [3].

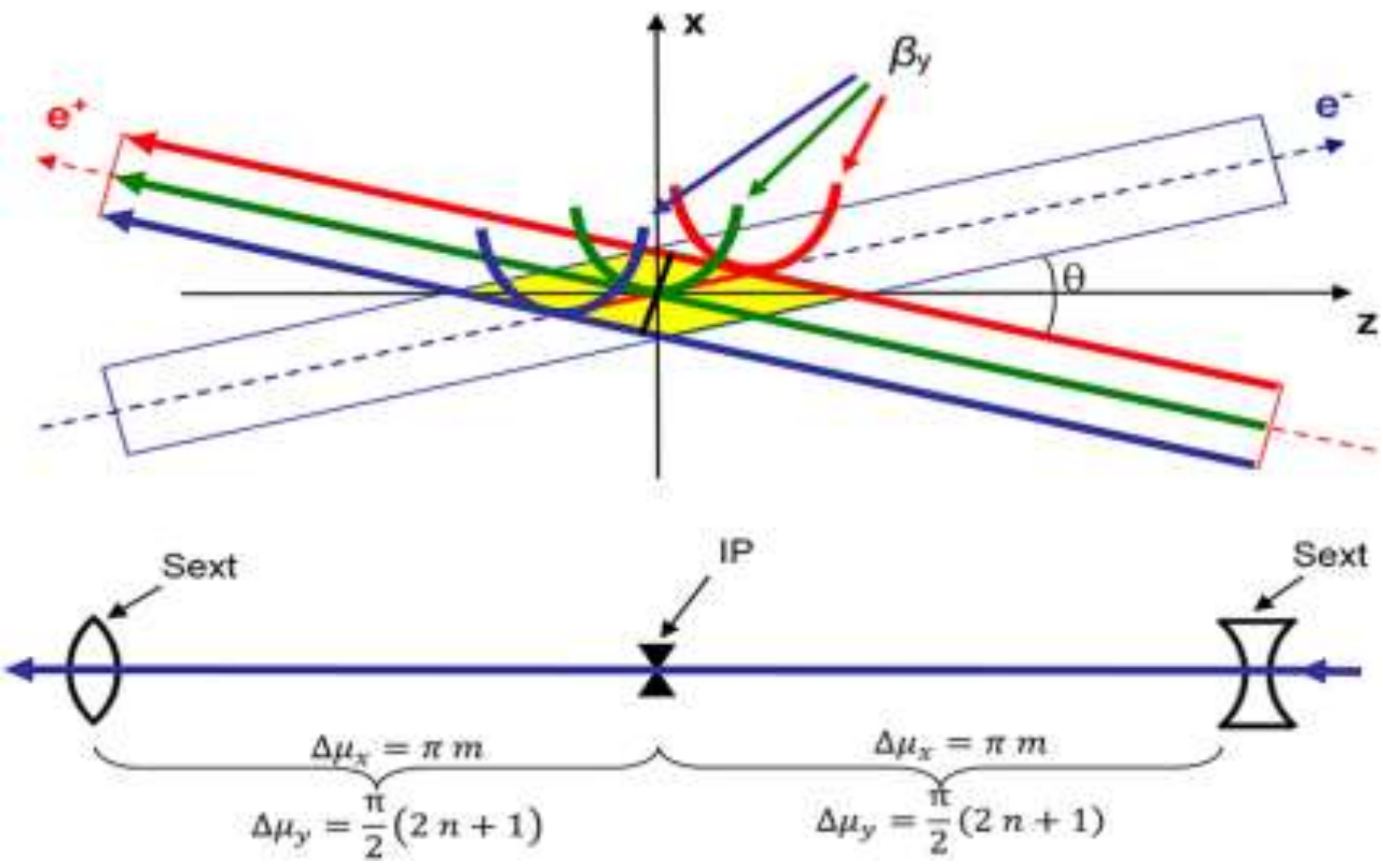

Figure 2.1-1: Illustration of the principle of the crab-waist scheme [3]

In addition, further measures to improve the luminosity of STCF include achieving an ultra-small $\beta_y^*$, moderately increasing the vertical beam-beam parameter $\xi_y$, and increasing the beam current. However, all of these are subject to certain limitations. For example, reducing $\beta_y^*$ to the sub-millimeter level is constrained by the technology of the final focus superconducting magnets at the IP, and it also results in larger local chromaticity and nonlinear effects. Increasing the beam-beam parameter $\xi_y$ may lead to uncontrollable beam dynamics and is generally not advisable to exceed 0.1. Raising the beam current can result in excessively short Touschek lifetime and excessive synchrotron radiation power, and a total circulating current of 3 A may represent the practical limit.

The crab-waist collision scheme requires strong coupling between the X/Y directions of the bunches near the IP via crab sextupoles. Combined with strong focusing at the IP, this leads to significantly reduced dynamic and momentum apertures of the collider rings. In addition to leading to serious difficulties for beam injection, it can cause a very short Touschek lifetime under high bunch currents and low emittances—for example, less than 300 s, which is significantly shorter than in other circular electron accelerators. Such a short Touschek lifetime not only increases the difficulty of designing and constructing the injector (as it would require significantly higher injection repetition rates) but also results in severe beam loss during injection, which in turn negatively impacts the experimental background. Balancing high-luminosity collisions with beam dynamics, beam lifetime, and stability is one of the major challenges faced in the physical design of the collider rings [4]. The physical design of the STCF collider rings must find a trade-off between the design goals and these critical parameters.

Based on analysis of the collision scheme proposed for STCF, selection of key parameters for the collider rings must be guided by detailed physics studies and iterative coordination with major hardware system designs, while also referencing results from other third-generation $e^+e^-$ colliders and synchrotron light sources. Considerations for several key parameter choices are summarized as follows:

- **Beam energy range:** 1.0-3.5 GeV, with an optimized energy point at 2 GeV, as determined by the overarching scientific goals of the project.

- **Luminosity:** ≥ 0.5 × $10^{35}$ $cm^{-2}s^{-1}$ @ 2 GeV beam energy, also defined by the project's scientific objectives. At the lower and upper ends of the energy range, lower luminosity is permitted but must not decrease by more than one order of magnitude.

- **Ring circumference:** 800-1000 m, based on international experience and with allowance for future expansion. A longer circumference provides space for idealized interaction region design and accommodates components that occupy long straight sections—such as damping wigglers, collimators, injection/extraction elements, and RF cavities—and also reserves space for future upgrades involving polarized beams (e.g., spin rotators or Siberian snakes). However, excessively increasing the circumference should be avoided to contain construction costs.

- **Circulating current:** ~2 A @ 2 GeV, based on international experience and preliminary STCF studies. High current is essential for achieving high luminosity, but further increases will severely shorten the Touschek lifetime. At lower energies, Touschek effects are more pronounced, and moderate reductions in beam current and luminosity are acceptable. At higher energies, synchrotron radiation power may impose additional constraints on the beam current.

- $\boldsymbol{\beta_y^*}$**:** 0.6–1.0 mm, based on international experience. The smaller the value, the better for luminosity, but excessively small values can lead to very large local chromaticity, higher-order dynamic effects, and reduced momentum aperture, all of which severely reduce Touschek lifetime. A working value of 0.8 mm is currently adopted, subject to further optimization.

- **Transverse emittance:** Horizontal emittance is about 5 nm·rad @ 2 GeV, comparable to third-generation synchrotron light sources. The transverse coupling is limited to 1%, corresponding to a vertical emittance of approximately 50 pm·rad, sufficient to achieve a vertical beam size of about 100 nm at the IP. Emittance is allowed to vary within a reasonable range at other energies (lower at low energies, higher at high energies). At 1 GeV and 1.5 GeV, the coupling ratio remains at 1%; at 3.5 GeV, it is set to 0.5%. At low energy, vertical dispersion can be increased to boost vertical emittance and thereby improve Touschek lifetime.

As the optimization of the collider ring physics design progresses, the overall parameter table for the collider rings will also be periodically updated. Table 2.1-1 shows the latest version of the current global design parameters.

Table 2.1-1: Preliminary STCF Collider Ring Global Parameters

| Parameter | Unit | Value | | | |
|---|---|---|---|---|---|
| Beam Energy, $E$ | GeV | 2 | 1 | 1.5 | 3.5 |
| Circumference, $C$ | m | 860.321 | | | |
| Crossing Angle, $2\theta$ | mrad | 60 | | | |
| $L^*$ | m | 0.9 | | | |
| Relativistic Factor, $\gamma$ | | 3913.9 | 1956.9 | 2935.4 | 6849.3 |
| Revolution Period, $T_0$ | μs | 2.87 | | | |
| Revolution Frequency, $f_0$ | kHz | 348.47 | | | |
| Ratio, $\varepsilon_y/\varepsilon_x$ | | 1% | 15% | 10% | 0.5% |
| Horizontal Emittance (SR/DW, IBS), $\varepsilon_x$ | nm | 8.79/4.63 | 2.20/5.42 | 4.94/3.82 | 26.9/26.91 |
| Vertical Emittance (SR/DW, IBS), $\varepsilon_y$ | pm | 87.9/46.3 | 330/813 | 494/382 | 134.5/134.55 |
| β Functions at IP, $\beta_x/\beta_y$ | mm | 60/0.8 | | | |
| Beam Size at IP, $\sigma_x/\sigma_y$ | μm | 16.67/0.19 | 18.03/0.81 | 15.14/0.55 | 40.18/0.33 |
| Betatron Tunes, $\nu_x/\nu_y$ | | 30.54/34.58 | | 30.555/34.57 | |
| Momentum Compaction Factor, $\alpha_p$ | $10^{-4}$ | 13.49 | 12.63 | 13.24 | 13.73 |
| Energy Spread (SR/DW, IBS), $\sigma_e$ | $10^{-4}$ | 5.72/7.82 | 2.86/6.18 | 4.29/6.93 | 10.01/10.02 |
| Beam Current, $I$ | A | 2 | 1.1 | 1.7 | 2 |
| Bunch Filling Factor | | 48% | | | |
| Number of Bunches, $n_b$ | | 688 | | | |
| Bunch Spacing, $\tau_b$ | ns | 4 | | | |
| Single Bunch Current, $I_b$ | mA | 2.91 | 1.6 | 2.47 | 2.91 |
| Particles per Bunch, $N_b$ | $10^{10}$ | 5.20 | 2.86 | 4.42 | 5.20 |
| Total Particles per Beam | $10^{13}$ | 3.58 | 1.97 | 3.05 | 3.58 |
| Charge per Bunch | nC | 8.34 | 4.59 | 7.09 | 8.34 |
| Energy Loss per Turn (SR/Total), $U_{0_sr}$ | keV | 159/543 | 10/106 | 50/267 | 1494/1494 |
| Synchrotron Radiation Power Loss per Beam (SR/Total), $P_{SR}$ | MW | 0.32/1.09 | 0.01/0.12 | 0.085/0.453 | 2.988/2.988 |
| Damping Times, $\tau_x/\tau_y/\tau$ | ms | 21/21/11 | 54/54/27 | 32/32/16 | 14/14/6.7 |
| RF Frequency, $f_{RF}$ | MHz | 499.7 | | | |
| Harmonic Number, $h$ | | 1434 | | | |
| RF Voltage, $V_{RF}$ | MV | 2.5 | 0.75 | 1.2 | 6 |
| Longitudinal Phase, $\Phi_s$ | deg | 167 | 172 | 167 | 166 |
| Synchrotron Tune, $\nu_z$ | | 0.0194 | 0.0146 | 0.0154 | 0.0228 |
| Natural Bunch Length, $\sigma_z$ | mm | 7.21 | 6.62 | 7.89 | 8.26 |
| Bunch Length (0.1Ω, IBS), $\sigma_{z_ibs}$ | mm | 8.7 | 9.46 | 10.01 | 8.79 |
| RF Energy Acceptance, $(\Delta E/E)_{max}$ | % | 1.68 | 1.44 | 1.35 | 1.88 |
| Piwinski Angle, $\Phi_{Piw}$ | rad | 15.66 | 15.74 | 19.84 | 6.56 |
| Beam-Beam Parameters, $\xi_x/\xi_y$ | | 0.005/0.095 | 0.005/0.023 | 0.004/0.033 | 0.003/0.032 |
| Hourglass Factor, $F_h$ | | 0.915 | 0.905 | 0.927 | 0.750 |
| Luminosity, $L$ | $cm^{-2}s^{-1}$ | 9.42E+34 | 6.19E+33 | 2.09E+34 | 4.48E+34 |
| Touschek Lifetime (SAD/Elegant) | s | 252/304 | 195/264 | 215/294 | 5400/3479 |

## 2.2 Lattice Design

### 2.2.1 Baseline Scheme (Two-Fold Symmetry)

The STCF collider rings adopt a double-ring layout, consisting of an electron ring and a positron ring. Both rings lie in the same horizontal plane and intersect at the IP in the collision region and the crossing point in the region opposite the IP. The layout is symmetric along the line connecting these two points. Each collider ring features a two-fold symmetric structure. Due to the crossing geometry of the double-ring configuration, the drift sections in the arc regions on opposite sides of the same ring differ slightly in length, forming inner and outer rings with an approximate spacing of 2 meters. Space is reserved on both sides of the interaction region for the future installation of spin rotators to enable beam polarization upgrades.

Each ring consists of one collision region, four large arc sections (each bending by 60°), two small arc sections (each bending by 30°), one crossing region, and several straight sections. The straight sections serve different functions, including injection/extraction, damping wigglers, beam collimation, RF cavities, and working point matching. The total circumference of the ring is 860.321 m.

The main features of this lattice design include:

- A symmetric lattice layout that preserves the option of upgrading to a second interaction point;
- Flexibility in the type and length of damping wigglers—currently adopting a well-established room-temperature damping wiggler scheme, with adjustments in the damping wiggler section not affecting the rest of the layout;
- Relatively small injection deflection angle—both electron and positron beams require a 60° bend to enter the collider rings;
- Reserved space for future upgrades, such as spin rotators needed in potential beam polarization;
- A medium-scale arc section with a maximum arc bending of 60°, which improves local chromaticity correction and allows better optimization of the dynamic and momentum apertures.

Figure 2.2-1 shows the layout of the collider rings, and Figure 2.2-2 presents the optical functions. This section introduces the optical design of the arcs, straight sections, and crossing

region in the two-fold symmetric lattice; the design of the interaction region optics is presented in Section 2.2.3.

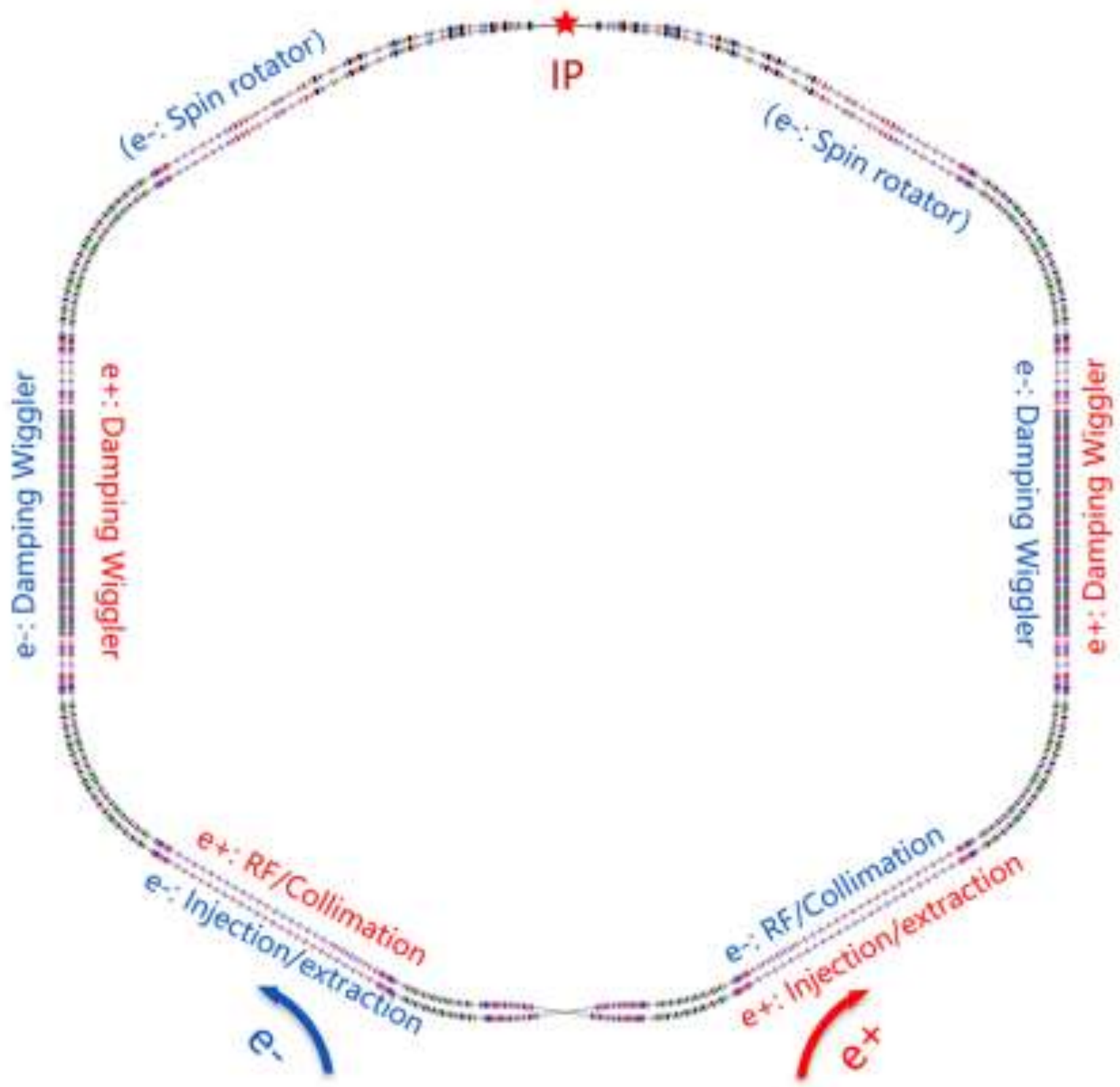

Figure 2.2-1: Layout of the STCF Collider Rings

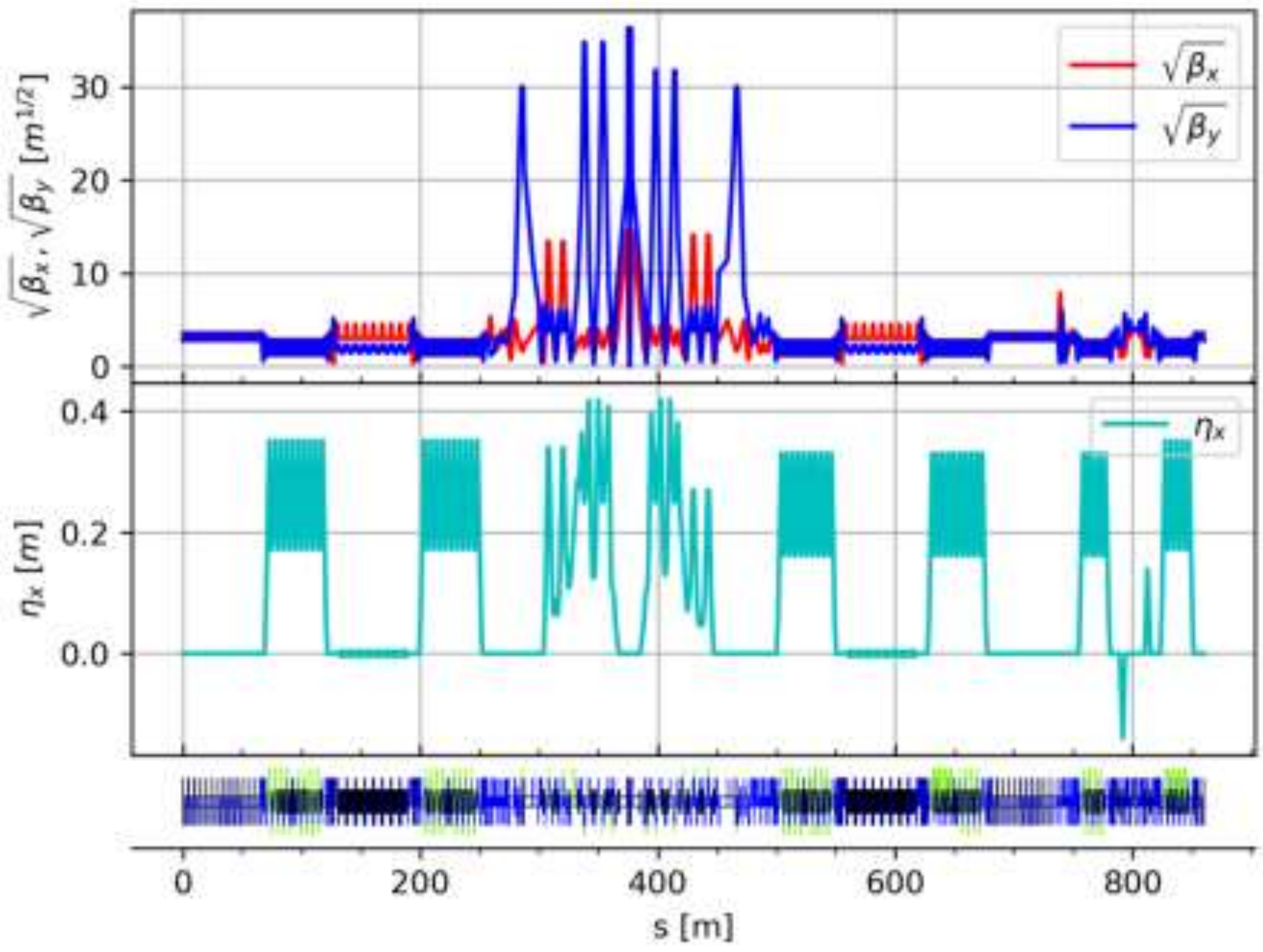

Figure 2.2-2: Optical functions of the full collider rings

#### *2.2.1.1 Arc Optics Design*

The large and small arc sections adopt the same standard FODO cell design. Each FODO cell is 4.7 m in length, providing a bending angle of 6°, a phase advance of 90°, and a momentum compaction factor of $4.9 \times 10^{-3}$. Each cell includes four 0.8 m drift sections for hosting sextupole magnets and beam collimators. Figure 2.2-3 shows the optical functions of a FODO cell. The ends of each arc section contain dispersion suppressor sections and optics matching sections. The large arc section consists of 9 FODO cells plus the dispersion suppression and matching sections, totaling 57.184 m in length and providing a total bending angle of 60°; the small arc section contains 4 FODO cells plus the dispersion suppression and matching sections, with a total length of 33.684 m and a bending angle of 30°.

Chromaticity correction and nonlinear optimization in the arc sections are achieved using sextupole magnets. These sextupoles are placed in pairs with a phase advance of 180°, satisfying the −I transformation condition, which effectively cancels first-order geometric resonance terms. The large arc section contains 8 groups of sextupoles (4 SD pairs + 4 SF pairs), and the small arc section contains 4 groups (2 SD pairs + 2 SF pairs). The field strength of each group of sextupole magnets can be adjusted independently. Figures 2.2-4 show the optical functions of the large and small arc sections, respectively.

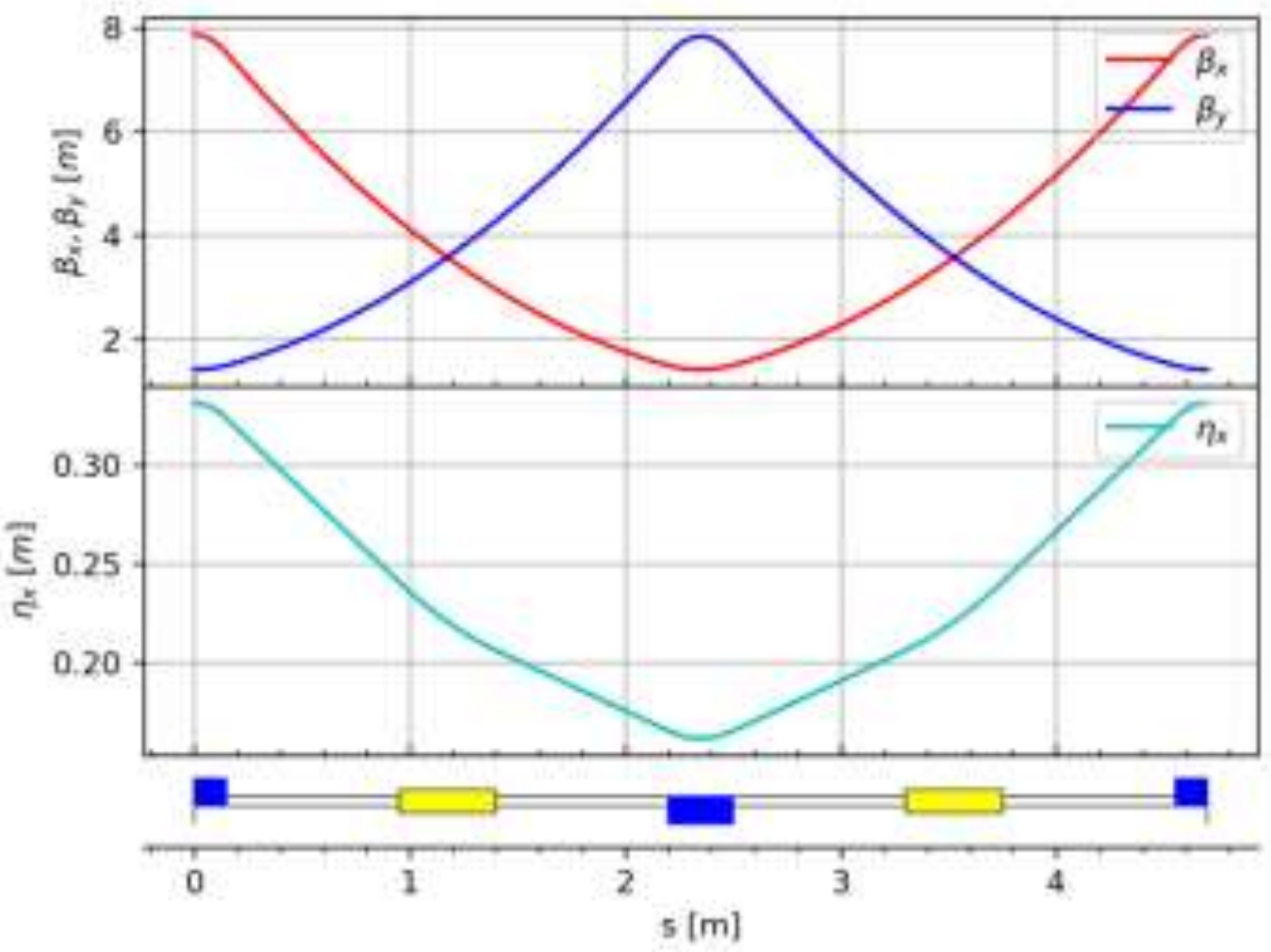

Figure 2.2-3: Optical functions of a standard FODO cell

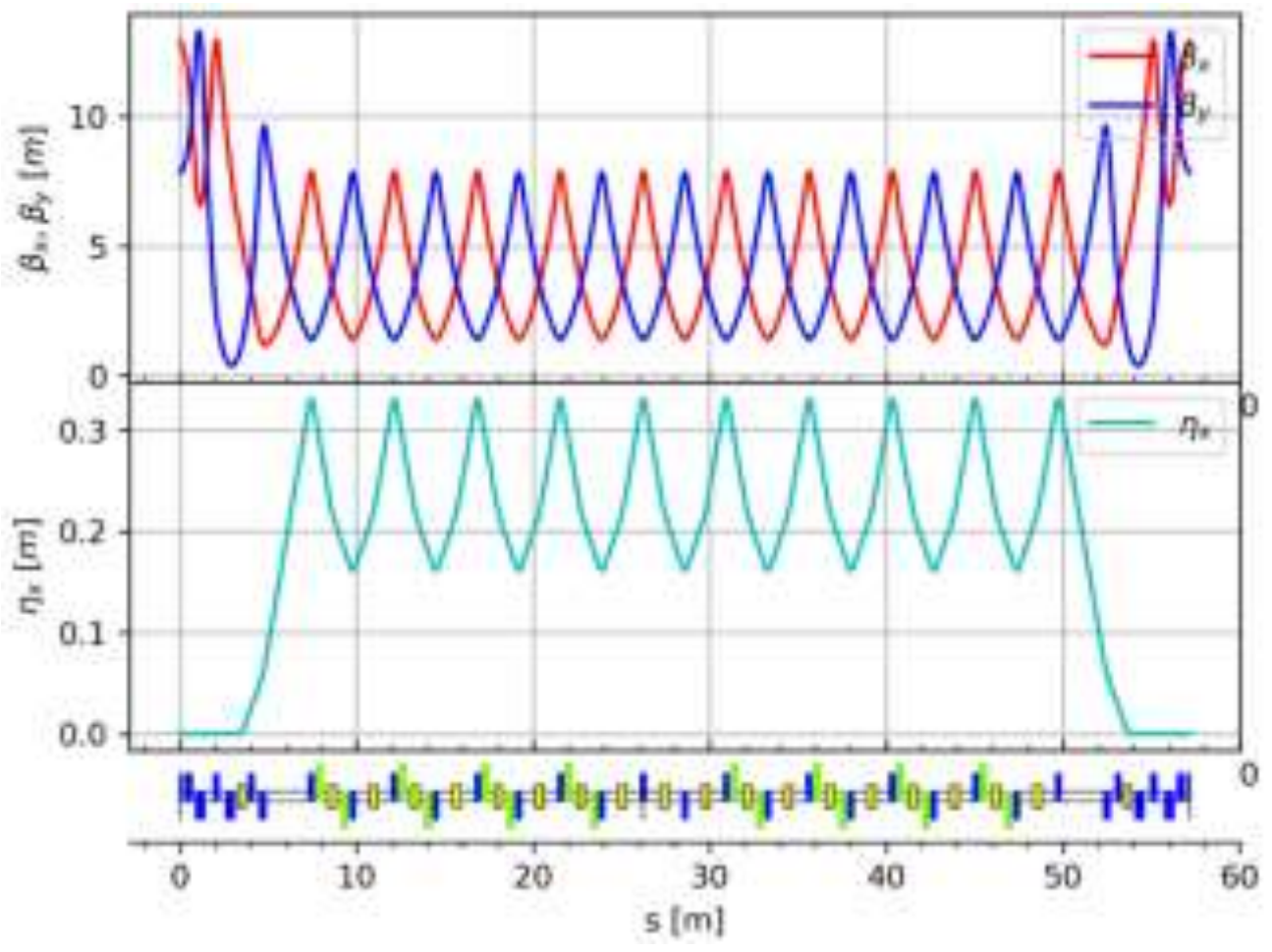

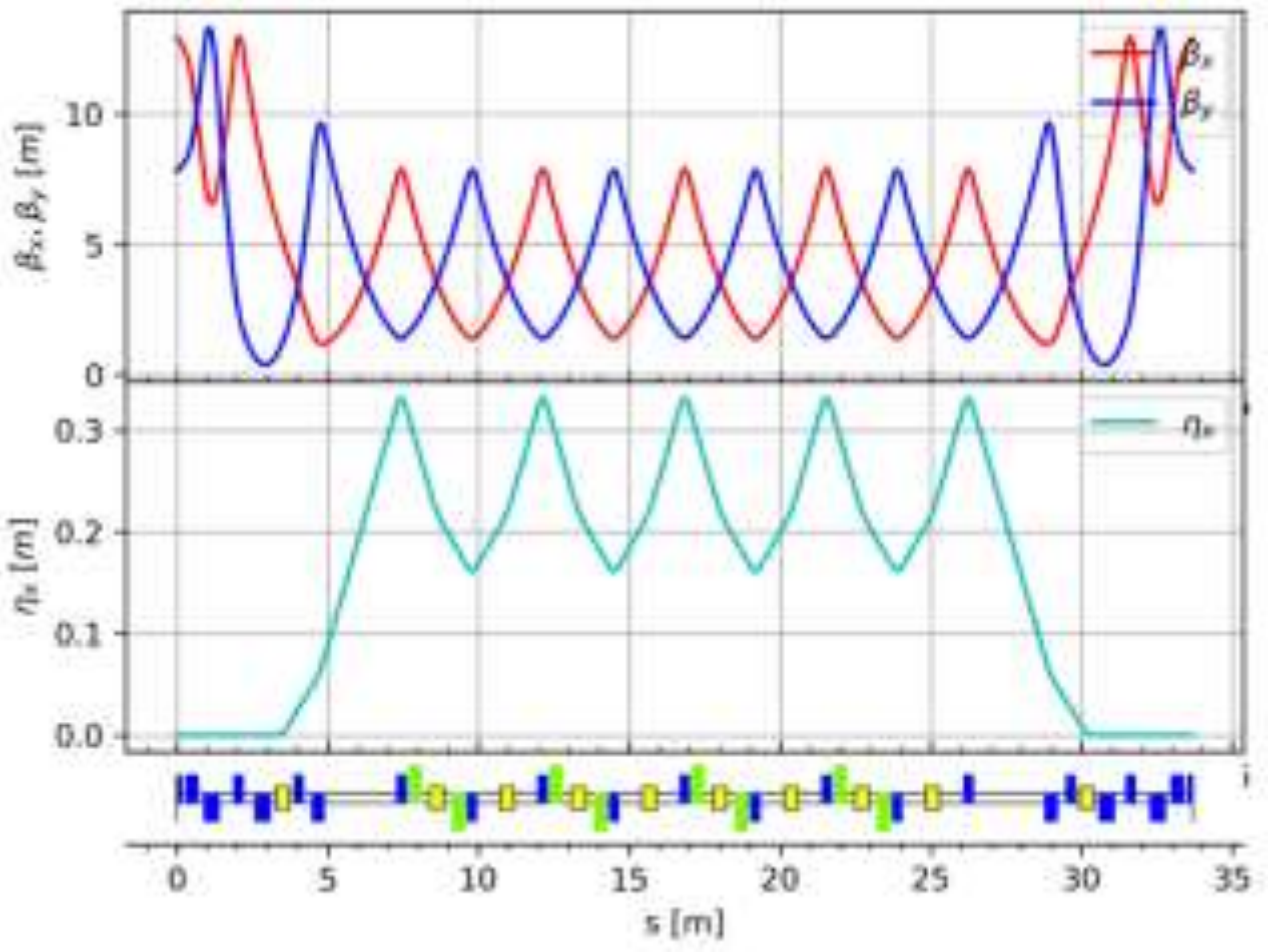

Figure 2.2-4: Optical functions of the large arc section (top) and small arc section (bottom)

### *2.2.1.2 Crossing Region Optics*

The crossing region employs two groups of dipole magnets to achieve the crossing and separation of the two beamlines. To achieve achromaticity, a triplet structure with a phase advance of π is used between the dipole magnets in each set. Additionally, two doublet structures are placed in the middle of the crossing region. To reduce the collision probabilities at the crossing point, the β functions in this section are designed to be relatively large, resulting in correspondingly larger beam sizes. The total length of the crossing region is 35 m, and the separation distance between the two rings ranges are about 2 m. Figure 2.2-5 shows the optical functions of the crossing region.

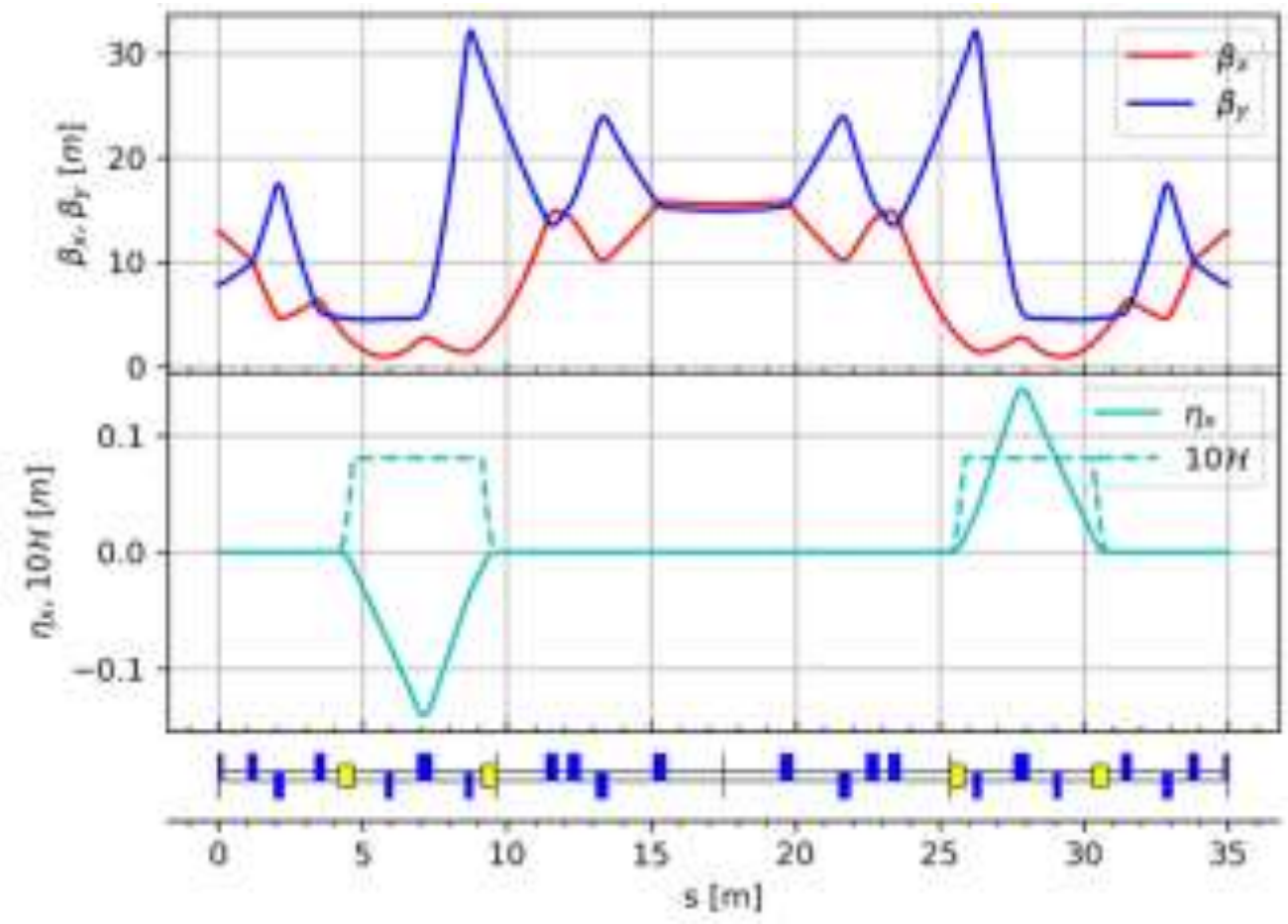

Figure 2.2-5: Optical functions of the crossing region

### *2.2.1.3 General Straight Section Optics*

In addition to generally designed straight sections that are just for connecting the arc sections, most of the straight sections in the collider rings are dedicated segments, such as the injection and extraction section, the damping wiggler sections, the RF section, and the beam collimation section, depending on their specific functions. Here, it describes only the design of the general straight section.

The general straight section adopts a standard FODO structure. Each FODO cell is 5 meters long with a phase advance of 30°, and contains two drift spaces, each 2.2 meters in length. Figure 2.2-6 shows the optical functions of the general straight section consisting of eight FODO cells.

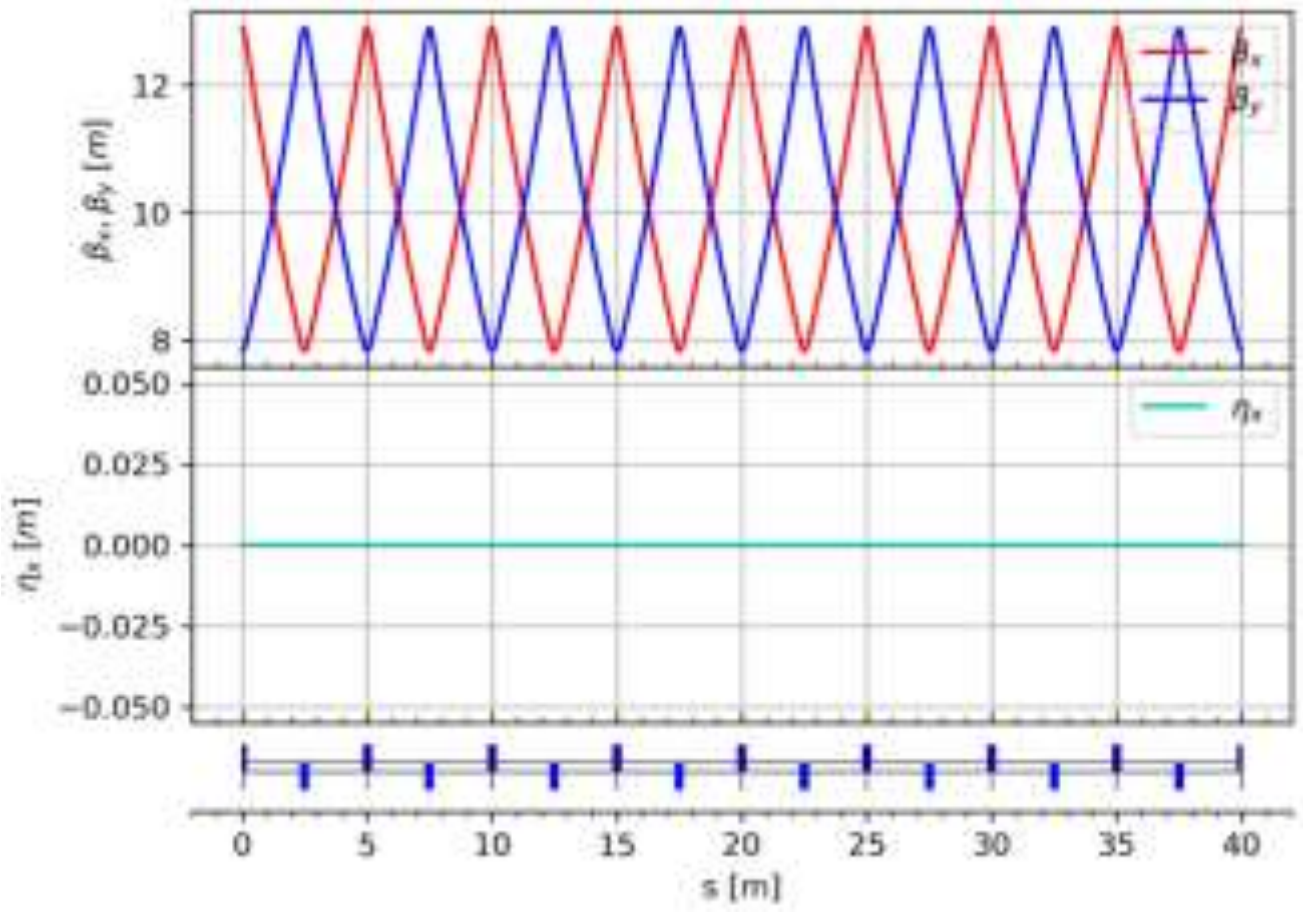

Figure 2.2-6: Optical functions of 8 FODO cells in the general straight section

### *2.2.1.4 Damping Wiggler Section Optics*

Each of the STCF collider rings includes two damping wiggler sections, which are used to reduce the damping time and to adjust the beam emittance at different beam energies. Each damping wiggler section contains eight damping wigglers, each 4.8 meters in length. To ensure relatively gentle variation of the β-function within the wigglers, the lattice design incorporates a specially optimized structure for the wiggler sections.

A triplet configuration is chosen as the primary lattice structure for this section due to its flexibility—it allows for long drift spaces to accommodate the wigglers while ensuring slow variation of the β-function at the wiggler locations. Since commonly used lattice design codes such as MADX and SAD do not include built-in damping wiggler models, the typical approach is to model the wigglers using a series of bending magnets and drift spaces that achieve equivalent radiation damping effects. Figure 2.2-7 shows the lattice functions of two focusing cells and the long drift section within the damping wiggler region.

Because damping wigglers influence the optical functions of the lattice, it is necessary to place several quadrupole magnets at both ends of each damping wiggler section. These are used to compensate for the optical distortions introduced by the wigglers and to match the Twiss parameters at the transition to adjacent straight sections. Table 2.2-1 lists the key parameters of the Lattice design for the damping wiggler section.

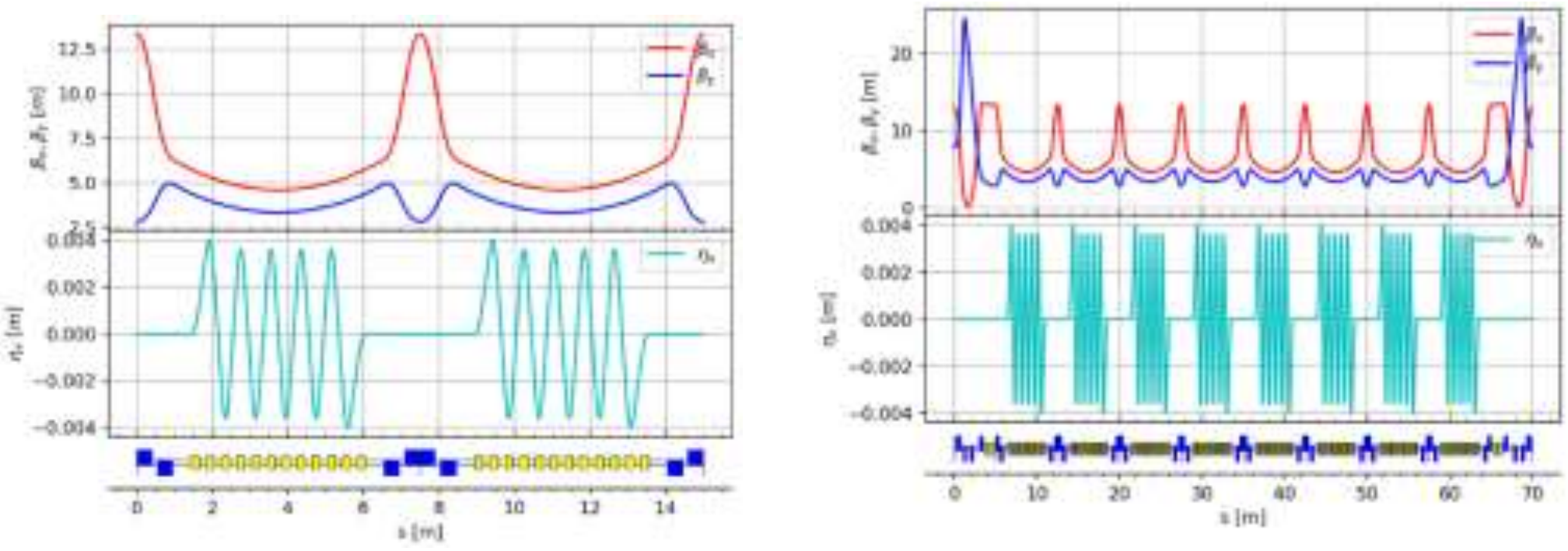

Figure 2.2-7: Lattice functions of two triplet units and long drifts in the damping wiggler section

Table 2.2-1: Key Parameters of the Damping Wiggler Section (per ring)

| Parameter | Value |
|---|---|
| Total wiggler section length (m) | 2 × 70 |
| Triplet unit length (m) | 7.5 |
| Wiggler unit length (m) | 4.8 |
| Number of wigglers per ring | 16 |
| Average β-functions at wiggler, $\beta_x/\beta_y$ (m) | 5.08 / 3.78 |

### 2.2.2 Scheme II (Single-Fold Symmetry)

At the same time, we have also studied a backup scheme (hereinafter referred to as Scheme II) that includes the reserved space for the installation of Siberian Snakes and a multifunctional long straight section.

The electron ring in this scheme consists of four arc sections, three Siberian Snakes (SS), four damping wigglers (DW), one interaction region (IR), and one multifunctional section. The geometric layout of the dual-ring and the distribution of different units are shown in Figure 2.2-8. This layout is realized by employing dipole magnets with different bending angles on both sides of the interaction region and arc dipole magnets with a bending radius difference of 2.14 meters between the inner and outer half-rings.

Figure 2.2-9 illustrates the full-ring lattice and its linear optical functions. Table 2.2-2 presents the main parameters of Scheme II for the collider rings, taking into account intra-beam scattering (IBS), synchrotron radiation, and the damping wigglers. Only the parameters at the optimal energy of 2 GeV are given.

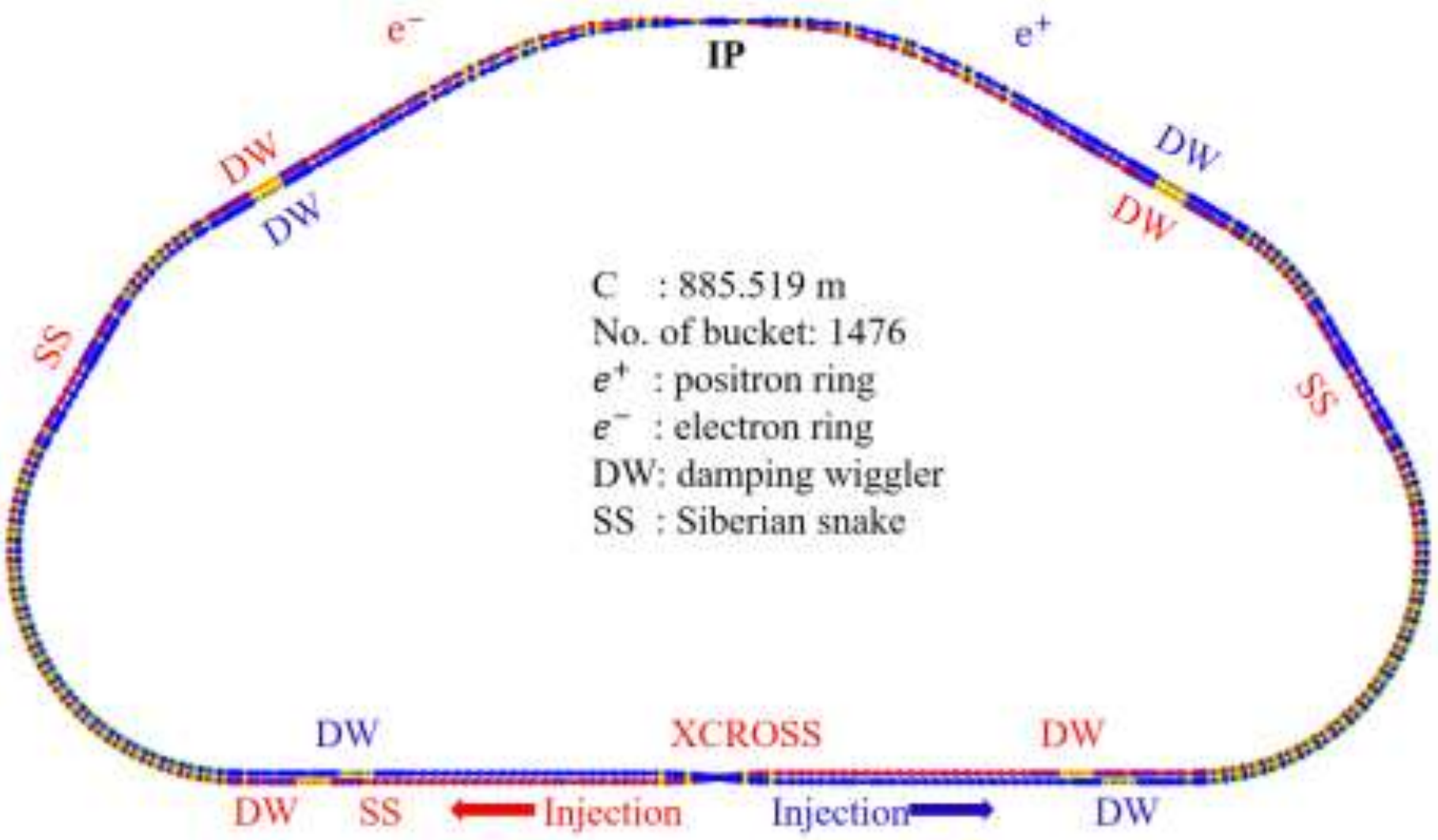

Figure 2.2-8: Layout of STCF collider ring lattice – Scheme II

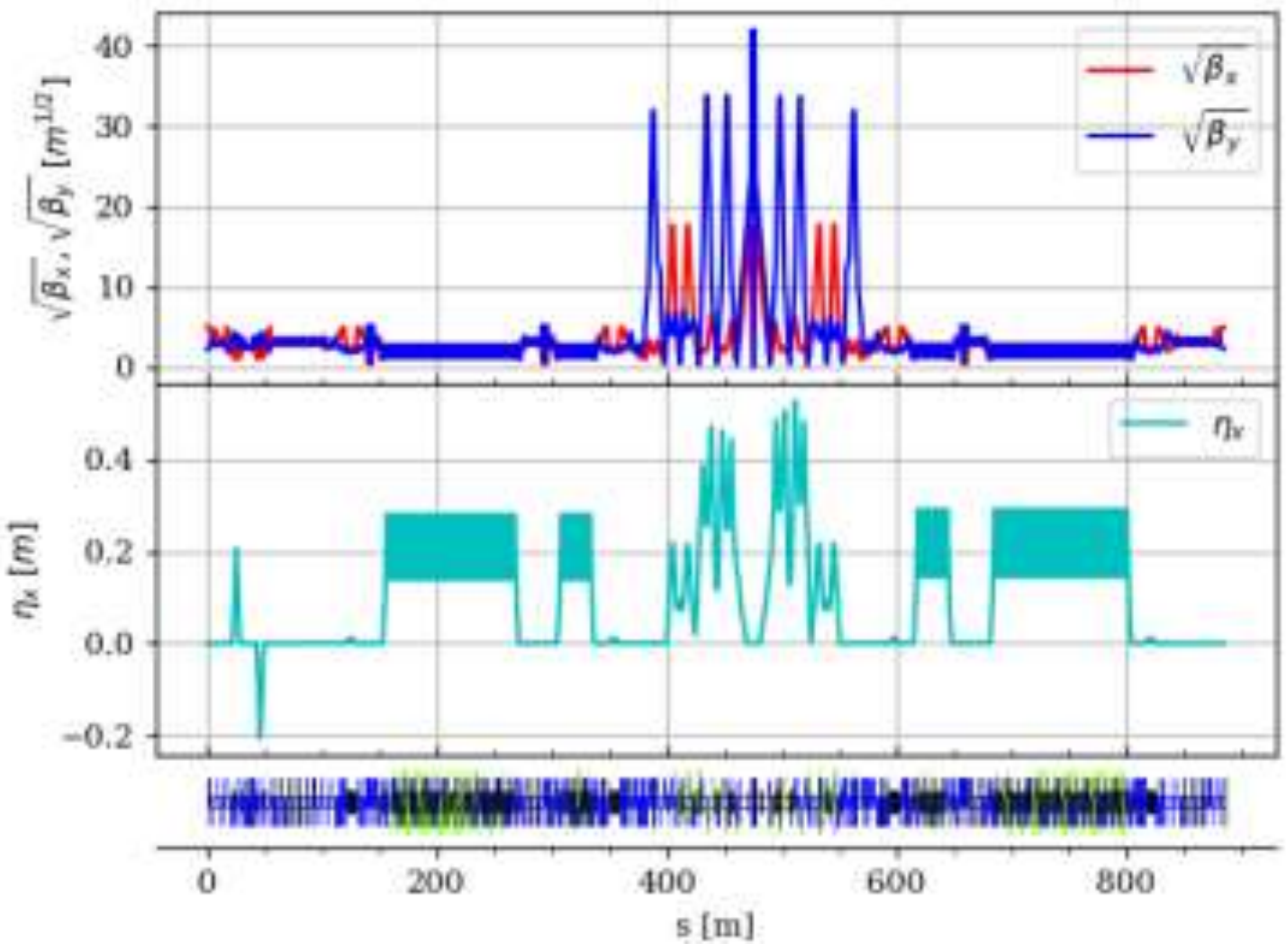

Figure 2.2-9: Full-ring lattice and linear optics of Scheme II

Table 2.2-2: Key Parameters of STCF Collider Ring – Lattice Scheme II

| Parameter | Unit | Value |
|---|---|---|
| Optimal beam energy, $E$ | GeV | 2 |
| Circumference, $C$ | m | 885.519 |
| Crossing angle, $2\theta$ | mrad | 60 |
| Revolution period, $T$ | μs | 2.953 |
| Horizontal / vertical emittance, $\varepsilon_x/\varepsilon_y$ | nm | 6.368 / 0.0318 |
| Coupling factor, $K$ | – | 0.5% |
| β-functions at IP, $\beta_x/\beta_y$ | mm | 40 / 0.6 |
| Beam sizes at IP, $\sigma_x/\sigma_y$ | μm | 15.96 / 0.138 |
| Betatron tunes, $\nu_x\ /\nu_y$ | – | 33.554 / 33.571 |
| Momentum compaction factor, $\alpha_p$ | $10^{-4}$ | 12.433 |
| Energy spread, $\sigma_e$ | $10^{-4}$ | 9.908 |
| Beam current, $I$ | A | 2 |
| Number of bunches, $n_b$ | – | 738 |
| Particles per bunch, $N_b$ | $10^{10}$ | 5.00 |
| Bunch charge | nC | 8.0 |
| SR energy loss, $U_0$ | keV | 383.77 |
| Damping times, $\tau_x/\ \tau_y/\ \tau_z$ | ms | 30.77 / 30.77 / 15.39 |
| RF frequency, $f_{RF}$ | MHz | 499.7 |
| Harmonic number, $h$ | – | 1476 |
| RF voltage, $V_{RF}$ | MV | 2 |
| Longitudinal tune, $\nu_z$ | – | 0.0169 |

| Parameter | Unit | Value |
|---|---|---|
| Bunch length, $\sigma_z$ | mm | 10.25 |
| RF acceptance, $\delta_{RF}$ | % | 1.58 |
| Piwinski angle, $\phi_{pwin}$ | rad | 19.26 |
| Beam–beam parameters, $\xi_x/\xi_y$ | – | 0.0024 / 0.081 |
| Hourglass factor, $F_h$ | – | 0.8804 |
| Peak luminosity, $L$ | $cm^{-2}\ s^{-1}$ | $1.03 \times 10^{35}$ |
| Touschek lifetime, $\tau_{Touschek}$ | s | >250 |

#### *2.2.2.1 IR Optics – Scheme II*

The interaction region adopts the crab-waist collision scheme, and its lattice structure is similar to that in Scheme I (see Section 2.2.3). The linear optics system of the right half (the left half has the same functional partitioning but with different parameters) consists sequentially of the Final Telescope (FT), the Local Chromaticity Correction System (LCCS), the crab sextupole section (CWS), and the Matching Transport section (MT), covering a total bending angle of 60°. At the IP, the $\beta$ functions are $\beta_x^* = 40$ mm and $\beta_y^* = 0.6$ mm.

In addition to sextupoles for correcting first- and third-order chromaticities [5], the interaction region is also equipped with weak sextupole magnets [6] to optimize nonlinear performance. Figures 2.2-10 and 2.2-11 show the geometric layout and linear optics functions of the interaction region, respectively.

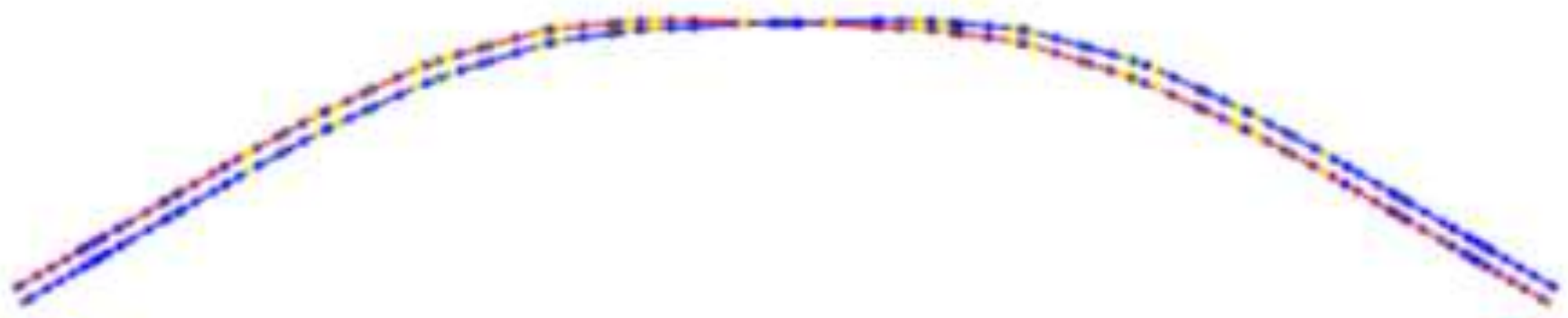

Figure 2.2-10: IR dual-ring layout

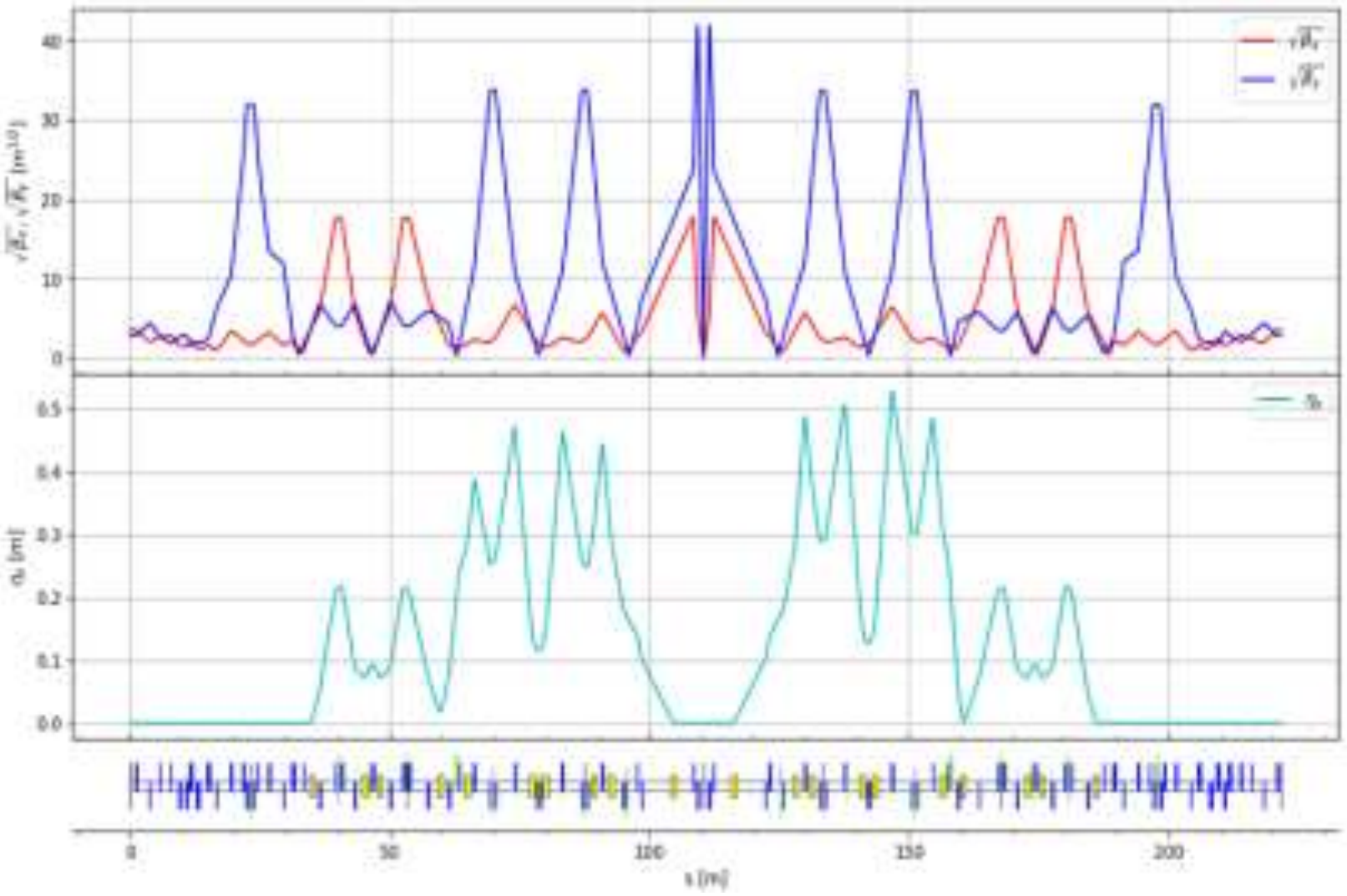

Figure 2.2-11: IR lattice functions

### *2.2.2.2 Arc Optics – Scheme II*

To implement the dual-ring layout, the arc sections are divided into inner and outer arcs, each comprising a long arc section with a 120-degree bend and a short arc section with a 30-degree bend. Both are constructed using FODO structures with horizontal and vertical phase advances of 90° and a bending angle of 5°. In the FODO structures of the inner and outer arcs, the bending magnets have a curvature radius difference of 2.14 m, while the quadrupoles, sextupoles, and drift sections have equal lengths.

Figure 2.2-12 shows the lattice functions of the long and short arc sections in the outer half-ring. In the short arc, non-interleaved sextupole pairs with a -I transfer are used for chromaticity correction. In the FODO cells of the long arc, a focusing sextupole follows each focusing quadrupole, and a defocusing sextupole follows each defocusing quadrupole. Every four identical FODO structures form one HOA (Higher-Order Achromat) unit, effectively canceling the first- and second-order geometric driving terms [7].

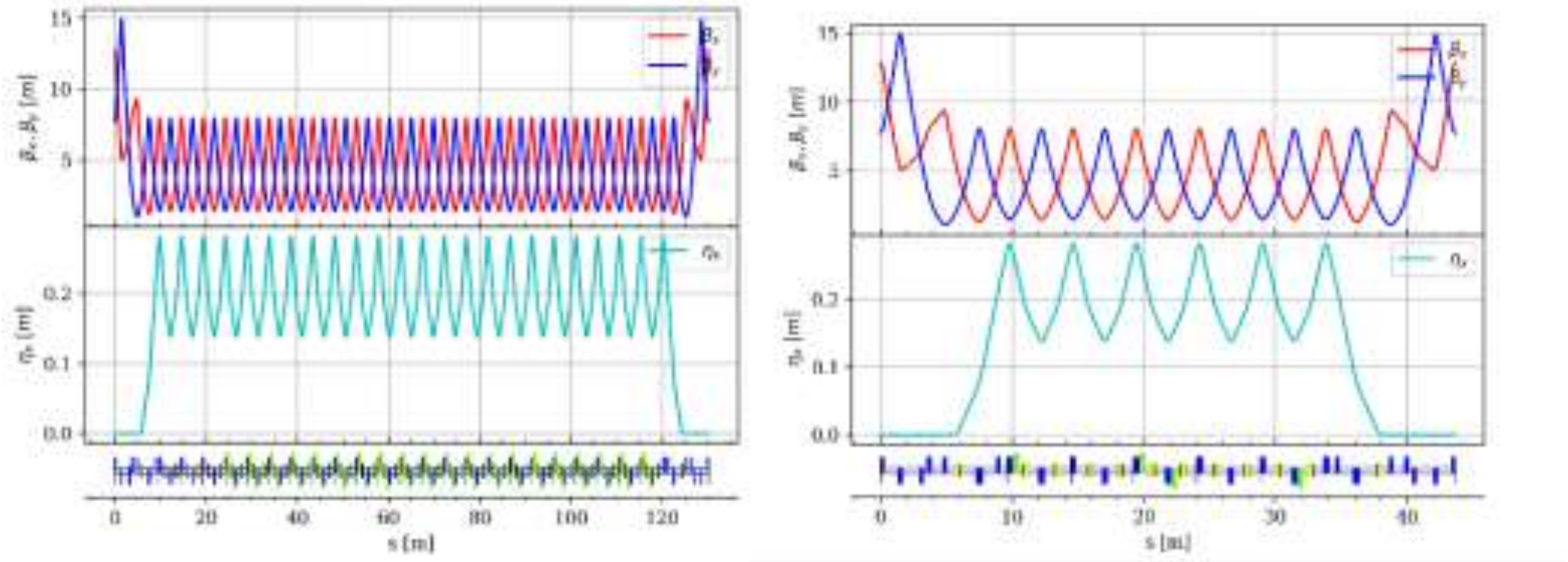

Figure 2.2-12: Lattice functions of the long (left) and short (right) arc sections

### *2.2.2.3 Straight Section Optics – Scheme II*

The straight sections include medium straight sections dedicated to the installation of Siberian Snakes (SS) and damping wigglers (DW), as well as the multifunctional region for hosting the injection system, RF cavities, dual-ring crossing, and phase adjustment. Their corresponding linear optical functions are shown in Figures 2.2-13 and 2.2-14.

SS refers to reserved space for Siberian Snakes, with an azimuthal angle of 120° between each pair. To regulate the damping time, two damping wigglers are installed in both the inner and outer half-rings. Each wiggler is positioned at the center of its respective damping wiggler section, where a non-zero dispersion function is present. By adjusting the magnetic field strength of the bending magnets, the dispersion function can be tuned, thereby enabling control over the damping time and emittance. Six quadrupole magnets are arranged to ensure the working point remains unchanged when the wiggler field strength varies.

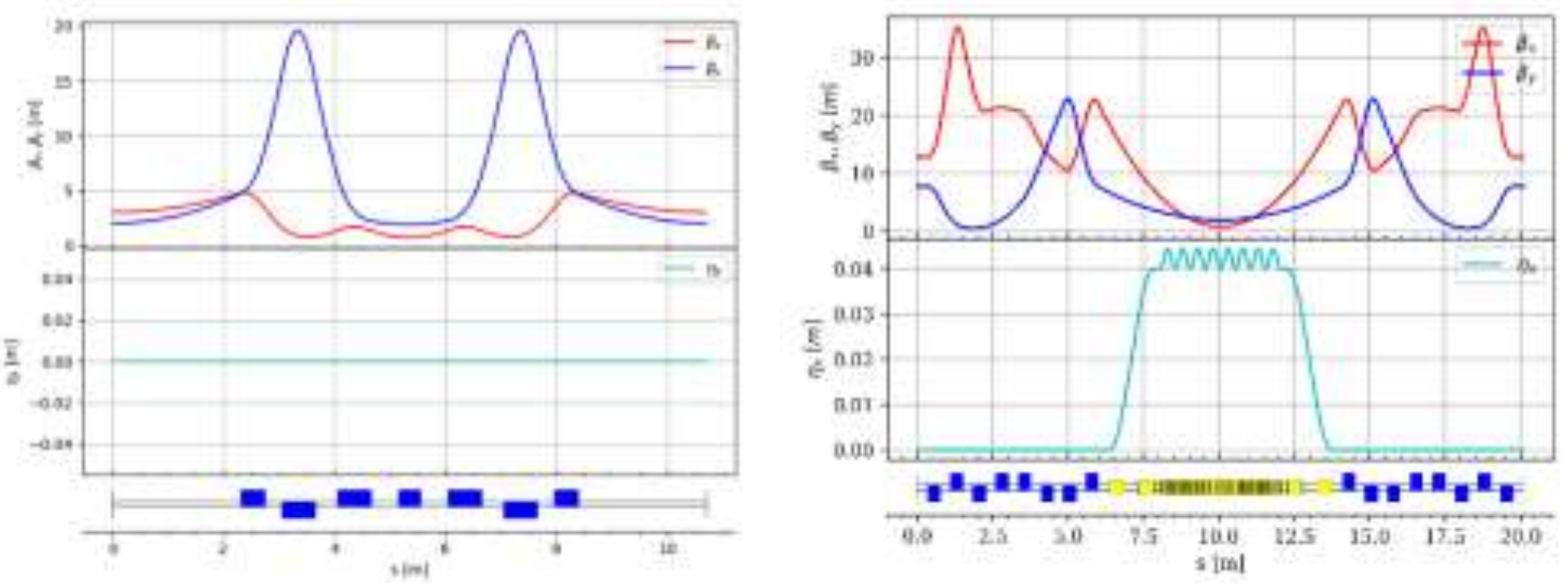

Figure 2.2-13: Lattice functions for the SS (left) and DW (right) sections

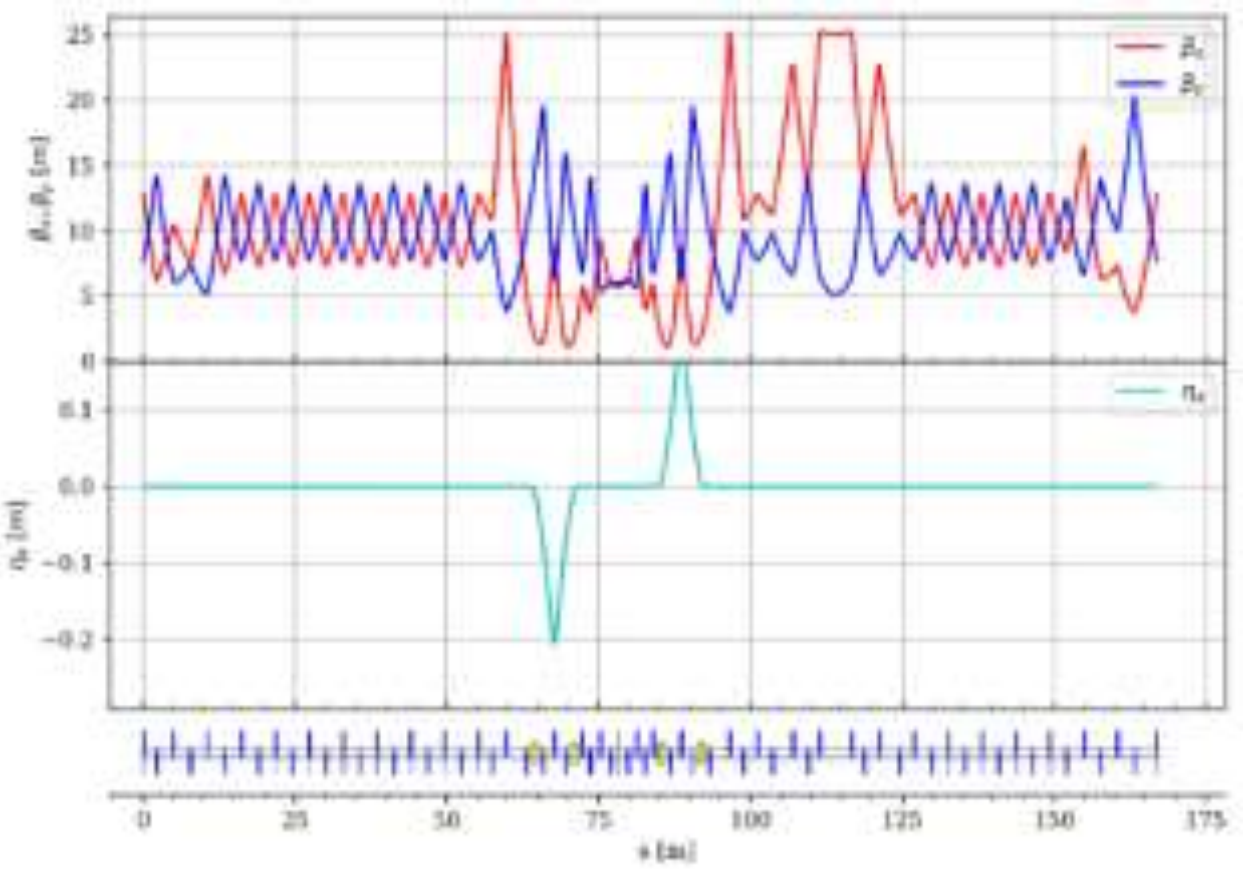

Figure 2.2-14: Lattice functions for the multifunctional straight section

### *2.2.2.4 Dynamic Aperture and Touschek Lifetime – Scheme II*

By employing phase tuners and sextupoles at the mirror position of IP [8] to separately optimize second- and third-order chromaticity, the W-function, and momentum acceptance, a large off-momentum dynamic aperture and momentum aperture have been achieved. At a design luminosity of $1.03 \times 10^{35}$ cm$^{-2}$s$^{-1}$, the off-momentum dynamic aperture reaches $12\sigma_{x,y} \times 13\sigma_e$, and the Touschek lifetime exceeds 250 seconds (as rigorously evaluated by SAD [9]). The target Touschek lifetime has been preliminarily achieved. However, the transverse dynamic aperture is still not sufficiently large, and the machine errors and corresponding orbit corrections have not yet been considered. As a result, the actual Touschek lifetime may be further reduced.

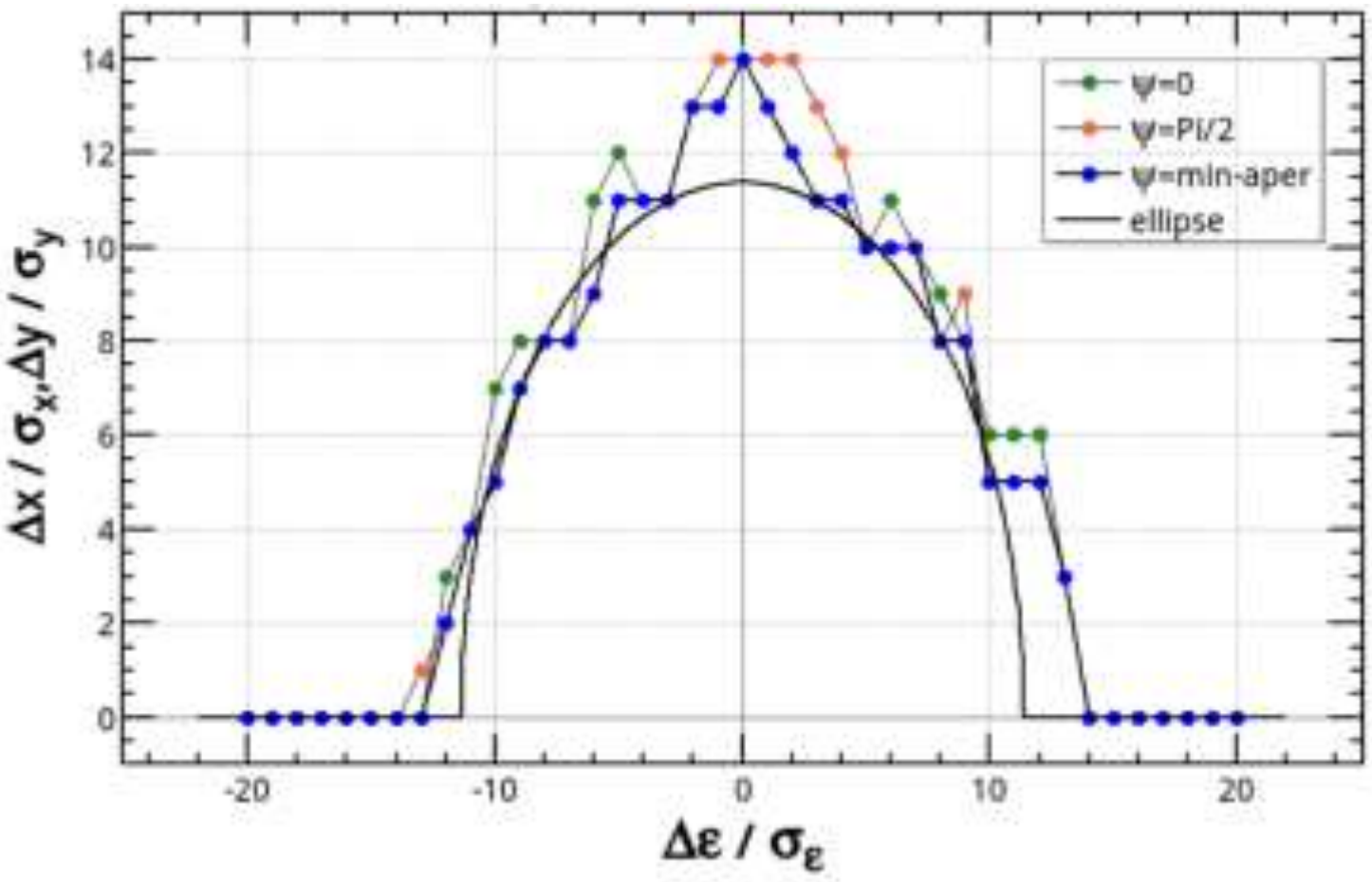

Figure 2.2-15: Momentum-dependent dynamic aperture in Scheme II

### 2.2.3 Interaction Region Optics Design

The key objective of the lattice design for the STCF interaction region is to compress the β function at the IP to enhance luminosity, while simultaneously ensuring sufficient momentum acceptance and dynamic aperture to maintain an adequate Touschek lifetime. This requires careful optimization of both the linear lattice and nonlinear optics corrections in the IR. The fundamental design principle is to coordinate the linear optics with the phase advances needed for nonlinear cancellation and to minimize the impact of nonlinear sextupole fields.

Following the design philosophy of other new-generation electron-positron colliders [1, 4, 10-12], the STCF IR lattice adopts a modular structure as illustrated in Figure 2.2.16. It consists of a final focusing telescope (FFT) to minimize the $\beta$ functions at the IP, a matching section (MCY) between FFT and the vertical chromatic correction section (CCY), local vertical/horizontal chromaticity correction sections (CCY/CCX), a matching section (YMX) from CCY to CCX, a dispersion suppressor (XMC), a crab sextupole section (CS), and a matching section (MS) connecting the IR to the long straight section. The linear optics functions of the STCF IR are shown in Figure 2.2.17.

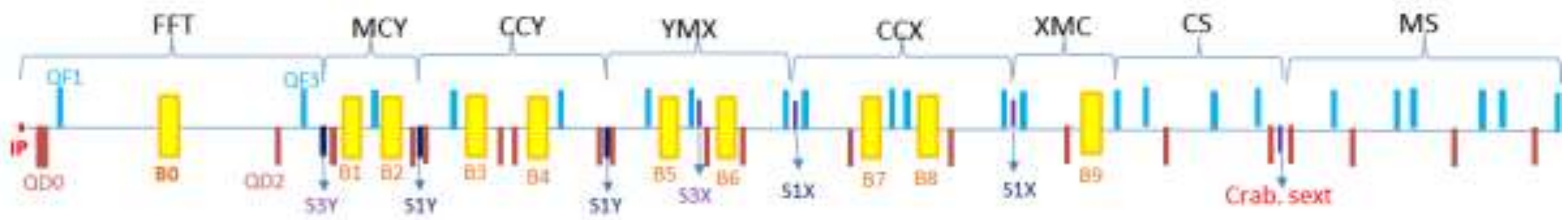

Figure 2.2.16: Lattice layout of the right half of the interaction region

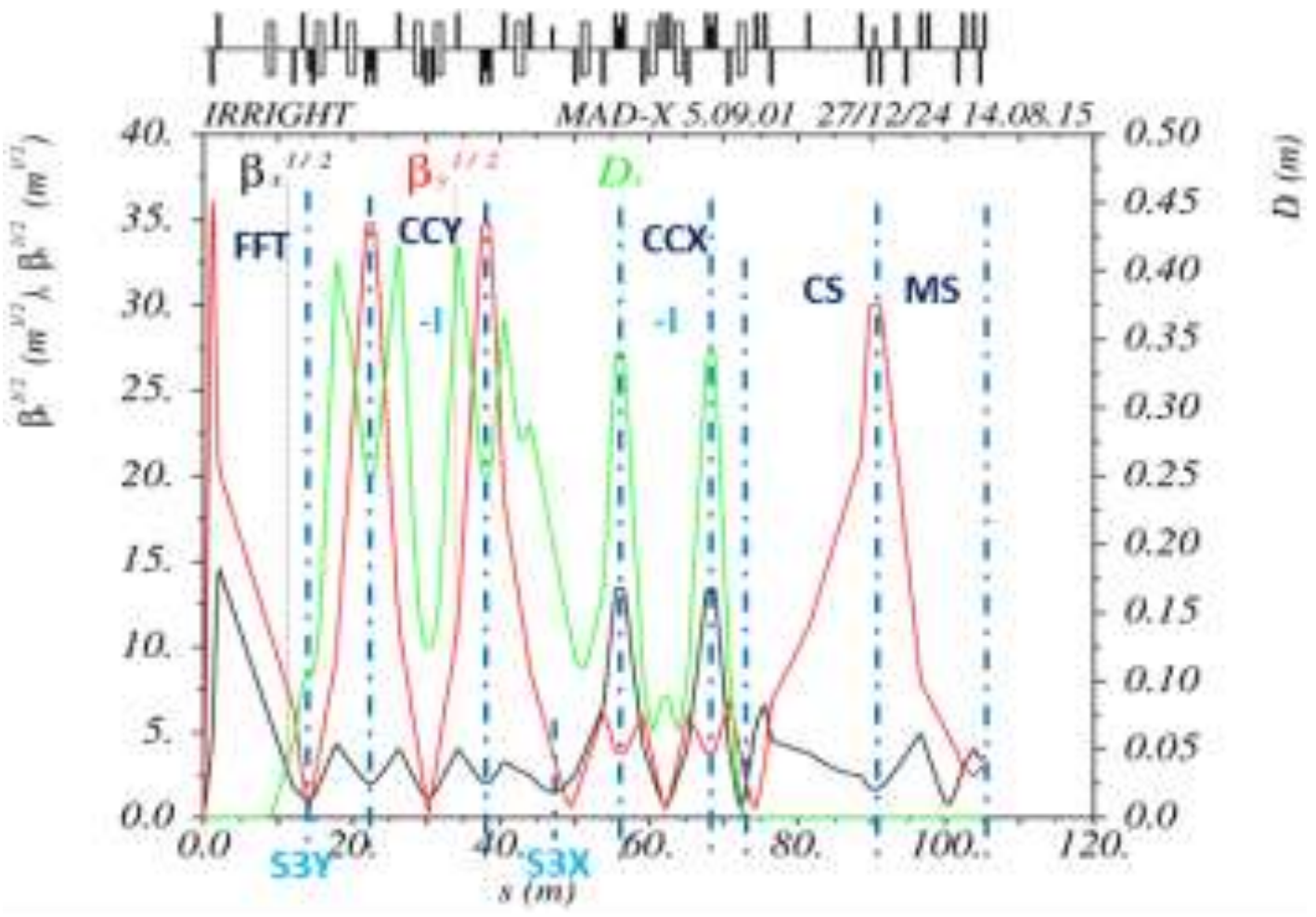

Figure 2.2.17: Optical functions of the right half of the interaction region

The Final Focus Telescope (FFT) section adopts the focusing structure composed of two groups of quadrupole magnets: one superconducting and the other room-temperature. The superconducting quadrupole doublet is used to achieve an extremely low $\beta_y^*$ at the IP, which is a key approach for enhancing luminosity in new-generation colliders. The room-temperature quadrupole doublet forms the mirror point of the IP. At this mirror point, both $\alpha_x$ and $\alpha_y$ are zero, and the phase advances to the IP in both the horizontal and vertical planes are π [13]. The advantage of the FFT lies in its ability to adjust the optical functions at the IP solely by tuning the optics at the mirror point, without modifying the components within the FFT.

A bending magnet (B0) is inserted into the FFT to generate the dispersion required for local chromaticity correction. Given its proximity to the IP, the strength of this bending magnet should be kept minimal to reduce synchrotron radiation background entering the detector.

The matching section from FFT to CCY (MCY) adopts a FODO-like cell starting with a defocusing quadrupole, enabling $\beta_y \gg \beta_x$ at the vertical chromaticity-correction sextupole S1Y. Additionally, the phase advance from S1Y to the final focusing quadrupoles (QD0 and QF1) is designed to be approximately π [14], with fine tuning required to correct second-order chromaticity. Two dipole magnets, B1 and B2, are included to increase dispersion at S1Y and thereby reduce the required sextupole strength.

The local vertical chromaticity correction section (CCY) must implement a –I transfer transformation between sextupole pairs to cancel out their nonlinearities. This can be realized using two FODO cells with both horizontal and vertical phase advances of π/2, arranged symmetrically about the center. Two identical dipole magnets, B3 and B4, are used to create symmetric dispersion.

The matching section from CCY to CCX (YMX) is responsible for converting the β-function from $\beta_x \gg \beta_y$ at the vertical chromaticity-correction sextupole S1Y to $\beta_x \gg \beta_y$ at the horizontal chromaticity-correction sextupole S1X. Additionally, the phase advance from S1X to the final focusing quadrupoles (QD0 and QF1) must be approximately 3π, with fine-tuning necessary to correct second-order chromaticity. At the midpoint of this section, where $\alpha_x = 0$, a second approximate mirror point of the IP is formed. Two dipole magnets, B5 and B6, are included to increase dispersion at S1X, thereby reducing sextupole strength.

The local horizontal chromaticity correction section (CCX) is designed similarly to the CCY section, requiring a –I transfer transformation between sextupole pairs to cancel nonlinearities. Two identical dipole magnets, B7 and B8, are used to form symmetric dispersion. The XMC section serves to cancel dispersion in the collision region, ensuring that the dispersion at the crab sextupoles is zero.

The crab sextupole section (CS) consists of six quadrupole magnets satisfying the phase advance constraint between the crab sextupole and the IP and the alpha function constraint at the crab sextupole: $\mu_x = 6\pi, \mu_y = 5.5\pi, \alpha_x = 0, \alpha_y = 0$. Additionally, $\beta_y \gg \beta_x$ at the crab sextupole locations, which helps to significantly reduce their required strength and associated nonlinearities [15]. The matching section (MS) consists of several quadrupole magnets, used to match the β/α functions required at the end of the long straight section, and also to tune the phase advance between the interaction region and the arc.

The total bending angle of dipole magnets in the interaction region is 60°, and all dipole magnets are set to 1 m in length. Except B0 that has a fixed bending angle of 1°, the strength of all dipoles must be chosen such that to adjust the dispersion function and maintain the dispersion invariant $\mathcal{H}_x$ below 0.02 m. To create a 60 mrad crossing angle at the IP between the two rings, the dipole magnets on the inner ring (beam out-going direction) are designed to bend 30 mrad less, while those on the outer ring (beam incoming direction) bend 30 mrad more. This approach facilitates a reasonable spacing (1.5–2 m) between the two rings, but results in asymmetry of the dispersion function about the IP, as shown in Figure 2.2-18.

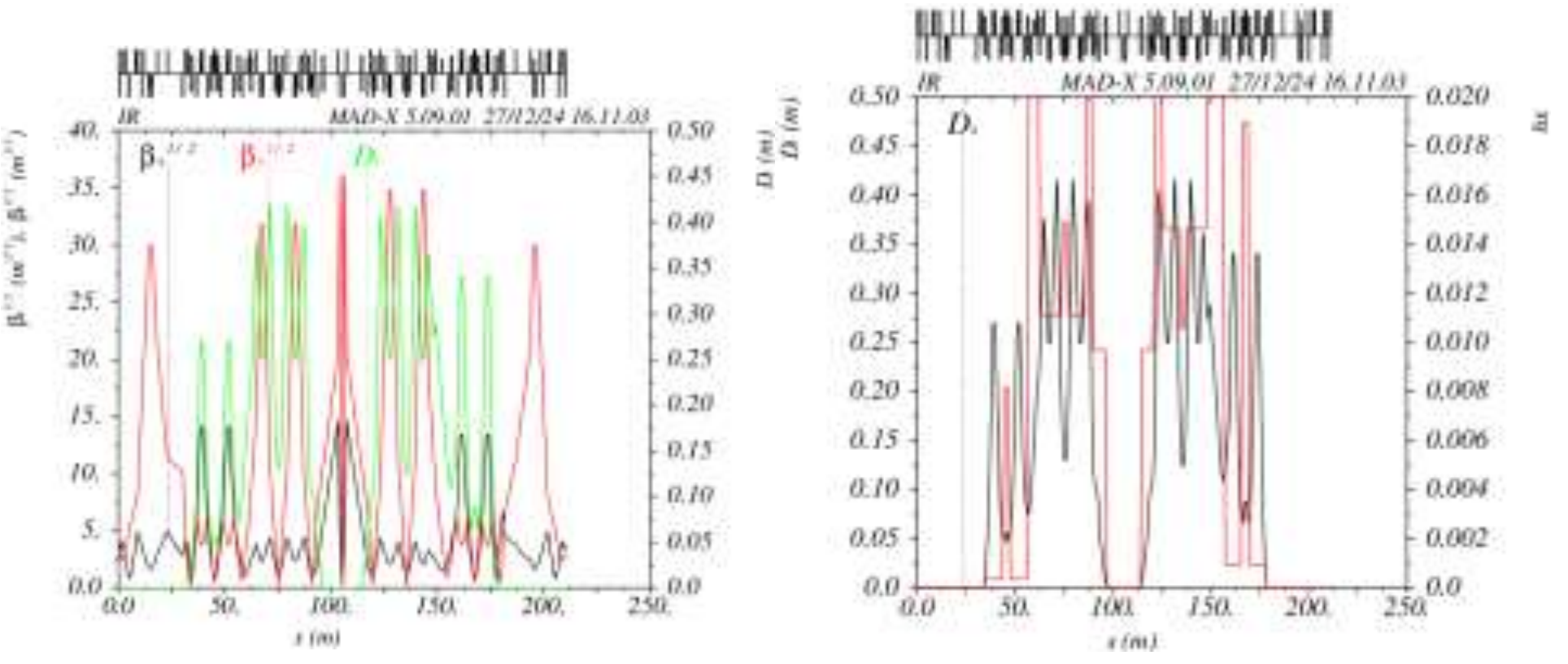

Figure 2.2-18: Optical functions of the entire STCF interaction region.

## 2.3 Transverse Beam Dynamics

The design goal of the transverse beam dynamics in the collider rings is to achieve a sufficiently large dynamic aperture and momentum acceptance to ensure high injection efficiency and long beam lifetime.

In the arc sections of the collider rings, strong quadrupole focusing is required to obtain a low emittance comparable to that of third-generation synchrotron light sources. A non-interleaved sextupole configuration was initially used for chromaticity correction in the arcs; however, due to the limited number of sextupole pairs, the required sextupole strength was relatively high, resulting in significant nonlinear effects. Therefore, based on the concept of second-order achromats proposed by K.L. Brown [13], an interleaved –I transfer transformation scheme was adopted for sextupole placement in the arcs. This configuration cancels the first-order geometric terms generated by the sextupoles, although it cannot cancel higher-order geometric terms. Nevertheless, the use of more sextupole pairs allows for lower individual magnet strengths, leading to weaker residual higher-order terms. The increased number of sextupoles also aids in the global nonlinear optimization of the collider rings.

However, the dynamic aperture and momentum acceptance of the collider rings are primarily limited by the interaction region (IR), which contributes the majority of the ring's nonlinearities [16, 17]:

- The IR must incorporate ultra-strong final focusing (FF) quadrupoles in a compact space to achieve extremely small $\beta_y^*$ at IP. This results in very high natural chromaticity and strong high-order kinematic and fringe field effects. A series of strong sextupoles must be installed in the IR for local chromaticity correction.
- The crab-waist mechanism imposes strict requirements on the phase advance between the crab sextupoles and the IP. Any lattice nonlinearity or imperfection can disrupt these phase constraints.
- The method of utilizing compensation solenoids to cancel the detector solenoid field has a direct impact on IR performance. For instance, overlap between the solenoid and quadrupole fields must be avoided, as it leads to vertical emittance growth.

All of these nonlinear and high-order effects degrade the dynamic and momentum apertures of the collider rings, thereby severely affecting beam injection efficiency and beam lifetime.

The optimization goal of the IR is to compensate as much as possible for the nonlinearities between the crab sextupole pairs. To achieve this, local horizontal and vertical chromatic sextupole pairs in the IR are designed to satisfy the –I transfer condition, which allows first-order chromaticity to be corrected while minimizing the nonlinear effects they introduce. The phase advance between the chromatic sextupoles and the final focusing (FF) quadrupoles is finely tuned to correct second-order chromaticity and the Montague function at the crab sextupole locations. Sextupoles are placed at the first and second IP mirror points to correct third-order chromaticity [6]. Since the $\beta$-functions at these mirror points are small, the geometric aberrations they contribute are relatively minor.

Due to the presence of numerous components and effects with significant nonlinear impacts in the STCF collider rings—such as the fringe fields of the FF quadrupoles, higher-order kinematic terms in IP drifts, chromatic sextupoles, and crab sextupoles—a comprehensive analytical treatment of these nonlinearities becomes extremely difficult. Consequently, deriving the dynamic aperture through traditional analytical or perturbative methods is no longer practical.

Instead, a common approach is to first construct simplified models to individually assess the influence of each nonlinear or high-order effect and propose targeted optimization strategies. Subsequently, professional software tools such as MADX, SAD, and Elegant are employed to perform 4D/5D/6D tracking and optimization of the dynamic aperture.

To optimize the dynamic aperture and momentum acceptance, advanced techniques such as multi-objective optimization algorithms are used for global nonlinear optimization of the collider rings, with the crab sextupoles always enabled during the process. A total of 48 chromatic sextupole pairs per ring are used as optimization variables (6 in the interaction region and 40 in the arcs). First, the program called PAMKIT [8], developed by project members, is employed in conjunction with intelligent multi-objective optimization algorithms to optimize sextupole strengths. First-order chromaticity and key resonance driving terms are used as constraints, while the dynamic and momentum apertures are treated as direct optimization objectives to determine the best sextupole configuration.

Then, the software SAD is used to precisely evaluate the dynamic aperture of the collider rings, as it accurately models the nonlinear effects mentioned above and has been successfully applied to KEKB and SuperKEKB. In SAD, six-dimensional canonical variables ($x$, $p_x$, $y$, $p_y$, $z$ and $\delta$) describe particle motion, where $p_x$ and $p_y$ are normalized transverse canonical momenta with respect to the design momentum $p_0$, and $\delta$ is the relative momentum deviation. Tracking simulations include longitudinal oscillations, synchrotron radiation in all magnets, high-order kinematic effects, finite-length effects of sextupoles, and crab sextupoles. Quantum excitation and Maxwellian fringe fields in quadrupoles are disabled. The number of turns tracked corresponds to one synchrotron radiation damping time.

After optimization, the dynamic apertures with the crab sextupoles turned off and on are shown in Figure 2.3-1. As can be seen, when the crab sextupoles are turned off, the horizontal dynamic aperture reaches nearly $36\sigma_x$, and at a momentum deviation of $10\sigma_\delta$, it still maintains a dynamic aperture of about $20\sigma_x$. When the crab sextupoles are turned on, the horizontal dynamic aperture reaches $34\sigma_x$ and even at $10\sigma_\delta$, the aperture remains at $17\sigma_x$. The variation of the collider ring's working point (fractional part) with respect to momentum deviation after optimization is shown in Figure 2.3-2, indicating that the momentum acceptance bandwidth reaches more than ±1.5%.

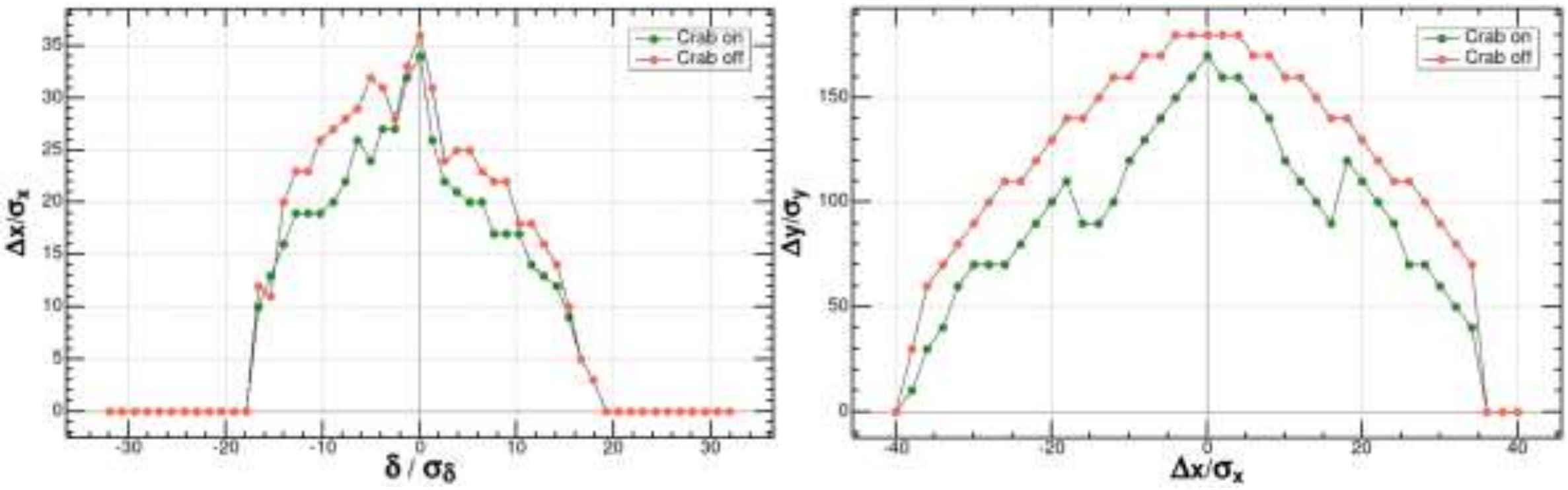

Figure 2.3-1: Off-momentum and on-momentum DA with crab sextupoles on and off

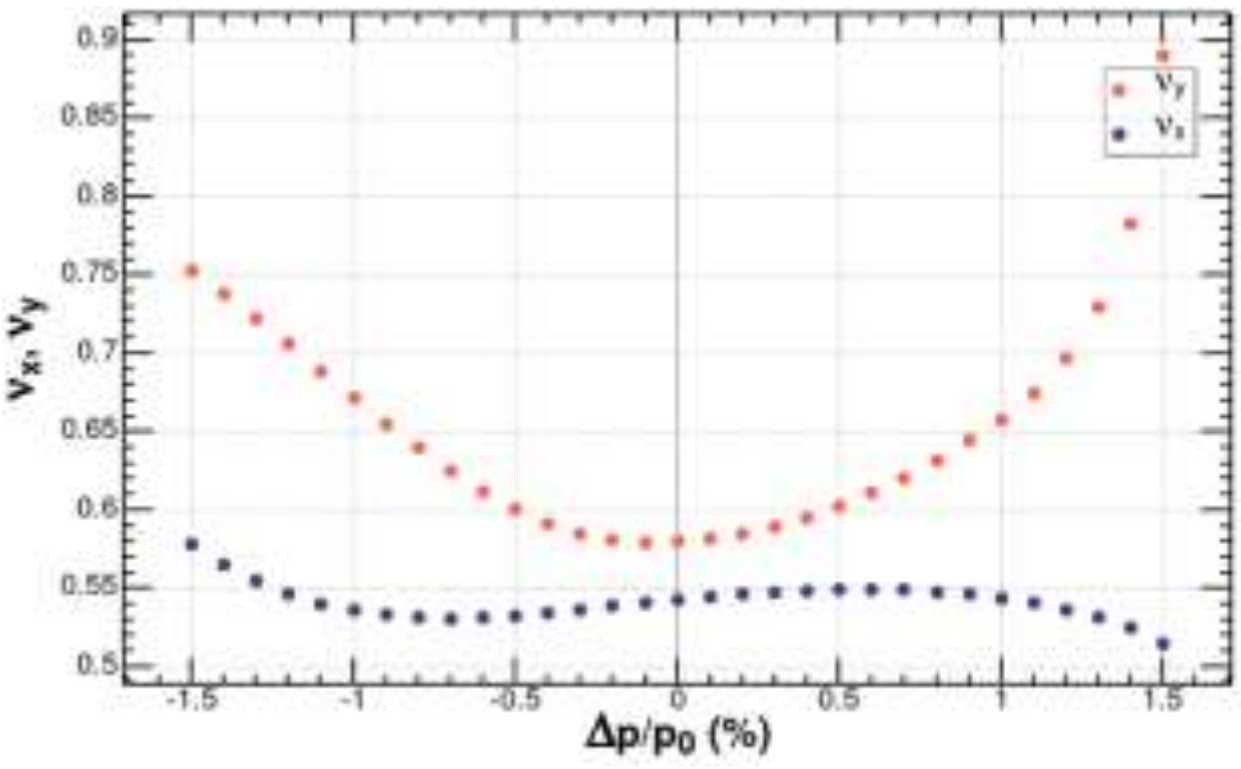

Figure 2.3-2: Tune (fractional part) versus momentum deviation $\delta$

Fringe fields exist in all magnets and are a source of nonlinearity that contributes to the reduction of the dynamic aperture. Using the SAD program to analyze the influence of Maxwellian fringe fields in dipole, quadrupole, and sextupole magnets, it was found that the fringe fields of the superconducting final focus (FF) quadrupole magnets have the dominant impact. Its effect on the dynamic aperture is shown in Figure 2.3-5.

The second term in the Hamiltonian of the quadrupole fringe field resembles the form of the Hamiltonian of a thin octupole magnet. Therefore, octupole coils can be installed outside the FF quadrupole magnets to compensate for the fringe field effects of the FF quadrupoles. Figure 2.3-3 shows that the off-momentum and on-momentum dynamic apertures for the cases with/without the FF quadrupole fringe fields and with/without octupole compensation. All the cases consider the crab sextupole on service. One can see that the dynamic aperture with the FF quadrupole fringe fields decreases significantly compared to the case without the fringe fields, however, it can recover to some extent with the compensation of additional octupole coils. The optimization of the octupole coils is still ongoing to further recover the dynamic aperture.

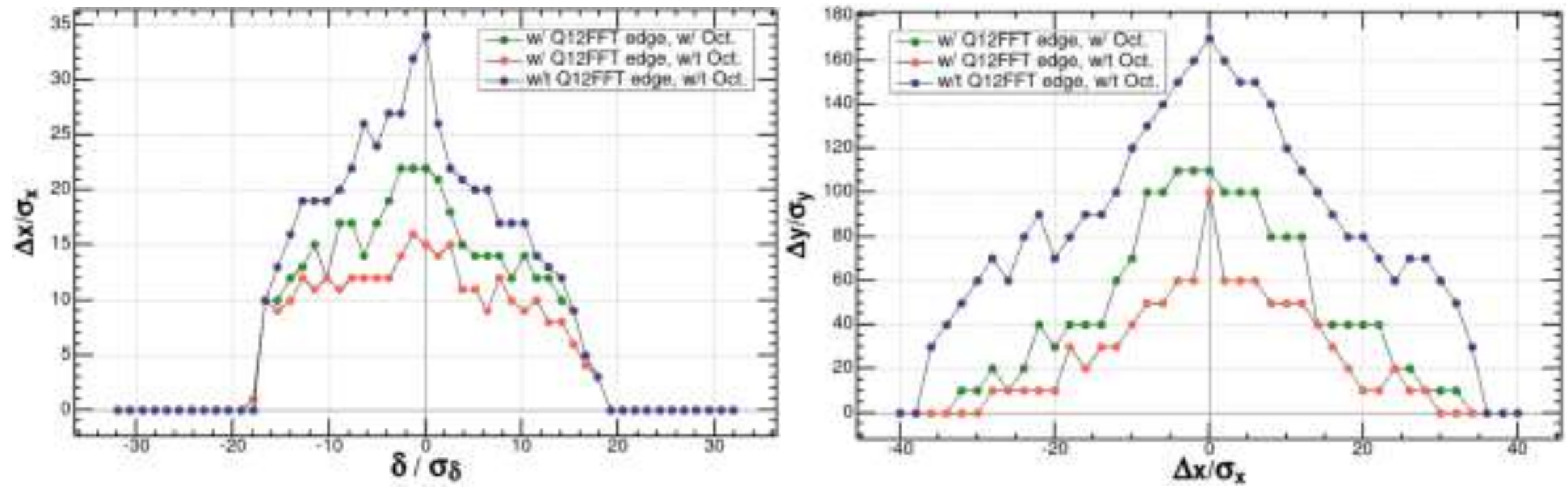

Figure 2.3-3: Dynamics aperture for the cases with/without the FF quadrupole fringe fields and with/without octupole compensation.

To further analyze dynamic aperture limitations, the Elegant program was utilized to track the local momentum acceptance (LMA) profiles around the entire ring for the cases with/without the FF quadrupole fringe fields and with/without octupole compensation. The results, as depicted in Figure 2.3-4, indicate that the integration of the FF quadrupole fringe fields significantly degrades the LMA performance compared to the idealized fringe-field-free scenario. With octupole compensation, the LMA demonstrates remarkable recovery. The minimum LMA under positive momentum deviation is about 0.7%, and about –0.6% under negative deviation. The interaction region (between 300 m and 450 m in the figure) remains the critical bottleneck for the ring's global momentum acceptance.

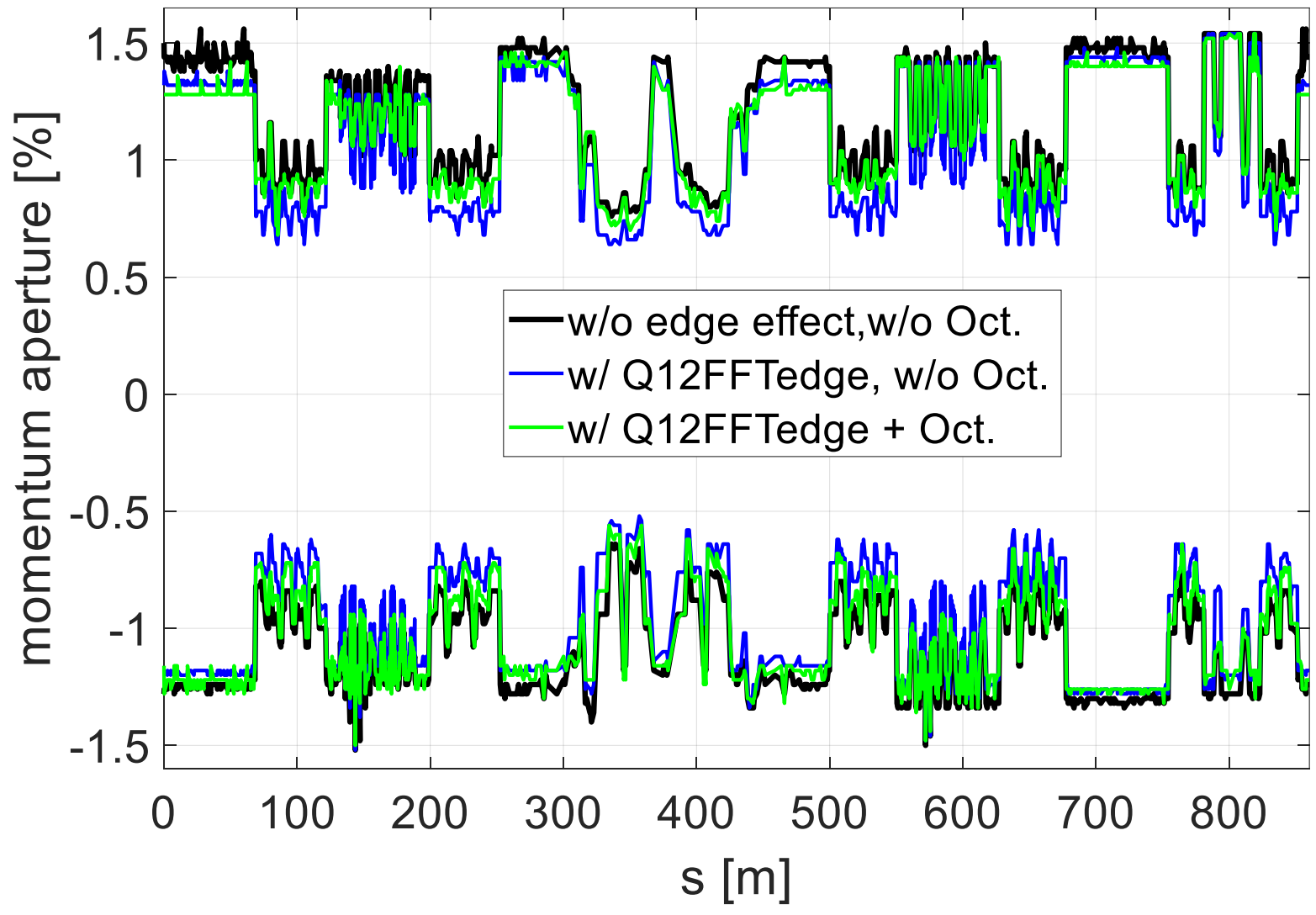

Figure 2.3-4: Local momentum acceptance along the ring for the cases with/without the FF quadrupole fringe field and with/without octupole compensation.

The solenoid field of the detector contributes to the vertical emittance growth in two main ways: the longitudinal $B_z$ field directly induces horizontal x-y coupling, while the fringe radial field $B_z$ introduces vertical dispersion. As a result, careful compensation is necessary. The general principles for compensation are as follows: (1) the integral of the longitudinal $B_z$ field must be zero; (2) within the region of the superconducting quadrupole magnets, the $B_z$ field must vanish;

(3) the integral of the radial fringe field of the solenoid should be compensated as much as possible. The focusing effect of the solenoid on the IR optics can be compensated by adjusting the strength of the final focusing magnets. Preliminary study shows that the dynamic aperture can be completely recovered for the case of perfect compensation for the detector solenoid field.

At the beam energy of 2 GeV and peak luminosity of $1 \times 10^{35}$ cm$^{-2}$s$^{-1}$ (with crab sextupoles activated), simulations using SAD and Elegant programs yield the Touschek lifetimes of 325 s and 304 s, respectively, when the fringe field effects are neglected. Both values meet the design goal of exceeding 200 s lifetime at the target luminosity of $5 \times 10^{34}$ cm$^{-2}$s$^{-1}$. When incorporating FF quadrupole fringe field effects, which typically introduce nonlinear distortions reducing dynamic aperture, the uncompensated Touschek lifetime drops to about 150 s. However, after implementing compensation with octupoles, the lifetime recovers to about 250 s, maintaining compliance with the 200-s operational requirement.

To further improve the Touschek lifetime, additional nonlinear optimization strategies are required to enlarge the local momentum acceptance. These include further optimization of nonlinearities in the collision region, the adoption of new nonlinear optimization strategies (such as those based on resonance driving terms and their fluctuations), and the inclusion of phase advances between different regions (such as between the interaction region and arc sections, or between adjacent arc sections) as tuning variables.

## 2.4 Longitudinal Beam Dynamics

For electron storage rings with target beam currents on the order of amperes, such as the STCF collider rings, the main challenges in longitudinal beam dynamics stem from longitudinal coupled-bunch instabilities caused by the fundamental mode and parasitic modes of RF cavities [18]. The RF cavity scheme must meet both specific voltage and power requirements and effectively suppress these coupled-bunch instabilities. STCF plans to adopt 500 MHz TM020-type room-temperature main RF cavities. Compared to conventional TM010 cavities, TM020 cavities feature relatively higher quality factors (Q) and lower R/Q values. Therefore, for the same cavity voltage and power requirements, the total R/Q can be reduced by approximately half [19], effectively mitigating coupled-bunch instabilities driven by the fundamental mode. This makes it feasible to suppress such instabilities through relatively simple means, avoiding overly complex low-level RF (LLRF) feedback system designs.

Table 2.4-1 presents the longitudinal beam dynamics parameters and corresponding RF parameters of STCF at three beam energies. For beams of 1 GeV & 1.1 A, 2 GeV & 2 A, and 3.5 GeV & 2 A, 2, 6, and 15 cavities are respectively required to meet both voltage and power demands, with some safety margins. Table 2.4-2 gives the fundamental mode parameters of the STCF TM020 cavity. These parameters are primarily used to evaluate the coupled-bunch instabilities induced by the accelerating mode.

Table 2.4-1: Longitudinal beam dynamics and RF parameters for STCF

| **Parameter** | **Unit** | **Value** | **Value** | **Value** |
|---|---|---|---|---|
| Beam energy, $E$ | GeV | 2 | 1 | 3.5 |
| Momentum compaction factor, $\alpha_p$ | $10^{-4}$ | 13.49 | 12.63 | 13.73 |
| Energy spread, $\sigma_e$ | $10^{-4}$ | 7.8 | 6.18 | 10.02 |
| Beam current, $I$ | A | 2 | 1.1 | 2 |
| Bunch filling factor | — | 48% | | |
| Energy loss per turn, $U_{0_sr}$ | keV | 543 | 106 | 1494 |
| Power loss per turn, $P$ | kW | 1086 | 117 | 2988 |
| Longitudinal damping time, $\tau_z$ | ms | 10.57 | 27.07 | 6.72 |
| RF frequency, $f_{RF}$ | MHz | 499.7 | | |
| Harmonic number, $h$ | — | 1434 | | |
| Total RF voltage, $V_R$ | MV | 2.5 | 1 | 7.5 |
| Number of RF cavities | — | 5 | 2 | 15 |
| Voltage per cavity | kV | 500 | | |
| Power per cavity | kW | 217 | 58.5 | 199.2 |
| Coupling factor | — | 4.71 | 2.33 | 5.5 |
| Detuning frequency | kHz | -64.1 | -57.6 | -99.0 |

Table 2.4-2: TM020 cavity fundamental mode parameters

| Parameter | Value |
|---|---|
| Frequency | 499.7 MHz |
| $R/Q$ ($V^2/(2P)$) | 47.5 |
| $Q_0$ | 60000 |
| R | 2.85 MΩ |
| $Q_L$ ($\beta_{opt}$=5.5) | 9230 |

### 2.4.1 TM020 Accelerating Mode Stability Analysis

To analyze coupled-bunch instabilities caused by the fundamental mode of RF cavities, it is necessary to consider the influence of low-level feedback, as it modifies the impedance seen by the beam [20]. Two common methods are used for evaluation: analytical calculations based on closed-loop cavity impedance and particle tracking simulations [21]. For the LLRF feedback, we only consider a PI (proportional-integral) feedback based on I/Q signals. The closed-loop impedance of the cavity can be expressed as:

$$Z_{clo} = \frac{Z_c(\omega)}{1 + \frac{k_p}{R_L} e^{-i\Delta\omega\tau_d} Z_c(\omega)} \tag{2}$$

where $Z_c(\omega)$ is the open-loop cavity impedance, $R_L$ is the load impedance, $k_p$ is the proportional gain of the PI feedback, and $\tau_d$ is the feedback loop delay.

For STCF, we find that by appropriately selecting the proportional gain and delay time of the PI feedback, the -1 mode can be effectively suppressed, and the -2 mode remains below the threshold determined by synchrotron radiation damping, thus avoiding instabilities caused by the fundamental mode. Using the 2 GeV & 2 A beam case as an example, Figure 2.4-1 shows the growth rates of major low-order modes obtained by both analytical and particle tracking methods for various proportional gains and delay times. Without PI feedback, the -1 mode has a growth rate of 368 $s^{-1}$, which is much higher than the damping rate of 93.5 $s^{-1}$. When $k_p = 1$ and the delay is between 700 and 900 buckets, the -1 mode is fully suppressed, and the -2 mode becomes the dominant instability, yet its growth rate still remains below the damping rate. This indicates that a low-R/Q TM020 cavity does not require a special feedback system for suppressing low-order coupled-bunch instabilities; using standard PI feedback with suitable gain and delay is sufficient to ensure stability. The same conclusion applies to the 1 GeV & 1.5 A and 3.5 GeV & 2 A beam conditions. To ensure this approach is effective, the LLRF feedback delay should preferably not exceed 1.5 μs.

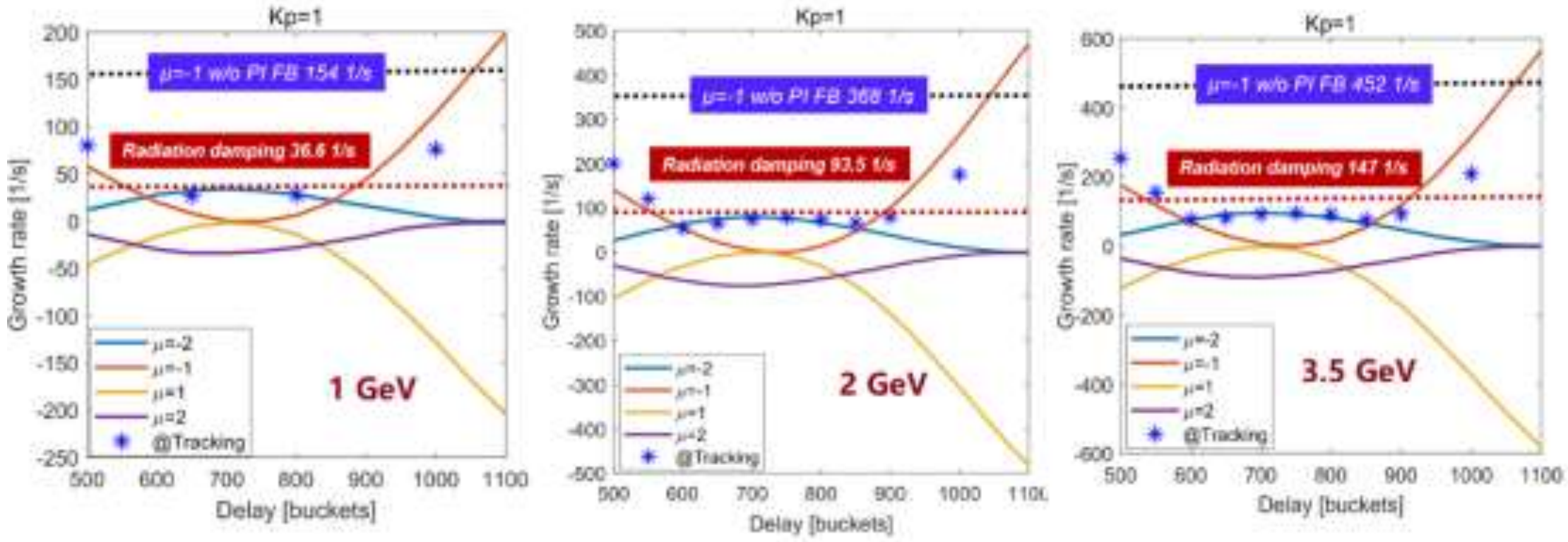

Figure 2.4-1: Growth rates of low-order coupled-bunch instabilities caused by the accelerating mode. Solid lines are from analytical methods; discrete points are from tracking simulations.

### 2.4.2 TM020 Parasitic Mode Stability Analysis

For the TM020 cavity, the TM020 mode is the accelerating mode, and modes with frequencies above or below its nominal frequency are referred to as higher-order and lower-order modes, respectively. Both types can cause significant coupled-bunch instabilities. For simplicity, we refer to both as parasitic modes. To mitigate these instabilities, the cavity design must incorporate strong suppression of parasitic modes from the outset. Assuming beam spectral lines coincide with the resonance frequencies of parasitic modes, the associated coupled-bunch instability growth rate is typically calculated by:

$$\frac{1}{\tau_{HOM}} = \frac{I_0 \alpha \omega_r}{4\pi \nu_s E/e} e^{-(\omega_r \sigma_\tau)^2} ReZ \quad (3)$$

where $\sigma_\tau$ is the rms bunch length, $\nu_s$ is the synchrotron tune, $\omega_r$ is the resonance frequency, and $ReZ$ is the real part of the impedance.

The synchrotron oscillation period of STCF is less than 0.2 ms. Assuming a conservative 1 ms damping time provided by the longitudinal bunch-by-bunch feedback system, the impedance thresholds for parasitic modes are shown in Figure 2.4-2. The threshold is lowest for 3.5 GeV, since it uses the most cavities and assumes all higher-order modes are aligned. Even so, the lowest single-cavity threshold is 1.7 kΩ @ 3.75 GHz, which is fully achievable for TM020 cavities.

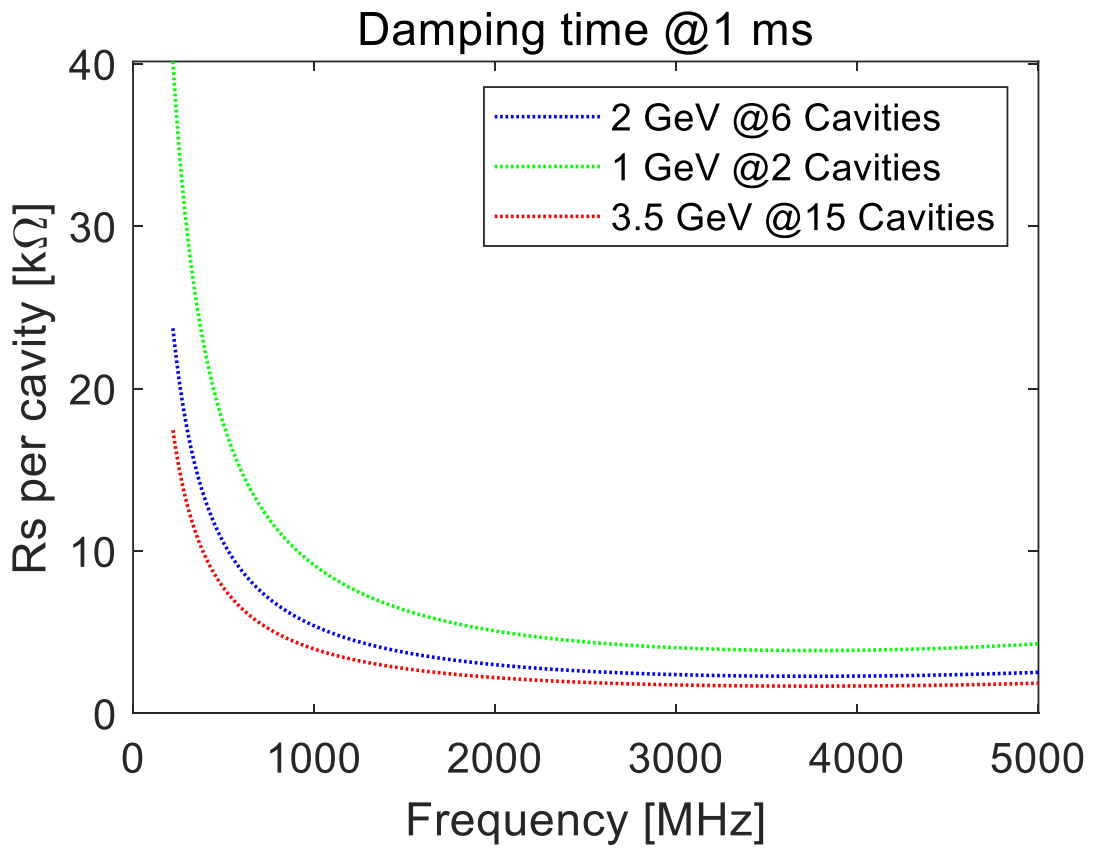

Figure 2.4-2: Parasitic mode impedance thresholds assuming 1 ms damping time from longitudinal feedback at three beam energies.

### 2.4.3 Transient Beam Loading Effects

The 5% empty buckets reserved to suppress ion trapping introduce transient beam loading effects [22], resulting in bunch-to-bunch variations in synchronous phase and bunch length. Tracking simulations of this effect show the resulting distributions of bunch centers and lengths, as illustrated in Figure 2.4-3. The effect on bunch length is minimal (within ± 1 ps), and the bunch center variation remains within ± 20 ps. This suggests the impact on luminosity is negligible.

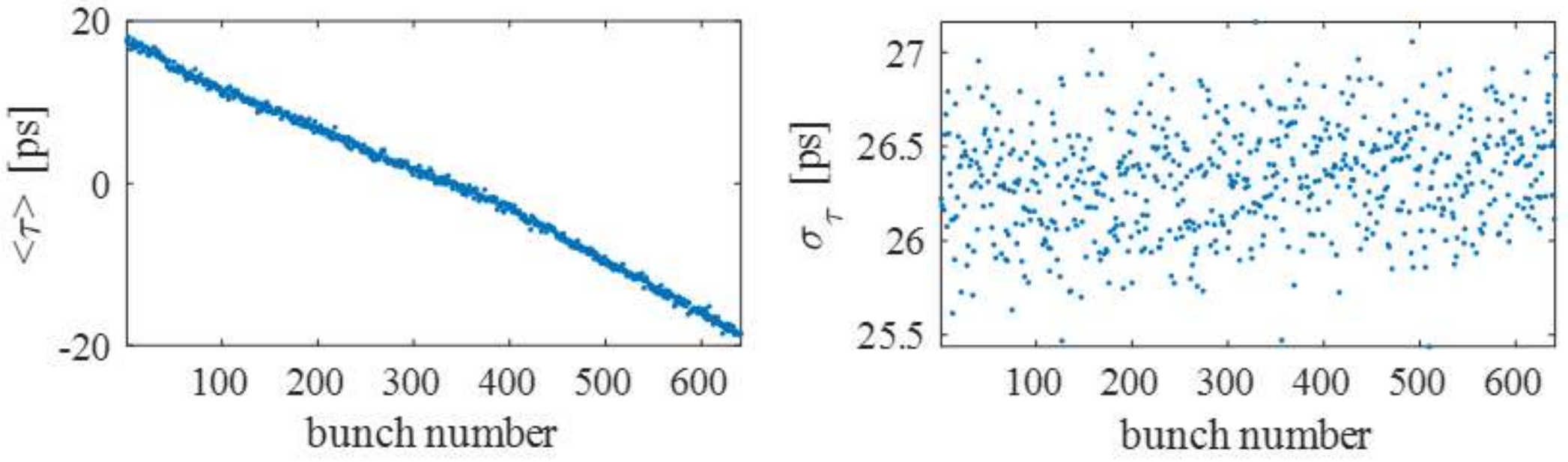

Figure 2.4-3: Bunch length and center distributions under nominal fill pattern

STCF intends to use the TM020-type room-temperature RF cavity scheme. Due to its relatively low R/Q, there is no need for complex feedback schemes such as those adopted by the PEP-II collider (e.g., one-turn delay feedback and direct feedback). The fundamental mode stability can be fully ensured by simply selecting appropriate PI feedback gain and delay parameters, thereby significantly simplifying the LLRF system. As for coupled-bunch instabilities caused by parasitic modes, radiation damping alone is insufficient; a longitudinal bunch-by-bunch feedback system providing at least 1 ms of damping is required. From the perspective of RF cavity design, suppressing parasitic modes below the instability threshold under such damping conditions is entirely achievable.

## 2.5 Beam-Beam Interaction and Luminosity Optimization

For a crab-waist scheme collider using flat beams (with transverse beam sizes at the IP satisfying $\sigma_y^* \ll \sigma_x^*$) and a large Piwinski angle (i.e., $\phi = \sigma_z \tan\theta / \sigma_x^* \gg 1$), the luminosity can be expressed as [23]:

$$L = L_0 R_H, \tag{4}$$

$$L_0 = \frac{n_b I_{b+} I_{b-}}{4\pi e^2 f_0 \sigma_x^* \sigma_y^* \sqrt{1+\phi^2}}, \tag{5}$$

$$R_H \approx \frac{\sqrt{\pi}}{\zeta_x} e^{\frac{1}{\zeta_x^2}} Erfc\left(\frac{1}{\zeta_x}\right). \tag{6}$$

where $Erfc(x)$ is the complementary error function and $\zeta_x = \sigma_x^*/(\beta_y^* \tan(2\theta))$. Typically, $\zeta_x \lesssim 0.5$ is chosen to avoid the degradation of luminosity and increase in emittance due to the hourglass effect. When $\zeta_x \lesssim 0.5$, $R_H \gtrsim 0.9$, and the luminosity can be approximated as $L \approx L_0$. The number of bunches $n_b$ mainly depends on the ring circumference and the RF system frequency. The beam-beam parameters in the horizontal and vertical directions can be approximately written as:

$$\xi_x = \frac{N_p r_e \beta_x^*}{2\pi\gamma\, \sigma_x^{*2}(1+\phi^2)} \approx \frac{N_p r_e \beta_x^*}{2\pi\gamma \sigma_z^2 \theta^2} \tag{7}$$

$$\xi_y = \frac{N_p r_e \beta_y^*}{2\pi\gamma \sigma_x^* \sigma_y^* \sqrt{1+\phi^2}} \approx \frac{N_p r_e \beta_y^*}{2\pi\gamma \sigma_y^* \sigma_z \theta} \tag{8}$$

where the single-bunch particle number $N_p$ is mainly constrained by collective effects from impedance.

The luminosity can also be written as [24]:

$$L = \frac{\gamma I}{2e r_e \beta_y} \xi_y^L \tag{9}$$

Based on operational experience, the vertical beam-beam parameter $\xi_y \approx \xi_y^L \approx 0.1$ can be achieved in present $e^+e^-$ storage ring colliders. While $\beta_x^*$ at the IP does not directly affect luminosity, choosing a small $\beta_x^*$ (i.e., small $\zeta_x$) is generally favorable. This reduces the synchro-betatron oscillation and coherent X-Z instability [25] induced by beam-beam interactions and lowers the required strength for the crab sextupoles. The required strength of the crab sextupoles is given by [26]:

$$K_2 = \frac{1}{\theta \beta_y^* \beta_y} \sqrt{\frac{\beta_x^*}{\beta_x}} \tag{10}$$

Beam-beam interaction is a key factor in determining collider luminosity, beam stability, and beam lifetime. In third-generation high-luminosity colliders using the crab-waist scheme, where the beam current is high and emittance is small, beam-beam effects are especially prominent. Experiences from colliders such as DAFNE and SuperKEKB, as well as numerous theoretical and numerical studies, have shown significant coupling between beam-beam effects and other processes like lattice nonlinearities and impedance, which limit overall machine performance. Therefore, the severity of beam-beam interaction depends jointly on key beam parameters and the machine operating mode and must be optimized carefully to enhance performance. In addition to qualitative theoretical analysis, the detailed impact of beam-beam interactions on accelerator performance is mainly assessed through simulations. Weak-strong simulations require fewer computational resources and are suited for wide exploration of the beam parameter space, such as tune, Twiss parameters at the IP, and machine tolerance to imperfections, serving as a foundation for strong-strong simulations. Strong-strong simulations, being computationally intensive, are used for localized studies of beam parameter space and to evaluate beam stability. Weak-strong simulations mainly address incoherent collective effects, whereas strong-strong simulations focus on coherent instabilities involving both beams.

### 2.5.1 Simulation Results and Analysis

The working point in $e^+e^-$ storage ring colliders is typically optimized near half-integer values to achieve maximum luminosity. The precise fractional tune values are optimized using beam-beam simulations. Figure 2.5-1 shows the luminosity versus working point calculated using the BBWS code for fractional tunes above the half-integer (beam energy is 2 GeV; other parameters refer to Table 2.1-1). Comparing the simulations with and without the crab-waist scheme reveals that the crab-waist can effectively suppress nonlinear betatron resonances induced by beam-beam effects, thus enlarging the high-luminosity region in the tune space.

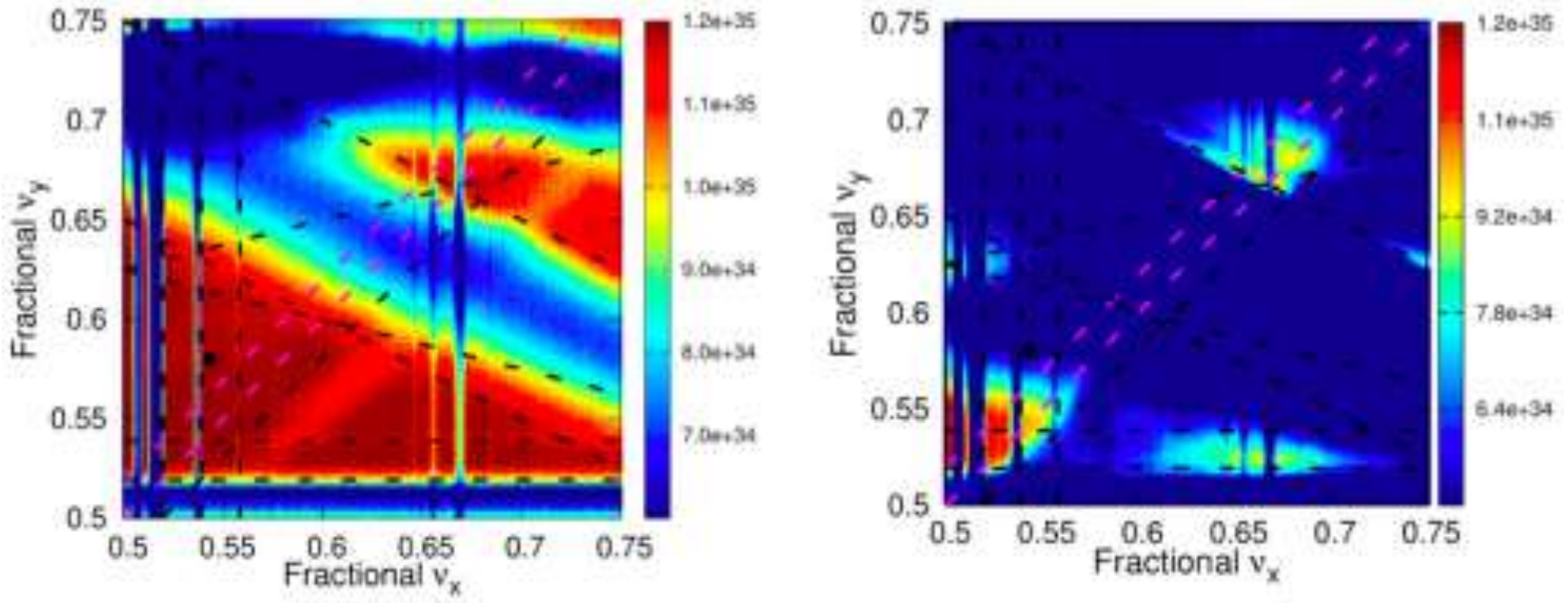

Figure 2.5-1: Comparison of luminosity maps with (left) and without (right) the crab-waist scheme. The black dot marks the nominal working point; dashed lines indicate resonance lines.

However, Figure 2.5-1 also shows that the crab-waist scheme cannot suppress synchro-betatron resonances induced by beam-beam interactions. Similar to SuperKEKB, STCF adopts the strategy of choosing a relatively large synchrotron tune $\nu_z$ and placing the transverse tune between $0.5 + n\nu_z$ and $0.5 + (n+1)\nu_z$ (n = 1 or 2). Since these resonance lines are separated by $\nu_z$, a large $\nu_z$ helps prevent the tune footprint from overlapping with resonance lines (see Figure 2.5-2), thereby avoiding emittance growth and luminosity loss. The general rule is $\nu_z/\xi_x \geq 3$. At 2 GeV, STCF chooses $\nu_z = 0.0194$, hence $\nu_z/\xi_x = 3.7$, which should be sufficient to mitigate synchro-betatron oscillations and coherent X-Z instabilities.

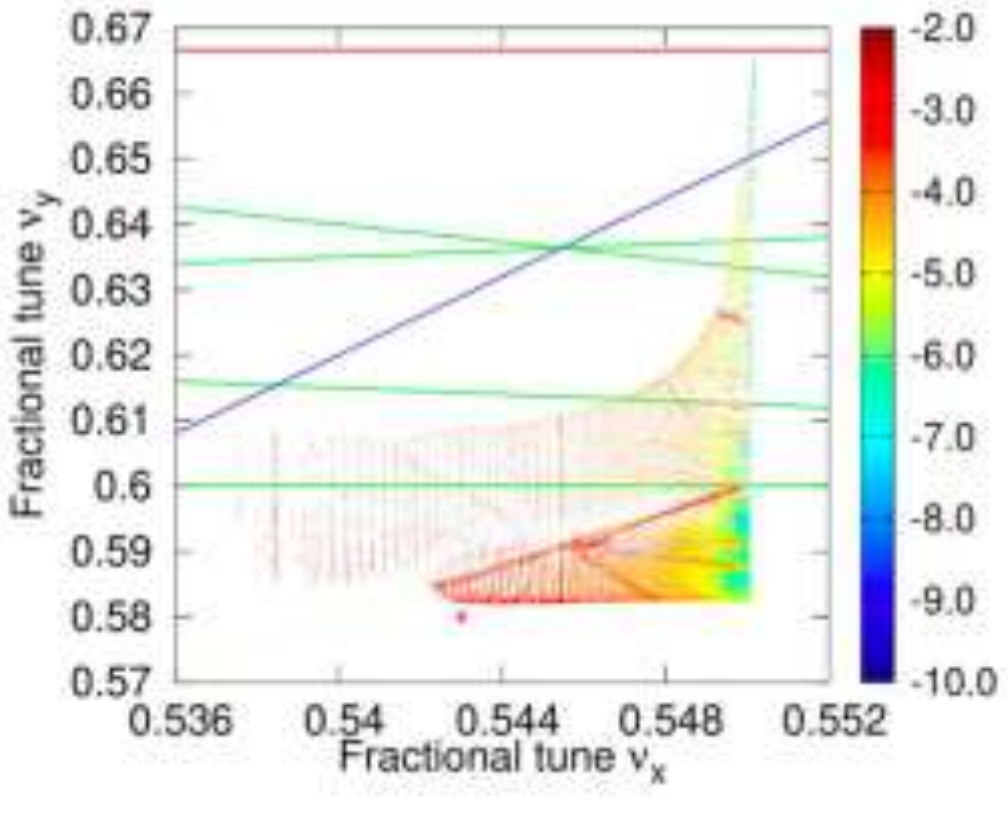

Figure 2.5-2: Tune footprint of STCF computed using the SAD code (including lattice). The red star indicates the nominal working point (0.543, 0.58). Solid lines in different colors represent various resonance lines.

Coherent X-Z instabilities are evaluated using strong-strong beam-beam simulations. Figure 2.5-3 shows the results of scanning the horizontal tune (with vertical fractional tune fixed at 0.58) using the BBWS and BBSS codes. The results indicate that a horizontal tune near 0.543 avoids coherent X-Z instability. The strong-strong simulations suggest that the range of horizontal tune values for achieving high luminosity is narrow and can be widened by increasing the synchrotron tune $\nu_z$ or reducing $\beta_x^*$ (Sec. 4.11 of [27]). Longitudinal impedance primarily causes bunch lengthening and synchrotron tune shifts. Bunch lengthening directly reduces luminosity, while tune shifts can introduce Landau damping that weakens synchro-betatron resonances, though it also broadens and shifts the oscillation peaks, complicating tune optimization.

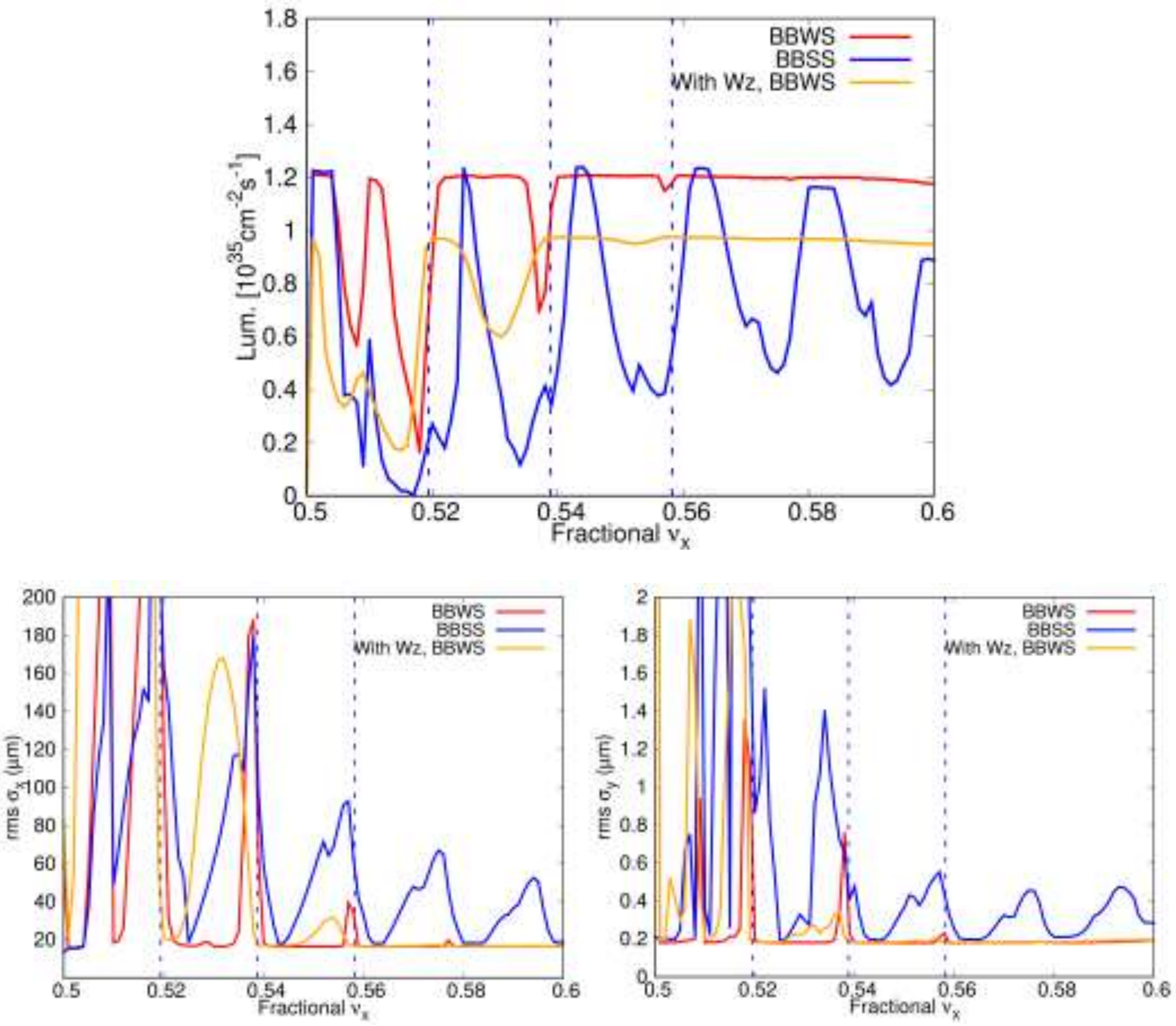

Figure 2.5-3: Luminosity (top) and horizontal bunch size (bottom) versus horizontal tune from BBWS and BBSS simulations. Brown curves show simulations including longitudinal impedance (modeled by the broadband resonator in Figure 2.6-1).

Fixing the horizontal fractional tune at 0.543, vertical tune scans were performed using BBSS. Figure 2.5-4 shows that the vertical fractional tune should be greater than 0.56, and that luminosity and transverse beam sizes are relatively insensitive to vertical tune within this range.

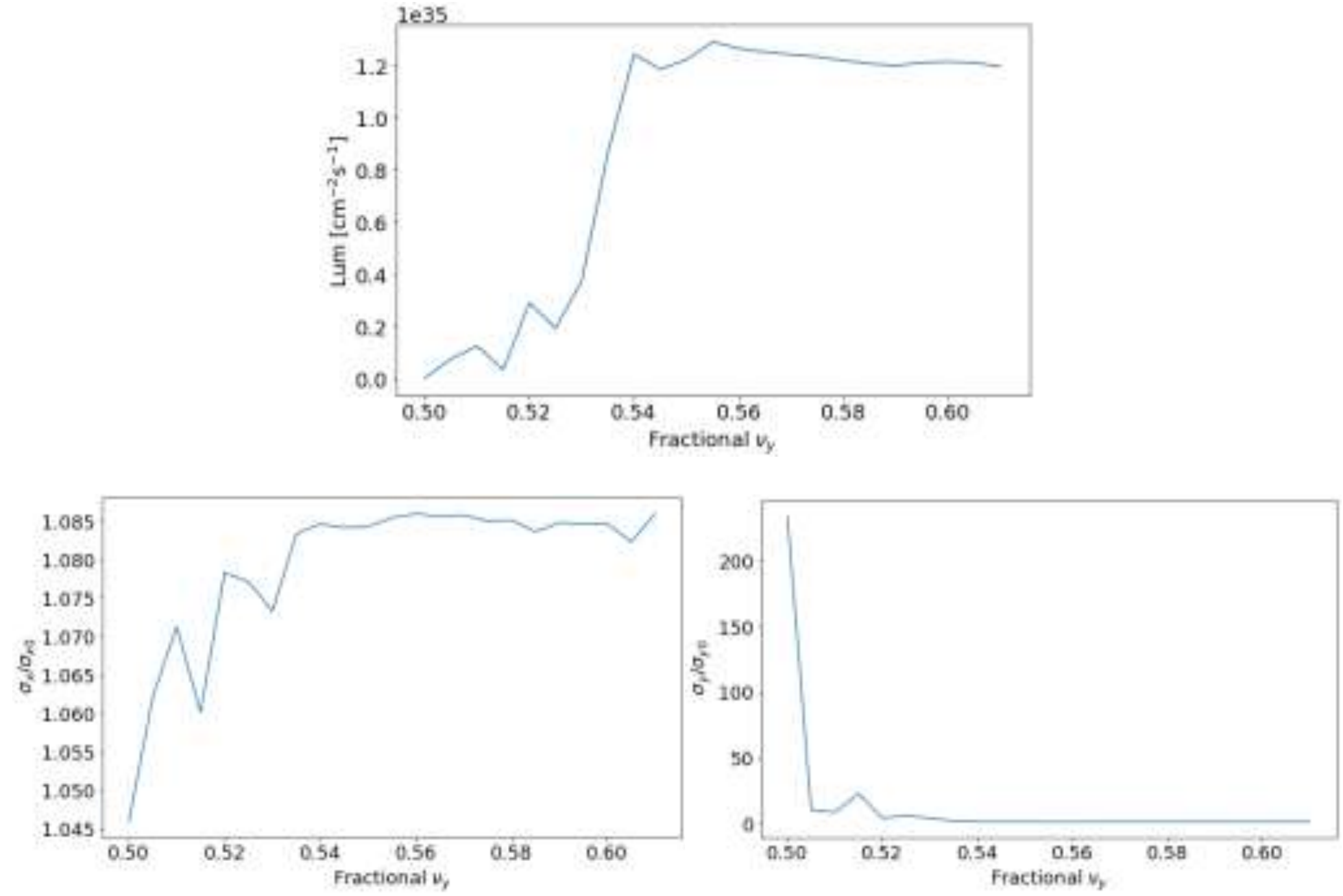

Figure 2.5-4: Luminosity (top) and transverse beam sizes (bottom) versus vertical tune from BBSS simulation.

Simulations considering lattice nonlinearities (for details of lattice design, see Section 2.2) were carried out using the BBSCL code and compared with the BBSS results. As shown in Figure 2.5-5, coupling between lattice nonlinearities and beam-beam effects affects luminosity, with the extent depending on specific lattice design details. This highlights the need for deeper lattice nonlinearity studies and optimization. Additionally, BBSCL was used to scan the single-bunch current with the lattice included. Figure 2.5-6 shows that coherent instability occurs when $N_p$ exceeds $6.2\times10^{10}$, above the nominal value of $5.2\times10^{10}$.

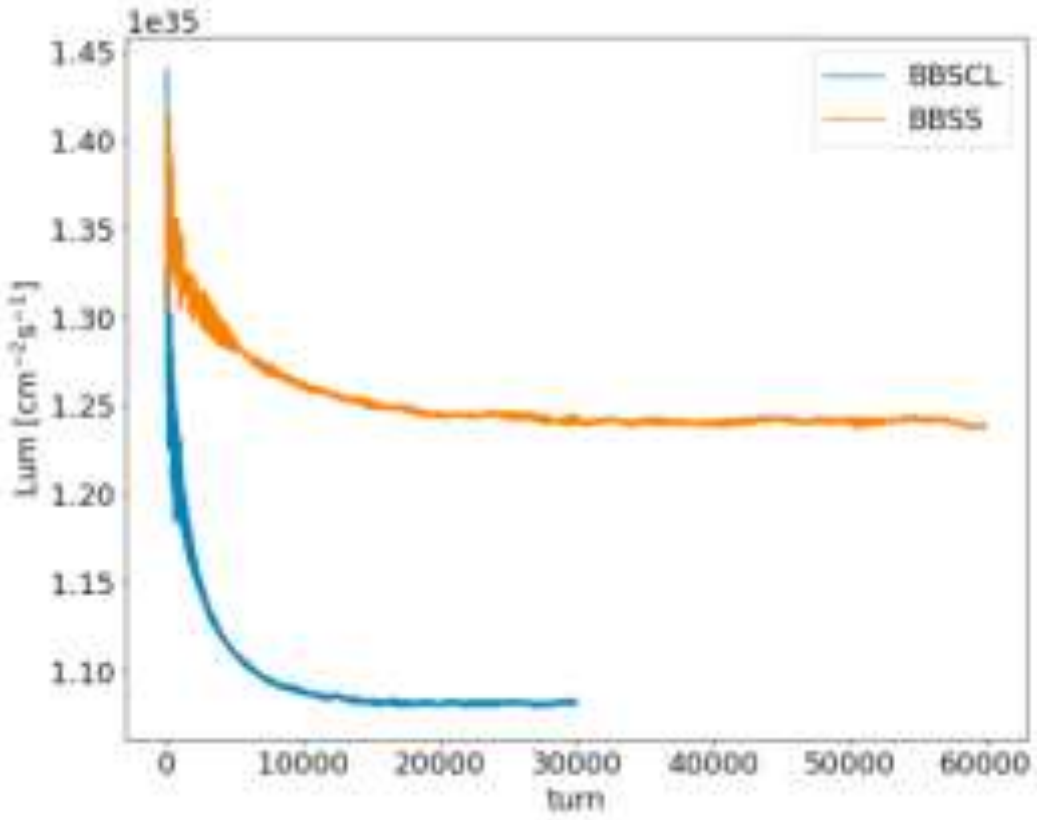

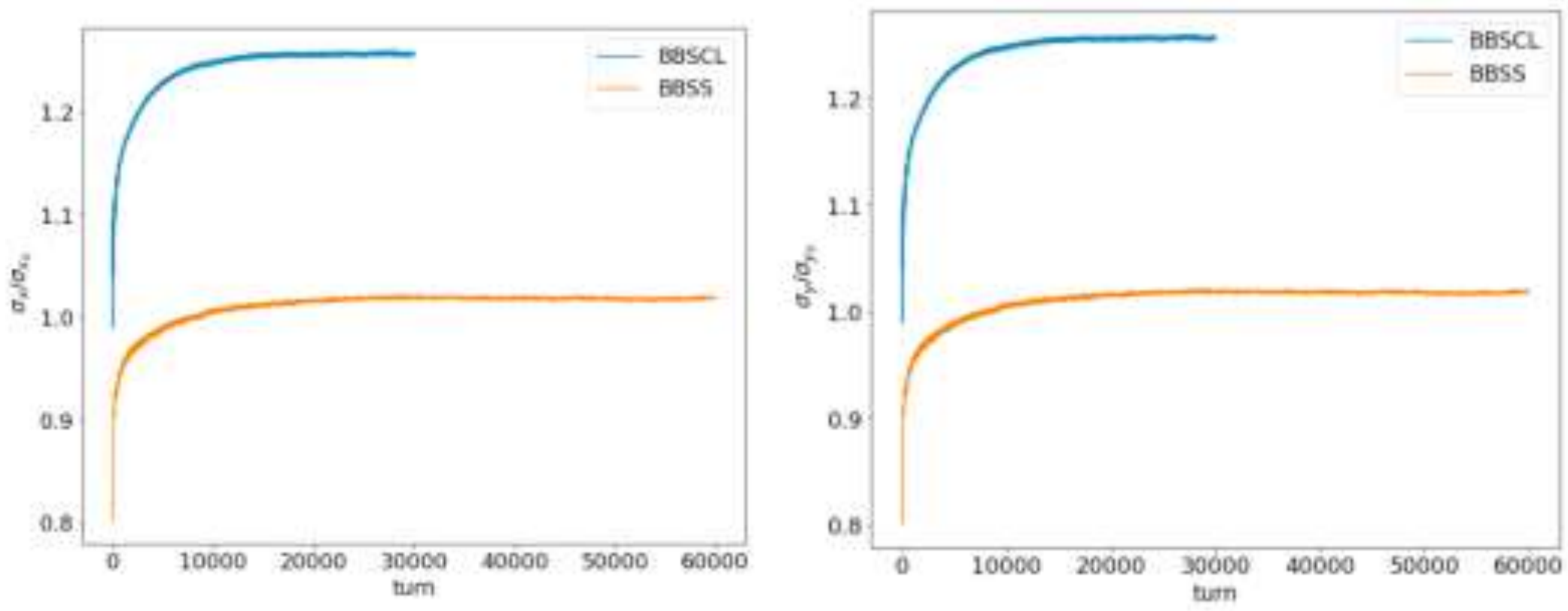

Figure 2.5-5: Luminosity (top) and transverse beam sizes (bottom) at the working point (0.543, 0.58) computed using BBSCL (with lattice) and BBSS (without lattice).

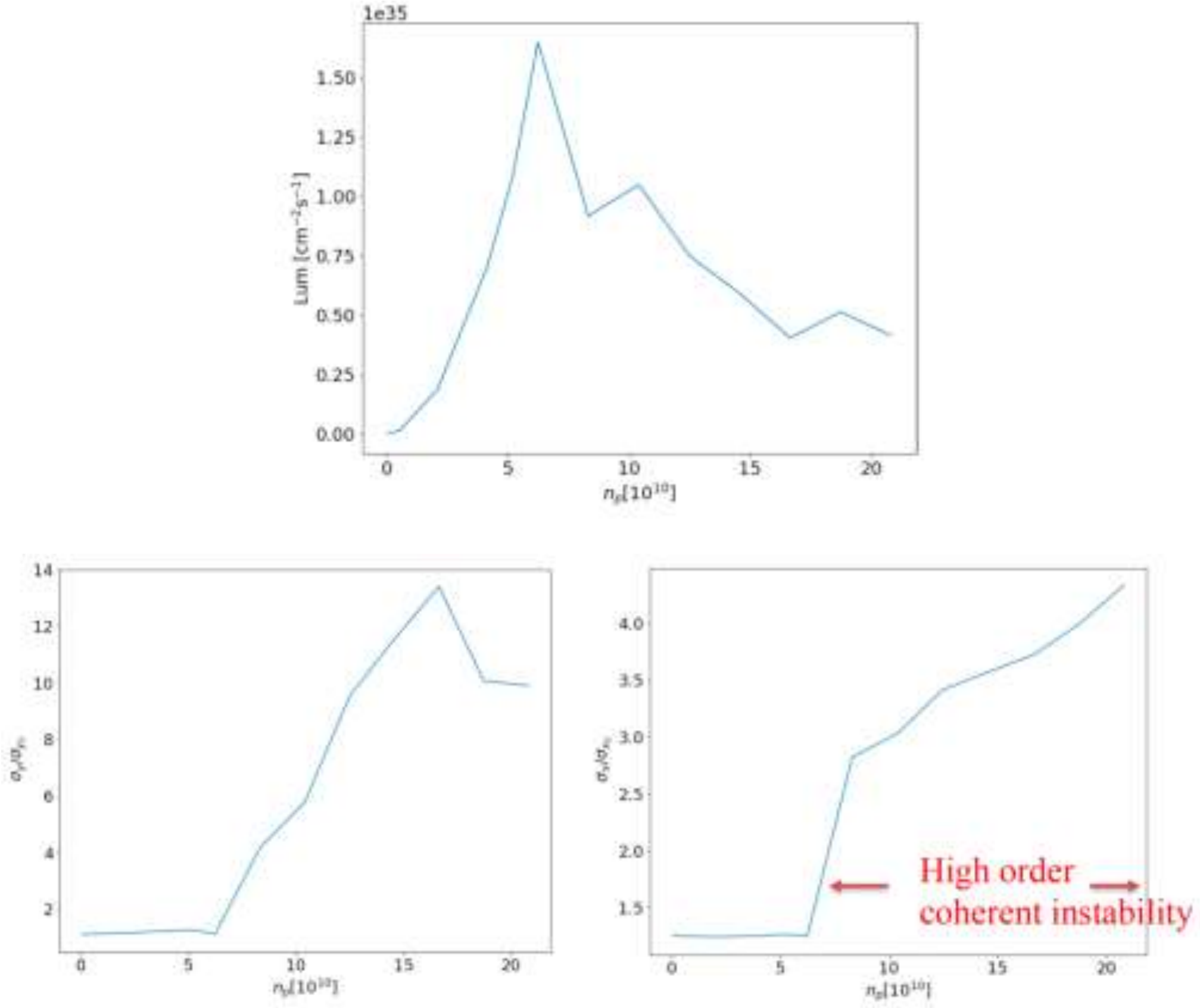

Figure 2.5-6: Luminosity (top) and horizontal bunch size (bottom) as a function of single-bunch population at the working point (0.543, 0.58) from BBSCL simulations.

### 2.5.2 Coupled Effects of Multiple Physical Mechanisms

The coupling of multiple physical mechanisms is particularly pronounced in colliders employing the crab-waist collision scheme, significantly increasing the complexity of machine design and the difficulty of achieving high-luminosity operation. In addition to the previously discussed couplings between beam-beam interactions, lattice nonlinearities, and impedance effects, several other factors warrant further investigation.

- **Space charge effects**: Under high-current conditions, the Coulomb field generated by the beam itself has a non-negligible influence on particle motion. According to the latest STCF lattice design and beam parameters, space charge effects induce vertical and horizontal tune shifts of approximately −0.043 and -0.0027 (see Fig. 2.5-8), which are opposite in direction to the shift caused by beam-beam interactions. As this effect is distributed along the entire ring and influenced by the alternating-gradient focusing structure, its overall impact cannot be simply canceled by beam-beam tune shifts and thus cannot be directly utilized to enhance luminosity. Beyond simplified particle tracking models, a more comprehensive assessment requires full-ring tracking simulations incorporating the actual lattice structure. Such simulations, performed using the BBSCL code with full-lattice and space charge effects, are currently underway. A preliminary result for luminosity under design beam parameters is shown in Fig. 2.5-7. Analysis of the simulation data reveals a notable luminosity degradation, attributed to a weak coherent X-Z instability. This instability is primarily associated with the horizontal tune shift induced by space charge (its absolute value is comparable with the horizontal beam-beam tune shift), rather than the vertical component. The horizontal tune shift brings the beams close to broad resonance peaks of the coherent X-Z instability, as illustrated in Fig. 2.5-3 under current machine conditions.

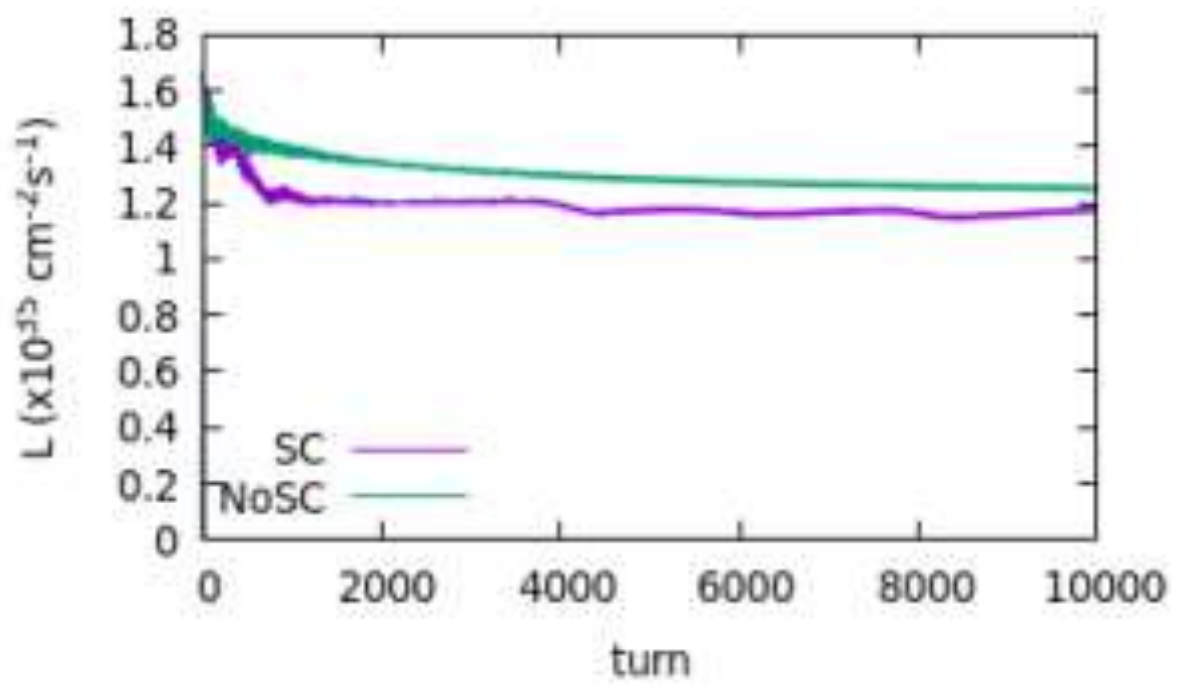

Figure 2.5-7: Simulated luminosity by the BBSCL code for STCF with and without the space charge effects

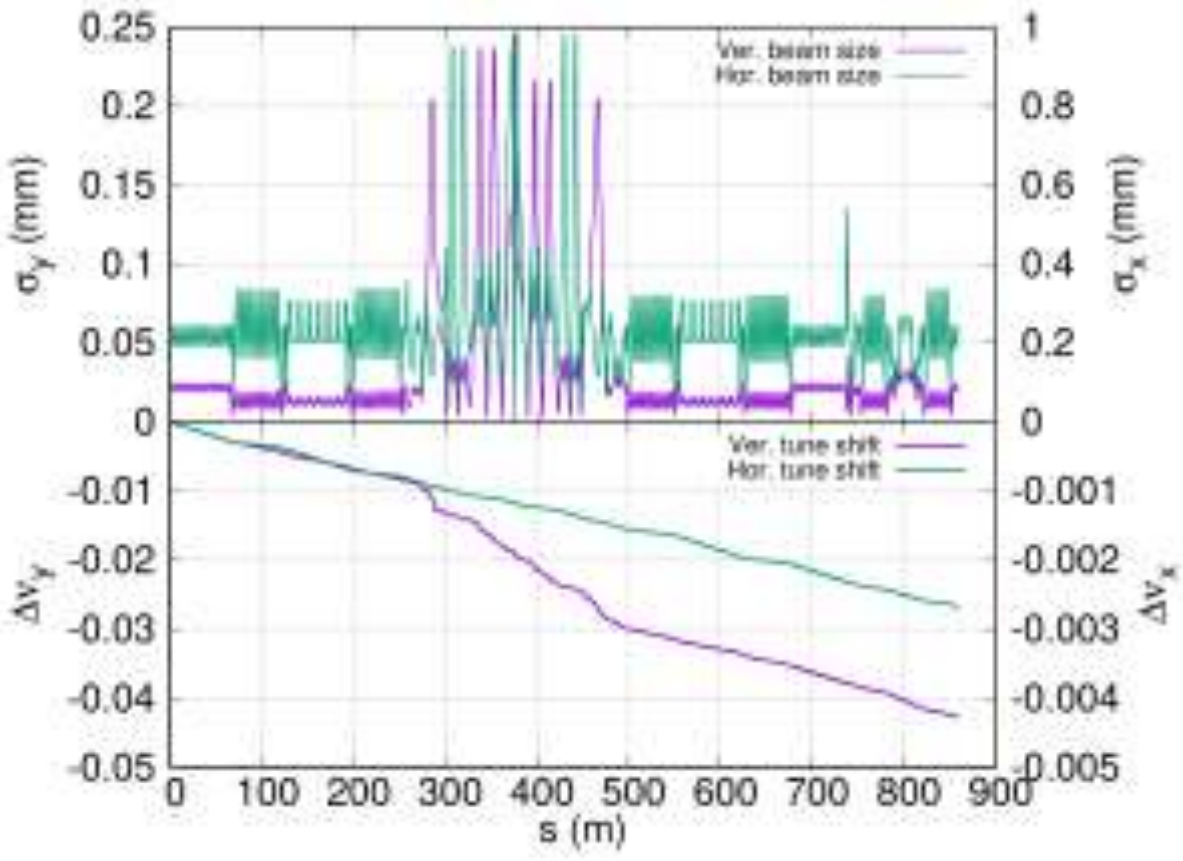

Figure 2.5-8: Space-charge tune shift as a function of the distance along the ring of STCF

- **Impedance effects**: In addition to the longitudinal impedance effects discussed earlier, transverse impedance can also couple with beam-beam interactions and induce coherent mode-coupling instabilities. These instabilities are closely related to the vertical tune and thereby constrain the accessible working point space. Development of a comprehensive impedance budget for STCF is currently in progress. In future studies, the combined impact of impedance and beam-beam interactions will be systematically investigated.
- **Imperfections in the IP optics**: These include non-zero coupling and dispersion at the IP, phase deviations between the IP and the crab-waist sextupoles, and orbit errors at both the IP and sextupole locations. Such imperfections can undermine the effectiveness of the crab-waist mechanism and consequently degrade the achievable luminosity.
- **Noise in the bunch-by-bunch feedback system**: Under high-current operating conditions, the machine's reliance on the feedback system for suppressing coherent instabilities becomes significantly greater. However, because of the extremely small transverse beam sizes at the IP, noises in the feedback system itself may excite beam motion and contribute to luminosity degradation. To evaluate this effect, simulations using the BBSS code were performed, incorporating turn-by-turn noise modeled as vertical collision offsets at the IP with a standard deviation of $\delta y = k_y \sigma_y^*$. The results, shown in Fig. 2.5-9, indicate that even modest noise levels (e.g., $k_y = 0.1$) can lead to a luminosity loss exceeding 30%, highlighting the need for stringent control of feedback-induced offset noise.

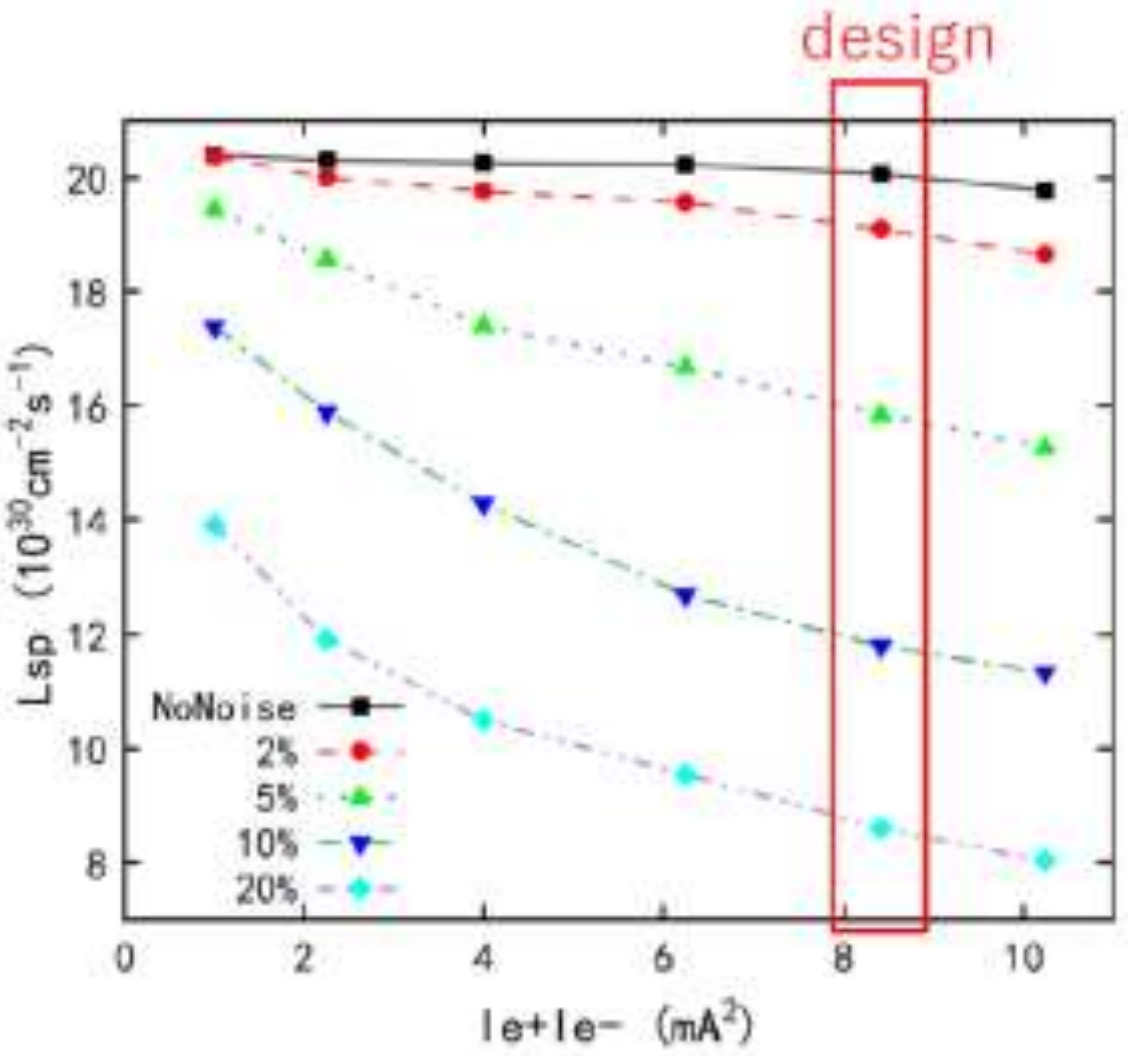

Figure 2.5-9: Simulated specific luminosity by the BBSS code for STCF, considering turn-by-turn noises in the vertical offsets of the beams at IP

- **Other collective effects**: Phenomena such as electron clouds and ion trapping may also pose potential performance limitations for the machine.

Recently, GPU-accelerated codes (e.g., BBSCL at KEK, APES-T at IHEP, Xsuite at CERN) have enabled more efficient simulation of coupled multi-physics effects. In the future, further development of STCF beam-beam studies will rely heavily on these advanced tools.

## 2.6 Impedance and Beam Collective Effects

### 2.6.1 Collective Effects

Beam collective effects encompass a variety of phenomena correlated with beam current. On one hand, they can degrade beam quality, such as by increasing emittance and energy spread; on the other hand, they can cause various instabilities that limit beam current and consequently reduce collision luminosity. Therefore, it is essential to carefully study the influence of these collective effects. The types of beam collective effects are diverse and include: intra-beam scattering (IBS) and Touschek scattering, which both are intra-scattering effects within the beam; various instabilities induced by the coupling of the beam with the impedance of the vacuum chamber environment, such as longitudinal microwave instabilities, transverse single-bunch instabilities, and coupled-bunch instabilities; instabilities caused by residual gas molecules, ions, and electron clouds within the vacuum chamber. To meet the high-luminosity requirements of a collider, it is necessary to reduce the beta function $\beta^*$ at the IP. This reduction implies the need for short bunch lengths and high total beam current, along with the additional requirement that the bunches remain stable, meaning that transverse and longitudinal oscillations are well suppressed. Therefore, for STCF, beam current limitation due to collective beam stabilities is a critical issue. During the design phase, it is crucial to carefully evaluate collective effects and propose effective mitigation strategies to ensure the stable operation of the STCF at high beam currents, thereby achieving the desired machine performance.

### 2.6.2 Impedance-Driven Single-Bunch Collective Effects

Single-bunch collective effects [28] are predominantly governed by the broadband impedance of the storage ring (from flanges, BPMs, bellows, collimators, etc.) and include bunch lengthening, microwave instabilities, and transverse mode coupling instabilities (TMCI).

To suppress these effects, strict control of broadband impedance is required. During engineering implementation, this requires optimization of the vacuum chamber structure through streamlines geometries and minimized discontinuities, combined with rigorous quality control in fabrication to mitigate impedance contributions – particularly from crucial vacuum components. Additionally, it is important to avoid structures that could trap higher-order modes, as they can lead to parasitic energy loss, local heating of vacuum chambers, and coupled-bunch instabilities.

When the single-bunch current remains below the threshold of microwave instability, the bunch lengthening is dominated by potential well distortion and can be described by:

$$\left(\frac{\sigma_z}{\sigma_{z0}}\right)^3 - \frac{\sigma_z}{\sigma_{z0}} = -\frac{cI_b}{4\sqrt{\pi}\alpha_p\omega_0\sigma_{z0}\sigma_{\delta 0}^2 E_0/e} Im\left(\frac{Z_\parallel}{n}\right)_{eff} \quad (11)$$

where $\sigma_{z0}$ is the rms bunch length, $\sigma_{\delta 0}$ is the rms energy spread, $\alpha_p$ is the momentum compaction factor, $E_0$ is the beam energy, and $\mathrm{Im}\left(\frac{Z_\parallel}{n}\right)_{eff}$ is the effective longitudinal impedance. For the STCF design, with a current target of 2 A and 50% filling factor, the required single-bunch current target is 2.8 mA. By optimizing impedance control strategies and referencing successful practices from existing storage rings, an effective impedance of 0.2 Ω is expected. Under these conditions, the bunch length would increase from the natural value of 6.8 mm to 8.9 mm at an RF voltage of 3 MV.

Microwave instability, a typical longitudinal single-bunch instability, arises when the current exceeds a threshold. While it does not cause particle loss, it degrades luminosity by amplifying the energy spread and elongating bunch length. A conservative threshold estimate is provided by the Keil-Schnell-Boussard criterion:

$$I_{th}\left|\frac{Z_\parallel}{n}\right|_{eff} < \frac{(2\pi)^{\frac{3}{2}} E_0 \alpha_p \sigma_\delta^2 \sigma_z}{C} \quad (12)$$

where $C$ is the ring circumference. Assuming $|Z_L/n|$ is 0.2 Ω, the estimated threshold current is 1.06 mA. Since the Boussard formula tends to underestimate the actual threshold, a more precise analysis should use macro-particle tracking simulations after the impedance model is built. Here, a simplified broadband resonator model is assumed with $f_r = 20\,\mathrm{GHz}$, $Q = 1$, $R_s = 11.6\ \mathrm{k\Omega}$, corresponding to an effective impedance of 0.2 Ω. The tracking simulation results for bunch length and energy spread versus single-bunch current are shown in Fig. 2.6-1.

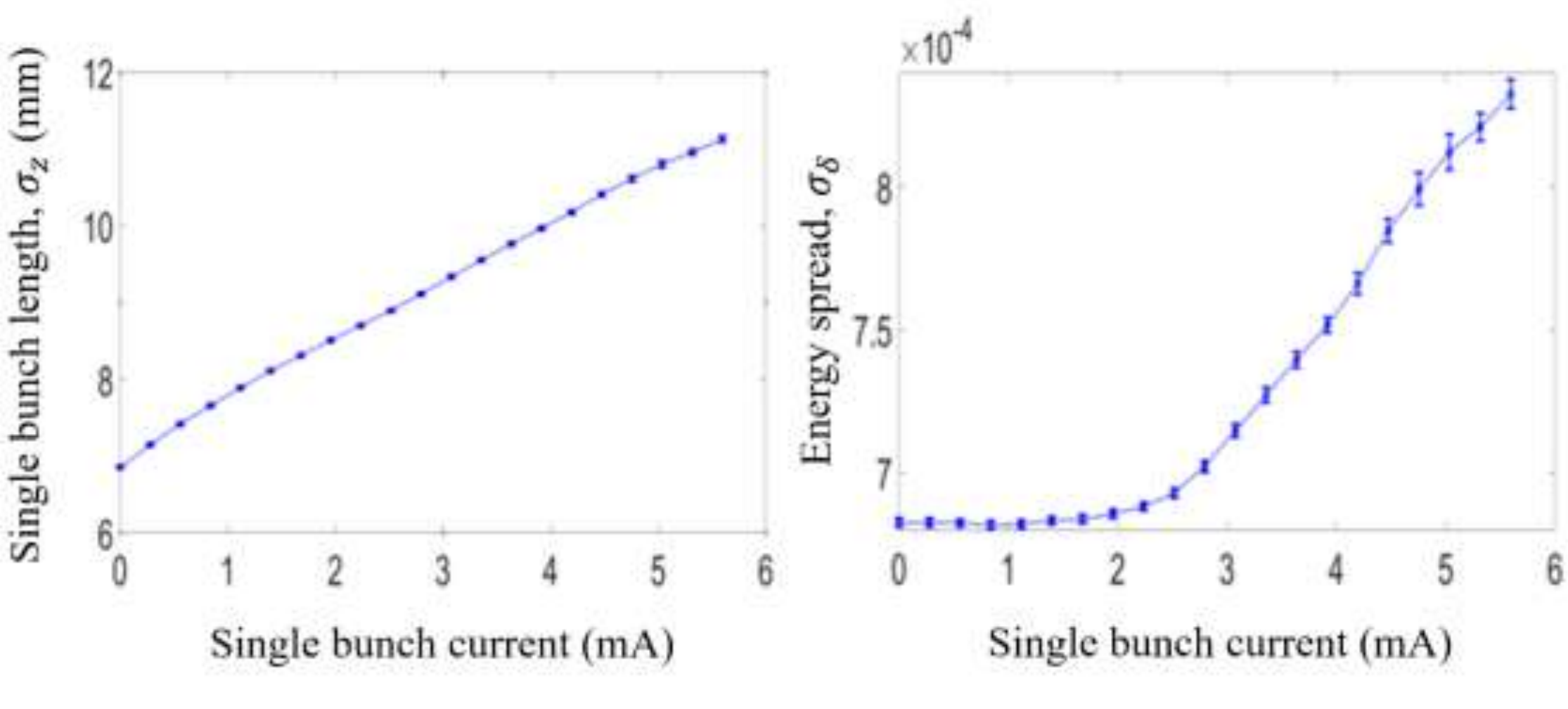

Figure 2.6-1: Variation of Single-Bunch RMS Length and Energy Spread with Bunch Charge

Simulations reveal the microwave instability threshold is about 2 mA, which is smaller than the STCF's target single-bunch current (2.8 mA). This proximity indicates potential operational risks. Increasing the momentum compaction factor $\alpha_p$ and the natural energy spread $\sigma_{\delta 0}$ and carefully managing the impedance budget in the design optimization can help raise the threshold.

Coherent synchrotron radiation (CSR) and coherent wiggler radiation (CWR) [29] can also drive microwave instability. Assuming a wiggler vacuum chamber with a gap height of 35 mm, a peak field of 1.6 T, a single wiggler length of 4.8 m, and a period length of 0.6 m, the simulations employing CSR impedance models obtained by the code CSRZ and macro-particle tracking yield a single-bunch instability threshold of about 0.6 mA at beam energy of 1 GeV, but the instability is weaker at 2 GeV with a threshold of 3 mA. Strategies to enhance the current threshold at 1 GeV include modifying the wiggler period length and optimizing the vacuum chamber geometry. One can also further optimize the lattice by increasing $\alpha_p$ and $\sigma_{\delta 0}$, as mentioned above. In addition, reducing the beam current in the operation to sacrifice the luminosity target at low energy is also an option if hardware modifications prove insufficient.

Head-tail instability and TMCI are critical transverse single-bunch instabilities. Negative natural chromaticity can trigger the 0-mode head-tail instability, which is conventionally suppressed by correcting chromaticity to positive values using sextupoles. While positive chromaticity may excite higher-order head-tail modes, these tend to grow slowly and can be suppressed by radiation damping and transverse feedback systems. At zero chromaticity, increasing beam current shifts the 0 mode toward the −1 mode, and once the threshold is crossed, the merging of the two modes will trigger bunch instability, which is the strongest of the TMCI, also called the strong head-tail instability. Once it occurs, the TMCI develops very fast and results in the rapid growth of transverse oscillation and particle loss. All efforts are needed to avoid it from happening.

Here we only discuss the transverse mode-coupling instability (TMCI) threshold under resistive wall impedance, assuming zero chromaticity, which is estimated using:

$$I_{th} = \frac{4\pi\nu_s(E_0/e)}{T_0 \sum_i \beta_{y,i}\, k_{y,i}} \tag{13}$$

where $I_{th}$ represents the threshold current, $\nu_s$ is the synchrotron tune, and $\beta_{y,i}$, $k_{y,i}$ are the vertical beta function and the transverse kick factor corresponding to the i$^{th}$ vacuum chamber element in the ring. Based on design experience from advanced colliders worldwide, collimators and interaction regions (IRs) are identified as the primary sources of transverse impedance. The former is due to their narrow apertures, and the latter is due to large beta functions. To ensure beam stability, in future design iterations, the total transverse impedance $\sum_i \beta_{yi} k_{yi}$ should be maintained below 78 kV/pC. For reference, at the low-energy collider ring of SuperKEKB, with a $\beta_y^*$ of 0.27 mm, the total $\sum_i \beta_{yi} k_{yi}$ is 138 kV/pC, and the collimators contribute more than half of this value. Thus, at STCF, the collimator design including apertures, locations and materials, will critically influence the impedance optimization.

To suppress the TMCI, correcting chromaticity to positive values can help to damp the head-tail 0-mode, though this has a risk of exciting higher-order head-tail modes. These higher-order modes, however, grow slowly and can be effectively suppressed by enhancing radiation damping through damping wigglers and implementing a transverse bunch-by-bunch feedback system, thereby raising the single-bunch transverse instability threshold current. In addition, the nonlinear collimation technique currently under development at SuperKEKB—designed to mitigate impedance effects—may offer further benefits. Its potential application to STCF will be explored.

### 2.6.3 Impedance-Related Coupled-Bunch Instability

This analysis focuses on the long-range wakefield effects, including narrowband impedance from RF cavities or vacuum structures, and low-frequency resistive wall impedance. Instabilities induced by the RF cavity modes have been already discussed in Section 2.4, and will not be repeated here. Geometric narrowband impedances, which can trap resonant modes, should be minimized through the optimization of the geometric structures of the devices.

Resistive wall instability mainly arises from the strong transverse impedance near zero frequency. These long-range wakefields can cause transverse coupled-bunch instability. For Gaussian beams, the instability growth rate is governed by [30]

$$\frac{1}{\tau_{\mathrm{n,m}}} = -\frac{1}{1+m}\frac{MI_b c}{4\pi\frac{E_0}{ev_\perp}}\frac{\sum_{p=-\infty}^{\infty} Re\, Z_\perp(\omega_{p,n,m})H_n(\omega_{p,n,m}-\omega_\xi)}{\sum_{p=-\infty}^{\infty} H_n(\omega_{p,n,m}-\omega_\xi)} \tag{14}$$

Here, the mode frequency is given by $\omega_{p,n,m} = (Mp+n)\omega_0 + \omega_\beta + m\omega_s$, where $M$ denotes the number of bunches, $n$ is the coupled-bunch mode number, $I_b$ is the single-bunch current, $\omega_\xi = \frac{\xi}{\alpha}\omega_0$ is the chromatic angular frequency, $v_\perp$ is the transverse betatron tune, $Z_\perp$ is the real part of the transverse resistive wall impedance, and the expression for the $H_n$ function is given in Equation (15).

$$H_n(\omega) = \frac{2\pi}{3\sqrt{\frac{\pi}{2}}\cdot\Gamma\left(n+\frac{1}{2}\right)}(\omega\sigma_t)^{2n}\, exp(-\omega^2\sigma_t^2) \tag{15}$$

Assuming a uniform aluminum beam pipe of 25 mm radius for the full ring circumference, Eq. (14) predicts an instability growth rate of 1.6 $ms^{-1}$ at zero chromaticity. Although modern bunch-by-bunch feedback techniques can achieve damping times of about 100 μs or better, the SuperKEKB experience [31] shows that the noise of the feedback system has an important impact to the luminosity stability. A slower feedback system combined with other measures such as correcting chromaticity to positive values to reduce the growth rate, are being considered. With these measures, STCF is expected to manage this instability effectively.

### 2.6.4 Beam Lifetime

The Touschek scattering effect is a dominant limitation for beam lifetime in the collider rings. It refers to large-angle Coulomb scattering between two electrons or positrons within a bunch, where, during collisions, one particle transfers its transverse momentum to the longitudinal one of another particle. Thus, some particles will potentially exceed the energy acceptance and will be lost. From the local momentum aperture (LMA) shown in Fig. 2.3-4, the calculated beam lifetime at 2 GeV is approximately 252 s.

Residual gas scattering includes elastic scattering and inelastic scattering. Elastic scattering leads to transverse oscillations, and the particle will be lost if its amplitude exceeds the dynamic aperture. Inelastic scattering causes the deceleration and energy deviation of the particle, and it will also lead to a loss if it is beyond the momentum acceptance. Based on the SuperKEKB experience, elastic scattering is the dominant factor for vacuum-related beam losses in this kind of the collider rings.

Assuming all residual gas molecules are CO, the vacuum-limited lifetime is:

$$\tau_C[\text{hrs}] = 11.1 \times \frac{E_0^2[\text{GeV}^2]\epsilon_A[\text{mm} \cdot \text{mrad}]}{\langle\beta_\perp[m]\rangle P[\text{nTorr}]} \tag{16}$$

$$\epsilon_A \equiv \min \frac{A^2}{\beta_\perp(s)} \tag{17}$$

where $A$ is the local physical aperture, "min" for taking the minimum over the whole ring. Assuming $\epsilon_A = 0.1$ μm·rad, to achieve a vacuum lifetime longer than 40 minutes, the vacuum system must maintain an average pressure below $1 \times 10^{-7}$ Pa.

### 2.6.5 Ion Effects and Electron Cloud Effects

Ion instabilities originate from the interaction between the ions trapped by the electron beam and the beam itself. The ions are generated from the ionization of residual gas in the vacuum and the desorption from the vacuum wall. While in modern electron storage rings, ultra-high vacuum systems can provide a pressure lower than $10^{-7}$ Pa, and can generally minimize these effects. During the design of the bunch filling scheme, it is usually required to introduce some large gaps between the bunch trains, which will help suppress the ion instability. In addition, transverse bunch-by-bunch feedback systems can effectively suppress fast ion-caused oscillations. Transient fast ion instabilities may still occur during the early machine operation under poorer vacuum conditions, but will diminish as the vacuum improves.

In the positron ring, the photons from synchrotron radiation strike the vacuum chamber wall and release photoelectrons, which are accelerated by the positron beam and can be multiplied with multiple reflections between the walls. The electrons captured by the positron beam form the so-called electron cloud [32]. The electron cloud interacts with the positron bunches, and can couple the motions of successive positron bunches and cause both the coupled-bunch instability and the single bunch head-tail instability. Using the codes PEI and PyECLOUD for simulations, at 2 GeV with an average chamber diameter of 50 mm and a secondary electron yield (SEY) of 1.3, the cloud density near the beam is $\rho_e$=$10^{13}$ m$^{-3}$ without applying additional mitigation measures. The density can be reduced to $5\times10^{12}$ m$^{-3}$ with antechambers. Using the code PEHTS, the single-bunch instability threshold is predicted to be $\rho_{e,th}$=1.4 $\times10^{12}$ m$^{-3}$. Thus, the vacuum system design should aim to keep $\rho_e$ below $10^{12}$ m$^{-3}$. Multiple mitigation measures to suppress the instability should be considered. This includes: antechambers to localize photon impact, TiN/NEG coatings to reduce SEY, chamber grooves, clearing electrodes, and external magnetic fields to disrupt electron accumulation, and bunch-by-bunch feedback for dynamic stabilization. Based on successful experience from other similar machines, these combined measures are expected to effectively suppress the electron cloud effect in STCF as well.

## 2.7 Beam Injection and Extraction

The primary objective of the beam injection design for the collider rings is to inject the positron/electron beams provided by the upstream injector into the collider rings with the highest possible efficiency. This ensures high integrated luminosity, minimizes disturbance to the circulating beam during injection, and reduces injection-related beam loss and its impact on detector background. The primary goal of the beam extraction design is to extract the high-energy, high-charge positron/electron bunch trains from the ring efficiently and transport them to the beam dump.

There are different injection methods are being used or studied for electron storage rings, the conventional off-axis injection with several variations, longitudinal injection, swap-out injection, etc. Considering that the momentum aperture of the lattice design for the STCF collider is quite small, and that STCF is not suitable for using a lower RF frequency than 500 MHz to increase the bucket length, longitudinal injection and off-momentum injection seem not be a good option. Currently, considering the design of the upstream injector and the overall cost of the accelerator, STCF will adopt a compatible injection scheme combining off-axis injection and bunch swap-out injection. The off-axis injection scheme [33] is chosen as the baseline, and will be employed in the construction. Different off-axis injection methods such as employing bump kickers, non-linear kickers, and anti-septum, are being investigated, but the one with bump kickers is described here. Thus, the injection system consists of a septum magnet and bump magnets. The injected beam and the circulating beam are transversely separated. Under the action of the bump magnets, the closed orbit of the circulating beam moves closer to the septum magnet, and after one or several turns—once the injected beam falls within the collider ring acceptance—it returns to the pipe center. Eventually, the injected beam merges with the circulating beam via synchrotron radiation damping. This scheme is mature and imposes relatively low requirements on the risetime of the injection elements but demands a large dynamic aperture.

In the future, depending on need, the system can be upgraded to a bunch swap-out injection scheme [34], while keeping use of the same injection section. With this scheme, circulating bunches are directly replaced one by one by newly injected bunches using ultra-fast pulsed kicker magnets. This scheme requires less dynamic aperture in the horizontal plane but imposes strict constraints on kicker pulse width—it must be less than twice the minimum bunch spacing. Additionally, injecting high-charge positron bunches places stringent requirements on the repetition rate of the upstream injector. To facilitate future upgrades, the quadrupole magnets in the injection section will be laid out in a unified design, requiring only the replacement of kicker components for the transition. Hardware parameters will be selected accordingly, and the injection process will be validated through tracking simulations.

The extraction system design for the collider rings also includes two operational scenarios. During bunch swap-outs, small angular deviations induced by ultra-fast pulsed kicker magnets are used, and the downstream septum magnet deflects the target bunch out of the ring. This method demands very high performance of the ultra-fast kickers, similar to the injection, requiring a pulse with a full width shorter than 6 ns. For beam dumping, the different kicker must have a rise time shorter than the gap between bunch trains and a flat-top width longer than the revolution period, enabling one-turn extraction of the entire bunches.

The luminosity fluctuation requirement forms the basis of the beam injection design. For example, assuming a 7% degradation in single-bunch luminosity and a typical beam lifetime of 250 s, the physical requirements for beam injection in the off-axis and swap-out injection schemes are summarized in Table 2.7-1. The goal of injection design is to meet the parameters listed in the table, particularly the injection efficiency.

Table 2.7-1: Beam Injection Requirements for the Collider Rings

| **Injection Scheme** | **Off-axis Injection** | **Swap-out Injection** |
|---|---|---|
| Collider Ring Circumference [m] | 860.321 | 860.321 |
| Bunch Spacing [ns] | 4 | 4 |
| Min. Single-Bunch Luminosity / Peak | 93% | 93% |
| Beam Lifetime [s] | ~250 | ~250 |
| Time to Complete Full Ring Injection [s] | 18.14 | 18.14 |
| Circulating Bunch Charge [nC] | 8.34 | 8.34 |
| Circulating Beam Emittance [rms, nm·rad] | 4.16 (0.5% coupling) | 4.16 (0.5% coupling) |
| Injected Bunch Charge [nC] | 1.0 (*) | 8.5 |
| Charge to Refill/Replace [nC] | 0.6 | 8.34 |
| Injected Beam Emittance [rms, nm·rad] | H: < 6; V: < 2 | H: < 30; V: < 10 |
| Injection Efficiency | > 60% | > 98% |

(*) For off-axis injection, the positron/electron injectors provide 1.0 nC per bunch at a repetition rate of 30 Hz.

### 2.7.1 Off-axis Injection

The key to the off-axis injection design is to utilize local orbit bumps to steer the injected beam into the ring acceptance while minimizing the transverse offset from the central orbit and reducing beam losses at the septum magnet [35]. First, the optics of the injection section must be configured with large β functions to minimize the effect of septum thickness or closed orbit errors on injection efficiency. Second, the beam clearance aperture must be optimized to balance between dynamic aperture and beam losses—a too-large clearance demands a very large dynamic aperture, while a too-small clearance leads to significant beam loss. Lastly, the design of the local bump orbit must carefully choose bump height and angle to bring the injected beam at the septum magnet exit as close as possible to the pipe center without causing excessive loss of the injected beam and by controlling the influence of the stray magnetic field of the septum magnet to the circulating beam.

The off-axis injection system in the STCF collider rings consists of three parts: a set of quadrupoles for optics matching, a group of bump magnets to form the local orbit, and septum magnets for large-angle bending of the injected beam. Table 2.7-2 lists the key parameters. Figure 2.7-1 shows a schematic of the off-axis injection scheme and its parameters. Figure 2.7-2 presents the layout and optics of the injection section that is located in one of the long straight sections in the collider rings, where 10 quadrupoles for optics matching are installed over a long straight section of 20 m in length. Figure 2.7-3 illustrates the local bump orbit, which is formed using four bump magnets. Each magnet provides a maximum bending angle of less than 3.5 mrad, and can be varied independently to change the orbit height and angle.

Table 2.7-2: Off-Axis Injection Design Parameters

| Parameter | Value |
|---|---|
| Circulating Beam Clearance [mm] | 1.68 (4.5σ) |
| Injected Beam Clearance [mm] | 1.18 (4.5σ) |
| βx at Injection Point (inner/outer ring) [m] | 33.34 / 11.53 |
| Injection Angle [mrad] | -2.45 |
| Orbit Bump Height at Injection Point [mm] | 3.86 |
| Orbit Bump Angle at Injection Point [mrad] | -2.50 |
| Position After Bump Recedes [mm] | 4.36 |
| Bump Duration [turns] | 1 |
| Bump Magnet Length [mm] | 350 |
| Max Bump Magnet Deflection Angle [mrad] | < 3.5 |
| Bump Magnet Field Strength [Gs] | < 1167 |
| Thin Septum Height [mm] | 5.54 |
| Thin Septum Thickness [mm] | 1.0 |
| Thin Septum Length (mechanical/effective) [m] | 0.65 / 0.6 |
| Thin Septum Field Strength [Gs] | 5500 |
| Thin Septum Deflection Angle [mrad] | 28 |
| Thick Septum Thickness [mm] | 2.0 |
| Thick Septum Length (mechanical/effective) [m] | 1.7 / 1.3 |
| Thick Septum Field Strength [Gs] | 9900 |
| Thick Septum Deflection Angle [mrad] | 110 |

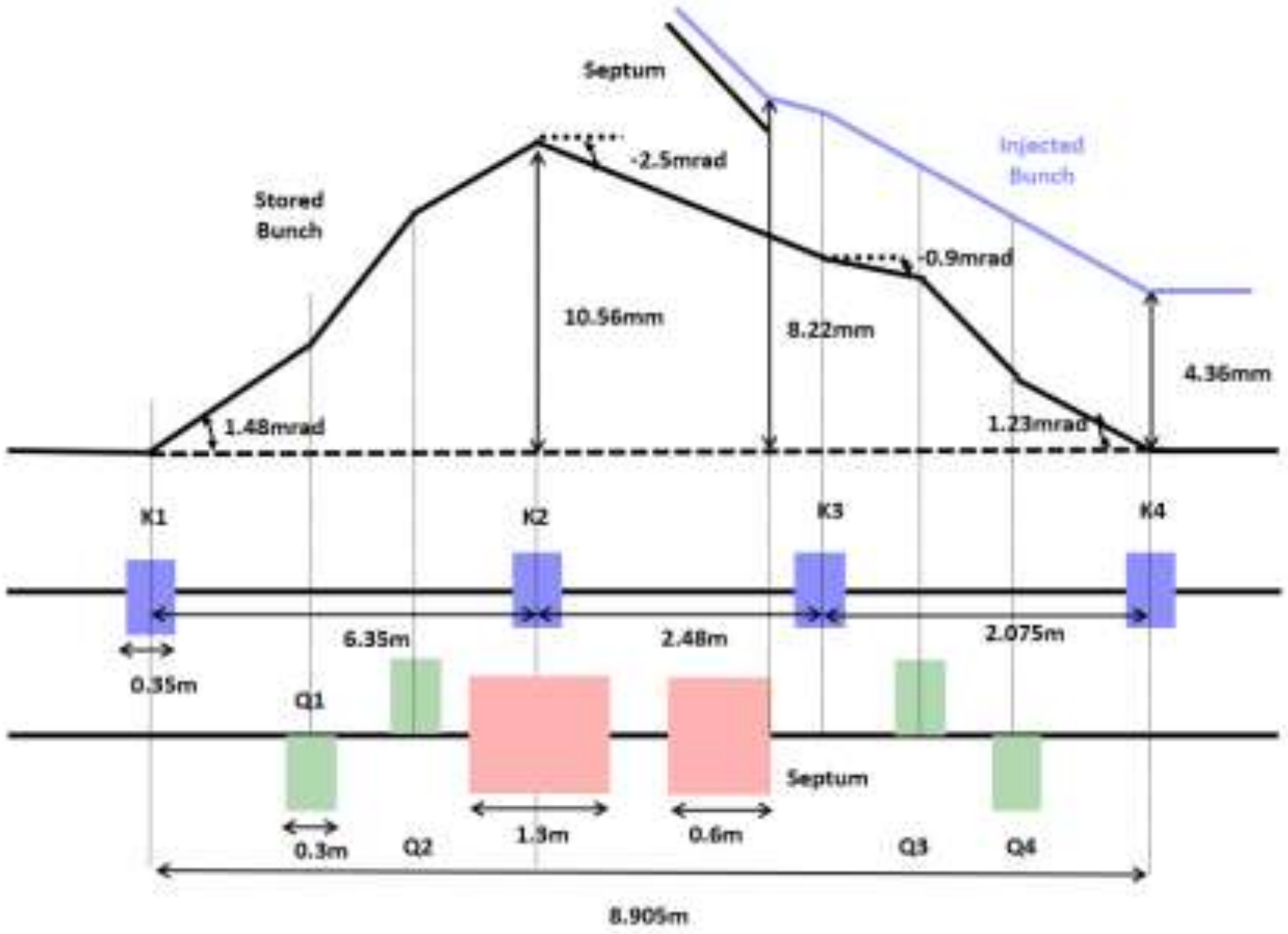

Figure 2.7-1: Schematic Layout of the STCF Off-Axis Injection Design

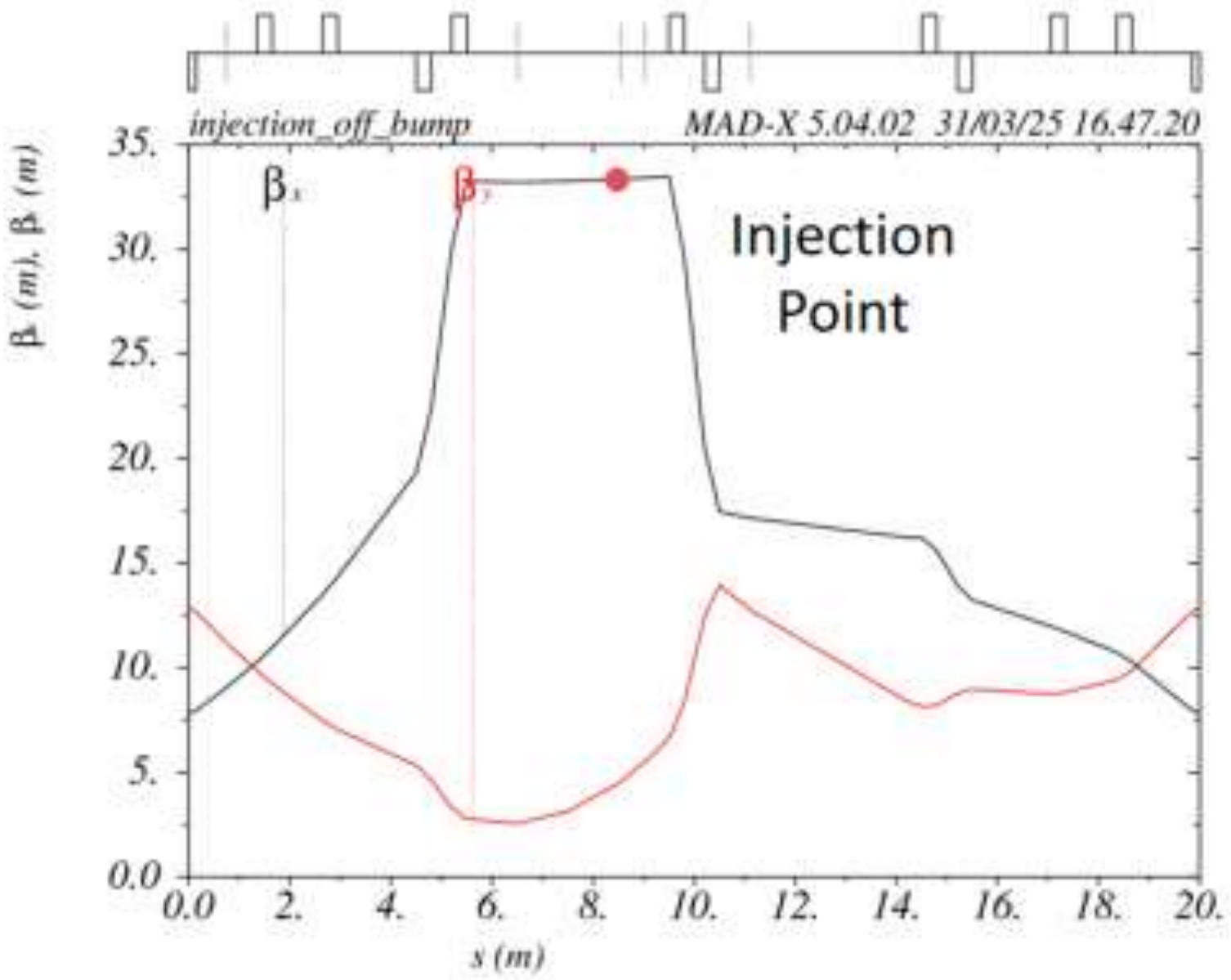

Figure 2.7-2: Injection Section Element Layout and Optics for the Off-Axis Scheme

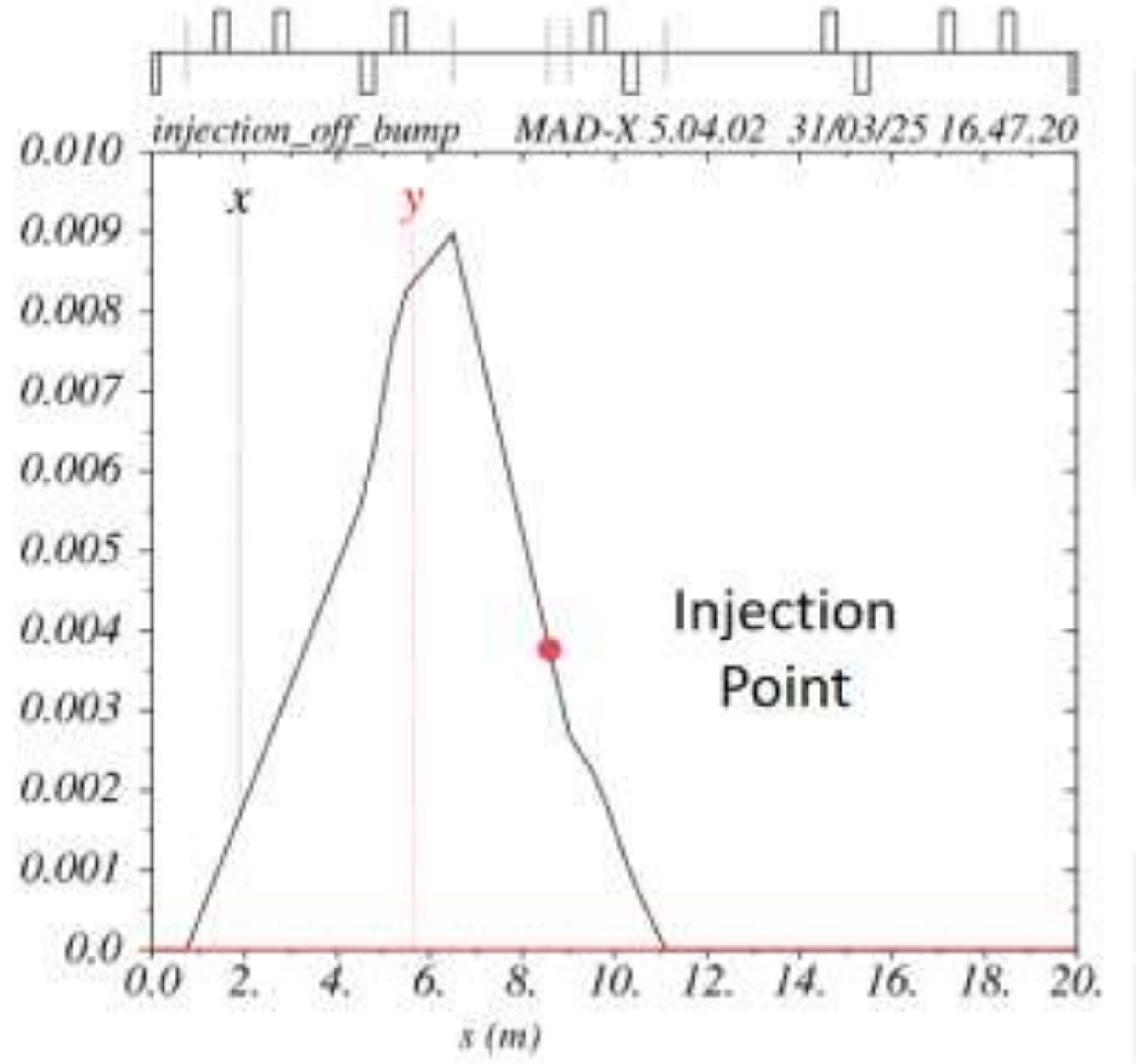

Figure 2.7-3: Local Orbit Bump for Off-Axis Injection Scheme

Figure 2.7-4 shows the evolution of horizontal geometric emittance (phase space area) of the injected bunch over five damping times as simulated by the code Elegant. Initially, the filamentation of the injected bunch dominates the emittance growth. Then, synchrotron radiation damps the off-axis oscillation of the injected beam exponentially. After five damping times, the emittance is reduced by more than an order of magnitude, merging completely with the stored bunch and reaching an equilibrium emittance of approximately 4.17 nm·rad. Considering the physical aperture of the septum and the dynamic aperture of the collider rings, the final injection efficiency reaches about 90%, which meets the design goals. The injection simulations with all the errors, which ensure the robustness of the injection physics design, are underway. More simulations with all kinds of errors and even with the combined beam-beam effect for merging bunches will be carried out.

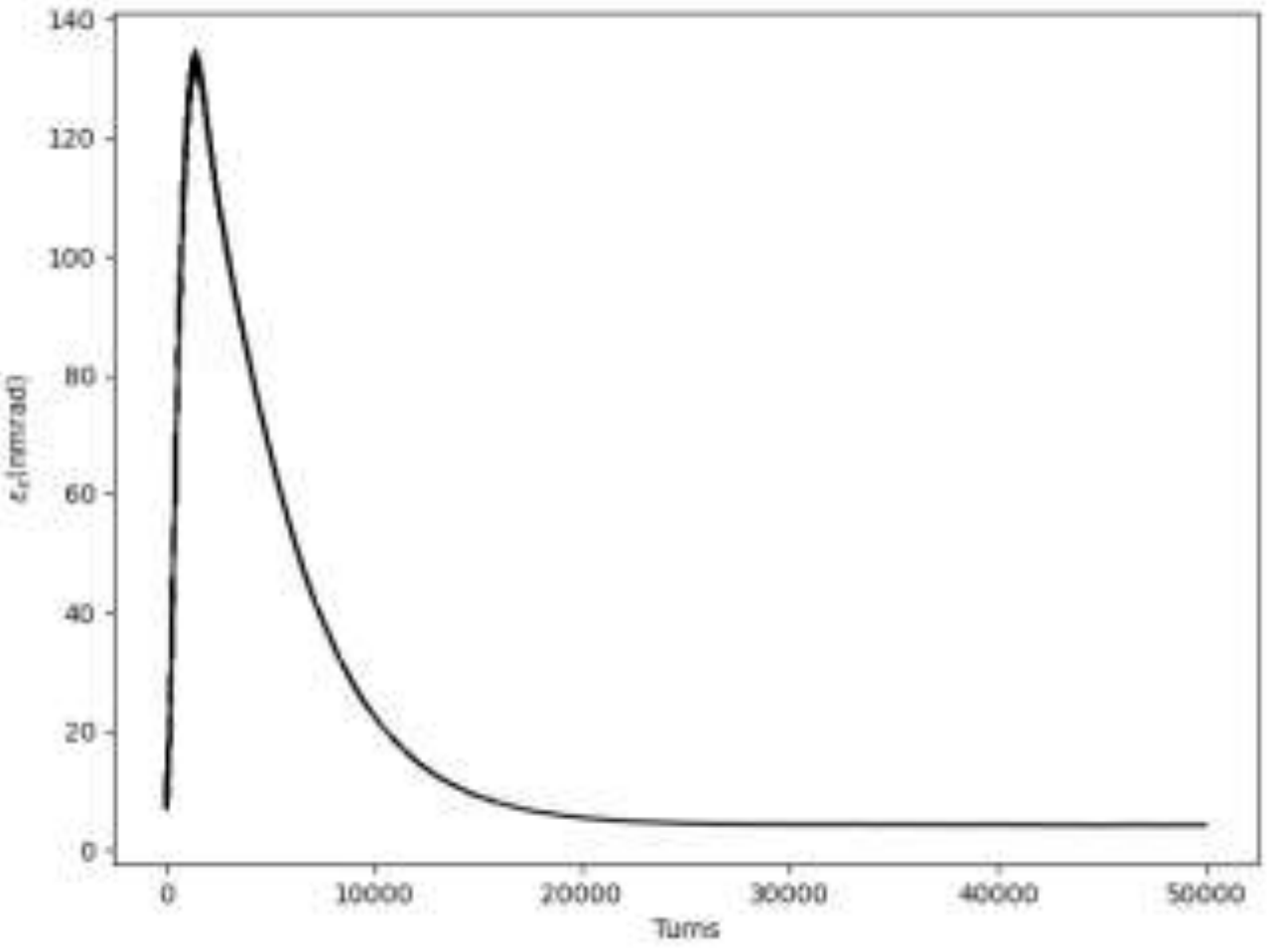

Figure 2.7-4: Evolution of Horizontal Emittance Oscillation for the Injected Bunch under the Off-Axis Scheme

### 2.7.2 Bunch swap-out Injection

The key to the physics design of the bunch swap-out injection scheme lies in arranging the components in a geometrically rational layout to ensure the injected bunch enters the closed orbit of the collider rings accurately, minimizing disturbance to the stored beam and reducing the required deflection angle of the kicker magnets [36]. The STCF bunch swap-out injection system primarily consists of a septum magnet for large-angle deflection and a set of ultra-fast horizontal kicker magnets. The septum magnet design is identical to that of the off-axis injection scheme. The kicker system comprises five identical modules spanning a 1.9-meter-long drift.

Table 2.7-3 lists the relevant design parameters. Figure 2.7-5 shows a schematic of the injection system. Figure 2.7-6 presents the layout and optics of the injection section, which is the same region for the off-axis injection spanning 20 meters and incorporates ten quadrupoles. To accommodate the deflection angle and good field region of the kicker magnets, about 6 meters of drift space is provided for the injected beam to travel from the septum magnet to the kickers. Figure 2.7-7 shows the beam trajectory and envelope for this region.

Table 2.7-3: Bunch Swap-out Injection Design Parameters

| Parameter | Value |
|---|---|
| Kicker Magnet Length [mm] | 300 |
| Kicker Magnet Deflection Angle [mrad] | 0.5 |
| Kicker Plate Gap [mm] | 12 |
| Number of Modules | 5 |
| Rise / Flat-top / Fall Time [ns] | 2 / 2 / 2 |
| Good Field Region [mm] | ±5 |
| Injection Point Position at Kicker Entry [mm] | 2.45 |
| Voltage Amplitude [kV] | > 17.5 |
| Injection Repetition Rate [Hz] | 30 |
| Thin Septum Height [mm] | 5.54 |
| Thin Septum Thickness [mm] | 1.0 |
| Thin Septum Magnet Length (Mechanical/Effective) [m] | 0.65 / 0.6 |
| Thin Septum Magnet Field Strength [Gs] | 5500 |
| Thin Septum Magnet Deflection Angle [mrad] | 28 |
| Thick Septum Thickness [mm] | 2.0 |
| Thick Septum Magnet Length (Mechanical/Effective) [m] | 1.7 / 1.3 |
| Thick Septum Magnet Field Strength [Gs] | 9900 |
| Thick Septum Magnet Deflection Angle [mrad] | 110 |

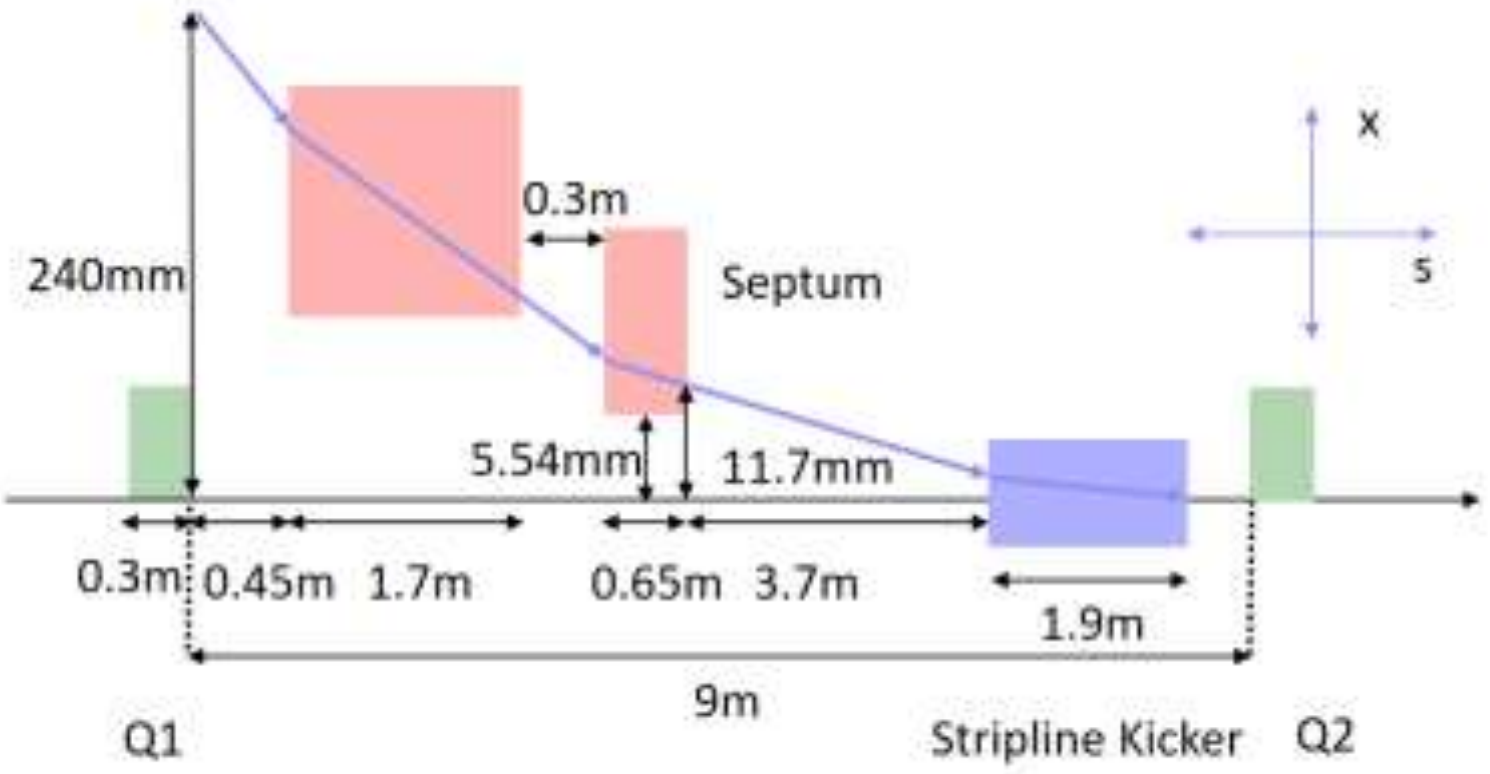

Figure 2.7-5: Schematic Layout of the STCF Bunch Swap-out Injection Design

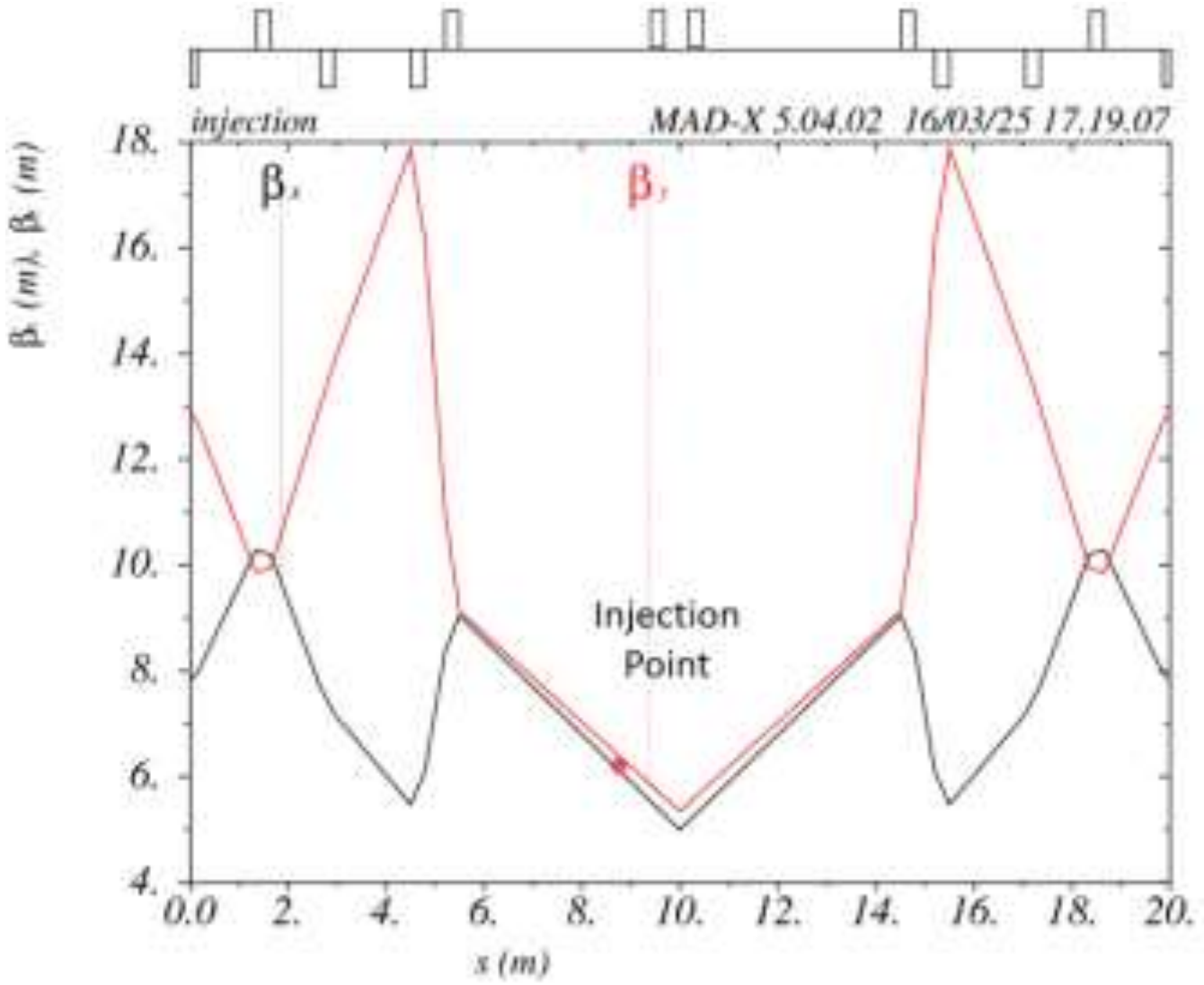

Figure 2.7-6: Layout and Optics of the Injection Section for the Bunch Swap-out Scheme

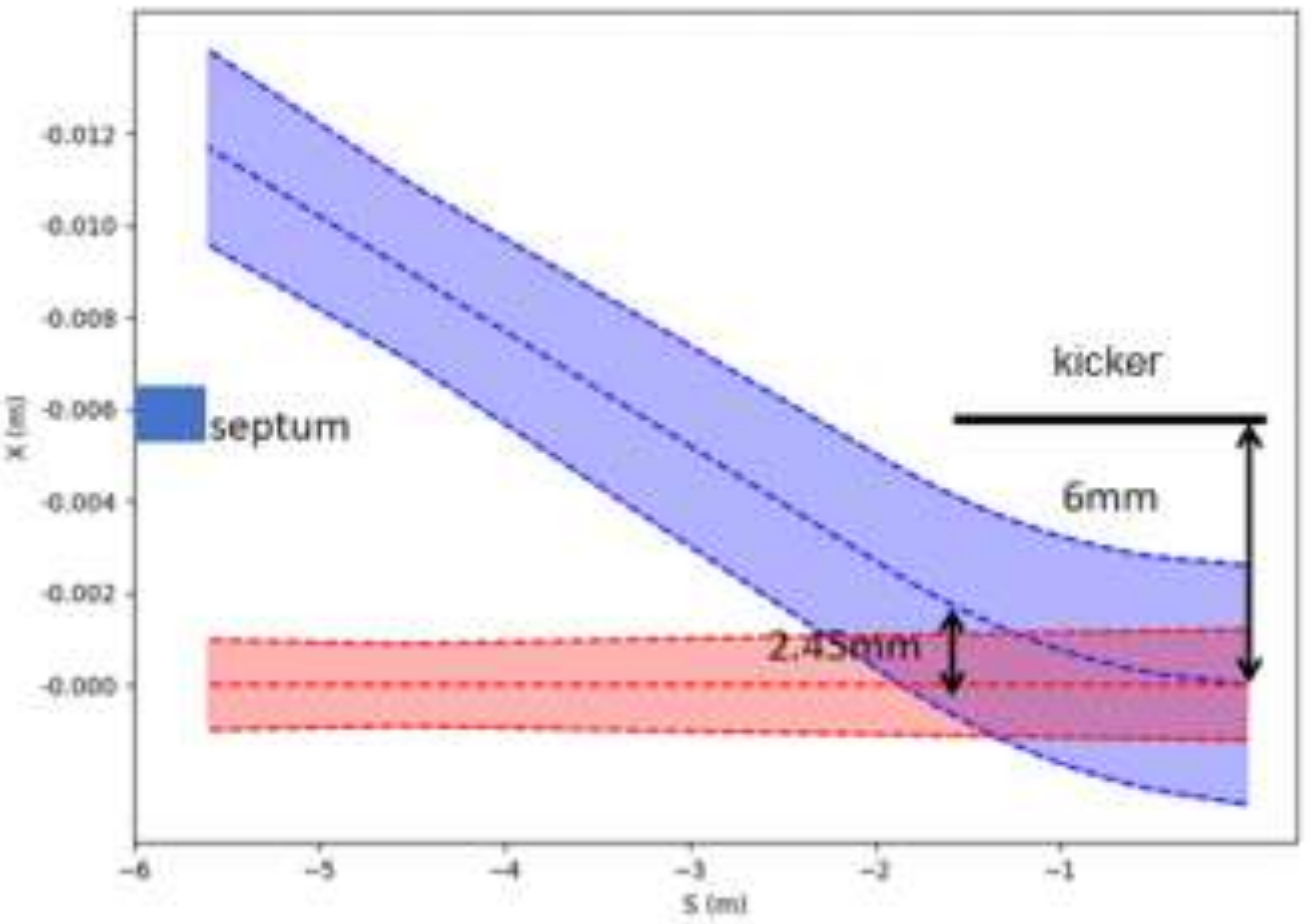

Figure 2.7-7: Beam Trajectory and Envelope from Septum Exit to Kicker Exit for the Bunch Swap-out Scheme

In the bunch swap-out injection scheme, the injected bunch has a relatively large emittance due to a very large bunch charge from the injector. Whether the injected beam can avoid major losses in the first few turns and be damped by synchrotron radiation to the equilibrium emittance of the collider rings is crucial for injection efficiency and final luminosity. Figure 2.7-8 shows the evolution of the horizontal beam emittance over five damping times, as simulated using the Elegant tracking code. After five damping times, the emittance converges to approximately 4.20 nm·rad, and the injection efficiency exceeds 98%, barely meeting the basic physical design requirements. In the future, smaller injection emittance will be applied to reduce the loss rate further.

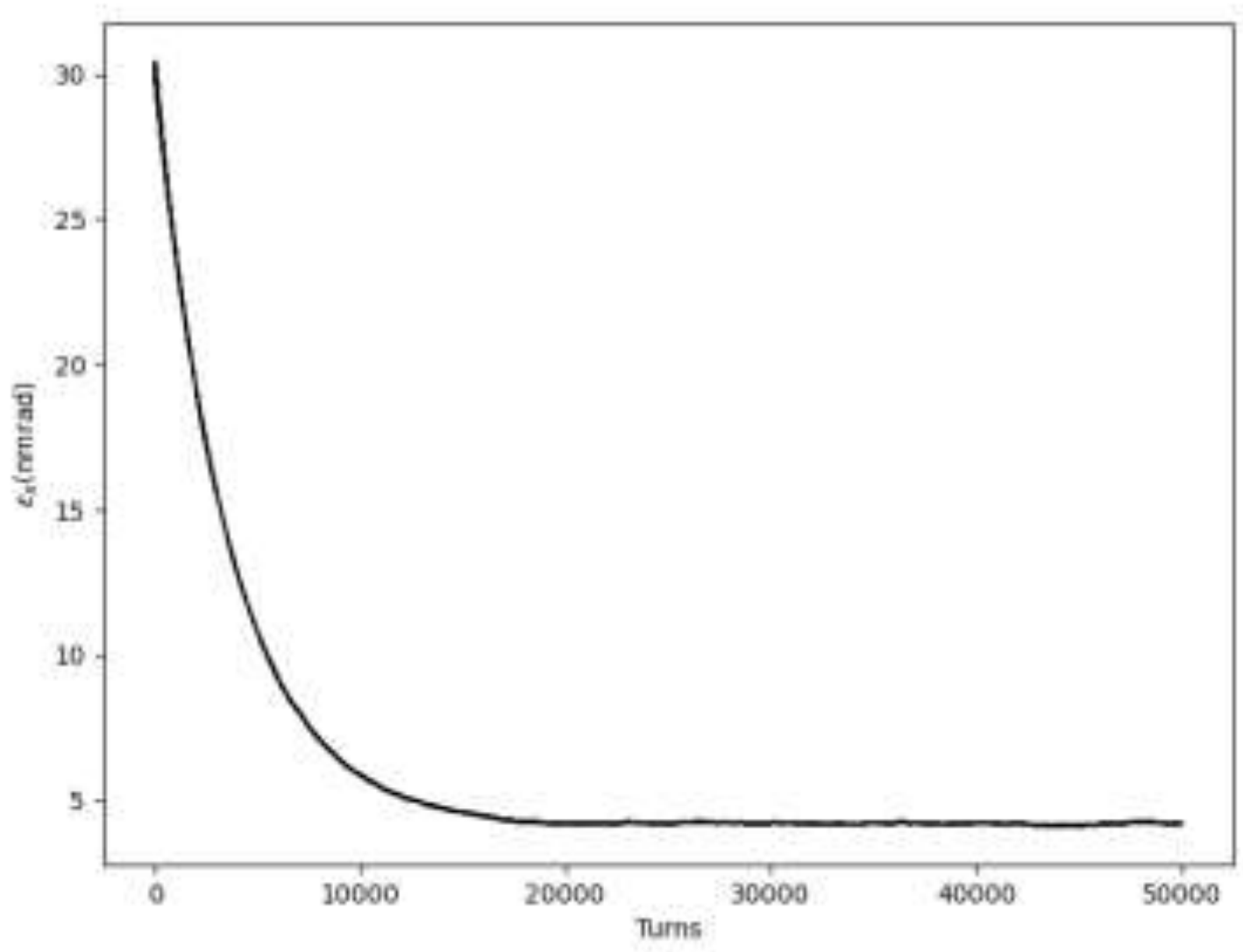

Figure 2.7-8: Evolution of Horizontal Emittance Oscillation for the Injected Bunch under the Bunch Swap-out Scheme

### 2.7.3 Beam Dumping and Extraction

Beam extraction can be considered the reverse process of bunch swap-out injection, which is accomplished by a set of pulsed kicker magnets and a downstream septum magnet. STCF requires two extraction systems to fulfill the following two beam extraction functions:

The first is during normal operation under the bunch swap-out injection scheme. Ultra-fast kicker magnets deflect the bunch to be replaced away from the equilibrium orbit, and the downstream septum magnet subsequently extracts it from the ring. Only one bunch is extracted every 33.3 ms to enable on-axis replacement. The hardware requirements are identical to those for swap-out injection, as shown in Table 2.7-3. The extraction section layout is also similar to Figure 2.7-6, except that the septum magnet is placed downstream in the beam direction. The main requirement for the extraction system in this case is to minimize disturbance to the circulating beam caused by the kicker magnets.

The second is for beam dumping during machine protection or experimental shutdown. This requires kicker magnets with a rise time shorter than the bunch train spacing and a flat-top width longer than the revolution period, so the entire ring can be emptied in a single extraction.

This process does not affect luminosity or experiments, and the design is relatively straightforward, with ample reference designs available. Therefore, it is not a technical challenge for the STCF beam injection/extraction design. However, considering the large stored beam energy in the collider rings, minimizing beam loss during extraction is critical. Additionally, kicker magnets with both short rise time and long flat-top width present certain technical challenges. A dedicated beam dump will be located outside the tunnel to receive the high-energy, high-charge (approximately 5.7 μC) dumped beam. Fig2.7-9 shows the beam orbits in the two extraction modes.

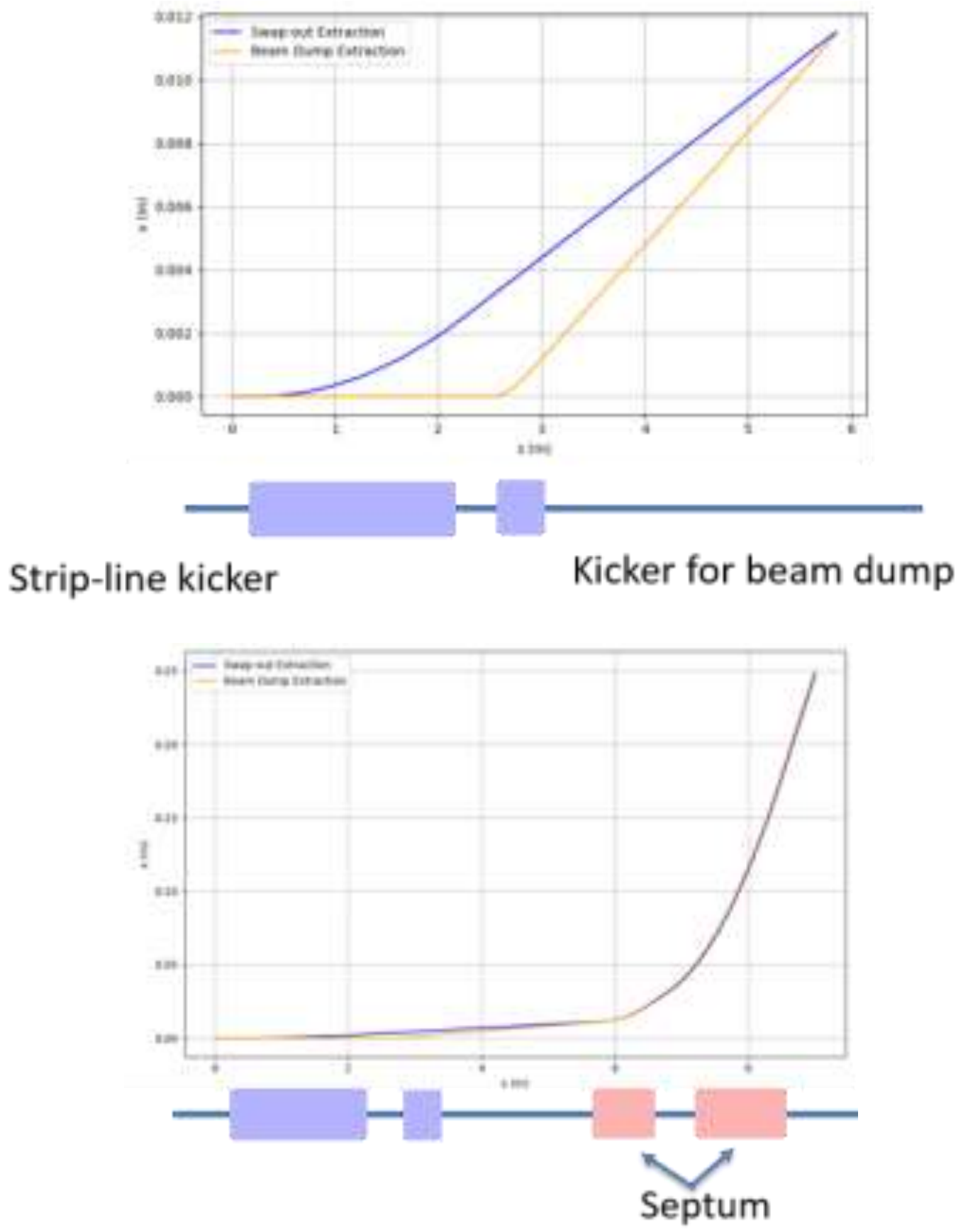

Figure 2.7-9: Schematic of the beam extraction (Upper: extraction trajectories for swap-out and for dumping; Lower: layout of the extraction section)

The off-axis injection scheme, widely used in positron/electron storage rings over the past decades, has proven effective and reliable in numerous experiments. Hence, it is entirely feasible to adopt the septum-and-bump-based off-axis injection as the baseline scheme for STCF.

The bunch swap-out injection scheme, which is considered ideal for some fourth-generation light sources and new-generation e+e− colliders, has gained much attention in recent years. Its main challenge lies in developing kicker magnets with ultra-short pulses and ultra-fast rise times. Preliminary numerical simulations confirm the feasibility of this scheme, with injection efficiency reaching as high as 98%. Therefore, the bunch swap-out injection scheme—based on fast pulsed kickers and septum magnets—will serve as the upgrade path for STCF. Since it is generally compatible with the off-axis scheme, especially in terms of injector requirements, it is advisable to adopt the off-axis method in the initial stage. After upgrading to the bunch swap-out injection scheme, the off-axis injection may still serve as a commissioning and operational fallback, ensuring the feasibility of maintaining peak luminosity via the top-up injection mode at STCF.

## 2.8 Error Analysis and Correction

In real operation, the collider rings inevitably deviate from the ideal design model due to various sources of errors, including alignment deviations caused by magnet installation and field errors from magnet manufacturing. These errors are primary contributors to closed orbit distortion, optical function perturbations, and transverse coupling. They significantly impact crucial machine performance metrics such as beam size, dynamic aperture, beam lifetime, and collision luminosity, and also affect the complexity of commissioning and operation of the collider rings. To assess the robustness of the magnetic focusing structure of the collider rings and to restore their nonlinear dynamic performance and luminosity, this section defines tolerances for static errors. Based on these tolerance levels, the error sensitivity of the collider ring lattice is analyzed, and corresponding corrections are applied to the orbit and optical functions.

Following the error settings of similar international facilities such as SuperKEKB [37], CEPC [12], and FCC-ee [38], the reference error levels are set as follows: magnet alignment error of 50 μm, angular misalignment of 0.1 mrad, and main field error of 0.02%. Under these assumptions, simulations of the closed orbit survival ratio of the STCF collider rings were performed to evaluate its error tolerance, as shown in the left panel of Figure 2.8-1, where the horizontal axis indicates the scaling factor of the error relative to the reference level. At the reference error level, the closed orbit survival ratio is approximately 25.2%; when the error level is doubled, the survival ratio drops sharply to only 1.7%.

Further investigation into individual error sources reveals that the transverse misalignment of quadrupole magnets has the most significant impact on the orbit survival ratio, especially in the vertical direction. Among all quadrupoles, those in the FFT region near the IP are the most sensitive and thus require higher alignment precision. The right panel of Figure 2.8-1 shows the impact on the orbit survival ratio when different alignment errors are introduced only to the doublet quadrupoles on both sides of the IP. The results indicate a noticeable drop in the survival ratio when the alignment error exceeds 20-30 μm. To reduce the orbit distortion and suppress the luminosity loss, the misalignments of the final doublet quadrupoles should controlled within 30 μm. Additionally, due to the high strength of sextupole magnets, their misalignments significantly affect the orbit, optics, and coupling. Therefore, the STCF collider ring design incorporates movers for the sextupole magnets to achieve precision alignment at the 10 μm level. Taking into account both STCF's design requirements and practical engineering capabilities, Table 2.8-1 summarizes the reference error values used for the current error analysis of the collider rings. The error levels are truncated at 3σ.

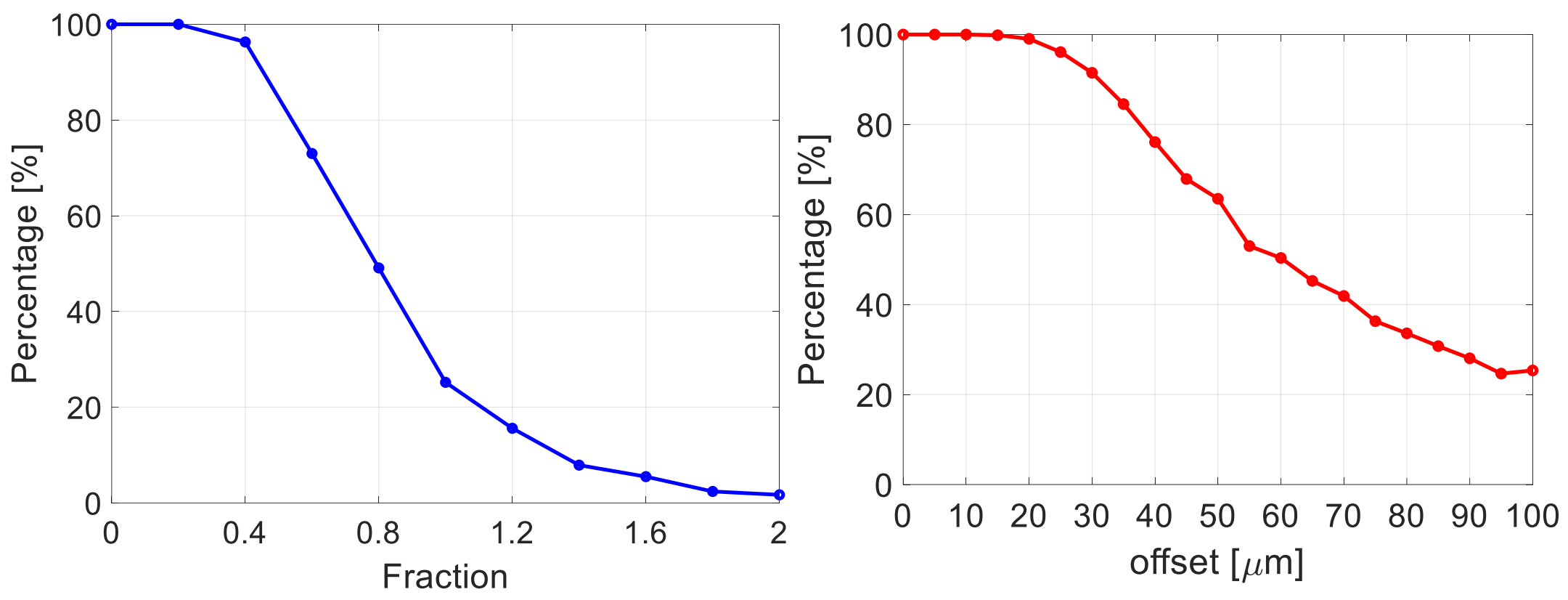

Figure 2.8-1: Closed orbit survival ratio of the STCF collider rings under different error levels (before correction)

Table 2.8-1: Reference Alignment Error Values (rms) for the STCF Collider Rings

| Component | Horizontal (μm) | Vertical (μm) | Longitudinal (μm) | Rotation (mrad) | Main Field Error |
|---|---|---|---|---|---|
| Dipole Magnet | 100 | 100 | 100 | 0.1 | 0.02% |
| Quadrupole Magnet | 50 | 50 | 100 | 0.1 | 0.02% |
| FFT Doublet | 30 | 30 | 100 | 0.1 | 0.02% |
| Arc/IR Sextupole | 50 / 30 | 50 / 30 | 100 | 0.1 | 0.02% |

Under ideal conditions, particles in the collider rings undergo $\beta$ oscillations around the ideal closed orbit. However, field errors in dipole magnets and alignment errors in quadrupole magnets distort the closed orbit, with the distortion magnitude depending on the size and distribution of field errors and the β value at the observation point. Orbit distortion degrades the dynamic aperture and alters the beam spot size and position, particularly near the IP, where orbit offset and beam size growth reduce luminosity. Therefore, correction is essential. Simulation results show that to limit luminosity loss to less than 10%, the horizontal and vertical orbit deviations at the IP must not exceed $12\sigma_x$ and 0.65 $\sigma_y$, respectively, where $\sigma_x$ and $\sigma_y$ are the nominal beam sizes at the IP. Field errors in quadrupole magnets, alignment errors in sextupoles, and magnet rotation errors can perturb optical functions. Calculations indicate that to limit luminosity loss to 10%, the increase in $\beta_y$ at the IP must be less than 30%, whereas $\beta_x$ is more tolerant. Rotational errors in dipoles and quadrupoles can induce transverse coupling and vertical dispersion; to keep luminosity loss below 10%, the vertical dispersion at the IP must be less than 0.1 mm. These effects not only reduce luminosity but also degrade nonlinear dynamics, reduce injection efficiency, and shorten beam lifetime. Effective measures are thus necessary for correction and recovery.

To avoid significant degradation in nonlinear dynamics and luminosity, effective error correction and compensation strategies must be implemented. The static error correction process for the STCF collider rings mainly includes the following steps: (1) first-turn trajectory correction, (2) closed orbit and dispersion correction, and (3) optical function and coupling correction. These steps are typically iterated until the required physics conditions are met. The correction system consists primarily of beam position monitors (BPMs), orbit correctors, and skew quadrupole magnets, with the layout guided by the following principles:

- A BPM is installed next to each quadrupole magnet and some sextupoles to monitor the local closed orbit and facilitate beam-based alignment (BBA), and also mitigate β-beating caused by the sextupole feed-down effects. A pair of BPMs is also installed near the IP to better correct the local orbit.
- A horizontal corrector magnet is placed next to each focusing quadrupole, and a vertical corrector is placed next to each defocusing quadrupole. These positions correspond to high β values, enhancing correction efficiency and reducing corrector strength. Additional independent correctors are installed near the IP to further improve orbit correction.
- Skew quadrupole coils are mounted on all sextupoles. Additional independent skew quadrupoles are placed near the IP and in certain straight sections.

The correction system layout for a single ring of STCF includes 405 BPMs, 287 horizontal correctors, 254 vertical correctors, and 102 skew quadrupoles.

During machine commissioning, the primary requirement is to achieve first-turn beam accumulation. Using 100 random seeds of error distributions, first-turn trajectories are corrected using the SVD method and trajectory response matrix. For *n* correctors and *m* BPMs, the corrector strength is given by:

$$X_m = R\Theta_n \qquad (18)$$

where $R$ is the m×n trajectory response matrix, and $X_m$ is the trajectory offset at BPMs. The corrector strengths $\Theta_n$ are calculated by:

$$\Theta_n = -R^{-1}X_m \qquad (19)$$

The pseudoinverse $R^{-1}$ is computed using the SVD method. Figure 2.8-2 shows the first-turn trajectories before and after correction for the 100 error seeds. After correction, the maximum horizontal trajectory deviation is within 2 mm, and the maximum vertical deviation is around 8 mm. All seeds achieve a closed orbit after correction.

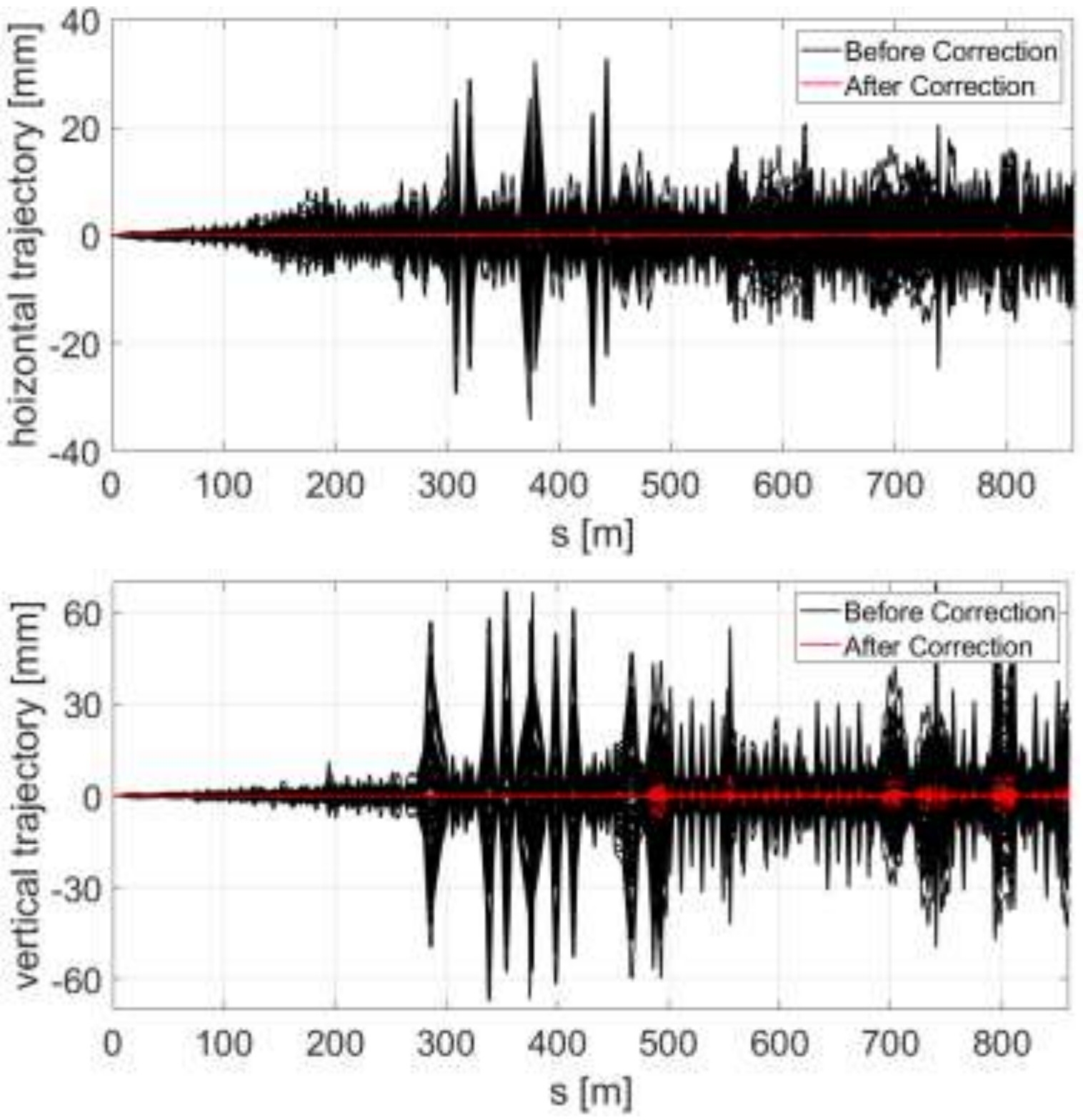

Figure 2.8-2: First-turn trajectories in the STCF collider rings before and after correction. Top: Horizontal direction; Bottom: Vertical direction

After successful first-turn accumulation and achieving a closed orbit, the Dispersion Free Steering (DFS) method [39] is used to further correct the closed orbit and dispersion function. The corrector magnet strengths remain as the tuning variables. Based on the orbit and dispersion response matrices, the goal is to minimize orbit and dispersion deviations. The corrector strengths $\Theta$ are calculated using the following equation:

$$d = R\Theta \qquad (20)$$

Here, $R$ is the response matrix, and $d$ is the deviation of orbit and dispersion at the BPMs:

$$R = \begin{pmatrix} 1-\alpha A \\ \alpha B \end{pmatrix} \qquad (21)$$

$$d = \begin{pmatrix} 1-\alpha u \\ \alpha D \end{pmatrix} \qquad (22)$$

where $A$ and $B$ are the orbit and dispersion response matrices, respectively, $u$ and $D$ are the closed orbit distortion and dispersion deviation vectors, and $\alpha$ is a weighting factor.

The orbit and dispersion at the IP are critical for achieving design luminosity and must be corrected to extremely low levels. A weighted SVD method [40] is applied to solve for the corrector strengths. The orbit and dispersion deviations at the BPMs on both sides of the IP are multiplied by a weighting factor $w$, and the corresponding rows in the response matrix are also scaled by $w$. The corrector strengths are then computed as:

$$\theta = -(wR)^{-1} \cdot wd \tag{23}$$

To improve the correction performance and control the magnitude of the corrector strengths, multiple iterative corrections of the closed orbit and dispersion functions are necessary, along with optimization of the number of singular values used in the SVD. The weighting factor $w$ influences the correction results: higher $w$ values enhance the correction at the IP but may worsen orbit and dispersion performance elsewhere. A value of $w = 1000$ is chosen to balance global and local correction performance.

Figure 2.8-3 shows the orbit distortion after DFS correction. The maximum horizontal and vertical orbit distortions across the entire ring are within 400 μm, and the rms values are under 50 μm. Notably, orbit distortions at the IP before and after correction are analyzed, as shown in Figure 2.8-4. After applying weighted DFS correction, the orbit distortion at the IP is reduced by 2–3 orders of magnitude, reaching the tens of nanometers level—sufficiently low to avoid any noticeable impact on luminosity. Figure 2.8-5 shows the dispersion function distortion before and after DFS correction. The correction significantly reduces horizontal and vertical dispersion distortions. The rms of the dispersion function across the full ring is about 0.25 mm after correction, while the rms values at the IP are about 0.4 μm (horizontal) and 0.5 μm (vertical). Figure 2.8-6 presents the required strengths of the corrector magnets during the DFS correction. The maximum corrector kick angles are about 0.3 mrad in both horizontal and vertical planes, with rms values around 32 μrad, all within feasible technical limits.

While DFS effectively restores the closed orbit and dispersion functions, it does not yet recover the beam emittance, β-functions, or tune. Approximately 25% of the seeds do not yield stable optical function solutions at this stage.

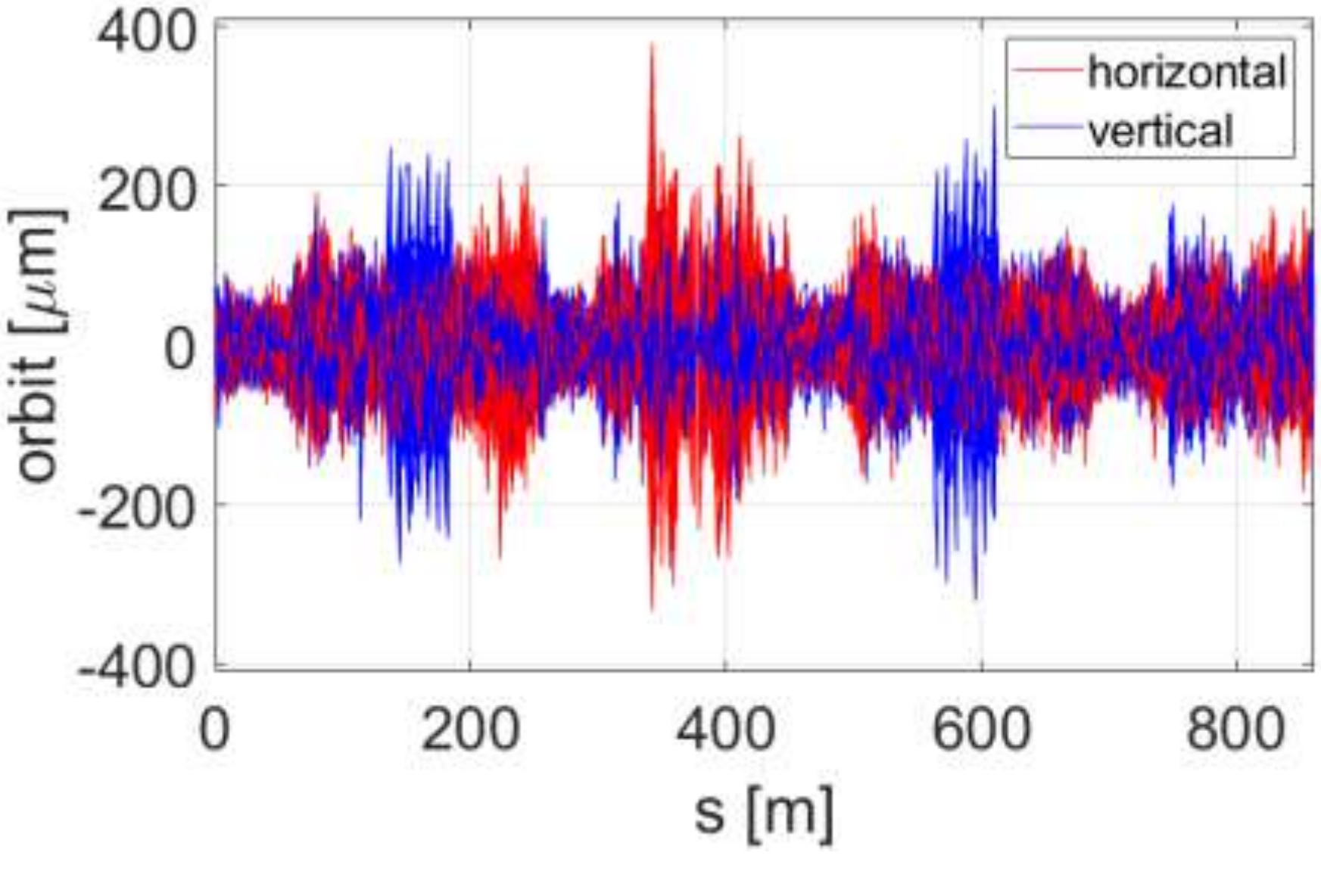

Figure 2.8-3: Closed orbit in the horizontal and vertical planes after DFS correction.

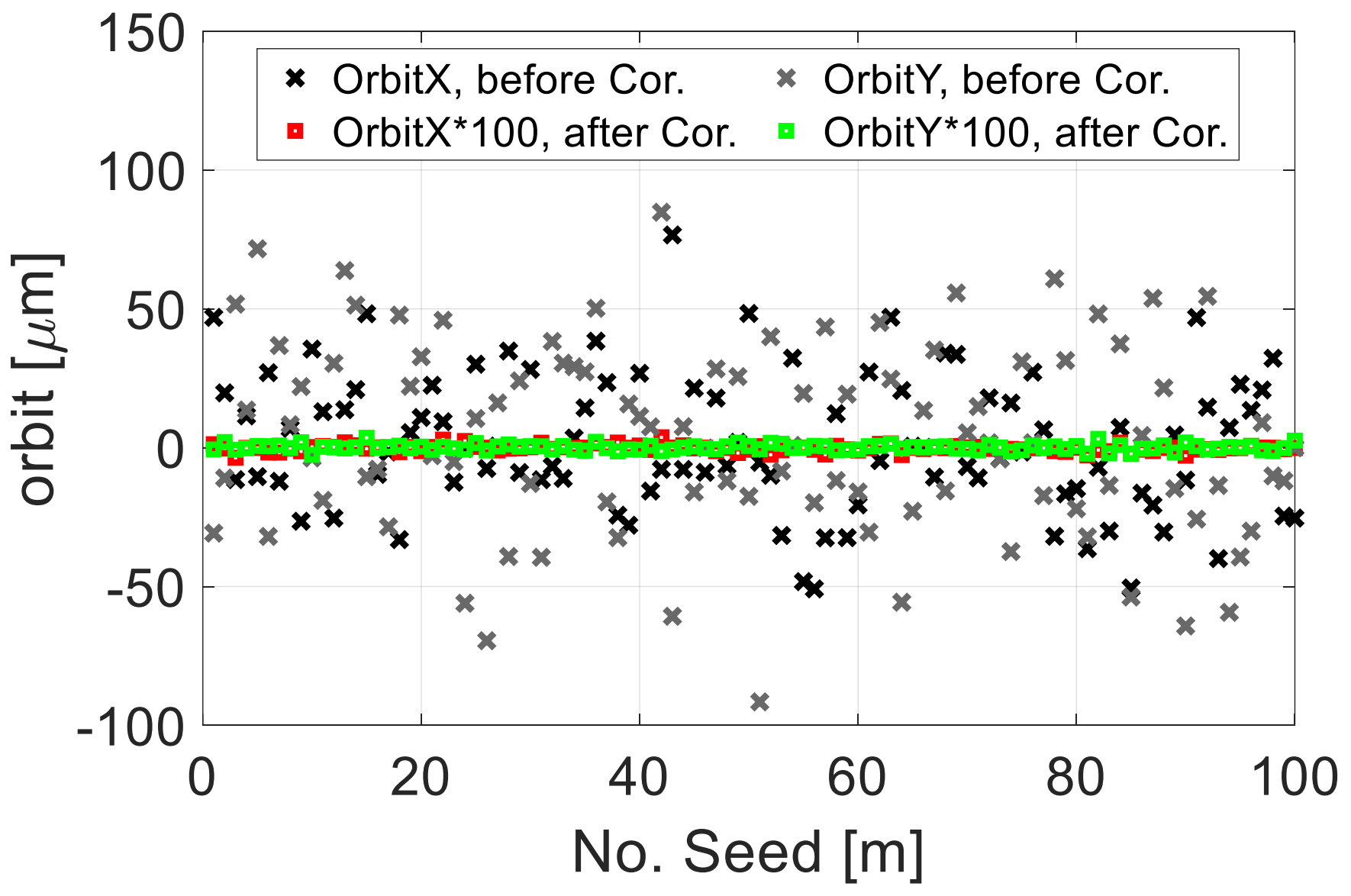

Figure 2.8-4: Closed orbit at the IP before and after DFS correction.

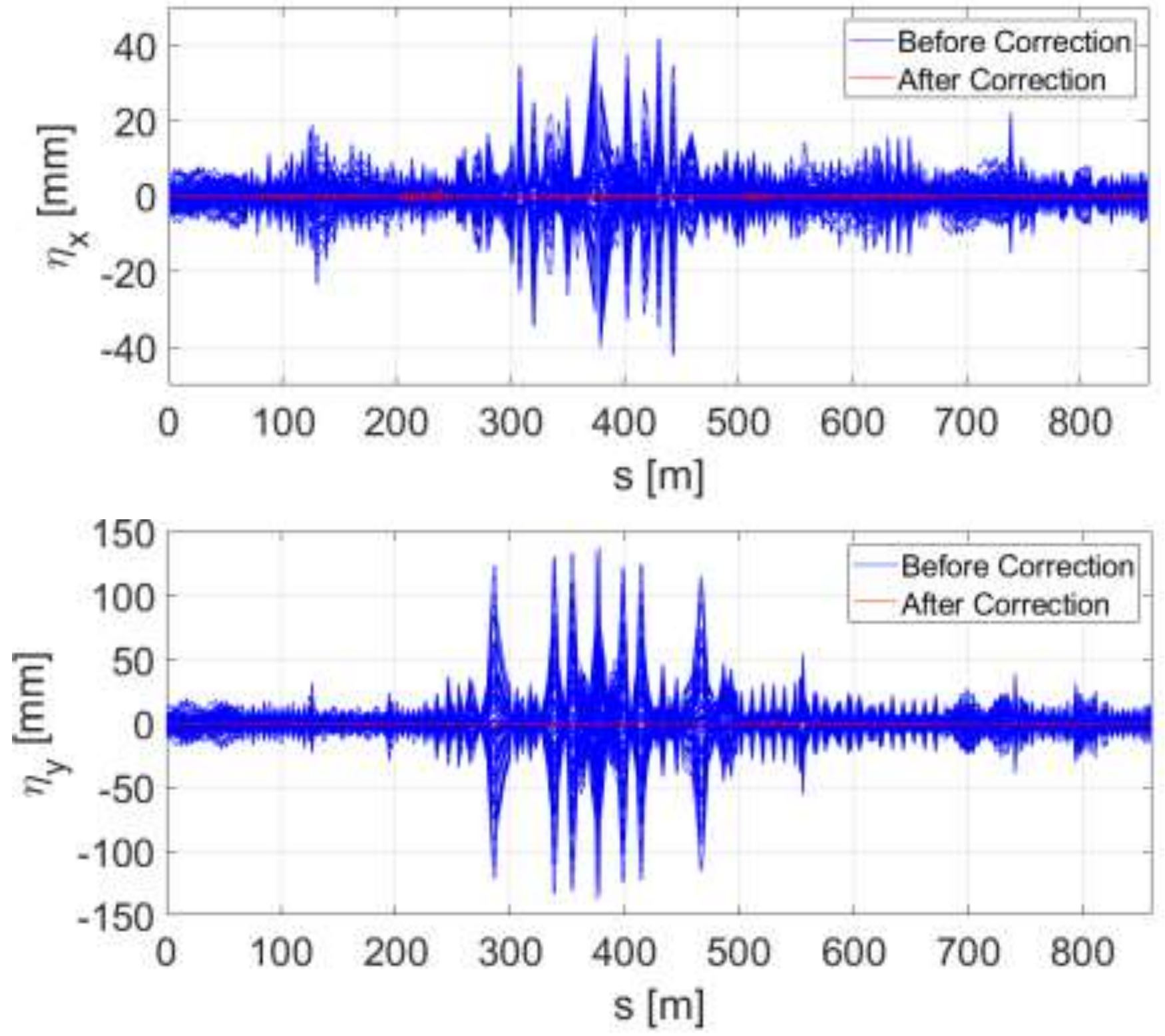

Figure 2.8-5: Dispersion function distortions before and after DFS correction; top: horizontal direction, bottom: vertical direction.

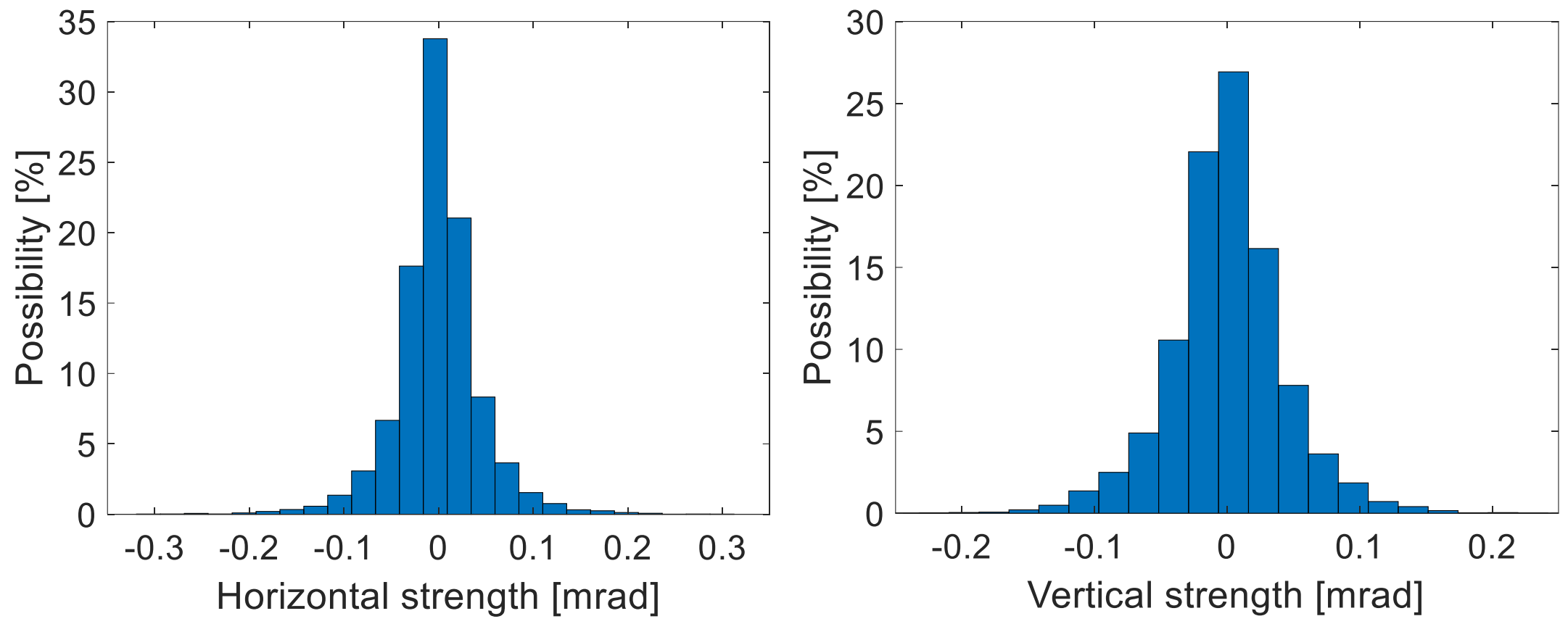

Figure 2.8-6: Corrector magnet strengths after DFS correction; left: horizontal plane, right: vertical plane.

To further restore beam emittance, tune, and dynamic aperture, the LOCO (Linear Optics from Closed Orbit) method [40] is employed for optical function correction. Skew quadrupole magnets are also used to correct transverse coupling and vertical dispersion. Figure 2.8-7 presents the statistical distribution of horizontal and vertical β-beating rms values before and after correction. Most seeds achieve good optical function correction, with β-beating rms values not exceeding 1% outside the IP region. However, $\beta_y$ distortion at the IP remains large, with rms and maximum values around 9% and 40%, respectively, leading to luminosity losses of approximately 3% and 13%. To recover the luminosity, the strengths of quadrupoles on both sides of the IP are retuned to rematch the optical functions, aiming to restore $\beta_y$ at the IP to near-design values. $\beta_x$ distortion is very small and has a negligible impact on luminosity.

After correction, the horizontal and vertical dispersion function rms values are under 0.5 mm and 0.2 mm, respectively, with maximum deviations not exceeding 5 mm. Since weighting was not applied at the IP during this stage, vertical dispersion at the IP increases, with rms and maximum values reaching approximately 15 μm and 46 μm, respectively—leading to luminosity losses of around 1% and 3%. Future iterations may apply additional weighting to reduce vertical dispersion at the IP. Horizontal dispersion distortion has a negligible effect on luminosity.

Figure 2.8-8 shows the statistical distribution of transverse coupling ratios before and after correction. Before correction, the transverse coupling ratio is very large, and the resulting vertical emittance would notably reduce luminosity. After correction, the emittance coupling ratio is reduced below 0.1%, ensuring minimal luminosity loss due to vertical emittance growth. During correction, the variation in quadrupole strengths remains below 3%, and the maximum strength of skew quadrupole magnets is approximately 0.3 T/m, which is within technical feasibility.

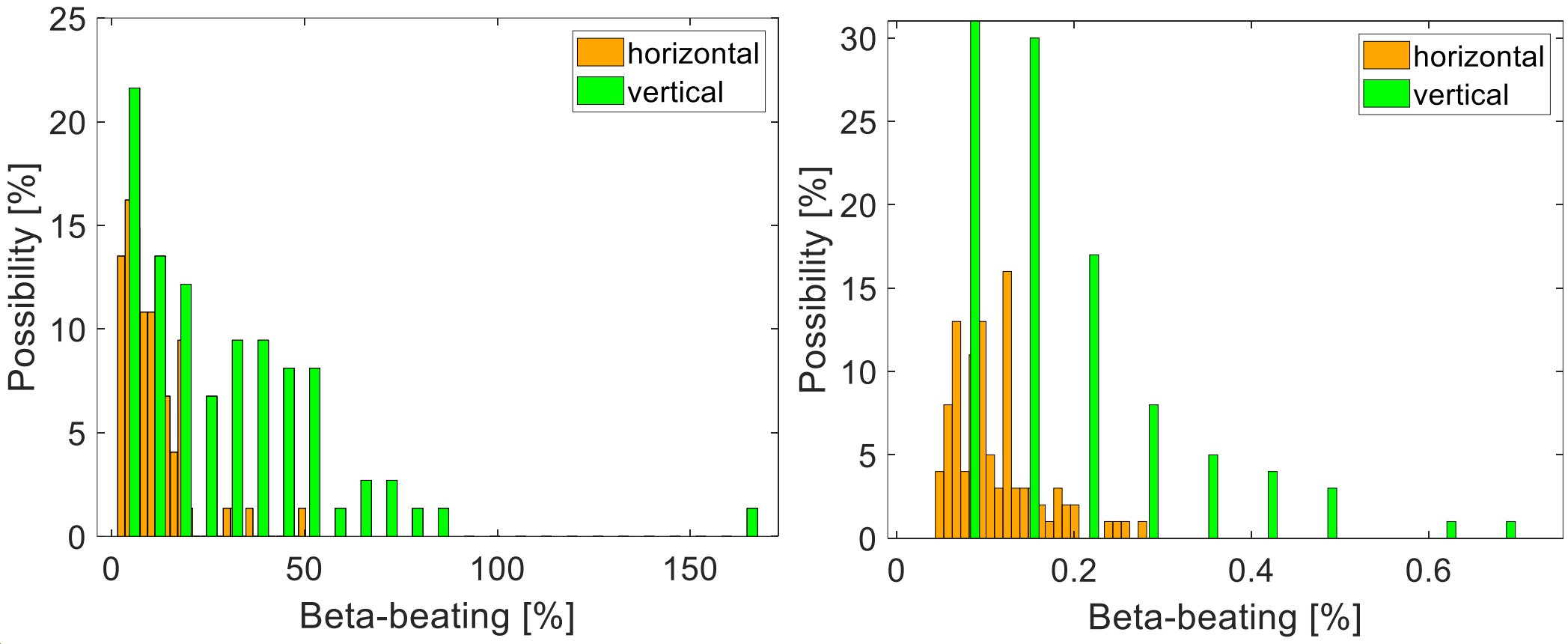

Figure 2.8-7: Statistical distribution of β-beating rms values before (left) and after (right) correction of optical functions and coupling.

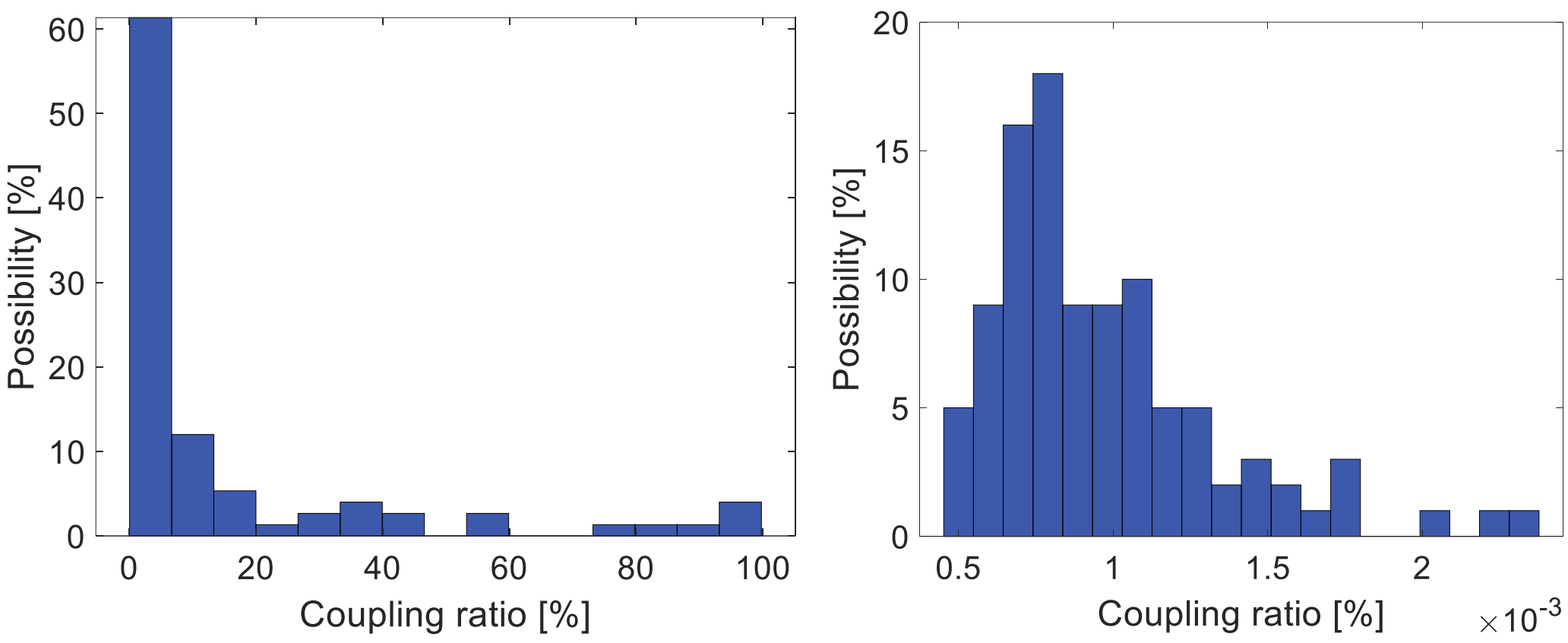

Figure 2.8-8: Statistical distribution of transverse coupling rate before (left) and after (right) correction.

After orbit, optical function, and coupling corrections, the dynamic aperture and momentum aperture of the collider rings are significantly restored. Figure 2.8-9 shows the dynamic aperture and momentum aperture with edge effect considered after correction, indicating that the dynamic aperture has recovered to near-ideal lattice levels. Momentum aperture is also well restored, with Touschek lifetime degradation due to current error sources estimated at less than 30%. The multipole field errors of the magnets will further affect the nonlinear dynamics. Simulation results show that the multipole fields of the doublet quadrupoles on both sides of the IP have the most significant impact on the dynamic aperture and momentum aperture. To mitigate the effects of multipole fields, the higher-order fields of these four quadrupoles should not exceed $2\times10^{-4}$, while those in other magnets should be controlled below $5\times10^{-4}$. This ensures that the reduction of dynamic aperture and Touschek lifetime remains within 10%.

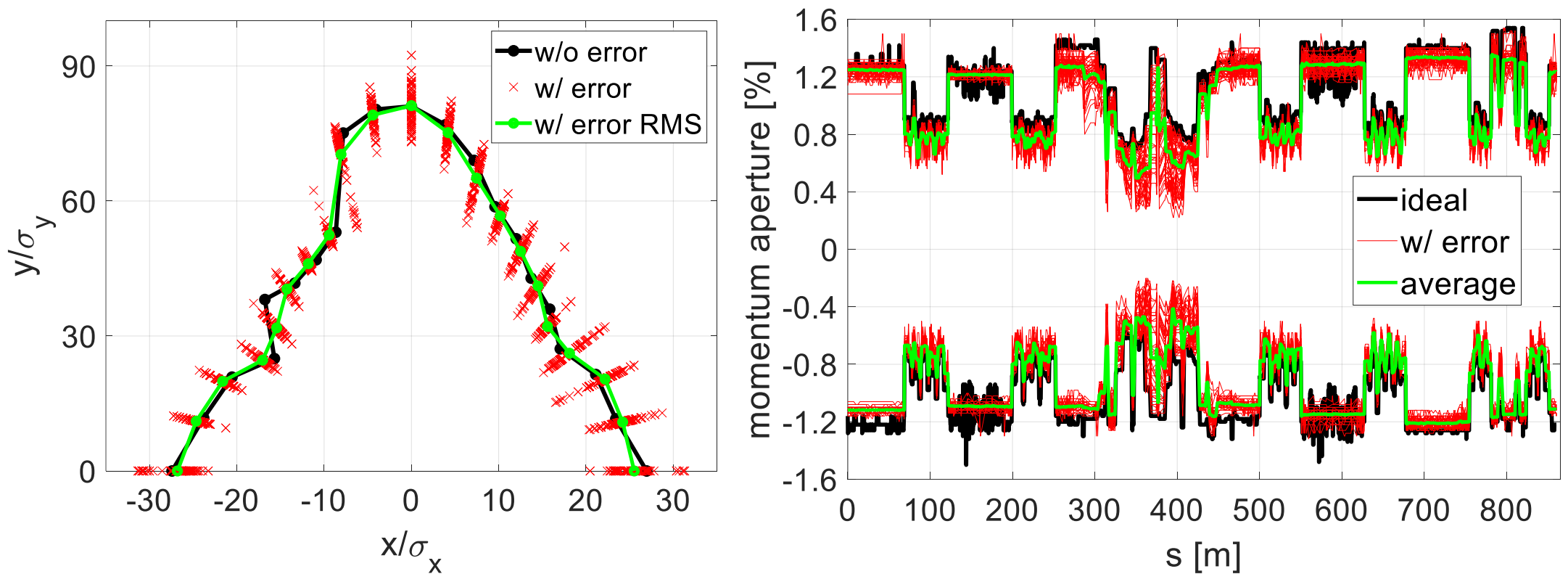

Figure 2.8-9: Post-correction dynamic aperture (left) and momentum aperture (right) of the collider rings.

## 2.9 Beam Collimation

Various beam loss mechanisms exist in the collider rings, such as Touschek scattering, beam-beam scattering, and beam-gas scattering, all of which limit the beam lifetime [41]. At STCF, Touschek scattering is particularly severe, reducing particle lifetime to around 200 seconds. Therefore, a sophisticated beam collimation system is required to mitigate particle losses.

The functions of the beam collimators in the collider rings are as follows: controlling experimental backgrounds in the interaction region, preventing uncontrolled particles from damaging accelerator components, and localizing beam losses to designated areas with additional shielding, thereby reducing radiation dose in the tunnel and easing the shielding requirements of the tunnel walls [42].

Collimation efficiency is defined as the ratio of particles intercepted by collimators (or their surroundings) to the total number of particles lost in a specific section of the collider rings. The design goal is a collimation efficiency greater than 90% outside the interaction region and greater than 80% within it.

Another requirement for the beam collimation system is to maintain as low impedance as possible, with collimators contributing no more than 50% of the total ring impedance.

### 2.9.1 Beam Collimation Methods

#### *2.9.1.1 Beam Loss Mechanisms*

During normal operation, there are multiple physical mechanisms responsible for beam losses in the collider rings, including Touschek loss, beam instabilities, injection losses, beam-gas scattering in vacuum, and beam-beam scattering at the IP [41]. It is estimated that the primary beam losses in the STCF collider rings arise from Touschek loss, beam-gas scattering in vacuum, and beam-beam scattering. The current estimated beam lifetime at STCF is around 200 seconds, with a total circulating particle count (based on a stored current of 2 A in the top-up mode) of about $3.6\times10^{13}$ $e^+/e^-$, resulting in a loss rate of approximately $1.8\times10^{11}$ $e^+/e^-$ per second. These losses exhibit no obvious time structure. Preliminary analyses of major loss mechanisms are as follows:

**Injection loss:** If STCF adopts the off-axis injection scheme, with an injection efficiency greater than 90%, the corresponding beam loss rate is around $1.8\times10^{10}$ $e^+/e^-$ per second. If the bunch swap-out injection is used, injection losses are negligible. Injection losses are time-structured, mainly occurring shortly after each injection (at 30 Hz or every 33 ms for both positron and electron rings), typically within 1 ms.

**Touschek loss:** This is the dominant beam loss mechanism at STCF, originating from intra-bunch scattering [41] and the very small momentum aperture of the collider rings. The most critical momentum aperture limitations occur in the interaction region (particularly at locations with high dispersion) and arc sections. A substantial number of collimators are placed in these regions to reduce losses in the interaction region and elsewhere in the ring. These losses are random and lack a time structure.

**Beam-beam scattering** [43]**:** This excludes physics processes that produce detectable events. Beam-beam **scattering** leads to two types of losses: instantaneous losses, which increase backgrounds at the IP, and emittance growth, which causes beam loss due to limited dynamic and momentum apertures throughout the ring. These losses are random in time, although coupling between off-axis injection damping and beam-beam interactions may introduce some time structure.

**Collective instability loss:** These are unpredictable losses. They may or may not occur, but if frequent, they will limit the achievable beam current and luminosity until new mitigation measures are implemented. Such losses tend to be distributed throughout the ring and impose demands on the collimation system. They are spontaneous and not time-predictable.

**Beam-gas scattering loss** [44]**:** Scattering of beam particles with residual gas molecules in vacuum causes both direct particle loss (e.g., scattered particles lost within a few turns) and emittance growth (which can be mitigated by radiation damping). Outside the interaction region, beam-gas scattering losses should be controllable due to well-designed vacuum systems. However, in and around the interaction region, especially inside the spectrometer, vacuum conditions are worse due to limited space for pumps, leading to relatively high beam loss rates. This part of the beam loss contributes directly to experimental background and must be suppressed. These losses are continuous but can occasionally spike if a sudden vacuum degradation occurs somewhere in the rings.

#### *2.9.1.2 Collimator Layout Design*

STCF plans to implement a more comprehensive beam collimation system than any previously constructed electron-positron collider. This is motivated by two factors: first, the exceptionally short beam lifetime in the STCF collider rings, which poses an unprecedented challenge; and second, as a newly built machine rather than an upgrade of an existing facility, STCF has the flexibility to allocate optimal collimator placements. This design approach is expected to better suppress backgrounds in the interaction region and also help manage the radiation dose distribution within the collider tunnel.

The beam collimators in the STCF collider rings can be divided into three categories: arc collimators, which are mainly responsible for cleaning particles from Touschek scattering; interaction region (IR) collimators, which block off-momentum particles (including those from vacuum scattering) from entering the interaction region and collimate Touschek-scattered particles within it; and straight-section collimators, which employ nonlinear collimation methods (i.e., nonlinear lattice designs) to remove particles with large emittance, thereby reducing the burden on the IR collimators.

**Arc Collimators:** Previous electron-positron colliders had relatively long beam lifetimes and thus did not require complex collimator systems. For example, BEPC/BEPCII had relatively few collimators [45]; KEKB and PEP-II employed more [46]; and SuperKEKB further increased the number, though it was constrained by space (collimators were only arranged on one half-arc for each of the positron and electron rings) [47]. As a new machine, STCF is well-positioned to provide more favorable collimator locations and space. Given that Touschek scattering is nearly uniformly distributed throughout the ring and contributes significantly to beam loss due to the short Touschek lifetime, collimators are planned for all arc regions, including those close to the IP (e.g., the small arcs). Collimators will be placed at positions with large dispersion and large beta functions, and it is expected that the majority of beam losses at STCF will be intercepted by these arc collimators.

**Interaction Region (IR) Collimators:** These require dedicated optimization to manage experimental backgrounds. In SuperKEKB, the IR collimators were overly burdened [48]; STCF seeks to relieve this pressure through a more comprehensive global collimation design. In addition, improvements will be made in terms of collimator materials and structural design to enhance their performance and effectiveness.

**Straight-Section Collimators:** These are a critical part of collimation systems in hadron colliders but are generally considered less important in electron colliders. However, SuperKEKB proposed a nonlinear beam collimation method [49] to reduce the burden on IR collimators, and STCF is even better positioned to implement this approach thanks to its advantageous spatial layout. In STCF, each collider ring ($e^+$ and $e^-$) will feature a dedicated nonlinear collimation section within the long straight segments. These collimators aim to remove particles with large transverse emittance, using relatively wide collimation gaps (to avoid generating excessive coupling impedance). In addition to large beta functions, nonlinear elements (such as sextupoles or octupoles) will be added to distort the beam in phase space, enhancing collimation effectiveness—an approach known as nonlinear collimation.

## 2.9.2 Calculation and Simulation Results

### *2.9.2.1 Beam Loss Simulation from Touschek Scattering*

Due to the very short Touschek lifetime, beam losses during routine STCF collider operation are primarily caused by the Touschek scattering effect. To better collimate Touschek-induced losses, the beam loss distribution in the absence of any collimators was first evaluated. The vacuum pipe radius was assumed to be 15 mm at the IP and the first quadrupole magnets on either side, transitioning uniformly through one drift section to a radius of 25 mm at the second quadrupole, and again transitioning uniformly to 33.5 mm in the main interaction region. The rest of the ring was assumed to have a vacuum pipe radius of 26 mm.

Under the condition of a single bunch charge of 8 nC and a transverse coupling ratio of 0.5%, the distribution of beam losses due to Touschek scattering—assuming scattering points at every quadrupole magnet—was simulated using the Accelerator Toolbox (AT). The result is shown in Figure 2.9.1. Analysis indicates that approximately 58.2% of the lost particles are lost in the interaction region (between 370 m and 380 m), and about 11.2% are lost in the CCX section preceding the interaction region (between 300 m and 320 m), and around 13.4% are lost near the crab sextupole section before the interaction region (between 270 m and 290 m). Most losses occur in the horizontal plane, though some vertical losses happen due to large $\beta_y$ at the crab sextupole region. Losses in the interaction region primarily occur in the transition drifts with changing pipe aperture, located within 10 m of the IP, where the β-functions are large and complex and the pipe aperture is relatively narrow.

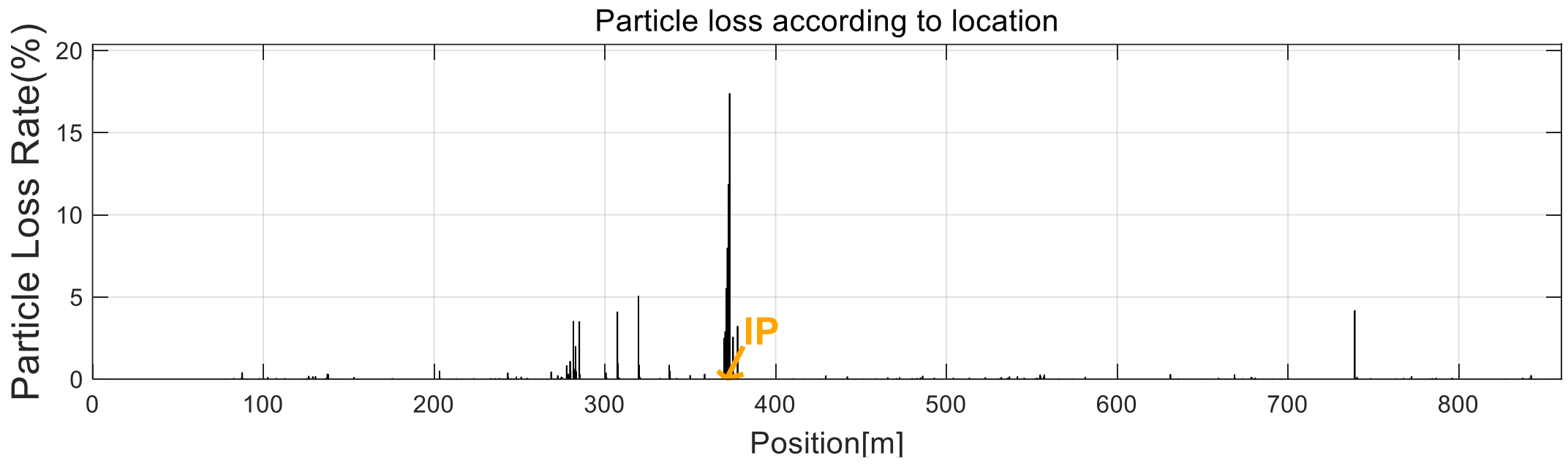

Figure 2.9-1: Beam loss distribution from Touschek scattering without collimators

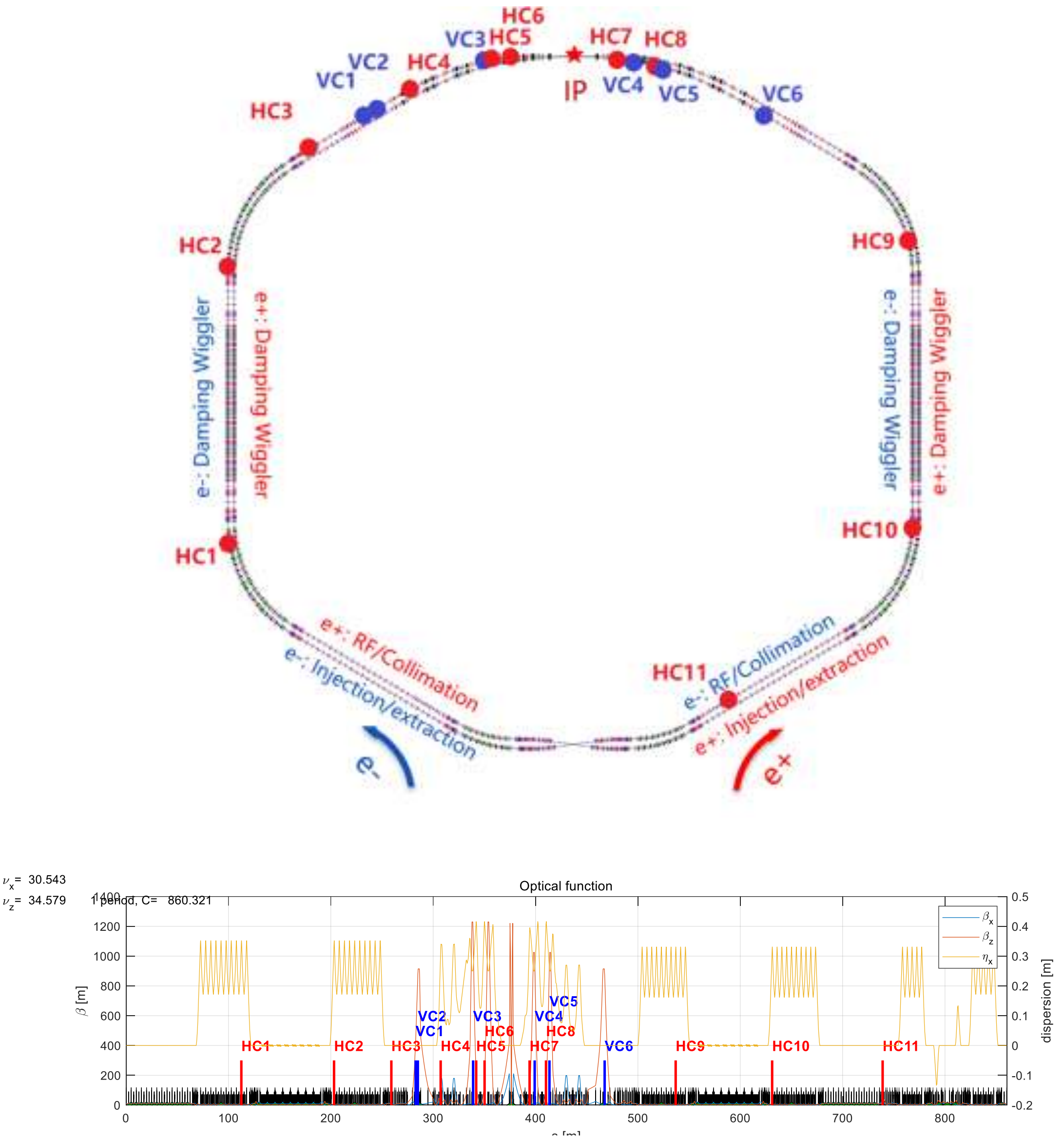

Figure 2.9-2: Full-ring collimator layout (Top: global beam collimator layout; Bottom: collimator layout with Lattice; example shown for one ring. VC = vertical collimator, HC = horizontal collimator)

Based on the loss distribution, global optical function map, and betatron function distortions for off-momentum particles, a layout of 11 horizontal collimators (HC) and 6 vertical collimators (VC) was determined after repeated simulation and optimization. The layout is shown in Figure 2.9.2. After collimators are placed, the Touschek-induced beam loss distribution is shown in Figure 2.9.3. The simulation indicates that the current collimator configuration achieves approximately 97.3% collimation efficiency. Within ±20 m of the IP, beam losses are about 1.36%, primarily occurring in the transition region upstream of QF1. As seen in Figure 2.9.3, placing horizontal collimators in the CCY section (HC5, HC6) and at locations with large horizontal β-function distortions (HC1) effectively intercepts horizontally scattered particles. Likewise, vertical collimators (VC1, VC2) placed at high-$\beta_y$ regions in the upstream of the interaction region effectively intercept vertically scattered particles.

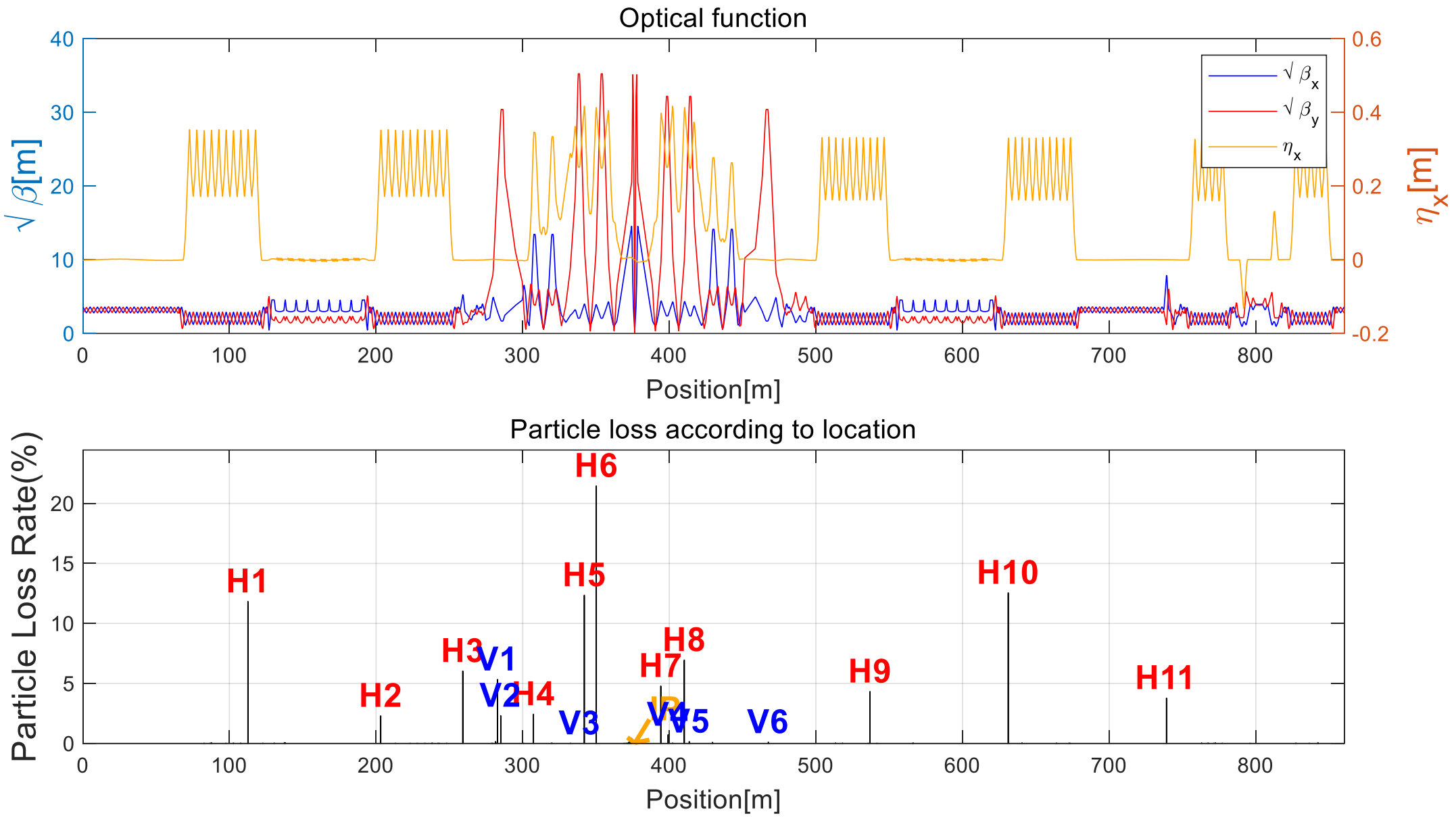

Figure 2.9-3: Interception ratio distribution of lost particles due to Touschek scattering after the collimators are placed.

With other loss mechanisms such as injection loss, beam-gas scattering and beam collision at the IP being included in the simulations, more collimators will be needed. A nonlinear collimation system as mentioned above is also foreseen to reduce the impedance of the collimators with very small gap.

#### *2.9.2.2. Preliminary Simulation of Lost Beam on Collimators*

When lost beam particles strike the collimators, they interact with the collimator material. The nature of this interaction varies depending on the choice of target material, affecting the ability to stop particles, the deposition of energy, and the production of secondary particles. Table 2.9-1 lists the properties of several commonly used collimator materials.

Table 2.9-1: Physical properties of candidate collimator materials

| Material | Density [g/cm³] | Specific Heat Capacity [kJ/(kg·K)] | Thermal Conductivity [W/(m·K)] | Melting Point [K] |
|---|---|---|---|---|
| Tungsten (W) | 19.3 | 0.132 | 173 | 3695 |
| Tantalum (Ta) | 16.69 | 0.14 | 57 | 3017 |
| Copper (Cu) | 8.96 | 0.385 | 400 | 1358 |
| Aluminum (Al) | 2.7 | 0.88 | 190 | 933 |
| Graphite (C) | 2.26 | 0.709 | 120 | 3915 |

Considering thermal properties and electrical conductivity, copper is preliminarily selected as the primary material for the collimators. Based on STCF's relatively wide beam energy range (1–3.5 GeV), an evaluation is underway to determine whether an absorbing-type or scattering-type collimator structure is more appropriate.

#### *2.9.2.3 Conclusions and Analysis*

The current collimator layout in STCF effectively reduces backgrounds in the interaction region while maintaining a reasonable loss distribution around the ring. Energy deposition from Touschek scattering losses appears relatively low and is not expected to cause severe thermal effects at this stage. As the lattice design evolves, the collimator layout will be adjusted accordingly, and simulation details will continue to be refined for higher accuracy. Simulations will also be extended to other beam loss mechanisms.

## 2.10 Synchrotron Radiation Damping

### 2.10.1 Requirements of the Collider Rings for Radiation Damping

The collider rings experience significant synchrotron radiation effects. On one hand, these effects lead to beam energy loss, impose heavy thermal loads, and contribute adversely to experimental backgrounds. On the other hand, synchrotron radiation exerts a damping effect on the circulating beams, causing shrinkage of transverse and longitudinal emittances. This damping effect is beneficial and crucial for beam injection, suppression of harmful collective effects, and enhancement of luminosity. Concurrently, the quantum excitation effect, also coming from synchrotron radiation, introduces randomness and competes with the damping; as a result, the horizontal emittance and energy spread eventually reach an equilibrium determined by the lattice design and beam energy.

The STCF's collider rings are high-current, low-emittance electron and positron storage rings operating over a wide energy range. The synchrotron radiation damping provided by the lattice alone is insufficient to achieve the required short damping times, low horizontal emittance, and relatively high energy spread, particularly at low beam energy. Therefore, additional damping wigglers must be installed in the dedicated long straight sections to enhance the damping effect.

The design requirements for radiation damping in the collider rings are as follows: at the optimal energy point (2 GeV), the radiation damping time should be less than 30 ms, the horizontal emittance should be around 5 nm·rad, and the energy spread should not be less than $6\times10^{-4}$. At the other energy points, these parameters may deviate slightly (either higher or lower) but must remain within an acceptable range.

### 2.10.2 Mechanism of Synchrotron Radiation Damping

Synchrotron radiation is emitted as a continuous spectrum of electromagnetic waves in the tangential direction when high-speed charged particles traverse curved trajectories in magnetic fields. In the STCF collider rings, synchrotron radiation is primarily produced by the dipole bending magnets and damping wigglers. The dipole magnets provide approximately uniform vertical magnetic fields, and electrons emit radiation as they bend in these fields. The wiggler magnets consist of alternating-polarity dipole magnets arranged periodically and perpendicularly to the beam's central plane. As electrons pass through this alternating field, their trajectories undergo periodic oscillations, producing synchrotron radiation.

The radiated energy $U$ per turn is expressed by Equation (24), where $I_2$ is the second synchrotron radiation integral, $C_r$ is a constant ($8.846\times10^{-15}$ m·GeV$^{-3}$), and $E$ is the electron energy [50]. When wigglers are added, $I_2$ includes contributions from both bending magnets and wiggler magnets. The contribution from the bending magnets $I_{20}$ is given by Equation (25), where $\rho_i$ is the bending radius and $L_i$ the effective length of the i-th bending magnet. The wiggler contribution depends on the magnetic field profile. Equations (26) and (27) describe $I_{2w}$ for sinusoidal and rectangular field wigglers, respectively, where $B\rho$ is the magnetic rigidity, $B_w$ the peak magnetic field, and $L_w$ the total effective length of the wiggler.

$$U = \frac{C_r E^4}{2\pi} I_2 \tag{24}$$

$$I_{20} = \int \frac{1}{\rho^2} ds = \sum_{i=1}^{N} \frac{1}{{\rho_i}^2} \cdot L_i \tag{25}$$

$$I_{2\mathrm{W,S}} = \int_0^{L_w} \frac{1}{\rho^2} ds = \frac{{B_w}^2 L_w}{2(B\rho)^2} \tag{26}$$

$$I_{2\mathrm{W,R}} = \int_0^{L_w} \frac{1}{\rho^2} ds = \frac{{B_w}^2 L_w}{(B\rho)^2} \tag{27}$$

In the collider rings, higher energy of the electron beam and positron beam will emit more energy by synchrotron radiation, and the oscillations of the particles will be more damped. The damping rates are different for different directions (horizontal, vertical, and longitudinal), as shown in Equation (28), where $J_i$ is the damping factor, $T$ for the revolution period, $E$ for the beam energy, and $U$ for the radiated energy in one turn. The average energy loss in the longitudinal direction will be compensated by the RF cavities.

$$\tau_i = \frac{2}{J_i} T \frac{E}{U} \qquad\qquad i = x、y、s \tag{28}$$

The radiation damping effect leads to a reduction in both the horizontal and vertical emittance, while the quantum excitation during the radiation process causes a slight increase in the horizontal emittance. When the quantum excitation and damping reach dynamic equilibrium in the horizontal plane, the horizontal equilibrium emittance is given by Equation (29), where $I_5$ is the fifth synchrotron radiation integral, $C_q$ is a constant (3.832×10$^{-13}$ m), $\gamma$ is the relativistic factor, and $J_x$ is the horizontal damping partition number [51].

$$\varepsilon_{x0} = \frac{C_q \gamma^2 I_5}{J_x I_2} \tag{29}$$

The beam energy spread $\sigma_{\mathrm{E}}$ is also affected by synchrotron radiation damping. When quantum excitation and damping reach equilibrium, the beam reaches a steady state, and the equilibrium energy spread is given by Equation (30), where $I_3$ is the third synchrotron radiation integral and $J_s$ is the longitudinal damping partition number [50, 52]. The contribution from the bending magnets to $I_{30}$, denoted as $I_{30}$, is expressed by Equation (31), where $\rho_i$ and $L_i$ are the bending radius and effective length of the $i$-th dipole magnet, respectively. Equations (32) and (33) describe the $I_{3w}$ contributions from sinusoidal-field and rectangular-field wiggler magnets, respectively, where $B\rho$ is the magnetic rigidity, $B_w$ is the peak magnetic field of the wiggler, and $L_w$ is its total effective length.

$${\sigma_E}^2 = \frac{C_q \gamma^2 I_3}{J_s I_2} \tag{30}$$

$$I_{30} = \int \frac{1}{|\rho|^3} ds = \sum_{i=1}^{N} \frac{1}{|{\rho_i}^3|} \cdot L_i \tag{31}$$

$$I_{3\mathrm{W,S}} = \int_0^{L_w} \frac{1}{\rho^3} ds = \frac{4{B_w}^2 L_w}{3\pi (B\rho)^3} \tag{29}$$

$$I_{3\mathrm{W,R}} = \int_0^{L_w} \frac{1}{\rho^3} ds = \frac{{B_w}^3 L_w}{(B\rho)^3} \tag{30}$$

Compared to the previous generation of colliders, the STCF collider rings have a significantly shorter beam lifetime, necessitating frequent beam injections and requiring a damping time of ≤30 ms. At lower energy (1 GeV), the radiation damping time is excessively long, necessitating the use of wiggler magnets to enhance damping and reduce the damping time. At higher energy (3.5 GeV), synchrotron radiation also affects the equilibrium emittance and energy spread of the stored beam; the introduction of wiggler magnets can also help tune these critical beam parameters.

### 2.10.3 Damping Wiggler Parameter Design

#### *2.10.3.1 Total Effective Length and Peak Magnetic Field*

At the optimized energy point of 2 GeV, the collider rings require a radiation damping time in the transverse planes shorter than 30 ms. At the beam energy of 1 GeV, the radiation damping from the lattice is much weaker, the required damping time should be less than 50 ms. Another radiation damping will be from the damping wigglers. By scanning the peak magnetic field strength $B_w$ from 1 T to 5 T, the theoretical total effective length of the wiggler magnets for achieving a 50-ms damping time at 1 GeV is calculated, as shown in Figure 2.10-1 for both sinusoidal and rectangular field configurations. The total effective length of a real magnet will lie between these two cases.

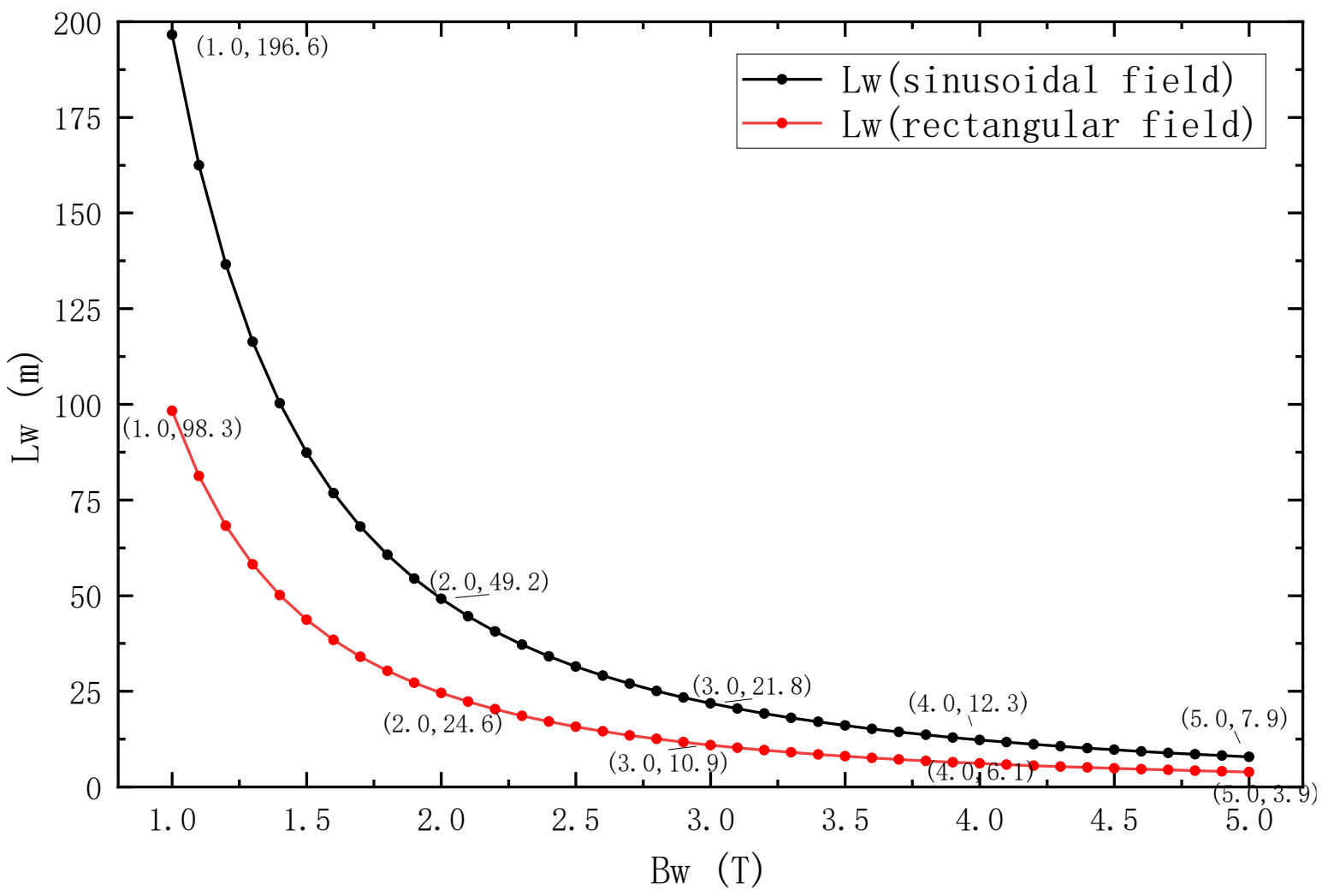

Figure 2.10-1: Total effective length of wiggler magnets required for a 50-ms damping time at 1 GeV versus peak magnetic field strength.

Currently used insertion magnet technologies in accelerators can generally be classified into three types: room-temperature magnets, cryogenic permanent magnets, and superconducting magnets [53]. Room-temperature magnets usually produce a magnetic field of 1 T to 1.5 T, cryogenic permanent magnets 1.5 T to 2 T [54, 55], and superconducting magnets 3 T to 5 T or higher.

Superconducting wiggler magnets are the most expensive and complex among the three types of magnets. However, for the same damping time, increasing the field strength from 3 T to 5 T yields diminishing returns in terms of reducing the total effective length. Given that the STCF collider rings operate over a wide energy range (1-3.5 GeV), the wigglers need to offer flexibility and tunability. While permanent magnets achieve magnet field tunability through mechanical gap adjustments, room-temperature magnets enable easier and broader tunability via coil excitation current control.

Thus, room-temperature wigglers have been chosen at STCF. Taking the cost, technical complexity, and ring layout into account, a peak field $B_w$ of 1.6 T and a total effective length $L_w$ of 76.8 m are used.

### *2.10.3.2 Selection of Wiggler Period Length*

Wiggler magnets are composed of a series of periodic dipole magnets. The period length $\lambda_w$ affects the field shape (more or less rectangular), electron trajectories, beam parameters and instabilities, and RF systems. Although the second synchrotron radiation integral (relevant for damping time) does not explicitly depend on $\lambda_w$, the damping time decreases with increasing $\lambda_w$, as longer periods lead to a more rectangular field profile, which provides stronger damping.

Table 2.10-1 shows damping times across the 1-3.5 GeV energy range with and without wigglers (with $B_w$=1.6 T, $L_w$=76.8 m). Rectangular fields are more effective than sinusoidal fields in reducing damping time. A realistic wiggler field profile is between sinusoidal and rectangular, with a longer period length being more rectangular. Considering the maximum trajectory deviations, maximum oscillation angles, and path length extensions to the ring circumference of the beams within the wigglers, a period length $\lambda_w$ =0.8 m is selected.

Table 2.10-1: Damping times (in ms) at various energies with and without wigglers

| Energy | No Wiggler | Sinusoidal Field | Rectangular Field |
|---|---|---|---|
| 1 GeV | 586.6 | 43 | 22.3 |
| 1.5 GeV | 173.8 | 26 | 14.2 |
| 2 GeV | 73.3 | 17.6 | 10.0 |
| 3.5 GeV | 13.7 | 6.7 | 4.5 |

### *2.10.3.3 Wiggler End Field Configuration*

The end field of a wiggler is crucial for ensuring that the beam trajectory returns to its original transverse position and angle. Ideally, the integral and double integral of the wiggler magnetic field should be zero, which requires special shaping of the end poles [7].

A simple method uses weakened end poles at both ends of a wiggler with a symmetric field arrangement, i.e., an odd number (2N+1) of poles are needed, where the end pole integral is half that of the central poles. This ensures that both position and angle of a beam are restored after its passage through the wiggler, although the center axis of the oscillatory trajectory inside the wiggler remains offset.

To ensure that the trajectory oscillates around the zero axis within the wiggler, two weakened end poles are used at each end. For an even number 2N of poles, the field pattern is: {+1/4, -3/4, +1, -1, ... +1, -1, +3/4, -1/4} for an odd number 2N+1: {+1/4, -3/4, +1, -1, ..., +1, -3/4, +1/4}. This configuration also relaxes the requirements for the good field region in the horizontal direction.

### 2.10.4 Impact of Damping Wigglers on Beam Parameters

Based on the current design parameters—room-temperature magnets with a peak magnetic field $B_w$=1.6 T, total effective length $L_w$=76.8 m, period length $\lambda_w$=0.8 m, and end fields arranged in an even-numbered pole configuration {+1/4, –3/4, +1, –1, ..., +1, –1, +3/4, –1/4}—each magnet has a length of 4.8 m, and a total of 16 wigglers are deployed in one ring. A more realistic wiggler field was calculated using the magnet design software OPERA. The vertical magnetic field $B_y$ along the central trajectory of a single wiggler is shown in Figure 2.10-2, illustrating that the actual field lies between sinusoidal and rectangular profiles and includes fringe field effects.

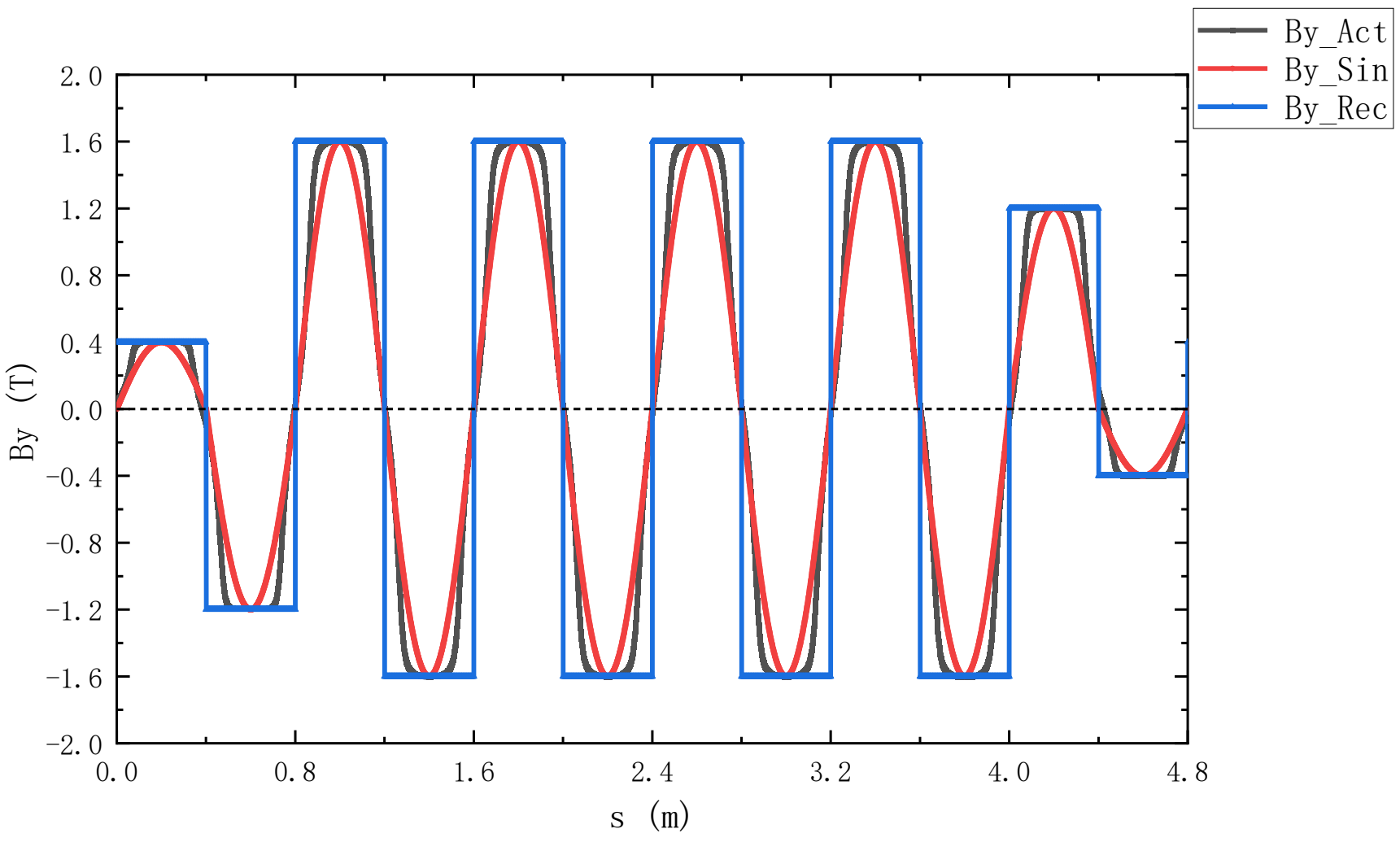

Figure 2.10-2: Vertical magnetic field distribution $B_y$ along the longitudinal axis for three types of wiggler field profiles in a single magnet

To accommodate STCF's broad energy range, the wigglers must support variable settings to meet beam dynamics requirements, such as damping time, horizontal emittance, and energy spread, across different operational energies. Since sinusoidal and rectangular wiggler fields yield different damping effects, and the real field lies between them, Table 2.10-2 presents simulation results for damping time, horizontal emittance, and energy spread at four beam energies (1.0, 1.5, 2.0, and 3.5 GeV) for the cases of no wigglers, sinusoidal wigglers, rectangular wigglers, and realistic wigglers.

The results in Table 2.10-2 show that adding wigglers reduces damping time and increases energy spread in all cases, with rectangular fields producing stronger effects than sinusoidal fields. Realistic fields fall between the two. Across all beam energies, the impact of wigglers on damping and energy spread follows the same trend.

Table 2.10-2: Damping time, horizontal emittance, and energy spread at four energy points in the STCF collider rings

| | $E$ (GeV) | $B_w$ (T) | $\tau_0$ (ms) | $\tau_u$ (ms) | $\varepsilon_0$ (nm.rad) | $\varepsilon_w$ (nm.rad) | $\sigma_{E0}$ ($10^{-3}$) | $\sigma_{Ew}$ ($10^{-3}$) |
|---|---|---|---|---|---|---|---|---|
| Sinusoidal Field | 1.0 | 1.6 | 586.2 | 54.5 | 2.4 | 0.7 | 0.29 | 0.52 |
| | 1.5 | 1.6 | 173.8 | 32.6 | 5.4 | 1.7 | 0.43 | 0.62 |
| | 2.0 | 1.6 | 73.3 | 21.3 | 9.7 | 3.6 | 0.57 | 0.71 |
| | 3.5 | 1.6 | 13.7 | 7.6 | 29.7 | 17.3 | 1.00 | 1.00 |
| Realistic Field | 1.0 | 1.6 | 586.2 | 42.4 | 2.4 | 0.9 | 0.29 | 0.55 |
| | 1.5 | 1.6 | 173.8 | 25.9 | 5.4 | 1.8 | 0.43 | 0.66 |
| | 2.0 | 1.6 | 73.3 | 17.4 | 9.7 | 3.5 | 0.57 | 0.75 |
| | 3.5 | 1.6 | 13.7 | 6.7 | 29.7 | 15.9 | 1.00 | 1.03 |
| Rectangular Field | 1.0 | 1.6 | 586.2 | 28.6 | 2.4 | 1.3 | 0.29 | 0.57 |
| | 1.5 | 1.6 | 173.8 | 18.0 | 5.4 | 2.2 | 0.43 | 0.69 |
| | 2.0 | 1.6 | 73.3 | 12.5 | 9.7 | 3.6 | 0.57 | 0.79 |
| | 3.5 | 1.6 | 13.7 | 5.3 | 29.7 | 14.0 | 1.00 | 1.06 |

Wiggler radiation is converted into substantial synchrotron light. At 2 GeV and 2 A, a single wiggler produces 64 kW of synchrotron radiation power. One straight section with 8 wigglers thus emits 512 kW in total. Figure 2.10-3 shows the power distribution of this radiation at the downstream end. To manage this radiation, photon masks must be placed along the beamline to prevent direct strikes on the vacuum chamber. The remaining radiation can be extracted near the first bending magnet in the downstream arc and absorbed by a high-power photon absorber. In the future, synchrotron light from the outer ring could also be extracted for high-flux applications.

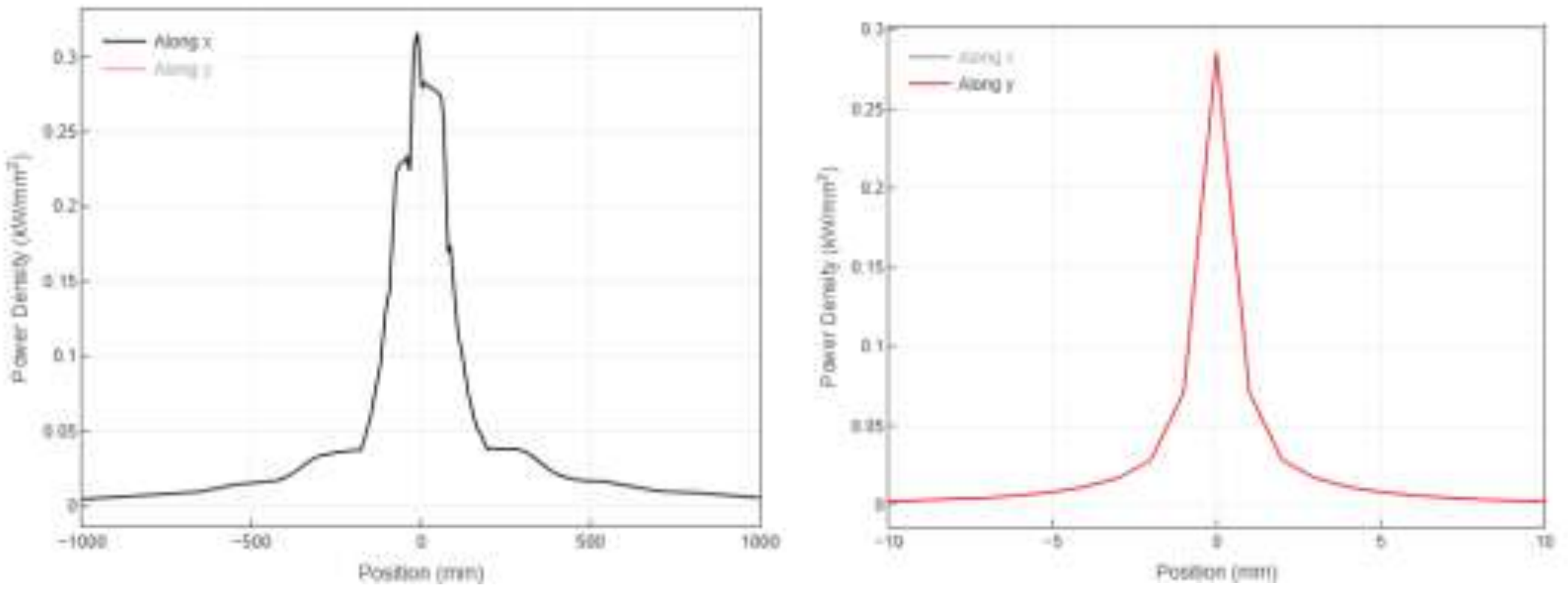

Figure 2.10-3: Superimposed power distribution from 8 damping wigglers at the end of a straight section (left: X direction, right: Y direction)

### 2.10.5 Evaluation of the Effects of Damping Wigglers on Linear Lattice and Nonlinear Beam Dynamics

#### *2.10.5.1 Impact on Linear Lattice*

Wiggler magnets with vertical dipole fields ($B_y$) exert additional vertical focusing forces on the beam, which can modify the lattice's optical functions and working point. The magnitude of this perturbation is proportional to $B_W^2 L_W \beta / E^2$, where $\beta$ is the beta function at the wiggler location. To mitigate such effects, the linear lattice design incorporates the wigglers from the outset, treating them as integral components of the storage ring. Triplet quadrupoles are placed adjacent to the wigglers to match the optical functions. These triplet quadrupoles provide sufficient tuning flexibility: as the beam energy in the STCF collider rings varies between 1 and 3.5 GeV, which affects the relative field strength of the wigglers, the quadrupole gradients can be adjusted accordingly to restore the desired optics.

#### *2.10.5.2 Impact on Nonlinear Beam Dynamics*

Wiggler magnets also influence the nonlinear beam dynamics of the lattice, primarily due to the roll-off effect in the magnetic field—that is, the non-uniformity of $B_y$ in the horizontal and vertical directions [57]. The perturbations exerted on the beam in the horizontal and vertical directions are proportional to $B_W L_W \lambda_W^2\, \partial B_y / \partial x$ and $B_W L_W \lambda_W^2\, \partial B_y / \partial y$, respectively, where $\partial B_y/\partial x$ and $\partial B_y/\partial y$ are the gradients of the vertical field in the horizontal and vertical planes. High-order field components, such as sextupole and octupole terms in the integrated wiggler field, introduce additional nonlinearities that can shrink the dynamic aperture.

When the wiggler's good field region is sufficiently wide and the field quality is high, these high-order components remain small, thereby minimizing their detrimental impact on nonlinear beam dynamics. Consequently, the design and manufacturing of wiggler magnets must ensure a large good field region and high magnetic field uniformity.

## 2.11 Physics Design Related to MDI

### 2.11.1 MDI Design Requirements

The Machine Detector Interface (MDI) is the interface region between the collider rings and the detector spectrometer, playing a key role in both routine accelerator operation and data collection for physics experiments. Since both beams must pass through the center of the detector and collide at the IP and a tightly focused beam spot is required to achieve high luminosity via superconducting quadrupole magnets, the inner part of the detector needs to be positioned very close to the IP. Thus, the available space for accelerator components becomes extremely constrained. For a third-generation e+/e- collider like STCF, the complexity of the MDI is significantly increased, mainly due to the ultra-high luminosity, large crossing angle, strong focusing, and short beam lifetime, all of which must be addressed in the design.

The accelerator's interaction region involves multiple technical systems, including physics design (Lattice design and complex beam dynamics such as nonlinear effects, chromaticity correction, beam-beam interactions, collective effects, etc.), superconducting magnets and cryostats, beam diagnostics, vacuum system, mechanical structures, collimation, and alignment. On the detector side, MDI-related systems include background simulation and shielding, spatial arrangement of the inner detectors and electronics, luminosity monitoring, and feedback. All of these aspects require coordinated study between accelerator and detector teams, with the design objective being to ensure spatial compatibility between accelerator and detector systems, enabling the accelerator to achieve its target luminosity and stable operation while minimizing background levels for collision experiments.

### 2.11.2 Impact of Interaction Region Design on MDI

The lattice design of the interaction region determines the layout and longitudinal space of accelerator components within the MDI. To achieve the maximum possible solid angle for physics experiments, the detector imposes tight constraints on the transverse space available to accelerator elements within the MDI. Figure 2.11-1 shows the layout of the STCF MDI, detailing the preliminary positions and dimensions of components within ±3.5 m of the IP, including detector boundaries, central beam pipe, vacuum pipe structures, superconducting magnets and cryostat, beam diagnostics components, mechanical supports, etc. Spatially, the detector requires that all accelerator components within the interaction region be confined within a conical region with a 15° opening angle centered at the IP.

The cryostat contains compensating solenoids, two superconducting quadrupoles, several correctors, and a shielding solenoid. The compensating solenoid upstream of the superconducting quadrupoles and closer to the IP, is used to cancel the integrated longitudinal magnetic field of the detector solenoid. The shielding solenoid, winding around the superconducting quadrupoles, ensures that the $B_z$ field is zero within the quadrupole region. The compensating solenoid after the superconducting quadrupole QF2 is used to counteract the long tail of the detector solenoid field.

The central beam pipe consists of the IP pipe and a transition pipe. The IP pipe is made of beryllium, with its inner wall coated in gold to shield against synchrotron radiation and reduce impedance. Its diameter is 30 mm. The transition pipe expands longitudinally from 110 mm to 500 mm, transitioning from a 30 mm round pipe to a racetrack-shaped cross-section measuring 60 mm wide and 30 mm high. At the Y chamber—Remote Vacuum Connector (RVC)—the beam pipe splits into two 30-mm diameter channels up to the end of QD0, then transitions over 0.5 m to a 50 mm diameter at the entrance of QF1, and finally increases within 0.5 m to 67 mm. The transition from the dual-aperture region in the superconducting magnets to the single-aperture central beam pipe considers impedance minimization as a key design focus.

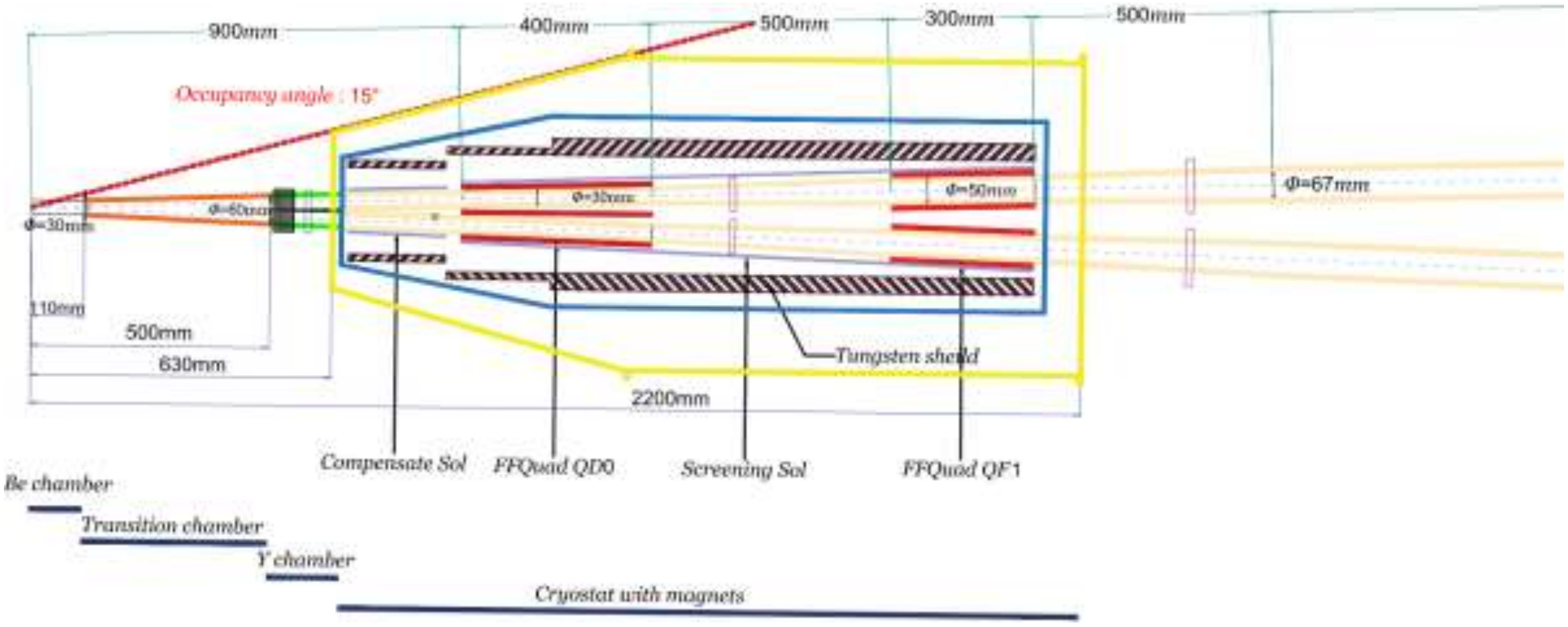

Figure 2.11-1: STCF MDI layout (right side only, layout is symmetric about IP)

The primary source of synchrotron radiation entering the MDI is the upstream bending magnet B0. In the lattice design, this magnet is deliberately placed away from the IP (about 8.5 m) and set to a relatively weak bending angle of only 1°. Synchrotron radiation distributions were analyzed for electron/positron beams entering the IP from both the outer ring and inner ring, as shown in Figure 2.11-2. The directly irradiated regions are marked in yellow. Only the positron beam case is illustrated for simplicity. The synchrotron radiation power deposited in various MDI regions is listed in Table 2.11-1. For beams entering from the outer ring, radiation does not directly strike the central beryllium pipe; for beams entering from the inner ring, it does. Therefore, we prefer the beams incident on the IP from the outer ring.

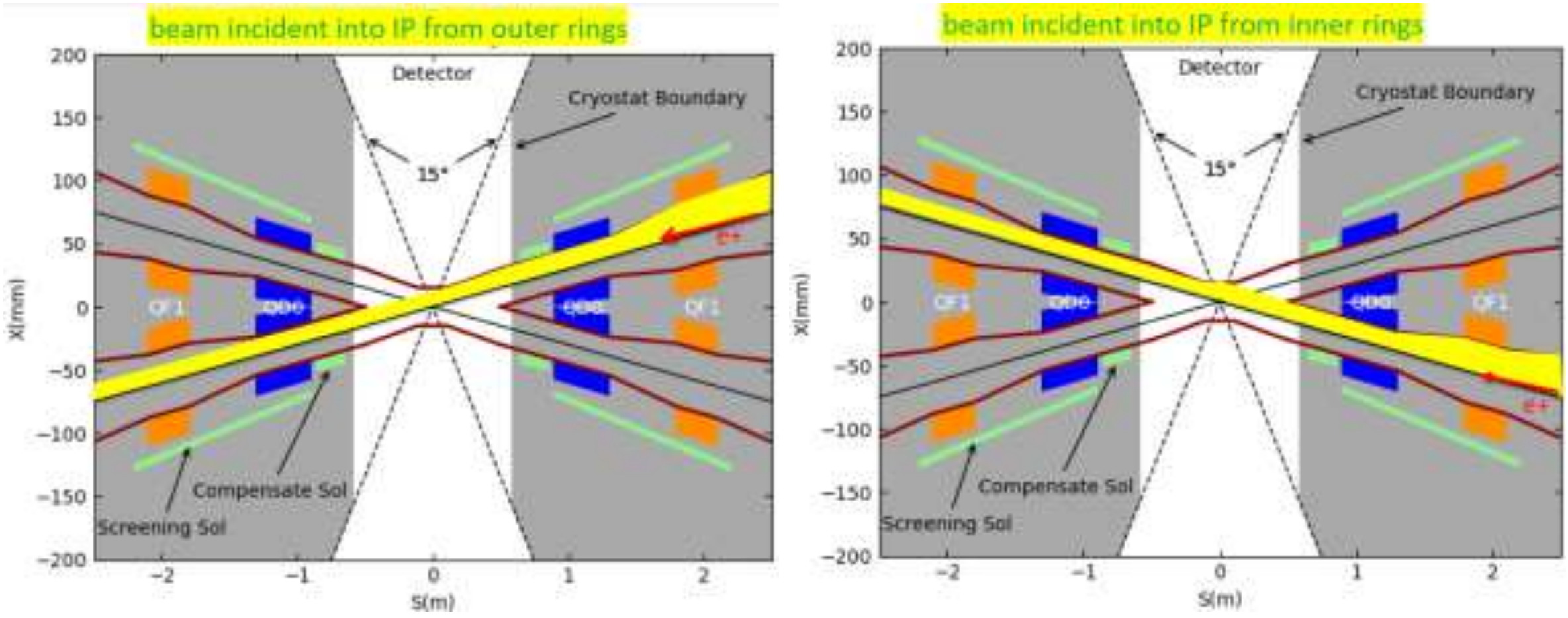

Figure 2.11-2: Synchrotron radiation distribution in the MDI for beam incident into IP from outer rings and inner rings

Table 2.11-1: Synchrotron radiation power distribution at the MDI for different beam injection directions.

| Synchrotron Power (W) | B0 | Central Be Pipe | Conical Transition Pipe | Y-section Pipe | Cryostat |
|---|---|---|---|---|---|
| Outer Ring @ 2 GeV | 137.2 | 0 | 3.64 | 0.43 | 17.8 |
| Outer Ring @ 3.5 GeV | 1287.3 | 0 | 34.1 | 4.02 | 167.2 |
| Inner Ring @ 2 GeV | 137.2 | 4.02 | 0 | 0.27 | 17.7 |
| Inner Ring @ 3.5 GeV | 1287.3 | 37.7 | 0 | 2.5 | 165.7 |

The definition of beam stay-clear regions in the interaction region is also critical for MDI magnet design. In second-generation $e^+/e^-$ colliders such as PEP-II and BEPCII, the stay-clear region is typically defined as 15–20 times the beam envelope $\sigma_x/\sigma_\gamma$ plus 1–2 mm of orbit deviation. For newer-generation colliders such as SuperKEKB, SuperB, and FCC-ee, due to the extremely small beam sizes at the IP, orbit deviations near superconducting quadrupoles must be strictly controlled (less than 100 μm), thus, the stay-clear region is defined without considering orbit deviation. CEPC follows the BEPCII definition.

Here we provide both interpretations. Figure 2.11-3 shows the beam stay-clear boundaries at 2 GeV and 3.5 GeV. The magenta and green dashed lines represent the stay-clear boundaries for the positron and electron beams, respectively. At 2 GeV, the horizontal stay-clear region is defined as $22\sigma_x + 2$ mm (or $24\sigma_x$), and the vertical as $22\sigma_y + 2$ mm (or $26\sigma_y$). At 3.5 GeV, due to the larger beam emittance, the stay-clear region becomes smaller: horizontal $9.5\sigma_x + 2$ mm (or $10.5\sigma_x$), vertical $9.5\sigma_y + 2$ mm (or $11.5\sigma_y$). In these calculations, the horizontal emittance $\varepsilon_x$ is taken as 5 nm at 2 GeV and 27 nm at 3.5 GeV. The vertical emittance $\varepsilon_y$ is assumed to be 5% of $\varepsilon_x$, i.e., $\varepsilon_y = 0.05\varepsilon_x$. The small beam stay-clear regions at 3.5 GeV can be improved by increasing the $\beta^*_{x/y}$ at IP. Table 2.11-2 lists the current superconducting quadrupole parameters near the IP that satisfy the IR optics requirements, including position, field gradient, effective length, beam pipe radius, and good field region.

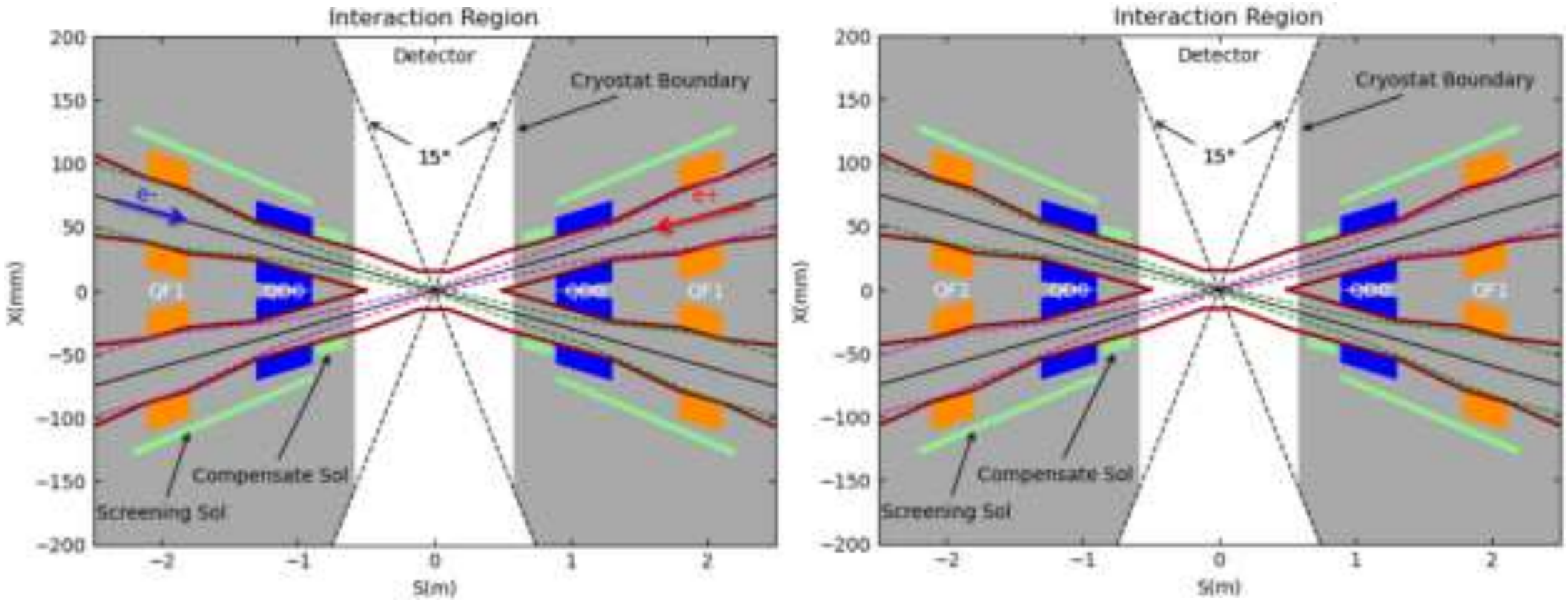

Figure 2.11-3: Beam stay-clear boundaries at 2 GeV and 3.5 GeV.

Table 2.11-2: Layout and Parameters of the Final Focus Superconducting Quadrupoles

| Superconducting Magnet | Distance from IP (m) | Beam Stay-Clear Region $2\times(22\sigma_{x,y}+2)$ (mm) | $e^+/e^-$ Beam Center Separation (mm) | Good Field Region H/V (mm) | Field Gradient (T/m) | Beam Pipe Inner Radius (mm) | Effective Length (m) |
|---|---|---|---|---|---|---|---|
| QD0 | 0.9–1.3 | Entrance: 15.2 / 24.5 | 54 | 24.6 × 27.3 | 50 | 15 | 0.4 |
| | | Middle: 18.7 / 27.3 | 66 | | | | |
| | | Exit: 24.6 / 26.3 | 78 | | | | |
| QF1 | 1.8–2.1 | Entrance: 43.2 / 19.2 | 108 | 48.3 × 19.2 | 40 | 25 | 0.3 |
| | | Middle: 47.3 / 17.6 | 117 | | | | |
| | | Exit: 48.3 / 17.0 | 126 | | | | |

**Note:** In the beam stay-clear calculations, the horizontal emittance is taken as 5 nm, and the vertical emittance is assumed to be 5% of the horizontal (i.e., 5% coupling).

### 2.11.3 Beam Loss and Experimental Background Simulation

The primary sources of experimental background in the interaction region are non-collisional beam losses, Bhabha scattering, synchrotron radiation effects, and beam-induced bremsstrahlung. In a low-energy $e^+e^-$ collider such as STCF, the latter two contribute less significantly, with beam loss and Bhabha scattering being the dominant factors. Bhabha scattering is an inevitable process accompanying collision experiments and is essentially unavoidable. However, while beam loss is also inevitable, its impact on experimental background can be mitigated through proper measures.

As a newly constructed accelerator, STCF will aim to minimize beam loss in the interaction region by using collimators placed at other locations in the collider rings to intercept particles from Touschek scattering and transverse beam halo, as described in the beam collimation section. Synchrotron radiation primarily affects the thermal load on the beam pipe in the interaction region, but its contribution to experimental background, particularly from hard X-ray components at the high-energy end, must also be considered. Beam-induced bremsstrahlung can affect background levels primarily when operating at the high-energy region near 3.5 GeV, which is also a major reason for maintaining ultra-high vacuum conditions in the interaction region.

Except for the central beam pipe section (particularly the beryllium pipe), which serves as the passage for reaction products entering the detector and thus cannot be shielded, all other beam pipe sections and components need to be equipped with shielding absorbers. These are designed to reduce the probability that secondary particles from upstream beam collimators, vacuum-induced bremsstrahlung, or synchrotron radiation reach the detector or the superconducting magnet coils. At the same time, care must be taken to minimize the impedance introduced by these shielding structures.

The simulation work related to collision-induced background is still ongoing. Preliminary results indicate that the background levels are within controllable limits.

## 2.12 Preliminary Considerations for Future Polarized Beams

The design of polarized beams involves two key considerations: achieving longitudinal polarization at the IP and satisfying the spin matching condition to mitigate depolarization effects. Longitudinal polarization at the IP can be achieved either by implementing Siberian snakes in the ring or by installing spin rotators at suitable locations near the IP [1-2]. To mitigate depolarization, the spin tune should be kept as far away from integer resonance as possible—ideally around half-integer. This can be done by using a Siberian snake to fix the spin tune at an integer value, or by employing a spin rotator to slightly adjust the beam energy such that the fractional part of $a\times\gamma$ equals 0.5, where $a$ is the anomalous magnetic moment ($a$=0.00115965218).

At STCF, the preferred approach is to use spin rotators to achieve both beam polarization and satisfy the spin matching condition. The advantage of this method is that it provides greater flexibility in the collider ring lattice design, without requiring dedicated structures to accommodate Siberian snakes. This enables the use of a two-fold symmetric lattice layout. Furthermore, since STCF plans to adopt a bunch swap-out injection scheme and the beam lifetime is around 200 seconds, the requirement on depolarization time is significantly relaxed, and the spin rotator approach is sufficient to achieve the desired longitudinal polarization and depolarization suppression. However, it is still unclear if the present lattice is compatible with polarized beam operation in the moment, and more serious study is needed in the future. Anyway, In Phase I of the STCF, polarized beams will not be considered, only some space for polarization components will be reserved in the lattice design.

# 3 Injector Accelerator Physics

## 3.1 Introduction to the Injector Design

### 3.1.1 Design Requirements and Specifications

As an integral part of the STCF accelerator complex, the injector is itself a sophisticated accelerator system. Its primary role is to provide high-quality, full-energy electron and positron beams for injection into the collider rings, serving as a key component to ensure the realization of high luminosity at STCF. The collider rings operate in the top-up mode, and at the moment, the beam injection design considers two different schemes: off-axis injection and bunch swap-out injection.

The off-axis injection scheme aims to replenish particle losses during collider operation. In this approach, the electron beam is accelerated and transported directly to the collider electron ring via a linac and beamline. For positrons, the beam that is produced by the high bunch-charge electron beam onto a target is first accelerated to 1 GeV, then damped in a damping ring, then further accelerated through the same main linac and transported into the collider positron ring. In contrast, the bunch swap-out injection scheme seeks to replace the bunches one by one in both rings whose charge has dropped below a threshold with newly injected high-charge bunches. In this scheme, high-charge electron bunches are provided directly from the electron gun and accelerated for on-axis injection into the collider electron ring. For positrons, bunch charge accumulation is required in an accumulator ring due to relatively low conversion efficiency from electrons to positrons, followed by emittance damping and acceleration before on-axis injection in the collider positron ring. The bunches to be replaced must be extracted in advance.

After comparison and evaluation, an injector scheme compatible with both off-axis and swap-out injection is proposed. This allows initial implementation with off-axis injection to control the construction costs, while enabling future upgrade to swap-out injection when needed.

Though the two schemes impose different design requirements on the injector, their main objective is the same: to maintain approximately 95% of the nominal electron and positron beam currents (2 A) in the collider rings, assuming a lower limit beam lifetime of 200 seconds. The off-axis injection scheme places higher demands on beam quality (e.g., emittance and energy spread), while the swap-out scheme requires high bunch charge and a higher repetition rate for part of the injector linac. The key injector design parameters are summarized in Table 3.1-1. The extremely short beam lifetime at STCF poses substantial challenges for beam injection at the collider rings and also for the design and construction of the injector.

Traditional off-axis injection achieves higher beam utilization, lower bunch charge per injection, and lower frequency requirements for the injector linac, making it relatively easier to implement. However, the associated larger beam loss during injection introduces significant background noise to the experiment and leads to emittance growth via coupling with the beam-beam effects.

To address these issues, the newly appeared swap-out injection scheme is also under study. It allows higher injection efficiency of individual bunches in theory and can suppress the experiment background from injection beam losses. However, this scheme results in a low beam utilization if the extracted bunches are not reused, which is in this case. In addition, the large bunch charge poses challenges in emittance control, positron bunch charge accumulation, and achieving a small vertical emittance to the injector design.

The swap-out injection scheme requires the beamline and accelerator components with large physical apertures, while ensuring strictly controlled emittance growth. Furthermore, swap-out injection demands that the collider rings pre-extract the bunches to be replaced before injection, adding complexity to the extraction system that needs only to extract the whole beam in case of machine protection and necessitating a more sophisticated beam dump. One can also consider the reuse of the spent high-charge bunches for other kinds of applications. A key advantage is that an injector designed for swap-out injection can also support off-axis injection with relatively little modification.

Table 3.1-1: Key Injector Design Parameters

| Parameter | Off-Axis Injection | Swap-Out Injection | Unit |
|---|---|---|---|
| Electron gun type | Photo-/thermionic cathode | Thermionic/PITZ cathode | – |
| Linac RF frequency | 2998.2 | 2998.2 | MHz |
| Bunch charge into collider | 1.5 | 8.5 | nC |
| Injection energy / nominal energy | 1.0–3.5 / 2.0 | 1.0–3.5 / 2.0 | GeV |
| Injected electron bunch geometric emittance (X/Y) | ⩽6/2 | ⩽30/15 | nm·rad |
| Injected positron bunch geometric emittance (X/Y) | ⩽6/2 | ⩽30/15 | nm·rad |
| Injected bunch energy spread (RMS) | ⩽0.1 | ⩽0.5 | % |
| Injected bunch length (RMS) | <7 | <7 | mm |
| Repetition rate (electron) | 30 | 30 | Hz |
| Repetition rate (positron) | 30 | 30 | Hz |
| Positron damping/accumulation ring | Damping ring | Accumulator ring | – |
| Bunch charge at ring entrance | 1.5 | 2.9 | nC |
| Entrance emittance (X/Y) | ⩽1400 | ⩽1400 | nm·rad |
| Entrance energy spread (RMS) | ⩽0.1 | ⩽0.3 | % |

| Parameter | Off-Axis Injection | Swap-Out Injection | Unit |
|---|---|---|---|
| Exit emittance (X/Y) | ⩽11 / 0.2 | ⩽30 / 5.2 | nm·rad |
| Ring RF frequency | 499.7 | 499.7 | MHz |
| Number of bunches in ring | 5 | 5 | – |

In the conceptual design phase, the injector physics designs for both off-axis and swap-out injections have been carried out. The injector design compatible with the two injections is also being developed.

Given the wide operating energy range and very short beam lifetime of the collider rings—especially under swap-out injection— the injector must deliver electron and positron beams with variable energy (1–3.5 GeV), high repetition rate (30/90 Hz), low emittance (<30 nm·rad), and high bunch charge (8.5 nC) for swap-out injection. This imposes substantial challenges on the injector design, such as:

- **Electron and positron linacs**: Including three-stage linacs for both injection schemes (see Figure 3.1-1). The low-energy sections must handle large bunch charges and control emittance growth. For positron linacs (PL and SPL), a larger positron emittance must be accommodated. The maximum linac repetition rate is expected to be 90–100 Hz.
- **Positron generation, capture, and pre-acceleration**: All schemes adopt conventional target-based positron production. Key tasks include efficient generation, collection, and pre-acceleration of positrons, involving adiabatic matching, e+/e− separation, and longitudinal compression, requiring detailed tracking simulations.
- **Damping and accumulator rings**: The damping ring (DR) reduces positron emittance via synchrotron radiation. In the swap-out scheme, the accumulator ring (AR) simultaneously accumulates and damps positrons to meet the collider's bunch charge and emittance requirements. The hybrid scheme uses both DR and AR in sequence.
- **Main linac**: Accelerates 1 GeV e+/e− bunches to 1-3.5 GeV for the e+/e- injection into the collider rings, operating alternately at 30 Hz for both the electron and positron beams.
- **Beam transport lines**: Includes the bypass beam line (for off-axis injection only), injection and extraction lines for the damping and accumulator rings, and transport lines from the main linac to the collider rings. The physics design is focused on the beam matching, emittance preservation, and high transmission efficiency.

The compatible injector scheme will be described separately in Section 3.6.

### 3.1.2 Overview of the Design Schemes

Based on the physics design requirements and specifications of the injector, and referencing layouts of similar international accelerators, schematic layouts of the injector for both the collider ring injection schemes are shown in Figure 1.2-1.

The linac system for each of the two injection schemes incorporates two electron sources—one corresponding to a low-emittance electron gun for direct injection into the collider electron ring, and the other a high-charge electron gun for positron production via target bombardment. For the off-axis injection scheme, the low-emittance electron source is a photocathode (abbr. as PC) RF gun producing 1.5 nC, and the high-charge electron source is a thermionic-cathode (abbr. as TC) RF gun producing 10 nC. For the swap-out injection scheme, the electron sources comprise a thermionic-cathode gun (8.5 nC) or an L-band PITZ photocathode gun (8.5 nC), along with an 11.6 nC thermionic-cathode gun. Different bunching schemes are used for beams from different types of electron guns: a chicane scheme is applied for beams from the RF photocathode guns, while for all three thermionic-cathode guns, a sub-harmonic buncher (SHB) and a fundamental harmonic buncher system are adopted.

In the off-axis injection scheme (Figure 1.2-1a), electrons from the two sources pass through respective bunching sections (BS) and pre-injector sections (PIS) for bunching and initial acceleration. They are then merged at 200 MeV at the entrance of the first main linac section (EL1), which accelerates the beam to 1.0 GeV. From there, the beam from the RF photocathode gun is delivered via a bypass line to the positron/electron main linac (ML) and accelerated to the injection energy required by the collider rings. The 10 nC electron beam from the thermionic gun is further accelerated to 1.5 GeV by a second linac section (EL2) before striking a tungsten target to produce positrons. The resulting positrons are collected to form a bunch charge of 1.5 nC and pre-accelerated to 200 MeV. In the main accelerator section of the positron linac (PL), the positrons are accelerated to 1.0 GeV and injected into a damping ring (DR) for emittance reduction. The DR uses a single-turn (on-axis) injection scheme, storing five bunches simultaneously. Each bunch undergoes damping from an initial emittance of 1400 nm·rad to below 11 nm·rad. The extraction repetition rate of the DR matches the injection of the collider rings. The damped positron bunches are then transported to the main linac ML and accelerated to the injection energy of the collider rings.

In the swap-out injection scheme (Figure 1.2-1b), 8.5 nC electron beams from a thermionic or PITZ L-band photocathode gun are bunched in Bunching Section 1 (BS1) and accelerated to 1.0 GeV in the linac section SEL1. They are then accelerated by the ML to the collider ring injection energy. The electron beam from the 11.6 nC thermionic is bunched in BS2, then accelerated to 2.5 GeV in SEL2 before striking a tungsten target to generate positrons. A positron beam of 2.9 nC in bunch charge is collected and pre-accelerated to 200 MeV, then accelerated to 1.0 GeV in the PL. These positrons are injected into an accumulator ring (AR) for bunch charge accumulation and emittance damping. The AR uses a multi-turn off-axis injection scheme, sequentially filling all stored bunches and allowing sufficient damping time before extraction. The injection repetition rate is a multiple of the extraction one, which is determined by the accumulation ratio. Once the accumulation and damping are complete, positron bunches are extracted and accelerated by ML to the required energy for injection into the collider rings.

Swap-out injection demands high bunch charges and stringent beam quality control for both electron and positron beams. Due to the positron production mechanism, the international community widely adopts the mechanism in which a high-charge electron beam bombards a production target. Given the cost sensitivity to electron beam energy and the technical limits of the bunch charge from the electron gun and linac, instead of a damping ring for the off-axis injection a positron accumulator ring together with a higher repetition rate and modest beam energy for the driving electron linac are adopted here, which will accumulate positron bunch charges by multiple injections and damp the transverse and longitudinal emittance simultaneously.

For electrons, taking into account technology maturity and the necessity to reduce cost, direct injection of high-quality and high-charge bunches into the collider electron ring to replace the degraded bunches is explored. The key issues are still with the production of high-charge bunches and the acquisition of low beam emittance. Two feasible approaches exist: (1) using conventional thermionic guns to generate bunches of larger than 8.5 nC. This approach is technically reliable and does not require a high-power laser system, but the relatively large emittance originating from the gun may necessitate beam scraping to reduce the beam emittance before injection into the ring. (2) Using an L-band photocathode gun to generate electron bunches of charge larger than 8.5 nC and high beam quality. Though technically promising, this approach places high demands on the driver laser, high vacuum, and lifetime of the semiconductor photocathode for the production of high bunch charge with a high repetition rate, which still requires engineering validation.

Despite differences in the injector physics designs corresponding to the two collider ring injection schemes, they share common requirements: strict control of beam emittance along beam acceleration, high capture efficiency of positrons, optimization design of the positron damping and accumulator rings, and optimization of the injection transport lines. Below physics design details for each scheme on how to meet the demands of bunch charge and beam quality are given.

## 3.2 Injector Physics Design for Off-axis Injection

### 3.2.1 Electron Source and Low-Energy Acceleration Section

Based on the electron charge requirements for direct injection into the collider electron ring and positron production via target bombardment, two electron guns are used to generate bunches of different charges. Consequently, the system is divided into a photocathode low-energy section and a thermionic cathode low-energy section. At the exit of the two low-energy sections, the beams from both electron guns are merged into the first main acceleration section through a combining segment. The layout of the dual-gun low-energy section is shown in Figure 3.2-1.

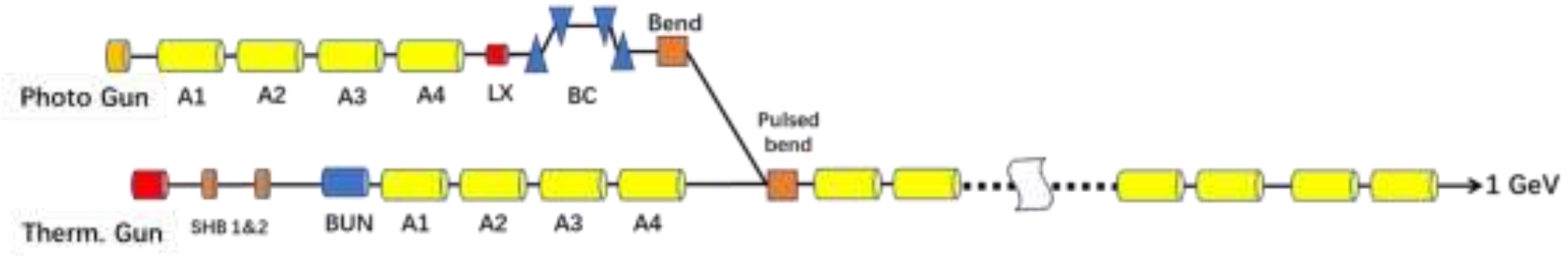

Figure 3.2-1: Layout of the dual-electron-gun low-energy section

#### *3.2.1.1 Photocathode Low-Energy Section*

The quality of the electron beam (e.g., emittance, current, bunch length, etc.) provided by the low-energy section is crucial in determining the final beam quality injected into the collider electron ring. To maximize the beam quality from the injector, a well-established S-band 2998 MHz photocathode RF gun, already proven at major international accelerator facilities and also in China, is adopted for STCF, which is capable of delivering the high beam quality that meets the stringent requirements of STCF. The basic configuration of this section is shown in Figure 3.2-2.

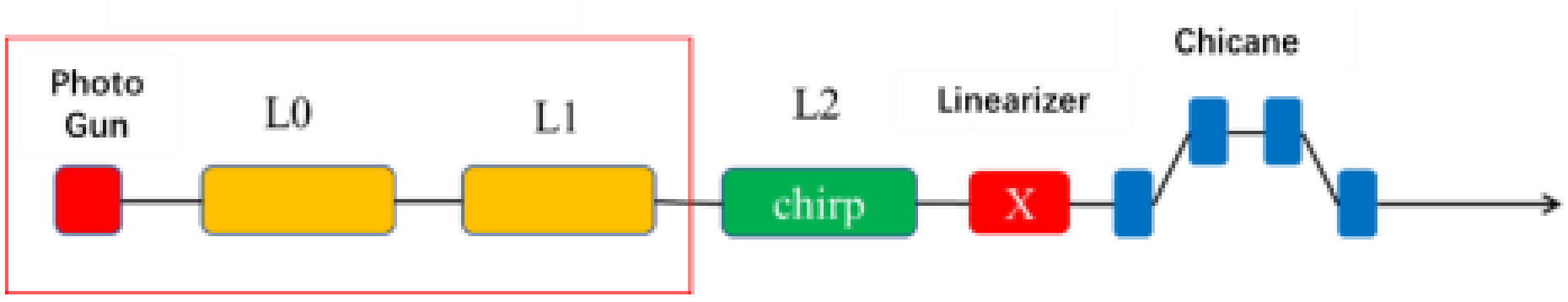

Figure 3.2-2: Layout of the photocathode low-energy section

The gun operates at a cathode gradient exceeding 100 MV/m and delivers a beam energy of ~4 MeV at the exit of the gun. The beam is then accelerated to 120 MeV using two S-band accelerating structures, L0 and L1. The next stage, L2, consists of two more S-band structures powered by a shared RF source with an energy doubler, accelerating the beam to over 200 MeV while imparting an energy chirp. Then it is followed by an X-band accelerating structure as the linearizer and a chicane structure as the magnetic bunch compressor that compresses the bunch length to below 1 ps (equivalent to 1° RF phase at S-band frequency). A Ti: Sapphire mode-locked laser system, synchronization control system, and various power supplies are also required.

Compared with a more conventional low-energy section based on a thermionic cathode, a key problem for the photocathode scheme is how to maintain low emittance at high bunch charge. Typically, an external solenoid is used to focus the beam after the RF gun. After a drift section, the linear space charge effects can be partially compensated, which mitigates the emittance growth.

According to the design requirements, the photocathode low-energy section must deliver bunches with a charge larger than 1.5 nC and a bunch length of less than 1 ps, to maintain low energy spread in the subsequent linac sections for direct and efficient injection into the collider electron ring. A global optimization of the parameters for the low-energy section is therefore necessary.

The electron beam properties are influenced mainly by the following four components:

1. **Accelerating structures** – provide energy gain and longitudinal focusing.
2. **Focusing elements** (e.g., solenoids) – provide compensation for the transverse emittance and transverse focusing to confine the beam size.
3. **Laser profile on the photocathode**– the transverse and longitudinal distributions of the drive laser affect space charge forces and nonlinearities.
4. **Photocathode material** – its thermal emittance has a direct impact on the beam brightness.

Given the complex interplay between these parameters and the spatial constraints of the injector, a manual or scan-based optimization approach is inefficient. Instead, multi-objective optimization algorithms, such as genetic algorithms, are employed and integrated with ASTRA beam dynamics simulations. Realistic 3D electromagnetic fields of the components are used in simulations to reflect true performance.

The optimization targets the bunch length and transverse emittance at the exit of the second S-band linac section. Three types of laser profiles are considered (see Table 3.2-1):

Table 3.2-1: Laser distribution cases used in optimization

| Case | Transverse Laser Profile | Longitudinal Laser Profile |
|---|---|---|
| Case 1 | Truncated Gaussian | Gaussian |
| Case 2 | Truncated Gaussian | Flat-top |
| Case 3 | Uniform | Flat-top |

Charge values of 1 nC, 1.5 nC, and 2 nC are studied. Results show that transverse beam quality is relatively insensitive to transverse laser shape, so a truncated Gaussian is used to avoid energy losses from laser shaping. However, a flat-top longitudinal laser profile significantly improves emittance. As a result, a flat-top longitudinal and truncated Gaussian transverse profile is chosen. From optimization and space constraints, the entrance to the first accelerating structure is located 1.5 m from the cathode, the second at 5.02 m, and the center of the second solenoid at 2.0 m.

Simulation results at the exit of the two S-band sections, including phase space, current, and energy spectrum, are shown in Figure 3.2-3.

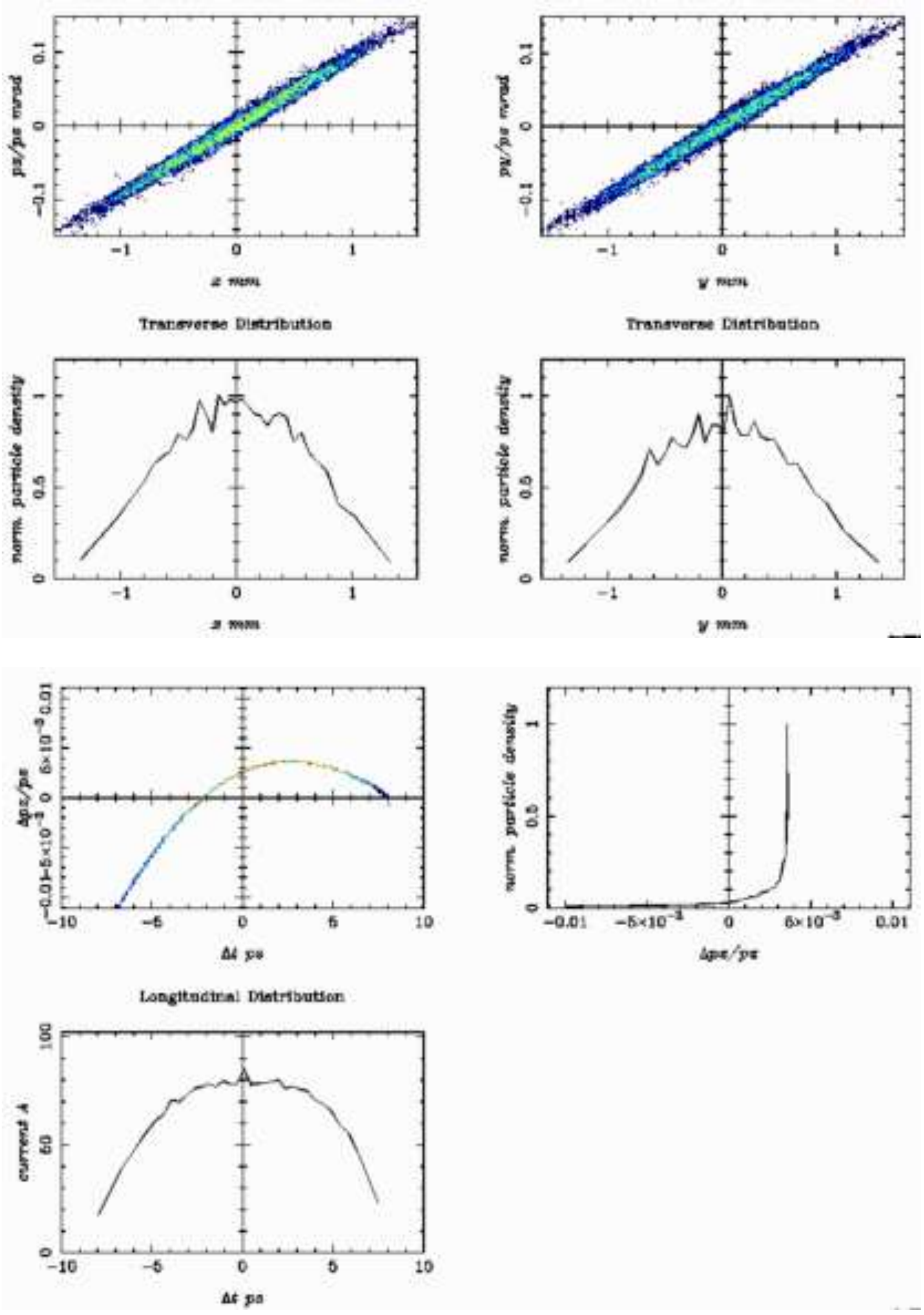

Figure 3.2-3: Beam distributions at the exit of the two S-band accelerating structures in the photocathode low-energy section

In the following L2 section, the RF phase is set before the peak to induce an energy chirp. The third and fourth S-band structures generate the chirp, while an X-band linearizer and a Chicane (providing negative R56) compress the bunch length to ~1 ps at ~200 MeV. At this point, both emittance and energy spread are optimized (<0.1%). With this setup, normalized emittance under 1.5 nC operation remains below 2 mm·mrad for all studied laser distributions. The final beam distributions in transverse and longitudinal directions are shown in Figure 3.2-4.

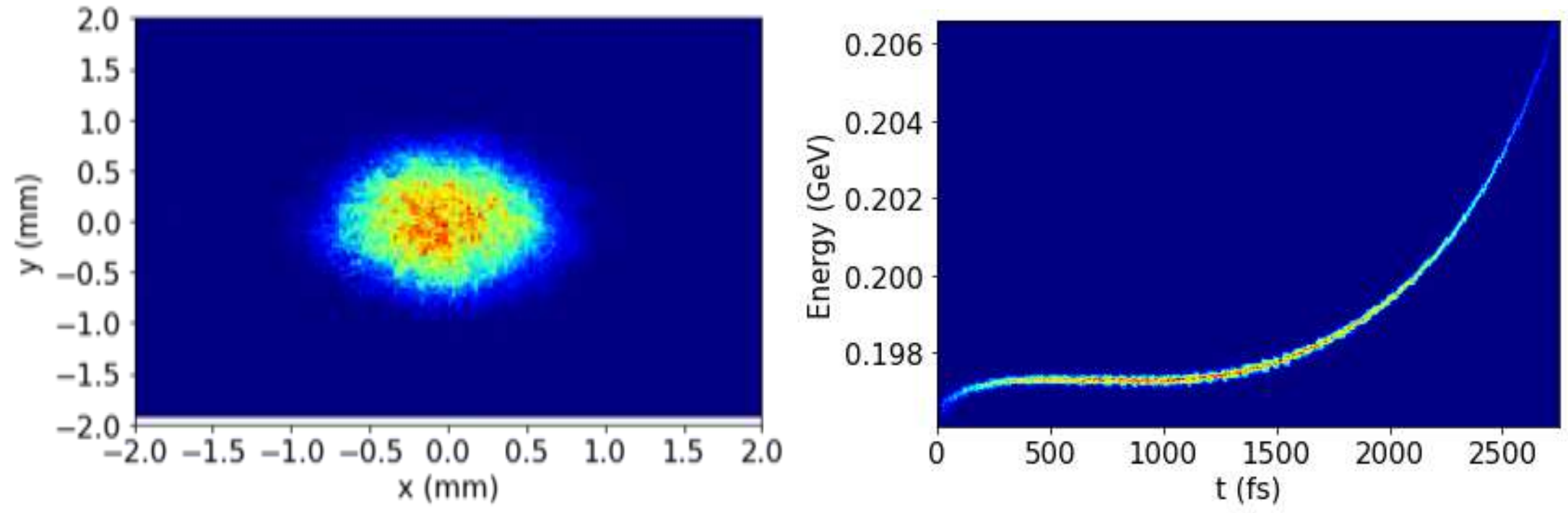

Figure 3.2-4: Transverse (left) and longitudinal (right) beam distributions at the exit of the photocathode low-energy section

### *3.2.1.2 Thermionic Cathode Low-Energy Section*

**Bunching Section**

The bunching section primarily consists of the electron gun and a bunching system. Its main function is to longitudinally compress the low-energy electron beam emitted by the gun to achieve a short bunch length (5-10 ps) while maintaining good capture efficiency, transverse phase space, and emittance to satisfy the requirements for further acceleration in the linac's high-energy section. To deliver a high-charge, high-energy electron beam for positron generation via a tungsten target, we choose a thermionic cathode electron gun capable of producing a bunch with an FWHM (Full Width at Half Maximum) greater than 1 ns. Considering capture losses, the total charge at the source should exceed 11 nC. Immediately downstream of the gun is the bunching system, which includes two sub-harmonic bunchers, one S-band fundamental buncher, and one S-band traveling wave accelerating structure, as shown in Figure 3.2-5.

The two sub-harmonic bunchers and one S-band fundamental buncher, located downstream of the gun, serve to longitudinally modulate the velocity of the non-relativistic electron beam and compress its bunch length, enabling effective injection into the main linac. A 1.3 ns FWHM beam from the gun is compressed to about 10 ps FWHM after passing through this bunching system.

After bunching, the 10-ps beam is injected into the accelerating structure L0-1 and further accelerated to 65 MeV, which is sufficient for subsequent acceleration to 1.5 GeV in the main linac for positron production via a tungsten target. The fundamental buncher and accelerator L0-1 share a klystron power source, while the two sub-harmonic bunchers are powered by independent solid-state amplifiers.

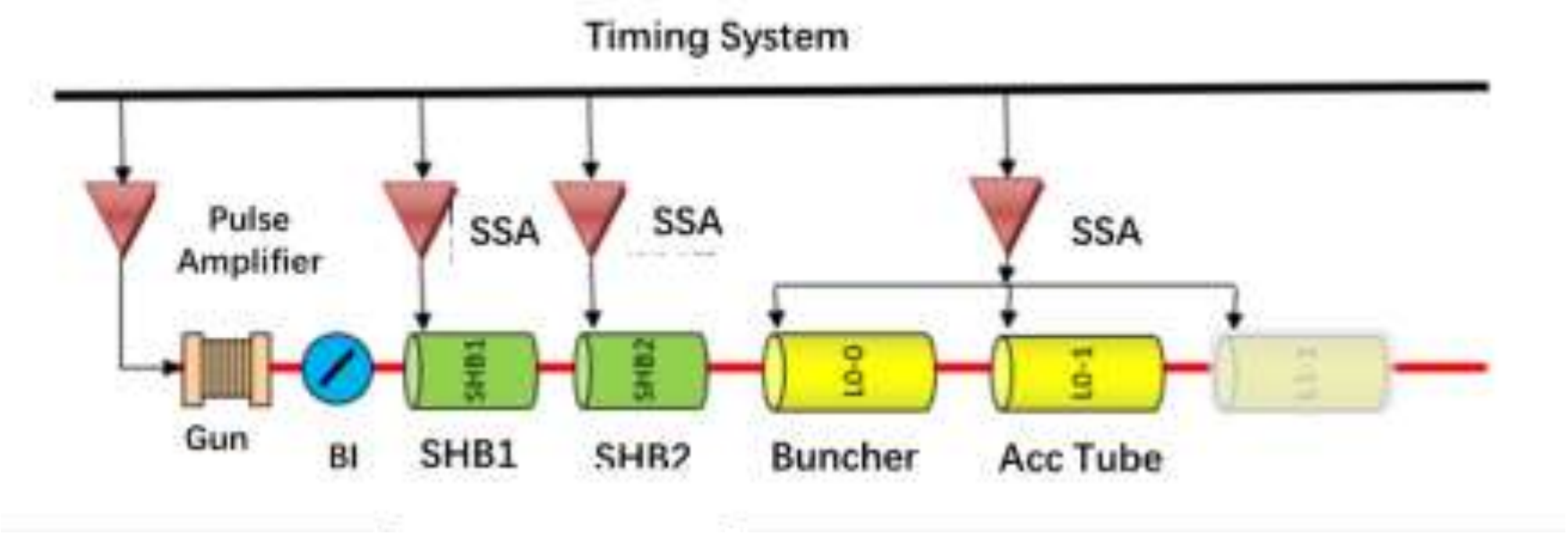

Figure 3.2-5: Thermionic cathode gun and bunching section

The beam dynamics of the bunching section are critical to overall accelerator performance. Simulations were carried out using the well-established Parmela code. A commercial Y-796 cathode-grid assembly with a maximum emission current of 12 A is used, paired with a 1.6 ns amplifier. The grid operates at 7 A, leaving a margin for current headroom. Figure 3.2-6 shows the initial phase space distribution of the thermionic electron beam used in the simulation.

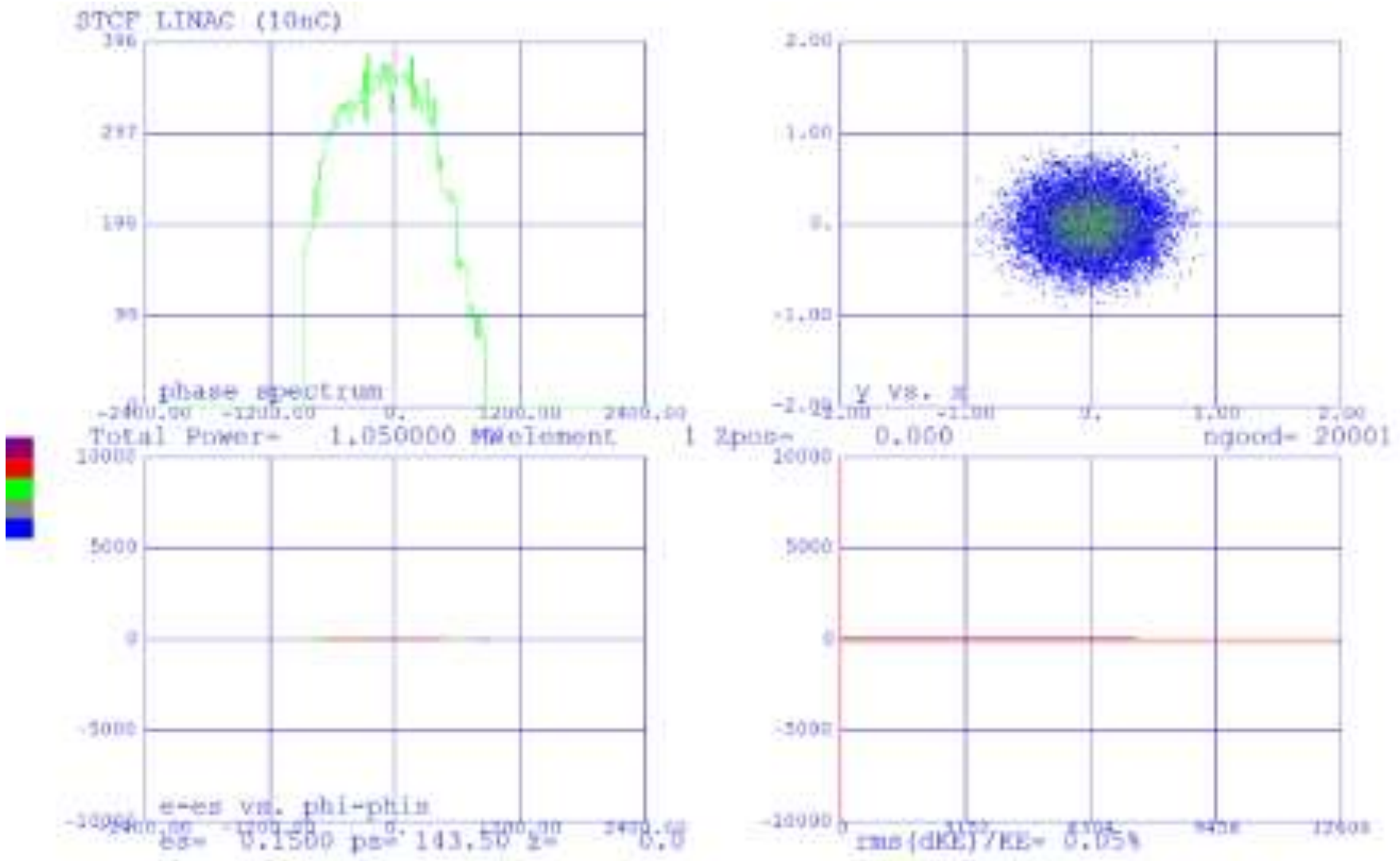

Figure 3.2-6: Phase space distribution at the exit of the thermionic cathode electron gun

After extensive optimization, especially tuning the cavity voltages, phases, and drift distances of the sub-harmonic bunchers, and the cavity length and voltage of the fundamental buncher, key beam parameters such as transverse and longitudinal envelopes along the beamline were obtained, as shown in Figure 3.2-7. The beam is well-focused both transversely and longitudinally, and a capture efficiency of 99.8% is achieved.

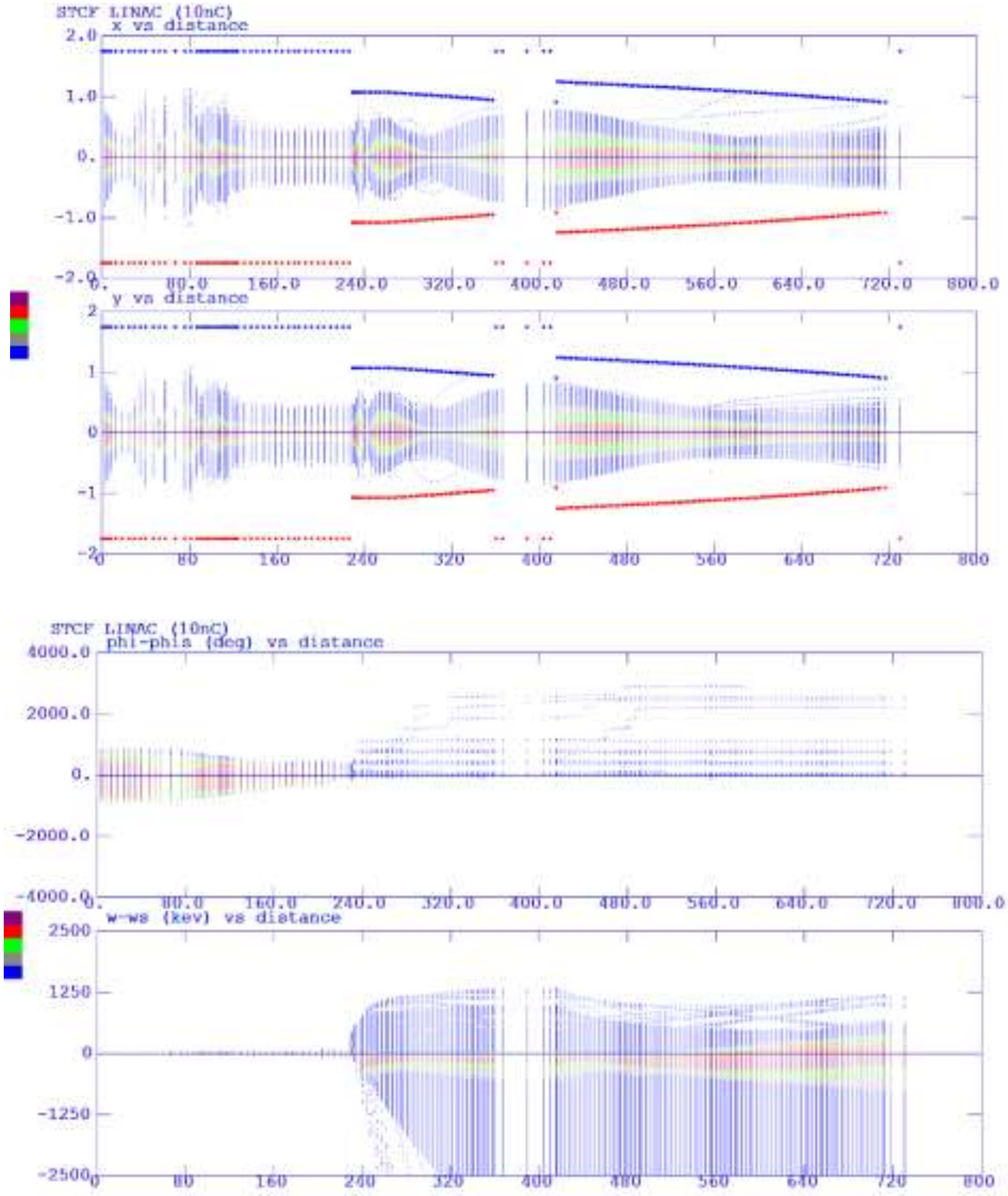

Figure 3.2-7: Transverse and longitudinal envelopes of the beam in the bunching section

The evolutions of normalized emittance and phase space distribution at the exit of the bunching section are shown in Figure 3.2-8. The normalized RMS emittance is about 50 μm·rad. There are about 5 to 6 satellite bunches following the main bunch, but their total particle count is only about 0.2% of the total. The main bunch contains about 10.9 nC, which meets the requirement of >10 nC for target impact as specified in Figure 3.1-1.

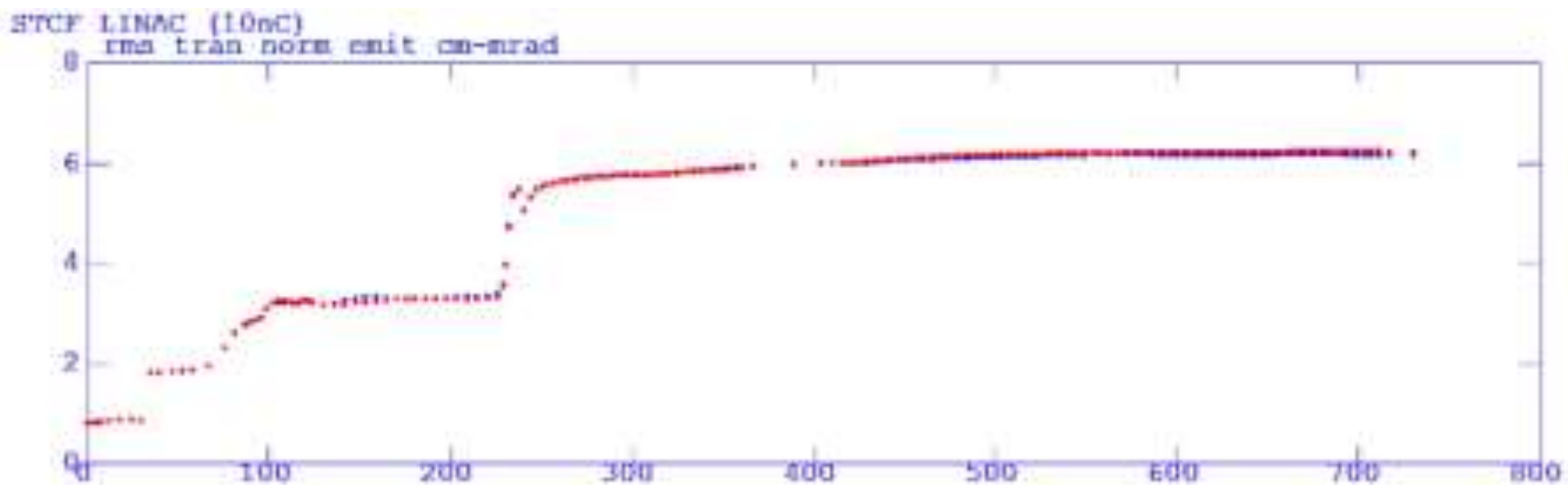

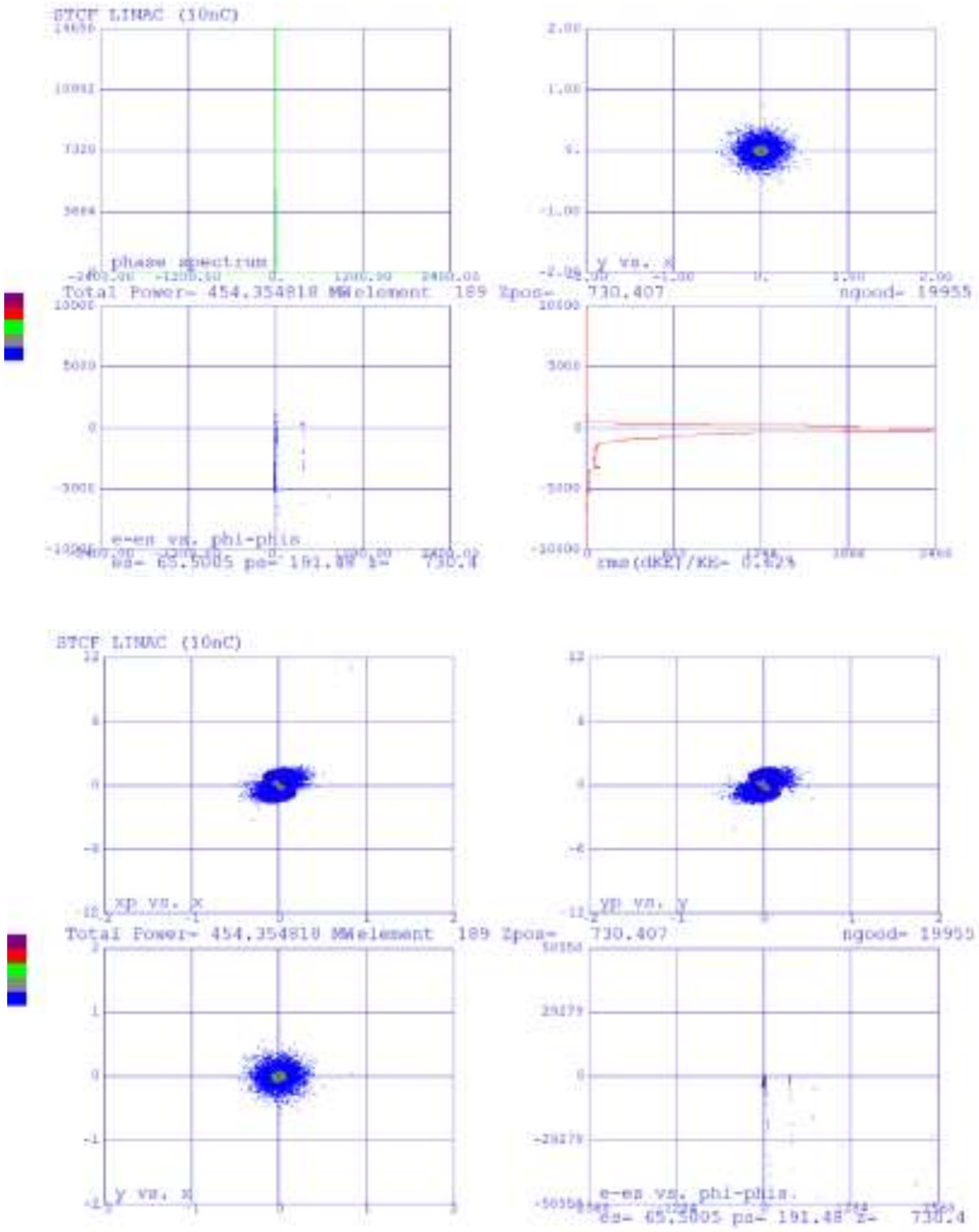

Figure 3.2-8: Evolution of normalized emittance and phase space at the exit of the bunching section

In the bunching section, solenoids and short magnetic lenses are effectively used for transverse beam confinement. This ensures good transverse parameters, such as envelope size and transmission efficiency, while preserving longitudinal compression. The magnetic field distribution of the bunching section is shown in Figure 3.2-9.

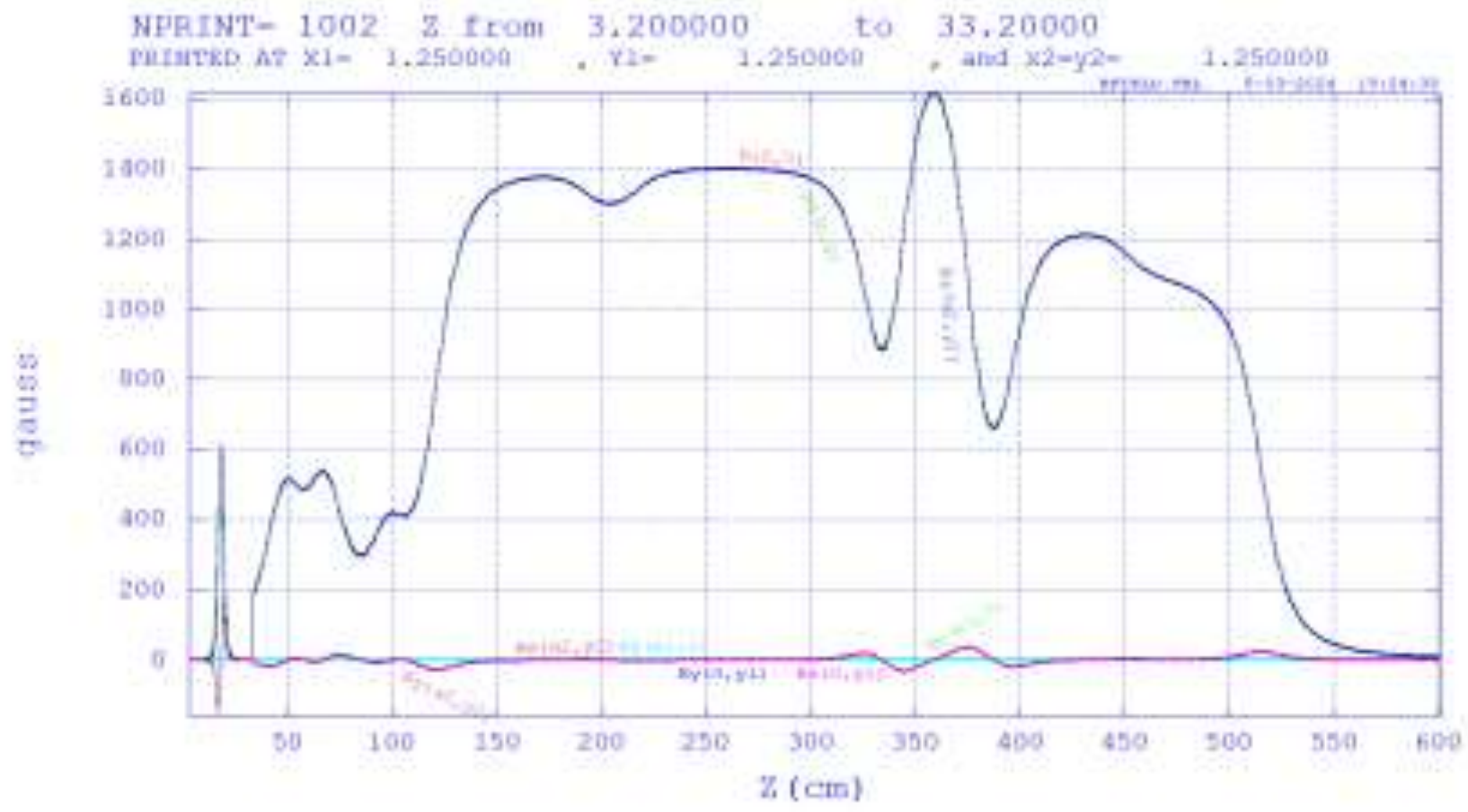

Figure 3.2-9: Magnetic field distribution in the bunching section of the thermionic cathode electron gun

**Pre-injector Section**

As shown in Figure 3.1-1, to enable the 10 nC thermionic cathode electron bunch to share the common linac section (EL1) with the 1.5 nC bunch from the photocathode source and preserve the beam quality for both modes, the thermionic beam must first be accelerated to 220 MeV. The pre-injector section (PIS) consists of a bunching section (BS), three 50 MeV accelerating structures, and two sets of triplet lenses, as illustrated in the top part of Figure 3.2-10.

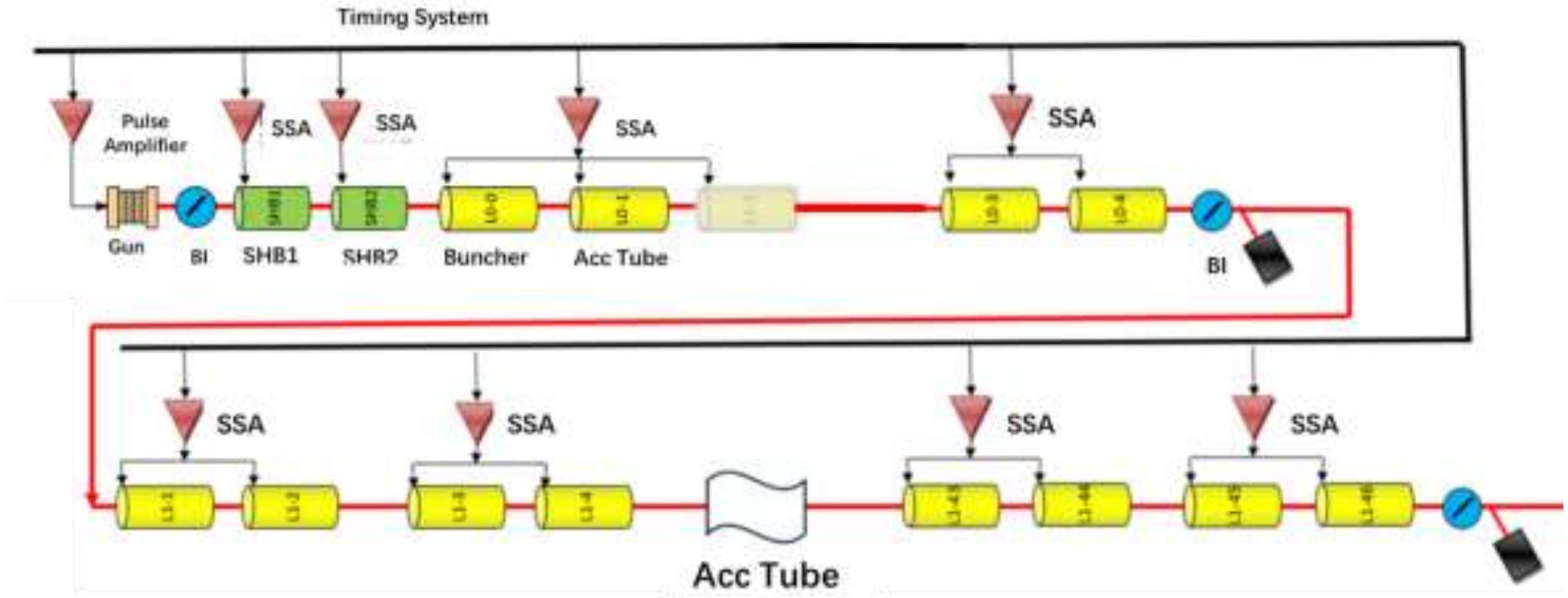

Figure 3.2-10: Layout of the thermionic electron pre-injector section and the full thermionic linac (EL1)

The fundamental buncher L0-0 and accelerating structures L0-1 and L0-2 share a single RF power source, while L0-3 and L0-4 are powered by another source. Two triplet lens assemblies are placed between L0-2/L0-3 and after L0-4 to provide transverse beam focusing.

Using the output distribution from L0-1 (shown in Figure 3.2-8) as the input, we optimized the phases of L0-2, L0-3, and L0-4, as well as the two triplets. The resulting transverse and longitudinal envelopes along the beamline are shown in Figure 3.2-11, indicating excellent transverse and longitudinal confinement.

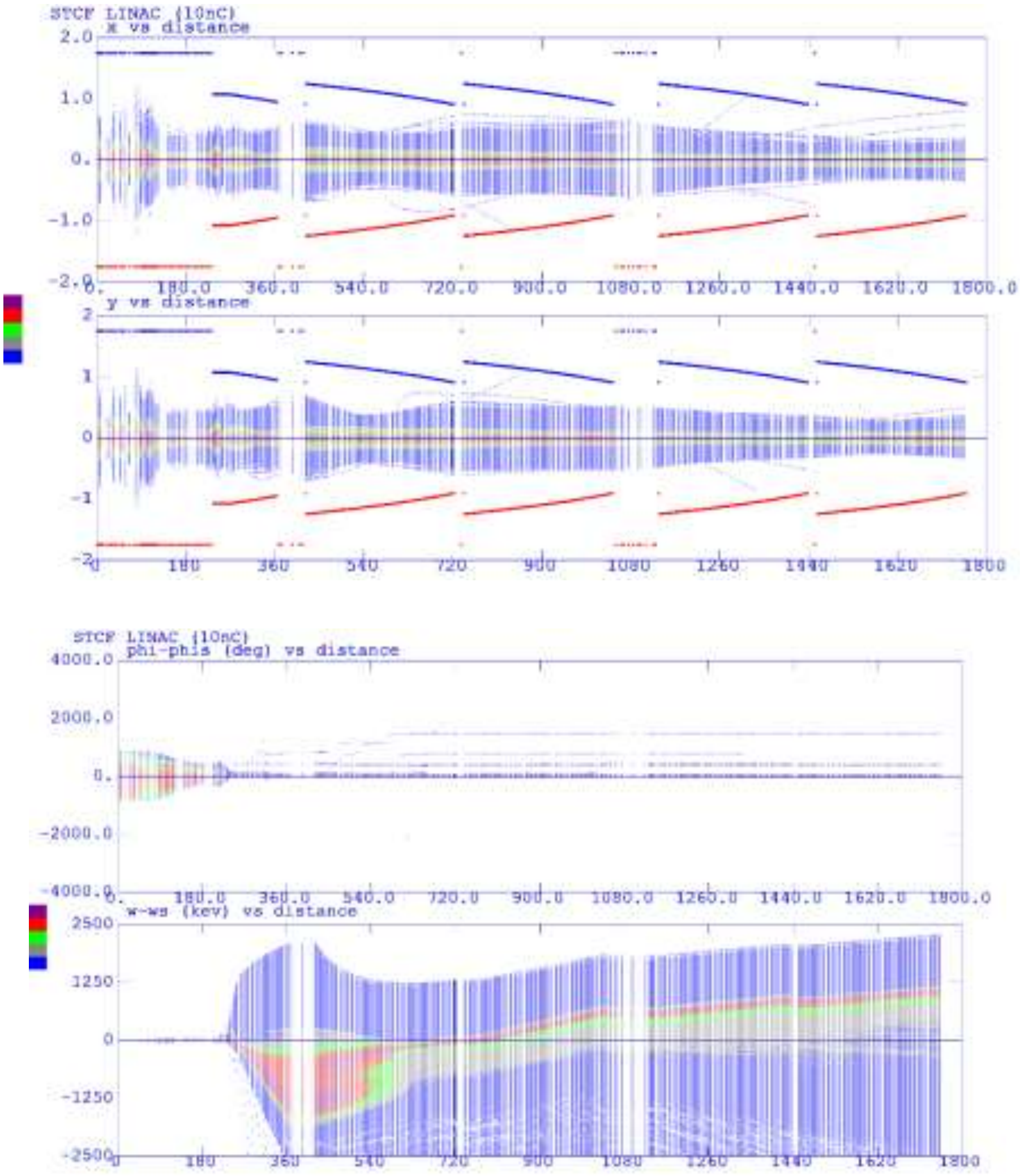

Figure 3.2-11: Transverse and longitudinal envelopes in the thermionic electron pre-injector section

The evolution of normalized emittance, phase space at the pre-injector exit, and energy gain curve are shown in Figure 3.2-12. The normalized RMS emittance and phase space are well controlled, satisfying the requirements for merger with the photocathode beam and further acceleration.

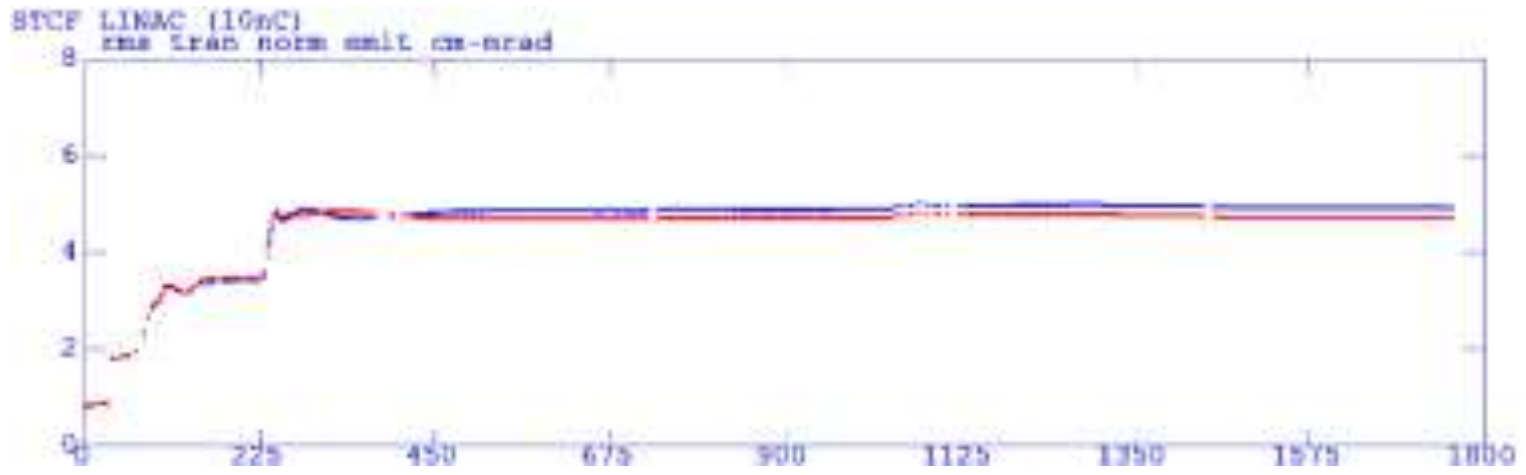

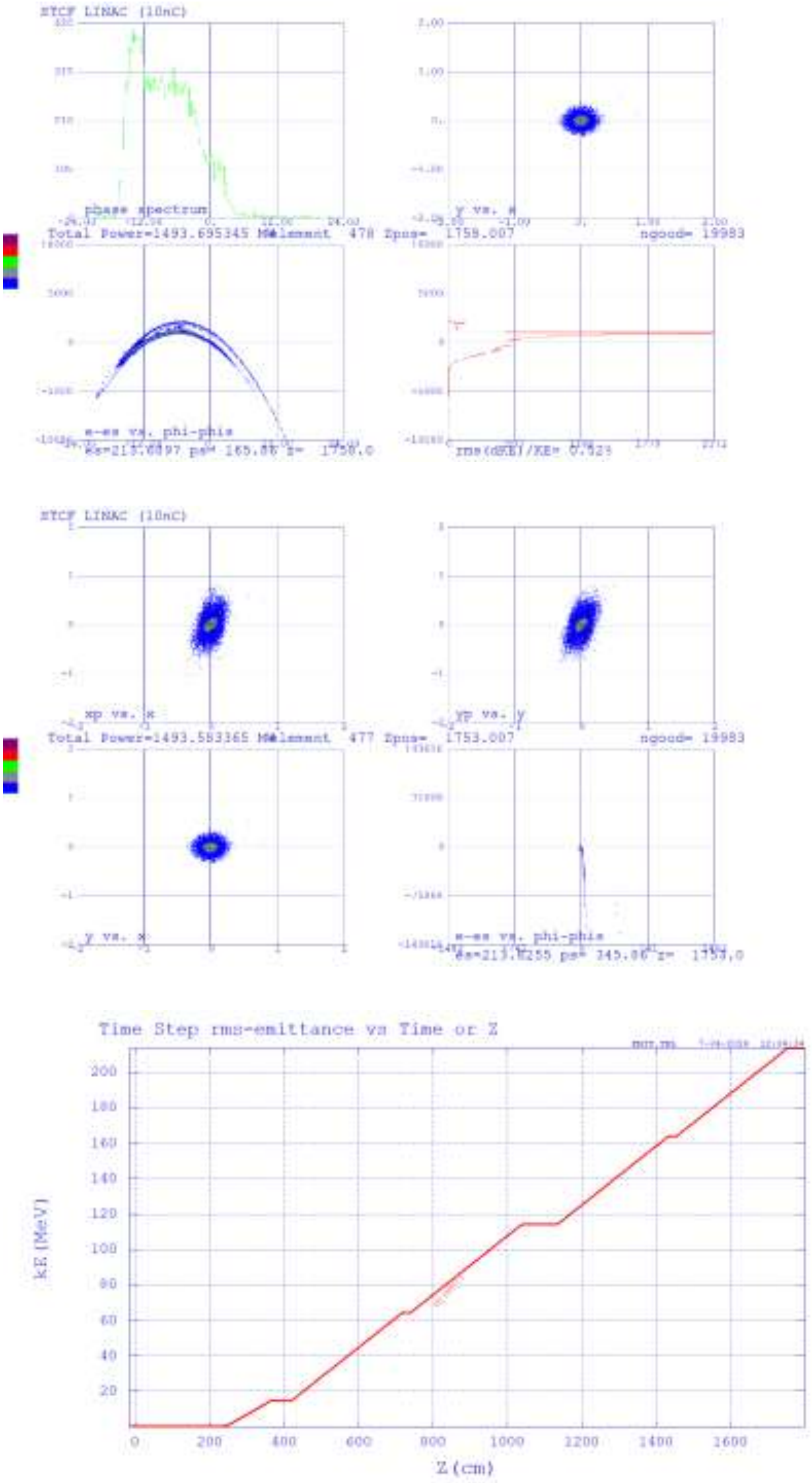

Figure 3.2-12: Emittance evolution, phase space, and energy gain in the thermionic pre-injector section

### *3.2.1.3 Electron Linacs EL1 and EL2*

High-charge electron beams for positron production and high-quality electron beams for direct injection into the collider electron ring are both pre-accelerated through two separate low-energy sections, and then merged into Main Acceleration Section 1 (EL1), where they are accelerated to 1 GeV.

## Key Design Considerations for the Merger Section

The layout of the merger section for the dual-electron-gun system in the STCF linac is shown in Figure 3.2-13. Considering the low charge and low emittance characteristics of the photocathode electron beam, it is preferable to deflect the photocathode beam into the main linac. This arrangement minimizes the impact of dispersion from dipole magnets on beam quality and facilitates beam optics design. The design goal of the merger section is to align the beam center of the photocathode electron beam with that of the thermionic beam while preserving the former's beam quality.

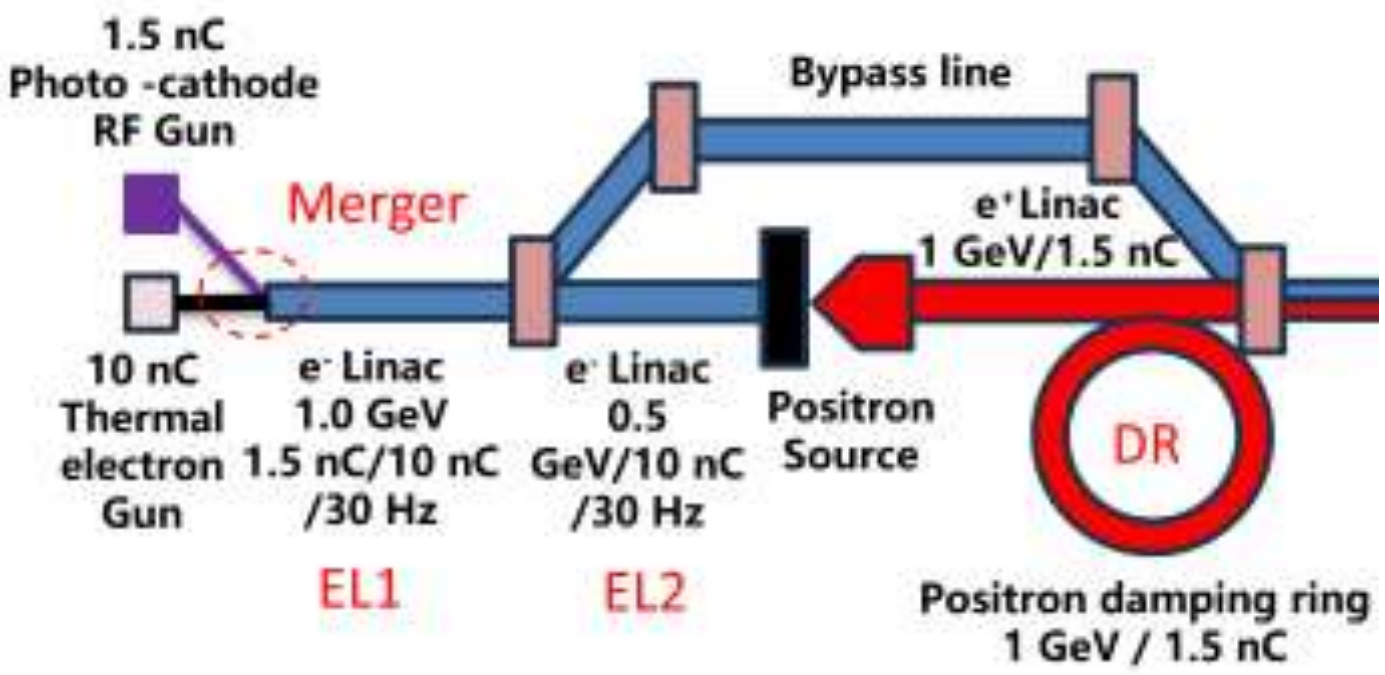

Figure 3.2-13: Schematic of the merger section for the dual-electron-gun linac

In our design, both the photocathode and thermionic cathode electron guns are placed on the same horizontal plane. Hence, we implement horizontal-plane deflection for the photocathode beam. The merger section adopts a dogleg structure, consisting of two dipole magnets with identical parameters deflecting in opposite directions, ensuring the beam propagates in the same direction before and after the dogleg while achieving lateral displacement to merge with the thermionic beam.

In practical terms, the photocathode beam must undergo bunch compression during transport to reduce energy spread. However, this compression introduces significant energy chirp/spread (1% RMS, 6% peak-to-peak). When passing through dipole magnets, this dispersion can cause up to two orders of magnitude increase in the horizontal emittance, severely degrading beam quality. Thus, the main challenge in merger beam optics design is not only to eliminate first-order dispersion but also to suppress higher-order dispersion terms, particularly the T_166 and T_266 matrix elements. Given space constraints, we fix the dipole length to 0.2 m and set the deflection angle to approximately 21.5°, resulting in a total merger length of about 7 m. The dogleg layout is shown in Figure 3.2-14.

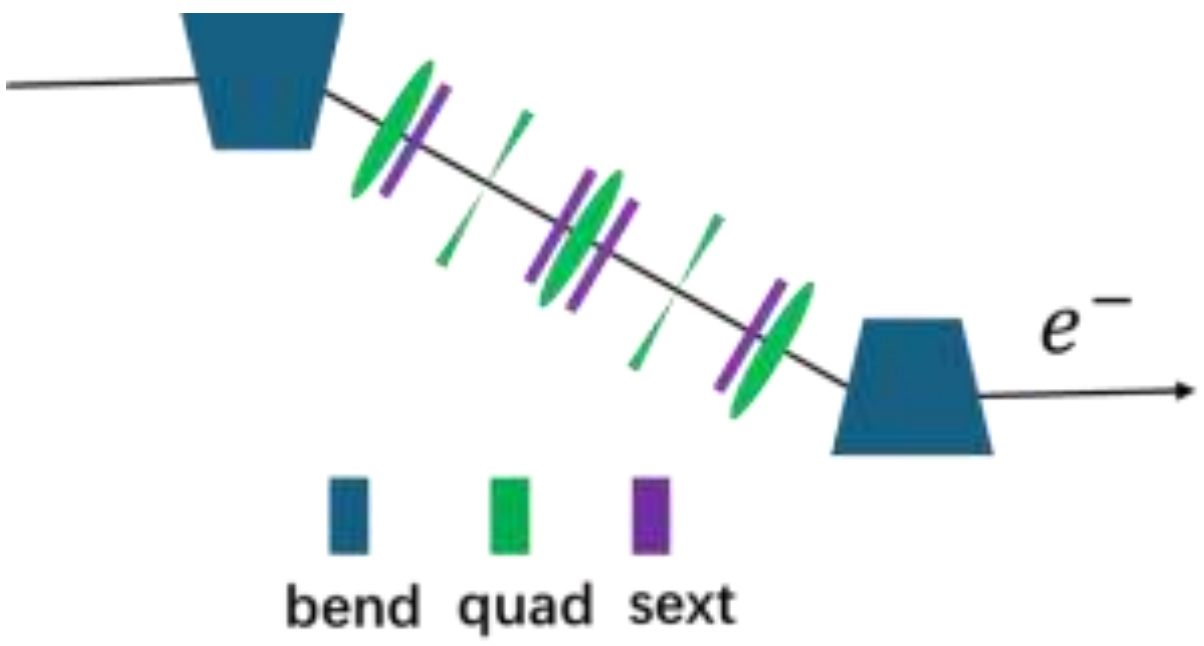

Figure 3.2-14: Layout of the merger section (dogleg)

We used the beam tracking code *Elegant* [60] for optimization. Quadrupole magnets were employed to match the first-order dispersion and transverse Twiss functions. To minimize higher-order dispersion, sextupoles were inserted. Inspired by the Elettra approach [61], sextupole windings were added to the quadrupoles to provide additional sextupole components, further optimizing the emittance. The optimization results show that reducing both first- and second-order dispersion below 0.01 effectively restores emittance to the desired level.

To reflect realistic conditions, the beam charge was set to 1.5 nC in the simulation. Effects including space charge (SC), coherent synchrotron radiation (CSR) in dipoles, and incoherent synchrotron radiation (ISR) at dipole edges were considered. Beam parameter optimization was performed using Simplex and second-generation genetic algorithms (NSGA2), with magnet parameters as variables. First, dispersion terms were minimized, followed by optimization of transverse Twiss parameters and emittance.

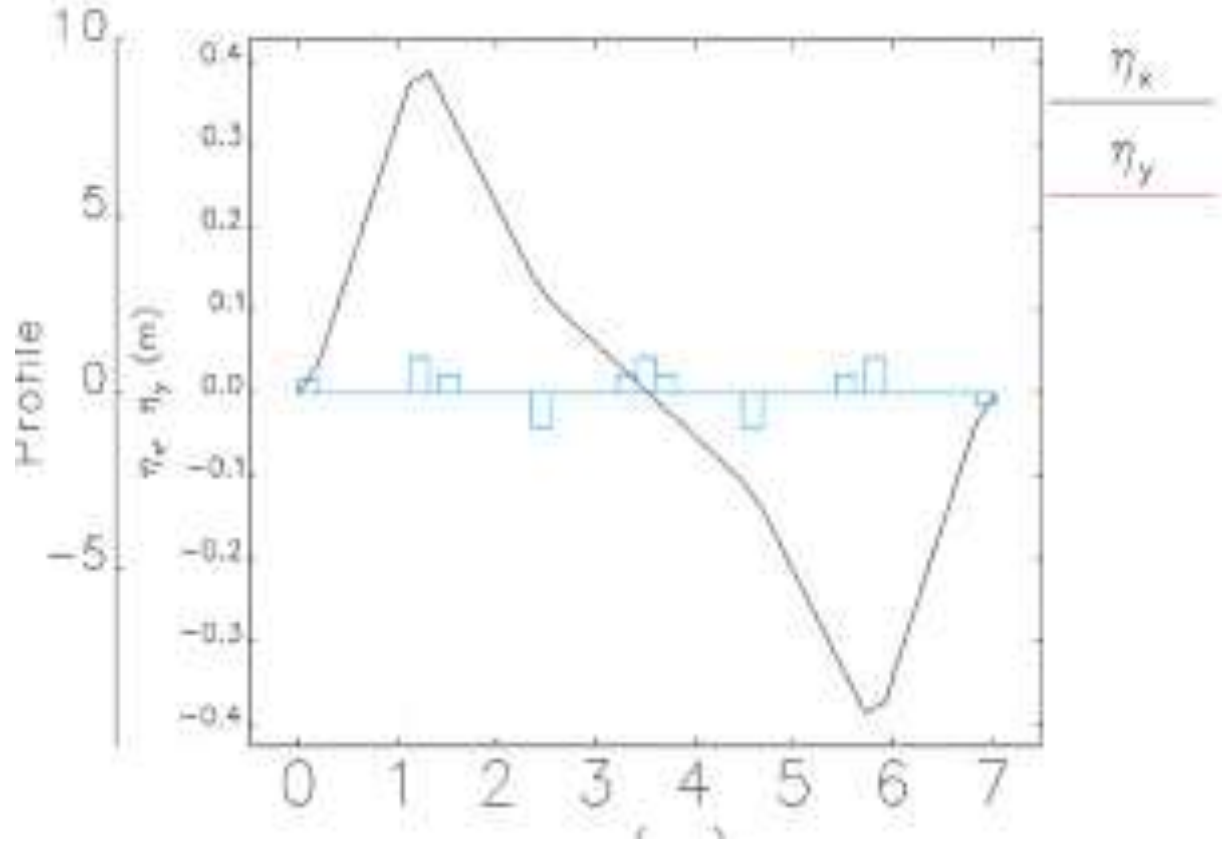

Figure 3.2-15: Variation of first-order horizontal dispersion $\eta_x$ with longitudinal position

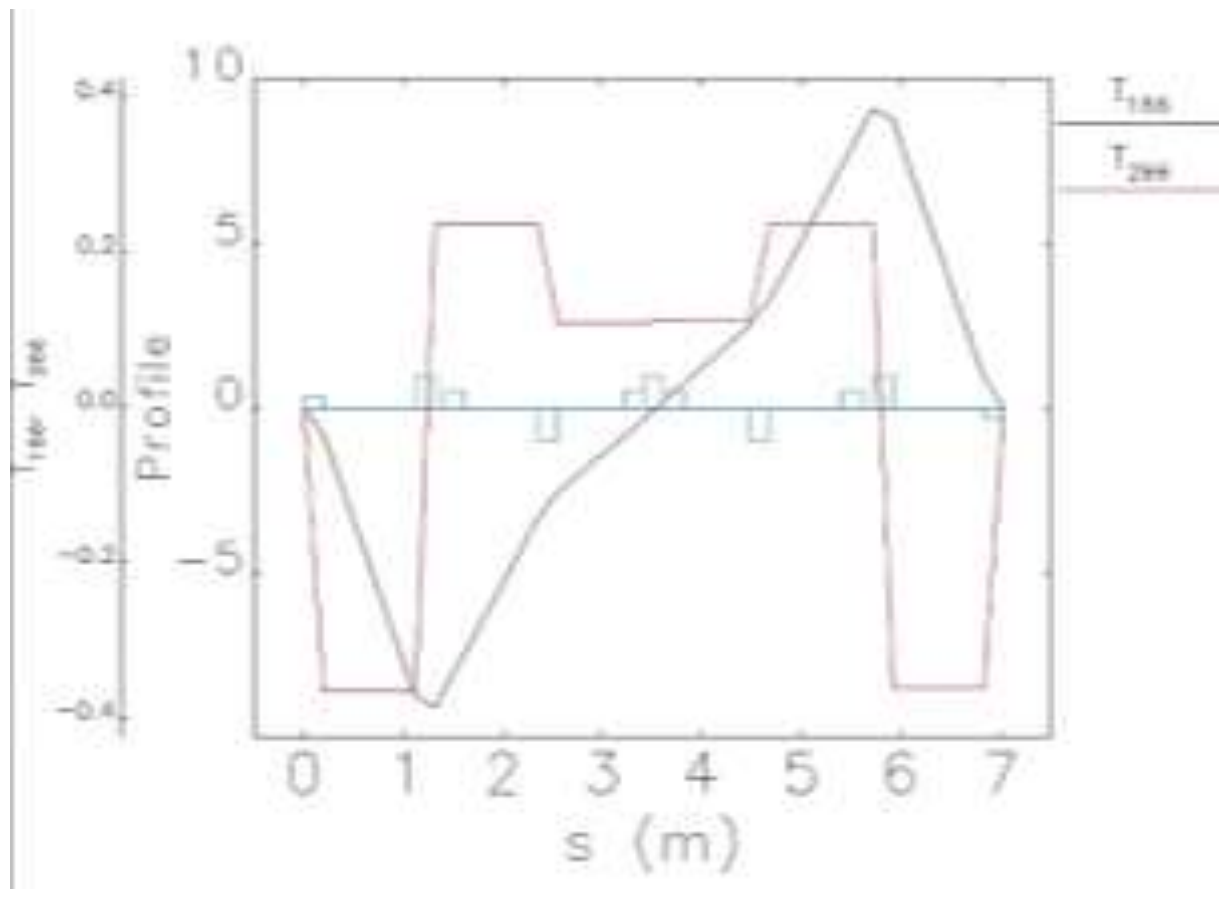

Figure 3.2-16: Variation of higher-order horizontal dispersion $T_{166}/T_{266}$ with longitudinal position

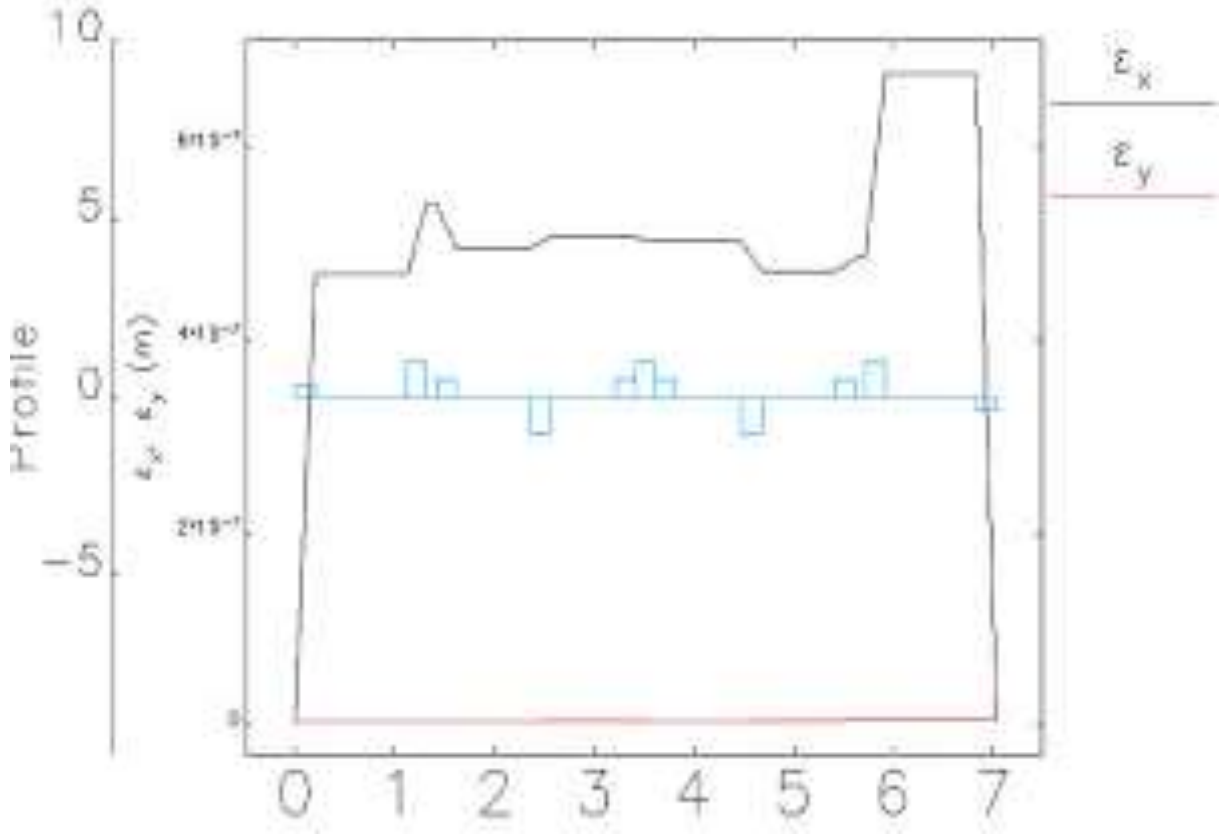

Figure 3.2-17: Emittance variation with longitudinal position

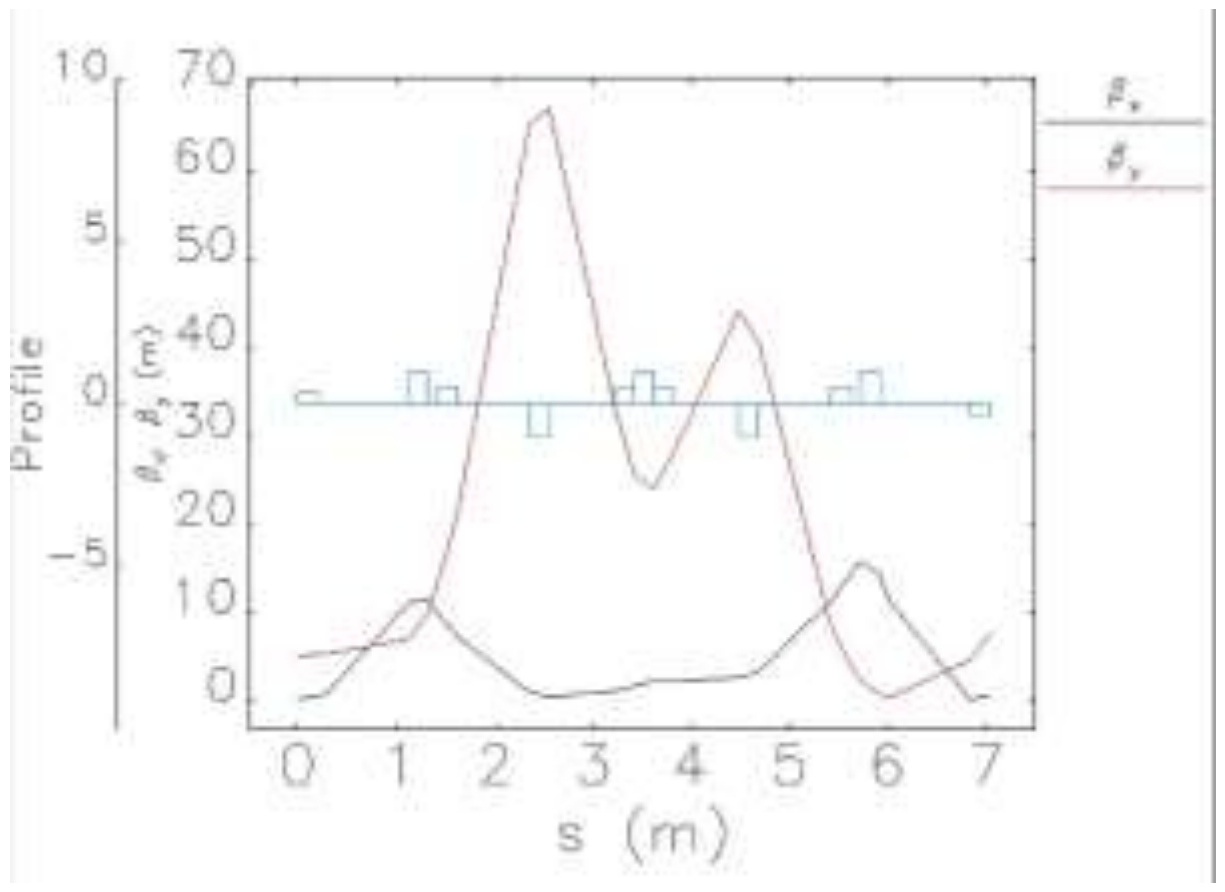

Figure 3.2-18: β-function variation with longitudinal position

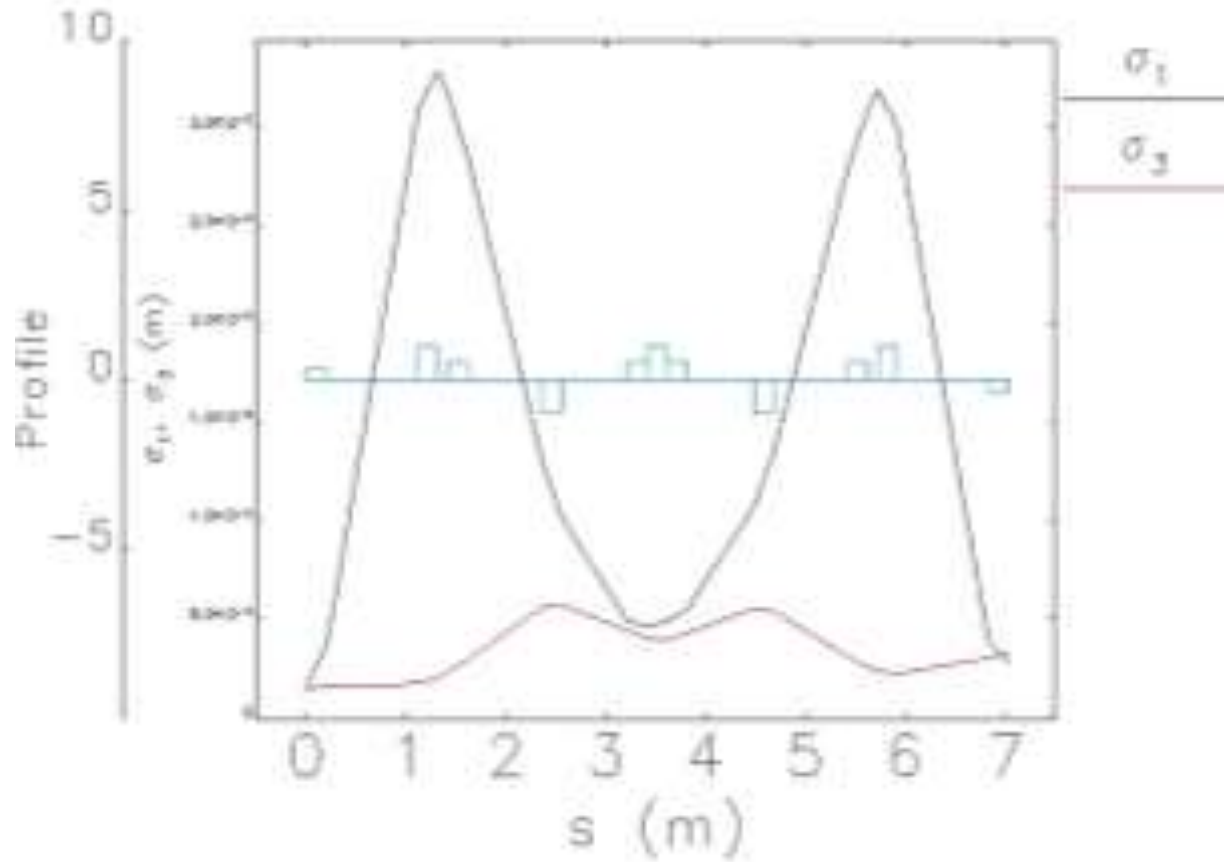

Figure 3.2-19: Beam size variation with longitudinal position ($\sigma_1$ and $\sigma_3$ represent horizontal and vertical beam sizes, respectively)

In conclusion, through the custom-designed dogleg, we successfully merged beams from the dual electron guns while restoring the horizontal emittance of the photocathode beam, even under significant energy spread, thus maintaining beam quality. The merged beams share the same beam center and can use the same main linac, optimizing cost and equipment usage.

**Key Design Considerations for EL1 and EL2**

The 1.5 nC beam from the photocathode is directly injected into the collider electron ring via the bypass after acceleration in EL1. Therefore, the beam quality requirements for this beam are more stringent than for the 10 nC beam that is further accelerated in EL2 for positron production. Hence, EL1's lattice is optimized for the 1.5 nC beam. The performance of the 10 nC beam in EL1 and EL2 was also simulated, with the resulting transverse and longitudinal envelopes, normalized emittance evolution, and phase space at the linac exit shown in Figure 3.2-20.

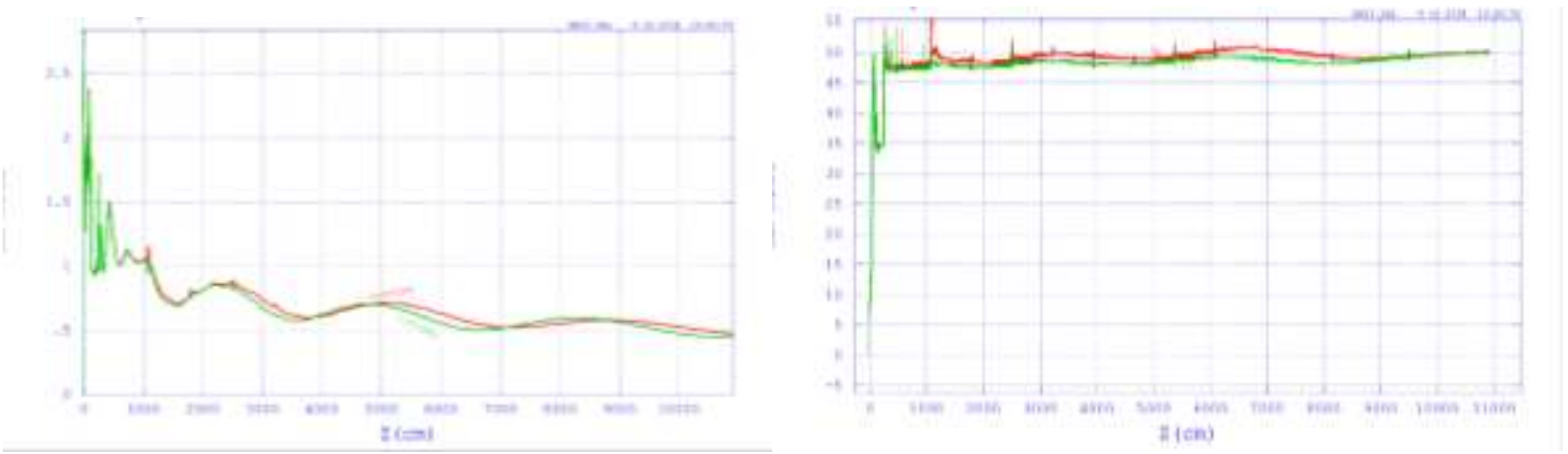

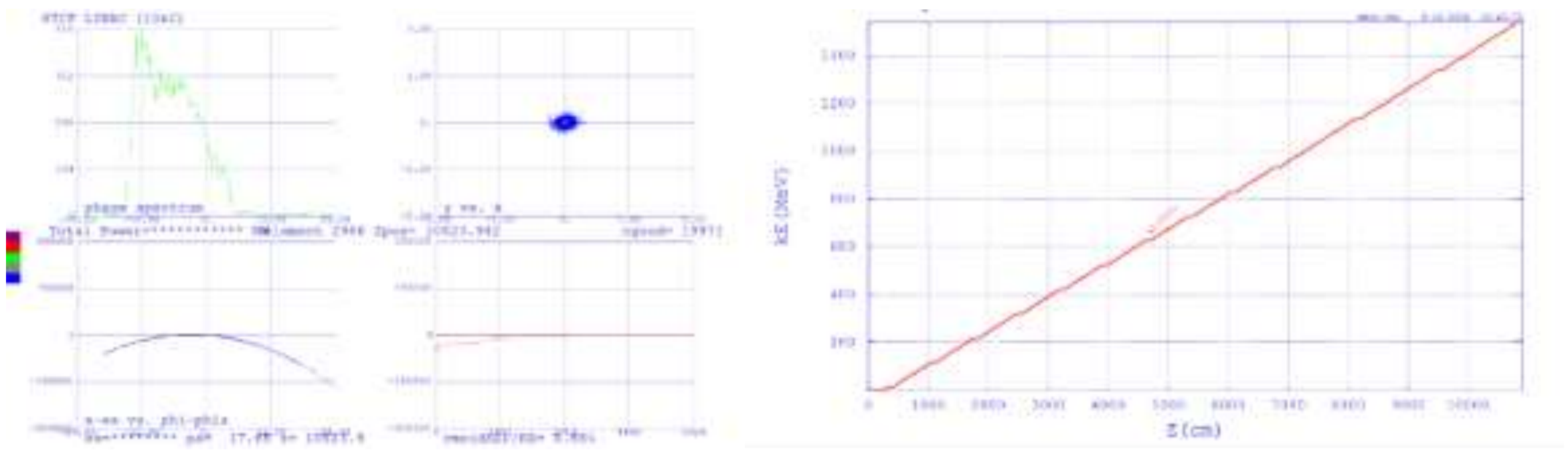

Figure 3.2-20: Beam parameter evolution along the linac for the positron target beam under the off-axis injection scheme

The plots show that the transverse RMS beam size at the linac exit is about 0.5 mm, the emittance is about 50 μm·rad, the energy reaches 1540 MeV, the transmission efficiency is 99.8%, the beam charge is 10.9 nC, and the RMS energy spread is 1.87%. These parameters meet the requirements for tungsten target positron production.

The EL2 section differs slightly in design. A triplet magnet is placed every four 50 MeV accelerating structures (i.e., after every 200 MeV energy increment). The total length of the thermionic cathode linac is about 110 m.

### 3.2.2 Positron Production, Capture, and Pre-Acceleration

As a high-luminosity electron–positron collider, STCF requires a high-performance positron source capable of generating positron bunches with sufficient quantity and quality. This source is one of the core components of the STCF injector. The positron source of STCF adopts the conventional method of positron production via electron beam bombardment of a high-Z target—currently the only viable approach for generating high-intensity positron beams.

STCF currently considers both off-axis and bunch-swap injection schemes for the collider rings. The off-axis injection scheme demands relatively low positron bunch charge and uses a lower primary electron energy of 1.5 GeV. This necessitates optimizing the subsequent capture and matching acceleration systems to ensure high positron yield from low-energy electron beams. In contrast, the swap-out injection scheme imposes significantly higher requirements on the positron yield and heat dissipation capacity of the target due to the need for high-charge positron bunches.

Table 3.2-2: Positron Source Parameters for STCF

| No. | Parameter | Off-Axis Injection | Swap-Out Injection |
|---|---|---|---|
| 1 | Injected Positron Charge | 1.5 nC/30 Hz | 2.9 nC/90 Hz |

| No. | Parameter | Off-Axis Injection | Swap-Out Injection |
|---|---|---|---|
| 2 | Incident Electron Beam | 1.5 GeV/10 nC/30 Hz | 2.9 GeV/11.6 nC/90 Hz |
| 3 | Injection Energy | 1.0 GeV | 1.0 GeV |
| 4 | Damping Ring Entry Emittance | ⩽1400 nm·rad | ⩽1400 nm·rad |

To develop a high-quality, high-current positron source that meets STCF requirements, two sets of target-converter systems, which are based on the two electron beam injection schemes, must be carefully designed and optimized. These include the positron conversion target, the adiabatic matching system (AMS), and the capture accelerating structure, all of which must be engineered to improve positron yield and meet the needs of both injection schemes.

#### *3.2.2.1 Positron Conversion Target*

The positron production process by electron beam bombardment is typically simulated using Monte Carlo codes such as EGS4, MCNP, FLUKA, or Geant4. These simulations determine the energy spread and angular distribution of positrons emerging from the high-Z target, providing input for the design of the downstream capture and matching systems. Figure 3.2-21 shows simulation results obtained using Geant4. It is worth noting that simulations performed using different codes often yield discrepancies in the positron angular and energy distributions, sometimes up to 10–20%. Since these results directly affect the design of the capture and matching systems, improving simulation precision and accuracy is crucial for the future development of positron sources for $e^+e^-$ colliders.

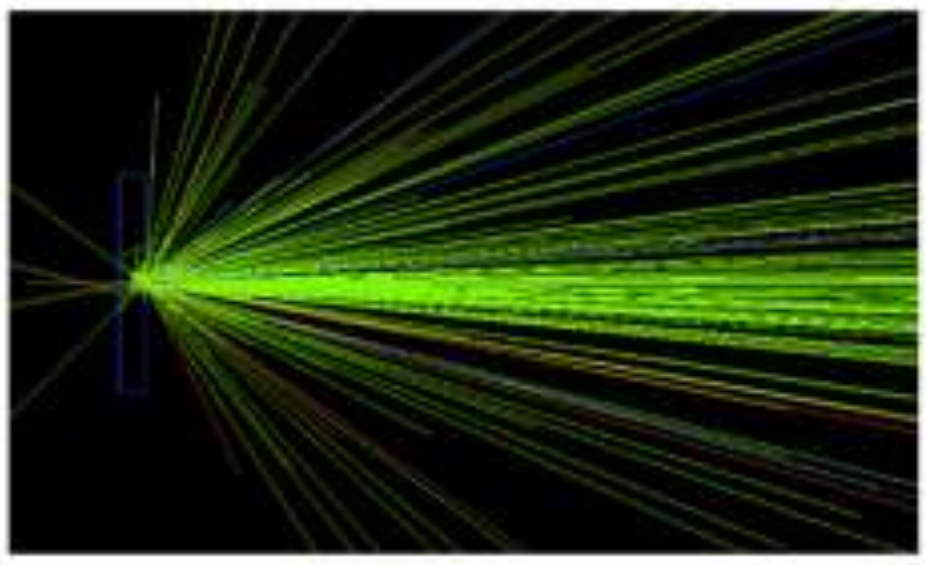

Figure 3.2-21: Geant4 simulation of positron production via electron beam bombardment

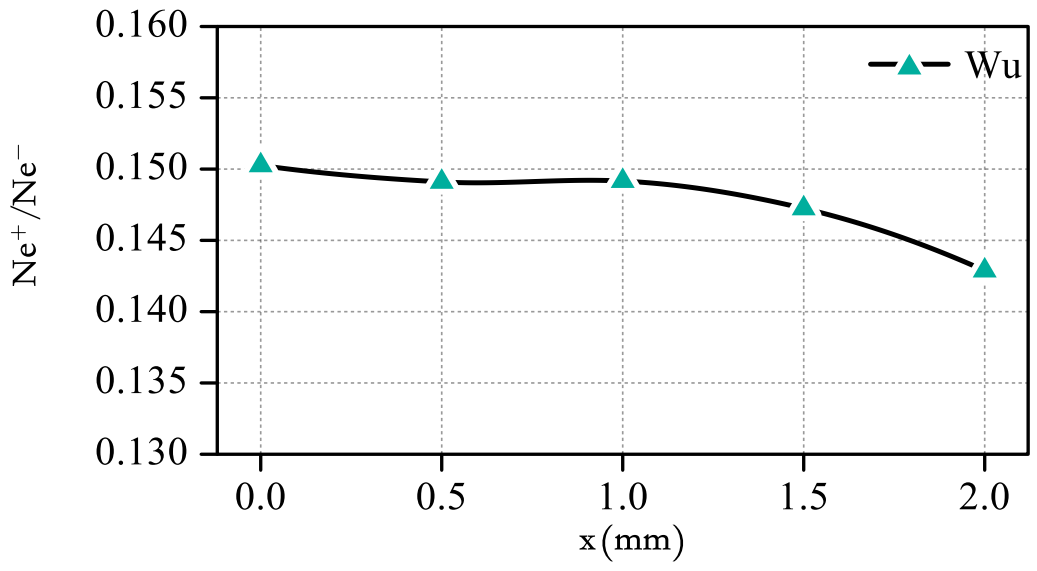

Figure 3.2-22: Impact of beam offset on positron yield

The emittance of the electron beam and its spot size on the target significantly influence the distribution and angular spread of positrons after the target. Smaller spot sizes generally yield higher positron collection efficiency. However, too small a spot size may cause excessive thermal damage to the target. STCF requires the electron beam spot size to be ≤1 mm. Moreover, the beam offset on the target also affects the positron yield. Figure 3.2-22 illustrates this effect for a 1.5 GeV electron beam; the result for a 2.5 GeV beam is similar. This requires of <1 mm beam offset for the primary electron beam.

For the off-axis injection scheme, the positron source must deliver 1.5 nC at 30 Hz. Figure 3.2-23 shows the simulated positron yield for a 1.5 GeV electron beam with a 1 mm RMS spot size striking conversion targets of different thicknesses and materials. Gold and tungsten yield the highest positron output. Considering cost-effectiveness, tungsten is selected as the target material. The optimal tungsten thickness is 13 mm, yielding a positron production efficiency of approximately 4.1 positrons per incident electron.

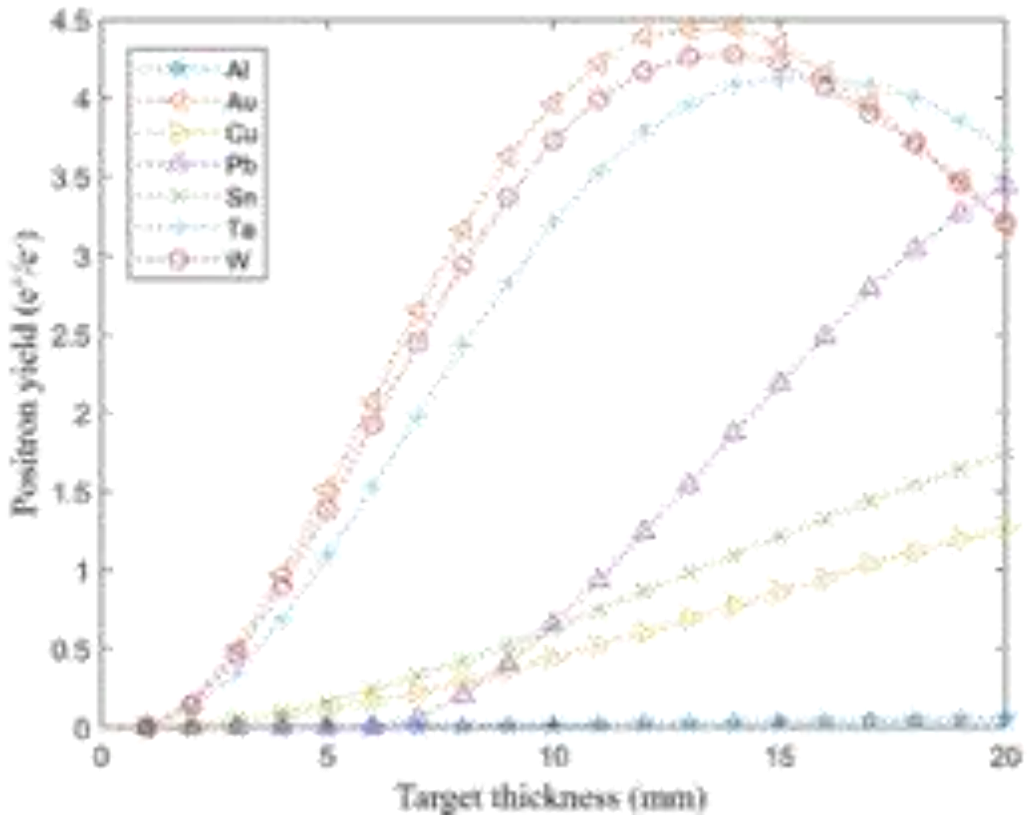

Figure 3.2-23: Positron yield from a 1.5 GeV electron beam hitting targets of different materials

Figure 3.2-24 presents the energy and angular distributions of positrons resulting from 1.5 GeV electron beams with and without a 5% energy spread. It shows that a 5% energy spread has an acceptable impact on the resulting positron characteristics.

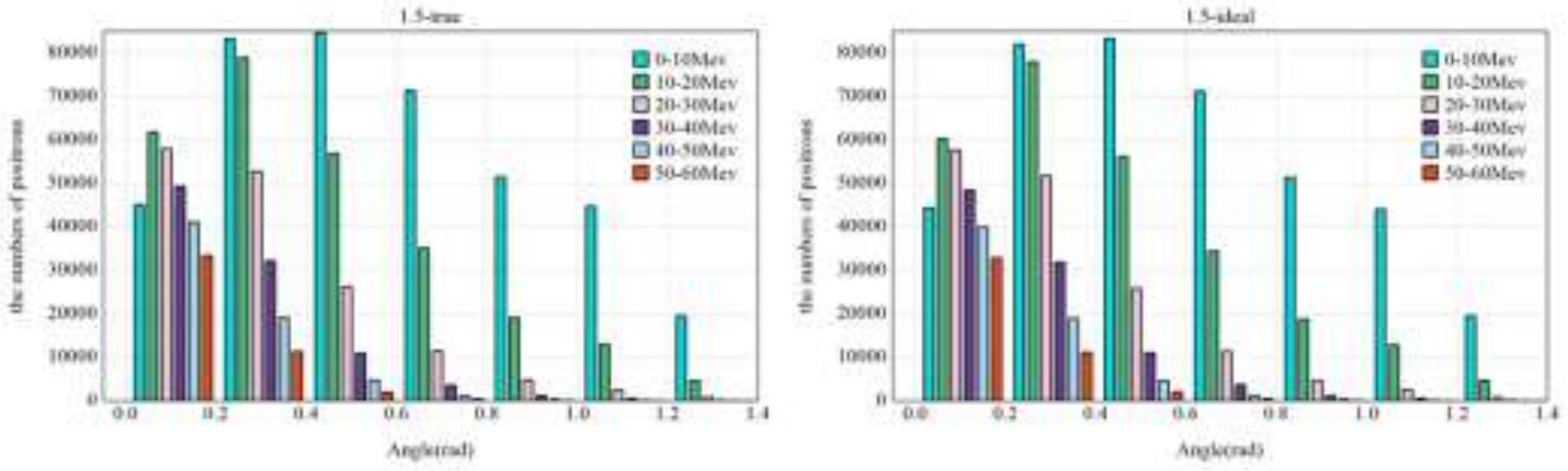

Figure 3.2-24: Energy and angular distribution of positrons after a 1.5 GeV beam strikes the target

(Left: with 5% energy spread; Right: zero energy spread)

### *3.2.2.2 Positron Accelerator PL Design*

The positron beam produced by electron bombardment of a target has a wide energy spectrum and large divergence angles. To match this with downstream acceleration structures, the positrons must be captured and converted into a beam with smaller divergence and larger transverse size. The mainstream solution is to use an Adiabatic Matching Device (AMD) within an Adiabatic Matching System (AMS), where the magnetic field gradually changes along the longitudinal axis as $B_z=B_0/(1+az)$. During transport through the AMD, the transverse phase space volume of the particle beam is conserved, allowing for the transformation from small transverse size and large divergence to large transverse size and small divergence. After passing through the AMS, the positron energy spectrum remains broad and is dominated by low-energy positrons.

Since the positron energy spectrum is broad with a large fraction of low-energy particles (1–20 MeV), a specialized accelerating structure is needed for capture. In the AMD, positrons follow helical trajectories due to the longitudinally varying magnetic field: low-energy positrons spiral outward, high-energy positrons inward. As they exit the AMD, the positron bunch lengthens and widens. Therefore, the first accelerating structure must compress the positron bunch, increase its energy, and perform longitudinal phase space matching. By selecting suitable RF phases, more positrons can be efficiently accelerated.

To increase transverse acceptance and improve positron capture efficiency, larger aperture accelerating structures are required. However, placing large-aperture L-band accelerators inside focusing solenoids would significantly raise the cost. As a cost-effective alternative, STCF adopts S-band large-aperture (30 mm) accelerating structures, with optimized RF phase settings for improved capture and acceleration. The initial stage comprises six special accelerating structures to boost positron energy to 200 MeV. This is crucial for positron beam intensity, emittance, and energy spread, impacting the damping ring, main linac, and injection system designs.

**Adiabatic Matching Device (AMD)**

The strength and profile of the AMD magnetic field are determined by the energy spread and angular distribution of positrons emerging from the target. Optimizing the AMD is critical for improving positron capture. Figure 3.2-25 shows the STCF AMD design on the electron–positron verification platform, where a 5 μs, 15 kA pulsed current is applied through 13 coils to generate the field.

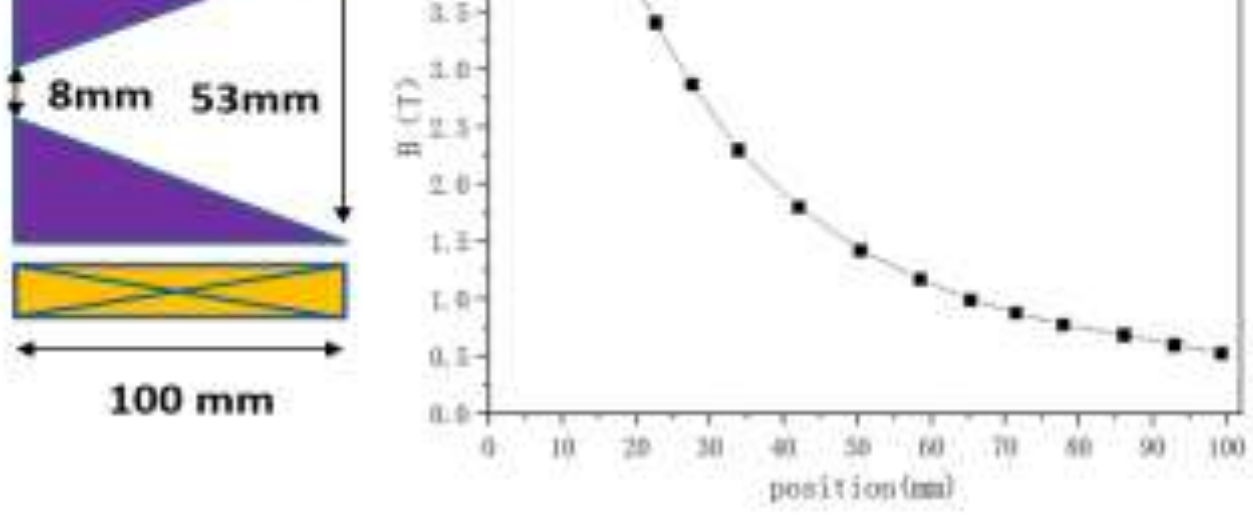

Figure 3.2-25: Schematic and magnetic field simulation of the AMD

Due to the large angular and momentum spread of positrons from the target, most would be lost without strong focusing. The capture system must have high acceptance. As positron beams have a small size but large divergence, while downstream accelerating structures allow larger sizes and smaller divergence, the AMD must perform the beam transformation shown in Figure 3.2-26, enabling proper matching. Figure 3.2-27 shows angular variation after AMD; Figure 3.2-28 shows energy spread and transverse beam size.

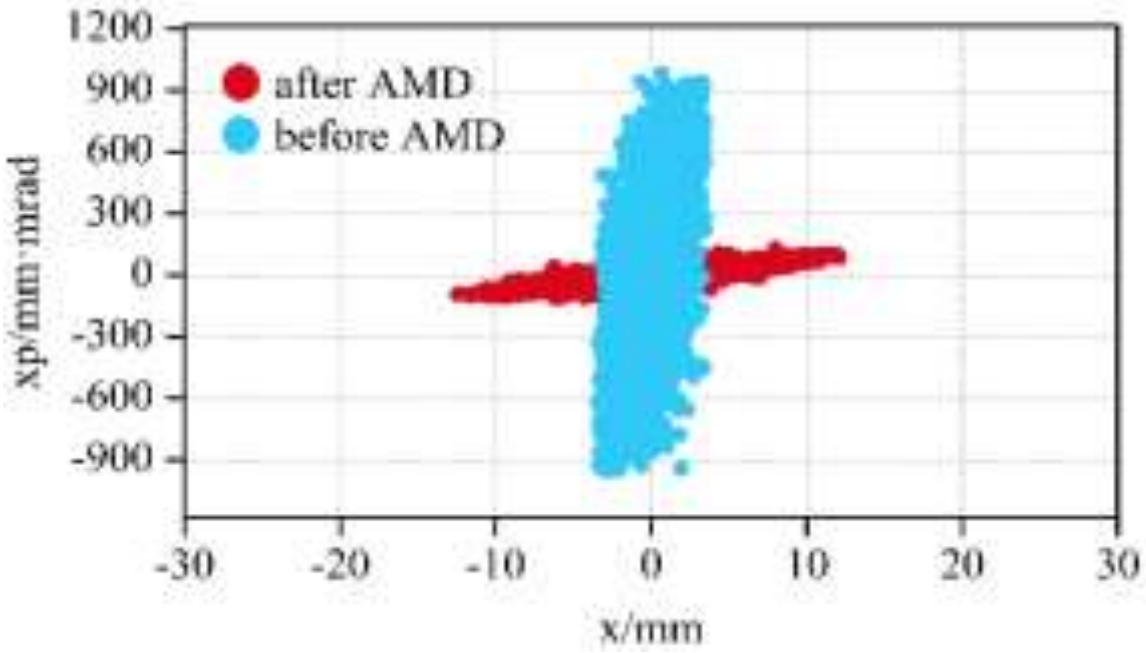

Figure 3.2-26: Phase space transformation of the positron bunch by the AMD

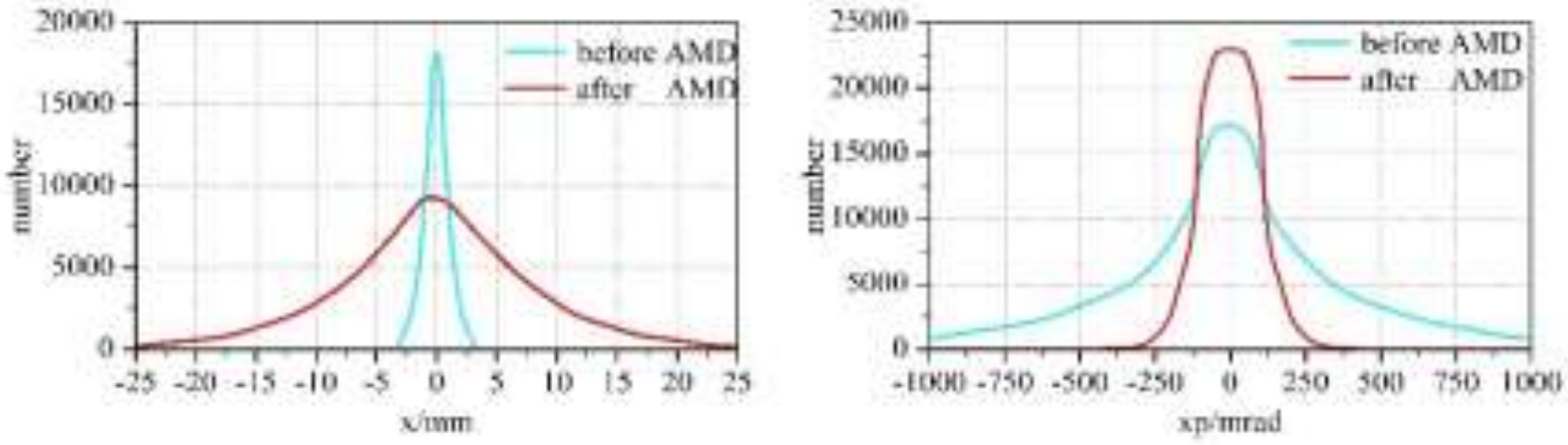

Figure 3.2-27: Positron beam distribution and divergence variation after passing through the AMD

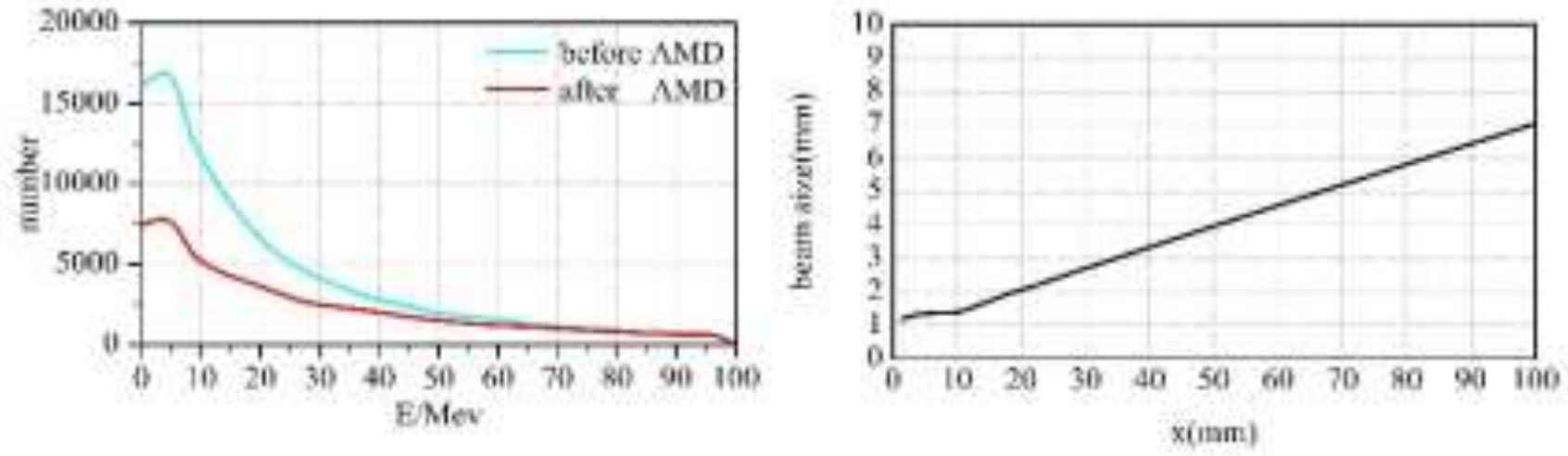

Figure 3.2-28: Energy spread and transverse beam size variation of positrons after passing through the AMD

**200 MeV Pre-Acceleration Section**

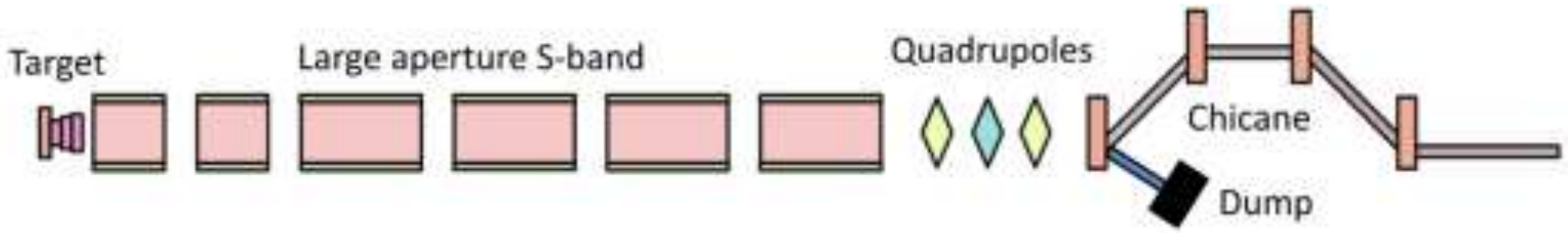

Figure 3.2-29: PAS-Pre acceleration section of the positron linac

Figure 3.2-29 illustrates the PAS-Pre section. Its structure includes the conversion target system, AMD, two 2 m large-aperture capture cavities, four 3 m capture cavities, solenoids for external focusing, a triplet lens for matching, and a Chicane for momentum collimation. Geant4 simulations provide energy and spatial distributions of 1.5 GeV positrons, which are captured in the AMD and two 2 m S-band accelerating sections. RF phase scans for these two sections yield the optimal phases of 20° and -40°, maximizing positron yield (see Figure 3.2-30). A final 3 m accelerating cavity boosts positrons to 200 MeV. Figure 3.2-31 shows particle tracking at this energy.

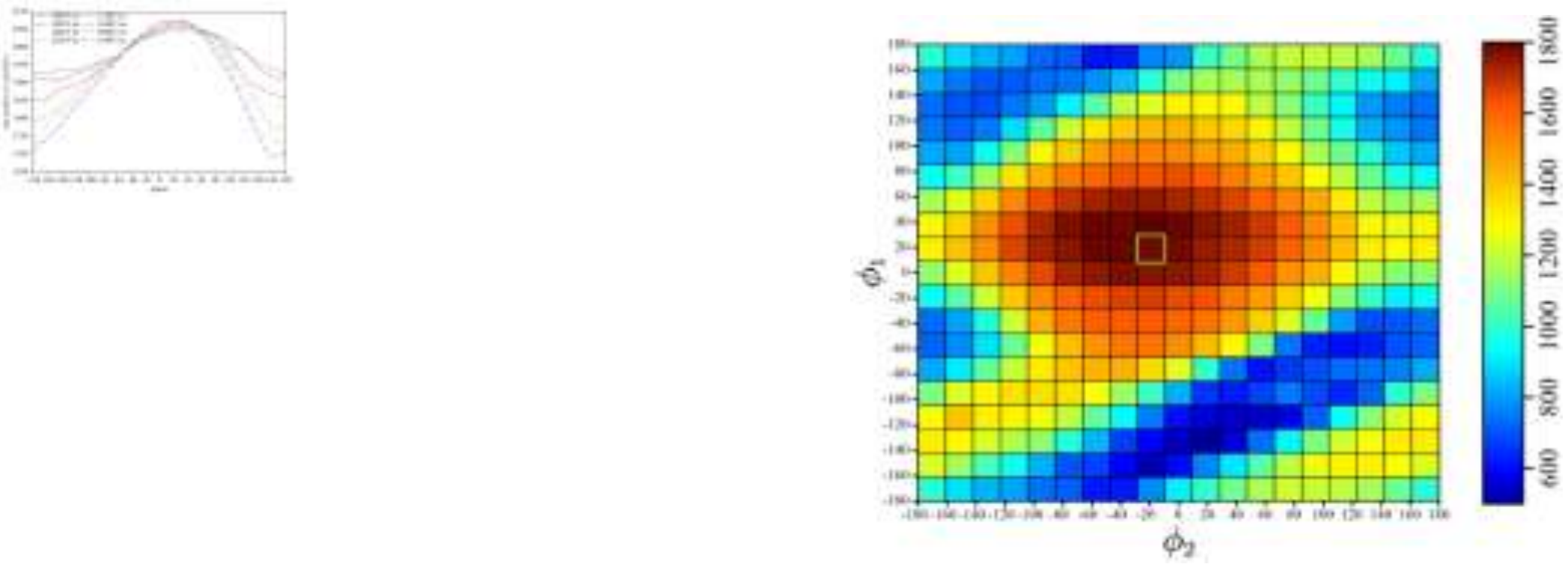

Figure 3.2-30: Phase optimization for the two 2-m capture cavities

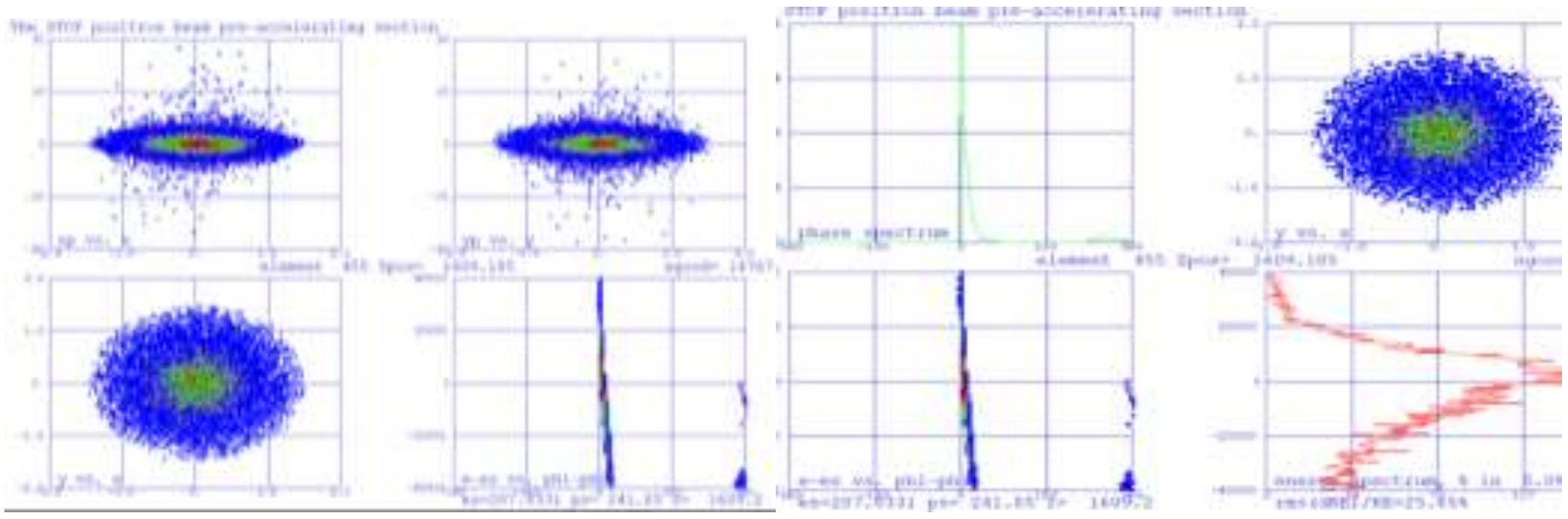

Figure 3.2-31: 200 MeV positron beam particle tracking results

After optimization, the overall positron yield reaches 52.7% with an RMS energy spread of 5.44%. Figure 3.2-32 shows the evolution of beam spot size, bunch length, and energy spread. Figure 3.2-33 presents the transverse and longitudinal dimensions. The main positron bunch is about 60 ps long and contains ~90% of the particles.

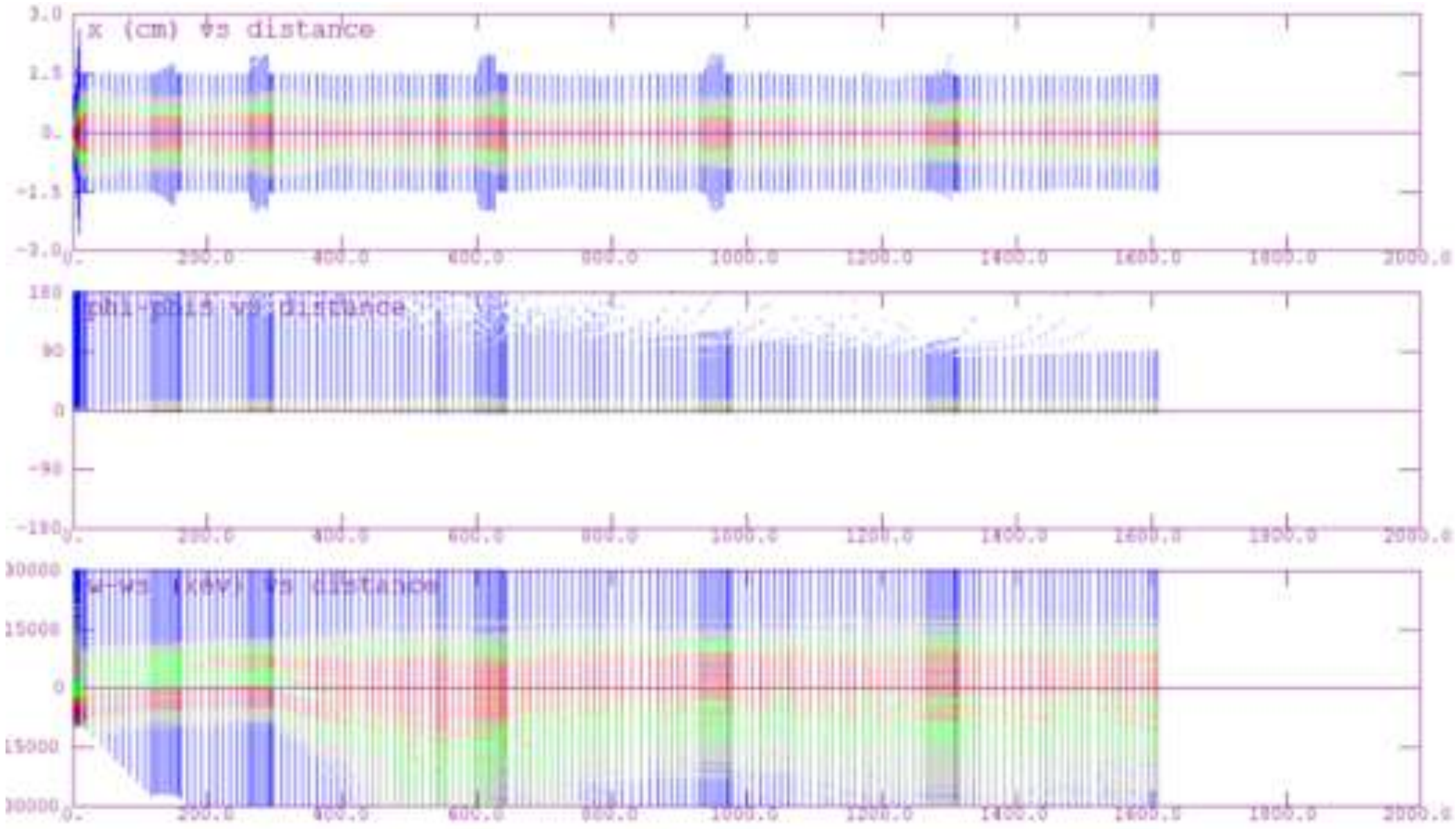

Figure 3.2-32: Variation of positron beam spot size, bunch length, and energy spread along the 200 MeV acceleration section

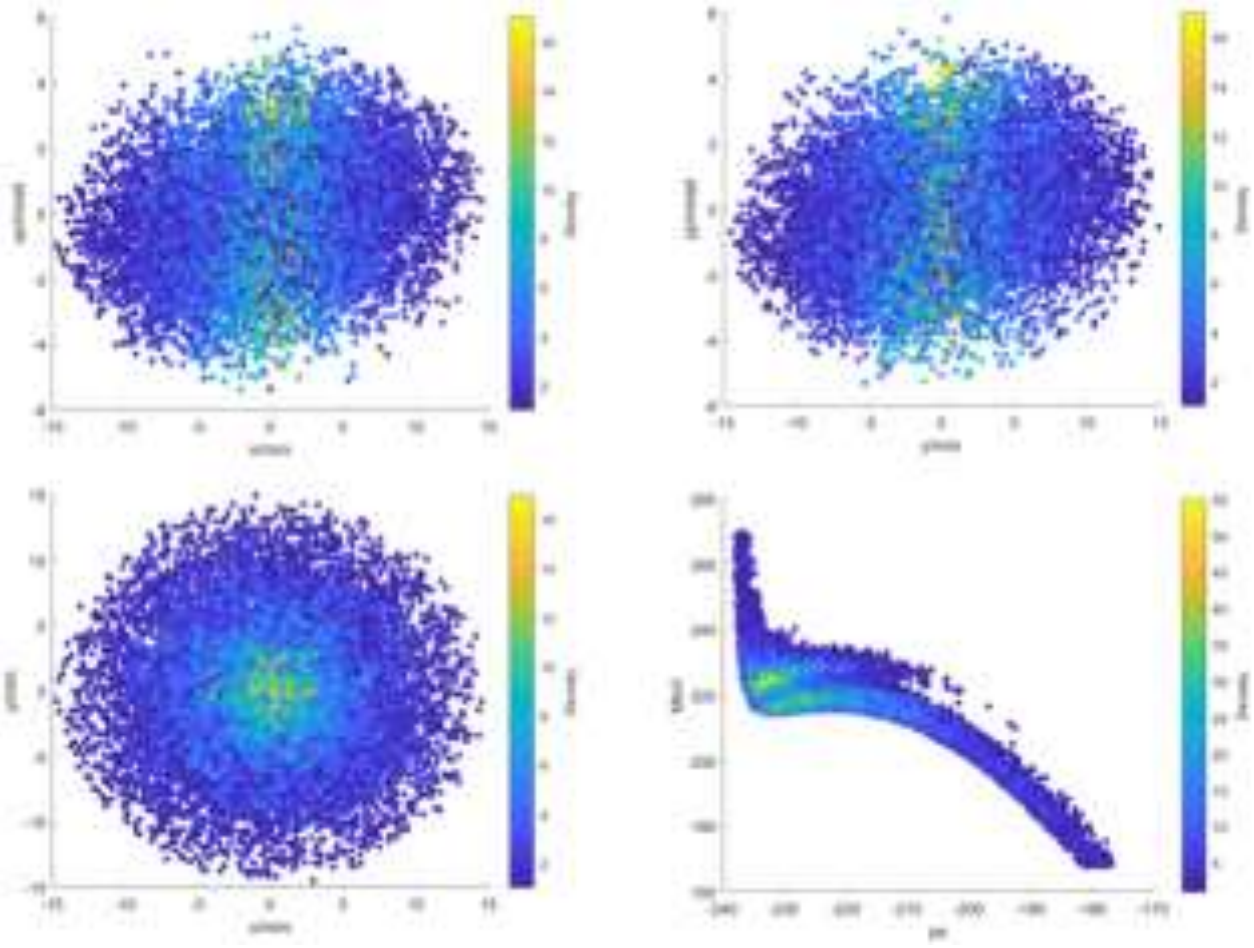

Figure 3.2-33: Phase space calculation results of the 200 MeV positron bunch

**Magnetic Compression Design**

Most high-energy electrons pass through the target and accompany positrons through the accelerator. To separate electrons from positrons, a Chicane is introduced, which also compresses the positron bunch. The layout is shown in Figure 3.2-34, with 8° bend angle and 27.83 mm offset, yielding $R_{56} = -2\theta^2\left(\frac{2}{3}L_{bend} + L_{drift}\right) = -0.05$ m. Optics are shown in Figure 3.2-35.

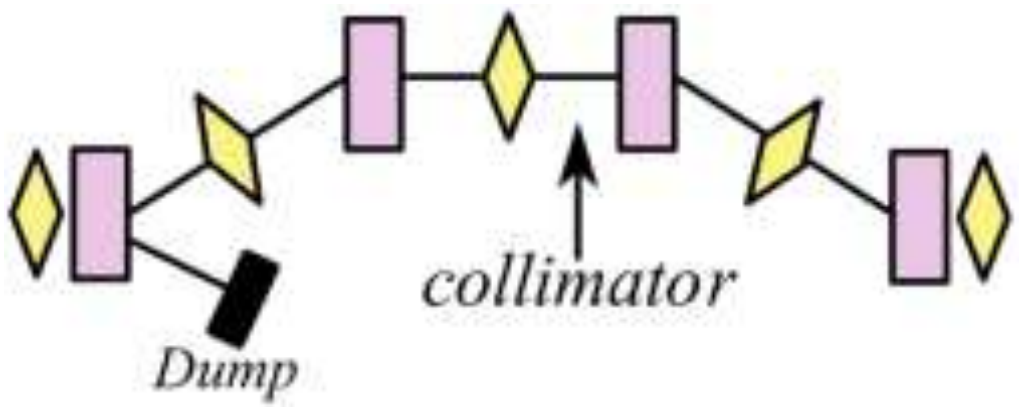

Figure 3.2-34: Chicane structure for positron magnetic compression

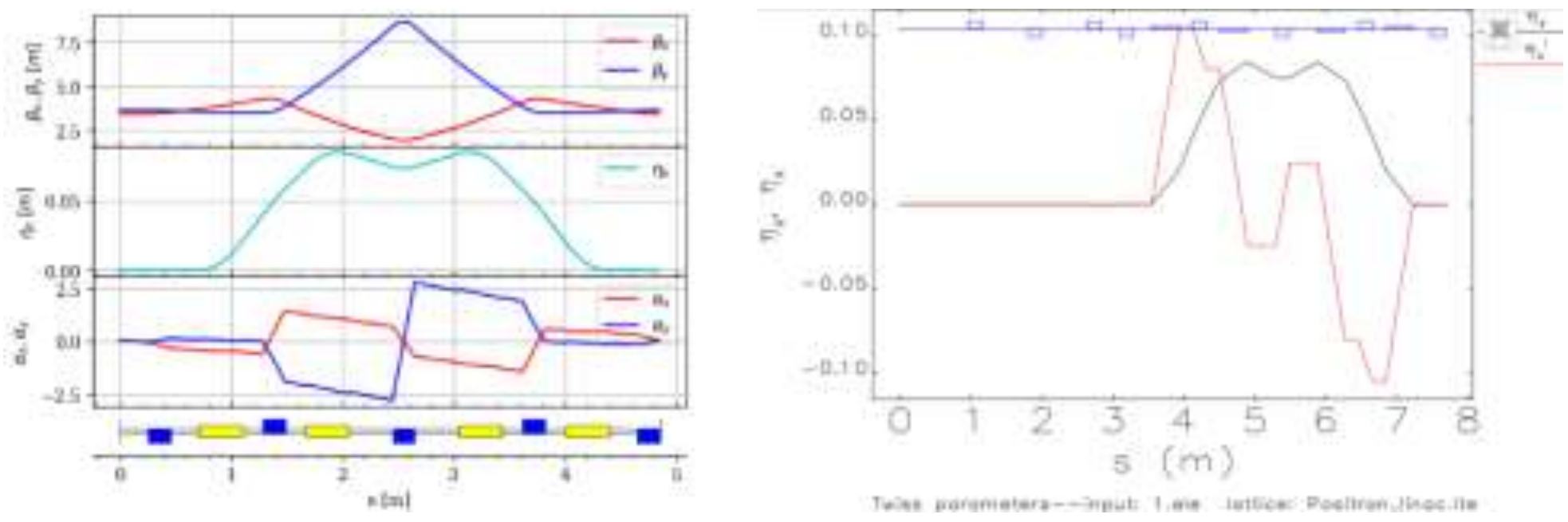

Figure 3.2-35: Optics functions of the Chicane section

To match the positron beam with the damping ring, a horizontal collimator is placed at the Chicane center to remove poor-quality positrons. For off-axis injection, the aperture is 10 mm. Figure 3.2-36 shows the filtering effect, reducing energy spread from 4.14% to 3.18%.

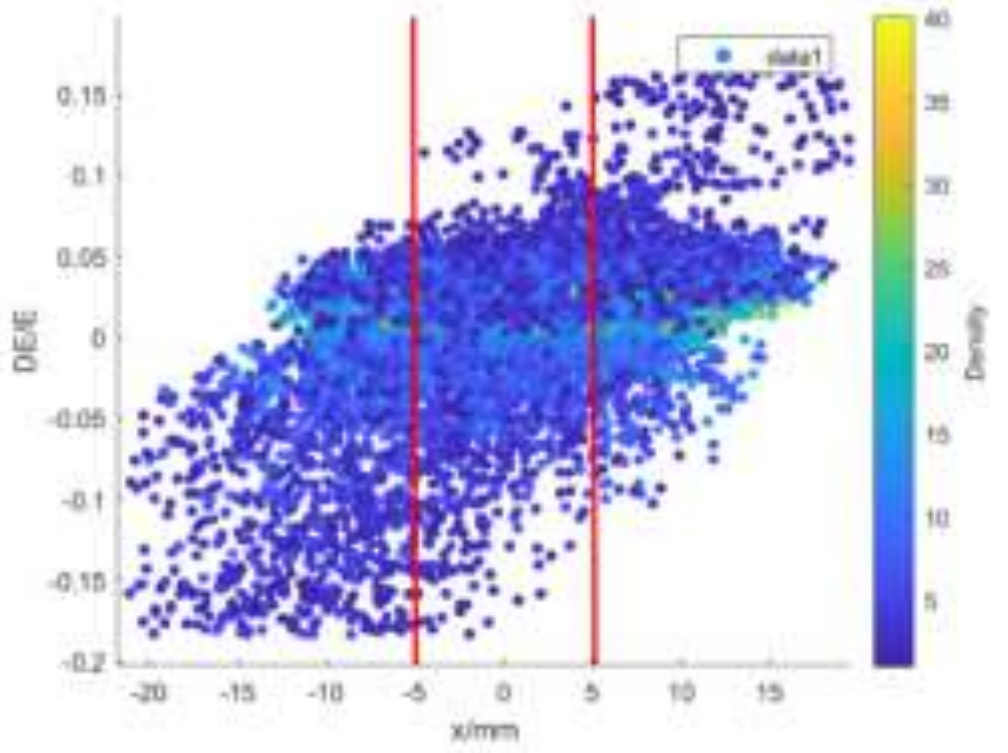

Figure 3.2-36: Schematic of positron beam collimation

**200 MeV to 1 GeV Positron Acceleration**

After initial capture and acceleration to 200 MeV, positrons are separated from electrons via the Chicane. Acceleration from 200 MeV to 1 GeV uses a Triplet-focusing structure (see Figure 3.2-37): from 200–600 MeV, 30 mm aperture S-band structures at 19 MeV/m; from 600 MeV–1 GeV, 20 mm aperture structures at 23 MeV/m.

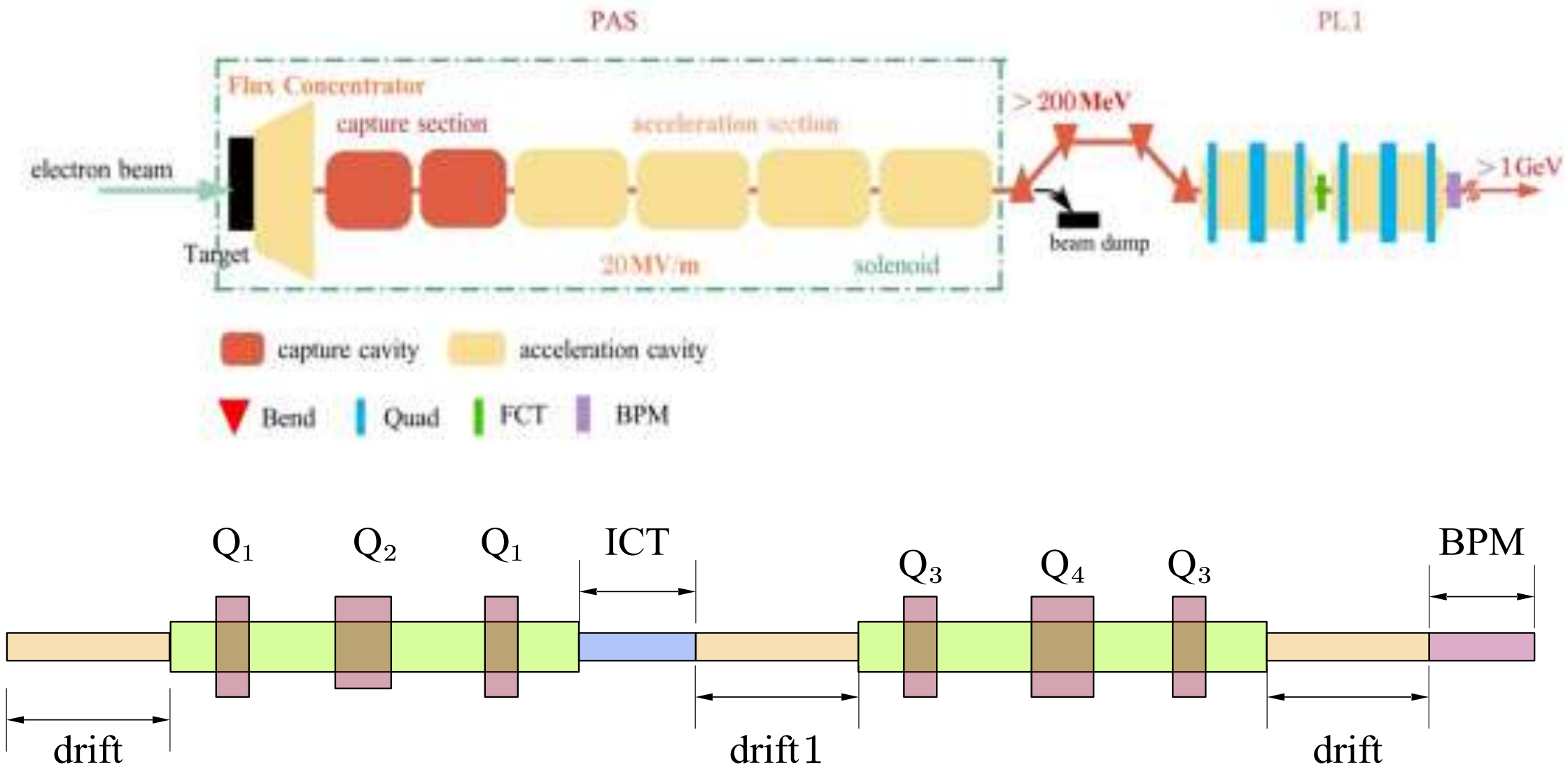

Figure 3.2-37: Layout of the 200 MeV–1 GeV PAS acceleration section

Figures 3.2-38 and 3.2-39 show the Twiss parameters and particle tracking results for the PAS section under off-axis injection. At 1 GeV, the RMS energy spread is 0.77%, positron yield 0.153 (i.e., 1.53 nC per pulse), RMS bunch length 5.62 ps, and geometric emittance $\varepsilon_x$=798 nm·rad, $\varepsilon_y$=1160 nm·rad —meeting the damping ring's injection requirements.

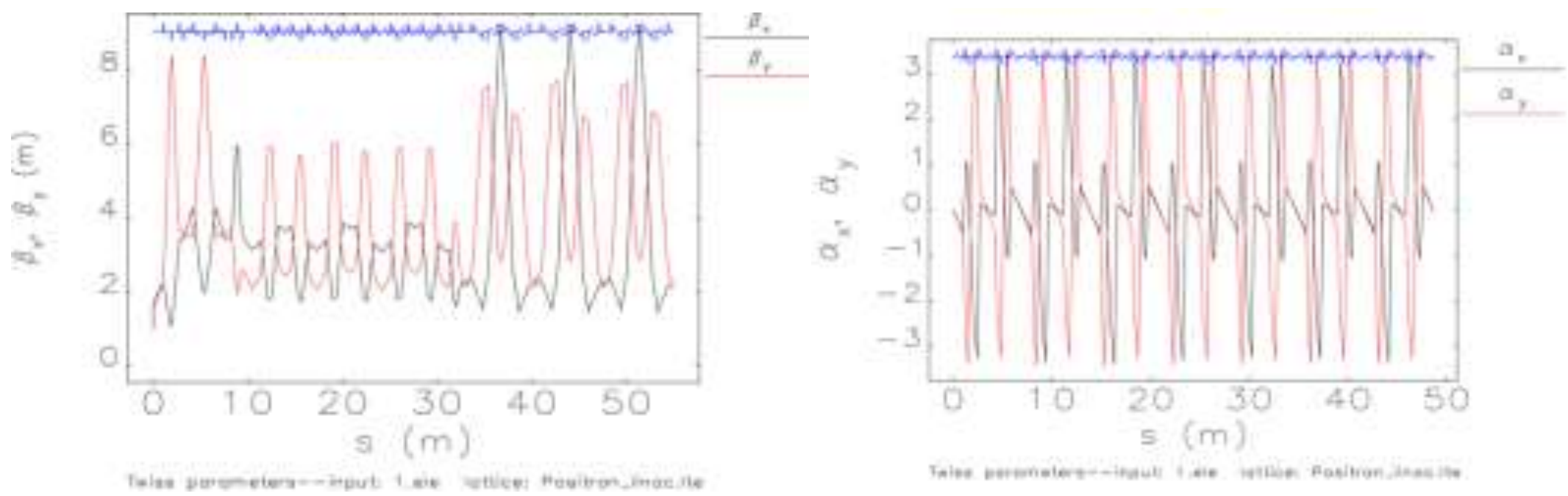

Figure 3.2-38: Twiss parameters of PAS under off-axis injection

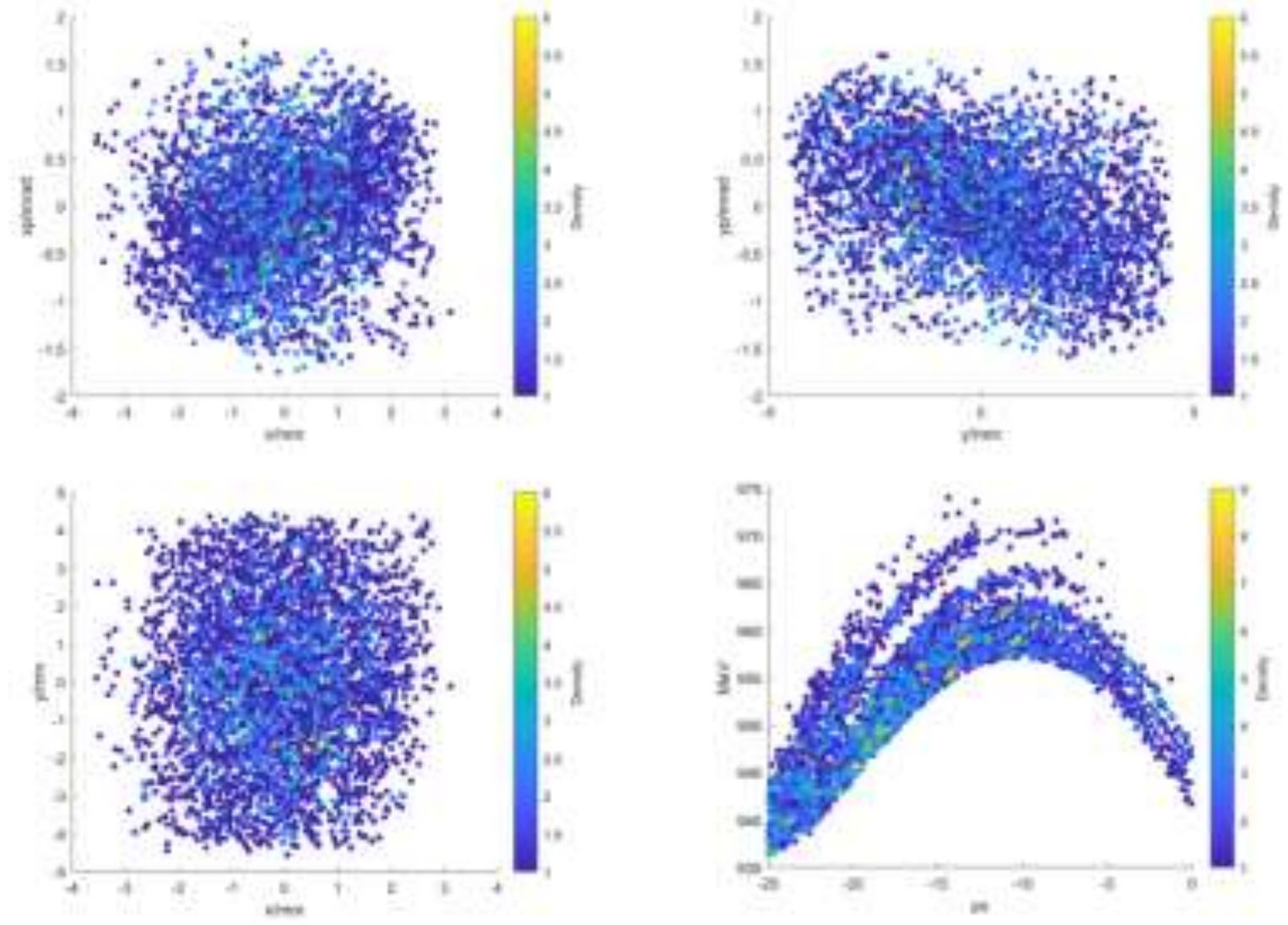

Figure 3.2-39: Particle tracking at 1 GeV PAS exit for off-axis injection

### 3.2.3 Positron Damping Ring

#### *3.2.3.1 Design Requirements*

In the STCF off-axis injection scheme, the injector includes a 1.0-GeV positron damping ring (DR) that serves to reduce the emittance of positron bunches—produced via electron beam target interactions—through synchrotron radiation damping, to levels suitable for injection into the collider positron ring.

The layout of the damping ring system is shown in Figure 3.2-40. It mainly consists of the injection and extraction beamlines and a circular synchrotron accelerator. The positron bunches generated by the positron source have a maximum charge of 1.5 nC and are injected into the damping ring via the injection line. The damping ring operates at an injection frequency of 30 Hz and stores five bunches simultaneously. Each bunch is stored in the ring for 166.7 ms. The extraction frequency is also 30 Hz, with one bunch extracted per cycle and sent to the subsequent linac section.

According to the collider ring design, the injected positron bunch emittance must be below 6 nm·rad at the nominal energy of 2 GeV. This corresponds to a maximum emittance of 12 nm·rad at the damping ring energy of 1 GeV. To satisfy the collider ring requirements while accounting for emittance growth during transport, the extracted bunch from the damping ring must have an emittance of less than 11 nm·rad. Currently, the full emittance (4 times rns) of the injected positron bunch is 1400 nm·rad. Table 3.2-3 compares the design parameters of the STCF damping ring with those of similar international facilities [12, 37, 62].

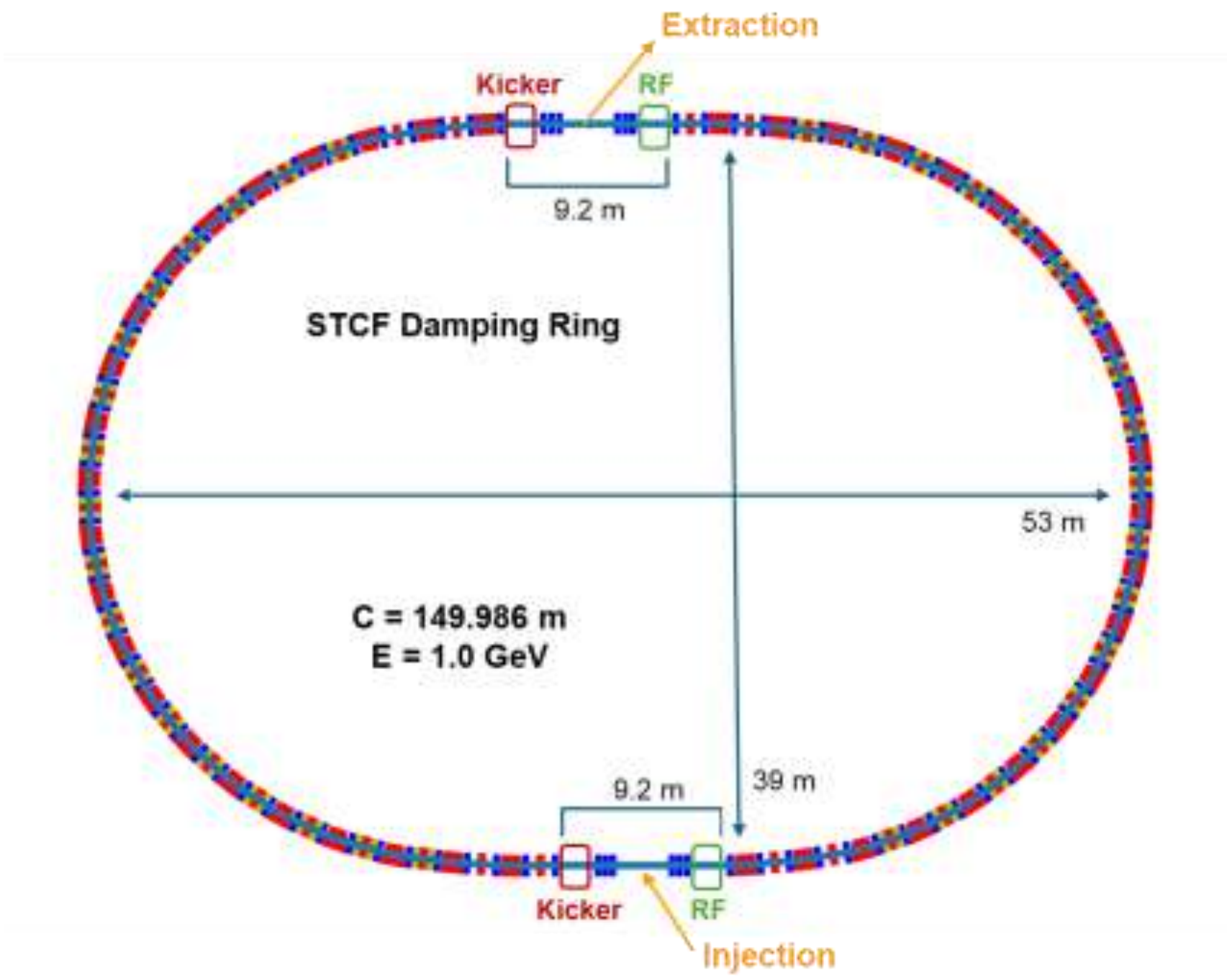

Figure 3.2-40: STCF Damping Ring Layout

Table 3.2-3: Comparison of Positron Damping Rings at STCF and International Facilities

| Parameter | CEPC | SuperKEKB | FCC-ee | STCF |
|---|---|---|---|---|
| Circumference (m) | 147 | 135.498 | 241.8 | 149.986 |
| Energy (GeV) | 1.1 | 1.1 | 1.54 | 1.0 |
| Natural Emittance (nm·rad) | 43.85 | 42.6 | 1.16 | 8.25 |
| Horizontal Injection Emittance (nm·rad) | 1161 | 1400 | 1290 | 350 |
| Horizontal Extraction Emittance (nm·rad) | 77.2 (44.9)* | 43.5 | 1.81 | 8.26 |
| Horizontal Damping Time (ms) | 11.4 | 10.87 | 10.6*** | 28.8 |
| Storage Time (ms) | 20 (40)* | 40 | 40 | 166.7 |
| Stored Bunches $N_{train} \times n_b$ | 2(4)*×1(2)** | 2×2 | 5×2 | 5×1 |

* Indicates CEPC operating in a 40-ms storage mode

** Indicates CEPC front-end Linac operating in dual-beam mode under Z-pole operation

*** FCC-ee utilizes two 6.64 m-long wigglers to reduce damping time

### *3.2.3.2 Damping Ring Optics Design*

The extracted emittance from the damping ring is calculated using the following formula:

$$\varepsilon_{ext}(t) = \varepsilon_{nat} + \left(\varepsilon_{inj} - \varepsilon_{nat}\right) \cdot e^{-2t/\tau} \quad (1)$$

where $\varepsilon_{inj}$、$\varepsilon_{ext}$、$\varepsilon_{nat}$ represent the injected, extracted, and natural emittances respectively, $t$ is the storage time in the damping ring, and $\tau$ is the damping time. If the injected emittance is 1400 nm·rad and the goal is to reduce the extracted emittance below 11 nm·rad, the beam must be stored in the ring for at least three damping times, and the natural emittance must be below 10 nm·rad. Since the natural emittance is proportional $\gamma^2/\rho$ and the damping time is proportional to $\rho/\gamma^3$, where $\rho$ is the bending radius and $\gamma$ is the Lorentz factor, it is difficult to achieve both a small natural emittance and a short damping time through optics design alone, without introducing additional synchrotron radiation sources like wigglers. Given that the STCF DR targets a much lower extracted emittance than CEPC and SuperKEKB, the damping time will necessarily be longer than those in those machines. Additionally, to reduce costs, wigglers will not be used. Therefore, the lattice design strategy is to first minimize natural emittance through optical design, then shorten the damping time as much as possible, and finally use a sufficiently long storage time to achieve the target emittance.

The damping ring will adopt a racetrack layout composed of two straight sections and two 180° arcs, minimizing the total circumference and hence the damping time. Both straight sections are dispersion-free and used for the injection/extraction elements and RF cavity. The ring will be based on a FODO lattice, which is sufficient for achieving a natural emittance in the nm·rad range. Although MBA lattices can yield even lower emittance, they reduce the dynamic aperture. The FODO structure here employs a Reverse Bend (RB) scheme, enhancing synchrotron radiation while achieving low emittance.

Several RB strategies have been proposed internationally. In 1989, CERN proposed a scheme in the CLIC design where dipoles with opposite bend directions were embedded within focusing and defocusing quadrupole pairs. However, this scheme offered limited flexibility for phase advance tuning. The ALS later introduced a reversed central dipole (Superbend) in a TBA structure. A similar "positive-negative-positive" dipole pattern was adopted by the U.S. EIC design, which is unsuitable for compact damping rings. KEK's 2005 SuperKEKB DR design proposed a "positive-negative" dipole arrangement within each FODO cell, which allows for damping time tunability and compact layout—STCF adopts this latter design.

With the RB scheme, the damping time $\tau$, natural emittance $\varepsilon$, and momentum compaction factor $\alpha_c$ are calculated as follows [63, 64]:

$$\tau_x = \frac{2ET_0}{J_x U_0}\frac{1-r}{1+|r|}, \quad \tau_y = \frac{2ET_0}{J_y U_0}\frac{1-r}{1+|r|}, \qquad \tau_E = \frac{2ET_0}{J_E U_0}\frac{1-r}{1+|r|}$$

$$\epsilon_x = C_q \frac{l\theta^2}{\rho}\gamma^2 F(r,\phi), \quad \alpha_c = G(r,\phi)\theta^2, \tag{2}$$

where:

- $E$: beam energy,
- $T_0$: revolution period,
- $r$: reverse bend factor (ratio of reverse to forward dipole angles),

- $U_0$: synchrotron radiation energy loss per turn,
- $J$: damping partition numbers,
- $\phi$: phase advance per cell,
- $\theta$: forward dipole bending angle,
- $F(r,\phi)$, $G(r,\phi)$: optical functions.

The explicit forms are:

$$G(r,\phi) = \frac{(1+r^2)(3+cos\,\phi)-8r}{16\,sin^2(\phi/2)}$$

$$F(r,\phi) = \frac{1}{24\,sin^2(\phi/2)sin\phi}\left\{\begin{array}{c} 1+5|r|+r^2 \\ +2(5-12r-2|r|+5r^2)cos^2(\phi/2) \\ +(1-|r|+r^2)\,cos^2(\phi/2) \end{array}\right\}$$

Figures 3.2-41 and 3.2-42 show the dependence of the damping time, natural emittance, and momentum compaction factor on the reverse bend factor *r*. Taking into account the requirement on damping time and feasibility of magnet fabrication, we choose *r*=0.3 as the design value.

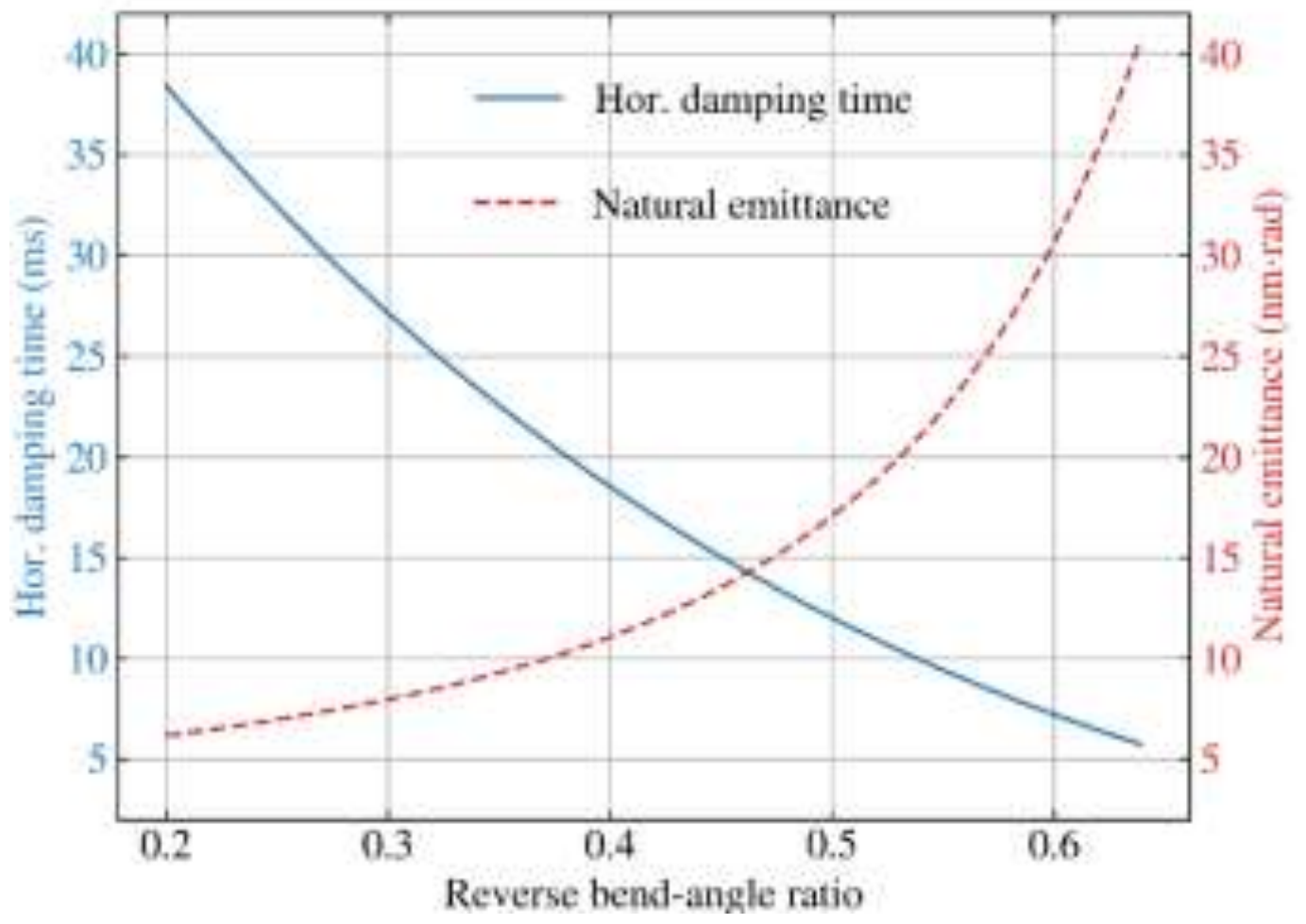

Figure 3.2-41: Horizontal damping time and natural emittance versus reverse bend factor

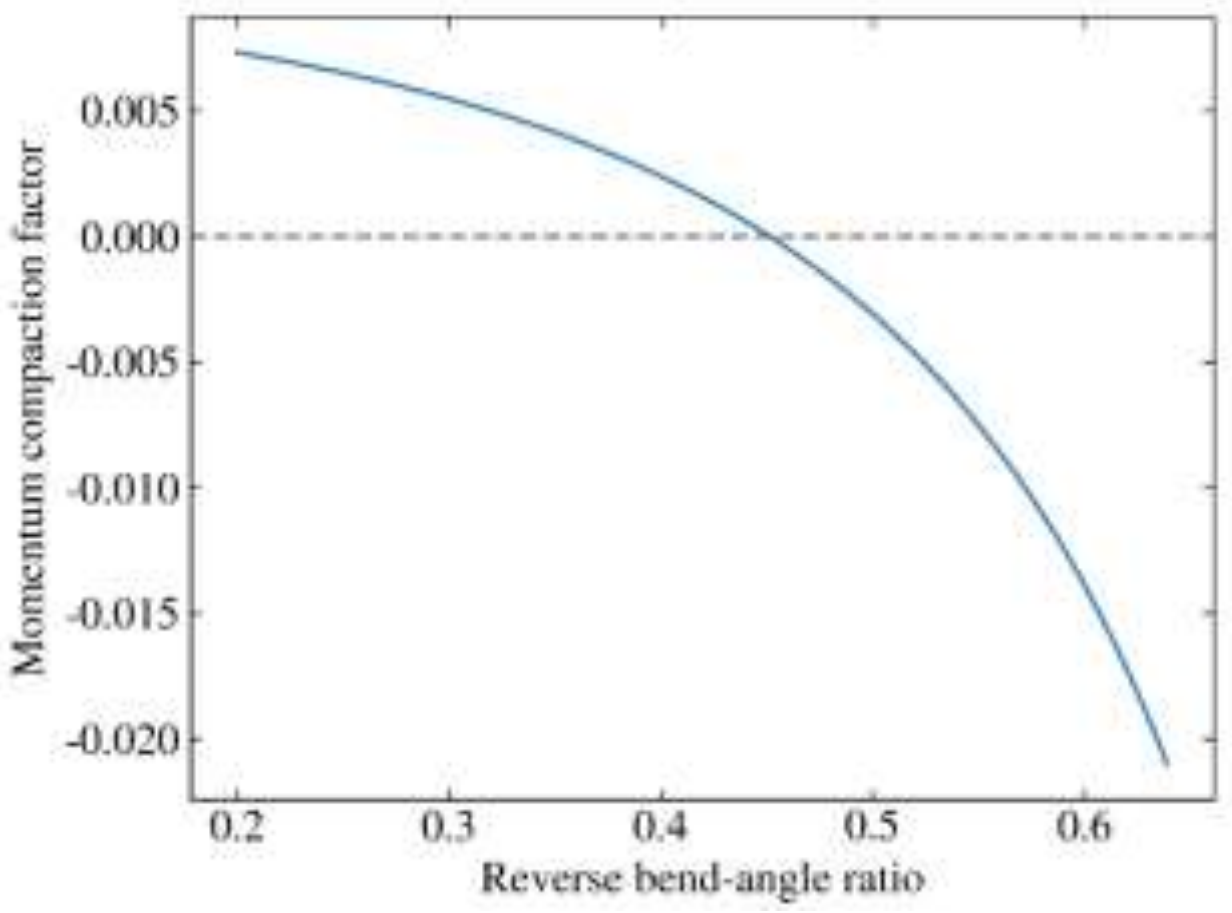

Figure 3.2-42: Momentum compaction factor versus reverse bend factor

Each arc contains 18 FODO cells (36 total in the ring). The phase advance per FODO cell is 90°. The forward and reverse dipoles are 1.0 m and 0.3 m long, respectively, both with a bending radius of 4.456 m and a field strength of 0.749 T. Each FODO cell is 3.0 m long. The Twiss parameters for the arc cell are shown in Figure 3.2-43.

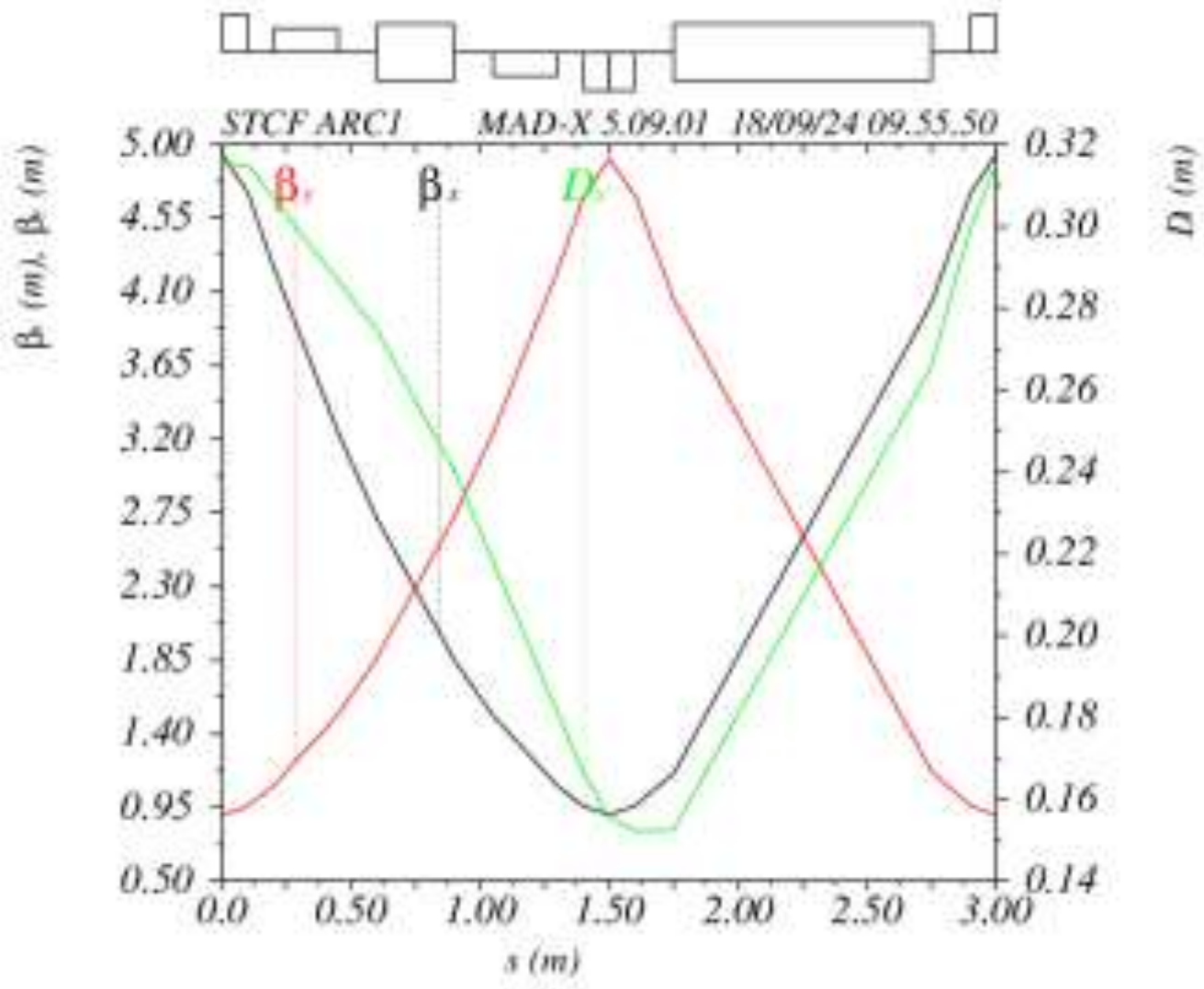

Figure 3.2-43: Twiss parameters for arc FODO cell in STCF damping ring

Each arc ends with two half-strength FODO cells for dispersion suppression. After that comes a straight section of about 9.2 m used for injection/extraction and housing the RF cavity. Twiss parameters for this region are shown in Figure 3.2-44. The injection/extraction point lies at the center of the straight, with horizontal and vertical beta functions of 9.5 m and 1.0 m, respectively. The RF cavity positions are indicated in the figure. Downstream of the injection/extraction point, three quadrupoles lead to the kicker location. Global Twiss parameters are shown in Figure 3.2-45.

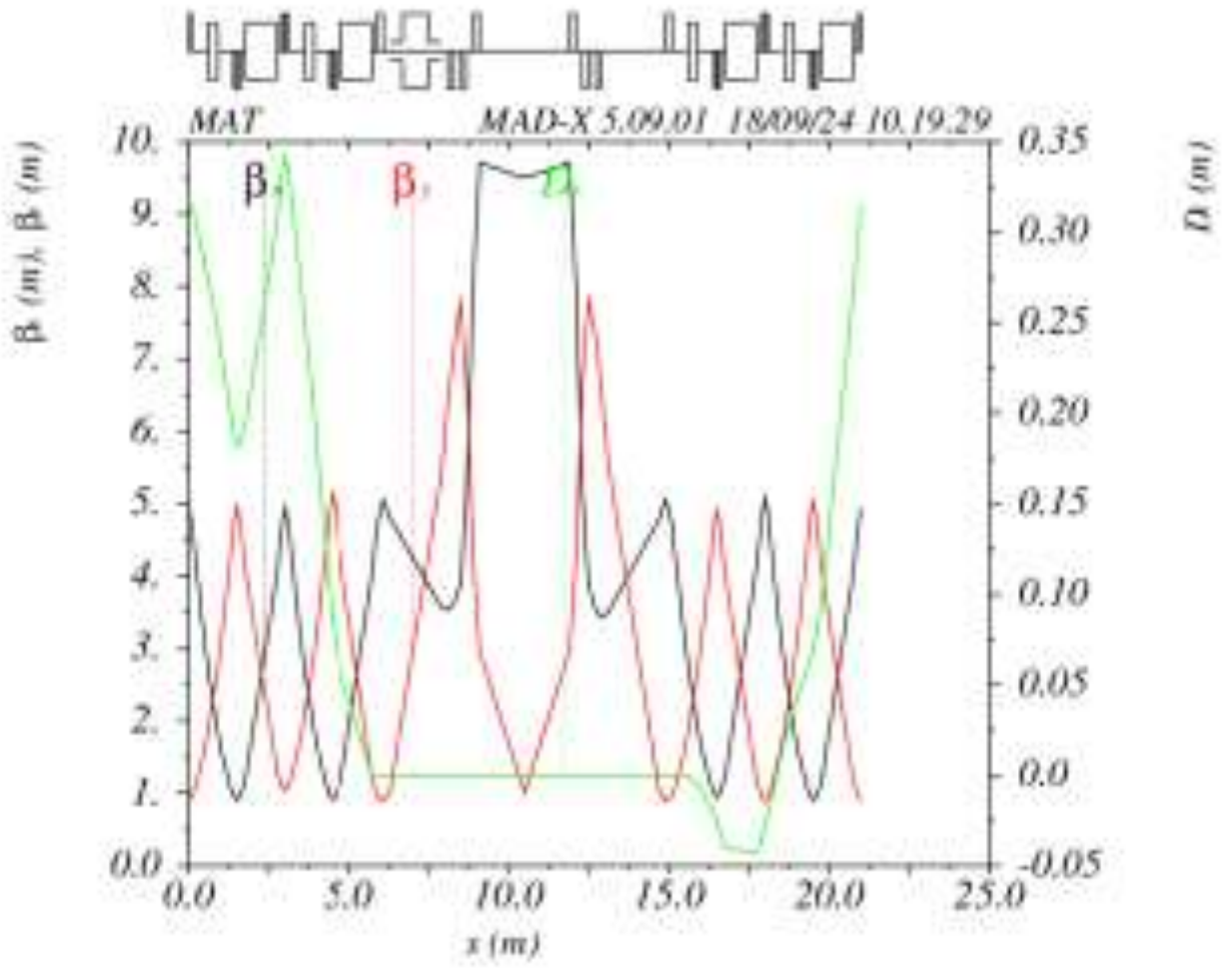

Figure 3.2-44: Twiss parameters of dispersion-free and straight section

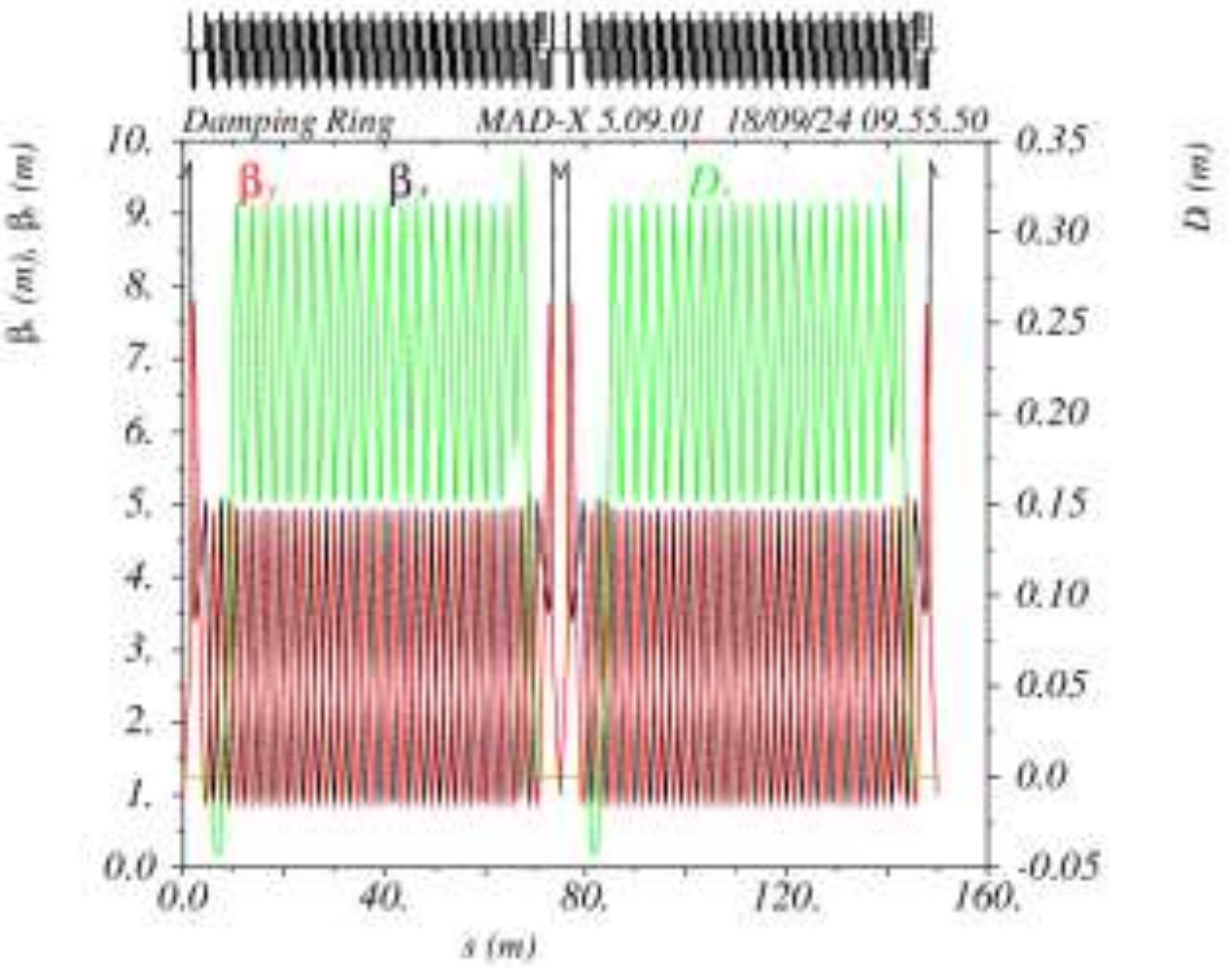

Figure 3.2-45: Global Twiss parameters of STCF damping ring

Figure 3.2-46 shows the dynamic aperture simulated with Elegant, which exceeds 10 times the RMS bunch size at the injection point in both transverse directions, fully meeting the injection requirement.

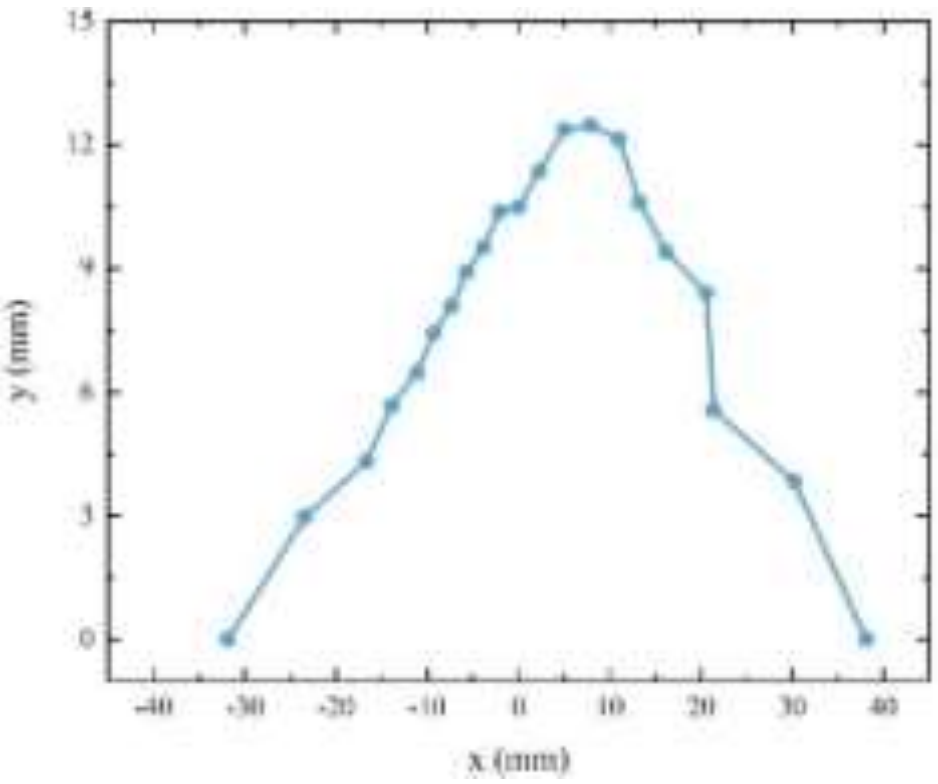

Figure 3.2-46: Dynamic aperture of the STCF damping ring

Figure 3.2-47 shows the emittance evolution of injected positrons as simulated using multi-particle tracking in Elegant. With a damping time of 28.8 ms and a design equilibrium emittance of 8.25 nm·rad, a 166.7 ms storage time ensures the emittance reduces from 350 nm·rad to about 8.4 nm·rad, which meets the extraction requirements.

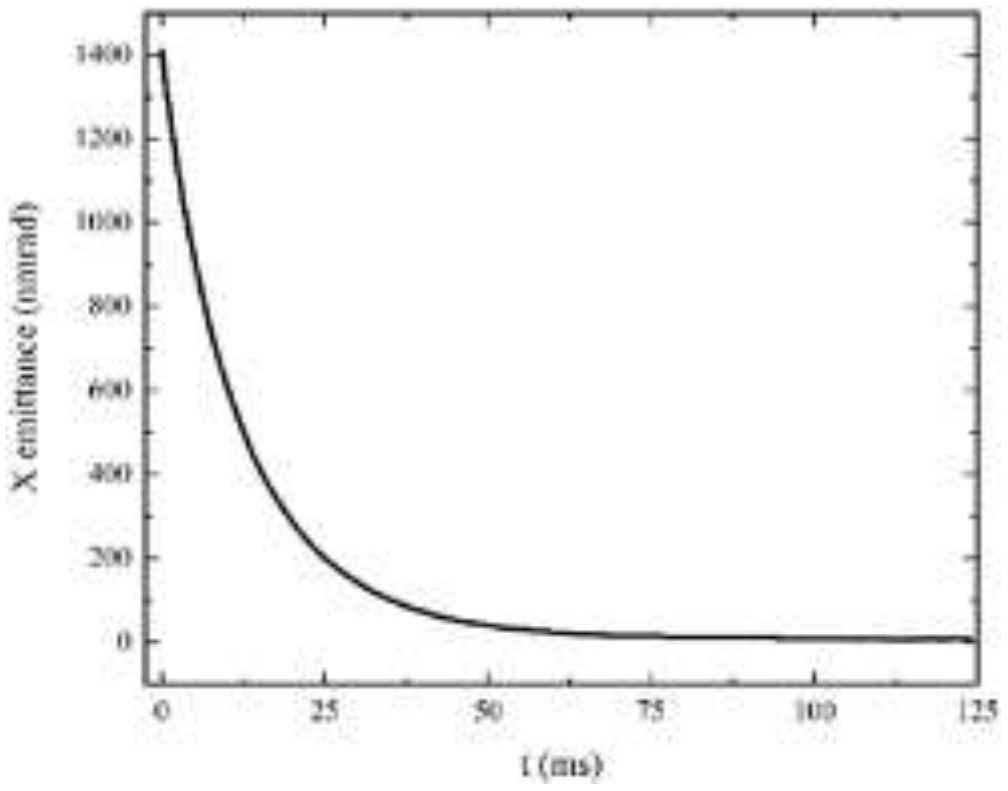

Figure 3.2-47: Injected positron beam emittance evolution in STCF damping ring

Table 3.2-4: Key Parameters of the STCF Damping Ring

| Parameter | Value |
|---|---|
| Energy (GeV) | 1.0 |
| Circumference (m) | 149.986 |
| Number of bunches | 5 |
| Max charge per bunch (nC) | 1.5 |
| Max current (mA) | 15 |
| Arc FODO phase advance x/y (deg) | 90 / 90 |
| Dipole bending radius (m) | 4.456 |
| Dipole field strength (T) | 0.749 |
| Reverse bend factor rrr | 0.3 |
| Synchrotron radiation loss per turn (keV) | 35 |
| Damping time x/y/z (ms) | 28.8 / 28.6 / 14.2 |
| Storage time (ms) | 166.7 |
| Natural emittance x/y (nm·rad) | 8.25 / 0.1 |
| Injected emittance x/y (nm·rad) | 350 / 350 |
| Extracted emittance x/y (nm·rad) | 8.26 / 0.2 |

| Parameter | Value |
|---|---|
| Natural relative energy spread (%) | 0.031 |
| Momentum compaction factor | 0.0064 |
| RF frequency (MHz) | 499.7 |
| RF voltage (MV) | 0.2 |
| Harmonic number | 250 |
| Natural chromaticity x/y | -15.0 / -15.3 |

#### *3.2.3.3 Injection and Extraction Design of the Damping Ring*

The damping ring stores a total of five positron bunches evenly distributed around the ring. After each bunch is extracted, it must be replenished via bucket-to-bucket single-turn injection. During each injection, a septum magnet deflects the incoming positron bunch from the transfer line into the acceptance of the damping ring. When the bunch center reaches the beam pipe center, a pulsed kicker magnet is used to remove the injection angle. The STCF damping ring adopts a horizontal single-turn injection scheme, consisting of one injection kicker magnet and one injection septum magnet. The layout of the injection system is shown in Figure 3.2-48.

The exit of the septum magnet is located 9 mm from the reference orbit of the circulating beam. The horizontal acceptance is 8100 nm·rad. The septum blade is 4 mm thick, and the injected beam envelope is 3.7 mm (corresponding to 1400 nm·rad). At the exit of the septum magnet, the injected beam center is 18 mm from the reference orbit, with an angle of -2.82 mrad. Parameters for the injection septum and kicker magnets are listed in Tables 3.2-5 and 3.2-6, respectively.

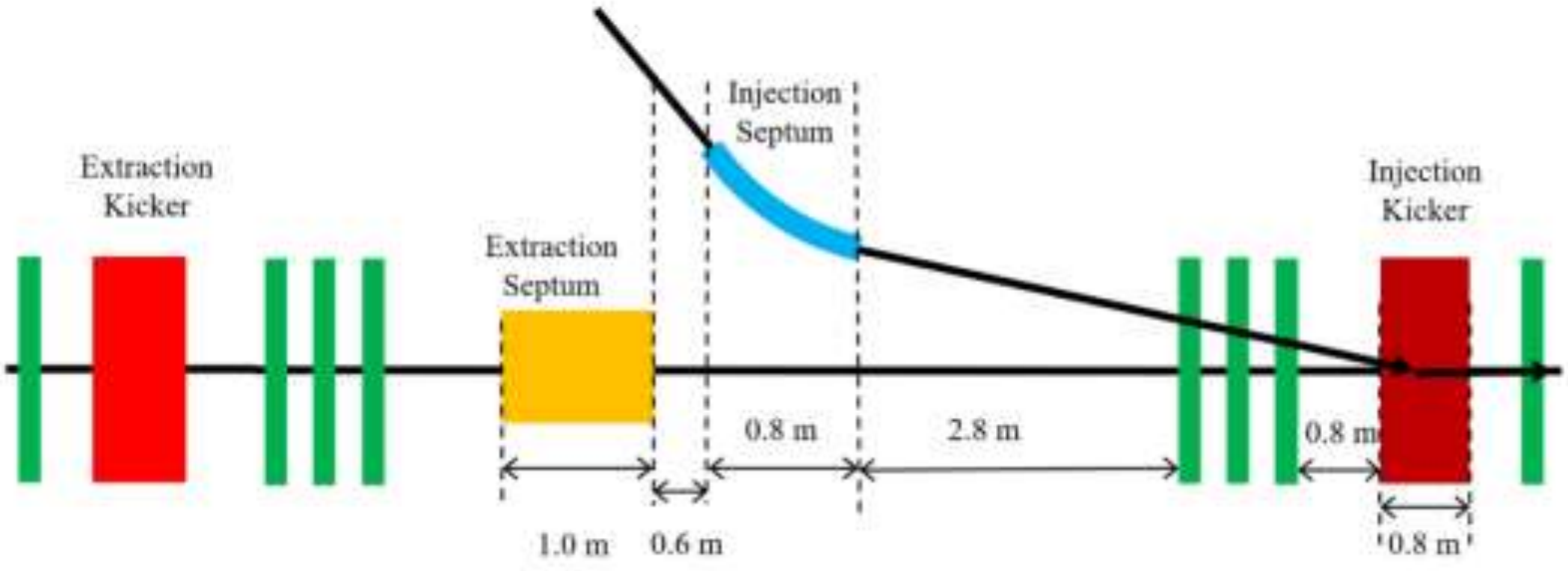

Figure 3.2-48: Layout of the horizontal injection system for the STCF damping ring

Table 3.2-5: Key Parameters of the Injection Kicker Magnet

| Parameter | Specification |
|---|---|
| Injection Direction | Horizontal |
| Number of Bunches | 1 |
| Kicker Type | Stripline |
| Deflection Angle | 4.04 mrad |
| Effective Magnet Length | 800 mm |
| Pulse Width | < 2 ns |
| Rise Time | < 35 ns |

Table 3.2-6: Key Parameters of the Injection Septum Magnet

| Parameter | Specification |
|---|---|
| Injection Direction | Horizontal |
| Deflection Angle | 10° |
| Effective Magnet Length | 800 mm |
| Magnetic Field Strength | 0.73 T |
| Septum Thickness | 4 mm |

The damping ring also employs a horizontal single-turn extraction scheme, using one horizontal extraction septum and one extraction kicker magnet. The magnet parameters are identical to those listed in Tables 3.2-5 and 3.2-6. The layout of the extraction straight section is symmetric to that of the injection section.

At present, the STCF positron damping ring achieves an extracted emittance below 11 nm·rad after a storage time of 166.7 ms, meeting the design requirements. The field strengths of the dipole, quadrupole, sextupole, and kicker magnets in the damping ring are all within feasible ranges. The engineering design of all components is technically achievable, with no apparent risks, indicating good overall feasibility.

## 3.3 Injector Physics Design for Swap-Out Injection

### 3.3.1 Electron Source and Main Linac SEL1 for Direct Injection

#### *3.3.1.1 Scheme 1 – High-Charge Thermionic Electron Source*

- **System Design Requirements and Specifications**

The beam parameters for swap-out injection are specified in Table 3.1-1. This design adopts a thermionic electron source with an RF frequency of 2998.2 MHz. The final output of the system is a single bunch with a charge of 8.5 nC, geometric emittance ≤ 30 nm·rad, RMS energy spread ≤ 0.5%, and RMS bunch length ≤ 2.0 mm.

- **Key Technologies and Design Strategy**

The linac scheme is determined by the beam injection requirements of the collider rings. Based on the overall performance goals, the design involves beam bunching and acceleration structures, combining theoretical modeling with simulation studies. The interaction between multiple physical fields and the driving beam is analyzed to establish an accurate theoretical framework. This guides the RF field and focusing magnet design. Multi-objective optimization algorithms are developed to enhance design efficiency. Parameters optimized by these algorithms are fed into 3D multi-particle beam dynamics simulations to evaluate bunching system performance.

Key technical challenges include the design of sub-harmonic bunchers, optimization of phase velocity distributions in non-uniform multi-cell structures, axial magnetic field design, longitudinal phase space optimization during acceleration, and transverse matching of beam parameters.

- **System Configuration**

The swap-out injection linac includes a bunching system and a relativistic main accelerating section, as shown in Figure 3.3-1. The bunching system consists of two sub-harmonic prebunchers with frequencies of 1/18 and 1/6 of the main frequency, an S-band traveling-wave buncher, and one section of velocity-of-light accelerating structure. The main linac consists of 68 SLAC-type 3-meter-long accelerating sections.

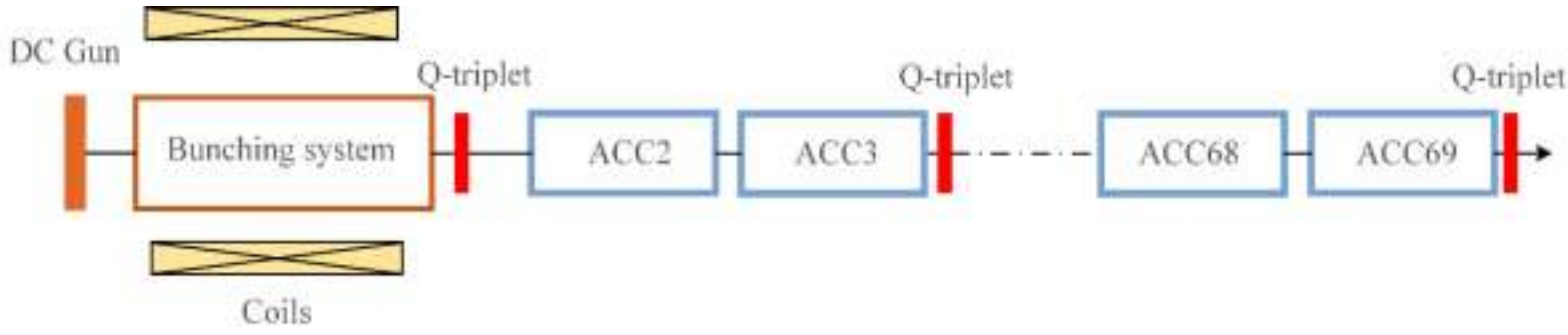

Figure 3.3-1: Layout of the swap-out injection linac (high-charge thermionic option)

- **Bunching System Design**

The thermionic electron gun produces a bunch with a full-width at half maximum (FWHM) of 1.3 ns, a single-bunch charge of 8.5 nC, and a kinetic energy of 200 keV. Its longitudinal distribution and projections in longitudinal and transverse directions are shown in Figure 3.3-2.

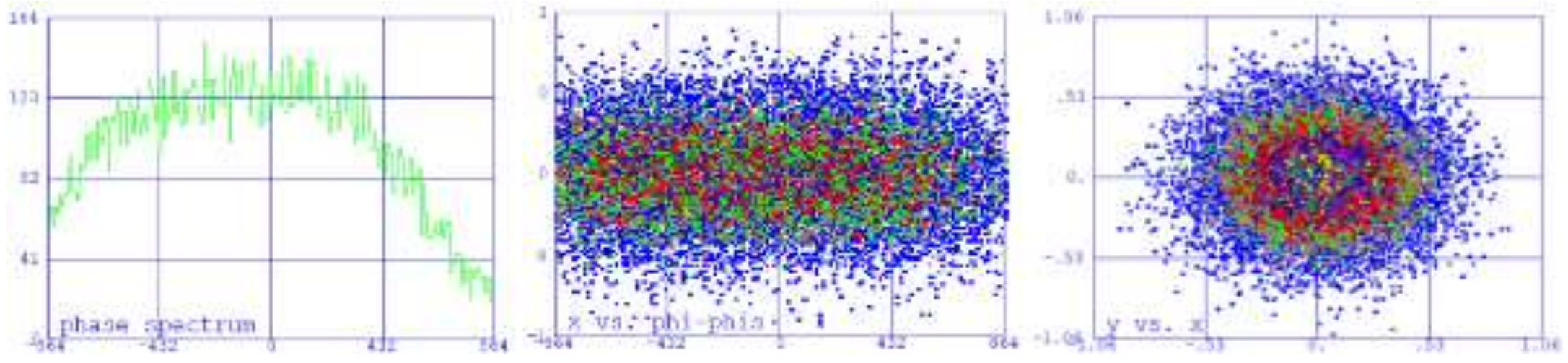

Figure 3.3-2: Input beam to the bunching system

After passing through the two sub-harmonic bunchers and the traveling-wave buncher (Figure 3.3-1), the bunch is compressed to 5 ps (RMS, 95% particles) and then accelerated by a 3-meter SLAC accelerating section to ~60 MeV with an energy spread of 1.39% (RMS, 95% particles). The traveling-wave buncher adopts a variable phase velocity and gradient disc-loaded waveguide structure. The accelerating field profiles for the buncher and the relativistic section are shown in Figure 3.3-3 (left and right, respectively).

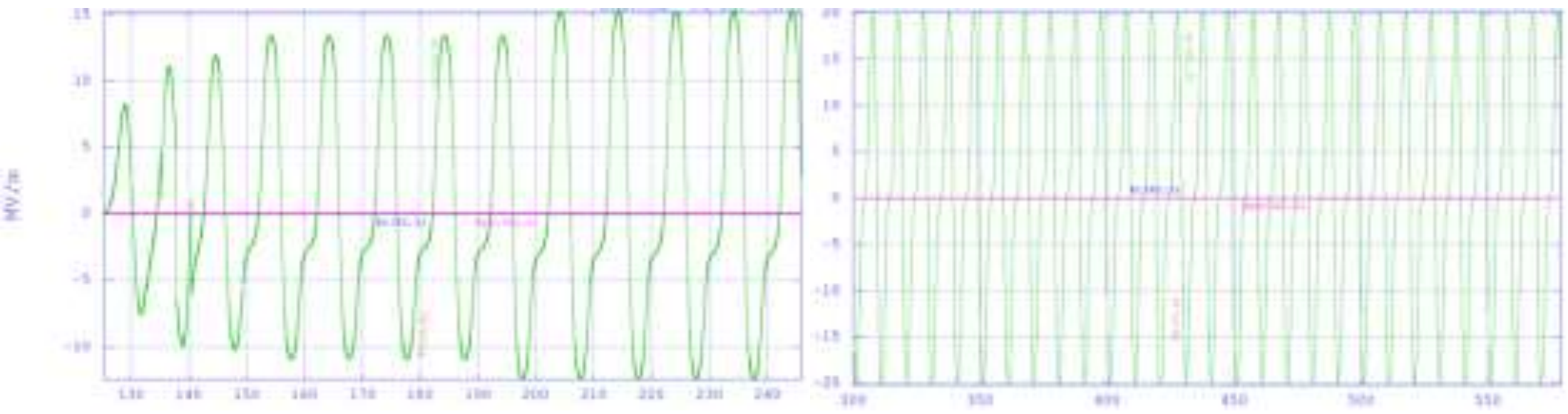

Figure 3.3-3: Field distributions of the traveling-wave buncher (left) and relativistic accelerating section (right)

At the output of the bunching system, the beam energy gain is shown in Figure 3.3-4, while the transverse and longitudinal phase spaces are shown in Figures 3.3-5 and 3.3-6.

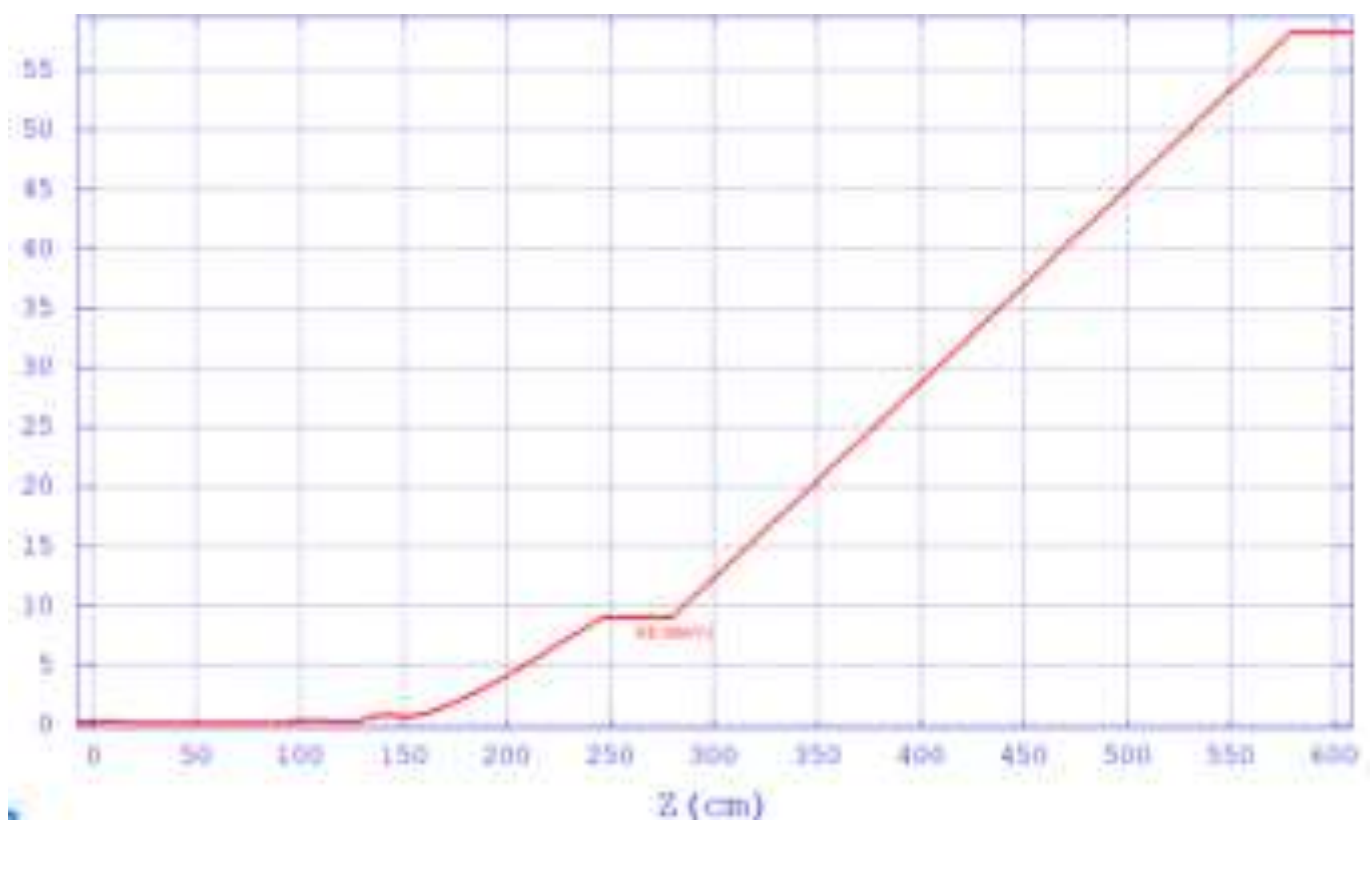

Figure 3.3-4: Energy gain in the bunching system

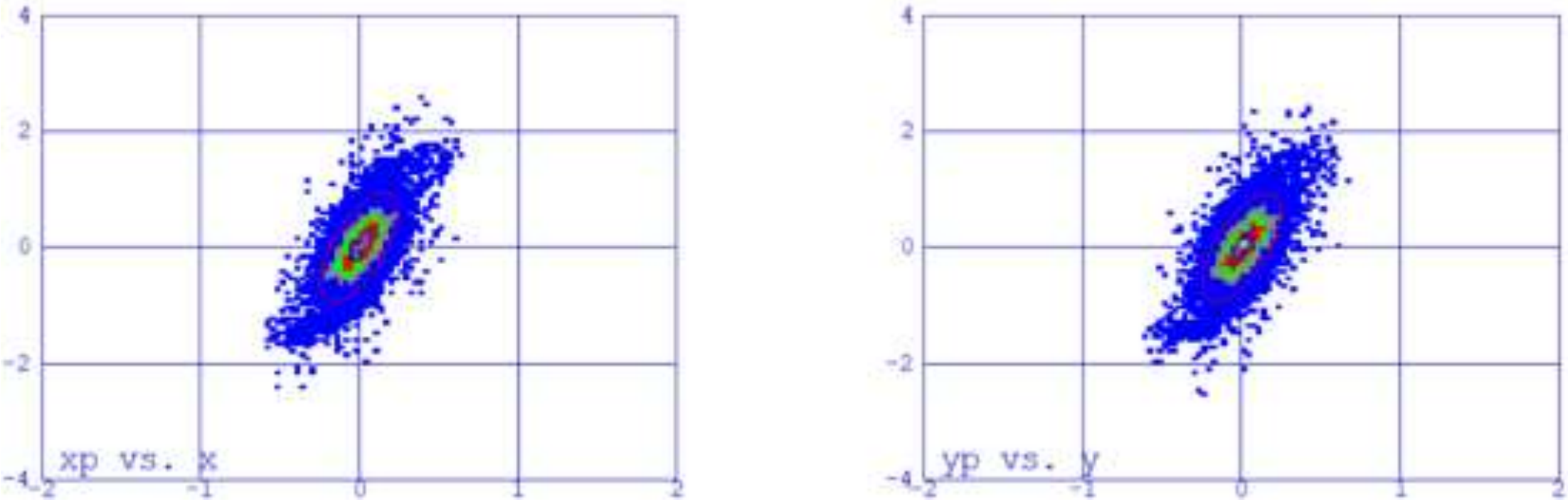

Figure 3.3-5: Transverse phase space at the bunching system exit

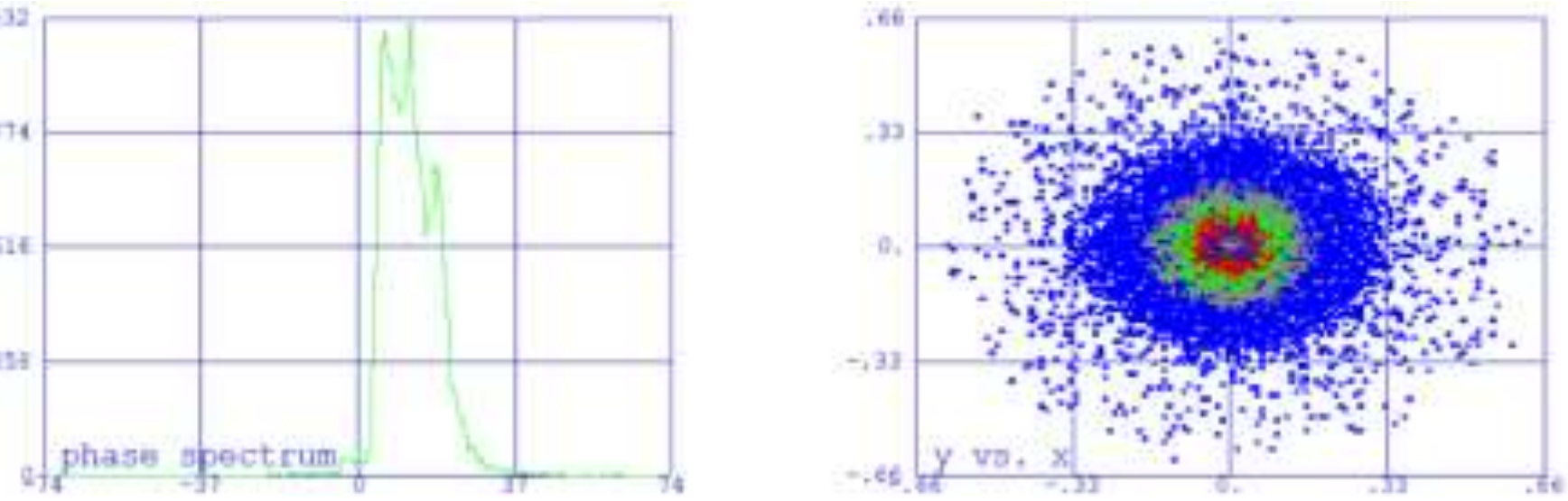

Figure 3.3-6: Longitudinal phase space and beam dynamics at the bunching system exit

To limit transverse beam loss, a solenoidal focusing system is used to constrain the beam size. The magnetic field distribution along the bunching section is shown in Figure 3.3-7. The evolution of transverse beam envelope, phase deviation, and energy deviation along the beamline is shown in Figure 3.3-8. The evolution of key parameters—including RMS bunch length, energy spread, normalized transverse emittance, and RMS transverse size—is presented in Figure 3.3-9.

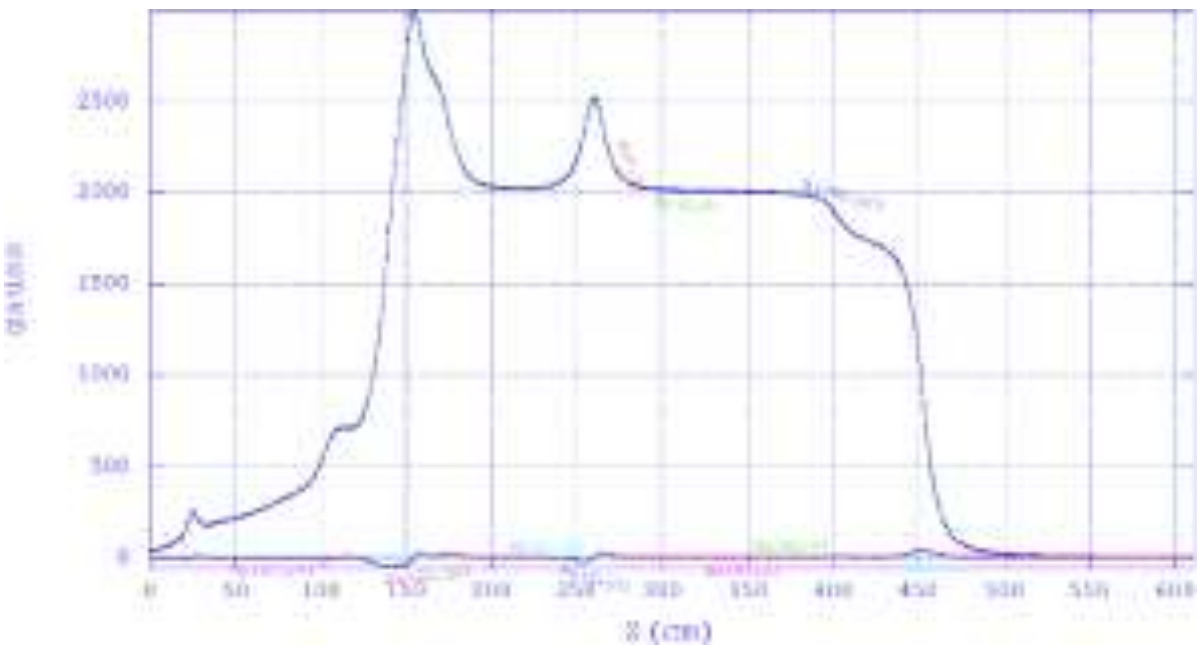

Figure 3.3-7: Magnetic field distribution along the bunching section

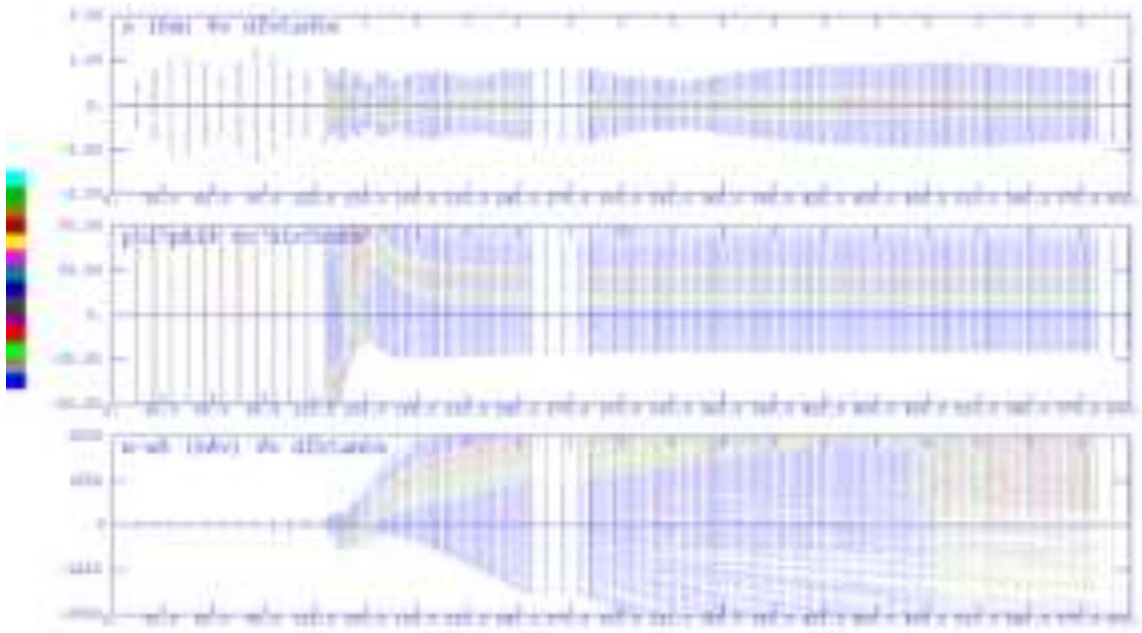

Figure 3.3-8: Evolution of transverse envelope, phase offset, and energy deviation

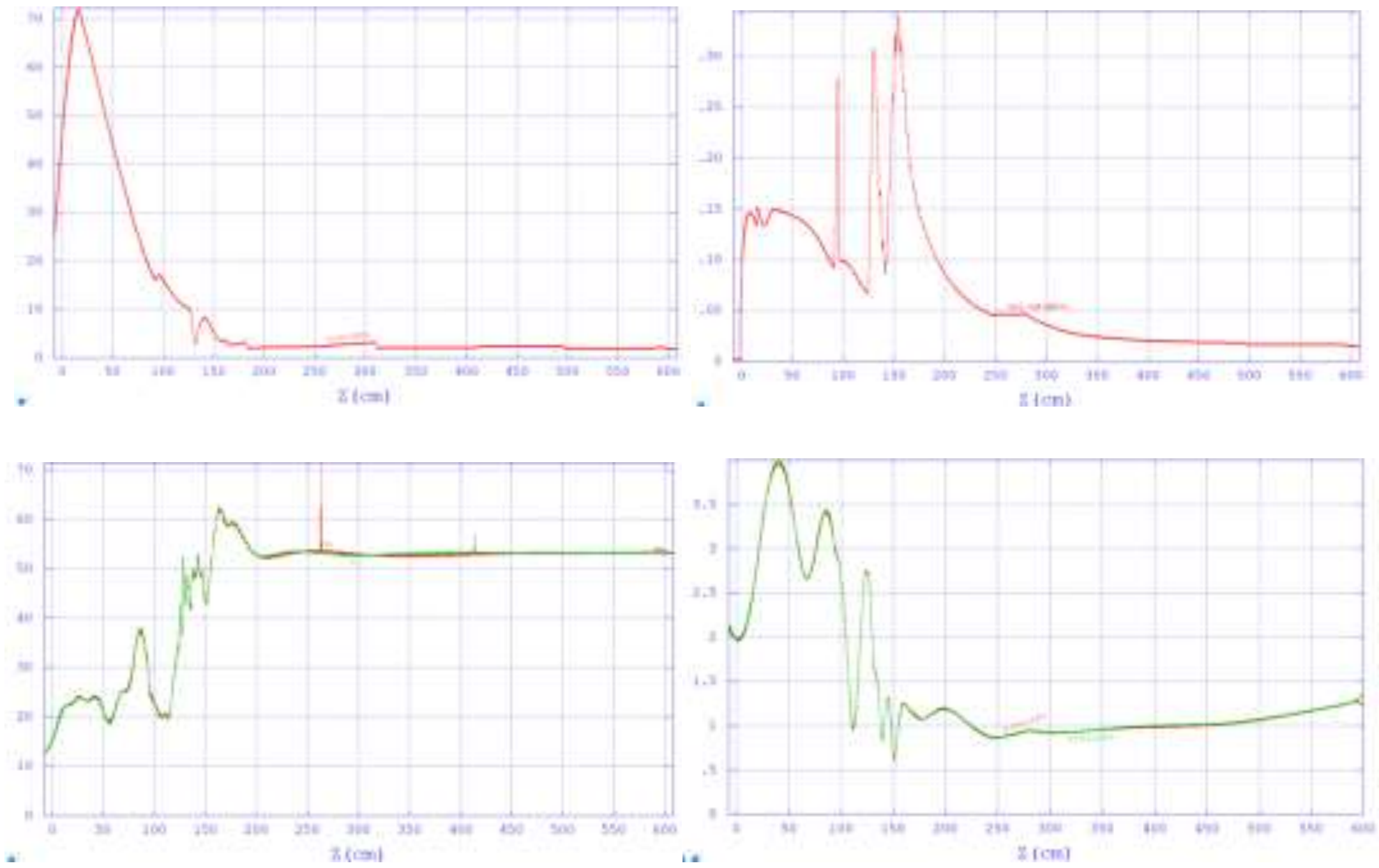

Figure 3.3-9: Evolution of bunch length (top left), energy spread (top right), normalized emittance (bottom left), and transverse beam size (bottom right)

- **Light-speed Linac Design**

The light-speed accelerating section consists of 68 SLAC-type 3-meter-long accelerating structures, increasing the beam energy from approximately 60 MeV to a maximum of 3.5 GeV. To control the transverse beam properties, one triplet (composed of three quadrupole magnets) is placed between every two accelerating sections for beam optics matching.

- **Longitudinal Performance**: The input beam energy to the relativistic linac is about 60 MeV, and it is accelerated to 3.51 GeV. The evolution of the energy spread and bunch length is shown in Figure 3.3-10. At the linac exit, the RMS bunch length is 4.87 ps (containing 95% of particles), and the RMS energy spread is less than 0.5% (95% of particles).
- **Transverse Performance**: After entering the relativistic linac, solenoids are no longer suitable for focusing high-energy beams. Therefore, triplet magnet systems are placed between every two accelerating structures to constrain the transverse beam size and match transverse optics by tuning quadrupole currents. After optimization, the RMS transverse beam size and normalized transverse emittance are shown in Figure 3.3-11.

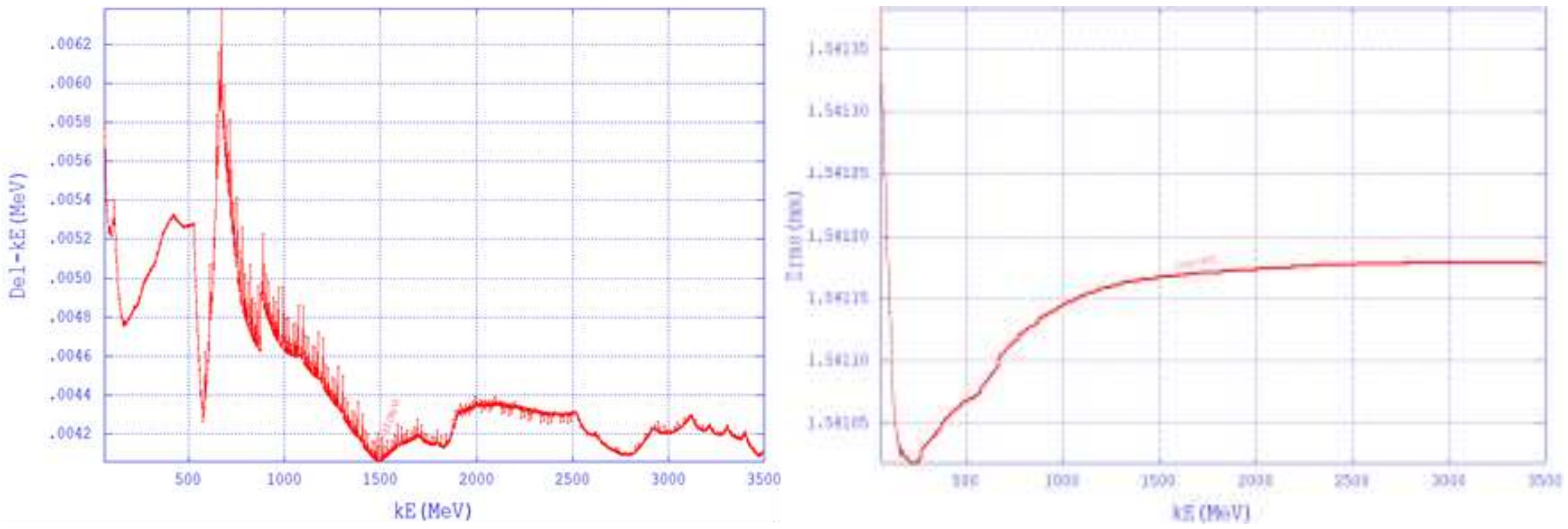

Figure 3.3-10: Evolution of energy spread and bunch length (RMS, 95% particles) in the relativistic linac

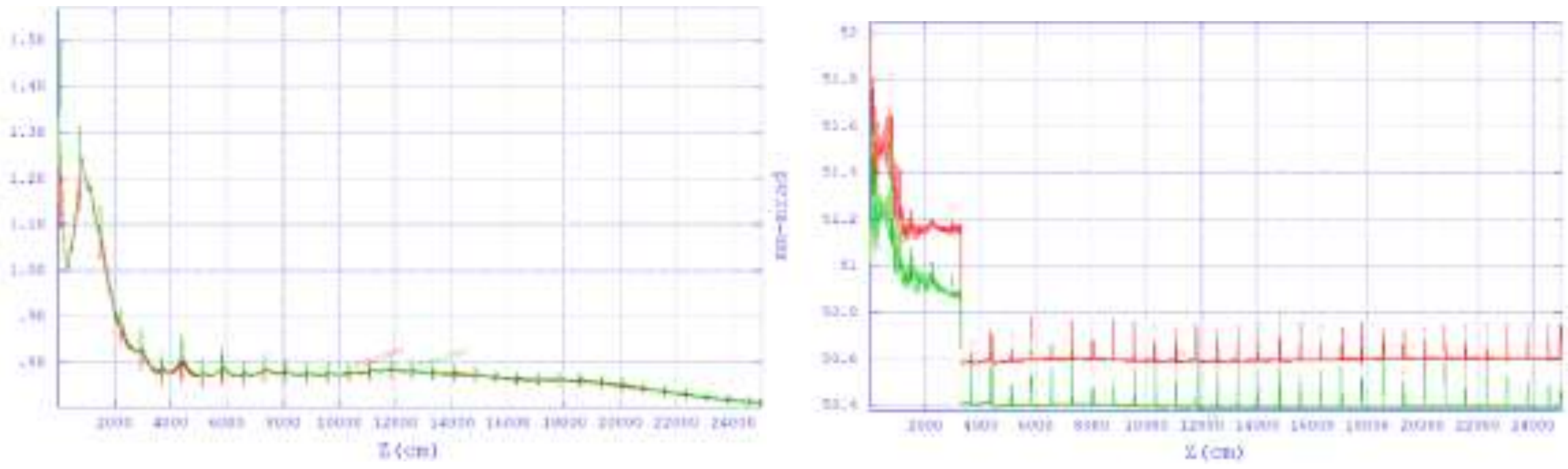

Figure 3.3-11: Evolution of RMS beam size and normalized emittance (95% particles) in the relativistic linac

The overall performance of the SEL1 linac for the swap-out injection is summarized in Table 3.3-1. It successfully accelerates an 8.5 nC high-charge electron bunch to 3.5 GeV while maintaining a transmission efficiency of 99.97% and energy spread below 0.5%, meeting the STCF injector design requirements.

Table 3.3-1: SEL1 Linac Design Results for the STCF Swap-Out Injection

| Parameter | Value | Unit |
|---|---|---|
| Injected electron bunch charge | 8.5 | nC |
| Injection beam energy | 1-3.5 | GeV |
| Transmission efficiency (main bunch) | 99.97 | % |
| Normalized emittance | $\leqslant$ 55 | mm·mrad |
| RMS energy spread (for $\geqslant$95% of particles) | $\leqslant$ 0.5 | % |

| Parameter | Value | Unit |
|---|---|---|
| RMS bunch length (for ⩾95% of particles) | ⩽ 4.9 | ps |

- **Performance at Different Output Energies**

As noted above, the 68-section linac can be operated with adjustable RF power to achieve 1 GeV and 2 GeV output energies. Figures 3.3-12 to 3.3-15 show the evolution of energy spread and normalized emittance for 1 GeV and 2 GeV cases.

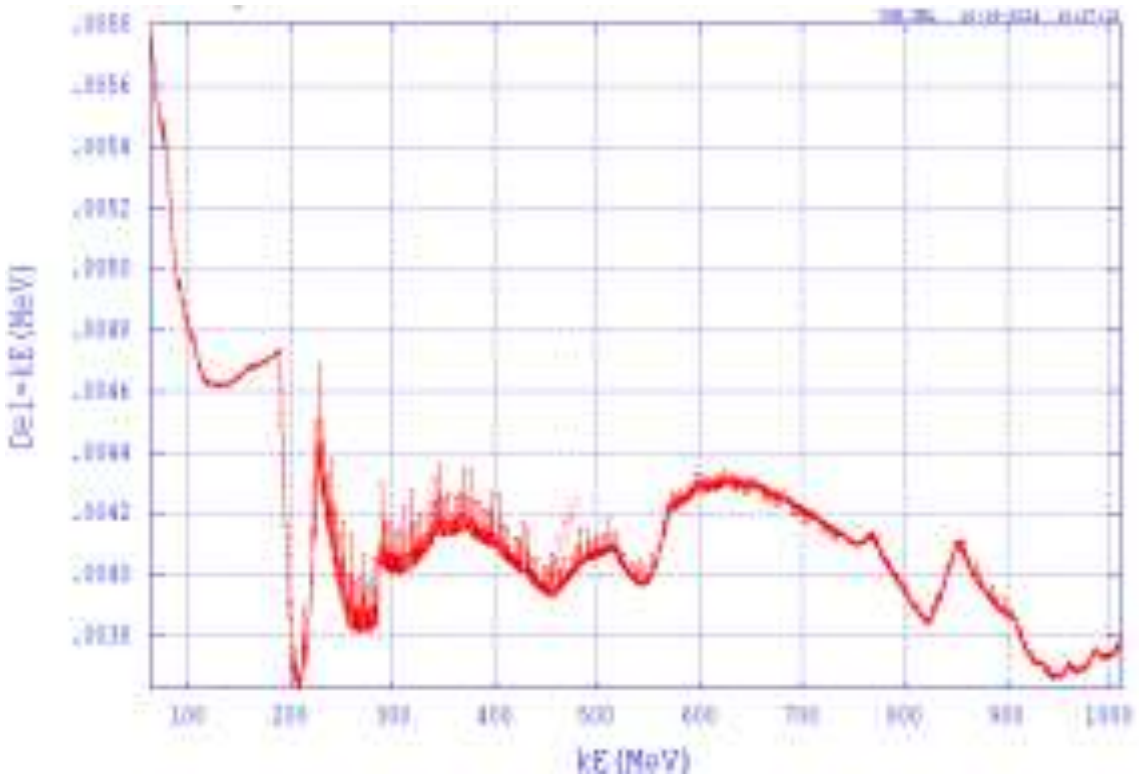

Figure 3.3-12: Energy spread evolution at 1 GeV

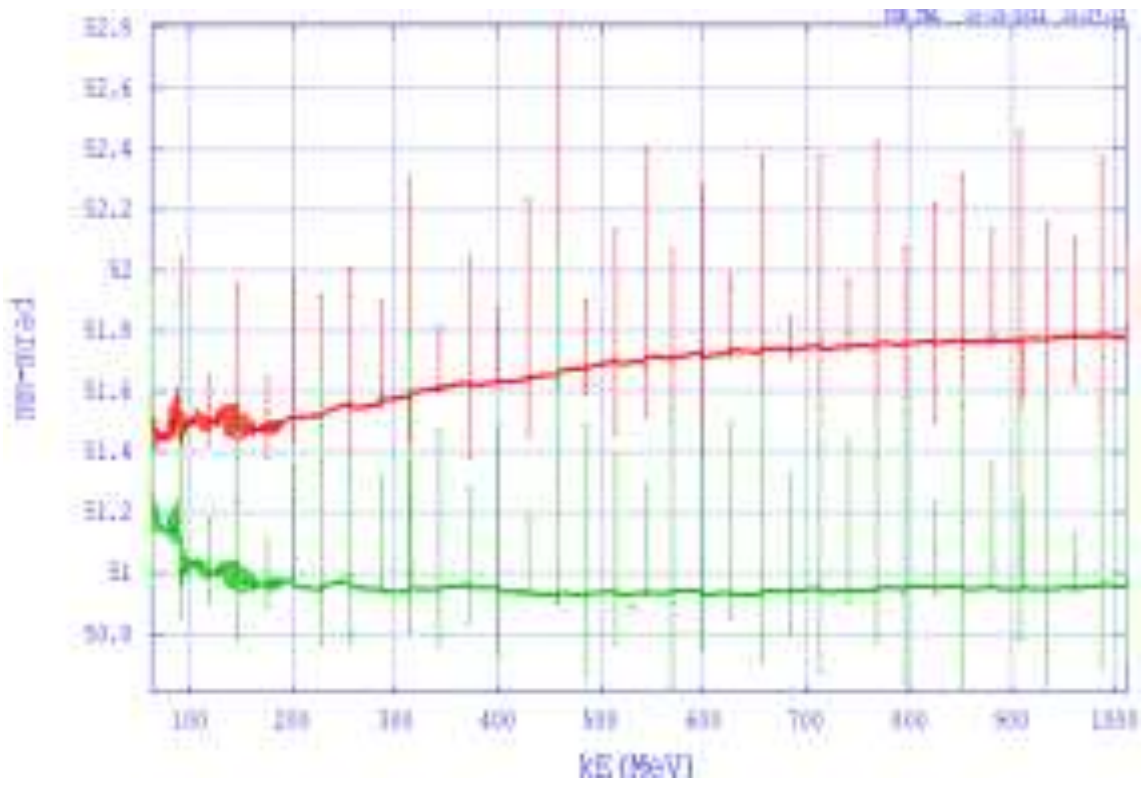

Figure 3.3-13: Normalized emittance evolution at 1 GeV

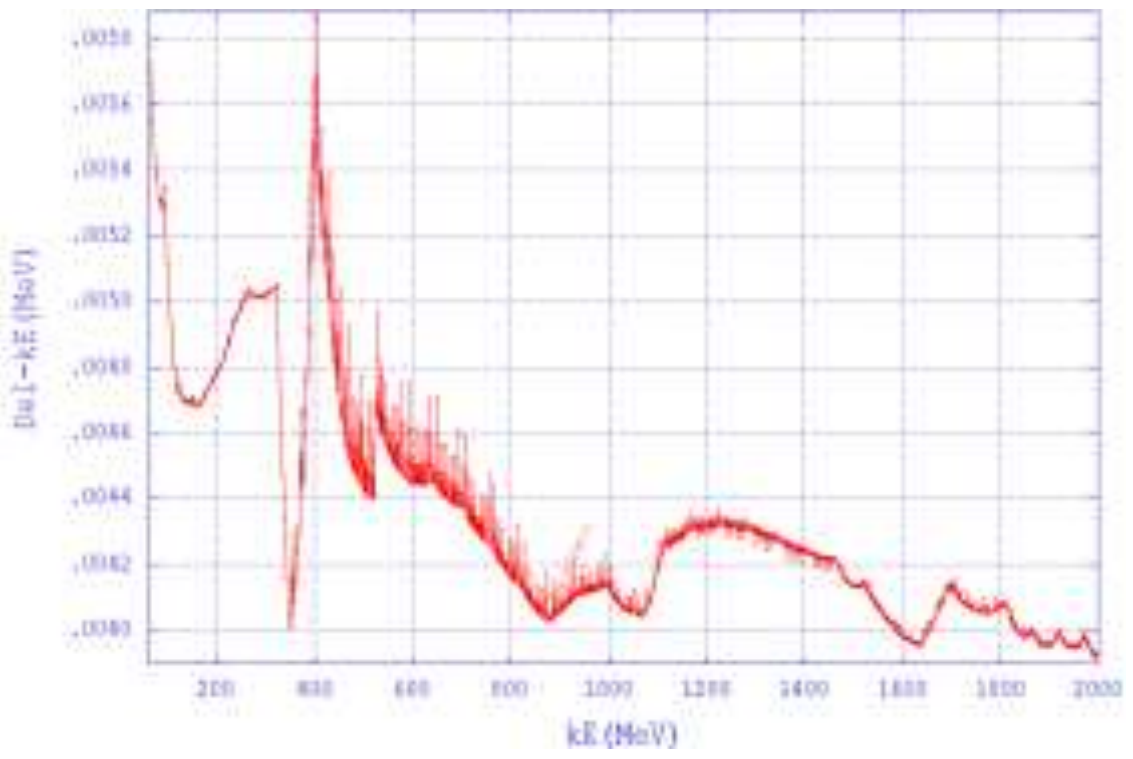

Figure 3.3-14: Energy spread evolution at 2 GeV

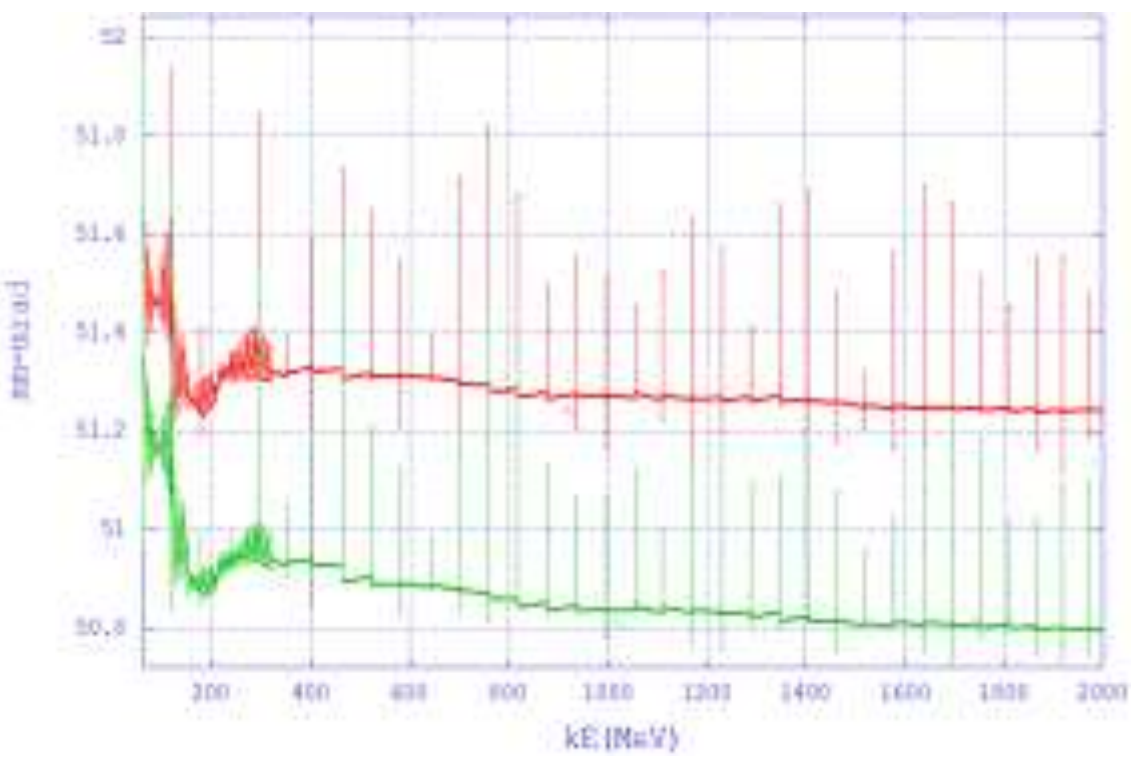

Figure 3.3-15: Normalized emittance evolution at 2 GeV

- **Feasibility Analysis**
  - The input beam distribution in this design is derived from the HEPS electron source. The simulation reflects a realistic upstream electron gun system and produces consistent output parameters.
  - The SuperKEKB injector in Japan uses a similar configuration with two SHBs + pre-buncher + buncher. The SHB frequencies are 1/25 and 1/5 of the main frequency. CEPC in China also uses two bunchers (SHBs + TW) with 1/18 and 1/6 frequencies. The current design is in line with successful international experience.
  - Beam dynamics simulations were conducted using Parmela. The results, shown in the previous section, confirm that the design satisfies the STCF collider's requirements for high-charge direct injection.
- **Conclusion**

For the swap-out injection scheme, a high-charge linac design based on a thermionic electron gun is proposed. Two SHBs and a traveling-wave buncher compress the large-charge electron beam, which is then accelerated through 68 accelerating structures to the injection energy. Beam dynamics simulations demonstrate that the beam reaches 3.5 GeV, with an RMS energy spread below 0.5%, an RMS bunch length below 4.9 ps, and an overall transmission efficiency of 99.97%, fully satisfying STCF injection requirements.

#### *3.3.1.2 Scheme 2 - L-Band Photocathode High-Charge Electron Source*

For the swap-out injection scheme, the STCF collider requires the injector to deliver high-quality electron bunches with a charge of at least 8.5 nC and stringent emittance constraints. Traditionally, generating bunches above this charge level relies on thermionic electron guns, which generally yield high emittance (~100 mm·mrad) and long bunch lengths (>100 ps). These lead to phase overlap in RF acceleration and require beam scraping, making it difficult to meet emittance and energy spread requirements.

In contrast, L-band photocathode electron guns, developed through years of research for FELs (Free Electron Lasers), offer a promising alternative. These guns can deliver both high bunch charge and high beam quality. The L-band photocathode gun technology has been validated at facilities like PITZ (Germany) and Tsinghua University [65, 66], with quantum efficiencies exceeding 5% and normalized emittances ~1 μm/nC. It has been adopted in projects like SHINE [67] and PWFA at BEPCII, making it a mature and reliable option for high-charge electron injectors.

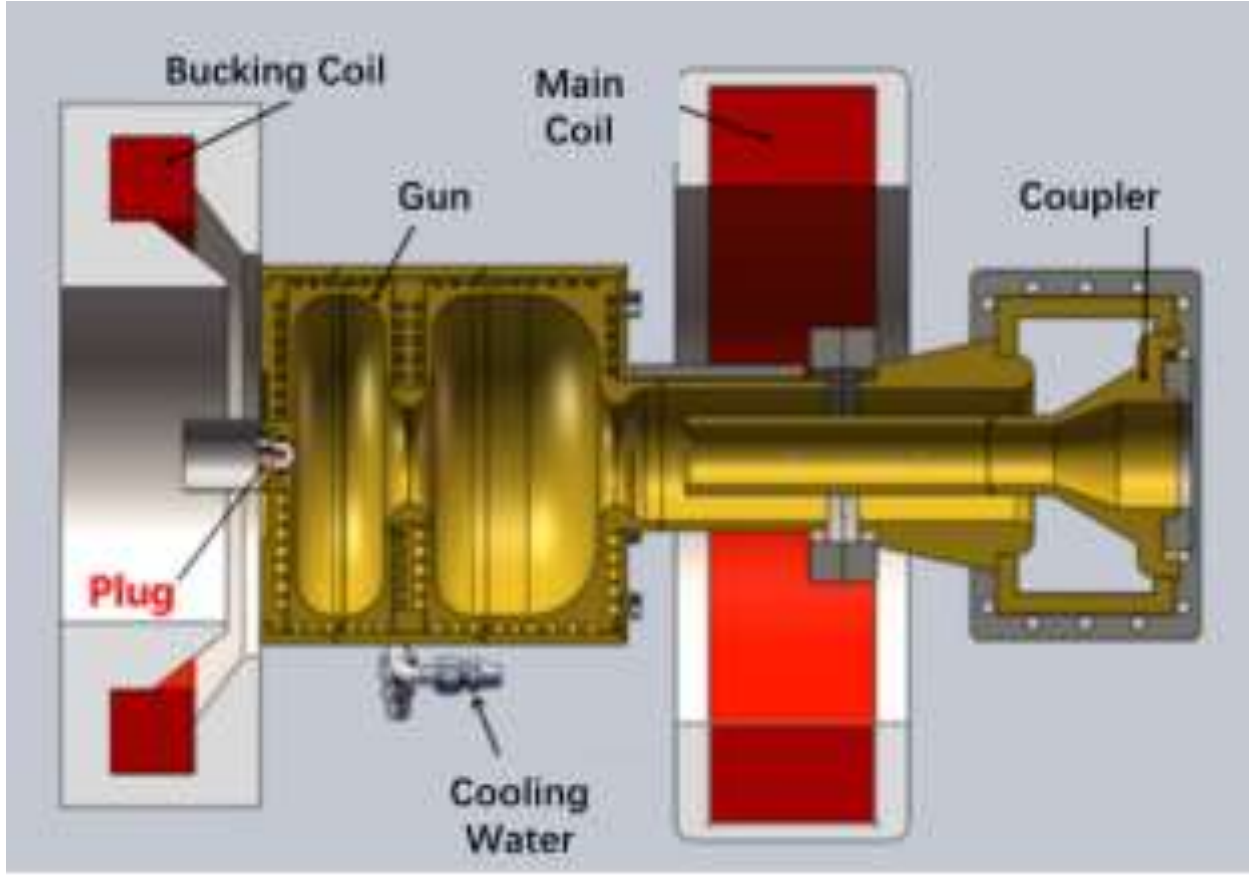

Figure 3.3-16 shows the major components of an L-band photocathode gun.

- **Layout Overview**

This scheme replaces the 8.5 nC thermionic gun in the injector layout (Fig. 3.1-1) with an L-band photocathode gun, while keeping the positron line similar to the off-axis injection scheme but raising the positron production energy (>2.5 GeV).

The L-band photocathode gun has been experimentally shown to operate stably at 5 nC @ 1 kHz. For STCF, a total yield of 240 nC/s is required (8.5 nC × 30 Hz), which this gun can meet. Given the RF power system constraints (≤100 Hz), the gun is configured for 8.5 nC @ 30 Hz operation. The design feasibility is thus being further studied.

- **Electron Bunch Generation and Acceleration**

The low-energy section of the L-band photocathode injector primarily consists of the photocathode RF gun, a solenoid (SOL), two L-band accelerating structures (L1), two S-band accelerating structures (L2), one X-band linearizer section, and a magnetic chicane compressor. It is also equipped with several magnets and beam diagnostics components. Together with the photocathode, drive laser, solid-state and pulsed power sources, vacuum, mechanical, power supply, control, radiation protection, and engineering systems, this forms the complete low-energy section of the injector. The overall layout is shown in Figure 3.3-17.

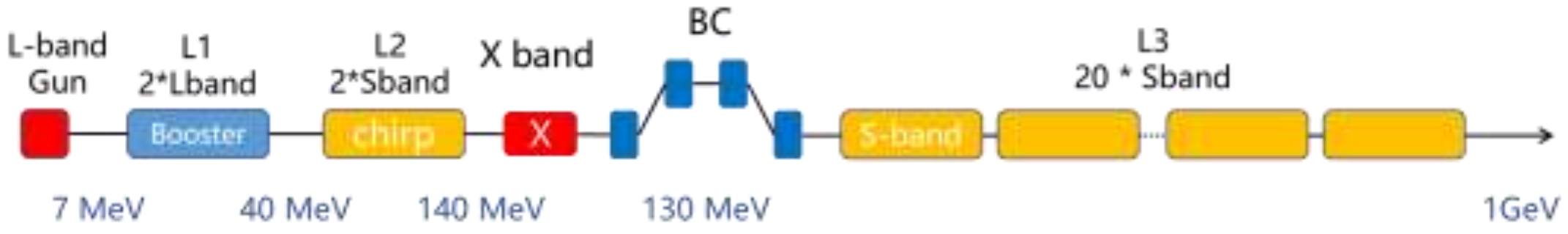

Figure 3.3-17: Schematic layout of the L-band photocathode low-energy section

Since the exit energy of the L-band RF gun is relatively high (>7 MeV), a traditional buncher cannot be used for compression. Instead, magnetic compression is employed to reduce bunch length and control energy spread. Based on international experience with L-band photocathode RF guns, for bunch charges as high as 8.5 nC, the bunch length at the gun exit is relatively long. Therefore, a pair of 1-meter-long L-band accelerating structures is used for pre-acceleration to increase the energy, followed by two 3-meter-long S-band accelerating structures (L2) and an X-band linearizer to introduce an energy chirp, which is then compressed in a magnetic chicane. Finally, the beam is further accelerated beyond 1 GeV by a section labeled L3, consisting of 20 S-band accelerating structures.

The photocathode gun is of the PITZ-type L-band design, with 10 water-cooling channels on the cavity body and a coaxial coupler for RF power feed. The solenoid employs a combination of a bucking coil and a main solenoid, also following the PITZ design. The performance parameters of the gun are listed in Table 3.3-2.

Table 3.3-2: Main parameters of the L-band photocathode RF gun

| Parameter | Value | Unit |
|---|---|---|
| Operating Frequency | 1300 | MHz |
| Operating Mode | Pulsed | |
| Max Repetition Rate | 1 | kHz |
| Duty Cycle | >0.4 | % |
| Cathode Gradient | >50 | MV/m |
| Dark Current | <200 | pC/pulse |
| Static Vacuum | $<1\times10^{-8}$ | Pa |
| Dynamic Vacuum | $<5\times10^{-8}$ | Pa |

As in the off-axis injection design, the optimization of this L-band photocathode low-energy section also adopts a multi-objective, multi-variable genetic algorithm, integrated with the ASTRA beam dynamics simulation software. The 3D electromagnetic field maps of actual components are used in the simulation, and the entire physical layout is modeled realistically.

During optimization, the longitudinal bunch length and transverse emittance of the electron beam at the injector exit are used as the main objectives. The tunable variables include the operational parameters and physical positions of the components mentioned above. The thermal emittance of the photocathode is set to 1 mm·mrad/mm. The transverse distribution of the drive laser is set to a flat-top profile, while the longitudinal profile is Gaussian. The main optimized results of the low-energy section are shown in Figures 3.3-18, 3.3-19, and 3.3-20.

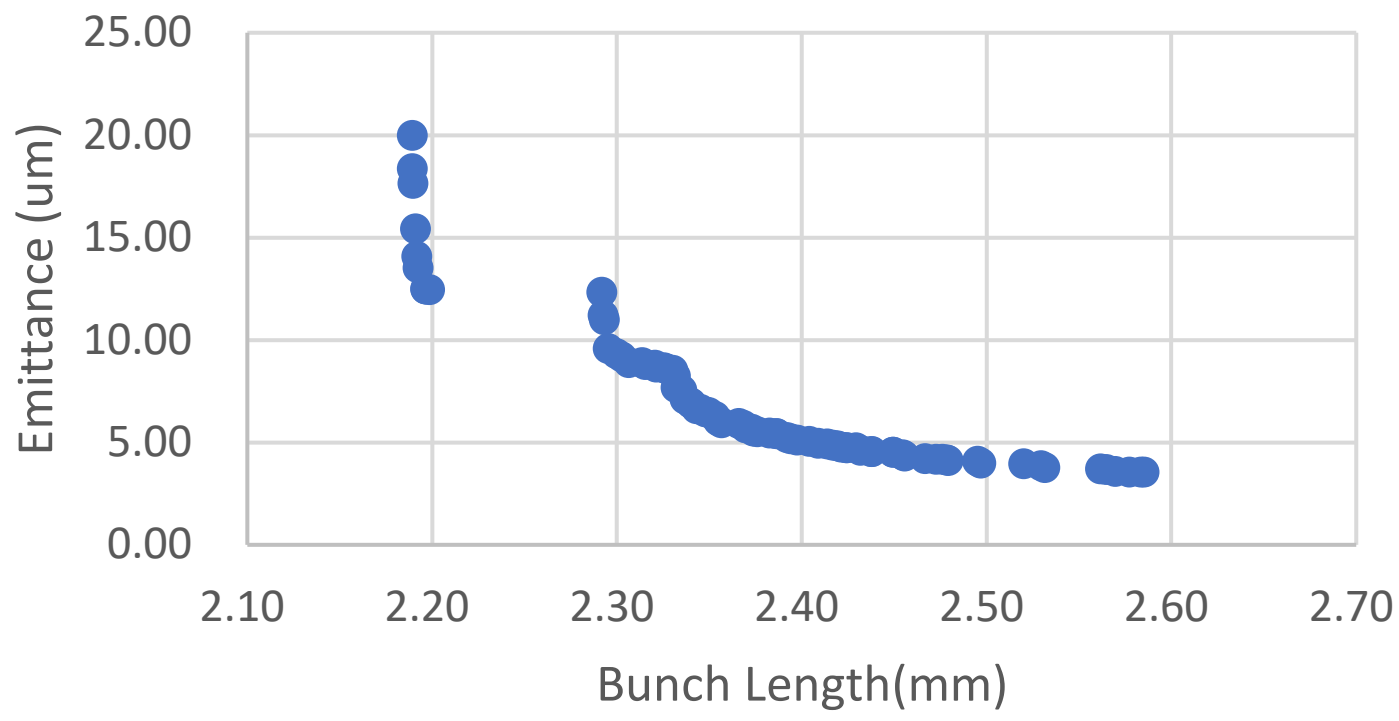

Figure 3.3-18: Multi-objective optimization results of the L-band photocathode low-energy section

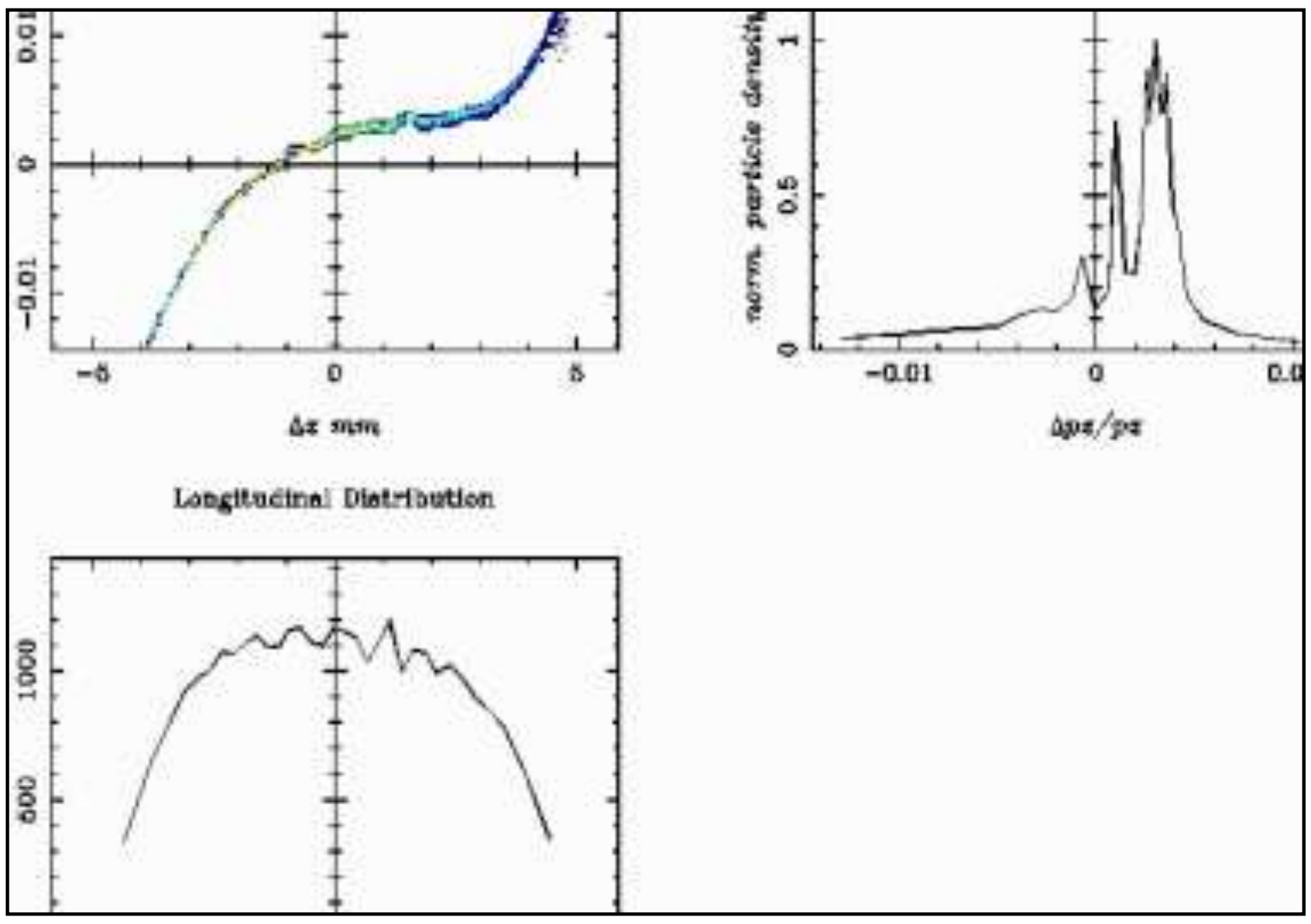

Figure 3.3-19: Phase space and current distribution at the L-band photocathode gun exit

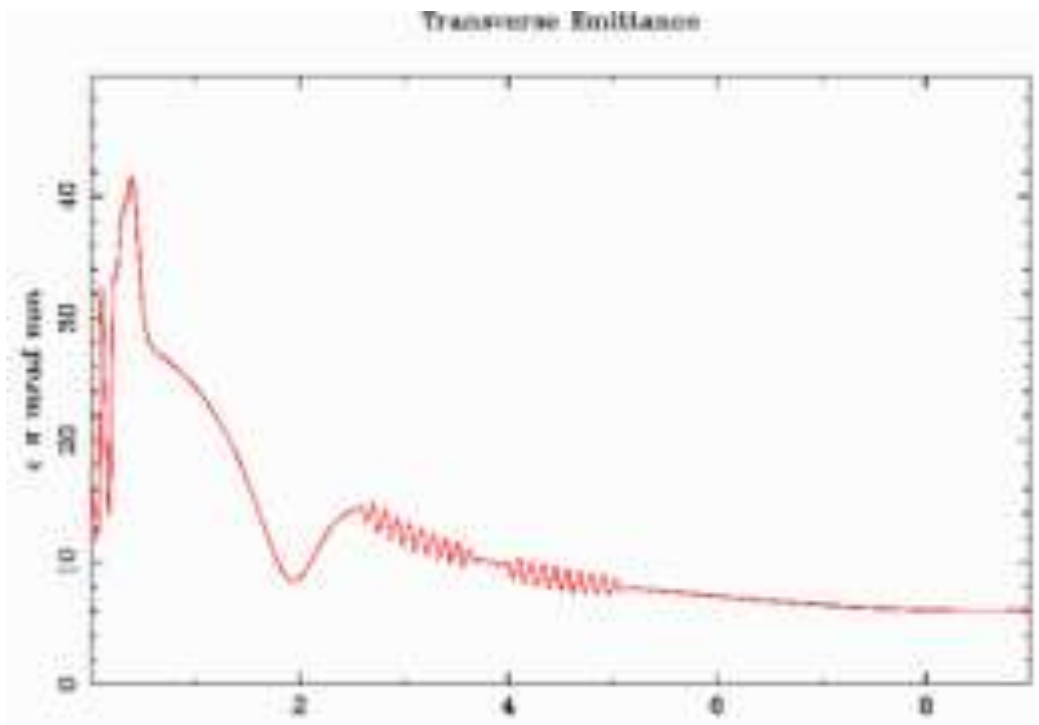

Figure 3.3-20: Emittance evolution along the L-band photocathode pre-injector

From the above results, it can be seen that at the exit of the L1 pre-buncher section, the beam energy is 38.77 MeV, the bunch charge is 8.5 nC, the bunch length is 2.2 mm, the normalized emittance is 4.8 μm, and the peak current reaches 300 A. The longitudinal profile approximates a Gaussian distribution. The low-energy beam then passes through the L2 and X-band linearizer sections, where it is compressed at around 200 MeV. The bunch length is reduced from 2.2 mm to 0.8 mm (corresponding to approximately 2° of the S-band RF phase), as shown in Figure 3.3-21, before being accelerated to over 1 GeV.

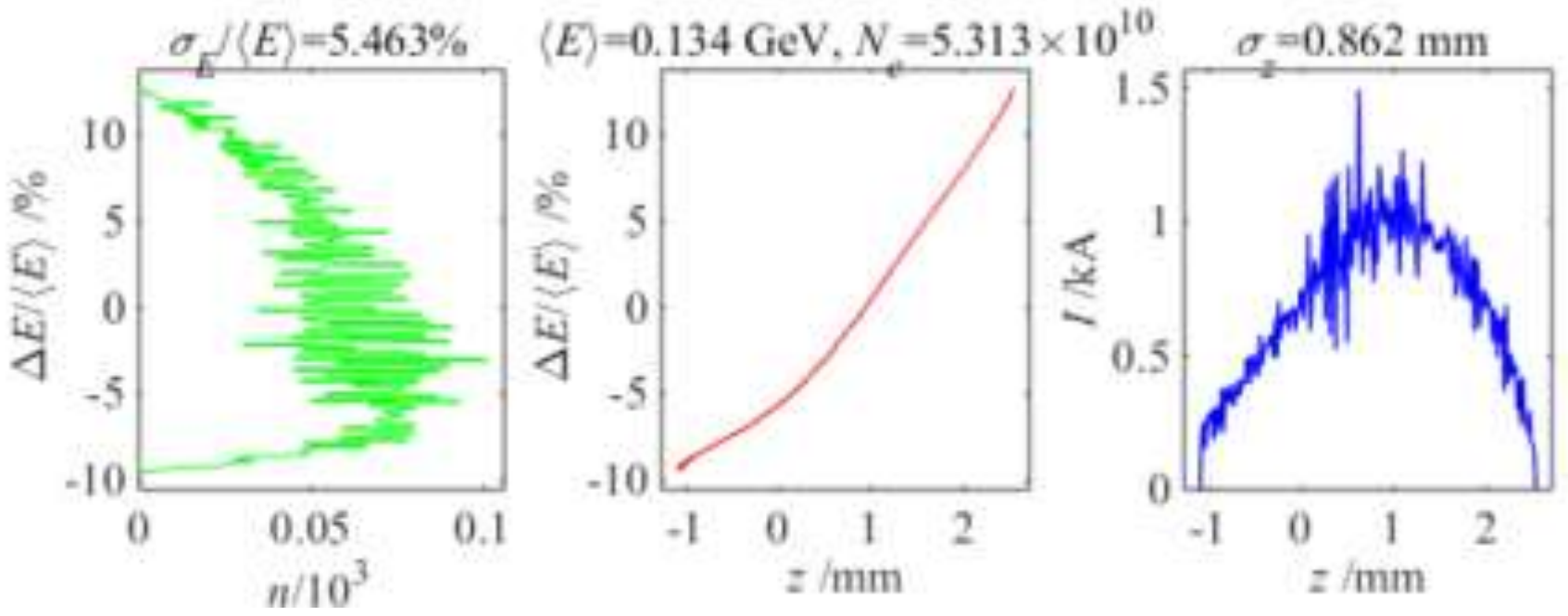

Figure 3.3-21: Phase space and current distribution at the exit of the magnetic compressor

At the exit of the SEL1 section, the energy spread is approximately 0.38%, with the phase space and current distribution shown in Figure 3.3-22.

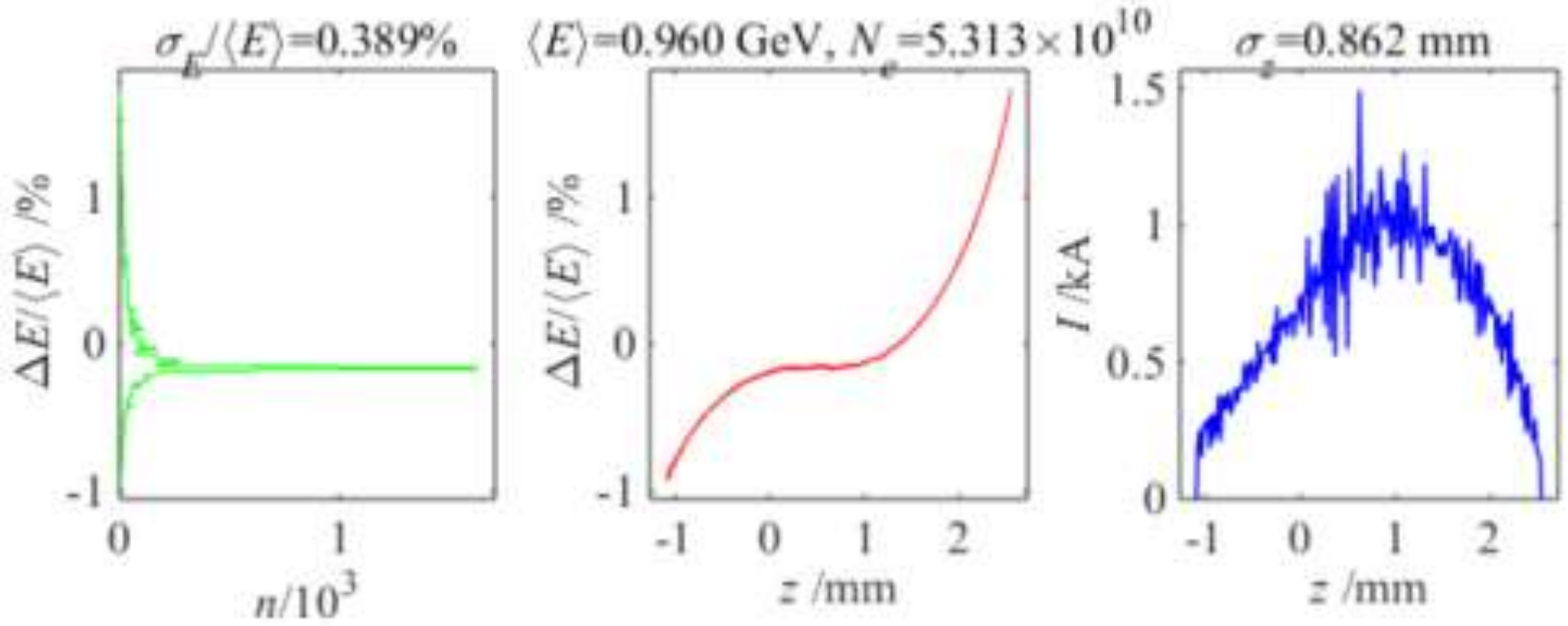

Figure 3.3-22: Phase space and current distribution at the exit of the SEL1 section

From these results, it is evident that the L-band photocathode low-energy section achieves shorter bunch length and lower emittance, making it suitable for direct injection. Moreover, both the cathode and drive laser offer sufficient operational margin, and the L-band photocathode gun shows potential to scale up to ~10 nC, fully meeting the requirements of single-bunch swap-out injection.

### *3.3.1.3 Electron Source and Main Accelerator Section SEL2 for Positron Production*

In the swap-out injection scheme, the thermionic electron source used for positron target bombardment is similar to that in the off-axis injection case, but with a higher bunch charge requirement (11.6 nC). The low-energy section design is also identical, consisting of a bunching section (BS) and a pre-injector section (PIS). The main accelerator section SEL2 boosts the beam energy to 2.5 GeV with an increased repetition rate of 90 Hz. A total of 46 accelerating structures are needed in SEL2, resulting in an overall linac length of approximately 180 meters, as shown in Figure 3.3-23.

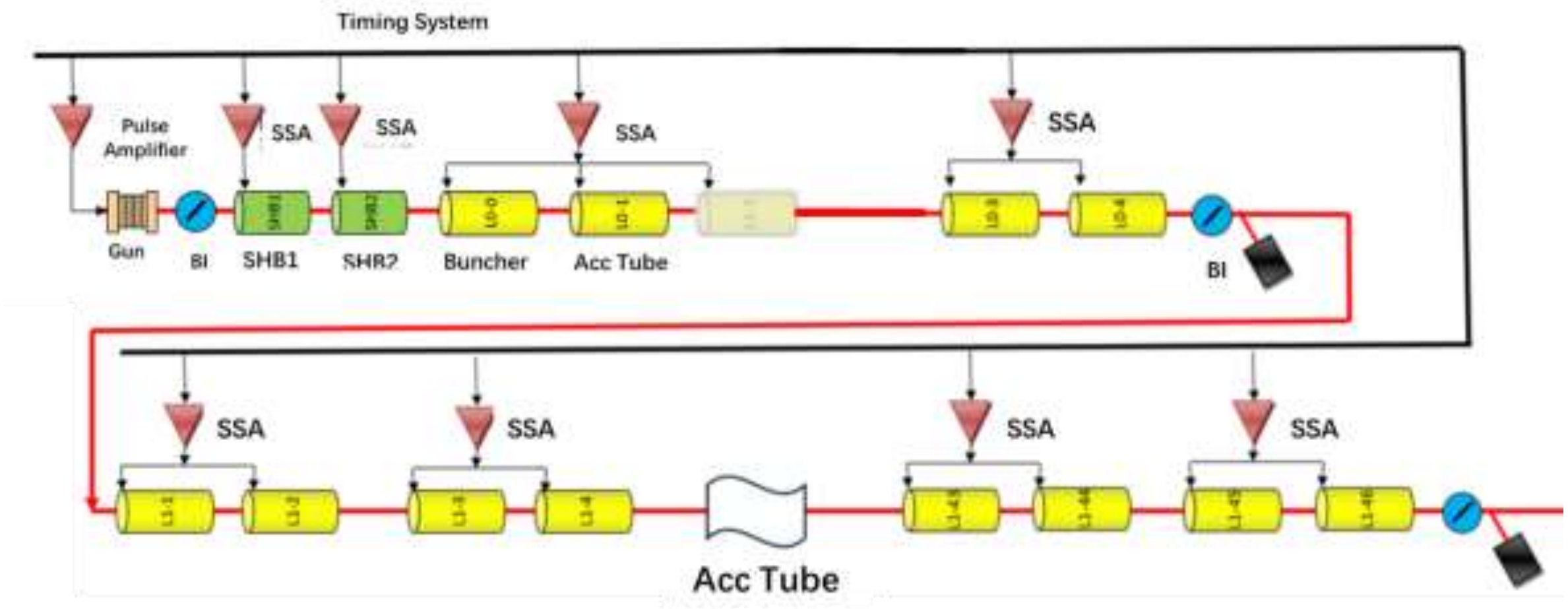

Figure 3.3-23: Layout of the linac for positron target bombardment in the swap-out injection scheme

Tracking of the 11.6 nC bunch through SEL2 yields the beam's transverse and longitudinal envelopes, normalized emittance evolution along the beamline, and the final phase space distribution at the linac exit, as shown in Figure 3.3-24.

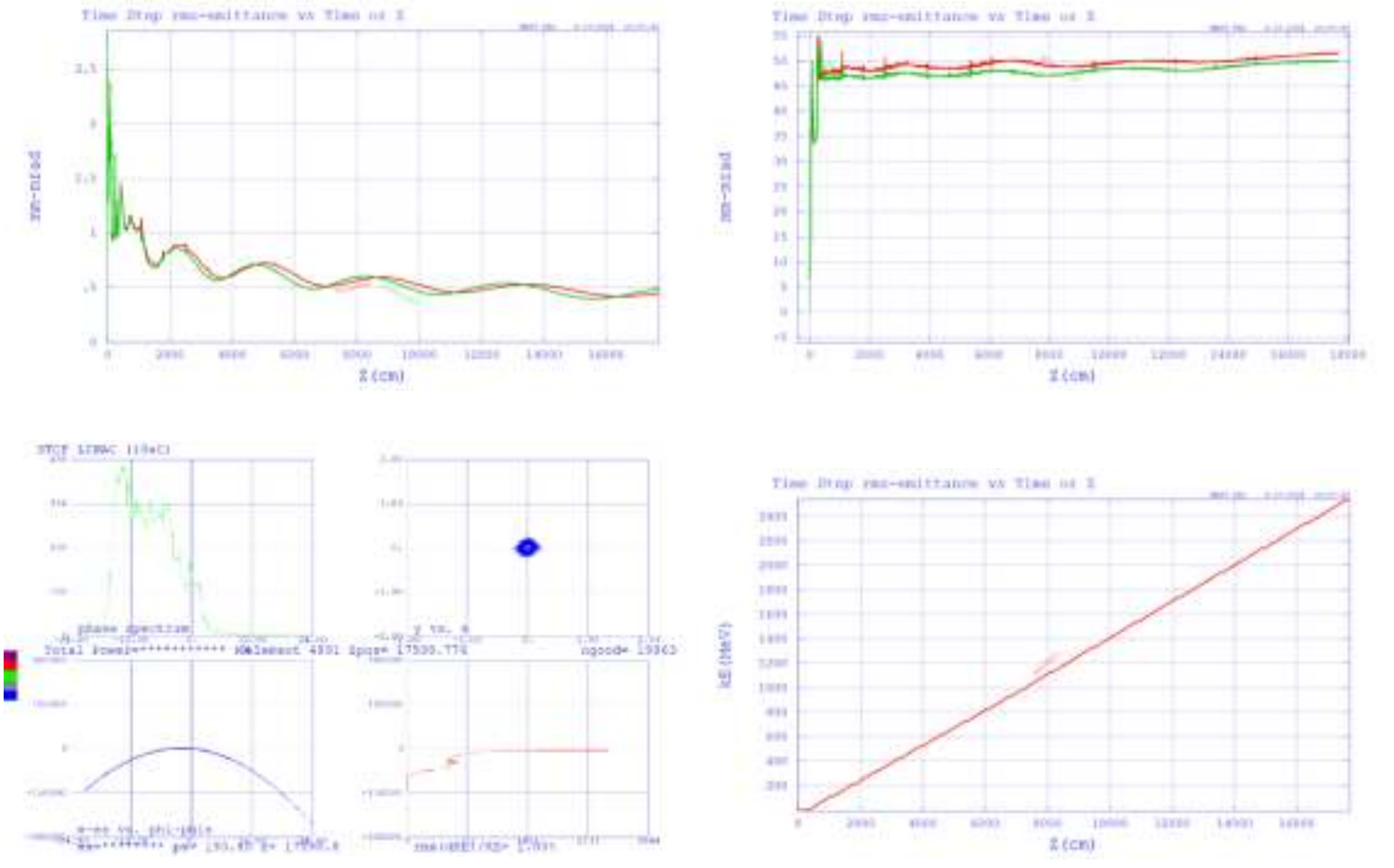

Figure 3.3-24: Beam parameters along the positron-target linac for the swap-out injection scheme

From the figure, the RMS transverse beam size at the exit is approximately 0.5 mm, with a normalized emittance of about 50 μm. The final energy is 2542 MeV, with a transmission efficiency of 99.8%, a bunch charge of 10.9 nC, and an RMS energy spread of 1.76%, all of which meet the requirements for positron production using a tungsten target.

As in the off-axis injection scheme, a Triplet lens is placed every two 50-MeV accelerating structures before 1 GeV, and every four 50-MeV accelerating structures after 1 GeV.

### 3.3.2 Positron Production, Capture, and Pre-Acceleration

The positron source is a key component of the STCF injector, responsible for generating positron bunches of sufficient quantity and quality. STCF will adopt the conventional method of positron generation via electron beam bombardment of a target, which remains the only viable approach for producing high-intensity positron beams. Both off-axis injection and swap-out injection schemes are considered. The off-axis scheme uses a relatively low bombardment energy of 1.5 GeV, requiring careful optimization of the subsequent capture and matching accelerator systems to ensure sufficient positron yield from low-energy electron beams. The swap-out injection scheme imposes a much higher requirement on the positron bunch charge, which demands greater positron yield and better heat dissipation performance of the target. The positron source parameters for the swap-out injection scheme are shown in Table 3.2-2.

#### *3.3.2.1 Positron Conversion Target*

For the swap-out injection scheme, the positron source must provide positron bunches with 2.5 nC at 90 Hz. Figure 3.3-25 shows the simulated positron yield for a 2.5 GeV electron beam hitting targets of different thicknesses and materials. Gold and tungsten produce the highest yields. Considering cost factors, tungsten is chosen as the target material, with an optimal thickness of 15 mm and a positron yield of 6.9 per incident electron.

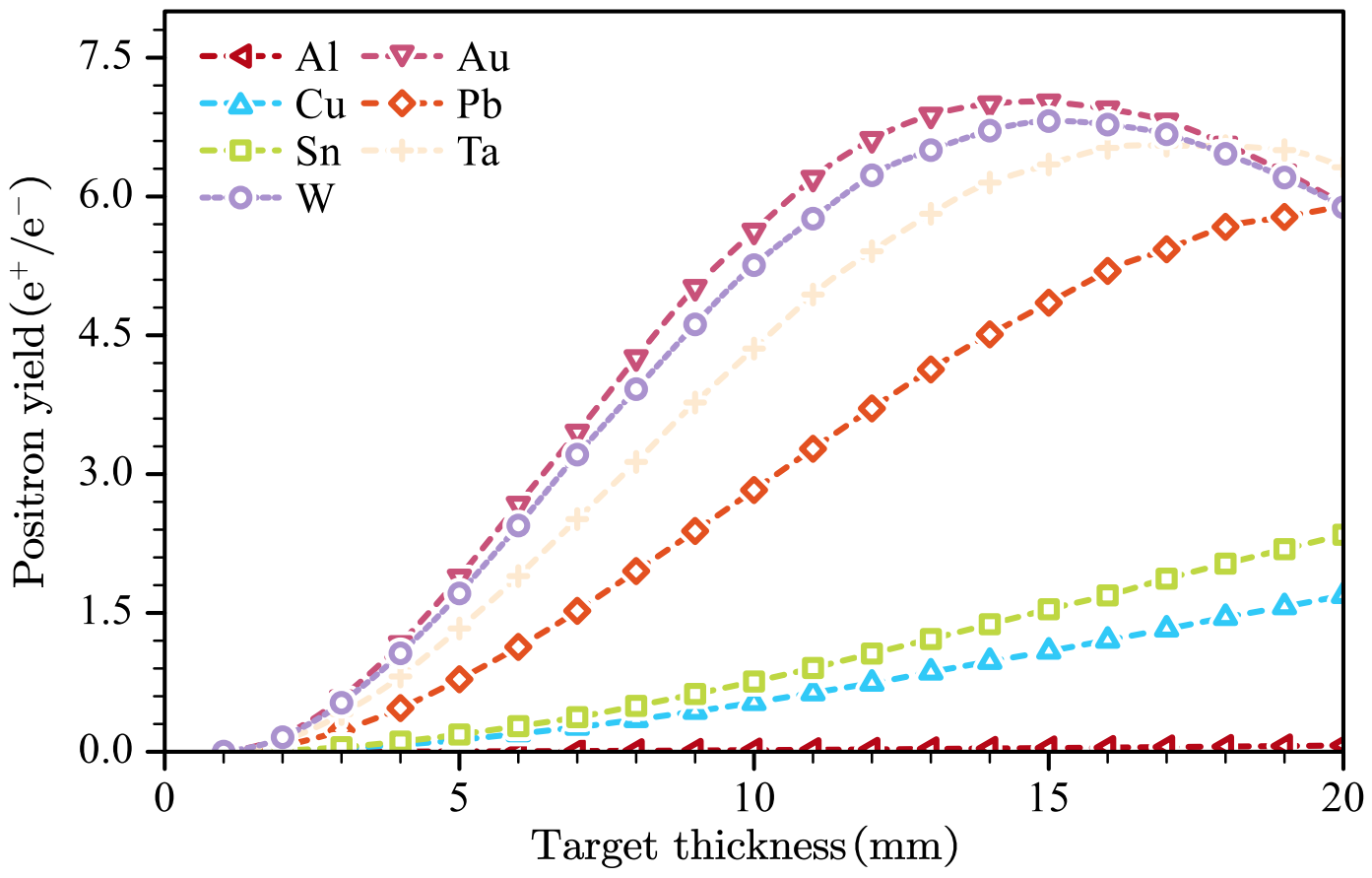

Figure 3.3-25: Simulated positron yields for a 2.5 GeV electron beam on targets of different materials and thicknesses

To optimize the downstream capture and acceleration of positrons, the energy and angular distribution of the positrons exiting the target must be known. Figure 3.3-26 compares the energy spectra and angular distributions of positrons produced by a 2.5 GeV electron beam with 5% energy spread and by an ideal monoenergetic beam. The impact of the 5% energy spread on the positron distribution is found to be acceptable.

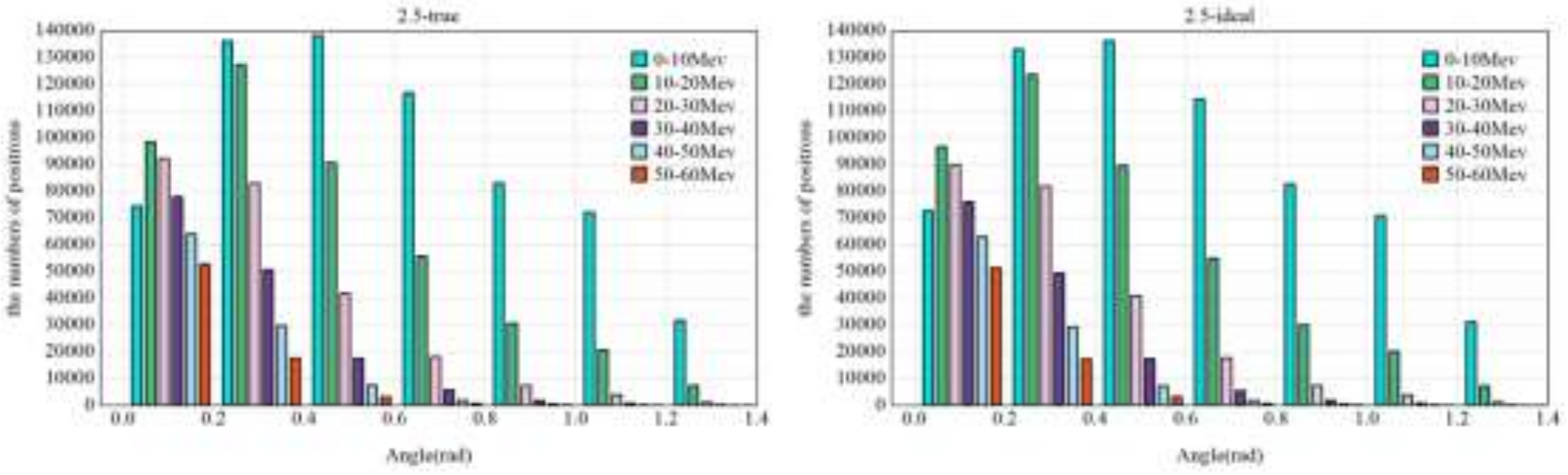

Figure 3.3-26: Energy and angular distribution of positrons produced by a 2.5 GeV electron beam with and without energy spread

### *3.3.2.2 Design for the Positron Linac (PL)*

This part, including the Adiabatic Matching Device (AMD), positron pre-accelerator, and main accelerator section, is nearly identical to the system used in the off-axis injection scheme. Therefore, it will not be repeated here.

## 3.3.3 Positron Accumulation Ring

### *3.3.3.1 Design Requirements and Specifications*

In the swap-out injection scheme of STCF, the injector is required to supply full-charge electron and positron bunches of 8.5 nC to replace those in the collider rings whose particle number has decayed below the level necessary for collisions. Due to the relatively short beam lifetime, the bunches must be prepared at a high repetition rate.

Assuming the injector delivers bunches to the collider rings at a repetition rate of $f_r = 30$ Hz, the accumulation ring must achieve a damping time constant τ ≈ 20~25 ms. It must store at least $n_b \geq 4$ bunches, with each bunch being accumulated $n_a \geq 3$ times. This leads to a required injector repetition rate of $n_a \cdot f_r \geq 90$ Hz. Each accumulated bunch must have a minimum damping time of $(n_b - 1)/f_r \geq 100$ ms $\geq 4\tau$ before extraction to ensure a final emittance below 30 nm·rad. To support multi-turn accumulation, the ring must have a sufficiently large dynamic aperture. The RF frequency of the accumulation ring is identical to that of the collider rings to allow phase synchronization. The $n_b$ bunches are evenly distributed within the ring, with a bunch spacing of $n_s = H_{AR}/n_b$ RF buckets, where $H_{AR}$ is the harmonic number of the accumulation ring and determines the ring circumference.

To ensure each bunch experiences sufficient damping before extraction, we adopt a rotational injection mode. During the initialization phase, each of the ($n_b$ buckets is filled via $n_a$ injections). Once all $n_b$ bunches are filled, the first bunch has experienced a damping time of $(n_b - 1)/f_r$, satisfying the extraction requirement. It is then extracted, and a new injection is immediately performed. This cycle of sequential injection and extraction continues to support the swap-out injection to the collider positron ring.

The bunch accumulation process uses multiple off-axis injections. The interval between injections is $1/n_a/f_r \approx 11$ ms, meaning previously injected particles are not fully damped before the next injection. This is similar to multi-turn injection, requiring a large phase space acceptance and dynamic aperture. Additionally, the closed orbit bump must be reduced after each injection to improve accumulation efficiency. This accumulation scheme is the core aspect of the accumulation ring design and requires careful optimization.

### *3.3.3.2 Lattice Design of the Accumulation Ring*

The accumulation ring adopts a structure similar to the damping ring, consisting of periodic bending sections, dispersion suppression sections, and straight sections. The full ring comprises 56 periodic bending cells, 4 dispersion compression sections, and 2 straight sections. The overall ring structure is shown in Figure 3.3-27, where blue indicates dipole magnets, red indicates quadrupole magnets, and green indicates sextupole magnets.

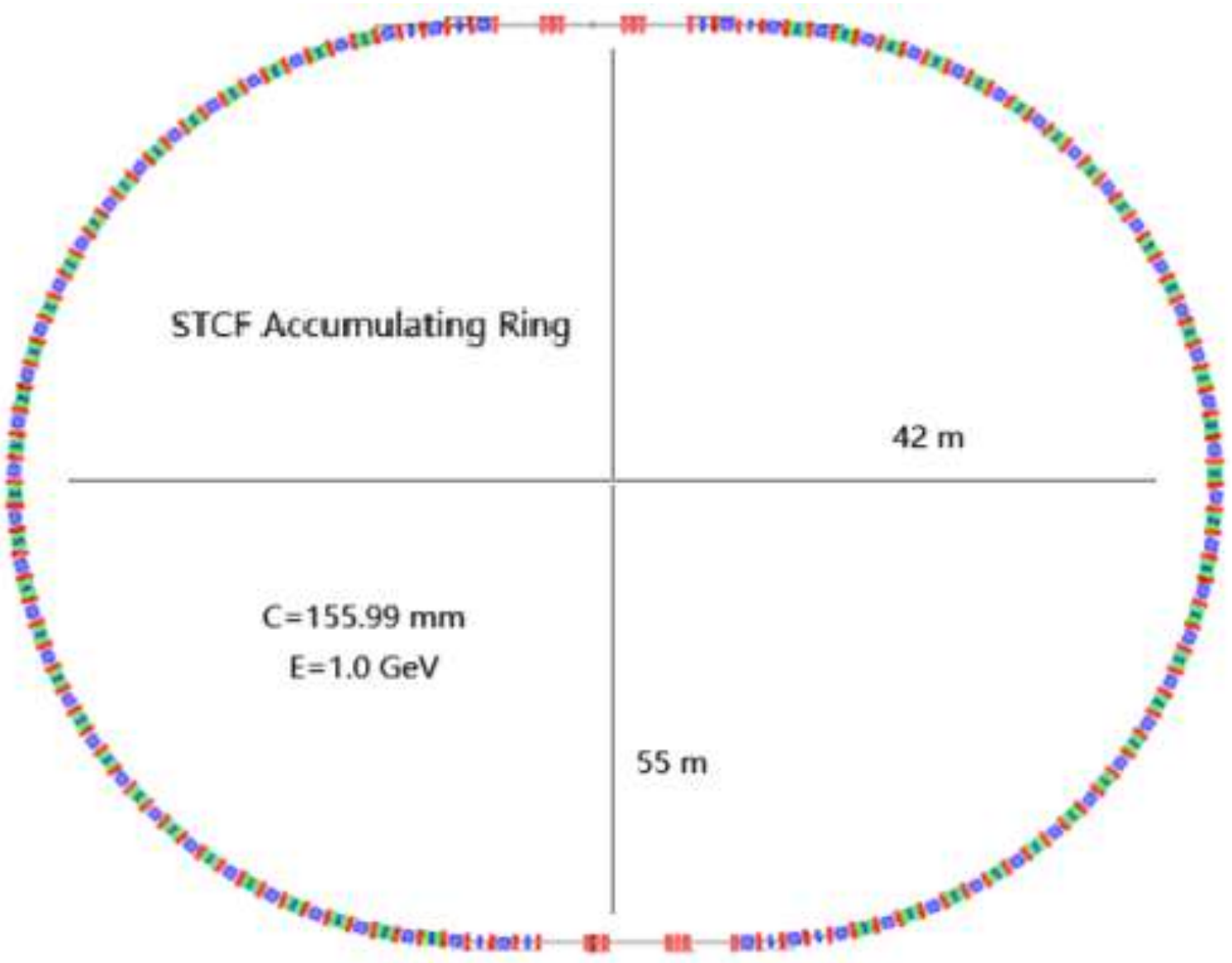

Figure 3.3-27: Layout of the accumulation ring

The periodic bending cell is illustrated in Figure 3.3-28. It adopts a “one normal + one reverse” dipole layout within a FODO structure [12][68], where the reverse dipole length is 0.3 times the normal dipole length and both have the same magnetic field strength. Each bending cell is 2.16 meters long and bends positrons by 6°. The normal and reverse dipoles bend by 8.57° and 2.57°, respectively. Each cell includes one focusing quadrupole, one defocusing quadrupole, one focusing sextupole, and one defocusing sextupole.

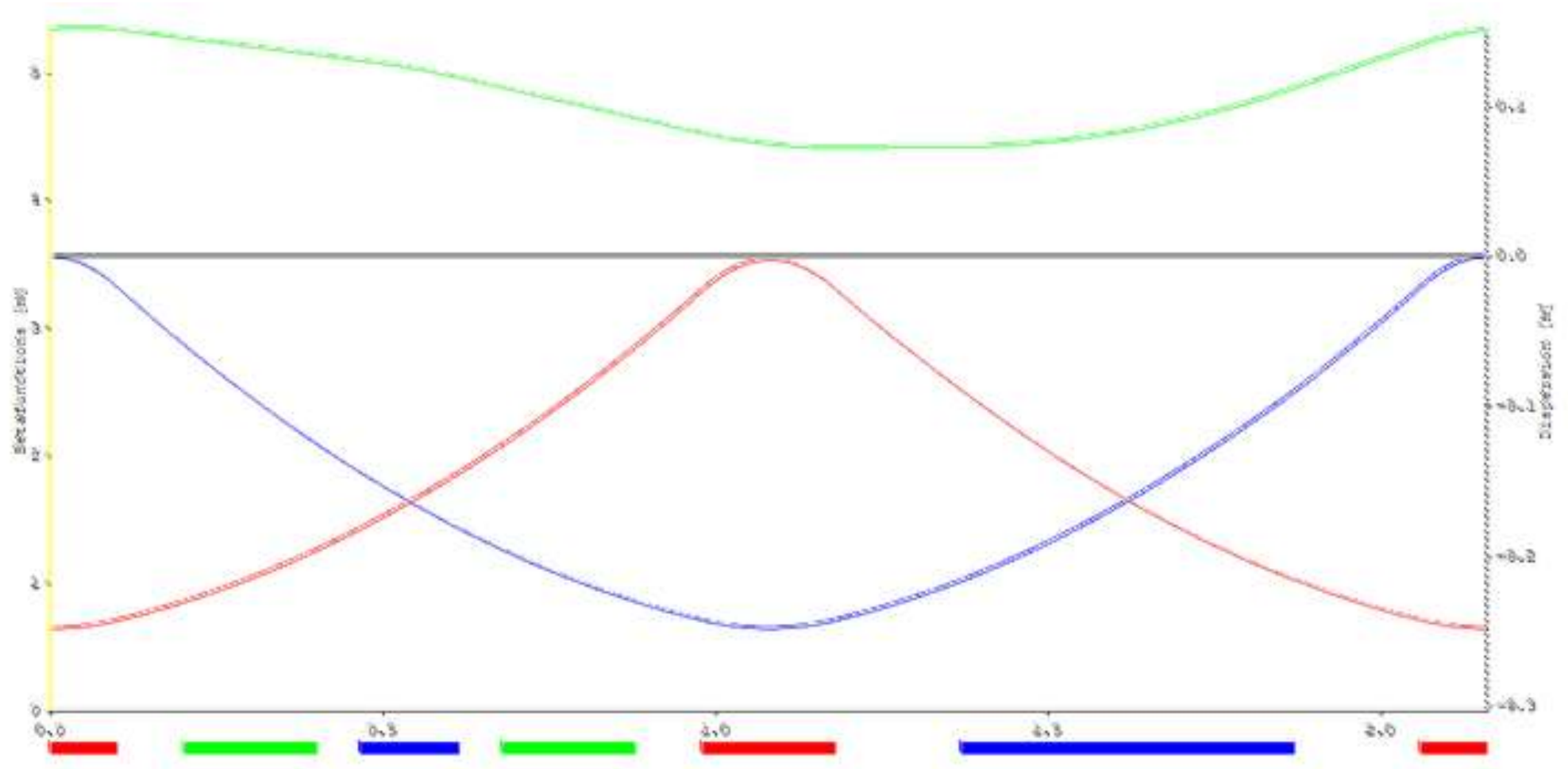

Figure 3.3-28: Periodic bending structure in the arc of the accumulation ring

The dispersion suppression section is shown in Figure 3.3-29. It follows a "half-bend" scheme, including two normal and two reverse dipole magnets, each with the same dimensions as those in the periodic bending cells but with half the magnetic field strength. This results in a total bend angle of 6°, matching the deflection in each bending cell.

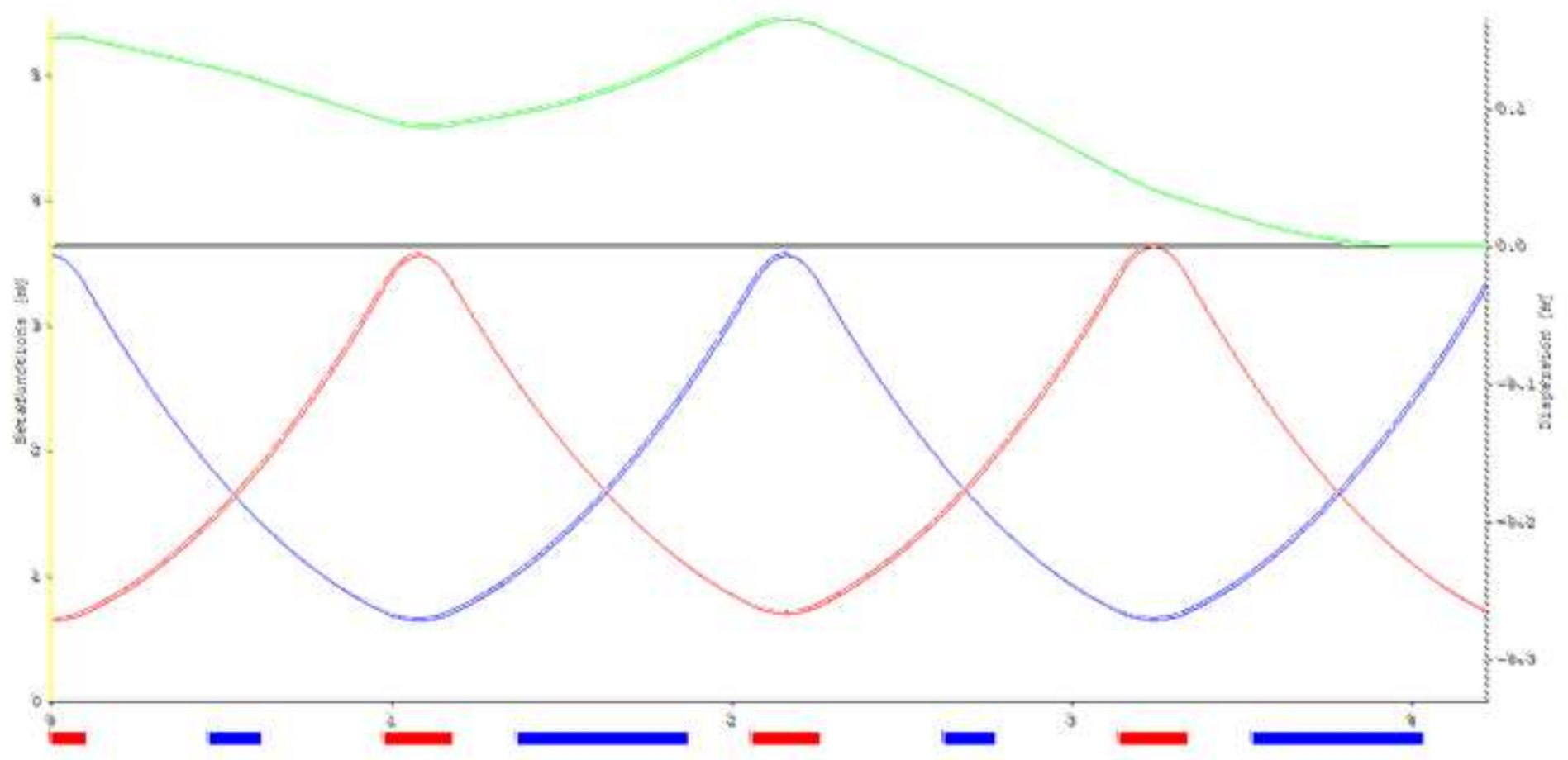

Figure 3.3-29: Dispersion suppression structure in the accumulation ring arc

The structure of the straight section is shown in Figure 3.3-30. Its primary function is to provide space for positron injection and extraction components, as well as RF cavities. The straight section consists of eight quadrupole magnets. After beam optics matching, the Twiss parameters at the injection/extraction point are set to $\beta_x$ = 9.5 m and $\beta_y$ = 1 m. Downstream of this point, after three quadrupole magnets, space is reserved for the injection/extraction kicker magnet. The beam optics matching is completed across the full ring, and the full-ring Twiss parameters are shown in Figure 3.3-31.

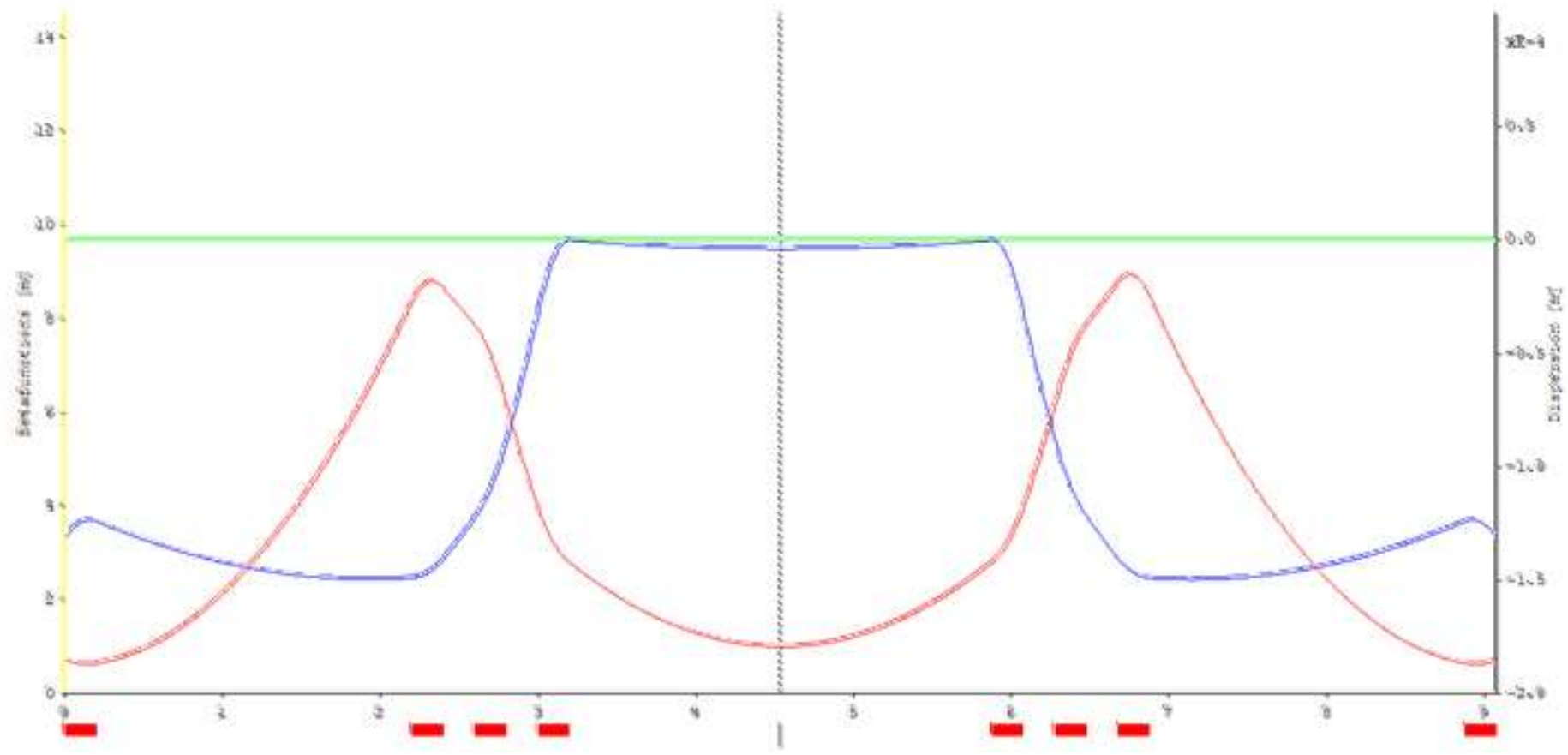

Figure 3.3-30: Structure of the accumulation ring straight section

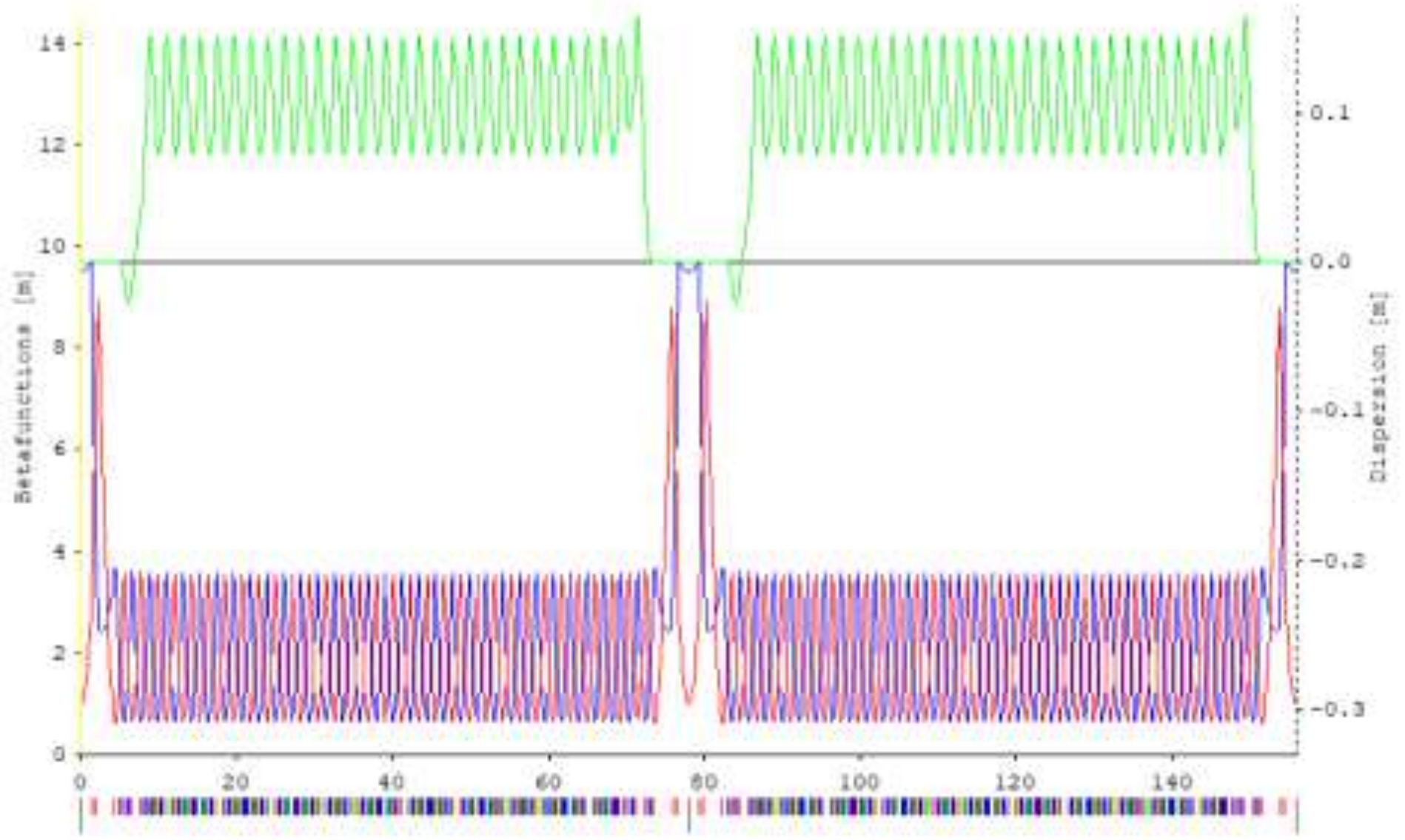

Figure 3.3-31: Twiss parameters of the full positron accumulation ring

In the arc region, each periodic focusing cell provides a horizontal and vertical phase advance of 0.25·2π. After the linear optics design is completed, sextupole magnets are added in each periodic cell to correct the first-order chromaticity. The final full-ring horizontal and vertical phase advances are 16.714·2π and 17.424·2π, respectively. Figure 3.3-32 shows the working point of the accumulation ring. It is well-separated from fourth-order resonance lines, although it is relatively close to fifth-order resonances.

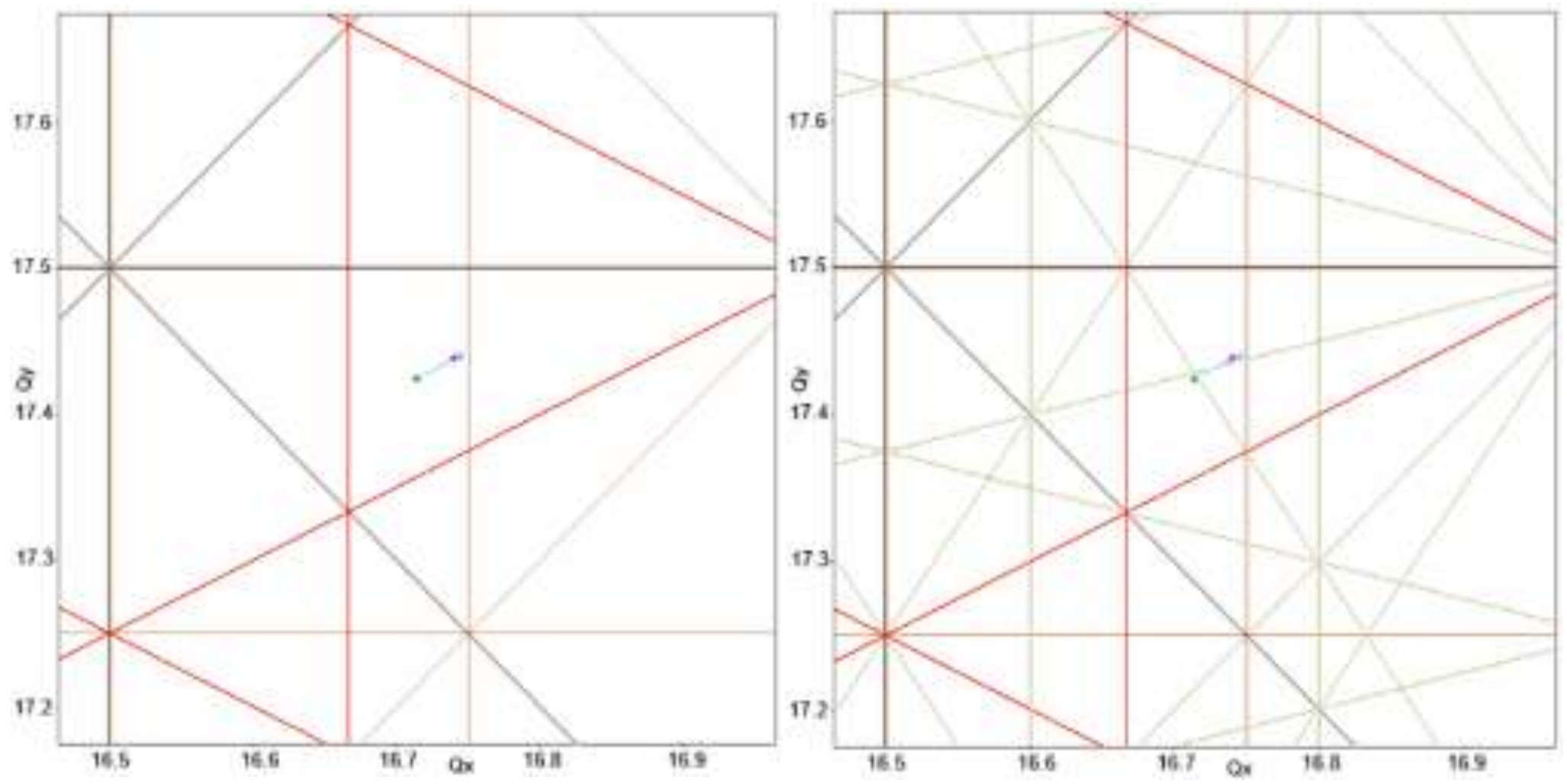

Figure 3.3-32: Working point of the accumulation ring; the left shows proximity to 4th-order resonances, and the right to 5th-order

To ensure injection efficiency and reduce beam loss, the dynamic aperture must be analyzed. The dynamic aperture is influenced by nonlinear elements such as sextupoles. After correcting chromaticity with sextupoles, the impact of their nonlinearities must be considered [69]. Figure 3.3-33 shows the dynamic aperture. During injection, the positron beam envelope is approximately 3.6 mm, while the dynamic aperture in the horizontal direction is about 15 mm, roughly four times the beam envelope, indicating a reasonably large acceptance.

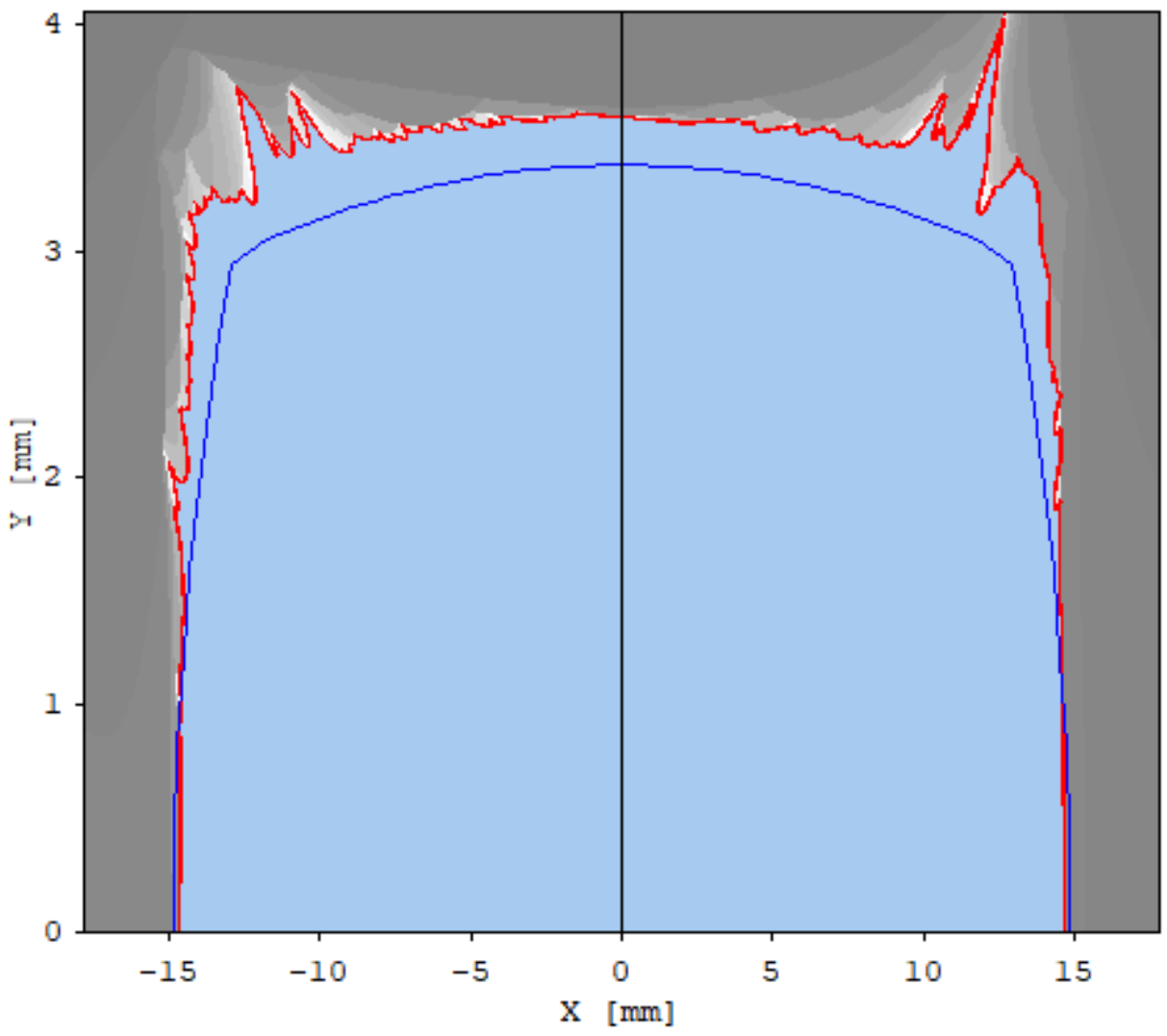

Figure 3.3-33: Dynamic aperture of the positron accumulation ring

Beam lifetime is also an important factor in the design. By adjusting RF cavity voltage and frequency, harmonic number, and gas species, the beam lifetime is estimated to be approximately 140 seconds. Figure 3.3-34 shows the momentum acceptance, which meets the requirements for positron injection.

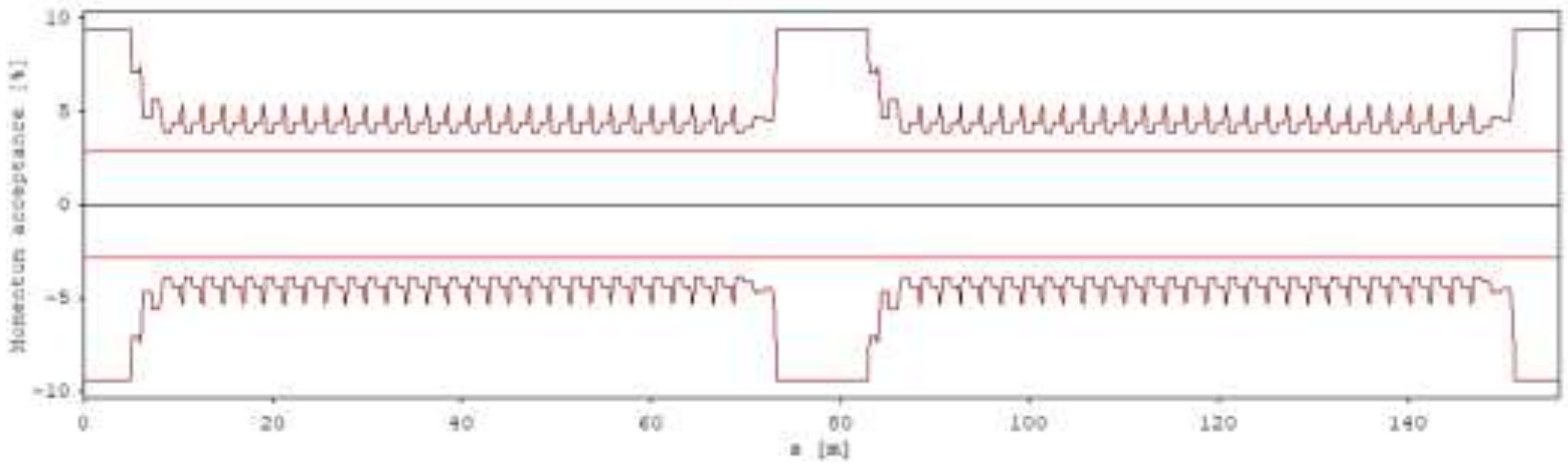

Figure 3.3-34: Momentum acceptance of the accumulation ring

The final injection into each stored positron bunch occurs approximately 100 ms before extraction, allowing damping to occur. The extracted emittance $\varepsilon_{\text{ext}}(\text{t})$ is calculated using the following formula:

$$\varepsilon_{\text{ext}}(\text{t}) = \varepsilon_{\text{nat}} + \left(\varepsilon_{\text{inj}} - \varepsilon_{\text{nat}}\right) \times \text{e}^{-\frac{2t}{\tau}}$$

Here, $\varepsilon_{\text{nat}}$ is the equilibrium emittance, $\varepsilon_{\text{inj}}$ is the injected emittance, and $\tau$ is the damping time constant. The ratio between reverse and normal dipole magnet lengths affects the relationship between damping time and equilibrium emittance. Figure 3.3-35 shows how this ratio affects horizontal equilibrium emittance and damping time. When the ratio exceeds 0.2, the emittance increases, but the damping time decreases. Thus, tuning this ratio can help minimize extracted emittance for a fixed damping time of 100 ms and initial emittance of 1400 nm·rad.

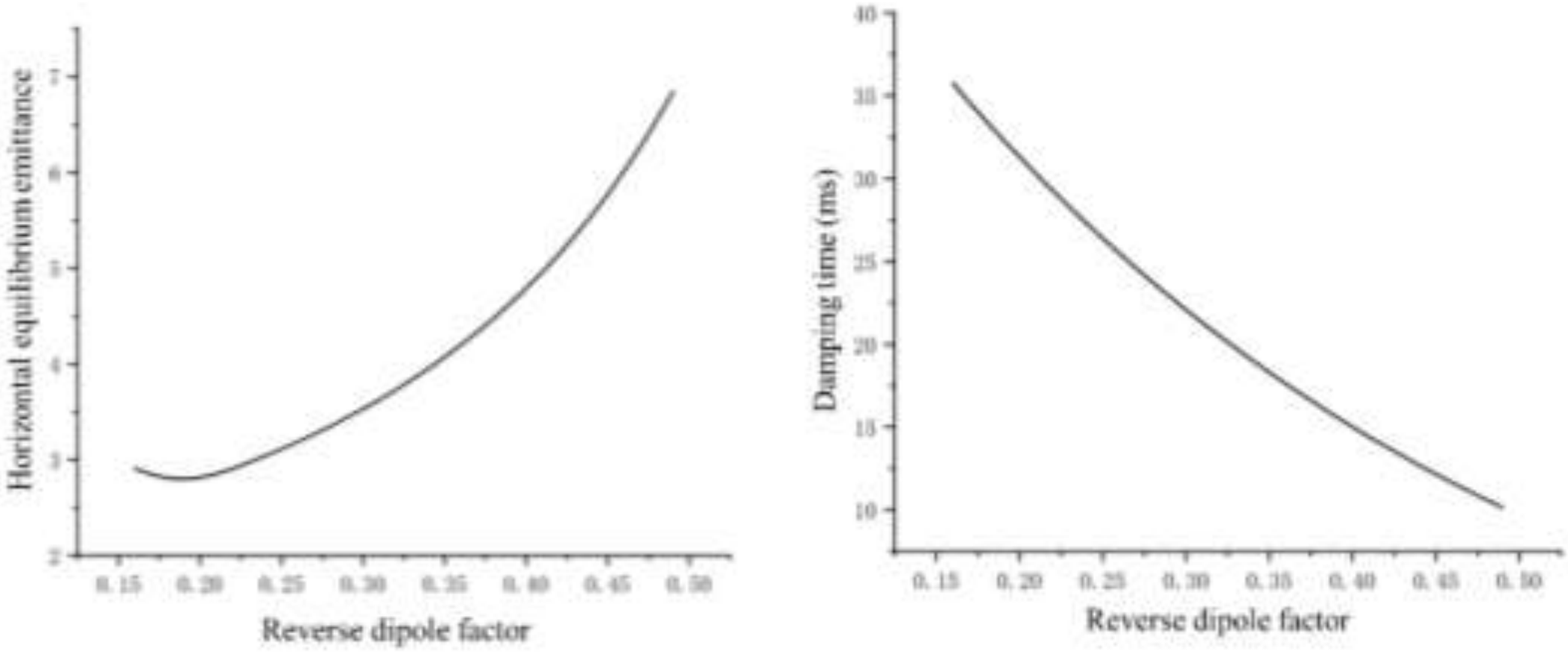

Figure 3.3-35: Relation between reverse dipole magnet ratio, horizontal equilibrium emittance, and damping time

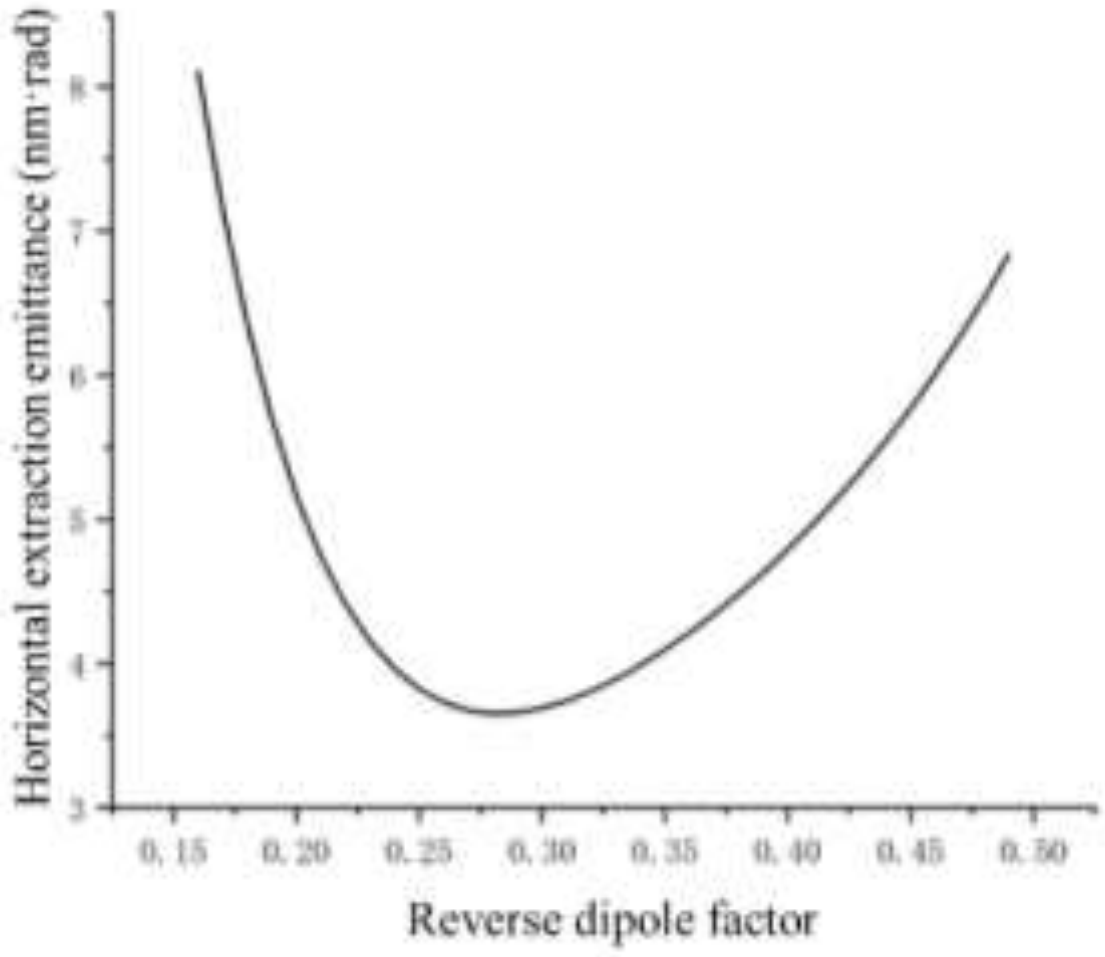

Figure 3.3-36: Relation between reverse dipole magnet ratio and extracted emittance

From Figure 3.3-36, a reverse dipole magnet ratio between 0.25–0.3 yields the lowest extracted emittance. Therefore, a ratio of 0.3 is adopted as the feasible design choice. The key parameters of the accumulation ring are listed in Table 3.3-3.

Table 3.3-3: Main Parameters of the Accumulation Ring

| Parameter | Value | Unit |
|---|---|---|
| Energy | 1.0 | GeV |
| Circumference | 155.99 | m |
| Number of bunches | 4 | |
| Max bunch charge | 8.5 | nC |
| Max current | 65.4 | mA |
| FODO phase advance (x/y) | 90 / 90 | degrees |
| Dipole bending radius | 3.337 | m |
| Dipole field strength | 0.998 | T |
| Reverse dipole magnet ratio r | 0.3 | |
| Energy loss per turn | 47.5 | keV |
| Damping time (x/y/z) | 21.9 / 21.9 / 11.0 | ms |
| Storage time | 100 | ms |
| Natural emittance (x/y) | 3.52 / 0.1 | nm·rad |
| Injected emittance (x/y) | 350 / 350 | nm·rad |
| Extracted emittance (x/y) | 3.56 / 0.23 | nm·rad |
| Natural relative energy spread | 0.0465 | % |
| Momentum compaction factor | 0.00273 | |
| RF frequency | 499.7 | MHz |
| RF voltage | 0.8 | MV |
| Harmonic number | 260 | |
| Natural chromaticity (x/y) | -21.4 / -22.3 | |
| Working point (x/y) | 16.7 / 17.4 | |

#### *3.3.3.3 Beam Accumulation and Extraction Scheme*

The accumulation ring uniformly stores four bunches, each of which reaches the target charge after three accumulation cycles. For a given bunch, the charge accumulation is performed via off-axis injection, as illustrated in Figure 3.3-37. The positron bunch is deflected by a septum magnet toward the injection point. At the injection point, a pulsed kicker magnet creates a local orbit bump, allowing the injected bunch to enter the acceptance of the accumulation ring.

Subsequently, the bump height is reduced, and the positron bunch undergoes synchrotron radiation damping to reduce its emittance.

During the first injection, the orbit bump is at its maximum height, and after injection, the bunch resides on the central orbit. For the second injection, since a circulating bunch already exists, the orbit bump must be reduced to prevent beam loss due to interaction with the septum. The newly injected bunch centroid will be offset from the central orbit, resulting in a significantly increased effective emittance. In the third injection, due to the short damping time, the circulating bunch still has a large effective emittance, and the bump height must be further reduced. After injection, the beam continues to damp in the accumulation ring to reach the design emittance. Table 3.3-4 lists the key design parameters of the pulsed kicker magnet used for injection in the accumulation ring.

Table 3.3-4: Key Parameters of Accumulation Ring Pulsed Kicker Magnet

| Parameter | Specification |
| --- | --- |
| Injection direction | Horizontal |
| Number of accumulations | 3 |
| Kicker type | Stripline |
| Max deflection angle | 11 mrad |
| Effective length | 800 mm |
| Pulse width | < 2 ns |
| Rise time | < 130 ns |

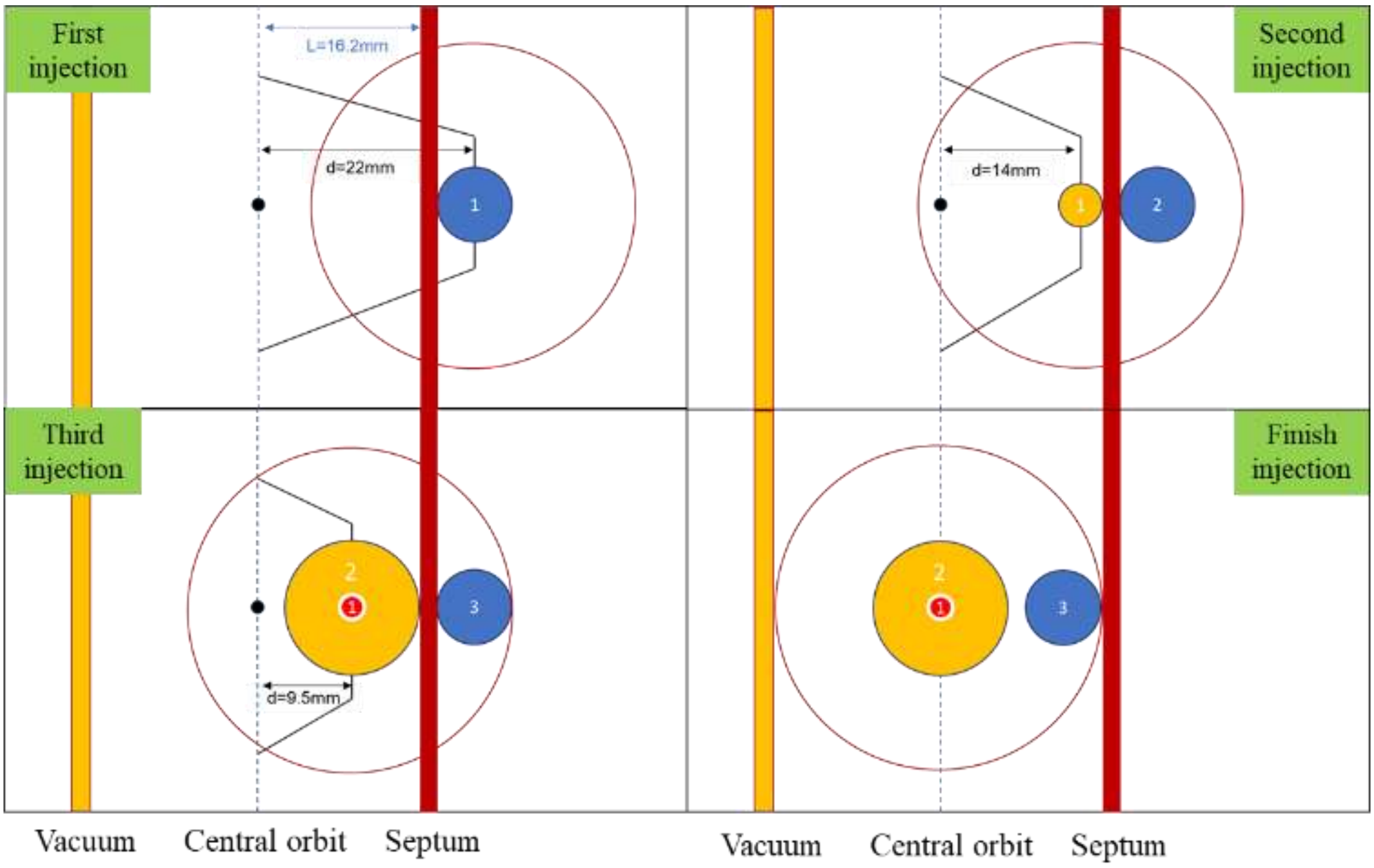

Figure 3.3-37: Schematic of the three-step accumulation process in the accumulation ring

In the current scheme, stringent requirements are imposed on the pulsed kicker magnet to minimize particle loss during accumulation. A new option under consideration is to split the accumulation ring into two separate rings: a damping ring and an accumulation ring. In this approach, the high-emittance positron bunch from the upstream system is first damped in the damping ring, and then the low-emittance beam is accumulated in the accumulation ring. This method not only relaxes hardware technical requirements but also simplifies the overall design complexity.

## 3.4 Main Linac

The main linac (ML) is responsible for boosting the beam energy from 1 GeV, provided by the upstream electron and positron linacs, to the injection energy required by the collider rings, which ranges from 1.0 to 3.5 GeV. Since the beam emittance from the preceding linacs is already determined, the primary design focus of this acceleration stage is to ensure sufficient energy gain.

### 3.4.1 General Physics Design

The ML uses the same RF frequency as the upstream linacs, 2998.2 MHz. A total of 40 RF power sources (klystrons) rated at 45–50 MW are employed. Each klystron, paired with a SLED energy multiplier, drives two 3-meter-long accelerating structures. The accelerating structures operate at a phase following the peak to compensate for the correlated energy spread introduced in upstream sections, ensuring low-energy-spread acceleration from 1.0 to 3.5 GeV before injection into the collider rings. The layout of the accelerating units is shown in Figure 3.4-1.

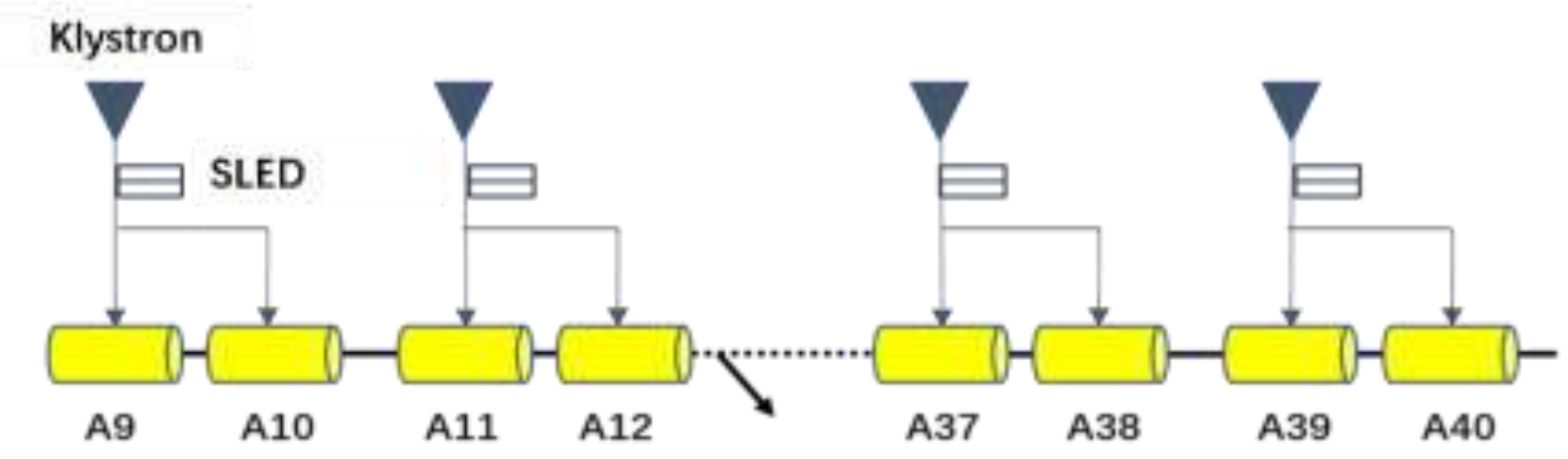

Figure 3.4-1: Layout of the accelerating unit in the main linac

To prevent excessive beam envelope divergence, one quadrupole lens is installed in each drift section between two accelerating structures. Beam diagnostic elements are placed between each power unit to monitor beam parameters. At the ML exit, provisions are made for quadrupole magnets and diagnostic tools between every two accelerating structures to facilitate matching with the downstream bending section and collision region, as shown in Figure 3.4-2.

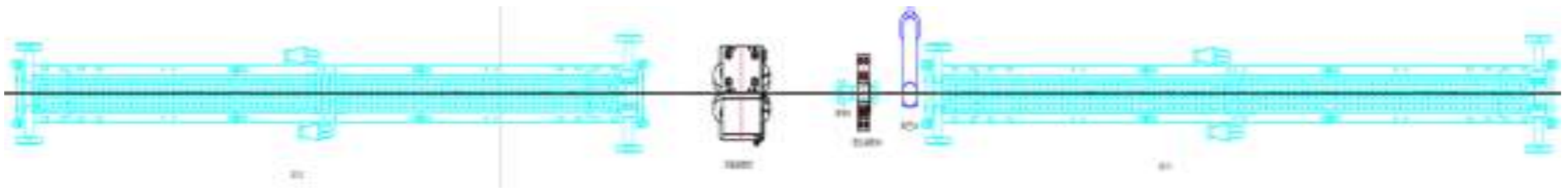

Figure 3.4-2: Distribution of quadrupole magnets and diagnostic units along the ML section

Quadrupole magnets before and after the magnetic bunch compressor must be carefully configured to control the coherent synchrotron radiation (CSR) effect during compression and to enable post-compression beam diagnostics. Based on the above lattice layout, the strength of the quadrupole magnets in the Triplet groups and FODO cells is optimized to ensure the beam's horizontal and vertical β-functions exhibit periodic oscillations in the L3 and L4 sections. Additionally, appropriate quadrupole magnets are reserved at the transport line entrance to perform Twiss parameter matching. The overall Twiss function distribution in the main linac is shown in Figure 3.4-3.

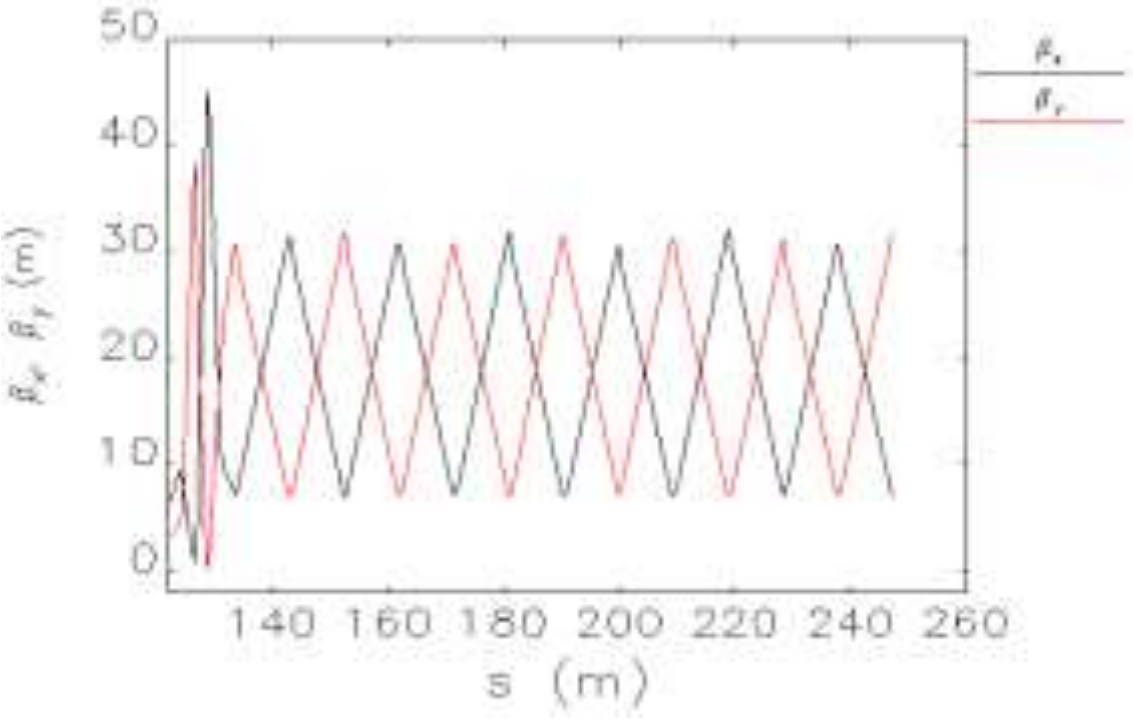

Figure 3.4-3: β-function distribution along the ML section

## 3.4.2 Error Analysis

### *3.4.2.1 Longitudinal Error Analysis*

Equipment parameter errors are the main contributors to variations in beam parameters, and they can generally be categorized into longitudinal and transverse components. The injected bunch charge and arrival time, amplitude and phase errors of the accelerating fields, as well as variations in the R56 of the bunch compressor, will affect longitudinal beam parameters such as bunch length, energy, and arrival time.

Installation errors and vibrations in dipole and quadrupole magnets, magnetic field errors, and high-order field components can cause orbit deviations, residual dispersion, Twiss parameter mismatches, and emittance growth. Similarly, misalignment and vibration of accelerating structures introduce transverse wakefields, leading to orbit deviations and further emittance growth.

These errors can be classified by their timescales and bandwidths into static, slow-drift, and fast-varying errors. Static and slow-drift errors can be corrected using dedicated algorithms and slow feedback systems. However, fast-varying errors must be mitigated by improving equipment stability to keep their impact within acceptable bounds.

Given the similarities in analysis methods across systems, this subsection focuses only on evaluating the effects of various errors and jitters on the beam in the S-band photoinjector under the 1 nC operation mode (prior to 1 GeV) in the off-axis injection scheme. The beam jitter requirements at the injector exit, as dictated by the collider ring physics goals, are listed in Table 3.4-1.

Table 3.4-1: Beam jitter requirements at the exit of the S-band photoinjector (off-axis injection)

| Parameter | Value | Unit |
|---|---|---|
| Energy jitter | < 0.2 | % |
| Peak current jitter | < 10 | % |
| Timing jitter | < 100 | fs |

These variations are mainly driven by injection charge and timing jitter, RF phase and amplitude errors, and magnetic field deviations in the bunch compressor. Due to multiple correlated sources and differing achievable error margins across systems, it is necessary to reasonably allocate tolerances to individual subsystems.

Simulations were performed at various operating points, analyzing the sensitivity of beam jitter to each type of equipment jitter. The normalized sensitivities and final results are summarized in Table 3.4-2.

Table 3.4-2: Beam parameter jitter sensitivity for the S-band photoinjector

| Jitter Source | Value | Unit | ΔE (%) | ΔI (%) | Δt (fs) |
|---|---|---|---|---|---|
| inj_dt | 0.25 | ps | 0.0000 | 0.4333 | 17.20 |
| inj_dQ | 5 | % | 0.0390 | 2.5133 | 49.80 |
| L1_pha | 0.1 | deg | 0.0525 | 0.0333 | 20.27 |
| L1_amp | 0.1 | % | 0.0207 | 5.3867 | 4.47 |
| Lx_pha | 0.4 | deg | 0.0062 | 0.0400 | 3.53 |
| Lx_amp | 0.04 | % | 0.0037 | 0.2667 | 3.20 |
| L2_pha | 0.1 | deg | 0.0093 | 0.0267 | 4.73 |
| L2_amp | 0.1 | % | 0.0140 | 0.0400 | 7.10 |
| L3_pha | 0.1 | deg | 0.0157 | 0 | 0 |
| L3_amp | 0.1 | % | 0.0166 | 0 | 0 |
| L4_pha | 0.1 | deg | 0.0157 | 0 | 0 |
| L4_amp | 0.1 | % | 0.0166 | 0 | 0 |
| BC1_R56 | 0.01 | % | 0.0047 | 0.0333 | 5.77 |
| Total | | | 0.08 | 5.97 | 57.75 |

As shown above, the beam stability at the S-band photoinjector exit meets the downstream collider ring injection requirements.

### *3.4.2.2 Transverse Errors and Correction*

Due to unavoidable manufacturing and installation imperfections, components in the main linac—especially magnets—may introduce mechanical alignment errors. These errors result in non-ideal magnetic fields, leading to orbit distortions. This can cause beam emittance growth and, in severe cases, beam loss due to excursions outside the dynamic aperture. Hence, precise control and correction of beam trajectory deviations are essential.

Since beam position monitors (BPMs) also suffer from alignment errors, Beam-Based Alignment (BBA) techniques are necessary to calibrate and eliminate these offsets. Track correction is then performed using either one-to-one, global, or dispersion-free correction methods to align the beam trajectory as closely as possible to the ideal orbit.

Numerical simulations were performed using the Elegant code's built-in BBA and global correction algorithms. Initial error settings are listed in Table 3.4-3. These simulations demonstrate that orbit distortions caused by such initial errors can be effectively corrected—horizontal deviations are reduced from the mm to sub-100 μm level. Future simulations will incorporate more realistic error budgets to refine the feedback and control system.

Table 3.4-3: Initial alignment error ranges used in orbit correction simulations

| Component | Δx/Δy (μm) | Δθ (mrad) | ΔK/K or ΔB/B (%) |
|---|---|---|---|
| Dipole magnets | 100 | 1 | 0.01 |
| Quadrupole magnets | 100 | 1 | 0.1 |
| BPMs | 100 | - | - |
| Accelerating tubes | 100 | - | - |

Using the global BBA correction method, Figure 3.4-4 shows the horizontal trajectories of various seeds before and after correction. Pre-correction orbit deviations reached several mm, with values exceeding 10 mm near the bunch compressor. After correction, orbit deviations were reduced to below 100 μm, meeting the design requirements.

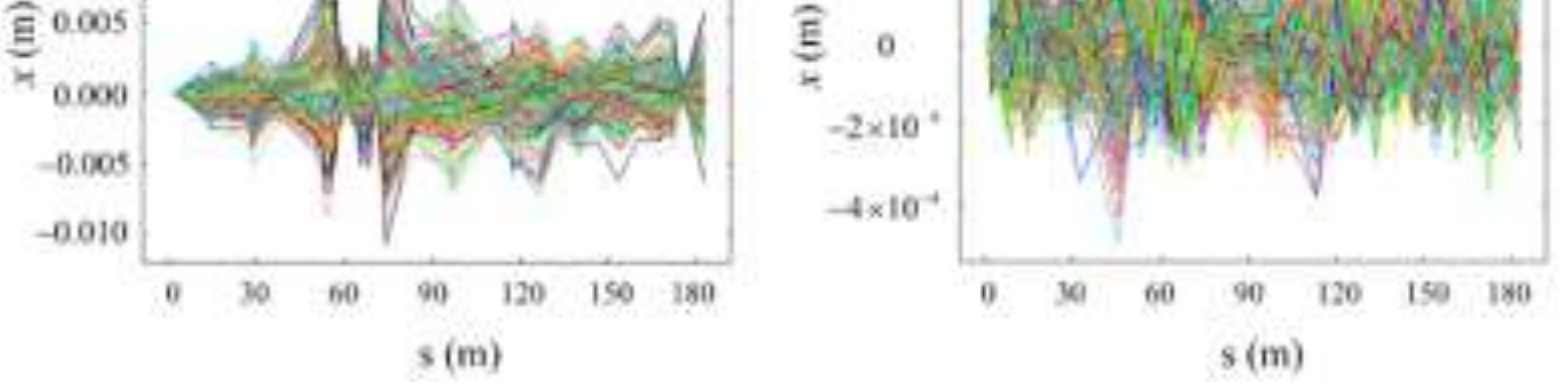

Figure 3.4-4: Beam orbit trajectories before and after correction (horizontal direction)

Off-axis beam injection and component misalignments introduce wakefield effects, leading to emittance growth. To evaluate this, simulations were run using the worst-case seed (i.e., largest pre-correction orbit deviation in Figure 3.4-4). The results in Figure 3.4-5 show that after BBA correction, the emittance growth due to wakefields is limited to within 10%.

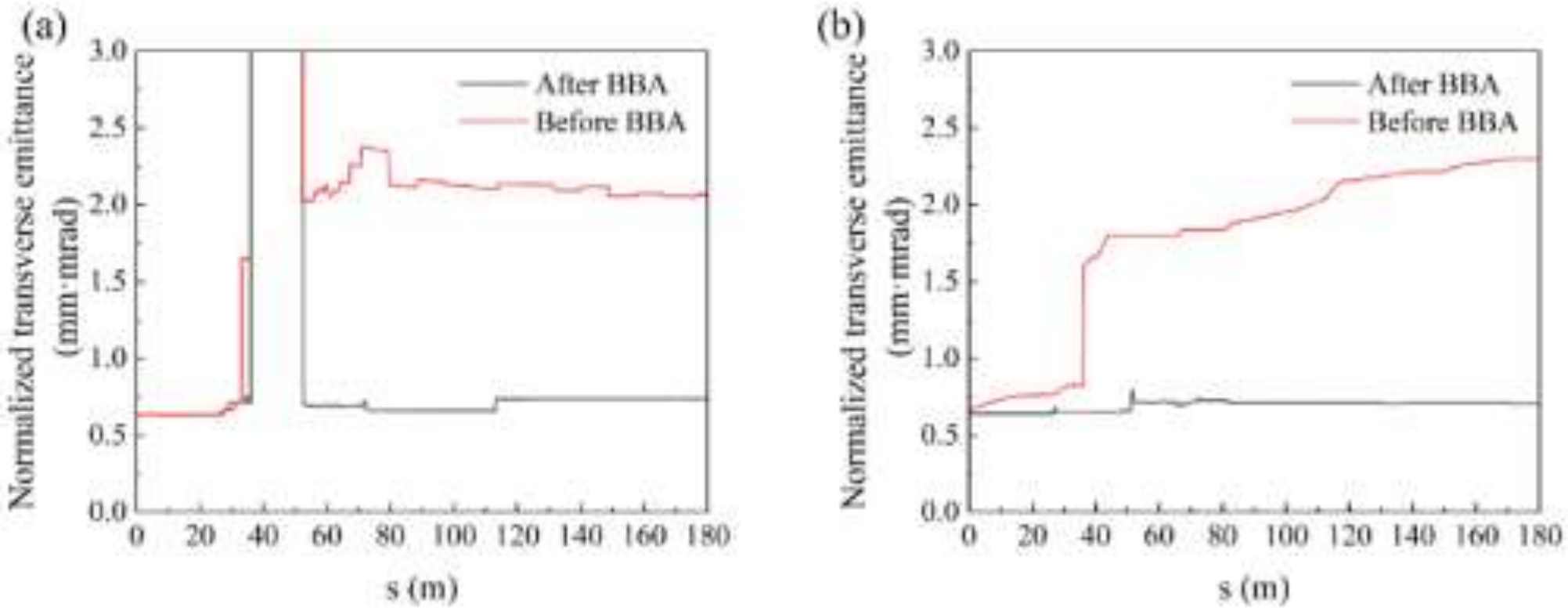

Figure 3.4-5: Evolution of normalized horizontal emittance before and after orbit correction

Error analysis and correction for the main linac follow the same procedures described above. Since the injection energy for the collider rings is adjustable from 1 to 3.5 GeV, the ML keeps its layout unchanged. When a lower-energy beam is required, some accelerators are switched off, but the wakefields induced by passing bunches still degrade beam quality. By tuning the RF phase of the active accelerators, the increase in energy spread from longitudinal wakefields can be effectively suppressed, and the growth in transverse emittance caused by transverse wakefields is likewise controlled after beam-based alignment (BBA) correction. The main beam parameters at ML exit energies of 1, 2, and 3.5 GeV are given below.

Table 3.4-4: Main beam parameters at the ML exit

| Energy | | 1 GeV | 2 GeV | 3.5 GeV |
|---|---|---|---|---|
| RMS energy spread (%) | | 0.0783 | 0.0534 | 0.074 |
| RMS bunch length (ps) | | 1.5669 | 1.5671 | 1.5667 |
| Nor. transverse emittance (μm.rad) | x | 2.187 | 2.175 | 2.166 |
| | y | 1.678 | 1.499 | 1.449 |
| RMS beam size (μm) | x | 113.66 | 81.09 | 60.91 |
| | y | 152.95 | 109.18 | 82.04 |

## 3.5 Beam Transport Lines

In the case where the collider rings adopt off-axis injection, the injector's beam transport lines are divided into five parts based on their regions and functions: the electron bypass section (Bypass), the damping ring transport section, the electron accumulation ring section, the positron accumulation ring section, and the injection transport sections to the collider rings. The Bypass and damping ring transport sections are used in the off-axis injection scheme, while the electron and positron accumulation ring sections serve the swap-out injection scheme. The injection transport sections to the collider rings are shared by both schemes.

In the transport sections of the positron damping ring and accumulation ring, each includes one injection line and one extraction line. The injection transport sections of the collider rings includes one transport line for the electron beam and one for the positron beam.

### 3.5.1 Bypass Section

To reduce the cost and footprint of the linac during positron production, a beam bypass drift section is typically used to extract electron beams that have been accelerated to a certain energy level, enabling them to either continue to higher energy stages or serve as drive beams for the positron source. This approach is commonly adopted in schemes such as SuperKEKB, Tau/Charm Factory, and CEPC. In this design, the Bypass section branches off from the 1 GeV electron linac, deflects horizontally to bypass the positron linac, and reconnects at the entrance of the 3.5 GeV positron/electron main linac.

A schematic layout of the Bypass section is shown in Figure 3.5-1. The total straight-line distance from start to end is approximately 160 m, with a horizontal offset of about 6 m from the linac. The Bypass section employs eight dipole magnets to steer the beam, 35 quadrupole magnets for transverse focusing, and energy slits in the bending regions for beam scraping to control energy spread. At the entrance of the Bypass, the electron beam parameters are: 1.0 GeV energy, 0.2% rms energy spread, 1.5 nC charge, and 2 mm·mrad rms normalized emittance.

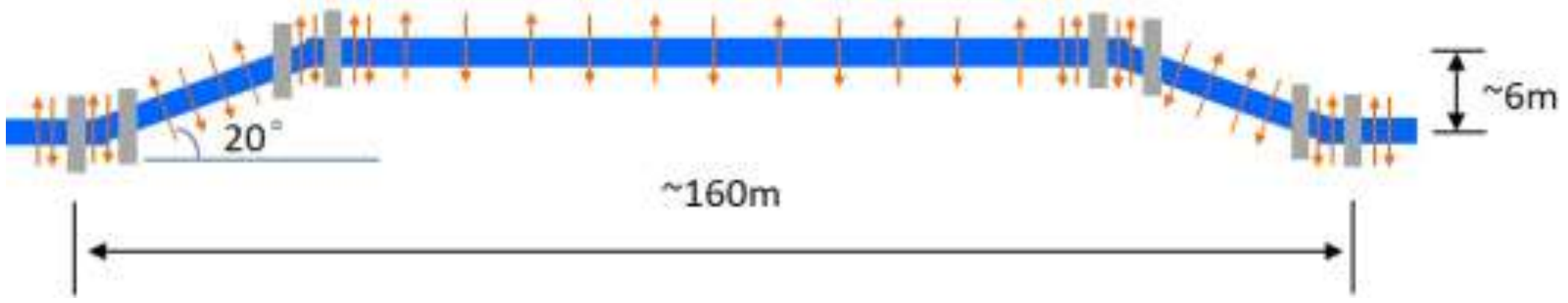

Figure 3.5-1: Layout schematic of the Bypass section

The beam optics of the Bypass section were matched using the MAD code. For initial Twiss parameters of $\beta_{(x,\gamma)}$ = 20 m and $\alpha_{(x,\gamma)}$ = 1, the lattice layout and optical design results are shown in Figure 3.5-2. The optimized Twiss parameters at the exit are $\beta_{(x,\gamma)}$ = 20 m and $\alpha_{(x,\gamma)}$ = –1. The β-functions are symmetrically distributed and controlled below 50 m throughout the transport line, oscillating between 10–40 m in the straight sections. The dispersion function reaches a maximum of 1.3 m in the bending section, sufficient for installing an energy slit, while both the dispersion and its derivative are zero in the straight sections. The transverse focusing system retains adjustment capabilities, allowing the lattice parameters to be tuned to match downstream optical requirements for the positron and electron main linacs under different initial conditions.

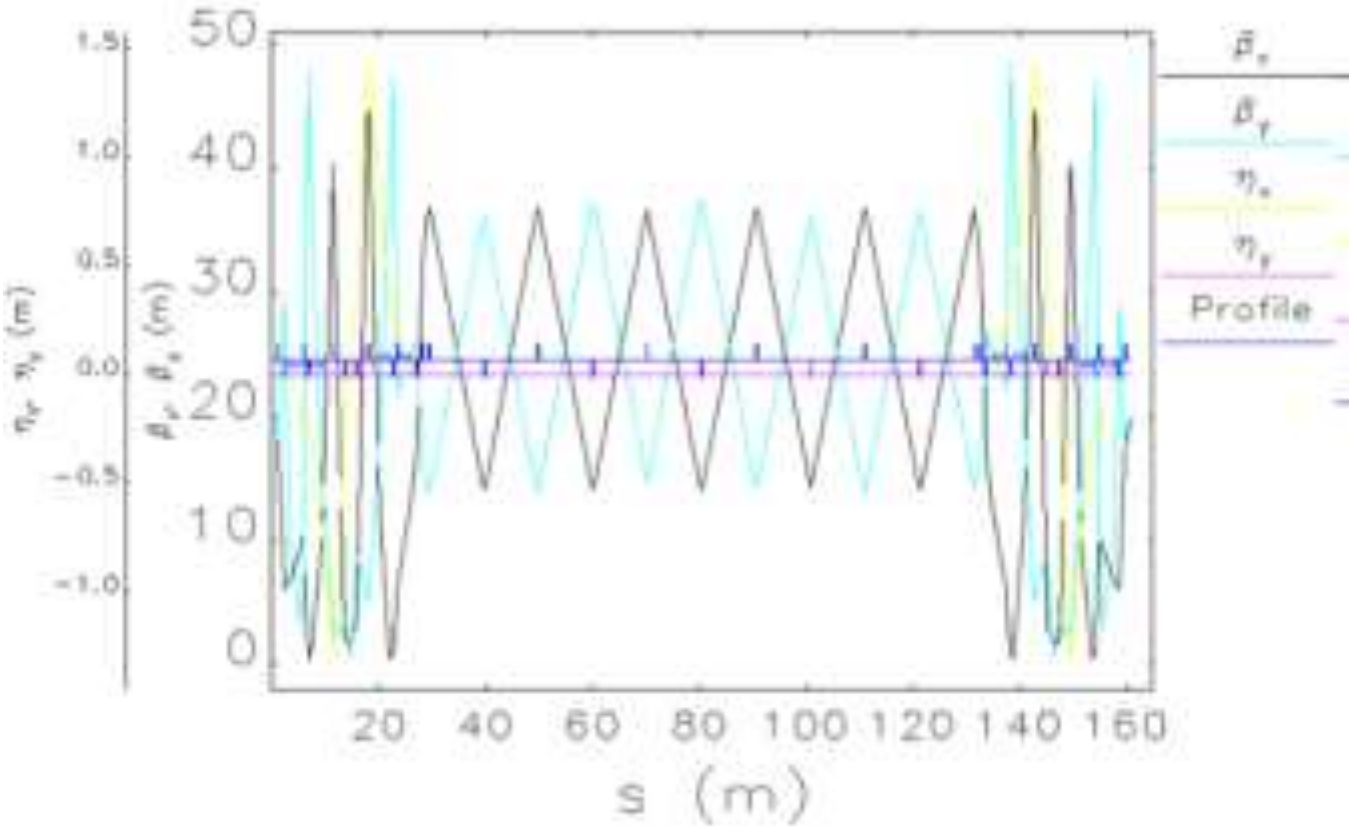

Figure 3.5-2: Optical design results of the Bypass section

Beam dynamics tracking was performed using the Elegant code. The simulated results for rms beam size and normalized emittance are shown in Figure 3.5-3. The beam size and emittance show symmetric distributions, with an rms beam size of approximately 0.2 mm at the exit. The normalized emittance is 2.14 mm·mrad horizontally and 2.0 mm·mrad vertically, with only a 7% increase in the horizontal direction, meeting the requirements of the main linac and the collider rings. Due to the Bypass transport line's R56 design value of 0.4 m, the bunch length will increase slightly, while the energy spread remains nearly unchanged.

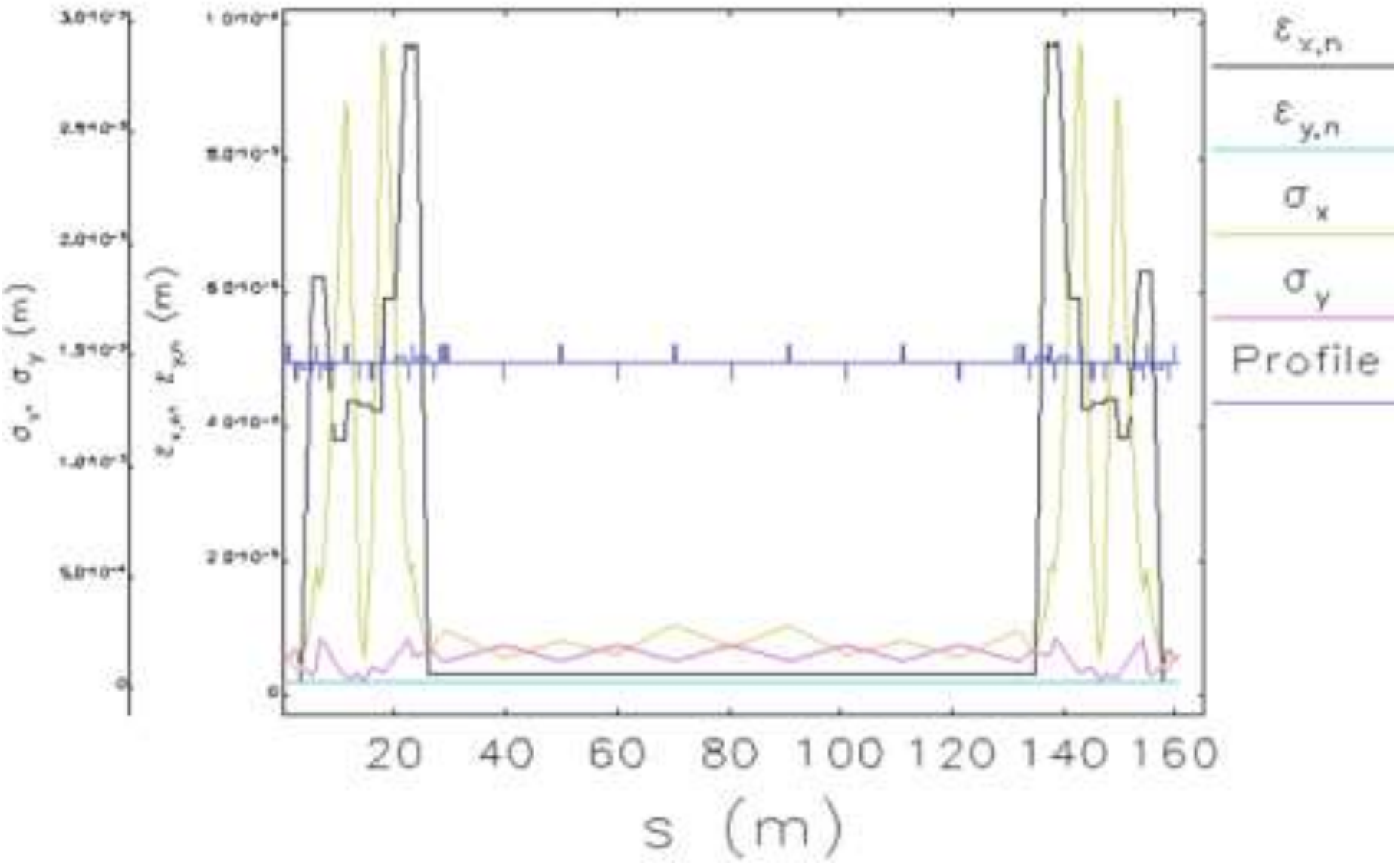

Figure 3.5-3: Transverse beam size and normalized emittance along the Bypass section

### 3.5.2 Damping Ring Transport Section

To meet the positron damping ring's requirements for bunch length, energy spread, and β-functions at injection and extraction, a dedicated design of the transport line connecting the damping ring and the positron linac is necessary. Before injection into the damping ring, the positron beam must pass through an Energy Compression System (ECS) to reduce the energy spread and meet the ring's energy acceptance requirements. After extraction from the damping ring, the positron beam needs to pass through a Bunch Compression System (BCS) to shorten the bunch length for proper matching with the subsequent linac.

A schematic layout of the damping ring transport section is shown in Figure 3.5-4. The light red and dark red lines represent the injection and extraction lines, respectively. The lengths of both lines are approximately 95 m. The injection line consists of two Chicane-based energy compression systems, a horizontal bending arc, and matching sections. In total, 14 dipole magnets are used for horizontal steering and the Chicane system, along with 30 quadrupole magnets for transverse focusing. Additionally, a ~1.3 m accelerating structure is used to provide an energy chirp for compression. The extraction line shares the same structure as the injection line, with only modifications to the Chicane bend angles to accommodate different compression requirements.

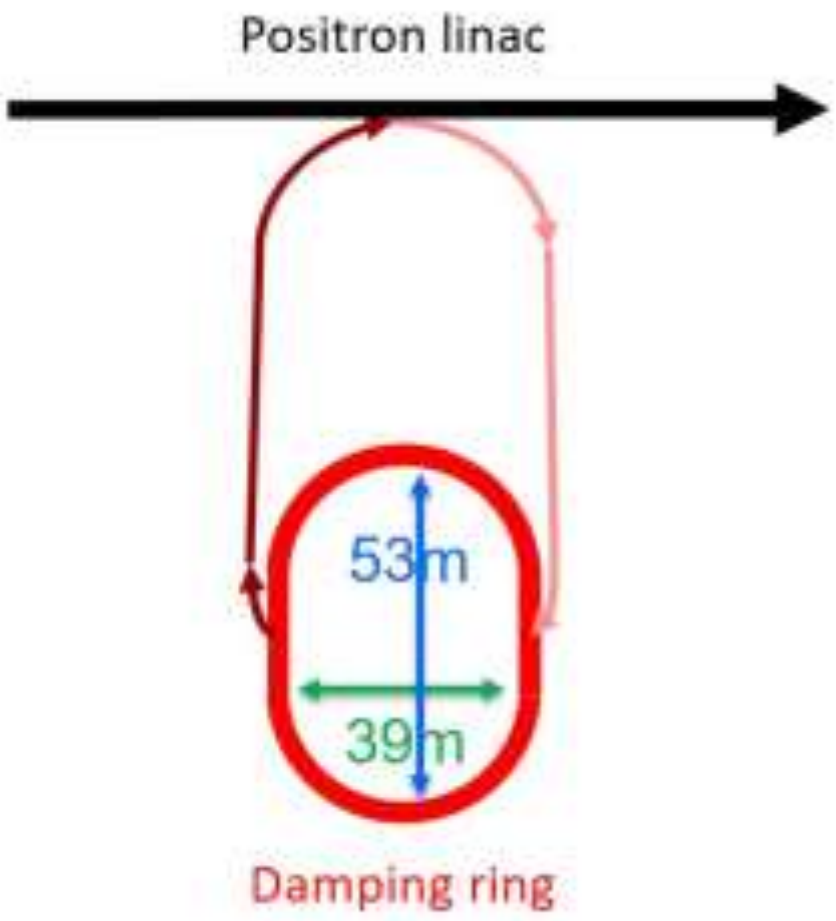

Figure 3.5-4: Schematic layout of the damping ring transport section

At the entrance of the injection line, the positron beam has an energy of 1.0 GeV, an rms energy spread of 0.6%, a bunch charge of 1.5 nC, an rms bunch length of 1.5 mm, and rms normalized emittances of 165 mm·mrad (horizontal) and 75 mm·mrad (vertical). The optics matching was performed using MAD. For initial Twiss parameters $\beta_x = 10$ m, $\beta_y = 55$ m, $\alpha_x = -1$, $\alpha_y = 2$, the lattice layout and optics design results are shown in Figure 3.5-5. The optimized Twiss parameters are $\beta_x = 2$ m, $\beta_y = 8$ m, $\alpha_x = 2$, $\alpha_y = 0$. The β-functions are symmetrically distributed, with the dispersion function reaching a maximum of 2.5 m in the Chicane section. The dispersion function and its derivative are zero in the matching section. The transverse focusing system provides sufficient flexibility for optics matching under different initial beam conditions to meet the damping ring's optics requirements.

Beam dynamics tracking was performed using Elegant, and the rms beam sizes and geometric emittances (with dispersion effects removed) are shown in Figure 3.5-6. Just before injection into the damping ring, the rms horizontal and vertical beam sizes are both 0.6 mm. The geometric emittances are 80 nm·rad (horizontal) and 39 nm·rad (vertical), with emittance growths of 4.4% and 13.3%, respectively, which satisfy the damping ring's requirements.

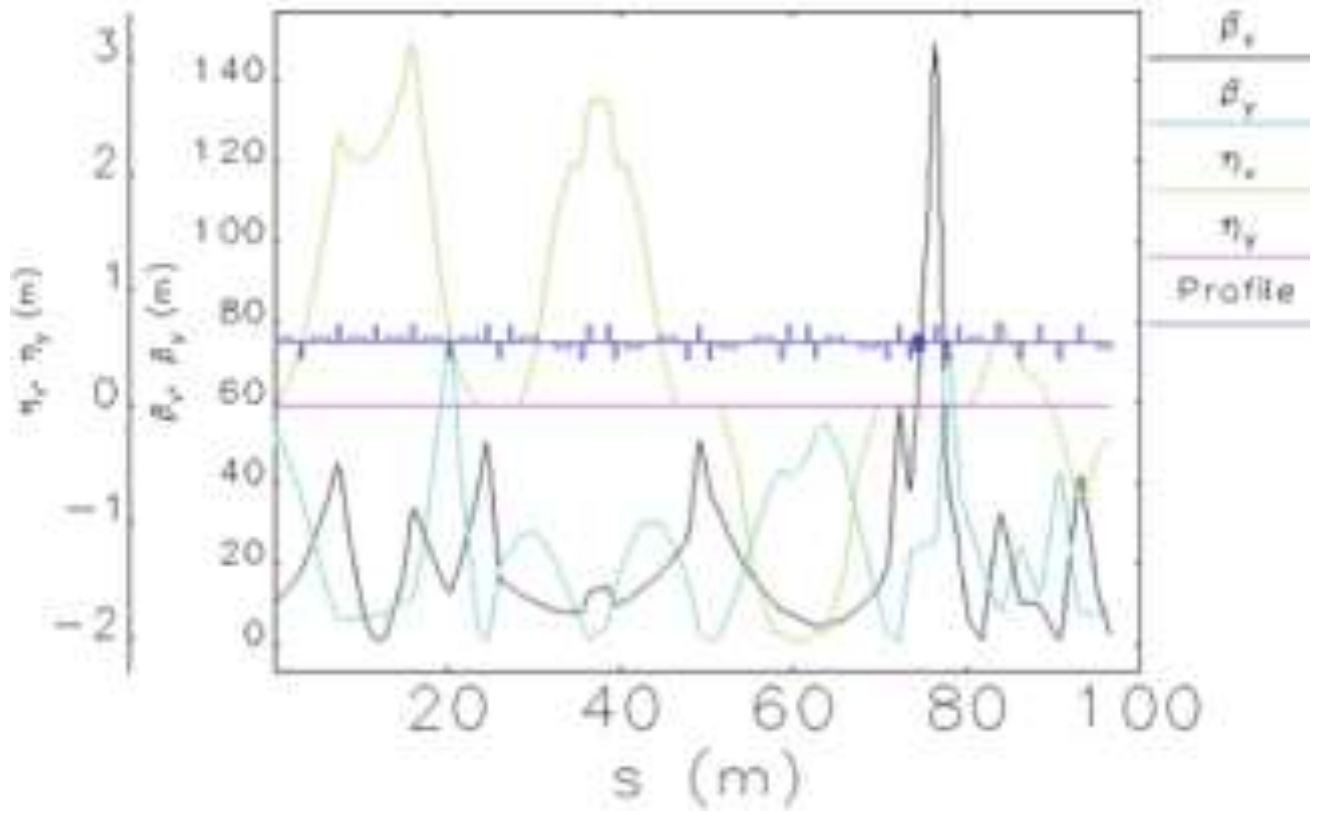

Figure 3.5-5: Optics design of the damping ring injection transport line

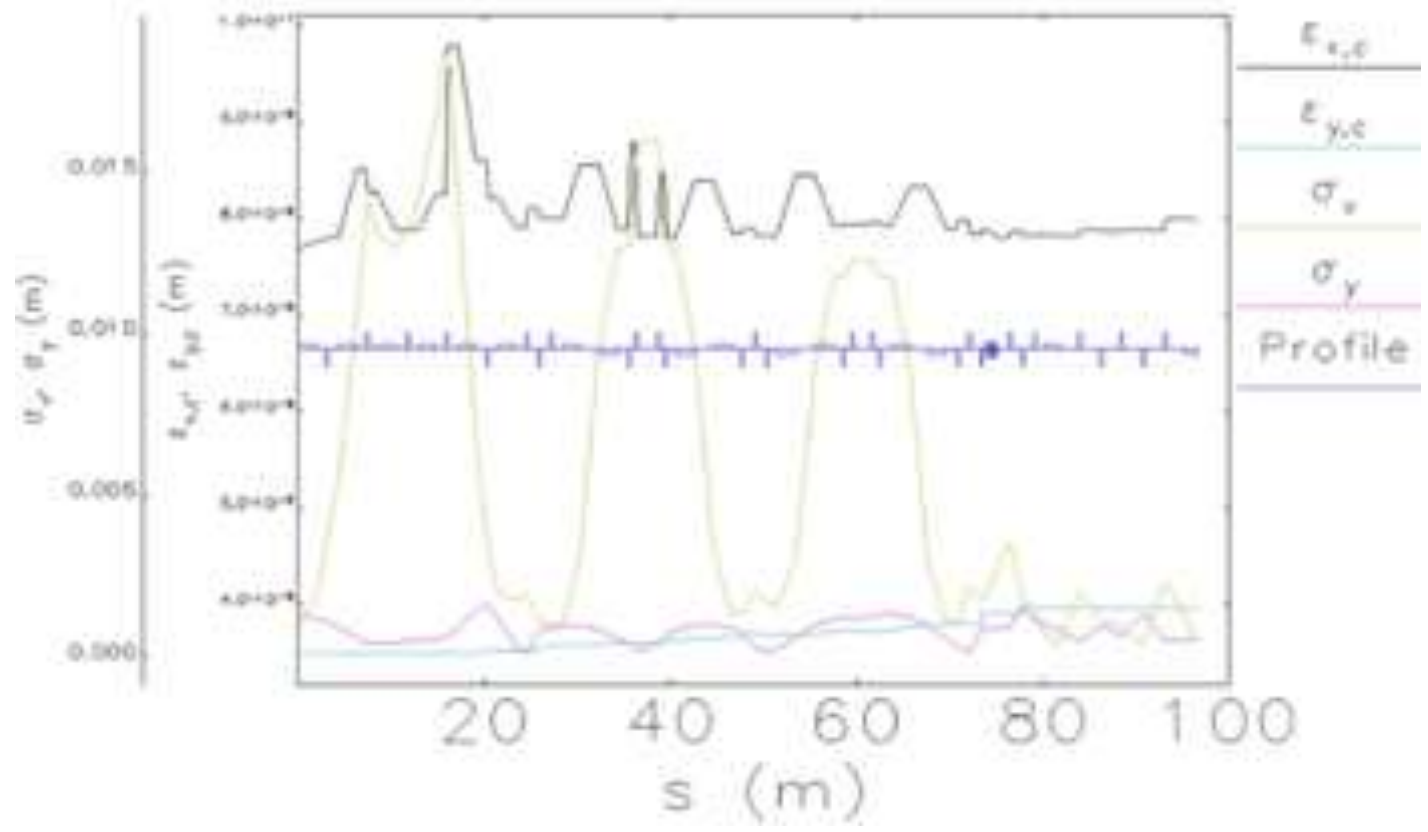

Figure 3.5-6: RMS beam sizes and geometric emittance variation in the injection line

The transport matrix element $R_{56}$ of the injection line is designed to be -1.48 m. As a result, the rms bunch length increases to 9.1 mm, while the rms energy spread is reduced to 0.16%. A comparison of the energy distributions at the entrance and exit of the transport line is shown in Figure 3.5-7, where a clear energy compression effect is observed.

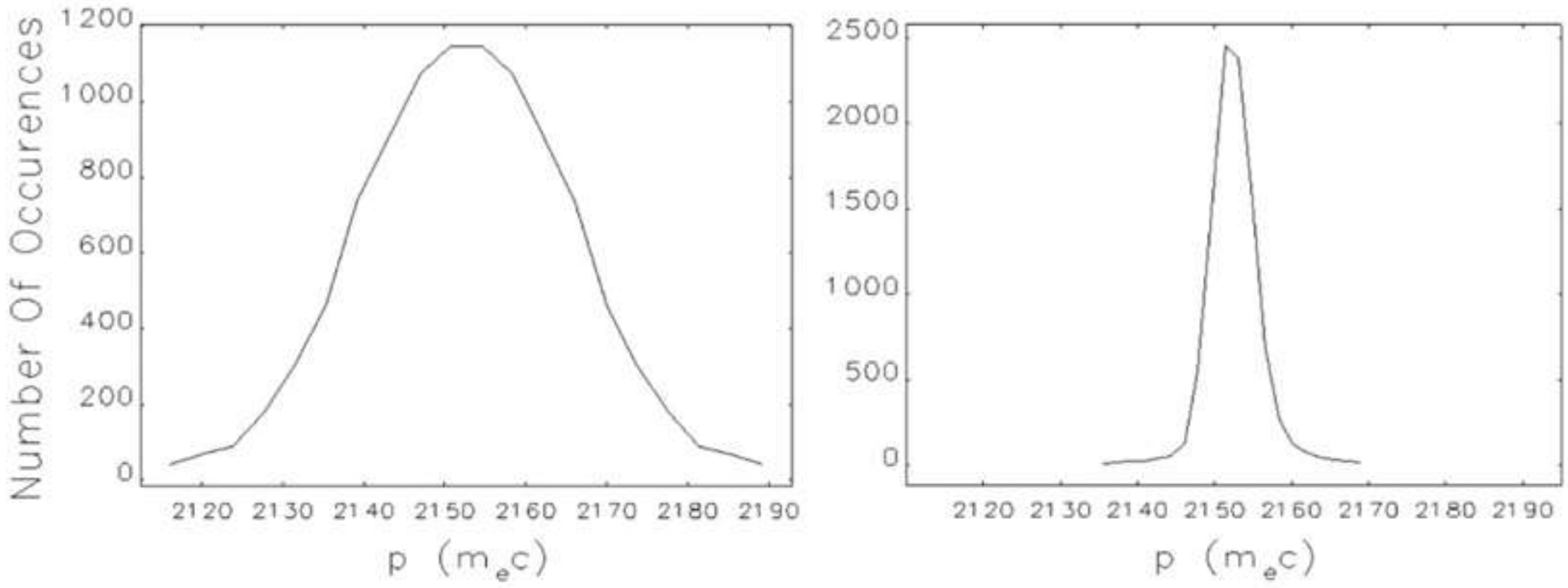

Figure 3.5-7: Energy distribution of the positron bunch at the injection line entrance (left) and exit (right)

### 3.5.3 Positron Accumulation Ring Transport Section

In the swap-out injection scheme, the positron beam must first be accumulated in the accumulation ring to reach the required bunch charge before being further accelerated and injected into the collider positron ring. The beam energy and quality in the positron accumulation ring transport line are similar to those in the damping ring transport section, and thus, the design can follow that of the positron damping ring transport system.

### 3.5.4 Collider Ring Injection Transport Section

The collider ring tunnel is assumed to be at ground level, while the injector is located partially underground, with a vertical height difference of 6.5 meters. As a result, the transport section connecting the injector to the collider ring consists of three main parts: a beam-splitting unit (±10° separation), a horizontal bending section, and a vertical bending section. The schematic layout is shown in Figure 3.5-8. In this layout, the blue line represents the electron transport line, and the red line represents the positron transport line. The electron and positron beams are bent from two different directions and injected into opposite straight sections of the collider rings, located symmetrically to the interaction point (IP).

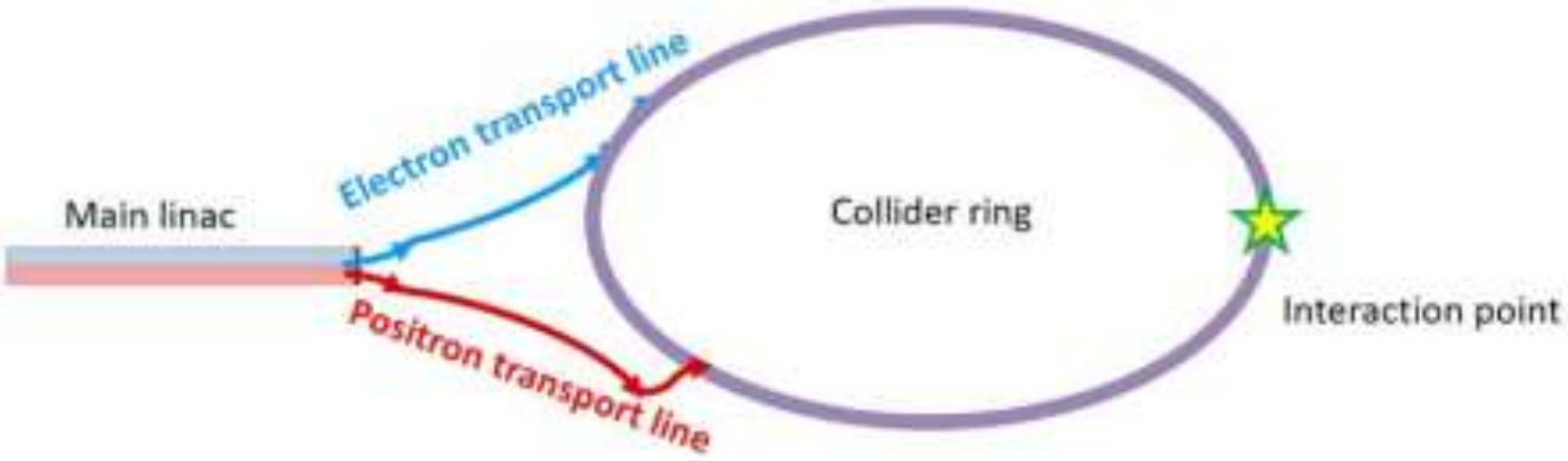

Figure 3.5-8: Schematic layout of the collider ring injection transport section

Taking the electron beamline as an example, the optical design of the horizontal bending section is shown in Figure 3.5-9. This section comprises five 10° dipole magnets and twelve quadrupole magnets. The dispersion function is designed to be zero. The horizontal and vertical β-functions are 12.88 m and 1.404 m, respectively, with both α-functions set to zero. The evolution of the normalized horizontal and vertical emittances (excluding the local growth due to energy spread in dispersive regions) is shown in Figure 3.5-10. Compared to the initial values, the horizontal and vertical emittances increase by 33.3% and 1%, respectively. The bunch length increases slightly, but the energy spread remains nearly unchanged.

The vertical bending section's optical design is shown in Figure 3.5-11, consisting of four 10° dipole magnets and eight quadrupole magnets. The dispersion function is also designed to be zero, with the horizontal and vertical β-functions again being 12.88 m and 1.404 m, and both α-functions equal to zero. The evolution of normalized horizontal and vertical emittances (again excluding dispersion-induced growth) is shown in Figure 3.5-12. Compared to initial values, the horizontal emittance increases by 1%, while the vertical emittance increases by 54%. At the end of the electron beamline, the horizontal geometric emittance is approximately 1.2 nm·rad.

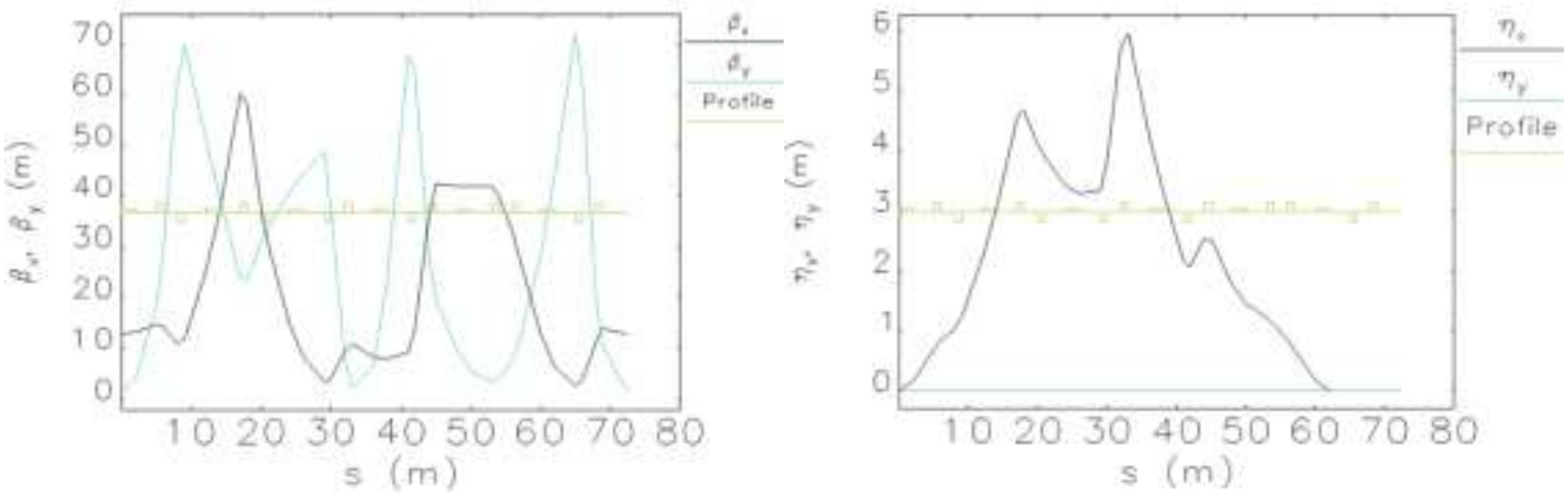

Figure 3.5-9: Optical design of the horizontal bending section in the collider ring injection line
(Left: β-functions; Right: dispersion function)

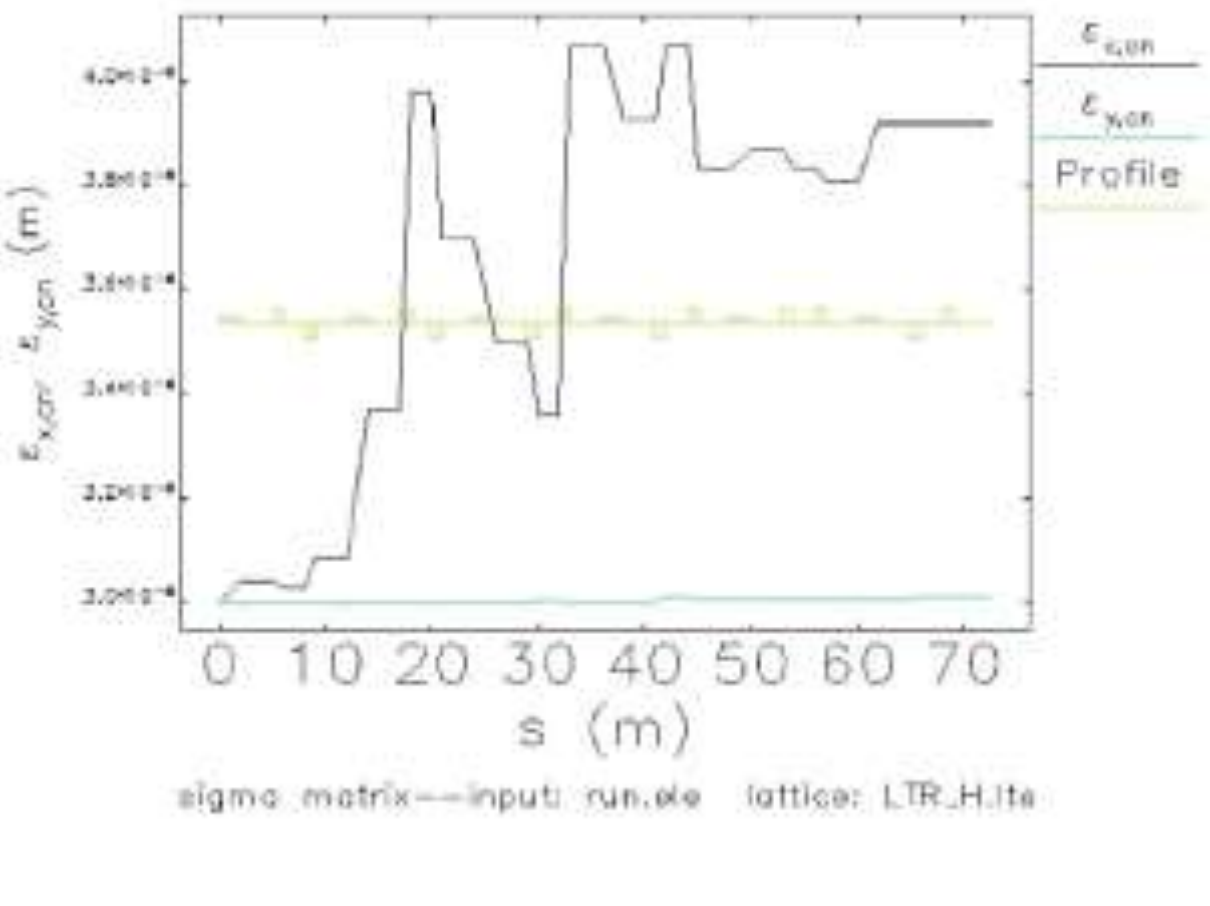

Figure 3.5-10: Normalized emittance tracking for the horizontal bending section in the collider ring injection line

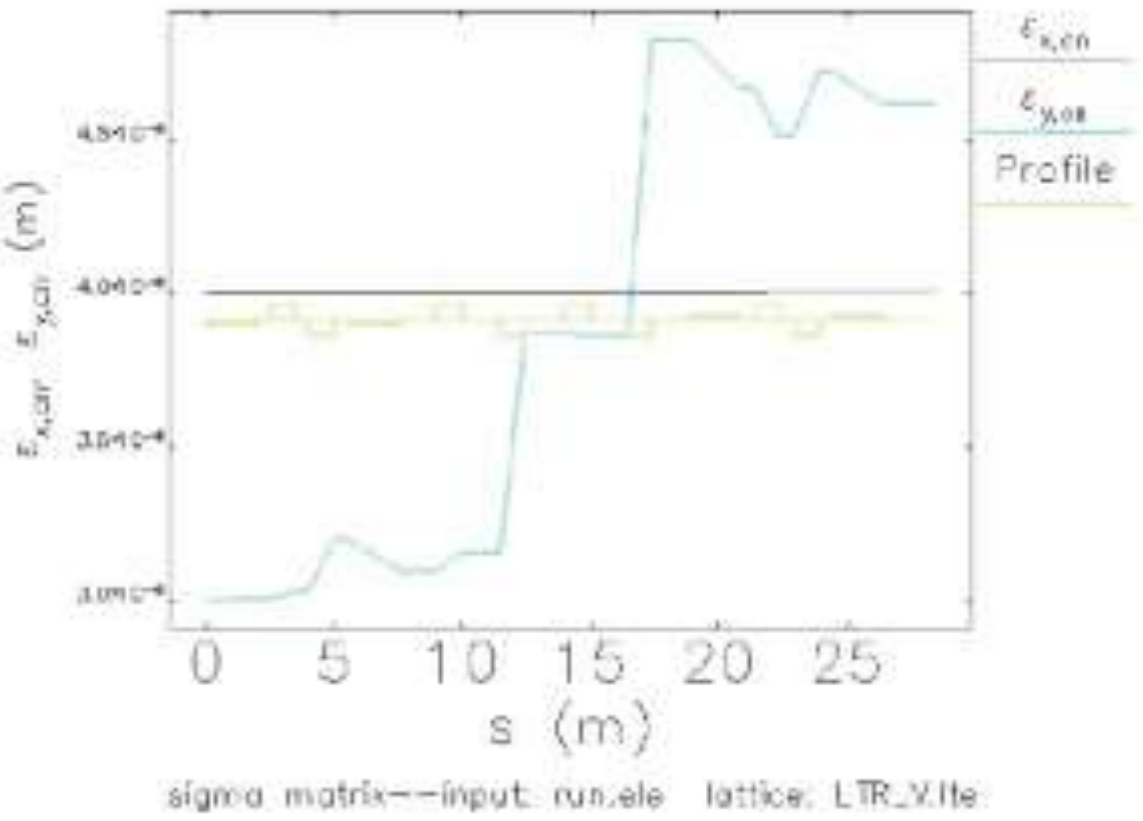

Figure 3.5-11: Normalized emittance tracking for the vertical bending section in the collider ring injection line

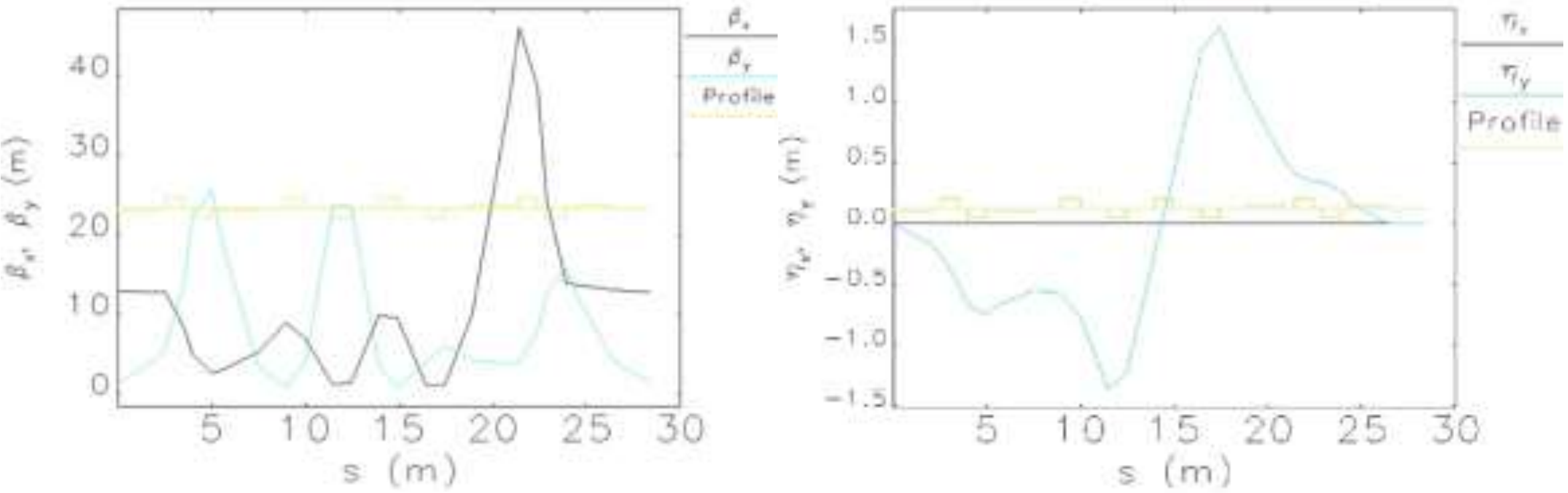

Figure 3.5-12: Optical design of the vertical bending section in the collider ring injection line (Left: β-functions; Right: dispersion function)

The collider ring injection transport lines consist of a total of eighteen 2-meter dipole magnets and forty 1-meter quadrupole magnets. Each transport line is approximately 100 m long and is designed with sufficient matching space and tuning capability. Particle tracking results confirm that the transverse emittances and Twiss parameters of the beams at the end of the line meet the injection requirements of the collider rings.

## 3.6 Compatible Injector Scheme for Off-Axis and Swap-Out Injections

This scheme is built upon the previously described off-axis injection and swap-out injection schemes. While recognizing the performance advantages of the swap-out scheme and the cost-effectiveness and technical ease of implementation of the off-axis scheme, the core objective here is to ensure that STCF meets its highly challenging performance targets, while allowing the project to begin with an off-axis injection system and be upgraded to the swap-out mode later with minimal structural changes. The main design parameters for the compatible injector scheme are shown in Table 3.6-1.

The layout of the compatible injector is shown in Figure 1.2-1c. The specific strategy of the compatible injector scheme is as follows:

1) The injector layout is designed to meet the requirements of the swap-out scheme, but the initial construction of accelerator hardware only needs to fulfill the off-axis injection specifications.
2) For the electron beam directly injected into the collider electron ring, an L-band photocathode electron gun is used (as described in Section 3.3.1.2). The off-axis mode uses a bunch charge of 1.0 nC, while the swap-out mode uses 8.5 nC.
3) For the electron linac used for positron production, the electron source provides 11.6 nC in both modes. In off-axis mode, 1.0 GeV / 30 Hz electron beams are used to strike the positron target; in swap-out mode, 2.5 GeV / 90 Hz beams are used, which requires upgrading a portion of the transport line to a new 1.5 GeV linac.
4) The positron target adopts a replaceable structure: a conventional fixed tungsten target is used in off-axis mode, while a rotating tungsten target is used for the swap-out scheme.
5) The positron linac operates at 30 Hz in off-axis mode and at 90 Hz in swap-out mode, both targeting 1.0 GeV output energy.
6) For positron damping and accumulation, a dual-ring structure is employed. Initially, only a damping ring is built for off-axis injection. When upgraded to swap-out mode, an accumulation ring of the same circumference is added 0.8 m above the damping ring. Positrons first reduce their emittance in the damping ring, then are transferred to the accumulation ring to increase the bunch charge to 8.5 nC. The extraction repetition rates of the damping and accumulation rings are 90 Hz and 30 Hz, respectively.
7) The main linac provides 2.5 GeV energy at a total repetition rate of 30 Hz + 30 Hz for electrons and positrons. The bunch charges for off-axis and swap-out injection are 1.0 nC and 8.5 nC, respectively.

Table 3.6-1: Main Design Parameters for the Compatible Injector Scheme

| Parameter | Off-Axis Injection | Swap-Out Injection | Unit |
|---|---|---|---|
| Electron gun type | L-band photocathode / thermionic | L-band photocathode / thermionic | — |
| Linac RF frequency | 2998.2 | 2998.2 | MHz |
| Electron bunch charge for collider e- ring | 1.0 | 8.5 | nC |
| Positron bunch charge for collider e+ ring | 1.0 | 8.5 | nC |
| Bunch energy / nominal energy | 1.0–3.5 / 2.0 | 1.0–3.5 / 2.0 | GeV |
| Electron bunch geometric emittance (X/Y) | ⩽6 / 2 | ⩽30 / 10 | nm·rad |
| Positron bunch geometric emittance (X/Y) | ⩽6 / 2 | ⩽30 / 10 | nm·rad |
| RMS energy spread | ⩽0.1 | ⩽0.25 | % |
| RMS bunch length | <7 | <7 | mm |
| Electron injection repetition rate | 30 | 30 | Hz |
| Positron injection repetition rate | 30 | 30 | Hz |
| Positron target beam energy | 1.0 | 2.5 | GeV |
| Positron target beam repetition rate | 30 | 90 | Hz |
| Positron preparation ring type | Damping ring | Damping + Accum. ring | — |
| Preparation ring input bunch charge | 1.0 | 2.9 | nC |
| Input emittance (X & Y) | ⩽350 | ⩽350 | nm·rad |
| Input energy spread | ⩽0.1 | ⩽0.3 | % |
| Output emittance (X / Y) | ⩽11 / 0.2 | ⩽30 / 0.2 | nm·rad |

# 4 Technical Systems Design

## 4.1 Conventional Magnets and Damping Wigglers

### 4.1.1 Technical Requirements and Scope

In the STCF accelerator complex, aside from the superconducting magnets in the interaction region, this system is responsible for the design of all other magnets, including those in the collider rings, damping (or accumulation) rings, linacs, and beam transfer lines. It also covers the damping wiggler magnets in the collider rings. The magnet parameters and requirements are provided by the collider ring accelerator physics group and the injector accelerator physics group, respectively. The objective is to design all magnets to meet the performance specifications defined by the beam physics.

Based on current beam dynamics designs, the collider rings use dipole, quadrupole, sextupole, and damping wiggler magnets (octupoles and correctors are not yet finalized). The damping/accumulation ring magnets include dipoles, quadrupoles, and sextupoles. The electron and positron linacs include solenoid steering coils, dipoles for magnetic chicanes, analysis magnets, quadrupoles, and bi-directional correctors. The positron linac also features adiabatic matching devices (AMD), large-aperture quadrupoles, and solenoids. Transfer lines include dipoles, quadrupoles, and bi-directional correctors.

Since the injector physics design spans multiple schemes with minimal variation in magnet requirements, this section focuses on magnet design for the off-axis injection scheme.

### 4.1.2 Key Technologies and Design Approach

Due to the wide beam energy range of STCF (1.0-3.5 GeV), the magnet design is particularly challenging. The magnet field performance will be optimized for 2.0 GeV, but must also ensure reliable operation at both low (1.0 GeV) and high (3.5 GeV) energies, and some reservations on the magnets are also needed to allow for the variations from the nominal lattice. Especially at 3.5 GeV, high field strengths may bring magnets near saturation. Therefore, beam dynamics groups will provide performance requirements to the magnet design across the full energy range.

Thanks to decades of accelerator development with the construction of numerous large accelerator facilities in China, the design and fabrication techniques [70, 71] for conventional magnets are highly mature. With the current requirements of STCF, no insurmountable technical risks are found. All the magnets—including damping wigglers—will adopt conventional electromagnet designs, balancing performance, technical maturity, and cost-effectiveness.

### 4.1.3 Design Strategy and System Composition

Magnetic designs use the codes Poisson and Opera-3D for field calculations, while damping wiggler magnets use Opera-3D and the code Radia. The goal is to achieve high field quality in constrained spaces by optimizing pole shapes, chamfers, and end shimming to suppress higher-order multipole components, to meet the requirements from beam dynamics. The technical design of the magnets is balanced among the requirements from the accelerator physics, the limitations from the fabrication and the cost effectiveness.

The system is organized into four subsystems:

- Magnets in the collider rings
- Magnets in the damping/accumulation ring
- Magnets in the linacs
- Magnets in the transfer lines

The magnets in the collider rings require the highest field quality due to large apertures and high fields. These will use silicon steel laminations of 0.5 mm thick (grade J23-50 from Wuhan Iron & Steel) with punched-lamination assembly. Damping ring magnets, though slightly less demanding in field quality, but still with a good number of units, will also use laminated construction.

The magnets in the linacs and transfer lines, which are usually fewer for each type and less demanding in field quality, will mostly use solid DT4 steel cores, and will be machined via CNC or wire-cutting machining technique that can also ensure field quality. However, for some types of magnets that need a large number of units and high field quality, the lamination technique may still be used.

### 4.1.4 Feasibility Designs and Analysis

#### *4.1.4.1 Collider Ring Magnets*

The STCF collider is a dual-ring machine with a circumference of about 860 m. Through iterations, the collider ring accelerator physics has defined parameters for four main magnet types for the CDR report: dipoles, quadrupoles, sextupoles, and damping wigglers. Following the specifications for each type of magnet, the magnet group has conducted feasibility studies and field modeling and has given the preliminary design results, including water cooling and electrical parameters.

**Dipole Magnets**

There are three types of dipole magnets (248 total for both rings), with gap sizes of 60 mm and 75 mm, and three types of effective lengths. An H-type yoke is proposed for its symmetric field; however, it requires top/bottom split assembly that is less convenient for installing vacuum chambers. Figure 4.1-1 shows a 3D model and transverse field distribution of a typical arc dipole (B-ARC).

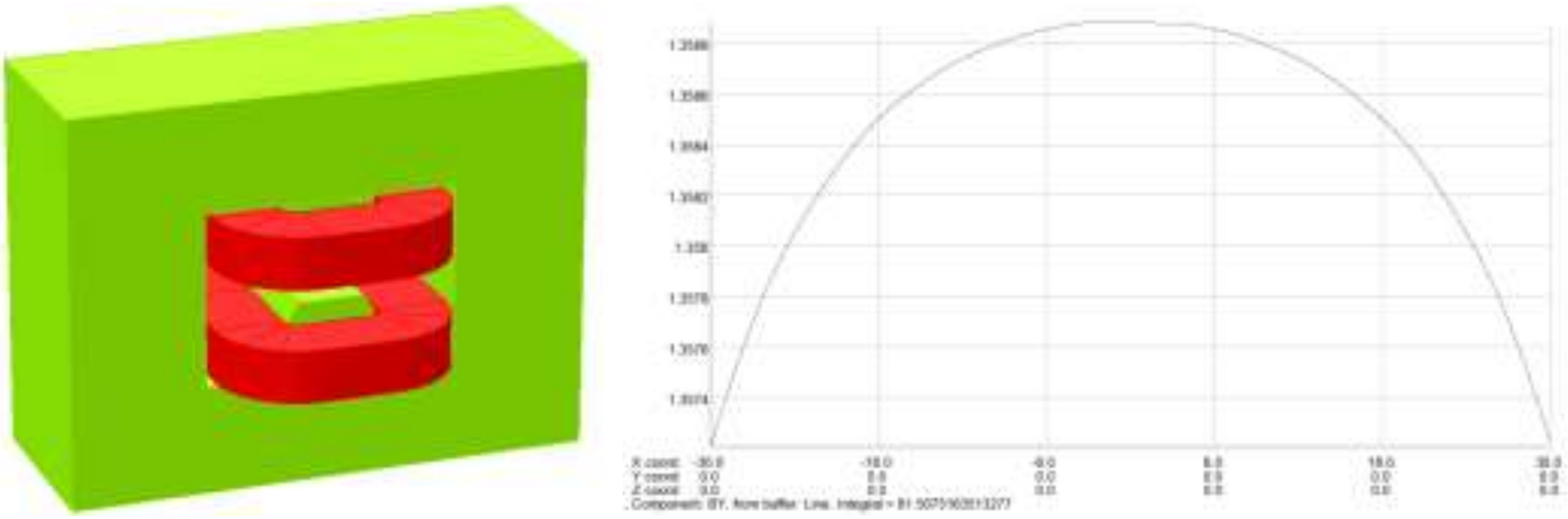

Figure 4.1-1: 3D Model and Field Distribution of Arc Dipole Magnet

**Quadrupole Magnets**

Seven types of quadrupoles (706 units in total) are defined, with bore diameters of 50 mm, 60 mm, and 75 mm, and four effective lengths. The spaces for vacuum ante-chambers have been reserved in the designs. Figure 4.1-2 shows the 3D model and field profile of one quadrupole type (Q1-IR).

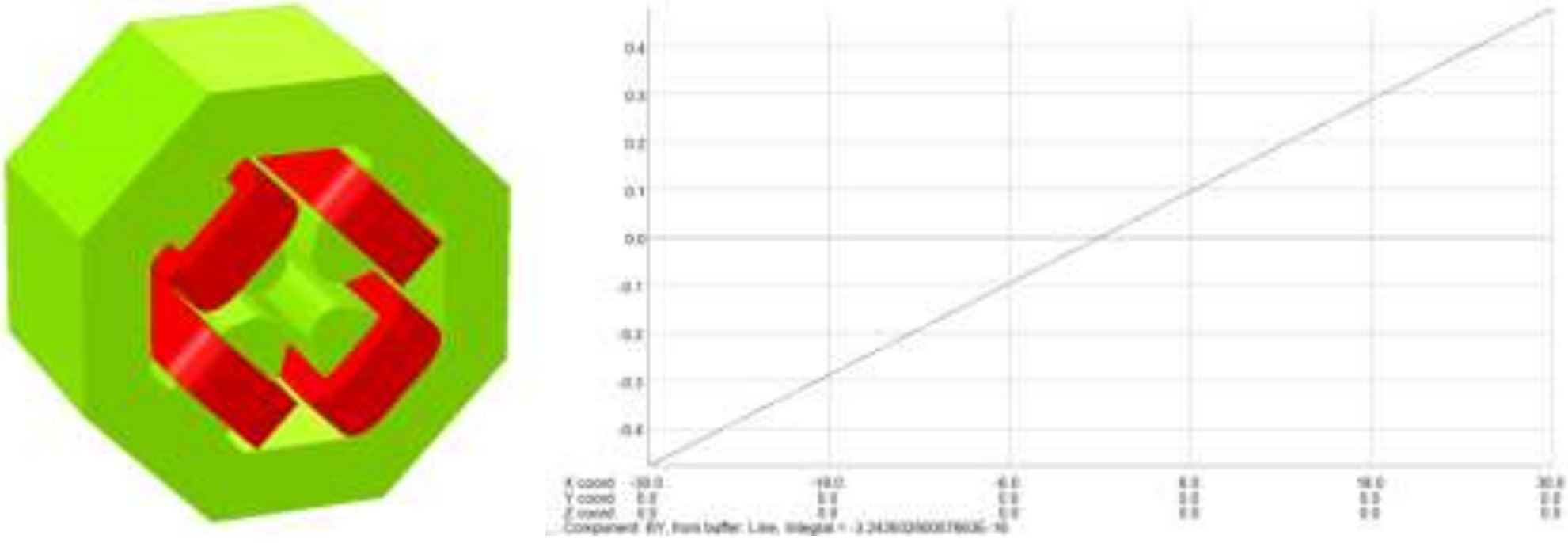

Figure 4.1-2: 3D Model and Field Distribution of Quadrupole Magnet

**Sextupole Magnets**

In addition to two large-quantity types (60 mm and 75 mm in bore diameter) for chromaticity correction, one pair of a special type of 75 mm aperture is needed as the so-called crab sextupoles. The total number of sextupoles for two rings is 192. Figure 4.1-3 shows the model and field profile for one type of sextupole (Sext-IR).

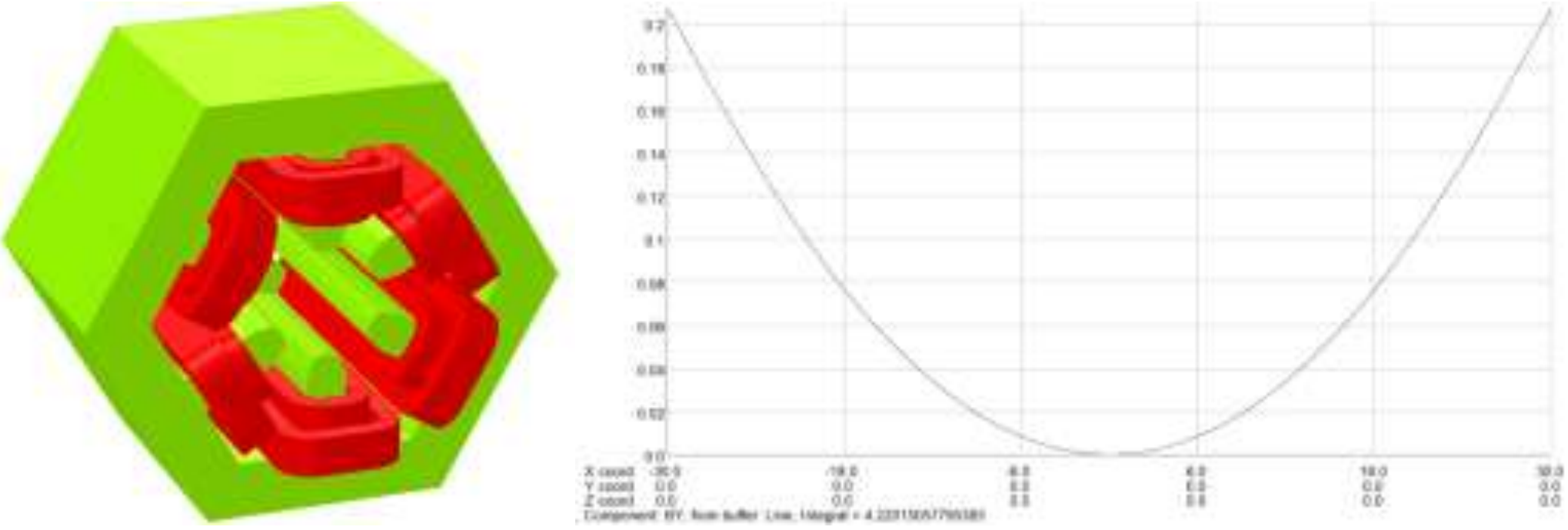

Figure 4.1-3: 3D Model and Magnetic Field Distribution of the IR Sextupole Magnet

**Damping Wiggler Magnets**

The primary function of the damping wiggler magnets in the STCF collider rings is to reduce the radiation damping time. Additionally, they contribute to the control of beam emittance and energy spread. As insertion devices in the storage ring, damping wigglers are designed to minimally perturb the beam. To this end, the relative angle and position of the particle trajectory at the entrance and exit of the wiggler should remain unchanged. Although the center axis of the oscillatory particle trajectory within the magnet (i.e., the reference trajectory) may deviate from zero, it is preferable to maintain a zero-centered trajectory to suppress transverse sextupole field components within the same pole width. To achieve this, the wiggler employs two pairs of end-pole field tapering structures at both ends.

The current wiggler design features an effective length of 4.8 m, a period length of 0.8 m, and a total of six periods (an even number). The peak magnetic field reaches 1.6 T with a vertical gap of 50 mm. A total of 32 such wigglers are planned for the double-ring configuration. The end pole configuration follows the sequence: (+1/4(①)、-3/4(②)、+1(③)、-1、...、+1、-1、+3/4、-1/4), corresponding to 12 pole pairs. Figures 4.1-4 and 4.1-5 show the 2D magnetic model and the longitudinal on-axis magnetic field distribution of the damping wiggler. Future magnet optimization will aim to further enhance the achievable peak field, as desired by the accelerator physics design.

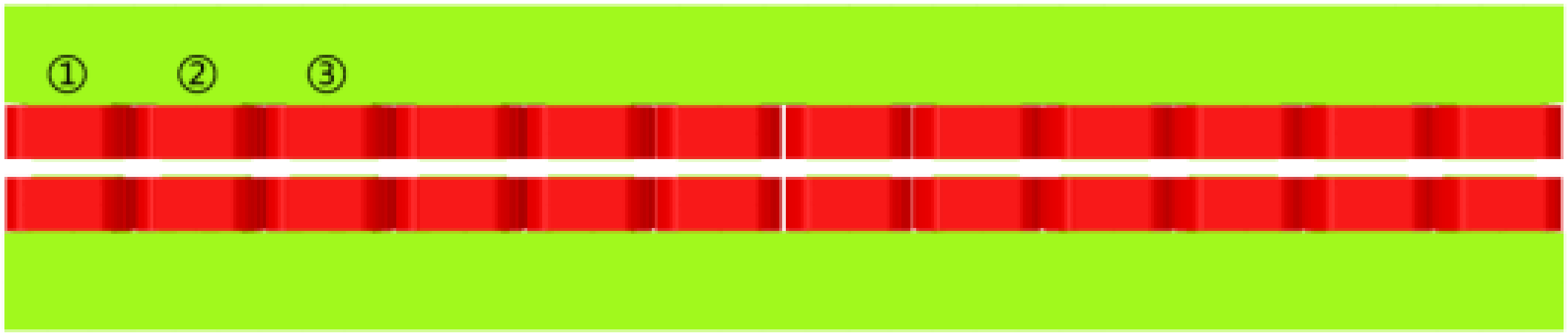

Figure 4.1-4: 2D magnetic model of the damping wiggler magnet.

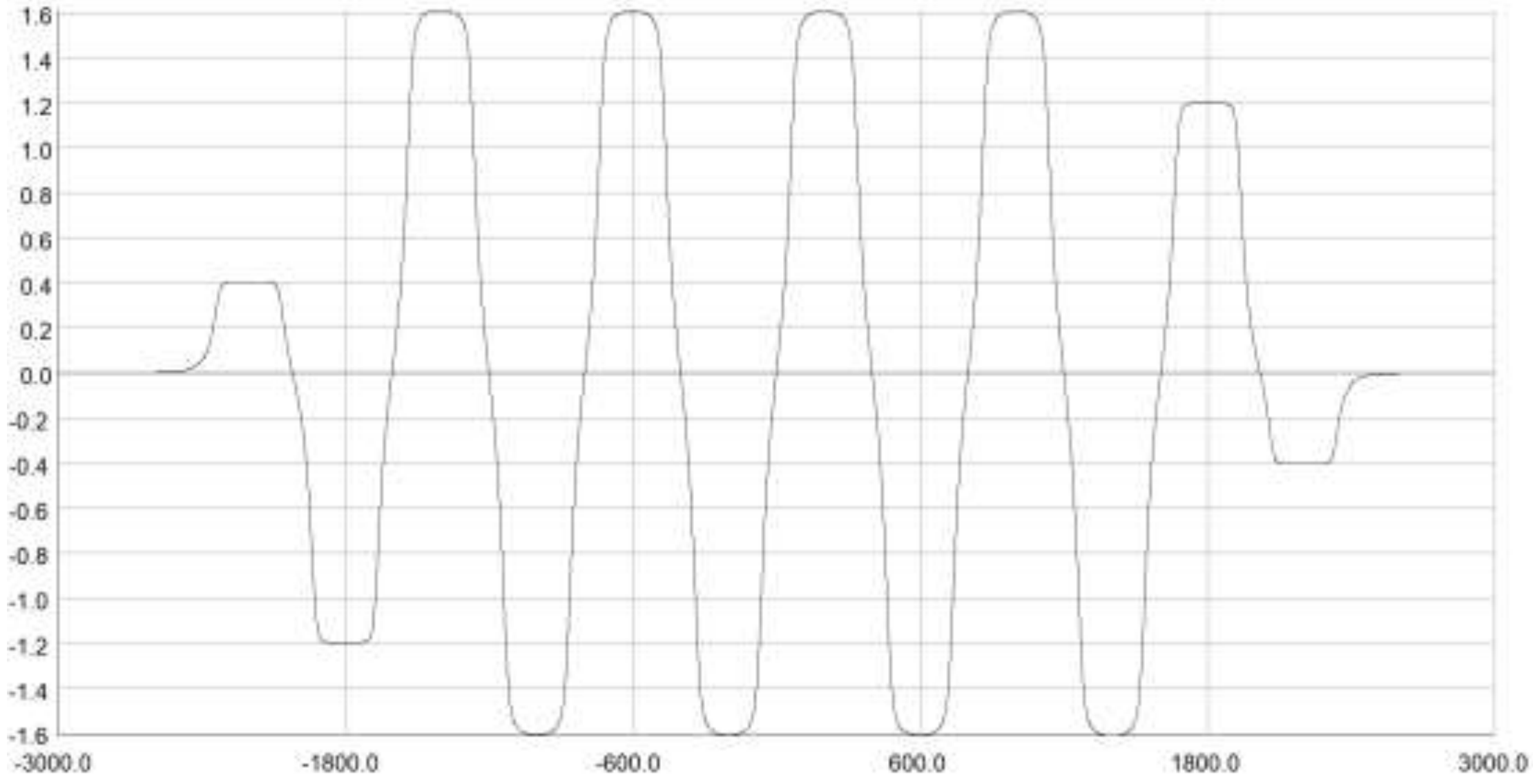

Figure 4.1-5: Longitudinal on-axis magnetic field distribution of the damping wiggler.

#### *4.1.4.2 Damping/Accumulator Ring Magnets*

The STCF damping ring has a circumference of approximately 150 meters and operates at an energy of 1.0 GeV. According to the current accelerator physics design, the required magnet types include dipole, quadrupole, and sextupole magnets. Compared with the collider ring magnets, the field strength and performance requirements for these magnets are relatively relaxed, so their design and fabrication are not expected to pose significant challenges. Preliminary modeling and estimates of the associated electrical and cooling parameters have been performed for each of the three types.

The magnets in the accumulator ring are similar to those in the damping ring, thus, they are not described here.

**Dipole Magnets**

Four types of dipole magnets are needed for the damping ring: normal bending magnets (Normal-bend), reverse bending magnets (Reverse-bend), normal dispersion-matching dipoles (Normal-bend DPS), and reverse dispersion-matching dipoles (Reverse-bend DPS), with a total number of 88. These dipoles also adopt an H-type structure and can be fabricated using either the laminated core technique or solid iron cores. The final choice will be determined during the engineering design phase. The 3D model is shown in Figure 4.1-6 (left).

**Quadrupole Magnets**

The damping ring requires three types of quadrupole magnets, totaling 102 units. Among them, the Q-arc quadrupoles account for 86 units and will be fabricated using laminated steel cores to ensure quality. The small-batch quadrupoles may be fabricated using solid iron cores. The 3D model is shown in Figure 4.1-6 (center).

**Sextupole Magnets**

The sextupoles in the damping ring include two types, SF and SD, totaling 72 units. These will be fabricated using solid DT4 iron cores with CNC machining for good precision. The 3D model is shown in Figure 4.1-6 (right).

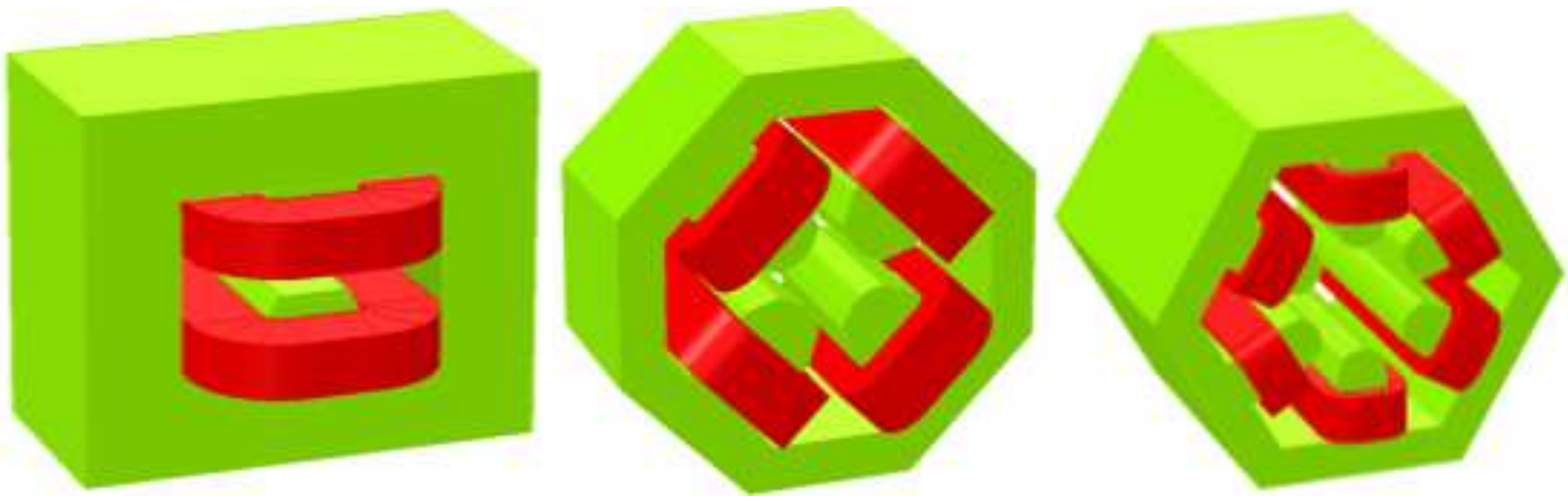

Figure 4.1-6: 3D models of damping ring dipole (left), quadrupole (center), and sextupole (right) magnets.

#### *4.1.4.3 Linac Magnets*

Magnet parameters for the linacs are provided by the injector physics design group and related to the magnets in the electron linacs, positron linac, and main linac. In the electron linacs, magnets include 26 dipole correctors, 4 movable magnetic chicane dipoles, 2 analyzing dipoles, 43 quadrupole magnets, and several solenoid coils. The positron linac includes a magnetic horn, an 18-meter-long solenoidal focusing coil, positron guiding solenoids, analyzing dipoles, 25 dipole correctors, 14 sets of combined-function quadrupole magnets, and 4 dipole magnets.

Although the linac magnets have a variety of types, in general, they are in small quantities of each type, except quadrupoles and dipole correctors, and the requirements for field strength and quality are low. They do not pose significant challenges to the design and manufacturing. However, the long solenoid focusing coil for positrons may present some fabrication difficulty due to its large aperture and relatively high field strength. To master the relevant technique, a similar solenoid is already under development as part of the electron-positron test platform within the frame of the STCF key technology R&D project, which will provide valuable fabrication experience for future STCF construction.

#### *4.1.4.4 Transport Line Magnets*

The magnets in the transport lines are mainly divided into three regions:

- Bypass Section: Includes 8 dipole magnets, 35 quadrupole magnets, and 2 bidirectional correction magnets.
- Transport Line Between the Damping Ring and the main Linac: Includes 14 dipole magnets, 30 quadrupole magnets, and 2 bidirectional correction magnets.
- Transport Lines Between the Main Linac and Collider Rings: Also includes dipole, quadrupole, and bidirectional correction magnets (quantities to be determined).

The two formers all operate in a fixed beam energy of 1.0 GeV, thus the magnets can be designed with less adjustability. However, all these magnets in the injection lines to the collider rings must operate under variable beam energy conditions ranging from 1.0 to 3.5 GeV, and thus the design must be comprehensively optimized to ensure performance stability across the full energy range.

### 4.1.5 Summary

A feasibility study and analysis of the conventional magnets and damping wiggler magnets for the STCF accelerator complex have been carried out. Starting from system design requirements, it is established that the performance of all magnets must meet the specifications defined by the accelerator physics design. These requirements cover magnets in the collider rings, the damping/accumulation ring, linac sections, and beam transport lines.

Given the wide energy span of the collider rings, magnet performance is optimized at 2.0 GeV, while also ensuring compatibility with 1.0 GeV and 3.5 GeV operation. Detailed modeling has been conducted for the magnets in the collider rings and the damping ring, and the preliminary results confirm feasibility. Initial estimates of electrical and cooling parameters have also been provided to support the power supply system design.

While the design and fabrication of conventional magnets pose no significant technical risks, owing to decades of mature domestic experience, the electromagnetic damping wiggler magnets present greater challenges. In particular, achieving higher peak fields with excellent field quality, while controlling fabrication costs, will be a key technical focus moving forward.

## 4.2 Magnet Power Supply System

The magnet power supply system provides stable or dynamically adjustable excitation currents for all types of magnets in the collider rings and the injector. These currents generate the magnetic fields required for guiding charged particles along designed trajectories inside the accelerator, thereby enabling precise beam control.

### 4.2.1 Design Requirements and Specifications

The STCF power supply system consists of a large number of highly stable DC current sources characterized by high stability, low ripple, and minimal electrical noise. The current stability is required to be within 10-50 ppm, and the current ripple within the order of $10^{-5}$. In addition to meeting stringent physical performance criteria, the power supply system must also conform to engineering requirements related to layout, electromagnetic compatibility, and long-term reliability.

According to their operational modes and power delivery schemes, STCF magnet power supplies are categorized into two main types:

- Main Magnet Power Supplies: These are unidirectional, high-stability static supplies serving dipole, quadrupole, and nonlinear magnets. The long-term current stability and ripple for dipole and quadrupole power supplies must reach the 10 ppm level.
- Corrector Magnet Power Supplies: These are bidirectional dynamic supplies used for beam orbit correction magnets. Based on bandwidth requirements, they are further divided into slow and fast corrector supplies. Fast corrector supplies must achieve a small-signal response bandwidth of at least 5 kHz while also maintaining low ripple, high output resolution, and excellent static stability.

### 4.2.2 Key Technologies and Design Approach

Based on both international engineering practices and domestic technical foundations, the system adopts a digitally controlled switched-mode power supply (SMPS) topology, as illustrated in Figure 4.2-1.

Each power supply consists of a converter and a controller. The converter is the core power-processing unit responsible for voltage/current transformation. The controller serves as the system's brain, using closed-loop feedback based on automatic control theory to regulate output with high precision. Digital management modules handle analog/digital signal acquisition and adjustment.

Power supplies are grouped into high, medium, and low power classes. By analyzing the load characteristics and physics requirements, power supplies are normalized and grouped to reduce the number of models and simplify manufacturing. Design priorities include high-frequency operation, modularity, digital control, thermal redundancy, and hot-swappable capability. Cost is strictly controlled under the premise of meeting all performance targets [72].

The key factor affecting power supply stability is the design of the digital controller and feedback loop. The system employs a high-precision digital current feedback loop, enhanced with output voltage feedback to suppress transients, and feedforward voltage compensation to mitigate grid fluctuations.

A classic digital PID algorithm is implemented for closed-loop control, where the tuning of proportional (Kp), integral (Ki), and derivative (Kd) gains is critical. A robust modeling and control framework is adopted, allowing for self-tuning of control parameters based on dynamic load conditions.

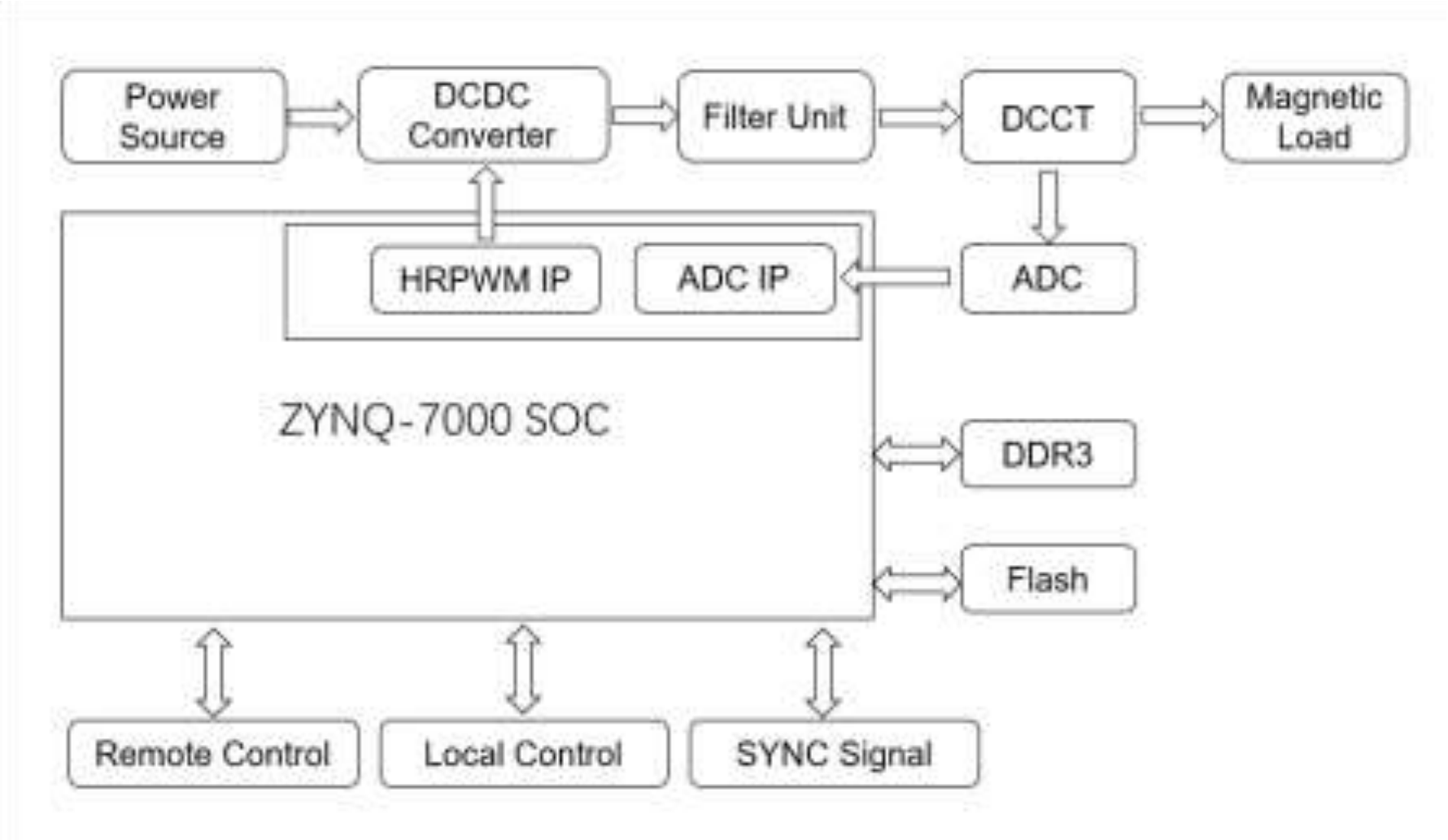

Figure 4.2-1: Schematic diagram of the digital power supply architecture

### 4.2.3 Design Scheme and System Configuration

The power supply system adopts a two-stage voltage conversion structure (see Figure 4.2-2). The front-end AC/DC stage provides isolation and stabilized voltage, effectively suppressing low-frequency voltage ripple. The rear-end DC/DC stage ensures stable current output for magnet excitation [73].

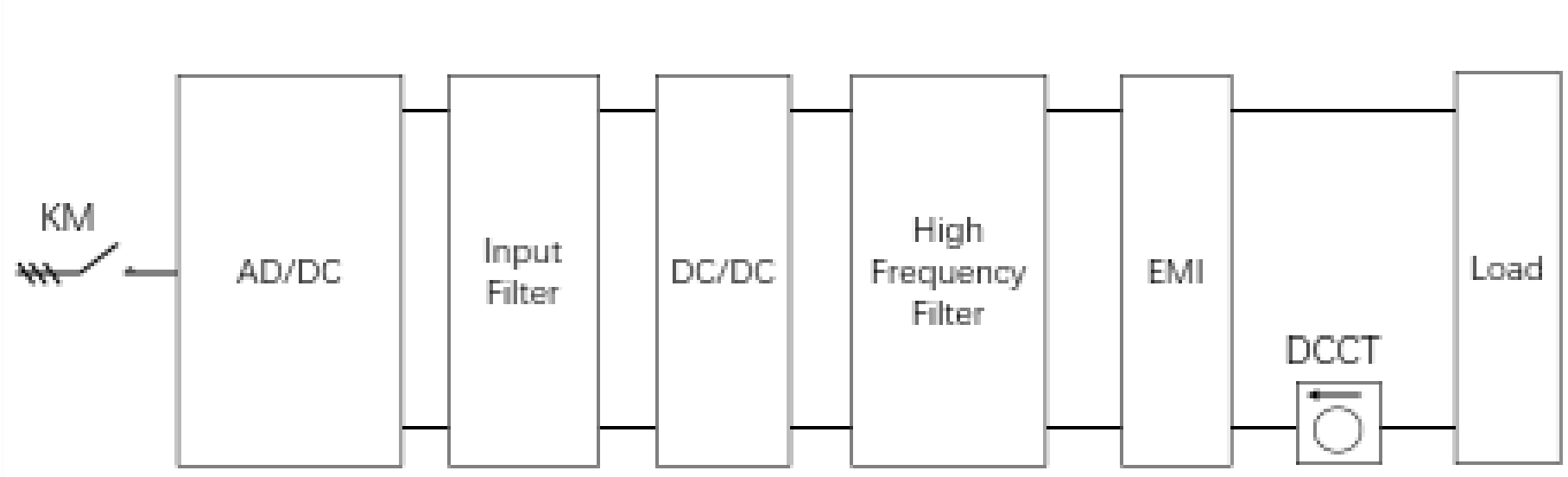

Figure 4.2-2: Main circuit block diagram of the magnet power supply system

The front-end AC/DC module can take several forms, including a phase-shifted full-bridge PWM converter, a single-ended forward converter, or a 50 Hz transformer combined with a three-phase full-bridge rectifier. The rear-end DC/DC module is selected based on the excitation requirements of different types of magnets. For unidirectional power supplies, a Buck chopper topology is used, which is simple and reliable. For bidirectional power supplies, an H-bridge converter composed of four power switches is adopted. By adjusting the conduction time of the switches, both the magnitude and direction of the output current can be continuously and smoothly regulated.

According to the magnet coil parameters and requirements defined by the accelerator physics design and magnet design, the collider rings include 248 dipole magnets, 706 quadrupoles, 192 sextupoles, and 32 damping wiggler magnets (DW), along with a number of corrector magnets. The arc dipoles are powered in series by type, while the dipoles in the interaction region and beam crossing region are powered independently. Each damping wiggler magnet is excited by one of three independently grouped power supplies matched to different current levels. Quadrupole, sextupole, and corrector magnets are powered individually. Excluding corrector supplies (to be defined), the collider rings require a total of 1062 power supplies.

The damping ring includes 88 dipole magnets in four specifications, 102 quadrupoles in three specifications, and 72 sextupoles in two specifications. All dipoles and sextupoles are powered in series, while quadrupoles are powered individually. The total number of required power supplies is approximately 108. Same as the magnet design, the accumulator ring is similar to the damping ring in size; thus, the magnet power supplies are not described.

For the linac, the initial physical design calls for about 241 power supplies. These include 52 for correctors, 28 for solenoid guiding coils, one for the chicane dipole magnet, two for analysis dipoles, and 43 for quadrupoles. In the positron linac, 28 supplies are used for solenoids, 18 for long solenoid focusing coils, one for the analysis dipole, 50 for correctors, 14 for quadrupoles, and four for dipoles.

In the transport lines, which include the electron bypass section, the section between the damping ring and the linacs, and the two sections from the main linac to the collider rings, approximately 364 power supplies are needed. This includes 37 dipole supplies, 283 quadrupole supplies, and 44 corrector supplies.

Power supplies are grouped into three power levels, and those of the same class share similar circuit topologies and control strategies [74]. High-power supplies, such as those for the arc dipoles in the collider rings, the damping ring dipoles, and some of the damping wiggler magnets, typically operate at power levels near 100 kW. These adopt a "12-phase rectification + dual interleaved chopper" topology, consisting of an EMI input filter, circuit breaker, AC contactor, 12-phase transformer, 12-phase rectifier, input filter, DC/DC converter, high-frequency output filter, and current sampling circuit (see Figure 4.2-3).

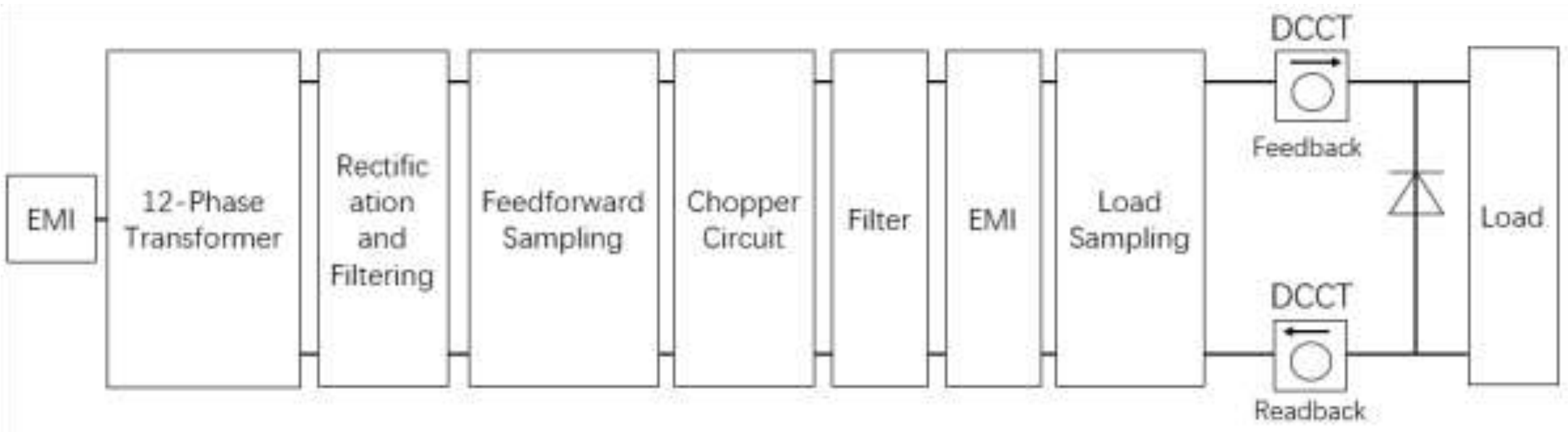

Figure 4.2-3: Main circuit block diagram of a high-power power supply

For power levels up to 300 kW, the main circuit is composed of four power modules in a series-parallel configuration, using multilevel control to meet ripple specifications and enhance efficiency. The system includes two transformers with dual secondary windings, enabling 24-pulse rectification to minimize disturbances to the grid.

For quadrupole and sextupole magnets in both the collider and damping rings, most power supplies fall in the several-kilowatt range. The preferred topology is a digitally controlled phase-shifted full-bridge converter, which supports soft-switching to achieve high efficiency and compact design. Each unit comprises a three-phase rectifier and filter, a DC/DC converter, an output filter, current and voltage sensing circuits, control loops, and driver and protection circuits.

For corrector magnets and some low-power quadrupoles, with outputs under 1 kW, the preferred circuit structure is a switching power supply combined with either a chopper or H-bridge converter. The front end uses standard switch-mode power modules.

The digital controller is the core of high-precision regulation for power supplies, capable of generating ultra-fine PWM signals with a minimum resolution of 78 ps, meeting the 10 ppm current stability requirement [75]. The controller, integrated into the power supply chassis, includes an FPGA-based control board and an ADC/DAC module. It communicates with the accelerator's IOC system via Ethernet or optical fiber, and interfaces with orbit feedback, local diagnostic interfaces, and synchronization signals from other supplies. The controller architecture adopts a modular layout—core daughterboard plus motherboard backplane—to facilitate future upgrades of logic chips and control algorithms [76]. The design concept is illustrated in Figure 4.2-4.

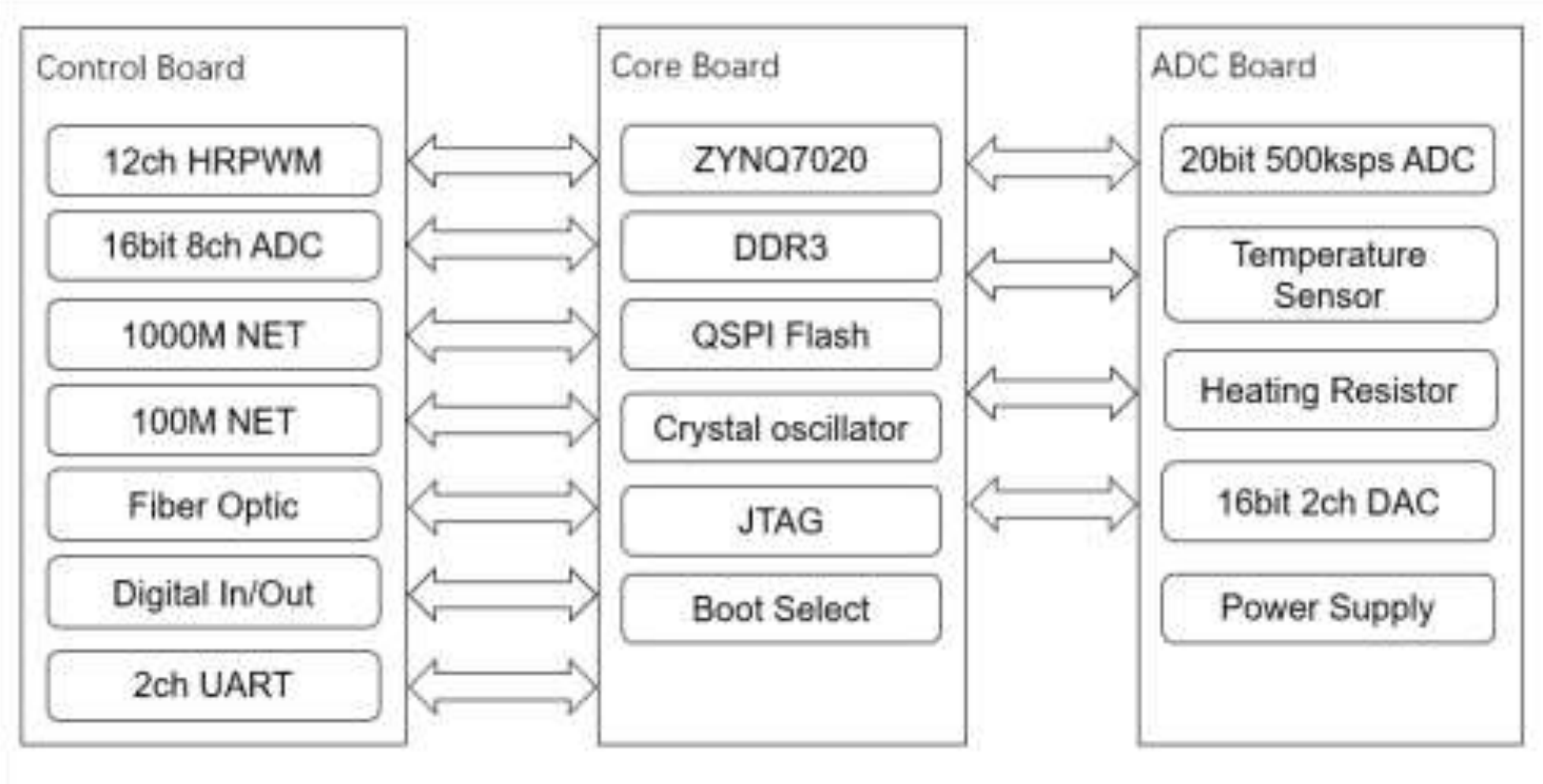

Figure 4.2-4: Schematic of the digital power supply control system architecture

### 4.2.4 Feasibility Analysis

Switch-mode power supplies (SMPS) are widely adopted for magnets in modern accelerators due to their high efficiency, wide operating range, and simple control. Digital regulation has become the mainstream solution for accelerator magnet supplies due to its excellent stability, configurability, and high integration.

The STCF power supply system shares similar design targets with facilities or projects such as Hefei Light Source (HLS) and Hefei Advanced Light Facility (HALF). Long-term R&D and construction experience from those facilities proves the technical feasibility and maturity of this design.

### 4.2.5 Summary

Based on the requirements from accelerator physics and magnet design, we have selected rational parameters and topologies for the magnet power supplies of the STCF accelerator complex. SMPS serves as the foundational architecture, supplemented by digital control, enabling high-precision software tuning of power loop parameters. Modular construction ensures that the system meets stringent performance goals, reliability, flexibility, and maintainability.

## 4.3 Ring RF System

### 4.3.1 Design Requirements and Specifications

The STCF ring RF system is responsible for compensating energy losses in the electron and positron beams in the collider rings and positron damping/accumulation rings, due to synchrotron radiation, vacuum chamber impedance, and other factors. As such, it is a core component affecting beam performance. The design of the RF system focuses on the selection of cavity technology and the definition of associated physical parameters.

Aligned with the 2998.2 MHz frequency of the linacs, the ring RF system operates at 499.7 MHz. In principle, either superconducting or room-temperature RF cavities could be used. However, since the damping/accumulation rings have relatively low RF requirements, the RF system of the collider rings can operate at reduced cavity voltage to also serve these rings—this approach is adopted at SuperKEKB, effectively reducing cost. Thus, separate studies for the RF system of the damping/accumulation rings are not needed.

When the STCF collider rings operate at 3.5 GeV beam energy with 2 A beam current, the synchrotron radiation power reaches 2954 kW. Additional power losses from vacuum impedance, cavity wall dissipation, and the higher-order modes (HOMs) also contribute and must be refined in collaboration with the accelerator physics design. For now, they are not included in the calculation. Therefore, the RF system must deliver 6 MV accelerating voltage and approximately 3 MW beam power to each of the electron and positron beams in the collider rings.

To reduce the number of RF cavities—and hence the fabrication cost—each cavity must provide an accelerating voltage larger than 0.5 MV and RF power to the beam larger than 250 kW. The cavity design targets are summarized in Table 4.3-1. A minimum of 12 cavities per collider ring is required to meet operational needs, and 2 or 3 more cavities are needed for redundancy. Such high beam power and current impose stringent demands on the RF input couplers and HOM dampers. Besides a dedicated RF power source, each cavity is also equipped with a low-level RF (LLRF) control system that stabilizes its frequency, amplitude, and phase against beam loading and environmental disturbances. Accurate beam control and synchronization are essential for the machine operation stability and the experimental data quality. The phase drift in reference signals can directly affect the beam energy stability. Therefore, a common reference signal must be distributed to all the cavities to ensure inter-cavity phase synchronization. The design requirements for the LLRF system are given in Table 4.3-2.

Table 4.3-1: RF Cavity Design Parameters for the Collider Rings

| Object | Target Specifications |
| --- | --- |
| 499.7 MHz Cavity | Accelerating voltage per cavity: ≥0.5 MV<br>Input coupler power: ≥300 kW (CW) |

Table 4.3-2: Low-Level RF System Design Parameters

| Parameters | Target Specifications |
| --- | --- |
| LLRF amplitude stability | 0.05% |
| LLRF phase stability | 0.05° |
| Reference signal frequency | 499.7 MHz |
| Integrated phase noise (jitter) | ⩽30 fs (rms) |
| Long-term synchronization | ⩽50 fs over 24 hours (rms) |

### 4.3.2 Key Technologies and Technical Approach

Both superconducting and room-temperature RF cavities are used in modern electron-positron colliders. Superconducting cavities offer higher accelerating gradients and better HOM damping, but require complex cryogenic systems, have higher maintenance costs, and are sensitive to single-cavity failures. In contrast, room-temperature cavities offer better operational stability, lower cost, and simpler maintenance, while still capable of sufficient HOM suppression.

Given the low-energy, high-luminosity nature of the STCF collider rings—with high beam current but modest cavity voltage requirements—a superconducting cavity would be over-specified and would be limited by coupler power constraints. Taking cost, operational reliability, and maintainability into account, STCF adopts a room-temperature RF cavity solution for both the collider rings and the positron damping/accumulation rings.

To compensate for beam loading from high current, all cavities must operate with intentional detuning. As shown in Table 4.3-3, for the four types of available room-temperature cavities, the single-cavity detuning frequencies of the ARES-type cavities used at SuperKEKB and the TM020-type cavities used at NanoTerasu are much smaller than the revolution frequency, making them less prone to coupled-bunch instabilities from the fundamental mode. The ARES type cavity uses a complex triple-cell structure (storage + coupling + accelerating cells), making it difficult to manufacture and expensive. In contrast, the TM020 cavity developed by KEK and RIKEN offers strong HOM damping, compact structure, and high cavity voltage, while achieving a total R/Q roughly half that of conventional cavities, thus significantly mitigating coupled-bunch instabilities. This cavity has been successfully used at the NanoTerasu light source [77, 78], achieving stable operation at ~250 kW per cavity.

Therefore, the TM020 cavity is selected for the RF system of the STCF collider rings. At 3.5 GeV operation, each collider ring will require at least 12 cavities to supply 3 MW of beam power and 6 MV accelerating voltage. Including operational margin, 15 cavities per ring will be installed.

Table 4.3-3: Comparison of Room-temperature RF Cavity Schemes for STCF collider rings (Single Ring)

| Cavity Type | ARES (Japan) | PEP-II (USA) | BESSY-II (Europe) | TM020 (Japan) |
|---|---|---|---|---|
| Beam power [kW] | 3000 | 3000 | 3000 | 3000 |
| Total voltage [kV] | 6000 | 6000 | 6000 | 6000 |
| Number of cavities | 10 | 15 | 30 | 12 |
| Voltage per cavity | 600 | 400 | 200 | 500 |
| Wall loss per cavity [kW] | 209 | 23 | 6 | 49.7 |
| Beam power per cavity [kW] | 300 | 200 | 100 | 250 |
| Input power per cavity [kW] | 509 | 223 | 106 | 299.7 |
| Detuning per cavity [kHz] | 12 | 283 | 557 | 75 |

The core challenges of the LLRF system include amplitude and phase stabilization of the cavity field, frequency control, and mitigation of longitudinal coupled-bunch instabilities. STCF adopts a direct sampling architecture with non-IQ demodulation algorithms, which enhance noise rejection (e.g., temperature drift, harmonic noise) while ensuring system stability [79].

The reference distribution system must deliver phase-synchronized, low-jitter signals to all the RF cavities. STCF chooses a continuous-wave optical carrier scheme, in which the reference signal from the RF master oscillator (RMO) is transmitted along the cable and partially reflected at the far end. The returning signal is compared with the original to generate a feedback loop, locking the receive-end phase.

### 4.3.3 Design Scheme and System Configuration

The RF system of the STCF collider rings consists of room-temperature RF cavities, solid-state power amplifiers, digital LLRF control systems, interlock and protection systems, and a reference distribution network. The room-temperature RF cavities supply the required accelerating voltage to the beam, compensating for radiation and other energy losses, while also ensuring a low higher-order mode (HOM) impedance environment. A 500 MHz solid-state RF amplifier with a maximum output power of 250 kW has been jointly developed by Chengdu Kaiteng Sifang Digital Broadcast Equipment Co., Ltd., the National Synchrotron Radiation Laboratory, and the Institute of High Energy Physics, CAS. With minor upgrades, this amplifier can meet the power demands of the STCF RF system. The digital LLRF system ensures amplitude and phase stability of the cavity field. The interlock protection system guarantees equipment safety, while the reference distribution network ensures phase synchronization among all system nodes.

#### *4.3.3.1 Design of the Room-temperature RF Cavity*

The TM020-mode room-temperature cavity with HOM damping operates at 499.7 MHz. A higher R/Q value often leads to a larger cavity detuning frequency for compensating beam loading, increasing the risk of coupled-bunch instabilities by the fundamental mode. Considering this, the optimized R/Q value is set to 77.2. The cavity adopts a nose-cone-shaped geometry. The electric and magnetic field distributions are shown in Fig. 4.3-1, and the optimized RF parameters are listed in Table 4.3-4.

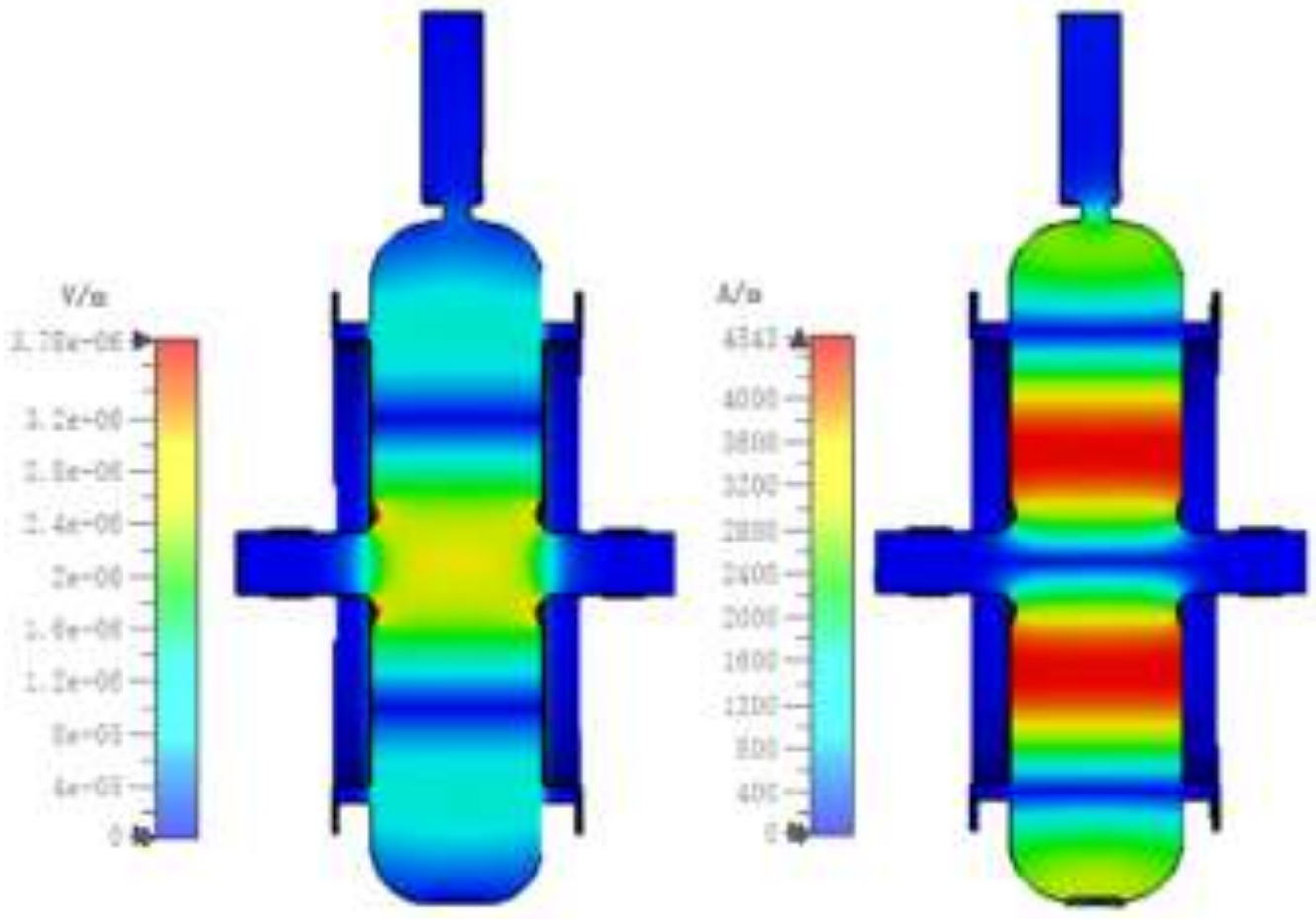

Figure 4.3-1: Electric field (left) and magnetic field (right) distributions in the TM020 cavity.

Table 4.3-4: RF Parameters of the TM020 Room-temperature Cavity

| Parameter | Value |
|---|---|
| Operating frequency [MHz] | 499.7 |
| Shunt impedance [MΩ] | 5.03 |
| Unloaded quality factor | 65110 |
| Leakage power of the accelerating mode to the HOM damping slot / total power | 1.38% |
| $E_p/E_{acc}$ | 2.12 |
| $B_p/E_{acc}$ [mA/V] | 2.55 |

The cavity is equipped with elliptical coupling slots and ferrite dampers located at the magnetic field radial node. This ensures that the fundamental mode field remains confined, while unwanted HOMs can propagate into the damping system. Simulation results confirm that all the longitudinal monopole and transverse dipole modes are effectively suppressed under 2 A beam current, as shown in Fig. 4.3-2, meeting the coupled-bunch instability (CBI) thresholds required by STCF.

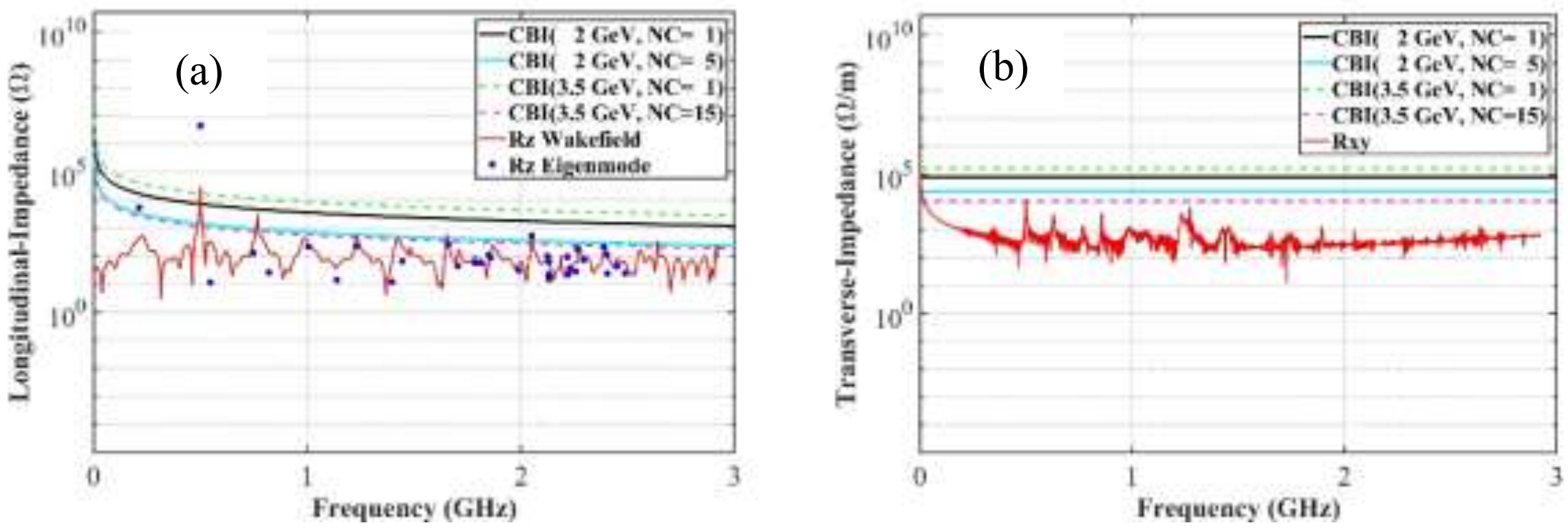

Figure 4.3-2: Impedance spectra for (a) longitudinal modes and (b) transverse modes (NC: number of cavities).

The coupling section adopts a half-height WR1800 rectangular waveguide. The coupler features an asymmetric, off-center rectangular port on the broad side of the waveguide. An adjustable tuner near the coupling port enables real-time tuning of the coupling strength by modifying the electromagnetic field distribution on both sides of the port. As shown in Fig. 4.3-3, with tuner insertion depth varying from 3 mm to 63 mm, the coupling coefficient ranges from 0.997 to 8.4, satisfying operational needs.

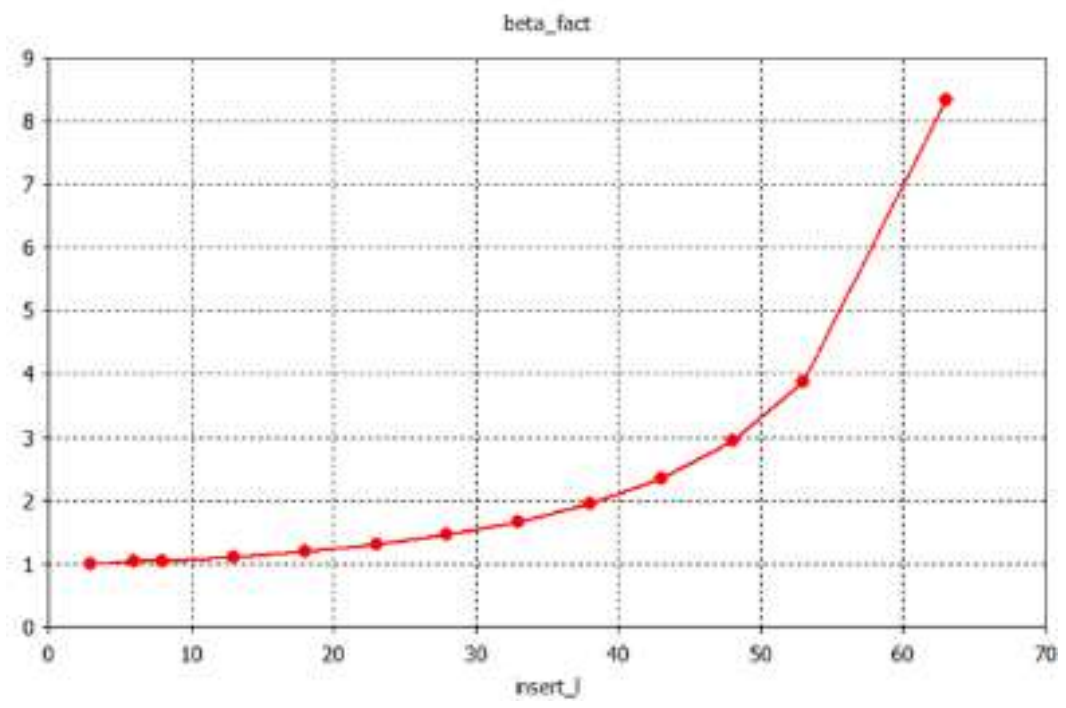

Figure 4.3-3: Coupling coefficient vs. tuner insertion depth.

A waveguide window isolates the vacuum inside the cavity from atmospheric pressure in the RF transmission system. The window is made of 99.5% high-purity alumina ceramic and matches the half-height WR1800 waveguide via the optimization of the length, width, and thickness of the window. The transition between the main WR1800 waveguide and the coupler's half-height WR1800 is realized via a step transformer. The final assembly model of the input coupler is shown in Fig. 4.3-4.

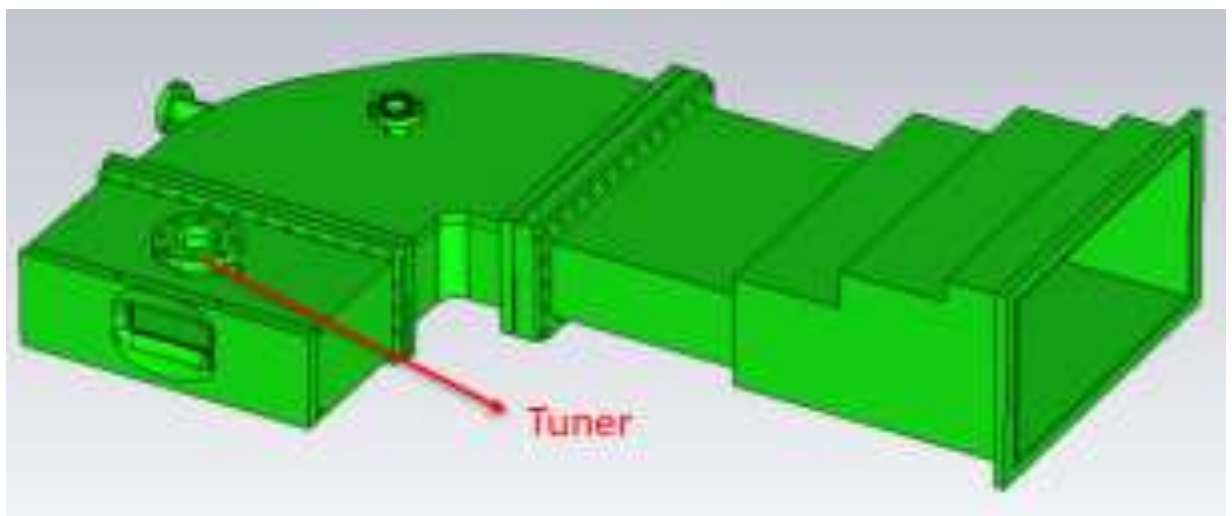

Figure 4.3-4: Assembly diagram of the input coupler.

To ensure stable operation of the high-power input coupler, the issues with thermal load, secondary electron multipacting, and stress must be evaluated. At 250 kW beam power per cavity, the cavity wall loss is 49.7 kW. Therefore, the input coupler is designed to stably operate at a CW power larger than 300 kW. For the waveguide couplers, it is complex and critical to suppress the secondary electron multipacting from the structure in the design phase. Simulations on the secondary electron yield (SEY) at the coupling port and waveguide window were performed. The results in Fig. 4.3-5 show that no structural multipacting occurs under the CW operation of 300 kW.

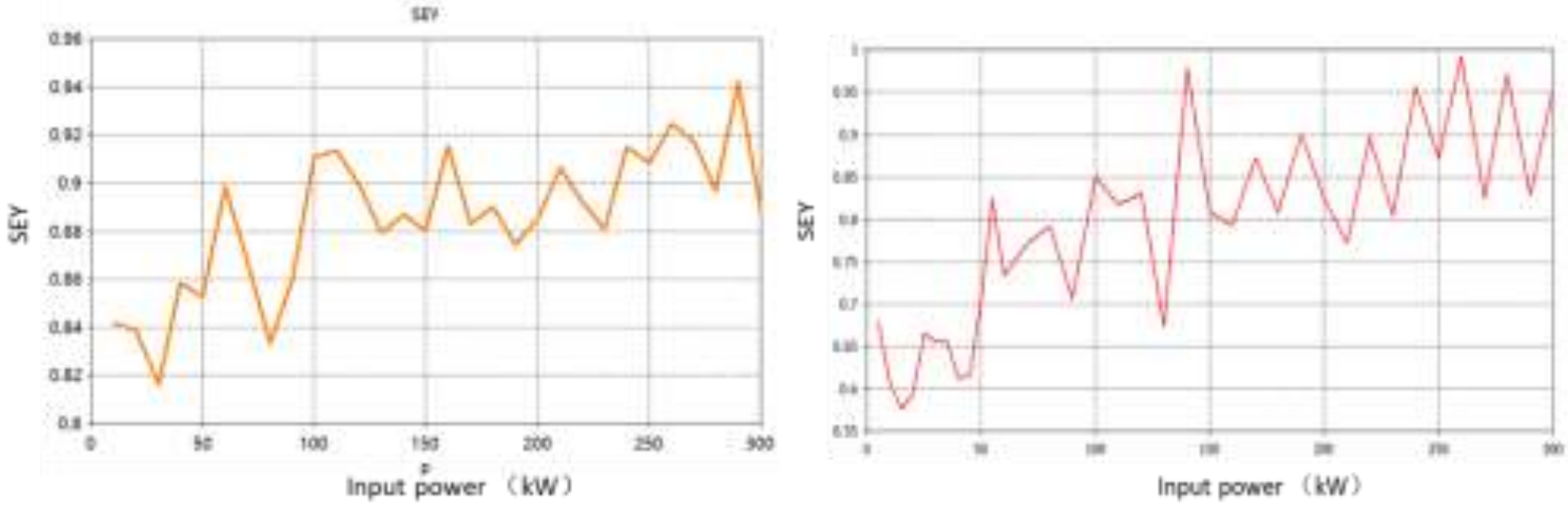

Figure 4.3-5: SEY vs. power for the coupler port (left) and waveguide window (right).

At a CW power of 300 kW for the RF forward pass through the coupler, the RF heating is significant. Therefore, active water cooling is applied to the coupling port, tuner, and waveguide window, together with air convection provided in the environment. The ceramic window of high-purity alumina helps resist the heat load, but water-cooling around its frame is still needed. Under 300 kW operation, the maximum temperature reaches 49.15°C, and both stress and deformation are within acceptable limits, as shown in Fig. 4.3-6.

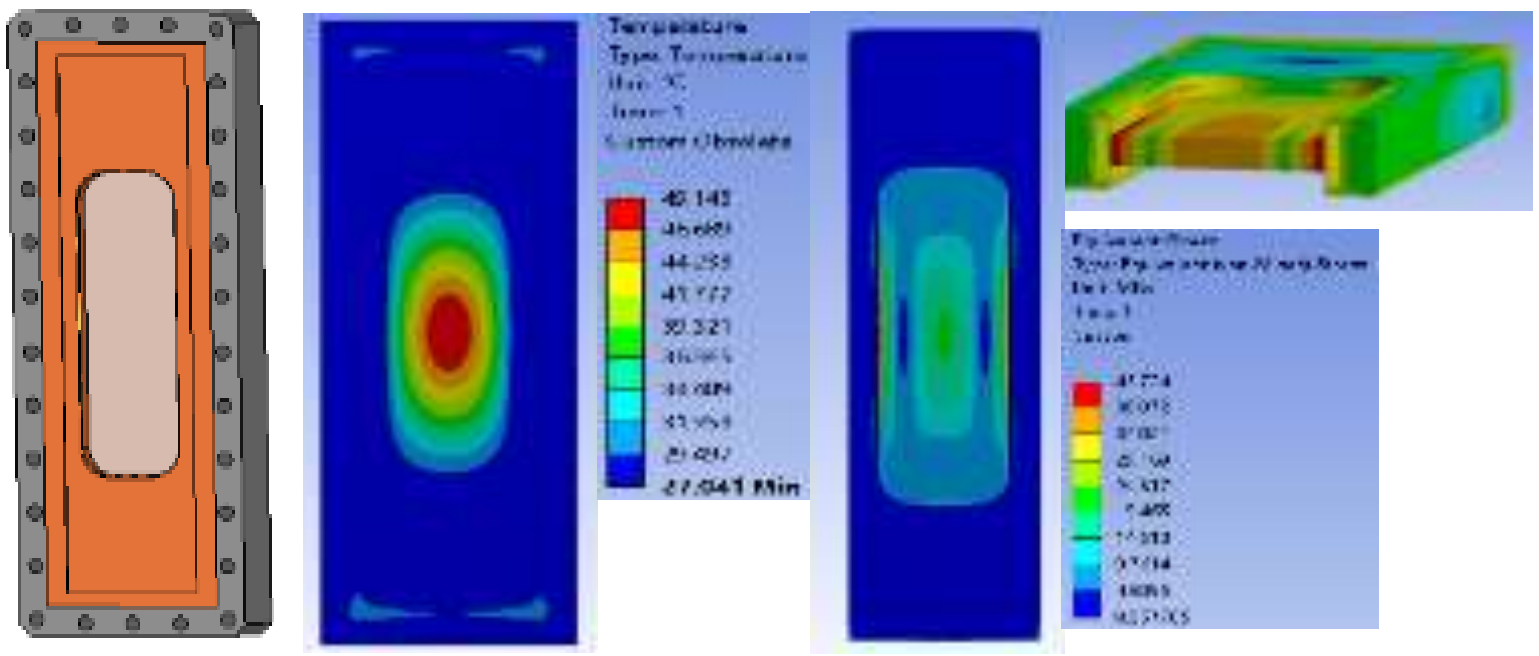

Figure 4.3-6: Mechanical model, thermal load, and stress distribution at the waveguide window

Combining the cavity and coupler designs, the full 3D model of the 499.7 MHz TM020 room-temperature RF cavity is shown in Fig. 4.3-7. Each cavity provides a cavity voltage of 500 kV and an RF power to beam of 250 kW (a total RF power of about 300 kW into the cavity). At least 12 such cavities per collider ring are required to meet performance goals. To ensure redundancy and prevent sustained operation at maximum capacity, additional cavities will be installed. In addition, designs with two input couplers per cavity are also under development to further reduce power demands per coupler.

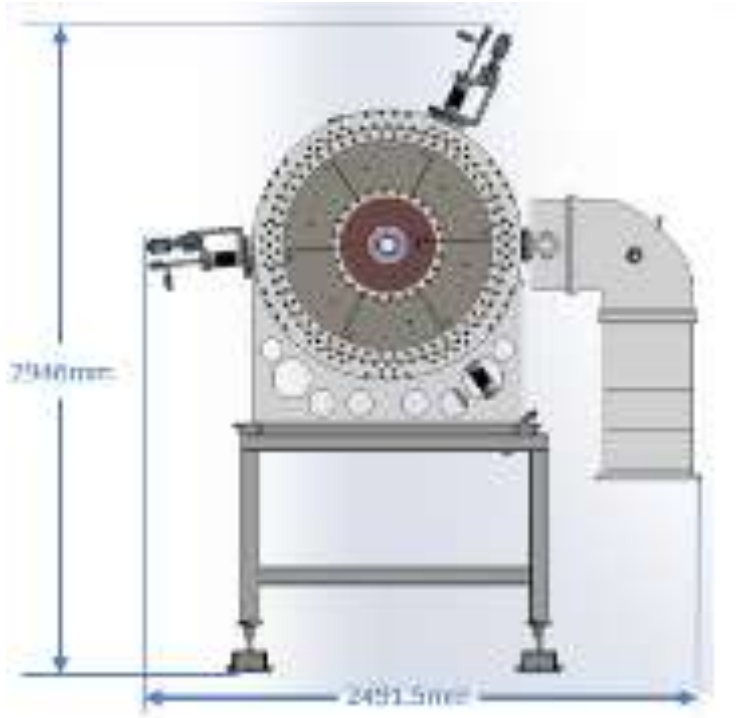

Figure 4.3-7: 3D model of the 499.7 MHz TM020 room-temperature RF cavity.

### *4.3.3.2 Digital Low-Level RF (LLRF) System*

The STCF consists of two separate collider rings for electrons and positrons, with each ring expected to house at least 12 room-temperature cavities. Each cavity requires its own LLRF control system. The overall design of the digital LLRF system is shown in Figure 4.3-8. We adopt a direct RF sampling hardware architecture, in which the 499.7 MHz signal is directly sampled by the ADC.

The LLRF system collects the cavity input signal, reflected signal, cavity field signal, and beam position monitor (BPM) signals. The reflected signal is used for monitoring cavity status. The cavity input and field signals are processed in real-time by an internal FPGA logic unit. One output controls the solid-state amplifier and the klystron to stabilize the RF field; another controls the tuner mechanism to ensure the frequency stability. Additionally, a third output is used to modulate the reference signal from the frequency synthesis system to suppress longitudinal coupled-bunch instabilities, particularly in the 0 mode.

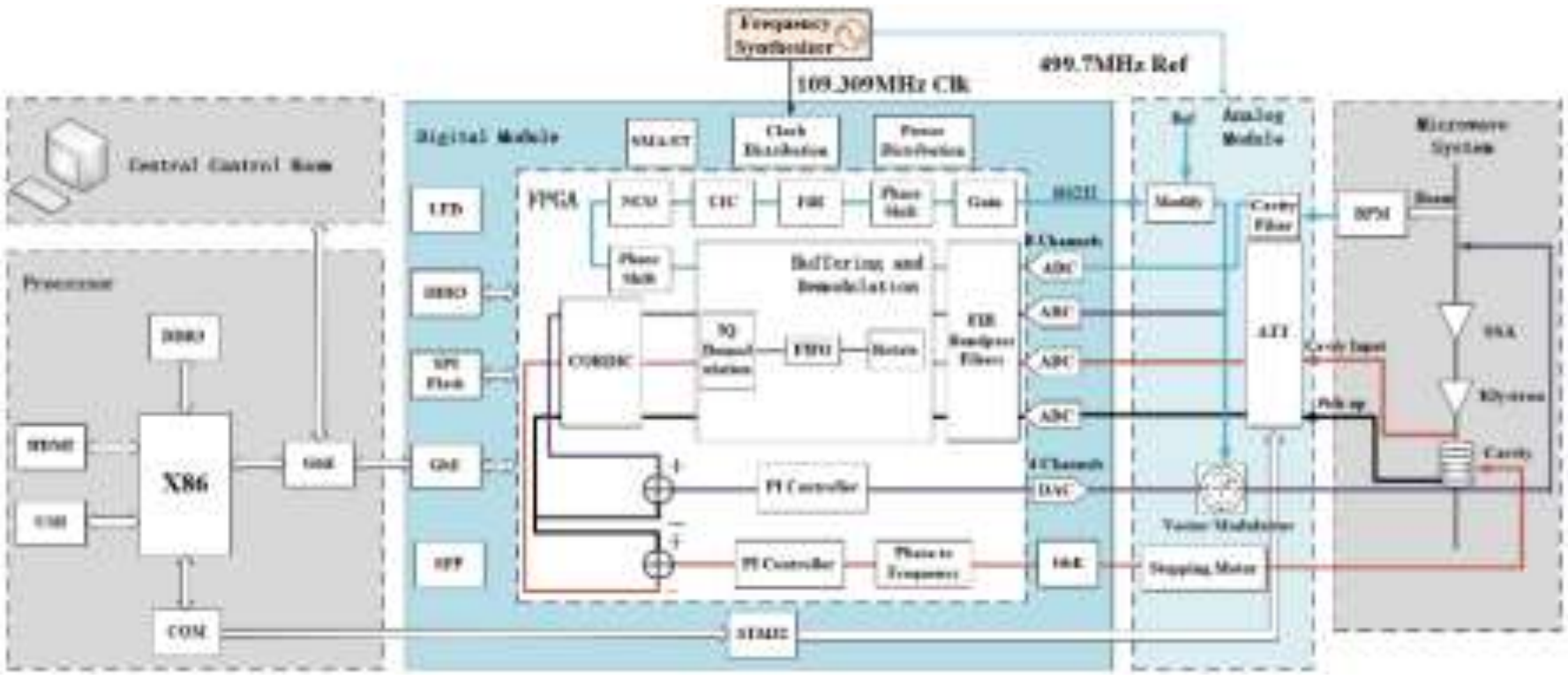

Figure 4.3-8: Design block diagram of the STCF ring's digital low-level RF system for room-temperature cavities.

#### *4.3.3.3 Interlock Protection System*

The interlock protection system ensures rapid response to equipment failures, providing protection to prevent damage to RF components. Implemented using FPGA technology, it offers fast response and high reliability. The data acquisition and interlock control software is developed using the Experimental Physics and Industrial Control System (EPICS), with data storage handled by Archiver Appliance.

To promote modularity and distributed control, the system is embedded in a Linux platform. EPICS is fully integrated, and automatic control is realized using the EPICS sequencer module. Specific applications include data acquisition, real-time analysis, and interlock event filtering.

#### *4.3.3.4 Reference Distribution System*

As shown in Figure 4.3-9, the synchronization system consists of a reference master oscillator (RMO), frequency synthesizer, synchronization modulator, transmitter, receiver, microwave phase compensation system, femtosecond phase comparator, and synchronization calibration controller.

The RMO generates the reference signal, which is amplified and modulated. Using a continuous-wave laser carrier, the reference signal is modulated onto optical fibers for distribution. A portion of the optical signal is reflected to the transmitter to enable phase comparison and compensation via a femtosecond-level phase comparator and an adjustable optical delay line.

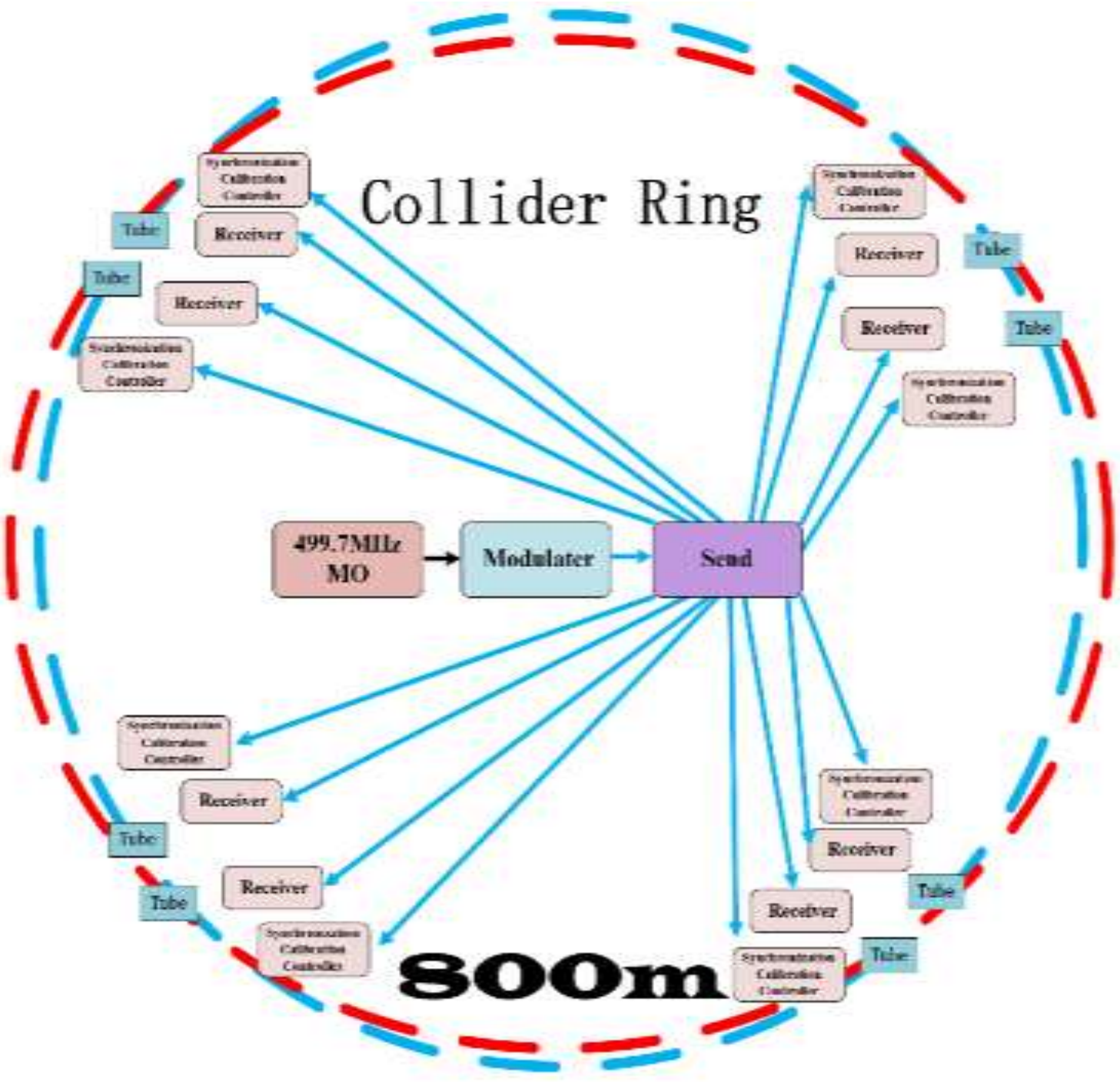

Figure 4.3-9: Schematic of the storage ring phase reference system.

### 4.3.4 Feasibility Analysis

In summary, although the TM020-mode room-temperature cavity offers promising technical advantages, it is a relatively new cavity type with limited operational experience worldwide and presents challenges in precision mechanical fabrication. STCF is actively pursuing technical development through prototyping and manufacturing process studies to ensure domestic production and technological independence.

To counteract beam loading effects, each cavity is detuned by 75 kHz. HOMs are deeply suppressed, thus avoiding coupled-bunch instabilities. The National Synchrotron Radiation Laboratory (NSRL) has extensive experience in the development and operation of digital LLRF systems. Independently developed LLRF systems have been running reliably at the NSRL microwave test facility and have been integrated successfully into the SSRF RF system in collaboration with the Shanghai Synchrotron Radiation Facility.

### 4.3.5 Summary

The conceptual design of the ring RF system—including the RF cavities, power sources, and LLRF systems—has been completed. Theoretical and simulation studies have been conducted for the key technologies. Some components are under prototype development and fabrication testing. Most devices will adopt proven, cost-effective technologies and products to reduce risk and ensure a successful implementation.

## 4.4 Injection and Extraction System

The STCF facility includes multiple circular accelerators: two collider rings (for electrons and positrons) and one positron damping ring (or accumulation ring). Each of these rings requires dedicated injection and extraction systems, which connect to the beam transport lines to ensure efficient injection and extraction of electron and positron beams.

### 4.4.1 Design Objectives

The overall objective is to meet the requirements of the injection physics design for each ring and to develop magnet systems that guarantee efficient injection and extraction of both electron and positron beams.

For the collider rings: The current design accommodates two different injection schemes—off-axis injection and swap-out injection. These two schemes have different requirements and technical challenges. For the off-axis injection scheme, the focus is on the structural design of nonlinear kicker magnets and the optimization of their magnetic field distribution. For the swap-out injection scheme, the emphasis is on the electromagnetic structure design of stripline-type ultra-fast kickers and prototype development. Regarding beam abort and single-bunch swap-out extraction, for full-bunch train extraction, kicker magnets with a long pulse-width are to be developed. For single-bunch extraction, the same ultra-fast kickers as used for injection are employed here. In addition, the design optimization of the thin pulsed septum magnet is required to eliminate perturbations to the circulating beams.

For the damping ring: The ring contains 5 bunches spaced 100 ns apart. An on-axis single-turn injection scheme is adopted to reduce the requirements on dynamic aperture. Ferrite-loaded transmission-line-type kicker magnets are employed to improve excitation efficiency, with a focus on achieving fast field rise times and impedance matching to suppress reflections. Extraction is single-bunch based and technically similar to injection.

For the accumulation ring: This ring stores 4 bunches with 130 ns spacing and uses local orbit bumps to allow for multiple injections to accumulate positron charge. The kickers used are similar to those in the damping ring and are relatively mature in technology. The extraction is also single-bunch, with technical requirements akin to injection.

### 4.4.2 Design Requirements and Key Technical Specifications

- ***Septum Magnets***

Designs are being developed for two types of septum magnets—eddy-current focused and reverse-field types. Key performance criteria include a septum thickness of ≤2 mm and residual magnetic field on the circulating beam orbit of <0.1% (with respect to the main field). These parameters may be adjusted based on physics design needs.

- ***Nonlinear Kickers***

Three types of nonlinear kickers are under design: air-core coil type, pseudo-octupole type, and shielded type. Key criteria include a central magnetic field amplitude of ≤5% (of the injection field) and a pulse width of ≤3 μs, adjustable per the physics design.

- ***Stripline-Type Ultra-Fast Kickers***

These are being designed and prototyped. Key criteria include magnetic field rise/fall time ≤2 ns and a total pulse width ≤6 ns, tunable to match injection requirements.

The key components of the injection and extraction systems for the three types of rings are the kickers and septum magnets. Their core technical requirements are summarized in Table 4.4-1. Other structural parameters will be adjusted according to specific design details.

Table 4.4-1: Technical Requirements for Key Components in Injection Systems

| | Kickers | Septum Magnets |
|---|---|---|
| Collider Rings | Off-axis: Central field ≤5%, pulse width ≤3 μs<br>Swap-out: Rise time ≤2 ns, pulse width ≤6 ns | Septum thickness ≤2 mm, residual field at orbit <0.1% |
| Damping Ring | Single-turn injection: Rise time ≤ 90 ns, pulse ≤190 ns | Same as above |
| Accumulation Ring | Off-axis injection: Rise time ≤120 ns, pulse ≤250 ns | Same as above |

### 4.4.3 Key Technologies and Technical Approaches

- ***Septum Magnets***

The main function of septum magnets is to bring the injected beam path as close as possible to the stored beam trajectory. This minimizes injection beam orbit disturbance and reduces beam loss in off-axis injection schemes. In on-axis injections, it helps reduce the required magnetic field strength of the kickers.

Technical approach at STCF: Based on the injection physics requirements, theoretical electromagnetic structure analysis and parameter optimization are first performed. Numerical simulation using the OPERA code validates and further optimizes the design. A comparative performance analysis of the two septum magnet designs is conducted under different application scenarios, and the final scheme is selected based on the STCF-specific injection/extraction needs.

- ***Nonlinear Kickers***

The core feature of a nonlinear kicker is a zero magnetic field at the center to ensure transparency to the stored beam, while producing a sufficient field in the offset region to meet the off-axis injection needs.

Technical approach: Injection requirements are translated into magnetic field profile specifications. Three structural types are considered, each analyzed electromagnetically and parametrically. OPERA-based numerical simulations are used for validation and optimization. A comparative study of field profiles is used to determine the final design according to injection/extraction physics.

- ***Stripline-Type Ultra-Fast Kickers***

The swap-out injection scheme requires ultra-fast, precisely timed electromagnetic pulses within the narrow interval between adjacent bunches to kick the injected bunch into the pre-emptied bucket of the stored beam without perturbing the neighboring bunches.

Technical approach: Electromagnetic parameters are analyzed theoretically, and the structure is designed according to the field waveform requirement from the physics design. CST simulations are used to optimize the waveform shape (pulse width and flatness), reduce impedance mismatch with adjacent components, and ensure the desired time structure. Prototyping and iterative testing are conducted to finalize performance.

### 4.4.4 Design Scheme and System Configuration

- ***Septum Magnets***

**Eddy-Current-Type Septum Magnet**: A simulation model of the eddy-current-type septum magnet [80, 81] was constructed using OPERA, as shown in Fig. 4.4-1(a). The physical length of the magnet is 650 mm. The excitation current is a half-sine waveform with a peak of 6270 A and a bottom width of 60 μs. The longitudinal distribution of the main magnetic field is presented in Fig. 4.4-1(b). Within the good field region, the peak field reaches 9832 Gs, the integrated field is 0.66686 T·m, and the leakage field at the circulating beam center is less than 0.01% of the main field.

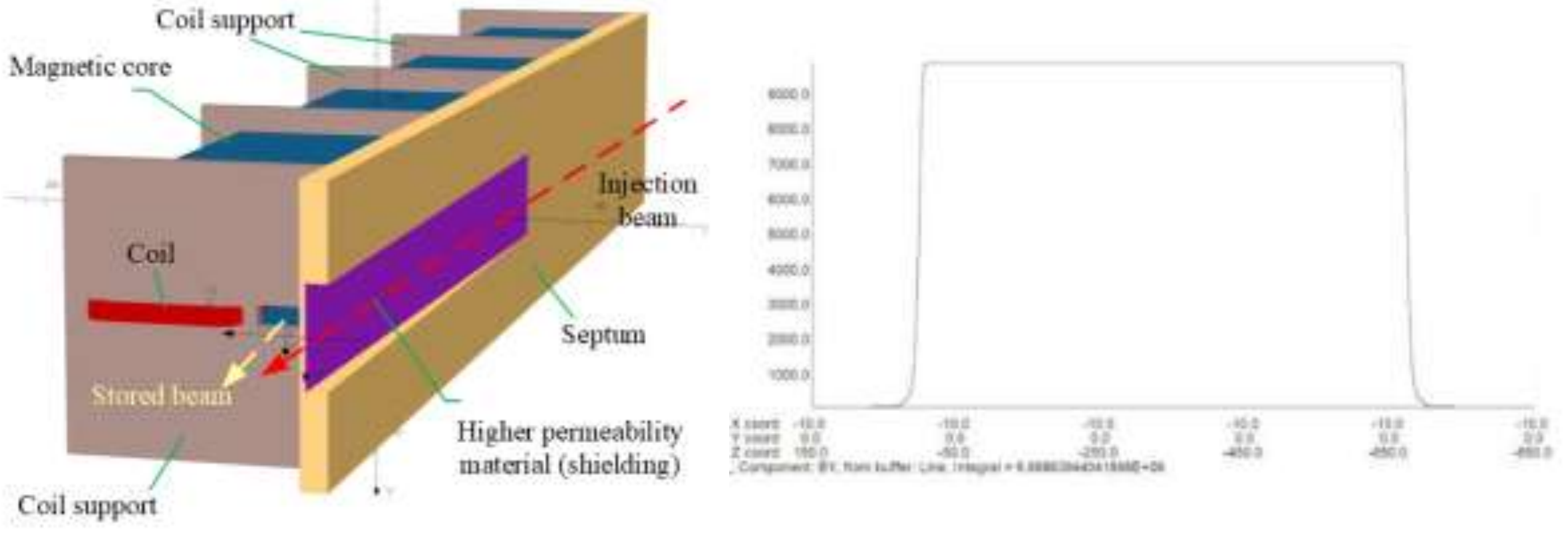

(a) (b)

Figure 4.4-1: Simulation model and magnetic field distribution of the eddy-current-type septum magnet: (a) simulation model; (b) longitudinal magnetic field distribution.

**Opposite-Field-Type Septum Magnet**: This design effectively comprises three dipole magnets [82]. The upstream and downstream compensating magnets are each 0.3 m in longitudinal length, and the central main septum magnet is 0.6 m long, as illustrated in Figs. 4.4-2(a)(b). The magnetic field distribution is shown in Fig. 4.4-2(c): the integrated magnetic field on the circulating beam side is canceled, while the integrated field on the injection beam side is doubled. Additionally, the electromagnetic forces on the septum plates cancel each other, reducing the structural design complexity of the coils.

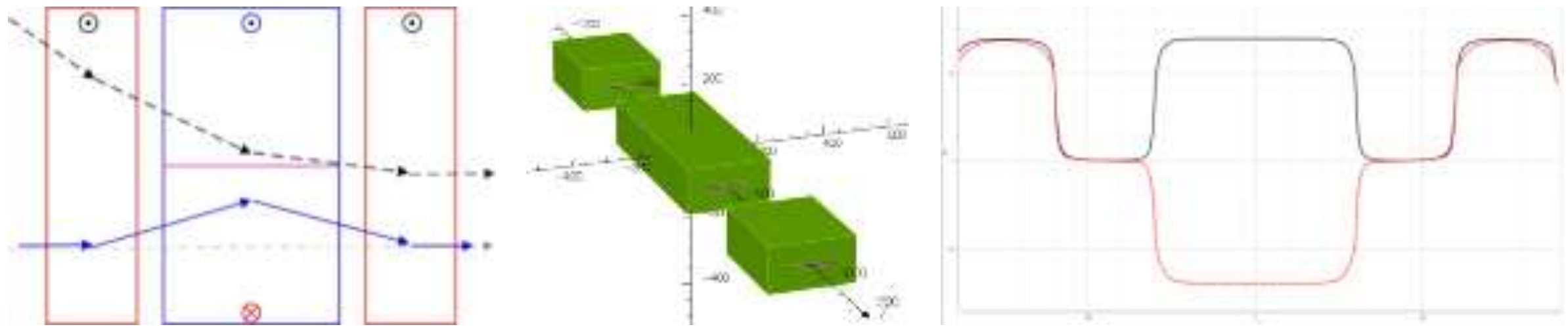

（a）　（b）　（c）

Figure 4.4-2: Simulation model and magnetic field distribution of the reverse-field septum magnet: (a) schematic; (b) simulation model; (c) longitudinal magnetic field distribution.

- ***Nonlinear Kicker Magnets***

**Air-Core Coil Type Nonlinear Kicker**: As shown in Fig. 4.4-3, the required nonlinear magnetic field distribution is formed by adjusting the relative positions of eight current-carrying conductors [83-85]. Both 2D and 3D simulation models were built in OPERA, with results shown in Fig. 4.4-3(c). At the injection point x = -5 mm, the field magnitude is 297.20 Gs. In the zero-field region (x = ±0.5 mm), the field is less than 0.28% of that at the injection point. The field distribution can be controlled by adjusting the conductor positions and excitation currents based on injection physics requirements.

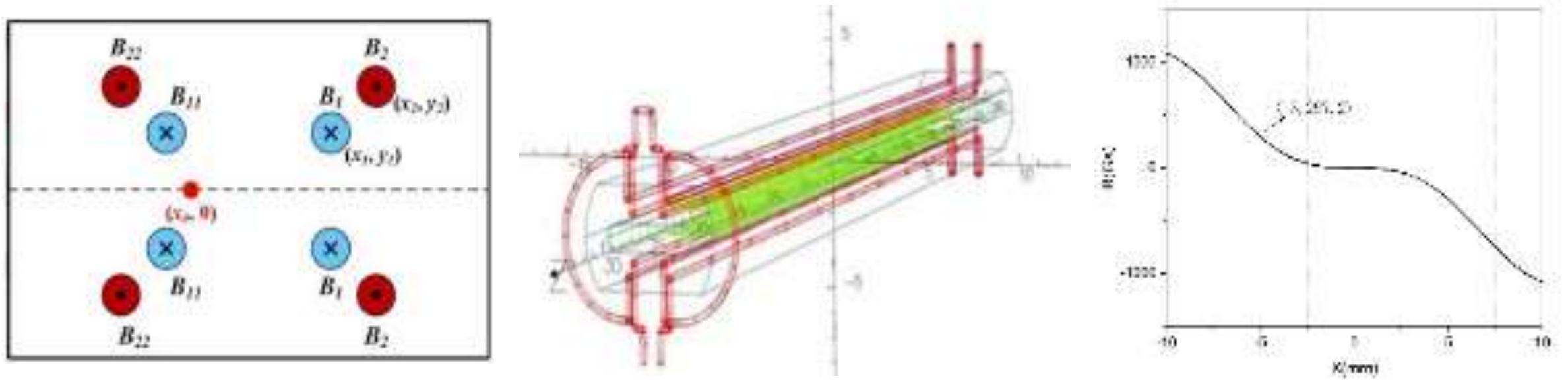

（a）　（b）　（c）

Figure 4.4-3: Air-core coil type nonlinear kicker: (a) principle; (b) structure; (c) integrated magnetic field distribution.

The engineering design for the air-core coil type nonlinear kicker has also been initiated. The design drawings are shown in Fig. 4.4-4. The supporting power supply is scheduled for testing by the end of April, and prototype manufacturing has been planned.

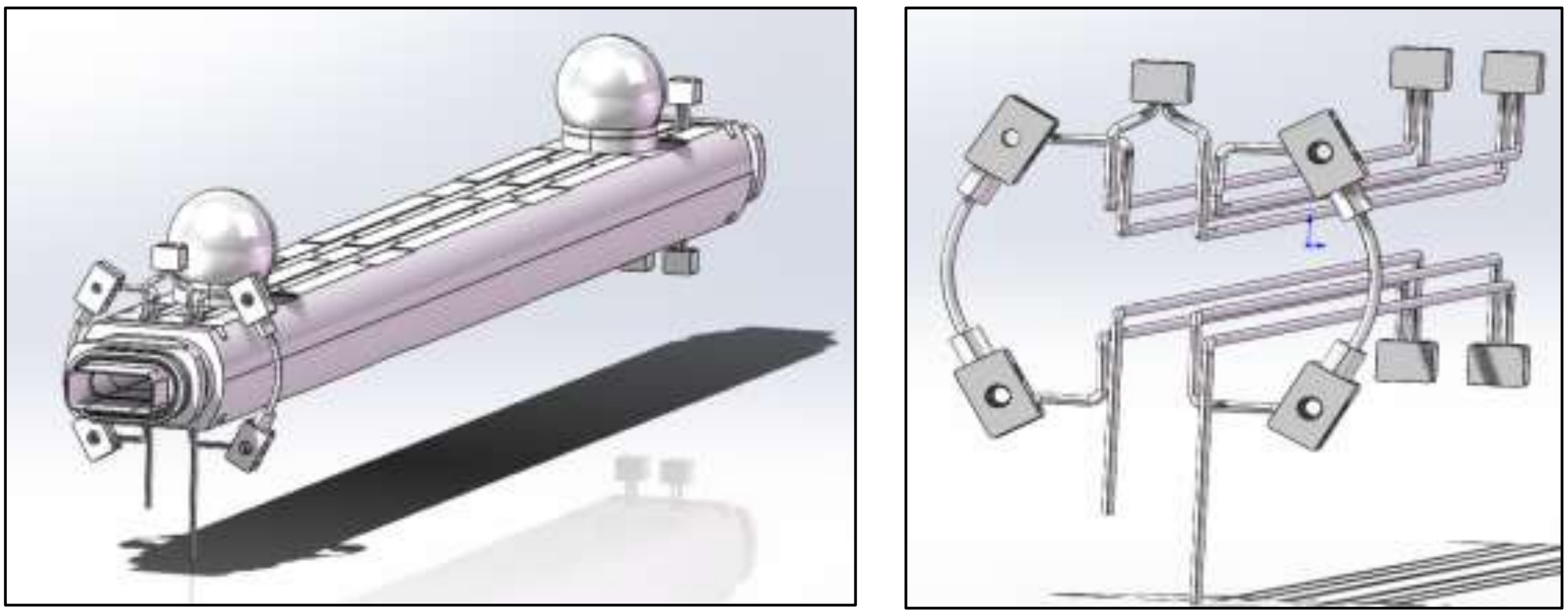

(a) (b)

Figure 4.4-4: Engineering design of the air-core coil type nonlinear kicker: (a) complete ceramic chamber structure; (b) conductor structure.

**Shielded-Type Nonlinear Kicker**: The structure of the shielded-type nonlinear kicker is shown in Fig. 4.4-5(a). The magnet consists of two C-type ferrite cores, separated by eddy-current plates to form opposing magnetic fields. Two copper arc-shaped shielding plates are placed in the magnet center to suppress the central field via induced eddy currents, creating a flat zero-field region. The peak field distribution at a specific time simulated in OPERA is shown in Fig. 4.4-5(b). A key advantage of this design is its uniform field distribution at the injection point.

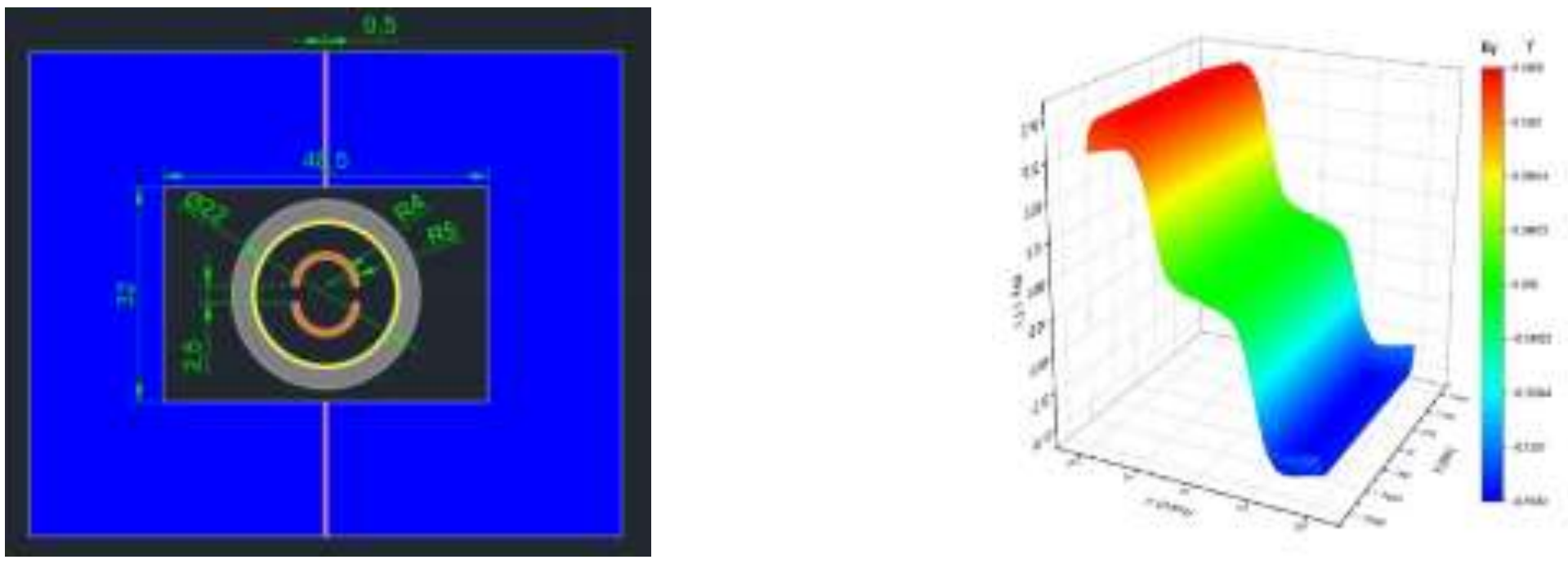

（a） （b）

Figure 4.4-5: Model and field distribution of the shielded type nonlinear kicker: (a) simulation model; (b) magnetic field distribution.

- ***Stripline-Type Kicker Magnet***

To meet the stringent time-structure requirements of pulsed electromagnetic fields for swap-out injection, the electrode structure of the stripline kicker magnet is optimized to match the impedance of the pulsed transmission cables, eliminating waveform distortion from distributed parameters [86]. Fig. 4.4-6(a) shows the simulation model of the kicker. Fig. 4.4-6(b) is a cross-sectional view of the electrode and vacuum chamber: the yellow region indicates the copper electrodes, which, together with the chamber, form a coaxial structure (odd-mode impedance of 50 Ω), achieving impedance matching with the pulsed cables and improving response time and pulse quality. The effective electrode length is 300 mm, and the gap between electrodes is 12 mm. Fig. 4.4-6(c) shows the tapered end transition structure, designed to reduce beam coupling impedance.

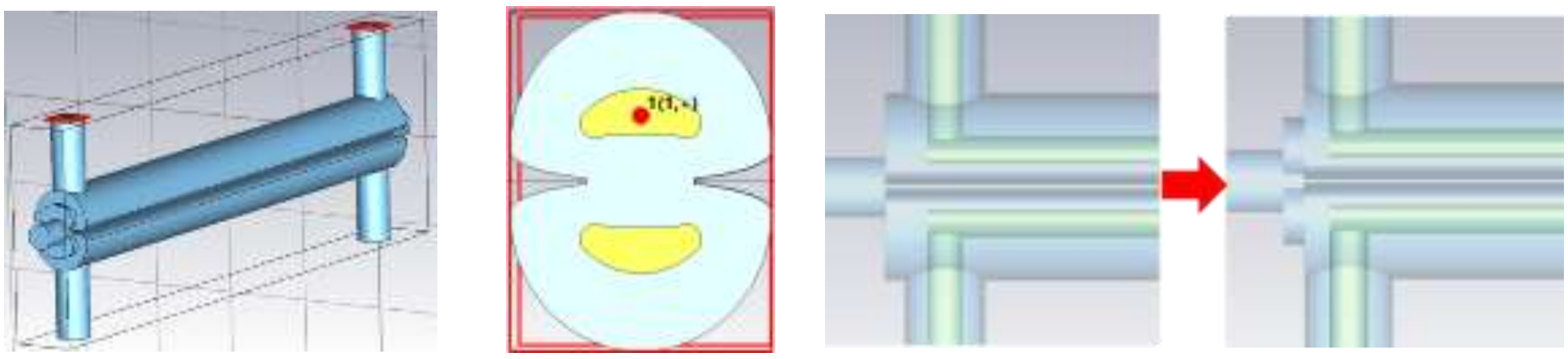

（a）　　　　（b）　　　　（c）

Figure 4.4-6: Simulation model of the stripline-type kicker: (a) full model; (b) cross-sectional structure; (c) tapered end transition.

When the excitation voltage reaches 17.5 kV, the electromagnetic field distribution at the kicker center is shown in Fig. 4.4-7. Within the range y = -5 mm to 5 mm, the electric field exceeds 1.44 MV/m, with a field uniformity of just 0.9%, and the magnetic flux density exceeds 48.2 Gs. These results indicate that the electric and magnetic fields have comparable effects on the electron beam.

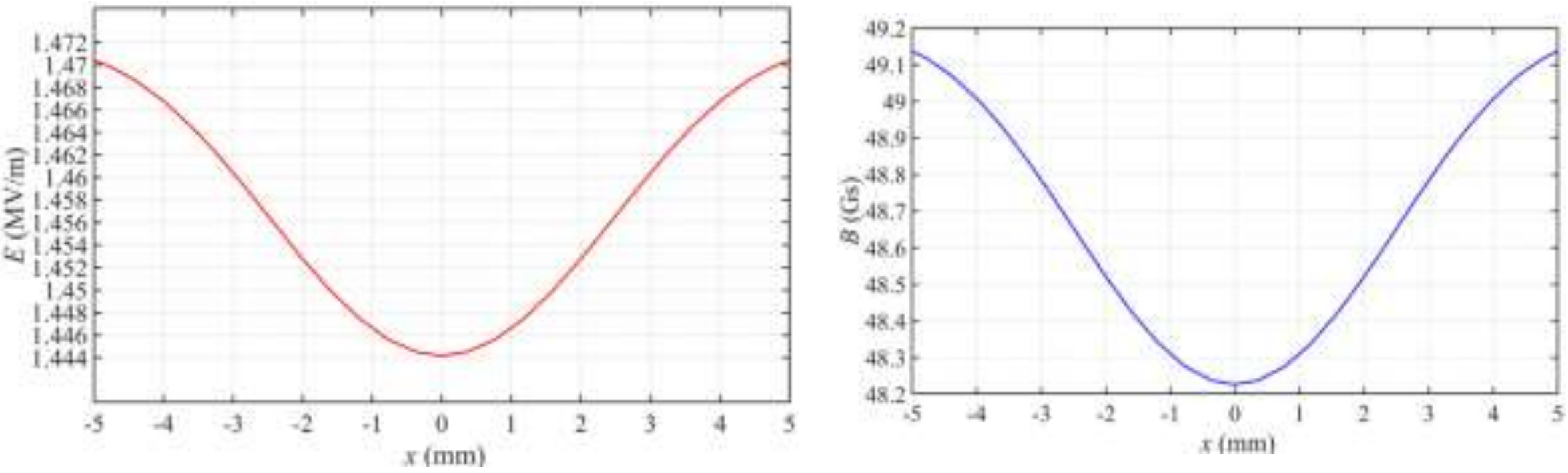

（a）　　　　（b）

Figure 4.4-7: Field distribution in the stripline-type kicker: (a) electric field; (b) magnetic field.

A comparison of beam coupling impedance before and after end-structure optimization is shown in Fig. 4.4-8(a). The peak impedance decreased from 856.8 Ω to 390.72 Ω. Fig. 4.4-8(b) shows the longitudinal impedance $Z_{||}/n$ per harmonic, where the effective impedance $(Z_{||}/n)_0$ dropped from 45.24 mΩ to 39.21 mΩ, a reduction of 13.33%. Fig. 4.4-8(c) presents the loss factor versus bunch length before and after optimization, with the optimized version always exhibiting lower losses. Fig. 4.4-8(d) shows power dissipation in the electrodes, where total power loss decreased from 89.2 W to 59.4 W, a reduction of 33.4%. These results demonstrate that the tapered end structure significantly reduces beam coupling impedance and thermal effects. A conical-tapered transition section will be designed in the future to further reduce the beam coupling impedance.

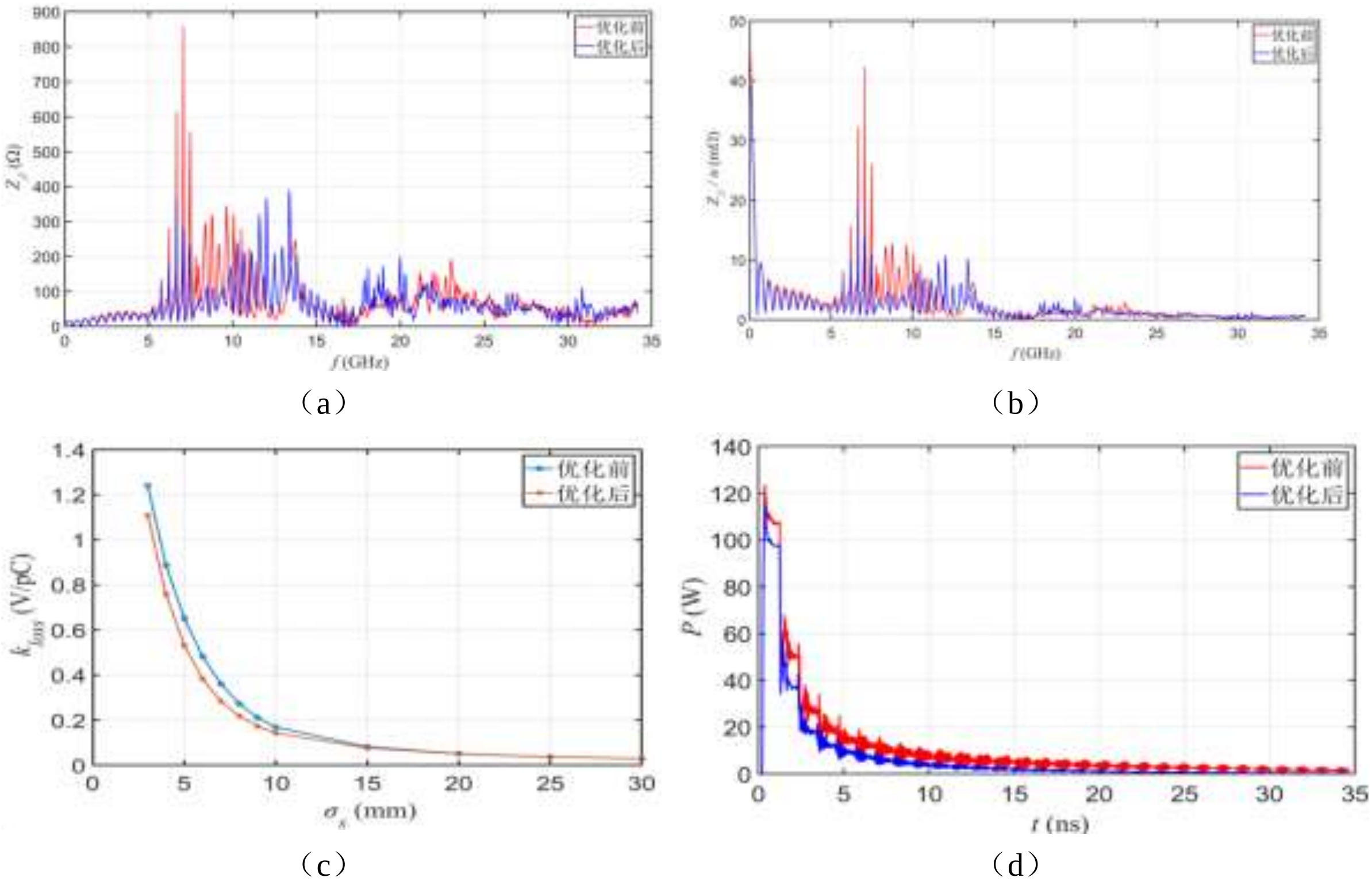

Figure 4.4-8: Performance comparison of the stripline kicker before and after end-structure optimization: (a) beam coupling impedance; (b) longitudinal impedance per harmonic; (c) loss factor vs. bunch length; (d) electrode power dissipation.

### 4.4.5 Feasibility Analysis

- ***Septum Magnets***

Eddy-current-type septum magnets are widely used in the accelerator field and represent a mature technology with successful domestic development experience. In contrast, the opposite-field-type septum magnet is a newly emerging design internationally and carries certain technical risks. To mitigate these risks, the team is concurrently advancing the design of both eddy-current and reverse-field septum magnets.

- ***Nonlinear Kicker Magnets***

The air-core-coil type nonlinear kicker magnet imposes stringent requirements on the positional accuracy of the excitation conductors. It still faces several engineering challenges, such as achieving the necessary machining precision of the ceramic vacuum chamber. These issues will need to be addressed progressively during engineering implementation. Compared with the air-core coil design, the pseudo-octupole and shielded-type nonlinear kicker magnets carry lower risk and are based on more mature technologies.

- ***Stripline-Type Ultrafast Kicker Magnet***

The stripline-type ultrafast kicker magnet must meet extremely strict timing requirements for its magnetic field structure. The geometry of its electromagnetic configuration, in conjunction with the pulse power supply, determines the response waveform of the electromagnetic field. The design is technically challenging and must be gradually refined during engineering development. The required nanosecond-level fast-response pulsed power supply is commercially available from international vendors such as FID, without export restrictions. In addition, the domestic team at IHEP (Institute of High Energy Physics, CAS) also has experience in developing such power supplies. Therefore, the technical risks associated with the power supply are considered manageable.

### 4.4.6 Summary

A high-efficiency injection and extraction system is one of the essential prerequisites for achieving the ultra-high luminosity goals of STCF. To meet the physical requirements of beam injection and extraction for the collider rings, the damping ring, and the accumulation ring, specialized magnet designs are necessary. The focus is placed on addressing the most technically challenging magnet designs to ensure the safe and efficient operation of the bunch swap-out injection, off-axis injection, and extraction processes.

## 4.5 Vacuum System

### 4.5.1 Technical Requirements and Design Objectives

In the STCF accelerator complex, the vacuum system covers all environments that require pressures below atmospheric levels, including both the collider rings and the injector components. The collider rings comprise the electron and positron rings, while the injector consists of multiple accelerator sections: the electron linacs, positron linac, positron damping ring (or accumulator ring), main linac, and various beam transport lines.

The vacuum requirements for the collider rings and injector are defined by the physics system. The goal of the vacuum system is to design vacuum chambers and configure vacuum pumps such that vacuum levels across all accelerator components meet the physical requirements of the STCF design.

According to specifications provided by the accelerator physics team, the dynamic vacuum pressure must be better than $1\times10^{-7}$ Pa for the collider rings and $6.5\times10^{-5}$ Pa for the injector. Corresponding static vacuum requirements are better than $2\times10^{-8}$ Pa and $1.3\times10^{-5}$ Pa, respectively.

The structural dimensions of the vacuum chambers for each part of the collider rings and injector are provided in the parameter tables from the accelerator physics design groups.

### 4.5.2 Key Technologies and Design Approach

The most challenging part of the STCF vacuum system lies in the collider rings. The primary challenges are twofold:

First, the high-current beams of 2 Amperes produce intense synchrotron radiation, which not only leads to high thermal loads but also generates large quantities of photoelectrons that trigger significant gas desorption. This effect is especially pronounced in the positron ring due to its sensitivity to the formation of the electron cloud. Moreover, the short bunch length and high bunch charge exacerbate the higher-order mode (HOM) effects, imposing strict requirements on the impedance of vacuum components.

Second, the beam–gas scattering in the interaction region (IR) contributes significantly to the experimental background, demanding particularly high vacuum levels in the IR where space constraints are very tight to add vacuum pumps.

International construction and operation experience from the previous high-current $e^+e^-$ colliders such as KEKB [33, 87-88] and PEP-II [89-91] provides valuable reference. The STCF vacuum design strategy includes:

- A carefully optimized global design.
- Vacuum chambers in the arc sections are designed with antechambers to deal with synchrotron radiation absorption and gas pumping.
- Appropriate material selection for vacuum chambers, and
- Efficient vacuum pump configurations.

In the positron ring, additional electron cloud mitigation measures will be employed, including TiN coating, localized magnetic fields, and grooved chamber geometries. In the interaction region, Non-Evaporable Getter (NEG) pumps will be used to improve local vacuum conditions, and low-impedance synchrotron radiation masks will be installed.

For the injector system—including the linacs, damping/accumulator rings, and transport lines—the vacuum requirements are conventional and well within the technological capabilities already mastered by several accelerator laboratories in China.

### 4.5.3 Design Scheme and System Composition

#### *4.5.3.1 Collider Ring Vacuum System*

The STCF collider rings are similar, both with a circumference of approximately 860 meters. Each ring is divided into arc sections, straight sections, and the interaction region. The vacuum requirements for the collider rings are defined by accelerator physics, primarily based on factors such as beam-residual gas scattering that contributes to experimental background near the interaction point (IP), beam lifetime, and emittance growth. The design must also consider fast ion beam instabilities in the electron ring.

The key aspect of the collider ring vacuum system is the vacuum chamber design. Drawing from the experience of SuperKEKB [92], the positron ring will adopt a dual-symmetric side chamber design, as shown in Figure 4.5-1. One side is used for absorbing synchrotron radiation, and the other for pumping. For the electron ring, a single-sided chamber may also be considered. All vacuum chambers in the damping wiggler sections must adopt dual-side symmetry. Flanges, bellows, and vacuum pump ports must all be equipped with RF shielding to minimize impedance contributions.

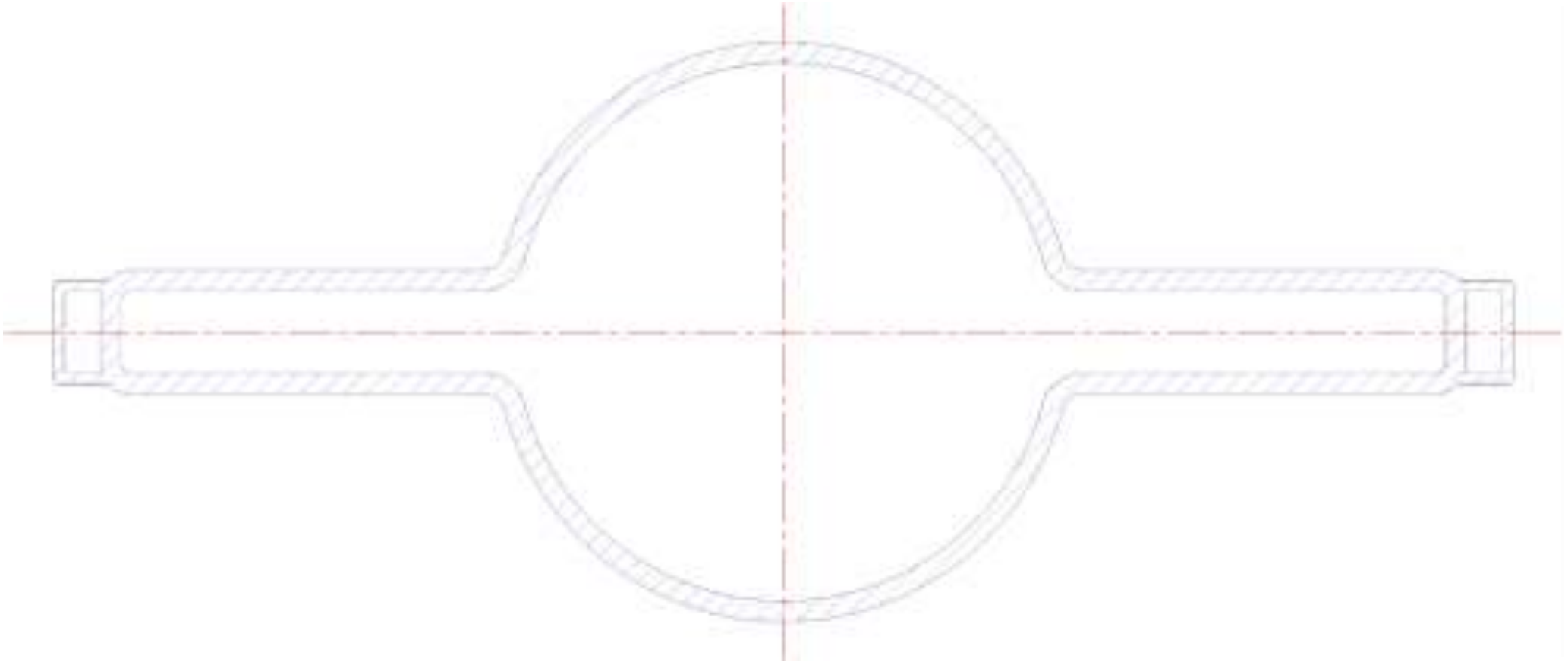

Figure 4.5-1: Cross-sectional schematic of the positron ring vacuum chamber

The pressure distribution in the collider ring vacuum system is determined by the gas load and the configuration of the pumping system. The design uses a vacuum chamber structure that combines the beam duct with side chambers. Synchrotron radiation is directed onto the cooled sidewalls of the vacuum chamber to reduce photoelectron entry into the beam channel and suppress the formation of electron clouds. The pumping configuration integrates both centralized and distributed pumping systems. These systems must effectively remove the substantial gas loads generated by synchrotron radiation to achieve the vacuum level required for acceptable beam lifetime. Once the distribution and rate of photon-induced desorption and thermal outgassing are determined, the pumping scheme can be finalized. The collider rings will employ a combination of distributed Non-Evaporable Getter (NEG) pumping and centralized ion pumping.

**Collision Ring Arc Section**

The vacuum pumping scheme in the arc section of the collision ring adopts NEG (Non-Evaporable Getter) coated vacuum chamber surfaces or distributed NEG pumping strips in combination with centralized vacuum pumps, see Fig. 4.5-2. The centralized pumps are configured as 400 L/s ion pumps paired with NEG hybrid pumps. For vacuum measurement in the arc section, a vacuum gauge and cold cathode gauge are installed at the entrance of each standard cell, corresponding to one set every 4.7 meters on average. Additionally, residual gas analyzers are placed at selected locations to monitor partial pressure distributions. The arc section is divided into 10 segments, with each segment separated by full-metal gate valves equipped with RF shielding. Each segment is equipped with a roughing valve and a venting valve.

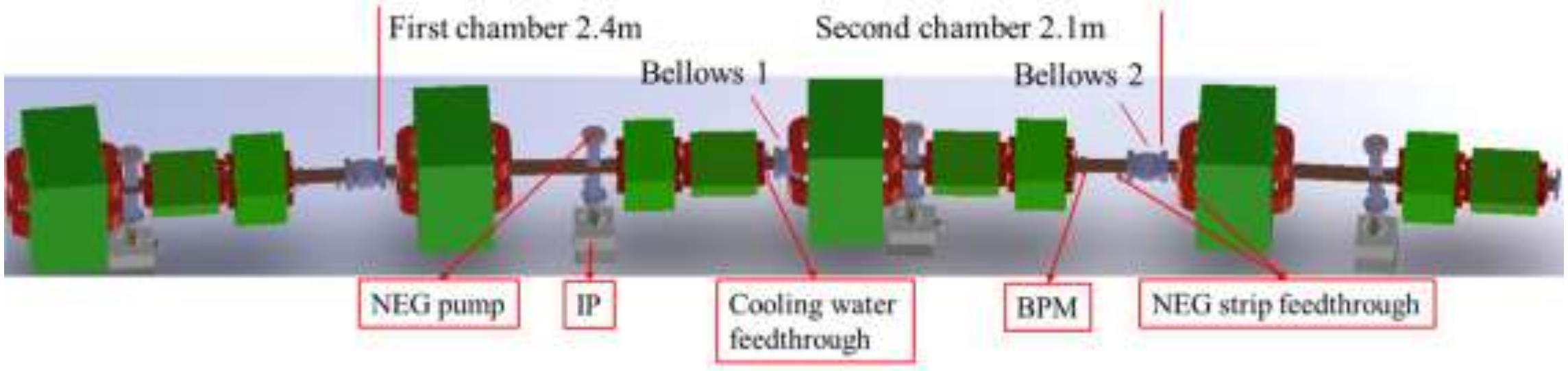

Figure 4.5-2: Vacuum layout for one arc cell of the collider rings

**Collision Ring Straight Section**

The vacuum pumping strategy for the straight sections also utilizes either NEG-coated vacuum chamber surfaces or distributed NEG pumping strips with centralized vacuum pumps, again configured as 400 L/s ion pumps with NEG hybrid systems. Vacuum measurement in the straight sections is implemented by placing a set of vacuum gauges and cold cathode gauges at the entrance of each standard cell, averaging one set every 5 meters. Residual gas analyzers are also installed in selected locations to monitor gas composition and partial pressures. Each straight section is similarly divided into 10 segments, each isolated by RF-shielded full-metal gate valves, and each segment includes a roughing valve and a venting valve.

**Interaction Region**

Based on the layout of beamline components near the interaction point, as shown in Figure 2.11-1, preliminary 3D modeling and simulation analysis have been conducted. The beam pipe located within the detector spectrometer relies on NEG coatings to maintain vacuum levels, providing an effective pumping speed of 500 L/s. In locations near the spectrometer, high-capacity NEG pumps and ion pumps are installed to extract gas loads stimulated by synchrotron radiation, delivering an effective pumping speed of 1000 L/s per meter.

The interaction region is divided into 12 sections, each separated by all-metal gate valves equipped with RF shielding. Each section is equipped with a roughing valve and a vent valve. The section containing the interaction point (IP) has limited space, extending from the IP to the end of the cryostat, leaving no room for vacuum pumps or related components. Therefore, high-speed vacuum pumps will be installed at more distant locations, along with a vacuum gauge and cold cathode gauge to monitor vacuum levels. Additionally, a residual gas analyzer will be installed to measure partial pressures. For the remaining sections, a vacuum gauge and cold cathode gauge will be installed approximately every 10 meters.

Based on the collider ring physics design, the preliminary list of major vacuum components is shown in Table 4.5-1.

Table 4.5-1: Major Vacuum Components in the Collider Rings

| Component | Quantity | Material |
|---|---|---|
| Beam position monitors | 105 | Stainless steel |

| Component | Quantity | Material |
|---|---|---|
| Various vacuum chambers | 458 | Oxygen-free copper |
| Ion pumps | 194 | — |
| NEG pumps | 158 | — |
| Vacuum valves | 102 | Stainless steel |
| Vacuum gauges | 96 | Stainless steel |
| Residual gas analyzers | 70 | — |

#### *4.5.3.2 Linac Vacuum System*

The vacuum system for the linacs in the injector primarily consists of accelerating tubes, various vacuum pipelines, and ion pumps. The vacuum system of the damping ring (or accumulation ring) is similar to that of the collider rings but can be somewhat simplified, relying on ion pumps for evacuation. The vacuum system for the transport lines is relatively straightforward and also uses ion pumps for pumping.

As part of the injector, the linac currently has two design schemes. In the first scheme, the electron beam is directly injected into the collider electron ring through the linac and transport line, while the positron beam passes through the damping ring and then through the final-stage linac and transport line before being injected into the collider positron ring; this is referred to as the off-axis injection scheme. In the second scheme, the electron gun provides high-charge electron bunches that are directly accelerated and injected into the collider electron ring, while the positron beam must be accumulated in the accumulation ring to reach the required bunch charge, then accelerated by a downstream linac and injected into the collider positron ring; this is referred to as the bunch swap-out injection scheme.

The main beamline components in the linacs include the electron gun, pre-buncher, buncher, accelerating tubes, vacuum components, beam diagnostics, magnets, and ancillary microwave equipment such as waveguides and loads for supplying RF power to the accelerating structures. The vacuum system components include the electron gun, S-band accelerating tubes, collimators, vacuum chambers for beam diagnostics, bellows, ion pumps, valves, vacuum gauges, and turbopump stations for vacuum startup. All vacuum valves are all-metal gate or angle valves.

Vacuum in the accelerating tubes is achieved via sputter ion pumps located at the pumping ports near the input and output couplers. Two 3-meter accelerating tubes form a 6-meter segment, with a 100 L/s sputter ion pump placed at each microwave input interface and another pump positioned at the output of the first tube, between the two. Shielded bellows are installed between adjacent 6-meter segments, and an all-metal gate valve is installed every two 6-meter segments for vacuum isolation. Outside of the accelerating tubes, sputter ion pumps are spaced approximately every 2 meters along the vacuum beamline.

A cold cathode gauge is installed at the input of the upstream accelerating tube in each 6-meter segment to measure the vacuum level.

Based on the parameters from accelerator physics and the corresponding injection schemes, the major vacuum components required for the two injection schemes are preliminarily listed in Table 4.5-2.

Table 4.5-2: Major Vacuum Components for Off-axis and Swap-out Injection Schemes

| Component | Off-axis | Swap-out | Material |
|---|---|---|---|
| Thermionic e-gun | 1 | 2 | – |
| Photocathode e-gun | 1 | 0 | – |
| S-band accel. tubes | 100 | 133 | OFHC Copper |
| X-band accel. tubes | 1 | 0 | OFHC Copper |
| Beam diagnostics | 40 | 45 | Stainless Steel |
| Vacuum chambers | 300 | 400 | Stainless Steel |
| Magnetic chicanes | 2 | 1 | – |
| Ion pumps | 343 | 473 | – |
| Vacuum valves | 27 | 36 | Stainless Steel |
| Vacuum gauges | 203 | 278 | Stainless Steel |

#### *4.5.3.3 Damping Ring (Accumulation Ring) Vacuum System*

In the off-axis injection scheme, the positron damping ring is used to reduce the emittance of the positron bunches generated by the target through synchrotron radiation damping, achieving the required emittance for injection into the collider positron ring. In the swap-out injection scheme, the positron accumulation ring is used to accumulate charge and optimize emittance. Both systems include injection lines, extraction lines, and a ring of about 150 m in circumference.

According to the damping ring design, the ring's vacuum system can be divided into 8 relatively independent sections by all-metal gate valves. Each section has roughing and venting valves. The vacuum system is first evacuated using oil-free turbopump stations, and then maintained by conventional sputter ion pumps. Ion pumps rated at 100 L/s are spaced approximately every 6 meters, each paired with a vacuum gauge and cold cathode gauge. The transport line vacuum system includes about 32 ion pumps, 27 vacuum gauges, and 9 VAT gate valves.

#### *4.5.3.4 Beam Transport Line Vacuum System*

The beam transport lines are divided into four parts based on area and function: the electron beam bypass line (Bypass section), the damping ring (accumulation ring) transport section, the positron-electron merging section, and the final injection section into the collider rings.

The main vacuum chambers in the transport lines have circular cross-sections, with rectangular cross-sections used in dipole magnet regions depending on magnetic aperture and deflection/splitting angles. Low-magnetic-permeability stainless steel is used for vacuum chamber construction.

The pumping system uses conventional 50 L/s sputter ion pumps, installed approximately every 5 meters along the beam path. Cold cathode vacuum gauges are installed every 10 meters, and all-metal gate valves are placed every 20 meters.

Based on the structural and design parameters provided by the accelerator physics team, the preliminary list of vacuum components required for the two injection schemes is shown in Table 4.5-3.

Table 4.5-3: Major Vacuum Components in Transport Lines for Off-axis and Swap-out Injection Schemes

| Component | Off-axis | Swap-out | Material |
|---|---|---|---|
| Circular vacuum chambers | 87 | 59 | Stainless Steel |
| Rectangular chambers | 38 | 30 | Stainless Steel |
| Ion pumps | 82 | 50 | – |
| Vacuum valves | 29 | 19 | Stainless Steel |
| Vacuum gauges | 47 | 29 | Stainless Steel |

### 4.5.4 Vacuum Control and Interlock

The vacuum control system primarily consists of ion pumps and their power supplies, radiation-resistant vacuum gauges, and isolation gate valves. Ion pump power supplies are equipped with built-in protection functions and interface with the EPICS control system via 0–5 V analog output signals, which are used for safety interlocks in accelerating tubes and for displaying ion current and voltage. The output signals from the gauges are transmitted through high-voltage cables to vacuum meters, which are then processed and monitored by the control system.

Gate valve operation is interlocked with vacuum gauge readings. When the vacuum at a specific location degrades to a predefined threshold, the two nearest gate valves automatically close to protect adjacent vacuum sections. Additionally, during accelerator maintenance, only the corresponding segment needs to be purged with nitrogen gas for service, avoiding exposure of the entire vacuum system to atmospheric pressure and preserving vacuum integrity in other regions.

### 4.5.5 Feasibility Analysis

The design, construction, and operation of the STCF collider ring vacuum system pose significant technical challenges, particularly in the positron ring, which requires additional measures to suppress electron cloud instabilities. Although the relevant technologies are largely mature internationally and the BEPCII experience provides a strong domestic reference, there is currently no precedent in China for the design or operation of an electron storage ring with circulating currents above 1 A.

In recent years, China has made substantial advances in ultra-high vacuum technologies thanks to the construction of several large-scale accelerator facilities and improvements in industrial capabilities. These include the production of high-temperature-tolerant copper materials, extrusion of non-standard structural profiles, and the widespread application of NEG coatings in new-generation light sources such as HEPS and HALF. TiN coating technology has been extensively used in the CSNS accelerator. For the collider interaction region, vacuum chamber design techniques such as remotely controlled vacuum connections (RVC) have also been applied domestically. Although further R&D is still required for some critical vacuum structures and technologies before construction begins, overall technical risks are considered manageable.

The injector vacuum system can be fully realized using mature and well-established technologies and presents no technical risk.

### 4.5.6 Summary

The STCF accelerator vacuum system is extensive and complex, with a total vacuum pipeline length of approximately 2500 meters. Each accelerator section requires different design approaches and vacuum technologies. Through comprehensive surveys and preliminary design studies, a basic structure and framework for the vacuum system have been established. The design plans and technical approaches for each subsystem have been preliminarily clarified, laying the foundation for continued technical development. Key vacuum technologies requiring further R&D have also been identified.

## 4.6 Beam Instrumentation System

### 4.6.1 Design Requirements and Specifications

The beam instrumentation system serves as a critical diagnostic tool for accelerator operation and tuning. It measures a wide range of beam parameters, providing essential support for machine commissioning, suppression of beam instabilities, and optimization of beam position/angle at the collision point. It also provides the data foundation for beam dynamics analysis and integrated luminosity performance evaluation. The STCF beam instrumentation system must support both the collider rings and the injector, enabling accurate measurement and stabilization of beam parameters. Its primary objective is to maintain stable beam operation at a high luminosity level of $5 \times 10^{34}$ $cm^{-2}s^{-1}$, with the time resolution capable of distinguishing bunch spacing at integer multiples of 2 ns (expected to be 4-6 ns), and delivering high-precision data for beam dynamics analysis. The technical specifications are listed in Table 4.6-1.

Table 4.6-1: Beam Instrumentation Technical Requirements for the Collider Rings

| Beam Parameter | Measurement Method | Specification | Value/Detail |
|---|---|---|---|
| Beam Position | Button electrodes and digital processors | Closed orbit resolution | SA: 50 nm @ 10 Hz; FA: 100 nm @ 10 kHz |
| | | Turn-by-turn resolution | 1 μm |
| | | Bunch-by-bunch resolution | 5 μm (transverse), 0.2 ps (longitudinal) |
| | | Dynamic range | -50 to 0 dBm |
| Beam Current | DCCT, PCT, data acquisition system | Measurement range | 0.01–3000 mA |
| | | Resolution | 2 μA |
| | | Accuracy (>200 mA) | 10 μA |
| Bunch Charge | BPM sum signal with fast digitizers or photodiodes | Measurement range | 0.1-50 mA |
| | | Resolution | 25 μA @ 50 mA |
| Betatron Tune | Sweep/FFT method | Update rate | 1 Hz |
| | | Resolution | $1\times10^{-4}$ |
| Beam Size & Emittance | Absolute (X-ray diffraction imaging) | Range | 10-50 μm |
| | | Resolution | 2 μm |
| | | Range | 20-80 μm |

| Beam Parameter | Measurement Method | Specification | Value/Detail |
|---|---|---|---|
| | Relative (X-ray pinhole imaging) | Resolution | 5 μm |
| | Relative (visible/X-ray interferometry) | Range | 5-50 μm |
| | | Resolution | 1 μm |
| Bremsstrahlung | Dedicated detector | Measure polarization | – |
| Bunch Length | Streak camera (visible light) | Range | 20-100 ps |
| | | Resolution | 2 ps |
| Bunch-by-Bunch Feedback | Transverse feedback (stripline, FPGA) | Bandwidth | 250 MHz |
| | | Drive power | 50-100 W |
| | | Damping time | 0.1-0.3 ms |
| | Longitudinal feedback (RF cavity, FPGA) | Bandwidth | 250 MHz |
| | | Drive power | 100-200 W |
| | | Damping time | 0.5 ms |
| IP Feedback | Slow and fast feedback | Stability | Better than 10% of beam size |
| BPM Drift Compensation | Capacitive sensor and temperature probe | Resolution | 50-100 nm |
| Beam Loss | Scintillator | Time response | ≤8 ns |
| | Optical fiber | Spatial resolution | ≤1 m |

### 4.6.2 Key Technologies and Development Strategy

In the collider rings, beam position monitors (BPMs) will be installed adjacent to quadrupole magnets, utilizing button electrodes with heterodyne detection and high-speed, high-precision electronics to enable turn-by-turn and bunch-by-bunch position measurements. Visible photon monitors will incorporate diamond-coated extraction mirrors to minimize thermal deformation induced by synchrotron radiation. An X-ray bunch size monitor based on coded aperture imaging will provide bunch-by-bunch measurements of transverse beam size, while an X-ray interferometry system will offer higher resolution diagnostics. At the IP, a large-angle bremsstrahlung radiation monitor will measure the polarization of emitted light to infer geometric characteristics of the colliding beams.

Bunch-by-bunch transverse and longitudinal feedback systems will integrate low-noise front-end electronics and high-resolution digital filters to enable instability suppression. The collision point feedback system will detect the orbit offsets in the interaction region that is linearly related to the experimental luminosity, and also take use of the luminosity monitor signal to maximize and stabilize the luminosity.

In the injector, beam position will be monitored using stripline pickups combined with high-speed, high-resolution electronics for bunch-by-bunch measurements. Bunch charge and length will be measured via passive resonant cavities to provide a large dynamic range and good signal-to-noise ratio. For high-resolution measurements of longitudinal beam structure, a transverse deflecting cavity will be taken into consideration.

Most of the beam diagnostics technologies required for the STCF collider rings and injector have been validated at the National Synchrotron Radiation Laboratory's Hefei Advanced Light Facility (HALF) [93], SuperKEKB at KEK [94], and the BEPCII/BEPCII-U collider in China [95]. Key techniques are also under development as part of the STCF Key Technology R&D project [96].

The development strategy for STCF beam instrumentation will follow the principles outlined below:

- Prioritize mature and proven technologies when performance specifications are met.
- Draw design references from the latest international projects of similar scope, subject to expert review.
- Undertake targeted development—via internal R&D or collaborations—for diagnostics tasks beyond existing technological capabilities, supported by pre-research programs.

### 4.6.3 Design Scheme and System Configuration

#### *4.6.3.1 Beam Instrumentation System of the Collider Rings*

**BPM and Displacement/Temperature Monitoring**

Each collider ring in the STCF is equipped with 402 button-type beam position monitors (BPMs), distributed as uniformly as possible adjacent to each quadrupole magnet. The number of the BPMS could be adjusted according to physical design updates, just like other beam diagnostics instruments. BPMs located near the beam injection points allow real-time monitoring of the injected beam condition, while those near the collision region provide precise position measurements prior to beam collisions, ensuring effective collisions. In addition, selecting uniformly distributed BPMs enables comprehensive beam position monitoring throughout the entire ring. Under the current STCF collider ring parameters, the optimized design of the button-type BPM probe has been completed. The resonant frequencies of the probe's modes are well separated from the operating frequency of 499.7 MHz, and are also shifted away from higher harmonics of 499.7 MHz that arise under multi-bunch operation. The structure of the BPM probe is shown in Figure 4.6-1.

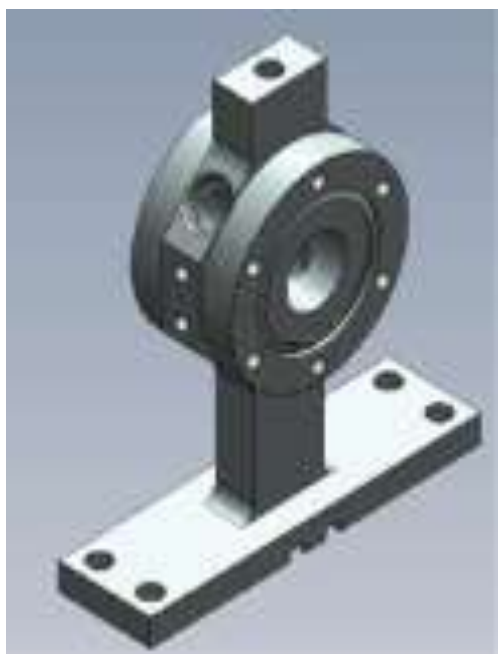

Figure 4.6-1: Schematic of BPM Probe

The four electrode signals from each BPM are transmitted via long cables to the BPM electronics, where analog signals are conditioned before calculating the 3D position of each bunch. For the signal acquisition system, the peak negative signal from each bunch is captured for transverse position measurements, while longitudinal phase measurements require sampling near the zero-crossing of the BPM signal. Based on the signal characteristics and measurement requirements, the system specifications for bunch-by-bunch 3D position measurement are listed in Table 4.6-2.

Table 4.6-2: Specifications for Signal Acquisition in Bunch-by-Bunch 3D Position Measurement

| Parameter | System Component | Minimum Requirement | Recommended Configuration |
|---|---|---|---|
| Number of channels | — | 4 | 8 |
| Analog bandwidth | Acquisition board | 500 MHz | >1 GHz |
| | Oscilloscope | 500 MHz | >6 GHz |
| Sampling rate | Acquisition board | 500 MS/s | >1 GS/s |
| | Oscilloscope | 5 GS/s | >20 GS/s |
| Sampling resolution | — | 8-bit | >10-bit |

Two acquisition methods are considered. One uses high-bandwidth, high-sampling-rate oscilloscopes [97], which provide high precision but involve complex data processing due to the oscilloscope's non-phase-locked nature and limited memory depth, making them less suitable for online real-time bunch-by-bunch monitoring.

To better serve the STCF project, a high-speed data acquisition board is selected for real-time, online bunch-by-bunch 3D position monitoring. The system block diagram is shown in Figure 4.6-2. Domestic signal processors have achieved 1 GSPS [98] sampling rates, enabling the development of an integrated signal processor for this purpose. This processor includes an analog front-end with eight ADC channels and a digital module based on an FPGA. The processor supports external clock and trigger inputs.

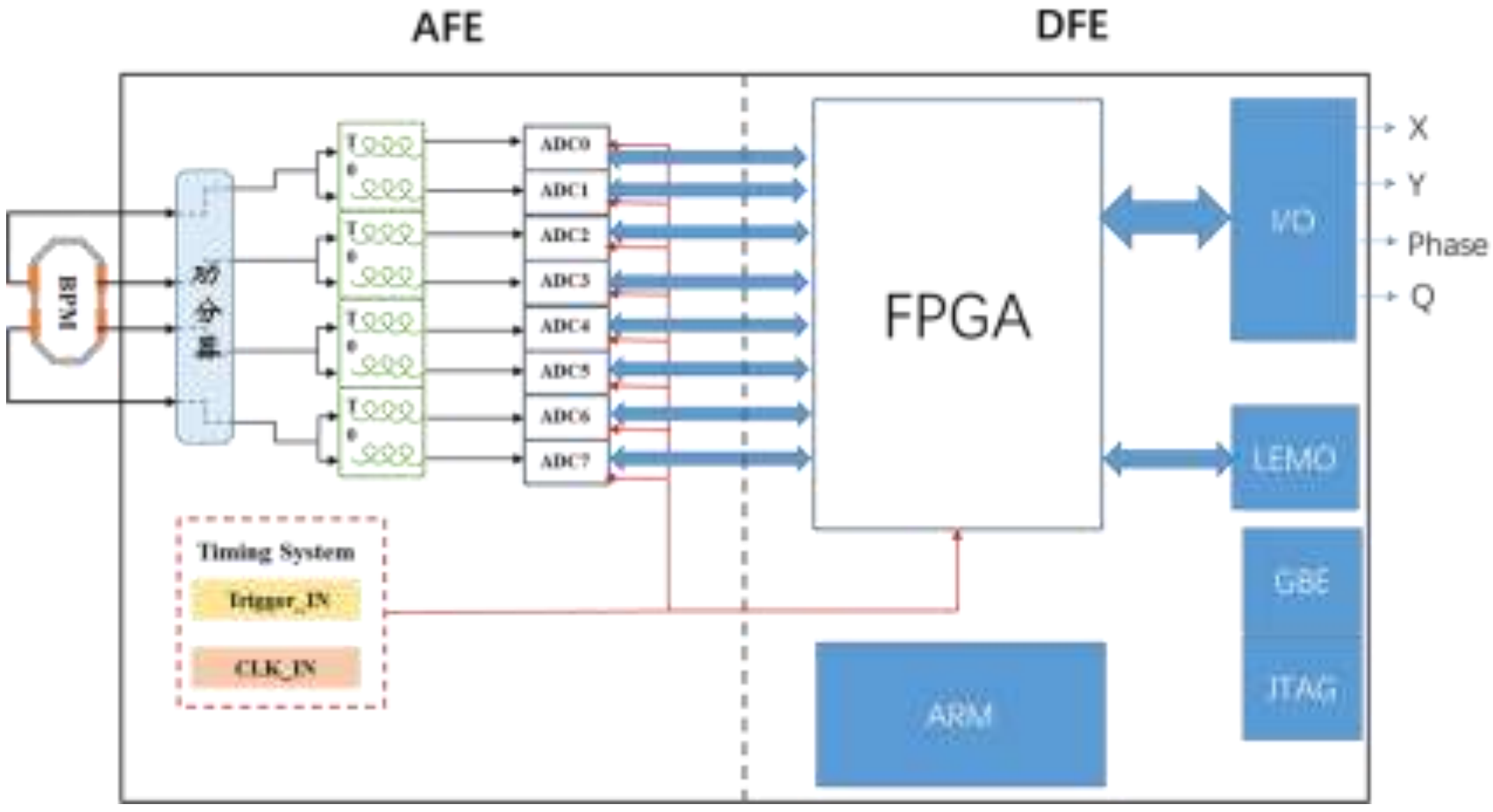

Figure 4.6-2: Integrated Bunch-by-Bunch 3D Position Signal Processor

As a high-current machine, STCF is subject to thermal variations due to synchrotron radiation and other effects, which may influence BPM measurements. Temperature monitoring and compensation techniques are required. Methods implemented at SuperKEKB [99] are referenced here. Temperature sensors are installed near BPM probes to monitor environmental fluctuations in real time. The BPM processor collects both position and temperature data, which are then used to build compensation models that correct temperature-induced displacement errors. Feedback control systems may later be incorporated to apply real-time corrections.

The design of the button-type BPM probe and the signal processing capabilities ensure effective operation across the full 1-3.5 GeV energy range of STCF. Although beam characteristics such as shape, intensity, and pulse width may vary with energy, these variations have minimal impact on the BPM measurement precision. Furthermore, the BPM electronics feature high-sensitivity amplification and filtering circuits that maintain a high signal-to-noise ratio and resolution under varying beam conditions.

**Bunch-by-Bunch Feedback and Tune Measurement**

The bunch-by-bunch feedback system is used to suppress coupled-bunch instabilities [100] and generally consists of three main components: oscillation signal detection, digital signal processing, and feedback implementation.

After acquisition by the front-end RF circuit, the oscillation signals are filtered, amplified, and down-converted to remove the DC component before ADC sampling and transmission to a digital signal processor. To extract beam oscillation information, appropriately ordered digital filters must be designed within the FPGA. A 90° phase shift can be implemented directly through FIR filtering algorithms. The signal is then converted by a DAC, delayed, and amplified by a power amplifier, and finally applied to the beam via a kicker to suppress instabilities [101].

High-frequency beam signals are mixed with $nf_{RF}$ to obtain an oscillation signal near $f_{RF}/2$ with better signal-to-noise ratio [102]. In order to capture beam oscillations at the fundamental frequency, the digital signal processor operates at a sampling rate of 500 MHz. High-order FIR filters are implemented in a high-performance FPGA to achieve both phase shifting and DC component removal.

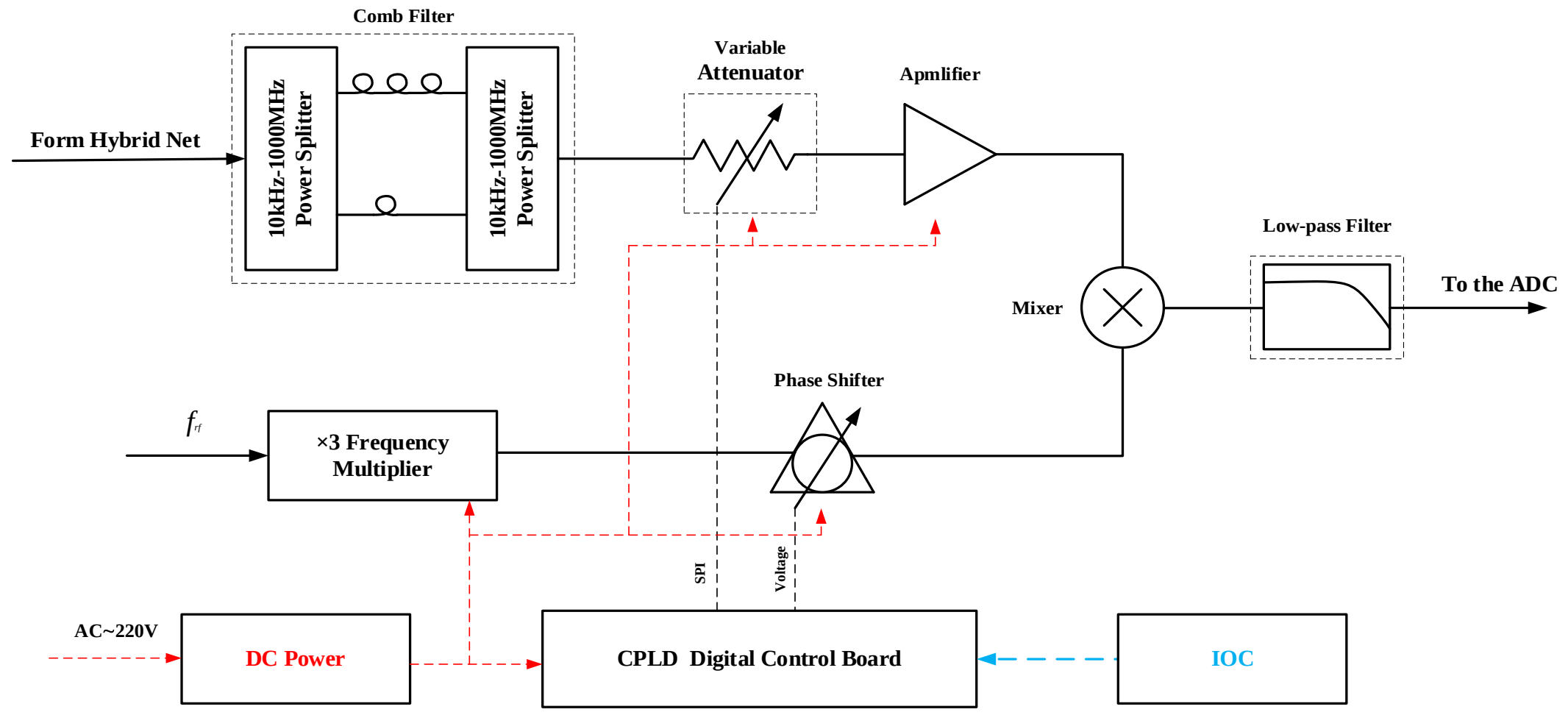

Figure 4.6-3: Block diagram of front-end electronics

The feedback actuator is the feedback kicker. The feedback signal, generated by the digital signal processor and amplified by an RF power amplifier, is applied to the kicker. For the longitudinal feedback system, the baseband feedback signal must first be modulated to the operating frequency band of the power amplifier and longitudinal kicker. The kicker then generates an electromagnetic field that applies a corrective kick to the bunch, damping its transverse or longitudinal oscillation. To suppress all coupled-bunch instability modes, the bandwidths of the power amplifier and kicker must not be less than $f_{RF}/2$. The central frequencies of the system differ between transverse and longitudinal feedback channels.

The SuperKEKB collider rings [103], with a circumference of 3016 m, are significantly longer than the STCF collider rings that are about 860 m and have a shorter revolution period. Furthermore, the low-energy ring of SuperKEKB operates at 4 GeV, while the beam energy at STCF is tunable from 1-3.5 GeV. Thus, the damping time in STCF is expected to be shorter than in SuperKEKB, which is beneficial.

High-energy beams impose more stringent requirements on the electronics of bunch-by-bunch feedback systems. Higher resolution and faster sampling rates are required for digital signal processors to provide low-latency, high-precision feedback signals. The entire feedback system also requires minimal latency, necessitating the use of low-latency power amplifiers.

The beam tune is a critical machine parameter that must be continuously monitored and recorded. It is directly related to beam quality and stability. Other indirectly measured parameters, such as the envelope function and chromaticity, can be calculated from tune values. Because the amplitude of free oscillations is very small during stable operation, excitation signals must be applied to the beam for measurement. To avoid disturbing the entire beam, one bunch can be selectively excited and not damped within the bunch-by-bunch feedback system, ensuring that all other bunches remain stable while the tune is measured.

**Interaction Point Feedback**

Beam orbit deviations at the interaction point (IP) can lead to significant luminosity losses and must be corrected. Sources of these deviations include vertical motion of final focus (FF) magnets caused by ground vibrations or mechanical fluctuations in cryogenic components such as cold boxes, which result in closed orbit distortions at the IP that cannot be fully compensated.

Collision point feedback based on beam-beam kick measurements exploits the linear relationship between beam-beam deflection and beam displacement at the IP. By measuring changes in BPM readings near the IP caused by beam-beam kicks, the orbit deviation can be inferred. The system calculates correction signals in real time to adjust particle trajectories using correction magnets. Operating in a high-frequency closed-loop mode, this feedback system dynamically compensates for orbit errors caused by mechanical noise or ground motion, ensuring precise alignment of the colliding beams at the IP, thus enhancing collision efficiency and luminosity [104, 105].

The IP feedback system primarily uses beam position signals, supplemented by luminosity and beam-induced radiation signals. To measure beam positions as close to the IP as possible, two eight-electrode BPMs are placed 60 cm to the left and right of the IP (distance will be updated according to the MDI design evolution), respectively. At this location, the electron and positron beams are still co-linear, and each BPM must simultaneously monitor both beams. Based on STCF beam parameters, maintaining the orbit deviation within 5% of the bunch size at the IP requires BPM resolution better than 5 μm. Because the horizontal beam size is very small, beam-beam kicks are no longer sensitive indicators of collisions in that direction.

$$\Delta y'^{*\,p,e} = \frac{2\pi}{\beta_y^{*\,p,e}} \xi_y^{p,e} \Delta y^* \qquad y_{BPM} = L \cdot \Delta y'^{*\,p,e}$$

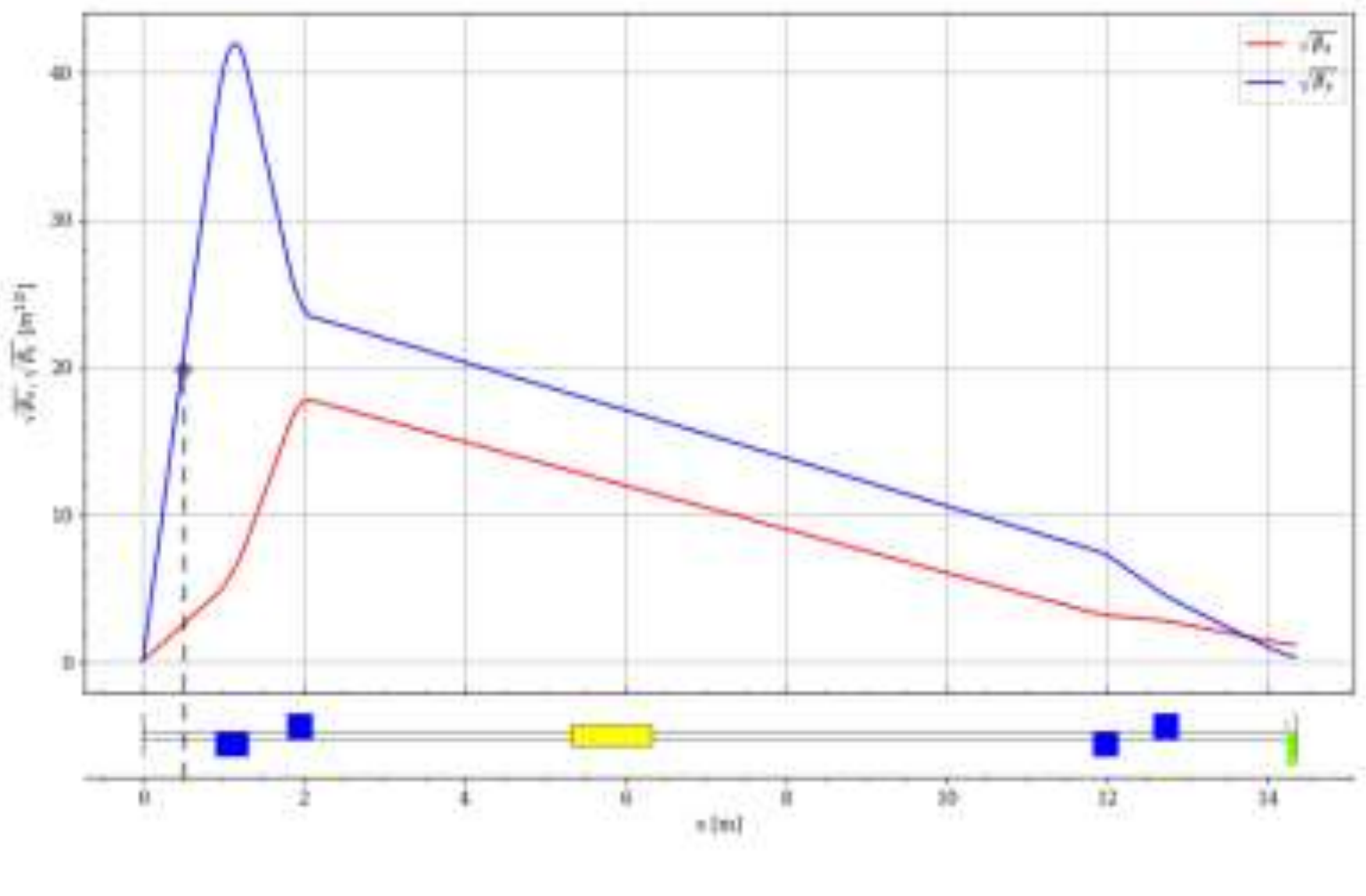

Figure 4.6-4: Location and β-function of the BPMs

BPM dimensions are optimized based on beam pipe diameter, signal strength, resolution requirements, and beam parameters. The chamber diameter matches the transverse beam size and vacuum chamber design. The diameter of the button electrodes balances signal strength and spatial resolution while meeting bandwidth demands. Because the beam pipe near the IP has a small aperture, the power output of the BPM near the IP is much higher than that of conventional BPMs, and specialized feedthroughs must be designed.

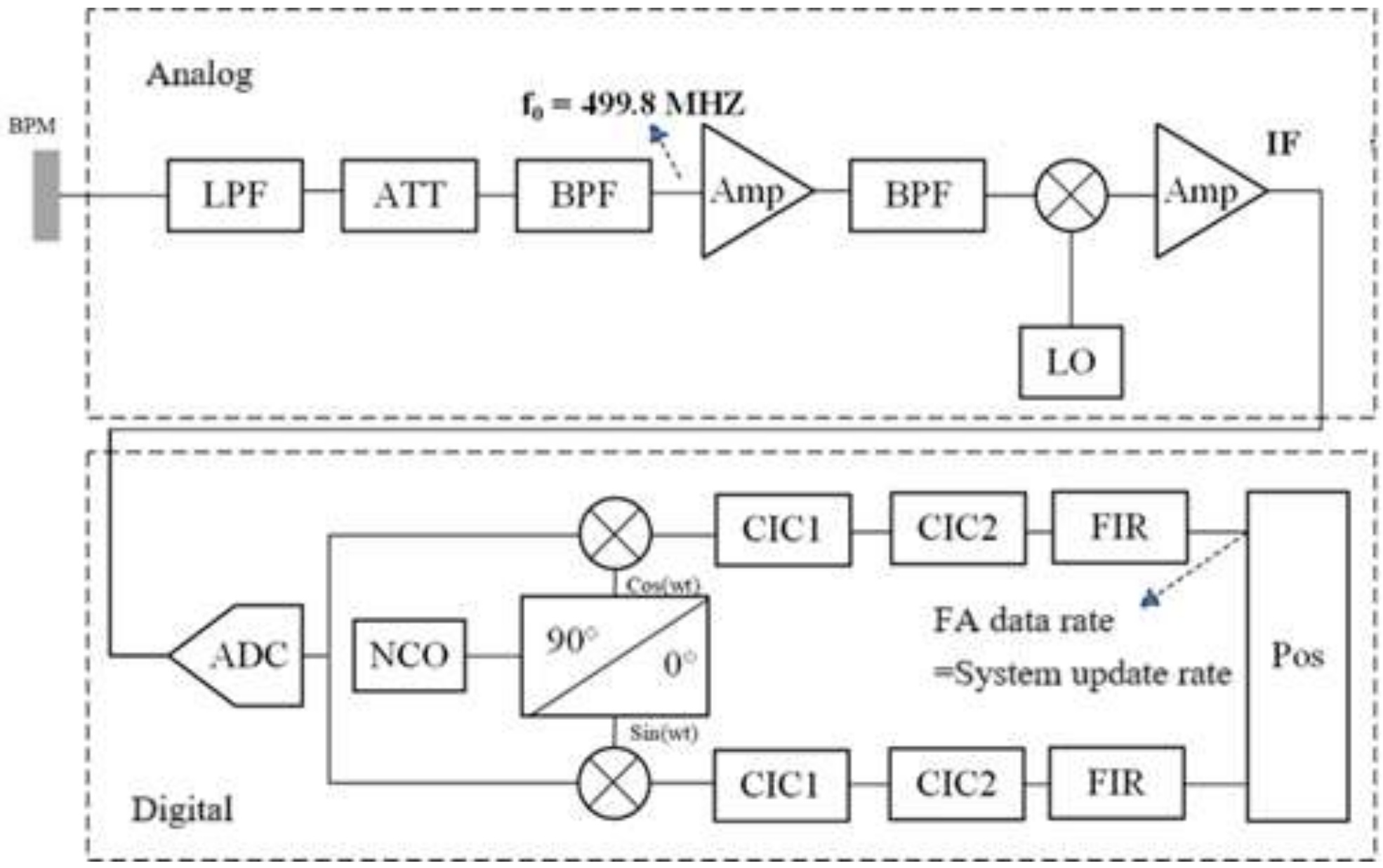

Figure 4.6-5: Schematic of the IP feedback detector system

The analog section uses low-pass and band-pass filters to convert beam signals into passband signals centered at the RF frequency of 499.7 MHz. An analog mixer then down-converts the 499.7 MHz component of the BPM signal to an intermediate frequency (IF) signal to reduce the impact of ADC clock jitter on the signal-to-noise ratio.

To ensure that the FOFB system operates with low latency and high bandwidth, the power supplies for fast correction magnets must also respond rapidly. The IP fast feedback system requires a correction magnet power supply with a bandwidth of at least 10 kHz. Fast corrector magnets must be capable of precise magnetic field control to minimize orbit correction errors and suppress feedback jitter around the target value. Improving power supply resolution requires reducing the quantization error of the DAC chips.

STCF may operate at energies ranging from 1 to 3.5 GeV. Beam-beam effects are energy-dependent and become more pronounced at lower energies. As energy increases, the same position offset results in smaller BPM signal changes, thereby raising the resolution requirements of the BPMs.

**Beam Diagnostics Based on Synchrotron Radiation**

High-energy charged particle bunches emit synchrotron radiation (SR) when deflected, with the SR characteristics reflecting the parameters beam at the light emission location, making it a valuable non-interceptive diagnostic tool. According to the beam characteristics of STCF, both X-rays and visible light from SR will be used to measure the transverse size and length of the electron and positron bunches.

The monitoring of beam transverse size and emittance can be achieved through X-ray diffraction imaging, X-ray pinhole imaging, and synchrotron light interferometry. Emittance is typically calculated by combining accurate measurements of the bunch transverse size with the beam's Twiss parameters.

X-ray diffraction imaging is a direct method for observing the beam spot. Figure 4.6-6 illustrates a beam size measurement system based on Fresnel zone plate (FZP) X-ray diffraction imaging [106]. The synchrotron light is first monochromatized by a monochromator, and then imaged directly through two FZPs. The system typically operates with a total magnification factor of 15–30. This method offers high spatial resolution but may be affected by diffraction effects, limiting its overall measurement range.

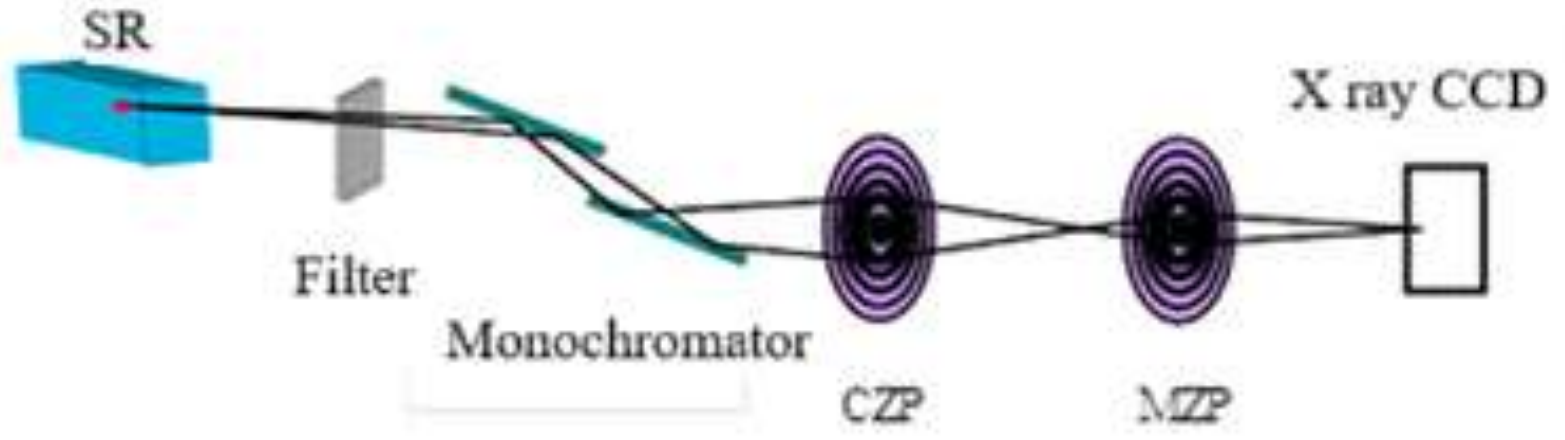

Figure 4.6-6: Schematic diagram of X-ray diffraction imaging with Fresnel zone plates

X-ray pinhole imaging uses a small aperture to project the transverse beam profile [107]. As shown in Figure 4.6-7, X-rays extracted from the vacuum chamber pass through a pinhole and illuminate a downstream fluorescent screen, where the image is converted to visible light and recorded by a CCD camera for profile reconstruction. This method is simple, monochromator-free, robust to thermal loads, and reliable. However, it has lower resolution than diffraction-based methods.

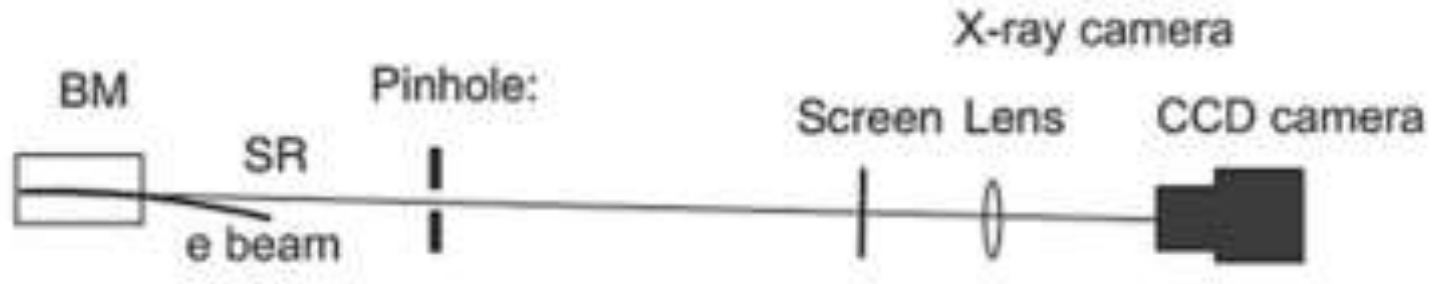

Figure 4.6-7: Schematic diagram of X-ray pinhole imaging system

Synchrotron light interferometry is implemented using a Michelson-type stellar interferometer [108]. After passing through a double slit, optical filters, and a focusing system, the light forms an interference pattern on a detector. The contrast of this pattern is used to deduce the beam spot size. This method can achieve sub-micrometer resolution but requires a complex optical path.

Bunch length is typically measured using a streak camera [109], which is capable of recording ultrafast optical phenomena. It can simultaneously provide information about light intensity, temporal structure, and spatial (or spectral) distribution.

In designing the STCF synchrotron light diagnostic system, beam energy variability must be considered. Beam energy affects both the synchrotron spectrum and radiation divergence: higher energies lead to an increased yield of high-energy (X-ray) photons and a narrower radiation cone. Therefore, optical elements must accommodate a wide dynamic range, and energy absorbers and related components must be designed with sufficient thermal capacity and safety margins.

**Beam Loss Monitoring**

According to the beam loss diagnostics requirements of STCF, a hybrid detector configuration combining optical fibers and scintillators is adopted. Optical fibers are installed along the damping ring and linac to monitor abrupt losses and localize the position of the loss event. Scintillator detectors are deployed along the dual-ring collider to provide a prompt response to frequent beam loss incidents. The entire beam loss monitoring system is designed to provide high spatial and temporal resolution.

In future upgrades, a self-developed scintillator detector with an overall response time below 2 ns is considered to enable bunch-by-bunch resolution. This would provide stronger support for maintaining high operational luminosity.

#### *4.6.3.2 Beam Diagnostics System in the Injector*

**Beam Diagnostics in the Linacs and Transfer Lines**

The beam diagnostics system in the linacs and transfer lines with an energy range of 1-3.5 GeV includes independent subsystems for beam position measurements, bunch charge measurements, beam transverse profile measurements, kinetic energy measurements, and energy spread measurements.

The beam position measurement system mainly consists of BPM probes (stripline and button types), signal transmission cables, digital BPM electronics, timing/trigger distribution systems, and frequency synthesizers. The BPM probes pick up beam position signals, which are processed by digital BPM (DBPM) electronics to extract beam position data, which are then published to the control network. This system is primarily used for position and orbit measurements and for beam-based alignment.

The beam charge measurement system mainly includes integrated current transformers (ICTs), signal transmission cables, digital oscilloscopes for current signal acquisition, and data processing terminals. The ICT signal is oversampled using digital oscilloscopes, and the collected waveform is processed to determine the bunch charge.

The beam transverse profile measurement system is based on view screen optical imaging. It includes beam spot screens, screen vacuum chambers, motion mechanisms, telecentric optical lenses, digital cameras, cabling, image processing terminals, motor drivers, and controllers. The local beam diagnostics station sends camera trigger signals via coaxial cables to the cameras. Control signals for stepper motors are sent via Ethernet to operate the screen actuators. Captured beam spot images are also transmitted via Ethernet to the server for remote control and image acquisition.

Beam energy and energy spread measurements are carried out using a spectrometer system. The beam is deflected by a bending magnet onto an energy-dispersive screen. By measuring the beam spot on the screen, beam energy and energy spread can be calculated. The imaging and motion control system for the energy screen shares infrastructure with the transverse profile system.

**Beam Diagnostics in the Damping Ring**

The positron beam of 1 GeV is injected into the damping ring to reduce its emittance through synchrotron radiation damping. Once the target emittance is achieved, the beam is injected into the collider. The diagnostics system in the damping ring monitors the evolution of beam parameters throughout this process, ensuring the beam meets injection requirements [110].

Beam position monitoring in the damping ring typically uses button-type BPMs. The four electrodes of a button BPM detect the bunch signal, which is transmitted through feedthroughs and coaxial cables to the readout electronics. The electronics compare the four signals (typically using difference-over-sum algorithms) to determine beam position. This system is simple, based on mature technology, easy to fabricate, and compatible with commercially available readout electronics.

A fluorescent screen is used during tuning phases to observe the beam spot position and shape. To ensure impedance continuity, a custom-designed beam shielding chamber is required, so that the vacuum chamber remains smooth when the screen is retracted.

A DC current transformer (DCCT) is employed to measure the DC current in the damping ring. The system includes a probe and its associated electronics. The probe is typically installed in the straight sections, and commercially available models with various apertures are available.

Beam loss in the damping ring is monitored using a combination of optical fibers and dual PIN diodes. Optical fibers cover the ring to enable real-time localization of loss events, while dual PIN diodes are deployed at key points for localized radiation monitoring.

Charge monitors such as FCT/ICT are installed at the injection and extraction points of the damping ring to assess overall transmission efficiency. An optical transition radiation (OTR) screen is installed at the extraction point to measure beam spot size and calculate the transverse emittance, ensuring the beam satisfies injection requirements.

### 4.6.4 Feasibility Analysis

Most subsystems, such as BPMs, bunch-by-bunch feedback, beam loss monitors, and synchrotron radiation monitors, are based on mature, standardized, and well-validated technologies already implemented at facilities such as HALF, HEPS, and SuperKEKB. These solutions meet the engineering requirements and are feasible for deployment, posing no significant technical risks. Some of the more demanding technologies, compared to those in existing storage ring light sources or colliders, have already been applied in FEL facilities and have undergone verification through dedicated R&D initiatives and beam test platforms.

Most technology subsystems will rely on domestically procured or independently developed equipment. For subsystems involving imported components—such as zone plates, streak cameras, and precision displacement stages—alternative strategies are under consideration, including diversifying import sources and developing domestic substitutes.

### 4.6.5 Summary

Based on the technical specifications and requirements derived from the STCF accelerator physics design, the beam diagnostics system will provide measurements of parameters including beam position and closed orbit, DC and bunch-by-bunch current, tune (working point), beam transverse profile and emittance, bunch length, and beam losses. These measurements support beam commissioning and accelerator physics studies.

The overall system is largely composed of well-established technologies that have been used at other facilities, with a few cutting-edge components expected to be validated within the next 2-3 years. Whenever possible, the system will utilize domestically produced and self-developed equipment. Future collaboration with leading domestic research institutes and instrumentation manufacturers will be pursued to promote localization and develop viable domestic alternatives.

## 4.7 Interaction Region Superconducting Magnet System

### 4.7.1 Design Requirements and Layout

To achieve extreme compression of the beam spot size at the interaction point (IP) and thereby enhance the luminosity of the STCF collider, a set of high-gradient superconducting quadrupole magnets is placed symmetrically on both sides of the IP, as close to the IP as possible. These magnets are installed within the experimental spectrometer, where space is extremely constrained. Furthermore, they must effectively cancel the solenoidal magnetic field of the detector to avoid adverse effects on the accelerator beam.

According to the accelerator physics requirements, the superconducting quadrupole magnets in the interaction region adopt a double-aperture structure, with a beam crossing angle of 60 mrad. The distance from the front end of the focusing quadrupole magnet to the IP (denoted as $L^*$) is 900 mm (see Section 2.1). A conceptual layout of the Interaction Region Superconducting Magnet (IRSM) is shown in Figure 4.7-1. Symmetrically placed on both sides of the IP, each IRSM system contains two sets of superconducting quadrupole magnets (QD0 and QF1), one anti-solenoid (AS), and two sets of corrector coils. These superconducting magnets operate within a background solenoidal field of 1.0 T provided by the detector spectrometer. The main physical parameters of these magnets are listed in Table 4.7-1.

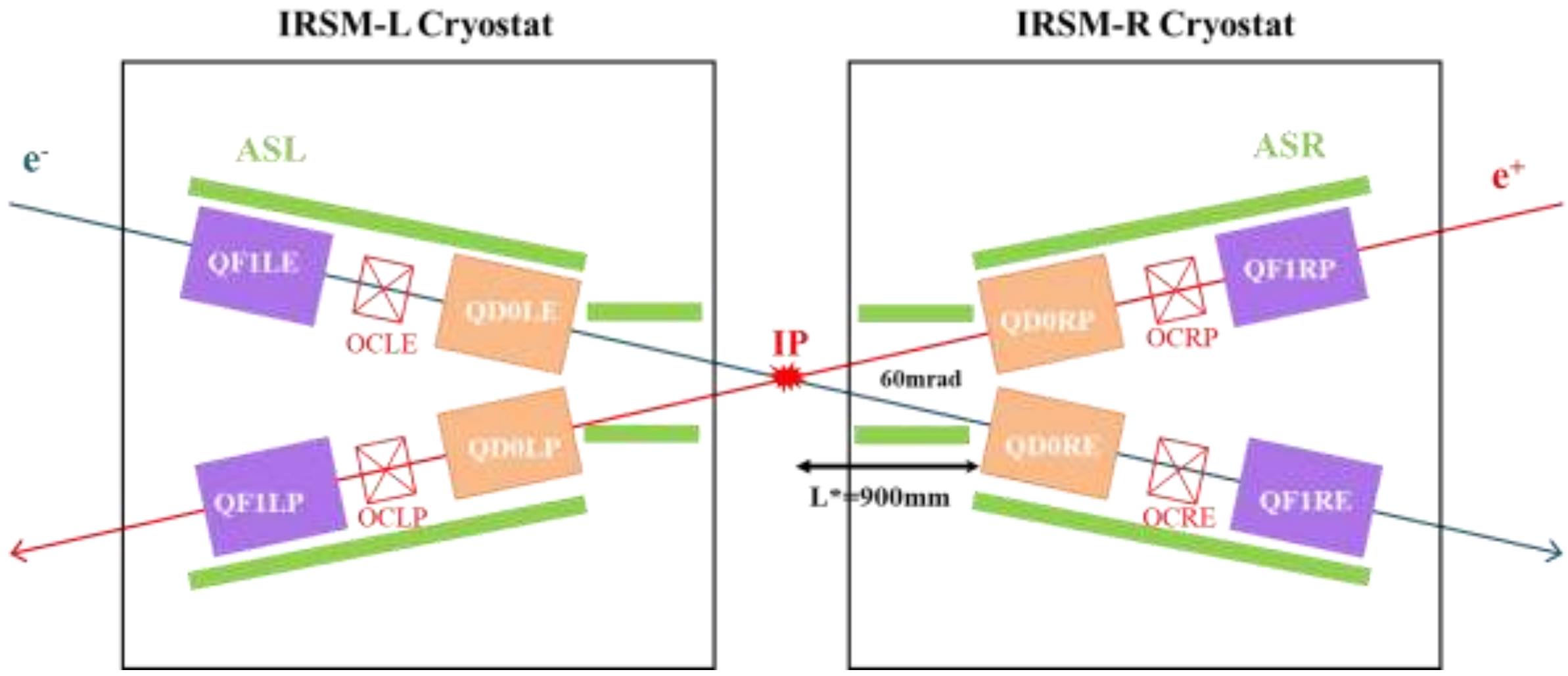

Figure 4.7-1: Conceptual layout of the IR superconducting magnet system (Note: IP = Interaction Point, ASL/ASR = left/right Anti-Solenoid, OCRP/E = Orbit Corrector for Right Positron/Electron beam, QD0RP/E = QD0 coil for Right Positron/Electron, others follow similar naming conventions.)

Table 4.7-1: Requirements for the STCF IR superconducting magnets

| Coil Type | Parameter | Unit | QD0_e+/e- | QF1_e+/e- |
|---|---|---|---|---|
| Quadrupole | Field gradient | T/m | 50 | 40 |
| | Integrated field harmonic (excluding b2) | | ⩽0.2 ‰ @R=10 mm | ⩽0.2 ‰ @R=15 mm |
| | Beam pipe inner radius | mm | 15 | 25 |
| | Magnetic length | mm | 400 | 300 |
| | Distance from IP | mm | 900 | 1800 |
| | Coil inner radius | mm | 20–22 | 20–22 |
| | Coil outer radius | mm | ⩽27 | ⩽27 |
| Correctors | Multipole orders | — | a1, b1, a2 | a1, b1, a2 |
| | Strength | @Rref | 0.016/0.016/0.6 | 0.03/0.03/0.3 |
| Anti-solenoid | Integrated field | — | Within ±2 m from IP, residual field less than 1% of detector solenoidal field | |

### 4.7.2 Key Technologies and Roadmap Selection

Double-aperture superconducting quadrupoles in the IR are a critical and distinguishing feature of new-generation electron-positron colliders. To date, only SuperKEKB in Japan has completed the engineering design and fabrication of such a system. Other projects like FCC-ee at CERN, SCTF in Russia, and CEPC in China are still in the key technology R&D phase.

The focusing quadrupole coils are the core components of the IR quadrupole magnets. Several technical options exist for their construction:

- **Cosine-theta (Cos2θ) coils**, as adopted by SuperKEKB [111]
- **Serpentine coils**, as used in BEPCII [112]
- **Canted Cosine Theta (CCT) coils**, adopted by FCC-ee [113] and SCTF at BINP [114]

Table 4.7-2 compares the IR quadrupole magnet technologies used in various international $e^+e^-$ collider projects. The key technical requirements include high magnetic field gradient, excellent field uniformity, compact structure, and light weight. After evaluation, the STCF has selected the **CCT coil** technology route for the prototype IR quadrupole magnet, due to its flexibility in design, superior magnetic field quality, lighter structure, and smaller end curvature.

Table 4.7-2: Comparison of IR quadrupole magnet technologies in new-generation $e^+e^-$ colliders

| Collider | SuperKEKB | FCC-ee | CEPC | SCTF | STCF |
|---|---|---|---|---|---|
| Country/Region | Japan | Europe | China | Russia | China |
| Beam energy ($e^-/e^+$) [GeV] | 7.0/4.0 | 104.5/104.5 | 120/120 | 1.3/1.3 | 3.5/3.5 |
| Beam crossing angle [mrad] | 83 | 30 | 33 | 60 | 60 |
| L* distance [mm] | 935 | 2200 | 2200 | 905 | 900 |
| Detector field [T] | 1.5 | 2.0 | 3.0 | 1.0 | 1.0 |
| Quadrupole coil tech | Cos2θ | CCT | Cos2θ | CCT | CCT |
| Max gradient [T/m] | 68.9 | 100 | 142.3 | 100 | 50 |
| Magnetic length [mm] | 334 | 1200 | 1210 | 200 | 400 |

## 4.7.3 Design Scheme and System Configuration

### *4.7.3.1 Double-Aperture Superconducting Quadrupole Magnets*

Each of the two quadrupole magnet assemblies includes two vertically focusing magnets (QD0) and two horizontally focusing magnets (QF1). These magnets are symmetrically arranged on both sides of the IP, with a total crossing angle ($2\theta$) of 60 mrad between the electron and positron beams. The front end of the QD0 is located 0.9 m from the IP, while that of QF1 is 1.8 m away. The relative positions of the eight quadrupole magnets are illustrated in Figure 4.7-1, and their design specifications are listed in Table 4.7-1.

Due to the high field gradient and the very compact spatial constraints, the QD0 magnet development is the most challenging, and thus, the main R&D work on superconducting quadrupole coils is focused on QD0. At the entrance of QD0, the center-to-center spacing between the electron and positron beam pipes is 54 mm. Given the end length of the CCT-type magnet and assuming an equivalent field length of 450 mm, the center spacing at a 60 mrad beam crossing angle reduces to 52.52 mm. This implies that the maximum outer diameter of each quadrupole coil is 26.26 mm.

Accounting for the beam pipe wall thickness of 1 mm, installation clearance of 3 mm, and helium channel thickness (installed directly on the inner surface of the QD0) of 1 mm, the minimum required inner diameter of the QD0 coil is 20 mm. The available radial space for coil construction is less than 6.26 mm, making this one of the most challenging components. A schematic of the radial space allocation is shown in Figure 4.7-2.

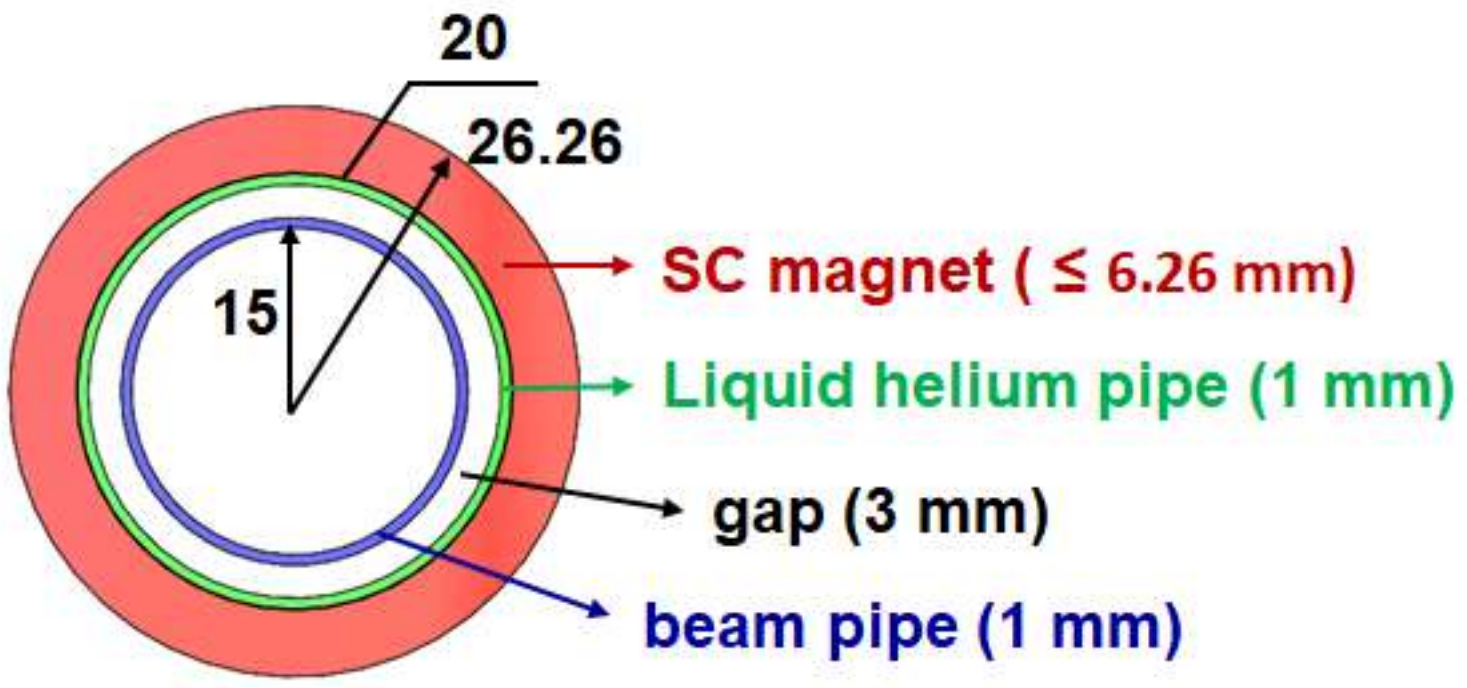

Figure 4.7-2: Schematic diagram of radial space allocation in QD0 magnet

To mitigate magnetic hysteresis loss and enhance field stability, superconducting wires with a fine filament diameter (<10 μm) are prioritized for coil design. The test prototype of the QD0 uses superconducting wire with filaments less than 10 μm in diameter. The main design parameters for the CCT QD0 magnet are listed in Table 4.7-3. The magnetic field contour and current load line are shown in Figure 4.7-3.

The maximum magnetic field on the quadrupole coil is 1.64 T, and the working point is at 67% of the load line. The QD0 coil is wound with superconducting wire of 0.9 mm diameter, using a pattern of 2 strands horizontally and 2 layers vertically, with 4 strands wound in parallel along pre-machined grooves on the CCT former.

Table 4.7-3: Design parameters for the CCT-type QD0 coil

| Parameter | Value |
|---|---|
| Good field region radius (mm) | 10 |
| Coil inner radius (mm) | 20 |
| Max QD0 outer radius (mm) | <26.23 |
| Operating current (A) | 638 |
| Peak field on coil (T) | 1.64 |
| Field gradient (T/m) | 50 |
| Magnetic length (mm) | 400 |
| Superconductor working point (at 4.2 K) | 67% |

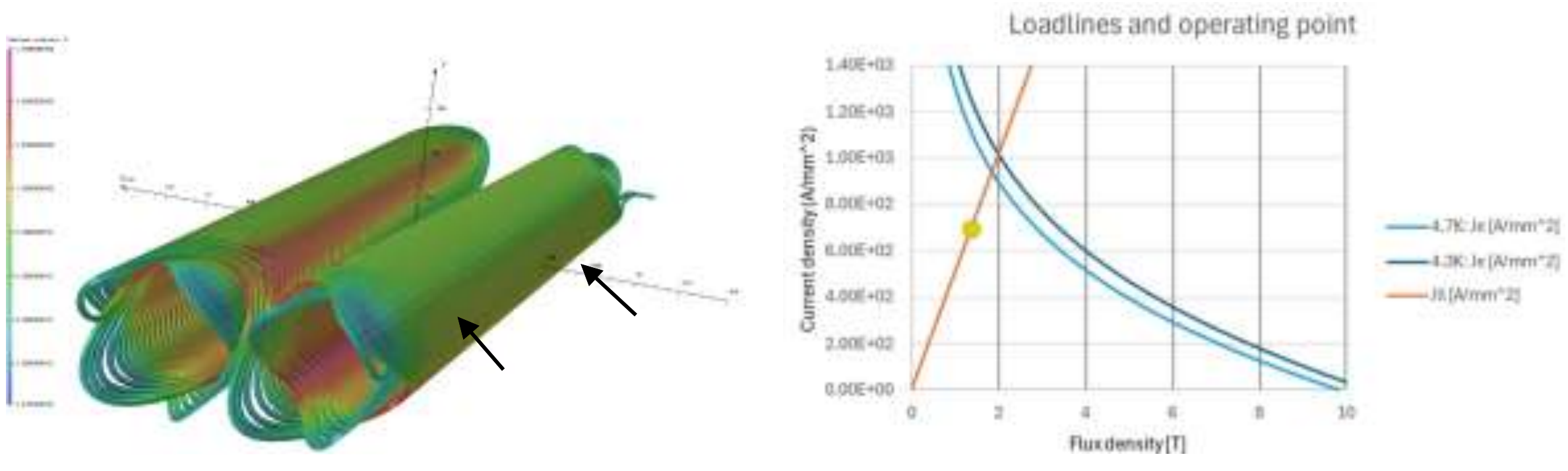

(a) (b)

Figure 4.7-3: (a) Magnetic field distribution on the QD0 double-aperture coil; (b) Current load line

Due to the electromagnetic coupling between the two apertures of the double-aperture quadrupole magnet, a cross-talk compensation technique is applied. This involves introducing reversed harmonic components and optimizing the overall coil trajectory.

$$z = \sum_{n_b}\left[K_n \frac{r\,sin(n_b\,\theta)}{n_b\,tan\,\alpha}\right] + \sum_{n_a}\left[P_n \frac{rcos(n_a\theta)}{n_a\,tan\,\alpha}\right]$$

Table 4.7-4 summarizes the optimized values of the 1st to 10th order harmonics of the QD0 double-aperture superconducting magnet. After optimization, the integrated harmonics of each order are less than 2 units (<0.2‰).

Table 4.7-4: Optimized field harmonics of QD0

| **Harmonic Order** | An (T·m) | Unit (An) | Bn (T·m) | Unit (Bn) |
|---|---|---|---|---|
| 1 | 3.41E-06 | 0.17 | -1.17E-06 | -0.06 |
| 2 | 2.80E-06 | 0.14 | 2.04E-01 | 10000 |
| 3 | -1.56E-06 | -0.08 | -4.27E-06 | -0.21 |
| 4 | -3.47E-06 | -0.17 | -2.64E-07 | -0.01 |
| 5 | -3.32E-06 | -0.16 | 3.87E-06 | 0.19 |
| 6 | 1.13E-06 | 0.06 | -2.98E-06 | -0.15 |
| 7 | 4.58E-07 | 0.02 | -2.94E-06 | -0.14 |
| 8 | -6.61E-07 | -0.03 | 2.99E-06 | 0.15 |
| 9 | 2.20E-07 | 0.01 | -2.70E-07 | -0.01 |
| 10 | -8.51E-08 | -0.00 | 2.97E-07 | 0.01 |

The groove-embedded former structure is designed based on the CCT coil parameters. The inner and outer formers are shown in Figure 4.7-4. Both formers are machined with helical grooves according to the CCT trajectory equations, with opposite winding inclinations. The inner and outer formers are aligned via positioning pins. Jumper and terminal structures are designed into the former to ensure continuous winding and allow for wire joint welding. The maximum mechanical stress in the CCT magnet coil is 1.6 MPa, within the allowable limits.

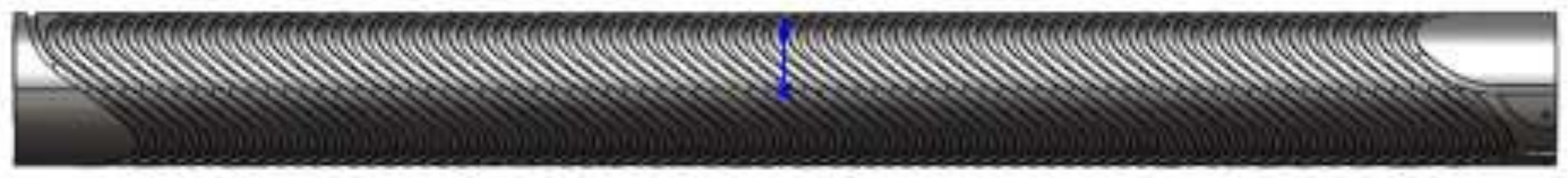

(a)

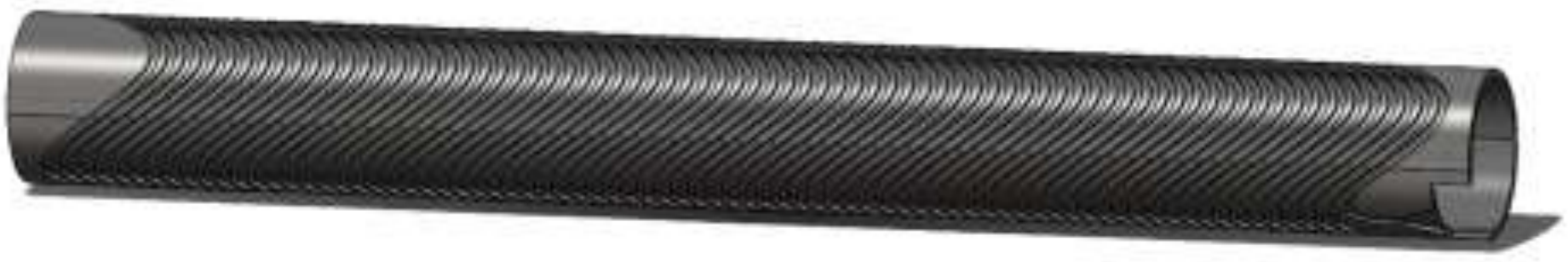

(b)

Figure 4.7-4: (a) Inner former of the CCT superconducting quadrupole magnet; (b) Outer former of the CCT superconducting quadrupole magnet

#### *4.7.3.2 Anti-Field Superconducting Solenoid*

The superconducting magnet system is embedded within the detector and operates under the background field of the detector solenoid of about 1.0 T. To eliminate the influence of the detector magnetic field on the beam trajectory, a set of anti-solenoid coils (field screening coils and field compensation coils) is used to compensate for the integral magnetic field. The specific technical requirements are as follows:

1) The residual integrated field within the beam pipe from the IP to QD0 and after QF1 must be ≤ 0.01 T·m;
2) The residual longitudinal magnetic field within the beam pipe from QD0 to QF1 must be ≤ 300 Gs;
3) The field distribution of the anti-solenoid must satisfy the vertical divergence constraint for the beams;
4) The axial layout and outer diameter must meet the mechanical detector interface (MDI) spatial constraints.

The anti-solenoid coils adopt a technical approach of "NbTi conductor low-temperature superconducting coil + vacuum impregnation process." The coils are distributed in a stepped arrangement along the central axis and are powered in series. An additional current lead is included in the central coils to allow fine adjustment of the integral field during beam commissioning and operation.

As a component of the combination magnet, the anti-solenoid coils for field screening are placed between the quadrupole magnets and the outer cryostat. Its structure consists of the NbTi anti-solenoid coils, coil former, mechanical supports, and current leads. Specifically:

1) The anti-solenoid coils provide a counteracting field to the spectrometer field, and their parameters are listed in Table 4.7-5;
2) The coil former ensures the geometric profile of the coils;
3) The support structure connects the coil former to the combination magnet cryostat, providing mechanical stability and thermal isolation;
4) End current leads power the anti-solenoid, while a center-tapped lead allows for integral field tuning.

Figure 4.7-5(a) shows the 1/2 model layout of the detector coil and the anti-solenoid (right side only), with spectrometer yoke iron hidden for clarity. Figure 4.7-5(b) shows the magnetic field distribution: the blue curve represents the spectrometer field, red represents the anti-solenoid field, and black shows the resulting field after superposition. The QD0 and QF1 locations (marked by black dots) exhibit nearly zero net axial field, confirming successful integral compensation.

Table 4.7-5: Design parameters for anti-field solenoid coils

| **Superconductor parameters** | | **Coil parameters** | |
|---|---|---|---|
| Bare wire size | 1.38×0.93 $mm^2$ | Operating current | 231 A |
| Insulated size | 1.48×1.03 $mm^2$ | Peak field | 4.11 T |
| Cu/SC ratio | 1.7 | Io/Ic | 20% |
| RRR | 150 | Number of coils | 15 |
| Filament count | 1500 | Coil length | 1.905 m |
| Filament diameter | 19 μm | Inductance | 3.0 H |
| Critical current (4.2K, 5T) | 1400 A | Stored energy | 0.081 MJ |

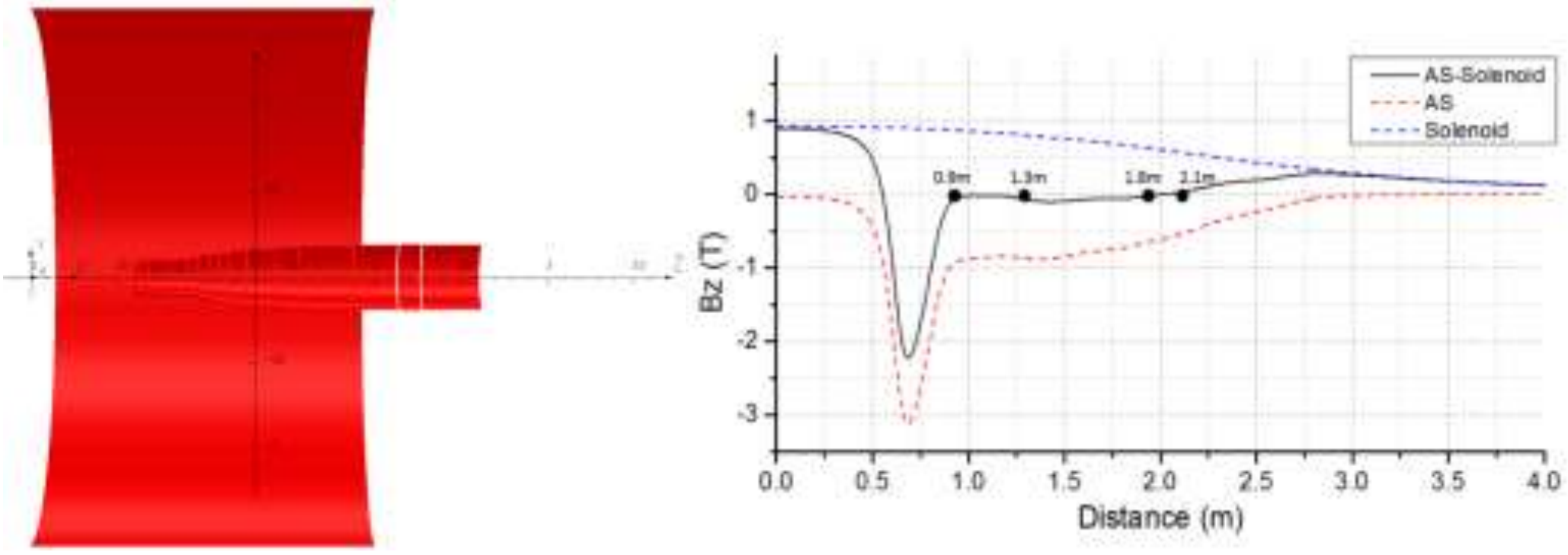

（a）　　　　　　　　　　　　　　　　（b）

Figure 4. 7-5: (a) Layout of anti-solenoid coils (yoke hidden); (b) Magnetic field distribution after compensation

#### *4.7.3.3 Corrector Magnets and Coils*

Among different accelerator magnets, harmonic correction coils (or harmonic trim coils) are essential components. They are typically categorized into field correction coils and orbit correction coils, as specified in Table 4.7-1.

Field correction coils are used to improve magnetic field quality and correct quadrupole misalignments. By generating higher-order fields that cancel the intrinsic multipole components of the magnet, they enhance field uniformity and stability, thereby ensuring stable particle trajectories.

Orbit correction coils mainly include low-order correction coils, which apply forces or coupling terms to adjust the dispersion function and correct motion coupling induced by alignment and magnetic errors. The a1 and b1 coils generate horizontal and vertical dipole fields, respectively, to correct beam orbits and control dispersion. The a2 coil provides a skew quadrupole field to mitigate x-y coupling.

Due to the space constraints imposed by the interaction region layout (e.g., crossing angle and *L**), only about 6 mm radial space is available, which is insufficient to install both the quadrupole and correction coils. As a result, a self-compensating harmonic design of the CCT-type quadrupole coil is employed, leveraging high-precision CNC machining of the coil former. This approach may eliminate the need for dedicated field correction coils. However, saddle-type coils are reserved as candidates for low-order beam-based corrections.

Currently, the initial specifications for orbit correction coils, including order and strength, are summarized in Table 4.7-6. The correction coils are positioned longitudinally between QD0 and QF1 and are wound in single-layer configurations with low current. The inner radius of these trim coils is the same as the quadrupole magnets (20 mm), and at a reference radius of 10 mm, the a1 and b1 design strengths for QD0 are 0.016 T·m. The design and simulated magnetic field distribution of the a1 (b1) coil is shown in Figure 4.7-6. The design parameters will continue to evolve as the machine optics and physics requirements become more refined.

Table 4.7-6: Specifications for the beam correction coils

| Magnet | Rref (mm) | A1 (T·m) | B1 (T·m) | A2 (T) | B2 (T) |
|---|---|---|---|---|---|
| QD0 | 10 | 0.016 | 0.016 | 0.6 | 0.6 |
| QF1 | 15 | 0.03 | 0.03 | 0.3 | 0.3 |

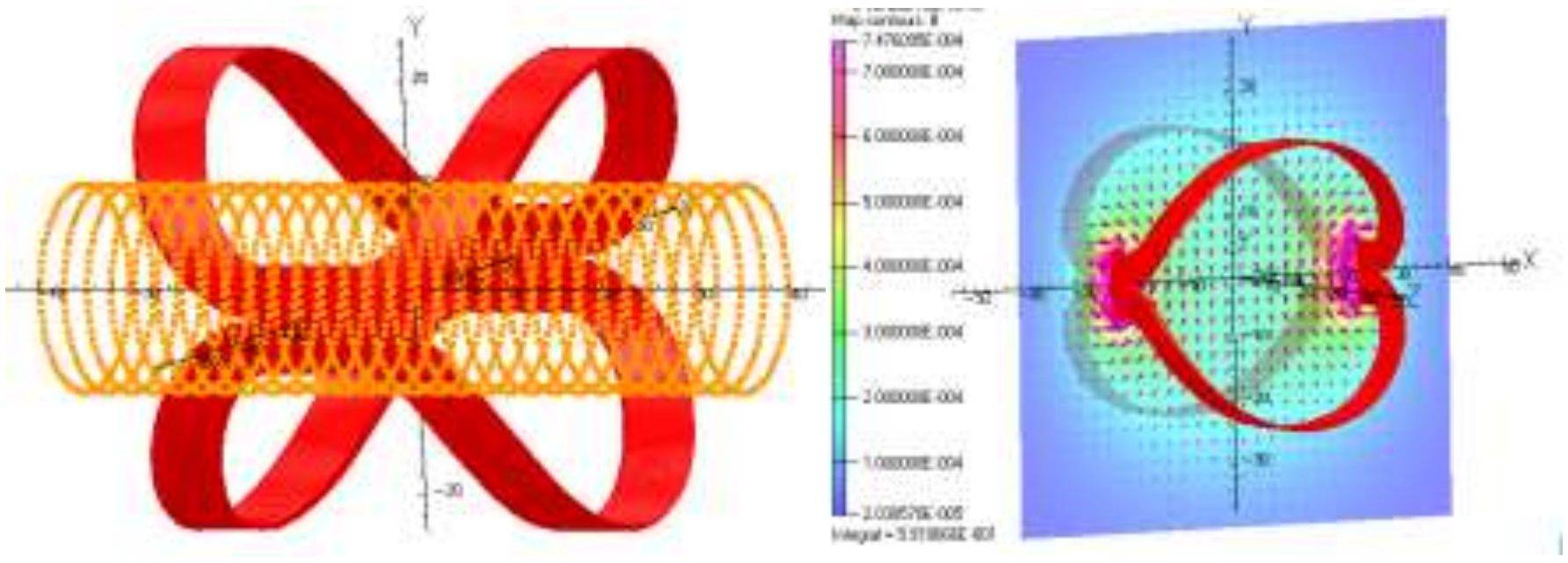

Figure 4.7-6: Preliminary design and field map of a1 (b1) correction coil

#### *4.7.3.4 Cryostat for the Interaction Region Superconducting Magnet System*

To meet the cryogenic requirements of the IRSM, the cryogenic system provides 4.5 K subcooled helium to cool the superconducting magnets and their current leads, and simultaneously offers sturdy mechanical support to extend the superconducting magnets into the detector. The system consists of cryostats, distribution valve boxes, and transfer lines. Each cryostat handles a thermal load of about 37.2 W and is cooled by a helium refrigerator with a total cooling capacity of 1000 W at 4.5 K. The refrigerator delivers up to 24 g/s of subcooled liquid helium to each IRSM system.

The core technical challenges for the IRSM cryogenic system are thermal insulation of components and the precise allocation and control of cryogenic working fluids. The technical route to address these challenges involves the following steps:

1) Based on the magnet and current lead configuration, the structure of the cryostat is determined and designed for thermal insulation. The static thermal load at 4.2 K is analyzed, and combined with the dynamic thermal load from magnet operation, the total cooling power requirement is defined.
2) From the cooling requirement, the subcooled helium flow rate and pressure are estimated. These parameters are then used to design the cryogenic transfer lines, whose insulation structure is also optimized to minimize transmission losses.
3) Using the determined flow and pressure requirements, the flow scheme of the distribution valve box is designed to allocate, regulate, and recycle the cryogen efficiently. Adequate monitoring and control components are installed to ensure precise system regulation.
4) Based on the valve box flow scheme, its structure is finalized and thermally insulated. The cooling losses in the distribution stage are then calculated, and the final total cooling requirement of the system is determined, which in turn defines the upper-level refrigerator capacity.

Referencing the cryostat design for SuperKEKB's interaction region magnets [115, 116], and adapted for the STCF IR superconducting magnet geometry, the cryostat is composed of an outer shell, liquid helium vessel, liquid nitrogen radiation shields, and a support structure. The internal cross-section is shown in Figure 4.7-7. The cryogenic structure of the magnet cryostat is illustrated in Figure 4.7-8. The liquid helium vessel is suspended by eight titanium alloy (Ti-6Al-4V) rods installed at inclined angles. This layout ensures that the relative positions of support points remain unchanged upon cool-down to 4.5 K.

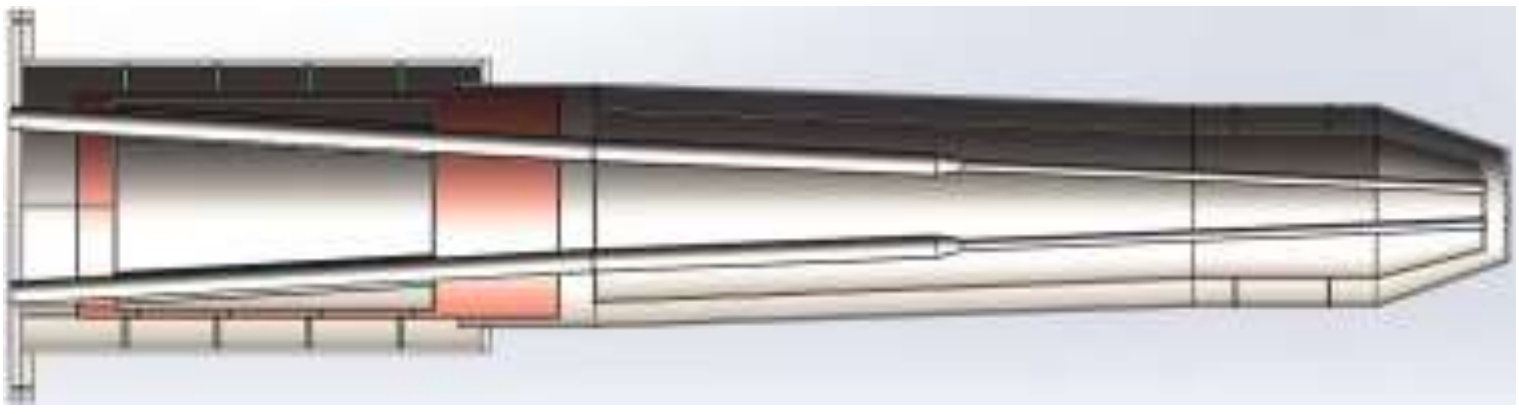

Figure 4.7-7: Cross-sectional view of the IR magnet cryostat

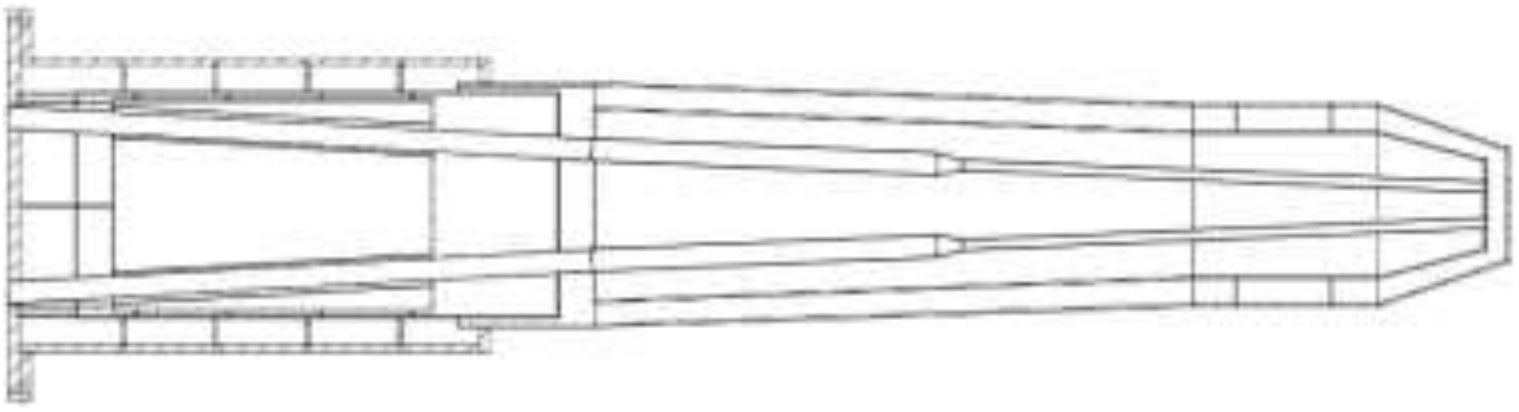

Figure 4.7-8: Structural diagram of the IR magnet cryostat

The cryogenic distribution system layout depends on the spatial configuration of the cryostats and the refrigerator. It consumes cooling power during operation, and thus, the structure, process, and scale of the valve boxes and transfer pipelines directly affect the overall system performance. The transfer and distribution system must meet the following technical specifications: room-temperature vacuum pressure $< 1.0 \times 10^4$ Pa and a total leak rate $< 5.0 \times 10^{-9}$ Pa·m$^3$/s. Within the valve box, helium flow is split, collected, regulated, and routed as required. The cryostat and valve box are connected by a 4-channel cryogenic transfer line, with a cross-sectional structure shown in Figure 4.7-9. The heat load of the 4.5 K line is less than 0.3 W/m. Liquid nitrogen at 77 K is split into two paths in the valve box, supplying the valve box radiation shield and the shields on the cryostat via the transfer line. Subcooled helium at 4.5 K flows to the cryostat, cools the magnet, and then returns via a helium vapor return line to the valve box.

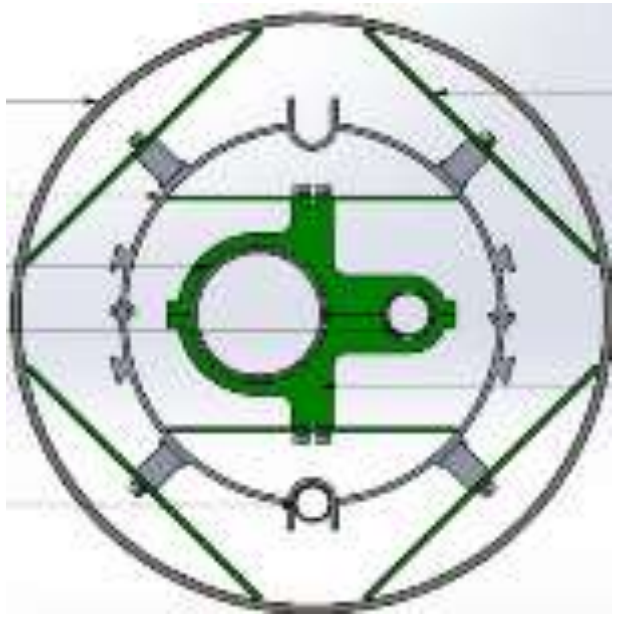

Figure 4.7-9: Cross-sectional view of cryogenic transfer line

### 4.7.4 Feasibility Analysis

All superconducting magnets will be based on domestically available NbTi superconducting wire, avoiding any risk of foreign export restrictions.

For the superconducting quadrupole coils, the design target has been set more stringently than the technical requirement to provide margin against possible magnetic field quality deviations during production. As the groove shape of the CCT coil support structure determines the conductor path and magnetic field quality, tight control over the machining precision of the coil formers is essential. Furthermore, careful control over coil fabrication and assembly tolerances is necessary to prevent degradation of magnetic field quality due to misalignments. The prototyping effort on the QD0 magnet will explore and verify the CCT-based technical path, providing the experience needed for full-scale development. The high-gradient, high-field-quality double-aperture quadrupole coils are the core of the IR superconducting magnet system. The CCT approach has already undergone prototype validation in the FCC-ee project, and the STCF team will draw on both international developments and domestic expertise through multiple expert reviews and prototype iterations.

The anti-solenoid will use an integrated stepped support structure, with pull rod lifting mechanisms and mounting holes at the ends. The support will be fabricated using a welded structure with full machining, ensuring good coaxial alignment across the multiple stepped coils. The anti-solenoid itself will be vacuum-impregnated to yield a compact, precise, high-strength coil capable of generating a high-quality reverse field and resisting the electromagnetic forces from coupling with the spectrometer magnet. The design, fabrication, and testing of NbTi-based anti-solenoid coils are mature technologies in China and pose no technical risk.

Orbit correction coils are designed based on SuperKEKB specifications. Their technical route is similar to the superconducting quadrupole coils, but the configuration is simplified using single-turn, single-layer saddle coils, making them highly feasible.

Cryogenic cooling for superconducting magnets in the interaction region is a mature technology with widespread deployment in both domestic and international accelerators. Particularly in the SuperKEKB and BEPCII-U projects, comprehensive simulation and experimental studies have clarified the thermal-hydraulic characteristics of subcooled helium and magnet behavior during cooldown, steady operation, and warmup. Proven technologies and protocols are already in place. As the technical requirements of STCF are consistent with those of these established projects, the cryogenic system and distribution infrastructure are considered mature and technically feasible.

### 4.7.5 Summary

The core component of the IR superconducting magnet system, the high-gradient quadrupole coil, has completed conceptual design and meets the STCF requirements in terms of field gradient and field quality. Prototype development of the CCT-type coil is underway, with work focused on electromagnetic optimization of the double-aperture layout and harmonics control. Other components, such as the anti-solenoid and correction coils, use mature technologies or off-the-shelf solutions, ensuring manageable technical risk for the full system.

## 4.8 Electron Source System

### 4.8.1 Design Requirements and Specifications

The primary options for the STCF electron source system are thermionic grid-controlled electron guns and photocathode RF electron guns. These are critical components of the STCF injector. The injector schemes for both the off-axis injection and bunch swap-out injection schemes in the collider rings have been studied and retained in the overall accelerator design.

In the off-axis injection scheme, the directly injected electron beam is generated by an S-band photocathode electron gun with 1.5 nC bunch charge, while the positron beam is produced by accelerating 10 nC bunches from a thermionic electron gun to 1.5 GeV for target bombardment. In the bunch swap-out injection scheme, the electron beam is generated by either a thermionic or an L-band photocathode electron gun with an 8.5 nC bunch charge, and the positron beam is produced by accelerating 11.6 nC bunches from a thermionic electron gun to 2.5 GeV before striking the target.

A compatible injector scheme supporting both the injection options has been finalized, with initial implementation using the off-axis scheme. In this configuration, the electron beam directly injected into the collider electron ring is generated by an L-band photocathode gun producing 1.0 nC per bunch. The positron beam is created using an 11.6 nC bunch from a thermionic gun accelerated to 1.0 GeV before striking the target. In the bunch swap-out scheme, the same L-band photocathode gun is used to provide 8.5 nC per bunch, while the positron beam is derived from the 11.6 nC per bunch from a thermionic gun accelerated to 2.5 GeV before striking the target. The bunch swap-out injection imposes significantly higher demands on the electron source system compared to the off-axis injection.

The design requirements and specifications for the electron guns in the compatible injector scheme with respect to both off-axis and swap-out injections are shown in Table 4.8-1.

Table 4.8-1: Main Specifications of Thermionic and Photocathode Electron Guns

| Injection Scheme | Electron Source | Expected Specification |
|---|---|---|
| Off-axis Injection | Thermionic | Beam current ⩾ 16 A, anode voltage > 150 kV, repetition rate 30 Hz |
| | Photocathode | L-band (1300 MHz) gun, static vacuum ⩽ $1\times10^{-7}$ Pa, bunch charge ⩾ 1.0 nC |
| Swap-out Injection | Thermionic | Beam current ⩾ 16 A, anode voltage > 150 kV, repetition rate 90 Hz |
| | Photocathode | L-band (1300 MHz) gun, bunch charge ⩾ 8.5 nC, repetition rate 30 Hz |

## 4.8.2 Key Technologies and Technical Roadmap

### *4.8.2.1 Thermionic Electron Gun: Physical Design and Optimization*

The thermionic electron gun system consists of the gun itself, the cathode-grid assembly, and a pulsed high-voltage supply up to 200 kV. The gun design must address factors such as beam energy, structural voltage hold-off capability, high-current stable emission, beam transverse size and divergence angle, ease of fabrication, assembly, and maintenance, long operational lifetime, and reliability. Achieving high current requires ultra-high vacuum, which must be given special attention.

Using the EGUN simulation software, the gun design was optimized. With a perveance of 0.2 μP and 200 kV voltage, it is capable of delivering electron bunches of up to 16 nC with beam quality suitable for the injector. Subharmonic buncher and accelerator system design were carried out in coordination with beam dynamics studies to ensure the feasibility and appropriateness of the electron gun parameters.

### *4.8.2.2 Photocathode Electron Gun: Core Technologies and Physical Design*

The generation of 1.0 nC per bunch using an L-band photocathode microwave gun is a mature technology and is applicable to the off-axis injection scheme. The bunch swap-out scheme requires at least 8.5 nC per bunch, and plans are in place to use an L-band gun to meet this requirement.

The core technologies involved include the structural design of the RF electron gun, photocathode preparation, and load-lock system, and laser shaping schemes. Optimization is also needed for the gun's acceleration, focusing, and laser beam shaping.

### *4.8.2.3 Technical Roadmap Selection*

For the thermionic gun system, EGUN simulations were used to perform design optimization. Subharmonic bunchers and accelerator sections were also designed with beam dynamics to validate performance.

For the photocathode source technology, the feasibility of using $Cs_2Te$ semiconductor cathodes in an L-band gun to produce 8.5 nC and 1.0 nC per bunch has been confirmed. In the future, to generate polarized electron beams, GaAs photocathodes may be considered.

### 4.8.3 Design Scheme and System Composition

The electron source system consists of both the thermionic electron gun system and the photocathode electron gun system.

#### *4.8.3.1 Thermionic Electron Gun Design Scheme*

CST simulations for the electron gun have been performed, and the results shown in Figure 4.8-1 indicate that the extracted beam charge and quality meet the requirements of the STCF injector.

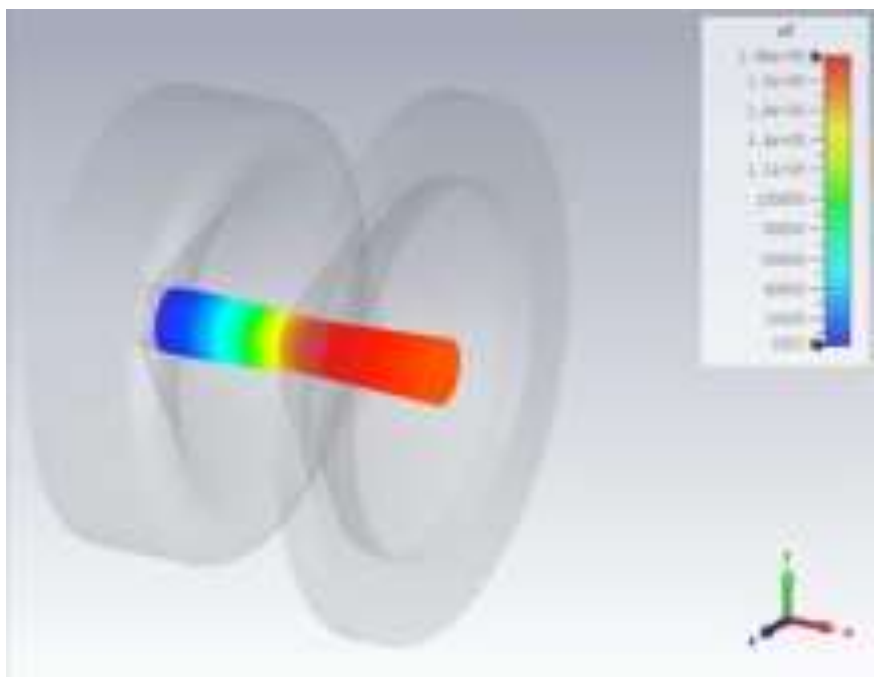

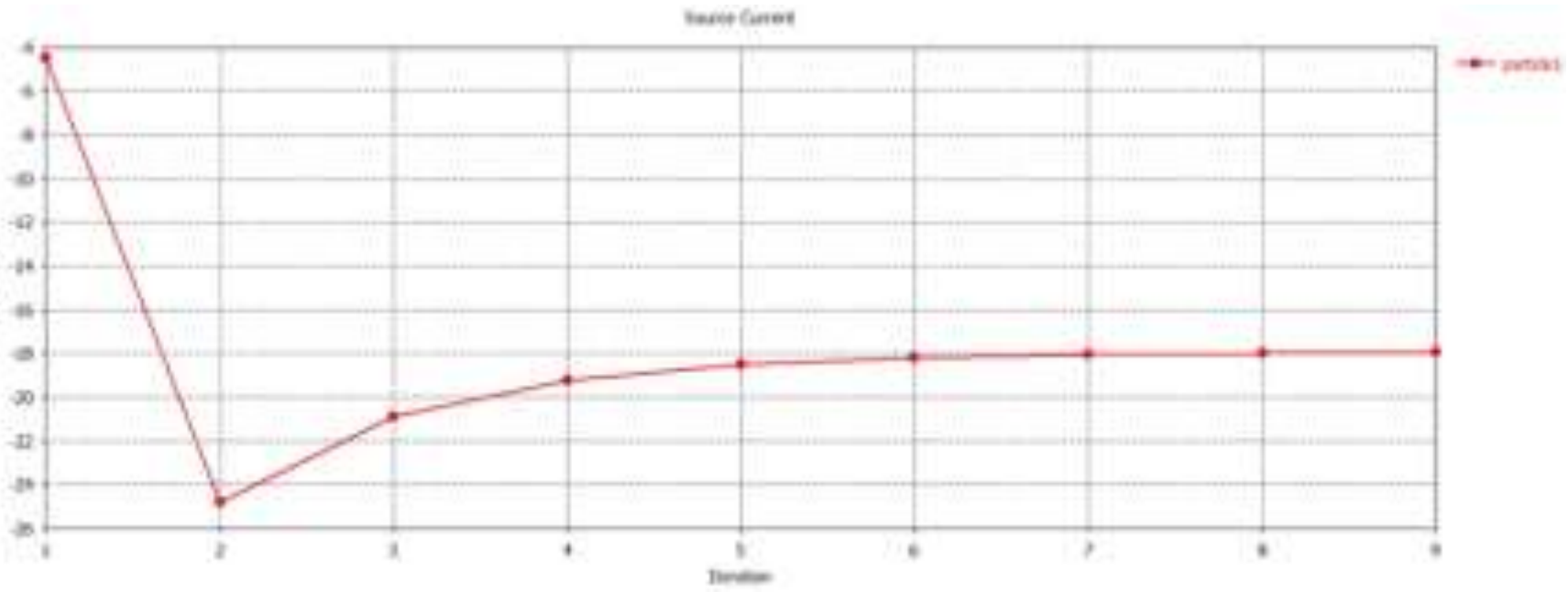

Figure 4.8-1: 200 kV Extraction Simulation Result for Thermionic Electron Gun

The gun cavity is primarily composed of the cathode, anode, focusing electrode, insulating ceramic tube, corona rings, and structural support. To prevent breakdown, corona rings are installed on both the vacuum (inner) and atmospheric (outer) sides of the welded area between the ceramic tube and metal. The 3D design of the thermionic electron gun is shown in Figure 4.8-2.

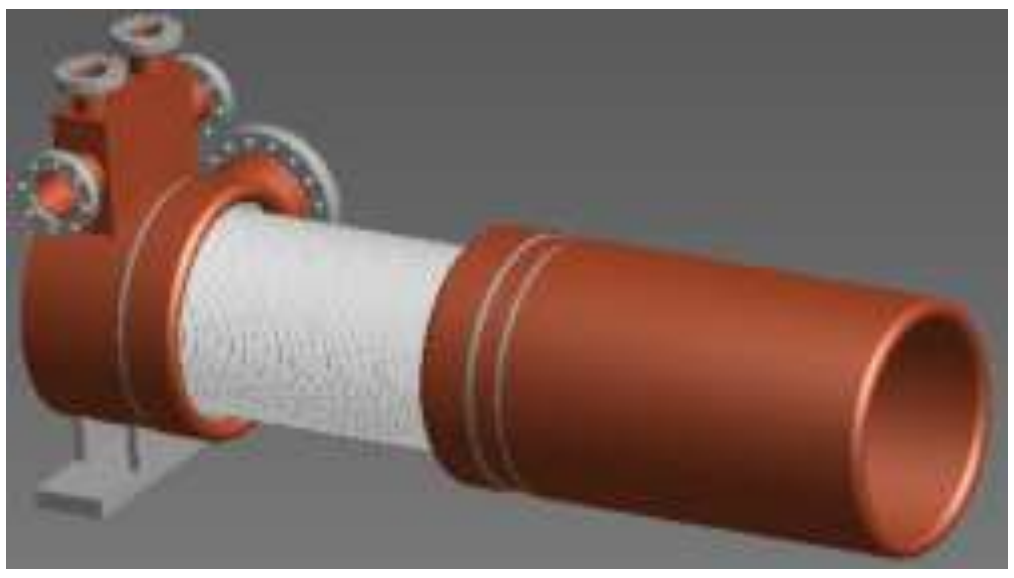

Figure 4.8-2: 3D Design of the Thermionic Electron Gun

This thermionic gun uses the EIMAC Y796 cathode-grid assembly or a component of equivalent specifications. The beam pulse generator for the electron gun includes a DC power supply, a control box, and a pulser. The pulser's DC supply is continuously adjustable from 0 to 1 kV, with the pulse having a full width at half maximum of approximately 1.1 ns and a base width of around 1.6 ns.

#### *4.8.3.2 Photocathode Electron Gun Design Scheme*

The photocathode electron gun used for direct electron beam injection under both the off-axis and bunch swap-out schemes is an L-band gun, as illustrated in Figure 4.8-3. This gun is compatible with high-quantum-efficiency semiconductor photocathodes. The L-band cavity is relatively large, and the aperture between the full and half cells provides sufficient space to support the generation of high-charge electron bunches.

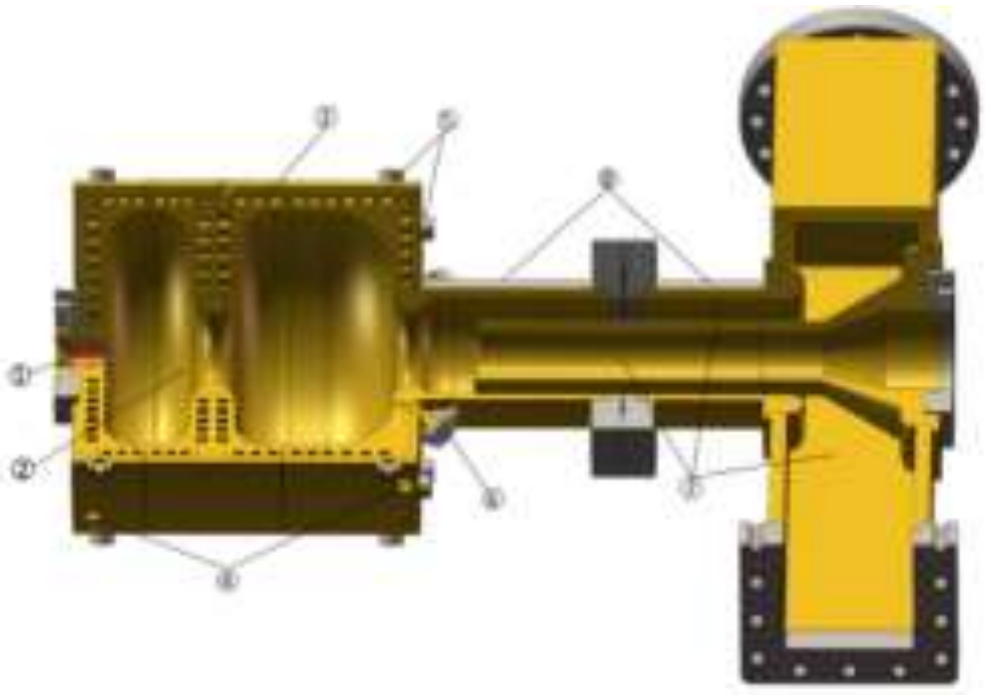

Figure 4.8-3: Structural Schematic of the L-band Photocathode Electron Gun

### 4.8.4 Feasibility Analysis

Thermionic electron guns are mature technologies that have been widely used in accelerator facilities such as SuperKEKB and BEPCII [1]. It is well established that at 150 kV, beam currents of approximately 12 A can be extracted. Simulations confirm that 16 A can be extracted at 200 kV, making the design feasible. The remaining key technologies are also mature, indicating a high overall feasibility.

For the L-band photocathode gun, Tsinghua University's accelerator laboratory has extensive experience in its design and operation. Internationally, groups such as one at PITZ, Germany, have conducted long-term and in-depth studies on L-band guns. These guns have been deployed in large-scale facilities such as the European XFEL. At Argonne National Laboratory, the AWA project developed L-band photocathode guns for the operation of tens-of-nC-level bunch charge. Tsinghua University has also designed L-band guns [118] for China Academy of Engineering Physics and the SHINE facility in Shanghai. There are established precedents for high-charge optimization of L-band guns to support the operation of a single-bunch charge of about 10 nC.

For polarized electron sources, design approaches can refer to international peers. For example, Jefferson Lab operates a polarized source [119], and the EIC project has proposed new-generation polarized sources capable of 7-16 nC per bunch. This project will continue to survey and analyze these validated and emerging technologies to support upgrade path decisions.

### 4.8.5 Summary

The thermionic electron gun system consists of the electron gun, the cathode-grid assembly, and a 200 kV pulsed high-voltage supply. Aside from the gun design itself, achieving high current relies heavily on ultra-high vacuum, which must be a key focus during system design. Subharmonic bunchers and accelerator sections are also designed with beam dynamics studies to verify and confirm the feasibility and reasonableness of gun parameters.

The L-band high-gradient RF gun is the core of the photocathode electron source and determines the upper limits of bunch charge, repetition rate, and beam quality. Innovations have been made in the structural design of the L-band photocathode gun to support large-charge semiconductor cathodes, including integration with the cathode load-lock system.

For the L-band gun, cavity geometry has been optimized to produce high-charge bunches with sufficiently good beam quality. The gun must also support long-term stable operation with high-quantum-efficiency photocathodes to produce large-charge, high-quality electron beams.

## 4.9 Linac RF System

### 4.9.1 Design Requirements and Specifications

The accelerating structures of the microwave systems not only provide the strong electric fields crucial for electron/positron beam acceleration and bunching but also act as vacuum components, ensuring the requisite vacuum environment during the beam’s transit. The reference signal synchronization system delivers low-phase-noise reference signals across different RF stations of the facility, synchronizing the microwave and clock systems from multiple power sources, and provides timing references for other time-sensitive subsystems of STCF. The digital low-level RF (LLRF) system ensures that the amplitude and phase of the accelerating field are automatically locked near the set values and forms a closed feedback loop with the beam pulse. The key parameters and design targets are summarized in Table 4.9-1.

Table 4.9-1: RF System Parameters for the Linacs

| Parameter | Value |
|---|---|
| Main operating frequency (MHz) | 2998.2 |
| Accelerating structure operating temperature (°C) | 35 |
| Effective gradient of accelerating structures (conventional for electrons) (MV/m) | $\geqslant$ 20 |
| Pulse mode | Single-bunch acceleration |
| Aperture of positron accelerating structures (mm) | 30 |
| Effective gradient of positron accelerating structures (MV/m) | $\geqslant$ 15 |
| Phase stability (rms, online closed loop) (°) | 0.09 |
| Amplitude stability (rms, online closed loop) | 0.04% |
| Phase noise jitter at reference signal input (rms) (fs) | $\leqslant$30 |
| Synchronization accuracy (long-term) (fs) | $\leqslant$50 @24h (rms) |

### 4.9.2 Key Technologies and Technical Path

Since STCF linacs operate predominantly in pulsed modes, the baseline design adopts room-temperature traveling wave accelerating structures and pulsed power operation (with pulse compressors in the waveguide system), which offers a high performance-to-cost ratio. Considering the high bunch charge of the injector, the S-band frequency is chosen as the principal frequency for the linacs, with 2998.2 MHz selected as the primary operating frequency, six times the collider ring RF frequency of 499.7 MHz.

### 4.9.3 System Design

The RF system is designed to match the two different injector configurations that are studied in parallel: the injector scheme for the off-axis injection scheme in the collider rings and the injector for the bunch swap-out injection scheme. Because for the swap-out injection, there is more than one linac scheme for the electron beam to be injected into the collider electron ring, the RF system design described here corresponds to the one using a thermionic electron source.

#### *4.9.3.1 Off-axis Injection Scheme Microwave System Design*

The microwave system is organized into acceleration sections. Table 4.9-2 lists the types of accelerator structures, number of structures, peak klystron power requirements, and power distribution strategies for each beamline segment. The final beam energy at the end of each segment and the overall topological layout are shown in Figure 4.9-1. In the off-axis injection scheme, the macropulse repetition rate of the klystrons is 30 Hz. According to the physical design, each 3-meter conventional accelerating structure contributes 50 MeV of energy gain.

Table 4.9-2: Microwave Power Distribution for the Off-axis Injection Scheme

| Beamline | Frequency (MHz) | Acc. Structure Type | Quantity | Peak Power per Klystron (MW) | RF Power Source to Acc. Structures Ratio | Number of RF Power Sources |
|---|---|---|---|---|---|---|
| PGun | 2998.2 | RF Gun | 1 | 10 | 1:1 | 1 |
| | 2998.2 | 3-m Std. Acc. Structure | 4 | 50 | 1:2 | 2 |
| | 11992.8 | X-band Acc. Structure | 1 | 6 | 1:1 | 1 |
| TGun | 166.57 | Subharmonic Pre-buncher (SHB1) | 1 | - | 1:1 | 1 |
| | 499.7 | Subharmonic Pre-buncher (SHB2) | 1 | - | 1:1 | 1 |
| | 2998.2 | Buncher | 1 | | Shared use of next RF power source | 1 |
| | 2998.2 | 3-m Std. Acc. Structure | 4 | 50 | 1:2 | 2 |
| EL1 | 2998.2 | 3-m Std. Acc. Structure | 16 | 50 | 1:2 | 8 |
| EL2 | 2998.2 | 3-m Std. Acc. Structure | 10 | 50 | 1:2 | 5 |
| PAS | 2998.2 | 2-m Large Aperture Acc. Structure | 1 | 50 | 1:1 | 1 |
| | 2998.2 | 3-m Large Aperture Acc. Structure | 14 | 50 | 1:2 | 7 |
| PL1 | 2998.2 | 3-m Std. Acc. Structure | 8 | 50 | 1:2 | 4 |
| ML | 2998.2 | 3-m Std. Acc. Structure | 50 | 50 | 1:2 | 25 |

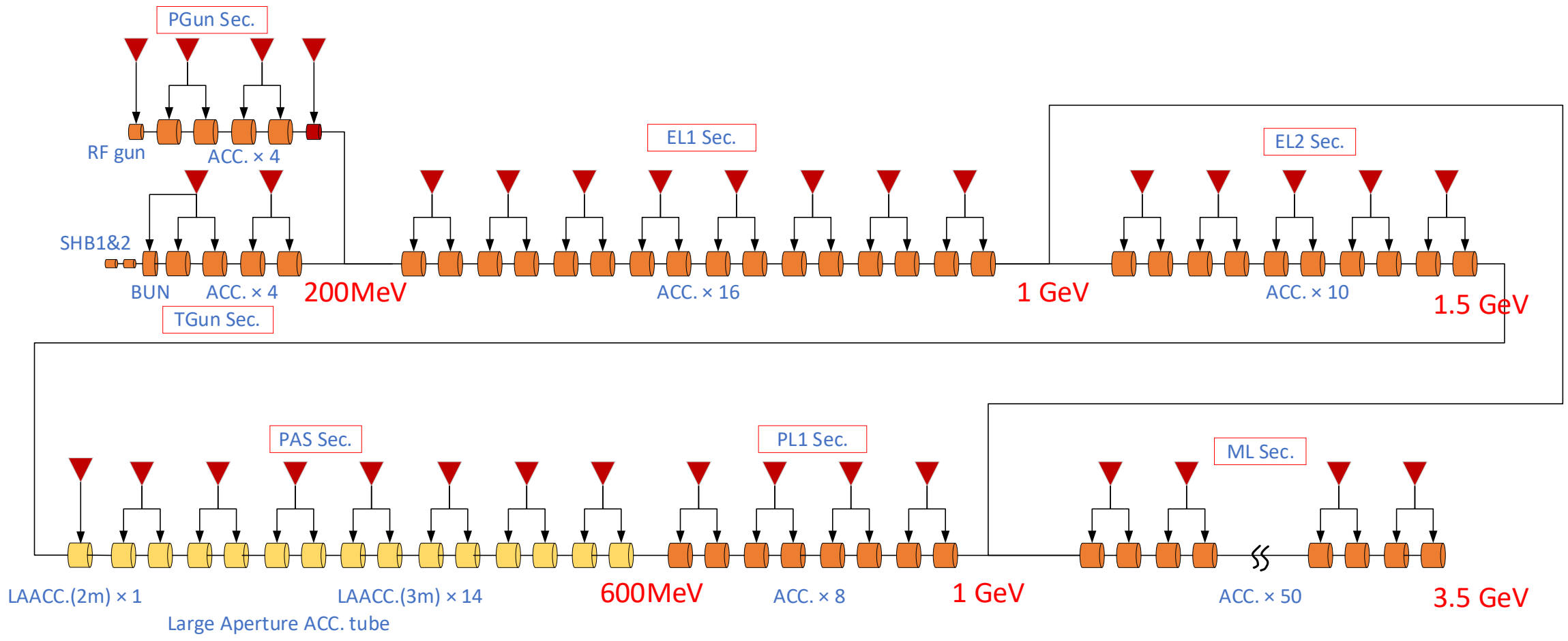

Figure 4.9-1: Layout of Accelerating Structures and RF Power Sources for the Off-axis Injection Scheme

The diagram referenced in Figure 4.9-1 illustrates the sequential arrangement of the accelerating structures and the distribution of RF sources across the various linac sections in the off-axis injection scheme.

#### *4.9.3.2 Swap-out Injection Scheme Microwave System Design*

The swap-out injection scheme requires more powerful linacs. The layout is similarly divided by acceleration sections, each with its corresponding number of accelerating structures, RF power sources, and macropulse repetition rates. Table 4.9-3 provides the detailed configuration. The beam energy at the end of each section and the overall topological layout are illustrated in Figure 4.9-2. According to the physical design of the injector, each 3-meter standard accelerating structure contributes 50 MeV of energy gain.

Table 4.9-3: Swap-out Injection Scheme — Accelerating Structures and Microwave Power Distribution

| Beamline | Frequency (MHz) | Acc. structure Type | Quantity | Peak Power per Klystron (MW) | Power Source to Acc. Structures Ratio | Repetition Rate (Hz) | Number of RF Sources |
|---|---|---|---|---|---|---|---|
| TGun | 166.57 | Subharmonic Pre-buncher (SHB1) | 1 | – | 1:1 | 90 | 1 |
| | 499.7 | Subharmonic Pre-buncher (SHB2) | 1 | – | 1:1 | 90 | 1 |
| | 2998.2 | Buncher | 1 | – | Shared use of next RF power source | 90 | – |
| | 2998.2 | 3-m Std. Standard Acc. Structure | 4 | 50 | 1:2 | 90 | 2 |
| EL1 | 2998.2 | 3-m Std. Standard Acc. Structure | 46 | 50 | 1:2 | 90 | 23 |
| EL2 | 166.57 | Subharmonic Pre-buncher (SHB1) | 1 | – | 1:1 | 30 | 1 |
| | 499.7 | Subharmonic Pre-buncher (SHB2) | 1 | – | 1:1 | 30 | 1 |
| | 2998.2 | Buncher | 1 | - | Shared use of next RF power source | 30 | 0 |
| PAS | 2998.2 | 3-m Std. Standard Acc. Structure | 20 | 50 | 1:20 | 90 | 110 |
| | 2998.2 | 2-m Large Aperture Acc. Structure | 14 | 50 | 1:2 | 90 | 7 |
| PL1 | 2998.2 | 3-m Large Aperture Acc. Structure | 8 | 50 | 1:2 | 90 | 4 |
| ML | 2998.2 | 3-m Std. Standard Acc. Structure | 14 | 50 | 1:2 | 30 | 25 |

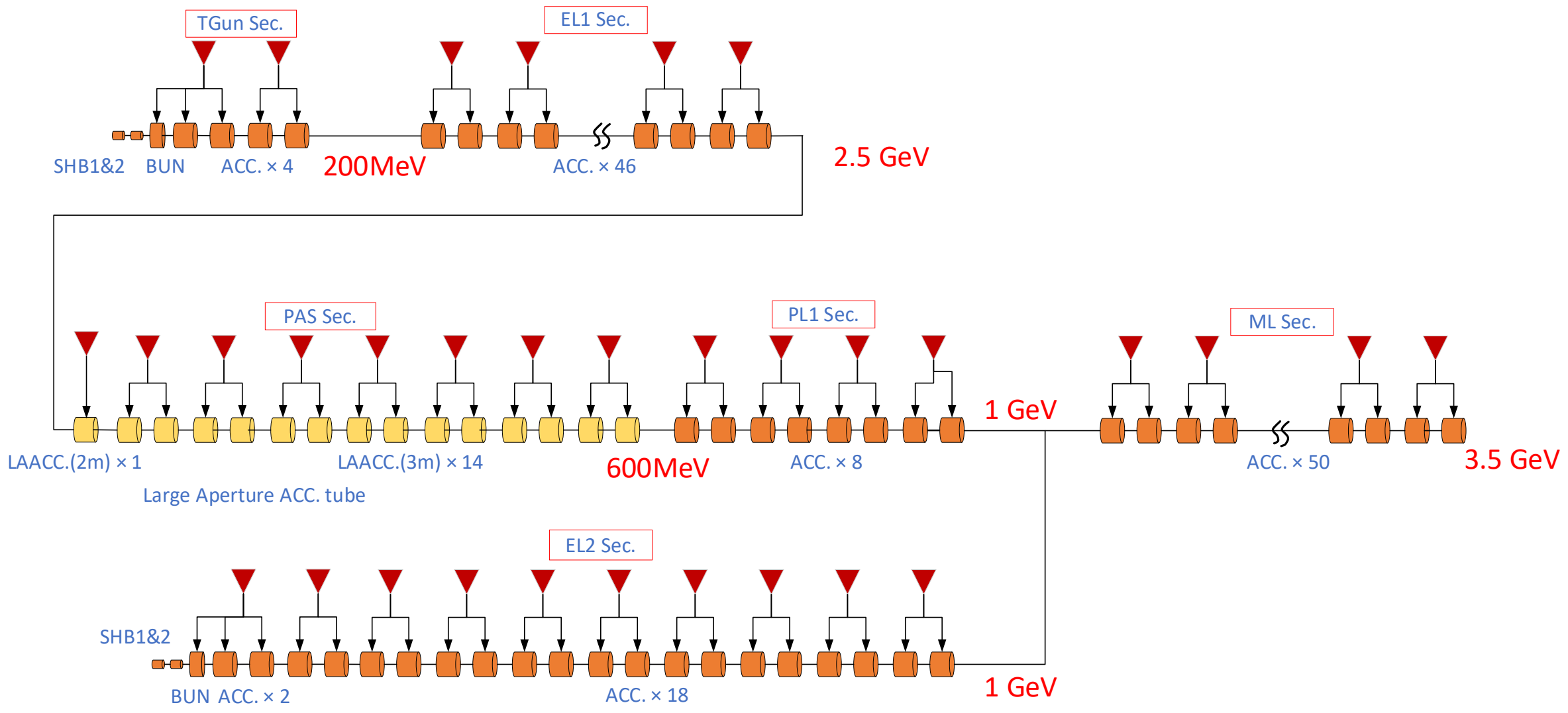

Figure 4.9-2: Layout of Accelerating Structures and Microwave Power Sources for the STCF Swap-out Injection Scheme

This figure depicts the sequential configuration of accelerating structures and associated RF power sources along the injector beamline designed for the swap-out injection scheme. The use of 2998.2 MHz S-band accelerating structures and subharmonic pre-bunchers ensures beam manipulation at high charge per bunch and varying repetition rates depending on beamline function. The layout aims for both compactness and high accelerating efficiency to support the performance requirements of the swap-out injection configuration.

### 4.9.4 System Composition

The microwave system primarily consists of accelerating structures, reference signal distribution and synchronization systems, digital low-level RF (LLRF) systems and their auxiliary equipment, solid-state amplifiers, and microwave transmission components. The solid-state amplifier and microwave transmission system adopt standard, mature schemes and are therefore not described in detail here.

#### *4.9.4.1 Traveling-Wave Accelerating Structures*

The accelerating structures for the STCF linacs include standard traveling-wave accelerating tubes, large-aperture traveling-wave accelerating tubes, and subharmonic pre-bunching cavities.

The standard accelerating structures are designed using elliptical-rounded irises and an internal water-cooling technique [120]. Based on the power source layout shown in Figures 4.9-1 and 4.9-2, these structures can achieve a maximum energy multiplication factor of 1.94 and an average effective accelerating gradient of up to 23.5 MV/m. The design parameters are shown in Table 4.9-4.

Table 4.9-4: Design Parameters for Standard Accelerating Structures

| Parameter | Unit | Value |
|---|---|---|
| Frequency | MHz | 2998.2 |
| Operating Temperature | ℃ | 30 |
| Number of Cells | – | 90 |
| Length | mm | 3165 |
| Cell Length | mm | 33.330 |
| Disk Thickness | mm | 4 |
| Iris Diameter (2a) | mm | 25.31–18.89 |
| Shunt Impedance | MΩ/m | 63.41–74.37 |
| Quality Factor | – | 15510–15403 |
| Group Velocity (vg/c) | – | 0.0264–0.0101 |
| Filling Time | μs | 0.57 |
| Accelerating Gradient (with pulse compressor) | MV/m | 23.5 |

The large-aperture accelerating structures are used for positron capture and initial acceleration. These use fixed iris diameters and adjust gradient distribution by modifying nose-cone lengths [121]. The structure dimensions and cavity distribution are optimized based on the uniform distribution of peak surface electric fields. According to injector design requirements, two types of structures are adopted: one 2-meter tube and fourteen 3-meter tubes. The 2-meter tube is located directly downstream of the positron target for capture and is powered by a single 40-MW klystron, with waveguide loss assumed as 0.6 dB. Each pair of 3-meter structures is powered by a standard microwave station. Design parameters are shown in Table 4.9-5. Subharmonic pre-bunching cavities follow standard standing-wave single-cell design.

Table 4.9-5: Design Parameters for Large-Aperture Accelerating Structures

| Parameter | Unit | 2-m Tube | 3-m Tube |
|---|---|---|---|
| Frequency | MHz | 2998.2 | 2998.2 |
| Operating Temperature | ℃ | 30 | 30 |
| Number of Cells | – | 60 | 90 |
| Length | mm | 2165 | 3165 |
| Iris Diameter (2a) | mm | 30 | 30 |

| Parameter | Unit | 2-m Tube | 3-m Tube |
|---|---|---|---|
| Cell Length | mm | 33.330 | 33.330 |
| Nose-Cone Length (d) | mm | 9.34–11.97 | 8.42-12.33 |
| Shunt Impedance | MΩ/m | 53.27–48.45 | 54.37-47.60 |
| Quality Factor | – | 13998–12877 | 14318–12695 |
| Group Velocity ($v_g/c$) | – | 0.027–0.020 | 0.030-0.020 |
| Fill Time | μs | 0.28 | 0.41 |
| Accelerating Gradient (with pulse compressor) | MV/m | 21.59 | 20.90 |

#### *4.9.4.2 Low-Level RF (LLRF) Processor*

The LLRF system uses a down-conversion sampling scheme. The detected signal is mixed with a local oscillator (LO) signal and passed through a band-pass filter to obtain an intermediate frequency (IF), which is then sent to the FPGA via ADC for demodulation [122]. A direct conversion scheme is used for signal output, generating VM signals with lower phase noise and eliminating the need for digital modulation via numerically controlled oscillators (NCOs) in FPGA or DSP, thereby improving reliability. Feedback control can be implemented using amplitude-phase or IQ control schemes. The amplitude-phase feedback scheme employs two independent loops to detect and regulate amplitude and phase separately [123].

#### *4.9.4.3 Reference Signal Synchronization System*

The reference signal synchronization system of the STCF injector provides low-phase-noise 2998.2 MHz signals for the lasers and microwave power sources. It ensures phase synchronization between microwave stations along the linacs and offers reference signals for other time-sensitive systems. The system maintains a phase drift of less than 100 fs (RMS) over 24 hours. The overall system is shown in Figure 4.9-3 and includes synchronized reference signals to LLRF units and calibration terminals near accelerating structures. The principle is based on modulating the RF reference signal from the RMO onto a continuous optical carrier transmitted via fiber. A portion of the signal is reflected at the remote end and returned to the transmitter, where it is compared with the reference. Phase locking between sender and receiver is achieved by adjusting fiber length based on this phase comparison [124].

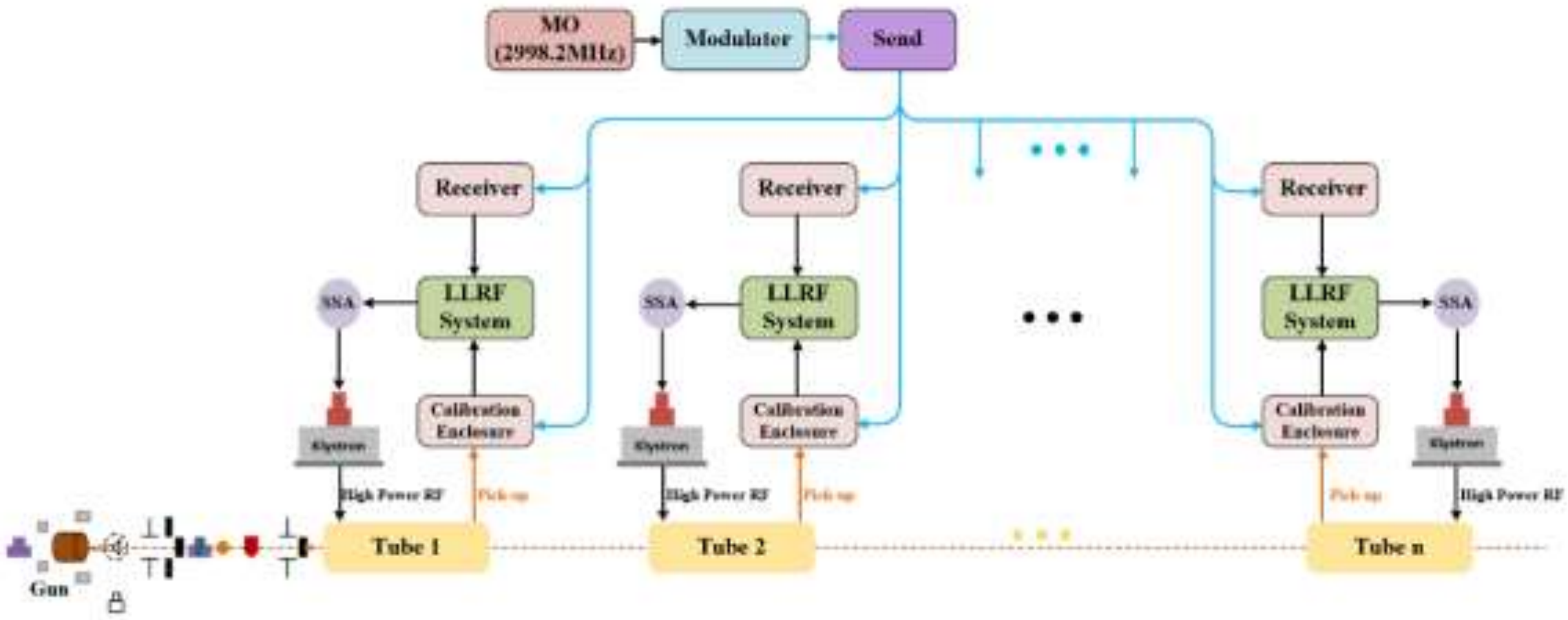

Figure 4.9-3: Block Diagram of the Linac Microwave Synchronization and Calibration System

### 4.9.5 Feasibility Analysis

The linac microwave system adopts mature and cost-effective technologies and products, with no insurmountable technical risks. The National Synchrotron Radiation Laboratory (NSRL) has extensive experience in designing and fabricating constant-gradient accelerating structures. The 3-meter 2π/3-mode traveling-wave accelerating tubes have been successfully applied in the HLS-II injector and have demonstrated excellent performance and reliability. The ongoing construction of the HALF linac will give us more experience. The previously untested large-aperture accelerating structures are currently being verified via the STCF key technology R&D project through the positron-electron beam test platform. NSRL also has rich experience in microwave transmission and LLRF systems and maintains close collaboration with IHEP, IMP, and Tsinghua University, ensuring smooth project development.

### 4.9.6 Summary

The design of the linac microwave system comprehensively considered performance requirements, technical feasibility, and cost-effectiveness based on the STCF injector's physical design. A detailed and rational technical route has been selected, prioritizing mature and proven technologies to reduce technical risks while meeting application demands. The design balances innovation and practicality, providing a solid foundation for subsequent system integration and optimization.

## 4.10 Linac Microwave Power Source System

### 4.10.1 Design Requirements and Specifications

The STCF injector consists of multiple segments of electron and positron linacs. The microwave power source system supplies S-band (2998 MHz) microwave power to the linac microwave system, enabling full-energy injection of electrons and positrons into the collider rings. The system primarily comprises high-power S-band klystrons and corresponding pulse modulators.

According to the collider's physics design, both off-axis and swap-out injection schemes are considered, and each requires a different injector configuration. The RF power system described here corresponds to the design using a thermionic electron source for direct electron beam injection, consistent with the microwave system.

In the off-axis injection scheme, the macro-pulse repetition rate is 30 Hz. The microwave power source system includes a total of 56 sets of pulsed power equipment: one X-band 6-MW source, one set of S-band 10-MW source, and 54 sets of S-band 45-MW sources.

In the swap-out injection scheme, the macro-pulse repetition rates are 30 Hz and 90 Hz. The microwave power source system includes 72 sets of pulsed power equipment, all using 45-MW S-band klystrons. The system is developed based on the selected injection scheme, with emphasis on microwave power source stability.

### 4.10.2 Key Technologies and Roadmap Selection

The linac microwave power source system comprises S-band and X-band klystrons and their associated high-voltage pulse modulators. It also integrates auxiliary systems such as microwave drive and low-level RF, vacuum, waveguide transmission, and water cooling.

For low repetition rates (<50 Hz), high-power S-band klystrons such as the commercial Canon E37302A are available and technically mature. Under support from the Institute of High Energy Physics (IHEP, CAS), domestic manufacturers such as Hanguang Tech (Hubei) and CETC No.12 Institute have also developed indigenous 2998 MHz klystrons. The Hanguang high-power klystron is currently under validation at the Wuhan Light Source.

Given the demand for large-scale deployment in major facilities and concerns about procurement restrictions, advancing the maturity of domestically developed high-peak-power klystrons is essential.

In terms of modulators, the "fractional turn ratio transformer" type solid-state modulator is selected due to its performance characteristics and alignment with trends in modulator technology [125]. This approach reduces insulation requirements, offers excellent pulse stability and uniformity, eliminates lifetime issues associated with thyratrons, enables low-voltage power supply operation, and provides improved electromagnetic compatibility and safety. These features support stable system operation and future upgrades [126].

## 4.10.3 Design Scheme and System Configuration

The microwave power source system provides microwave power with defined stability and performance for accelerating electrons and positrons in the STCF linacs. Different injector configurations or schemes require tailored microwave system integration.

### *4.10.3.1 Microwave Power Source Configuration for Off-Axis Injection Scheme*

For the off-axis injection scheme, the system includes 56 pulsed microwave power units: one unit of 6-MW X-band, one unit of 10-MW S-band, and 54 units of 45-MW S-band. The system layout is shown in Figure 4.10-1.

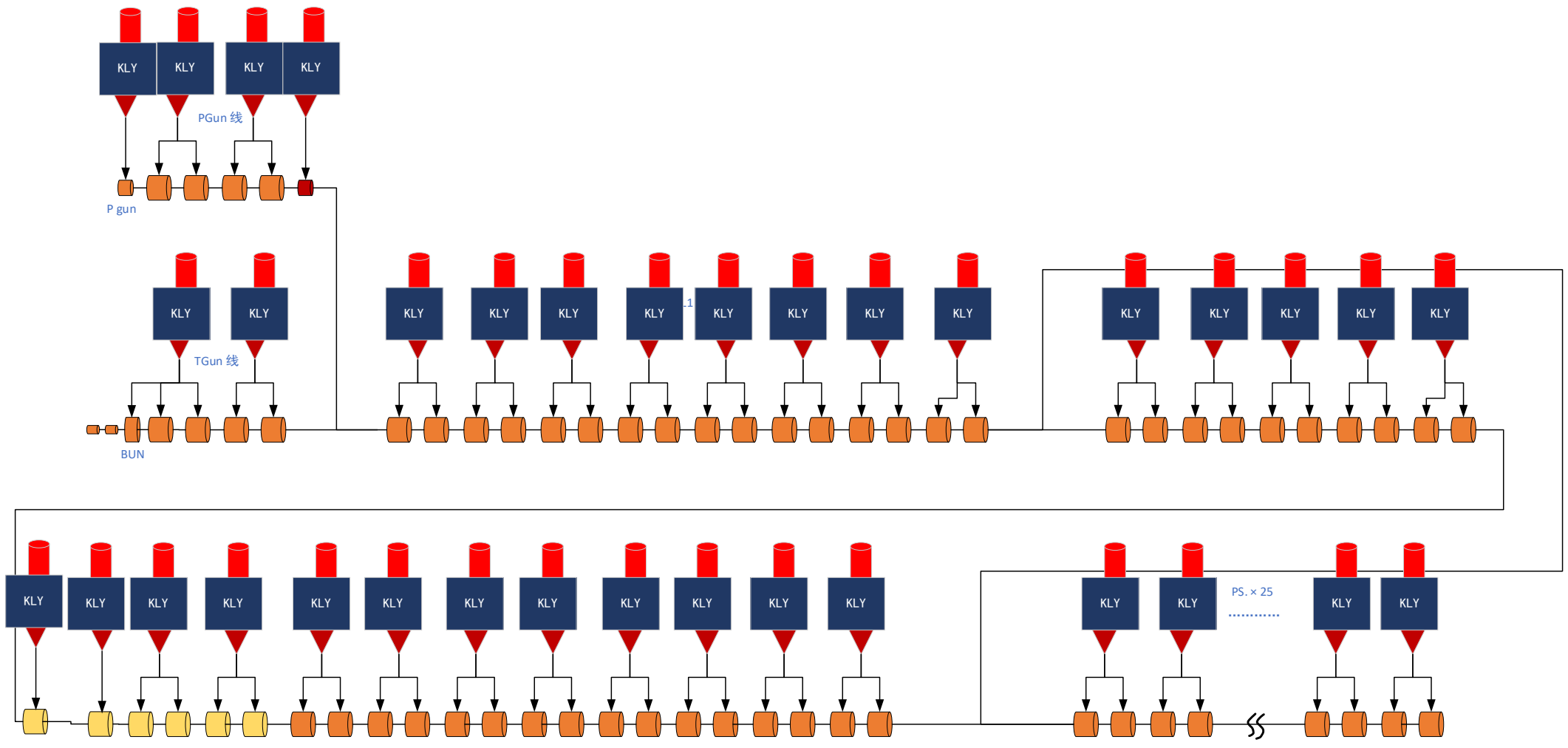

Figure 4.10-1: RF Power Source Layout for Off-Axis Injection Scheme

In this scheme, the macro-pulse repetition rate for the klystrons reaches up to 90 Hz (main linac section, ML). The specific performance requirements of the involved RF power sources are listed in Tables 4.10-1 and 4.10-2.

Table 4.10-1: S-Band RF Power Source Technical Specifications

| Parameter | Specification |
|---|---|
| Microwave Frequency | 2998.2 MHz |
| Maximum Repetition Rate | 90 Hz |
| Klystron Peak Power | 45 MW |
| Modulator Peak Power | 130 MW |
| Pulse Flat-Top Width | ⩾ 4.0 μs |
| Pulse High Voltage Stability | Jitter ⩽ 0.03% (RMS) |

| Parameter | Specification |
|---|---|
| Output Pulse Rise Time Jitter | ⩽ 10 ns (RMS) |

Table 4.10-2: X-Band Microwave Power Source Technical Specifications

| Parameter | Specification |
|---|---|
| Microwave Frequency | 11992 MHz |
| Maximum Repetition Rate | 30 Hz |
| Klystron Peak Power | 6 MW |
| Modulator Peak Power | 20 MW |
| Pulse Flat-Top Width | ⩾ 4.0 μs |
| Pulse High Voltage Stability | Jitter ⩽ 0.03% (RMS) |
| Output Pulse Rise Time Jitter | ⩽ 10 ns (RMS) |

#### *4.10.3.2 Microwave Power Source System Configuration for Swap-Out Injection Scheme*

For the swap-out injection scheme, the system consists of 72 sets of pulsed power equipment, all utilizing S-band 45-MW RF sources. The macro-pulse repetition rate of the klystrons reaches up to 90 Hz in the EL1, PAS, and PL1 beamlines. The system layout is illustrated in Figure 4.10-2.

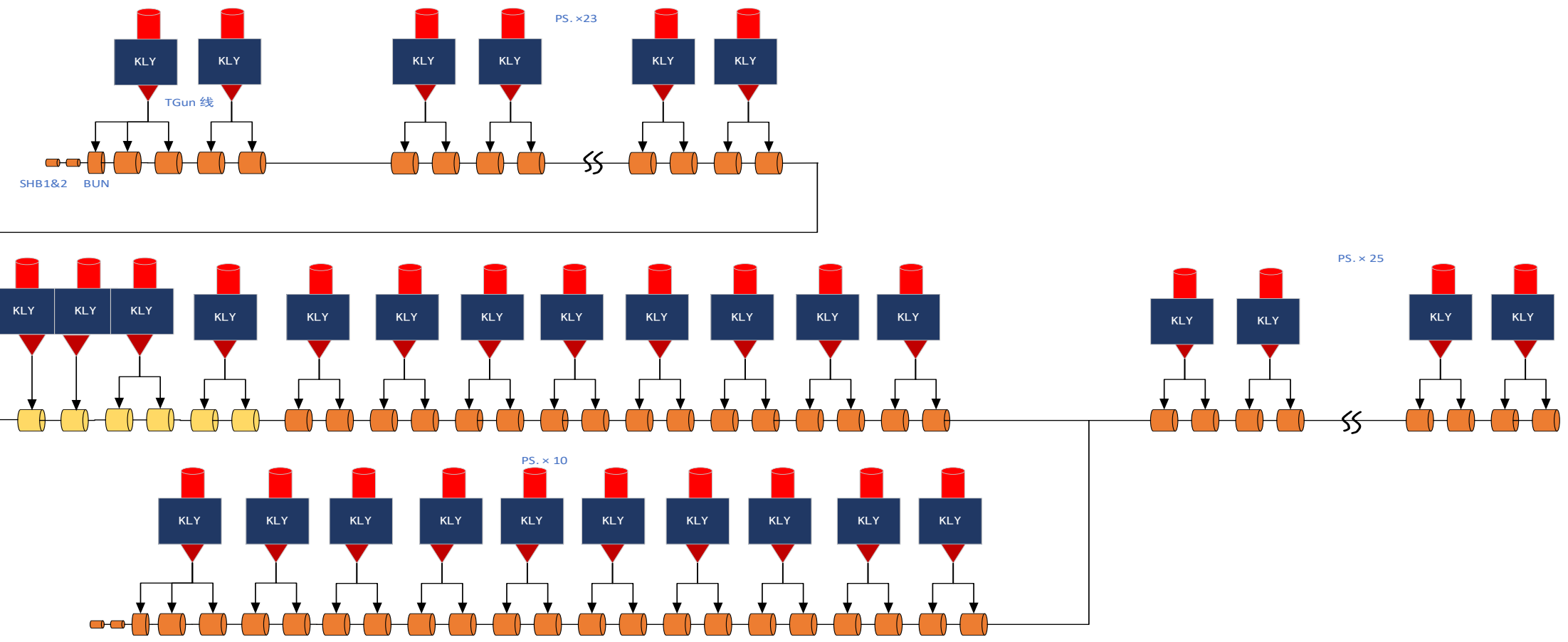

Figure 4.10-2: System Layout of the Pulsed Power Source for the STCF Swap-Out Injection Scheme

According to the operating parameters of the klystrons and the injector physics requirements, and with a margin considered in the design, the fractional-turn transformer-type solid-state modulator paired with the 45-MW klystrons has the parameters shown in Table 4.10-3.

Table 4.10-3: S-band Klystron Modulator Design Parameters for Swap-Out Injection

| Parameter | Design Value |
|---|---|
| Pulse voltage | 335 kV |
| Pulse current | 400 A |
| Modulator pulse power | 135 MW |
| Pulse width | 4 μs |
| Maximum repetition rate | 90 Hz |
| Pulse amplitude stability (RMS) | < 0.03% |
| Output pulse rise jitter (RMS) | < 10 ns |

Given the large number of RF sources required by the STCF, the design of the microwave power source system must follow a modular design philosophy. The system is standardized as much as possible with a unified design across multiple frequency bands, simplifying high-power system complexity and improving maintenance efficiency. This modular, highly integrated, compact, and reliable approach applies to both the X-band 6-MW and S-band 45-MW klystrons and their associated high-voltage pulse modulators.

Each microwave power unit includes a klystron, a modulator, and auxiliary power supplies. The solid-state modulator consists of a pulse transformer and oil tank, solid-state modulation components, high-voltage power supply, auxiliary power units, and measurement devices. While the structure of klystrons varies across bands and power levels, the modulation components are designed in a unified modular fashion, and the auxiliary power supplies are standardized as much as possible to facilitate engineering and production.

Taking the main 45-MW klystron power unit as an example, the system diagram is shown in Figure 4.10-3.

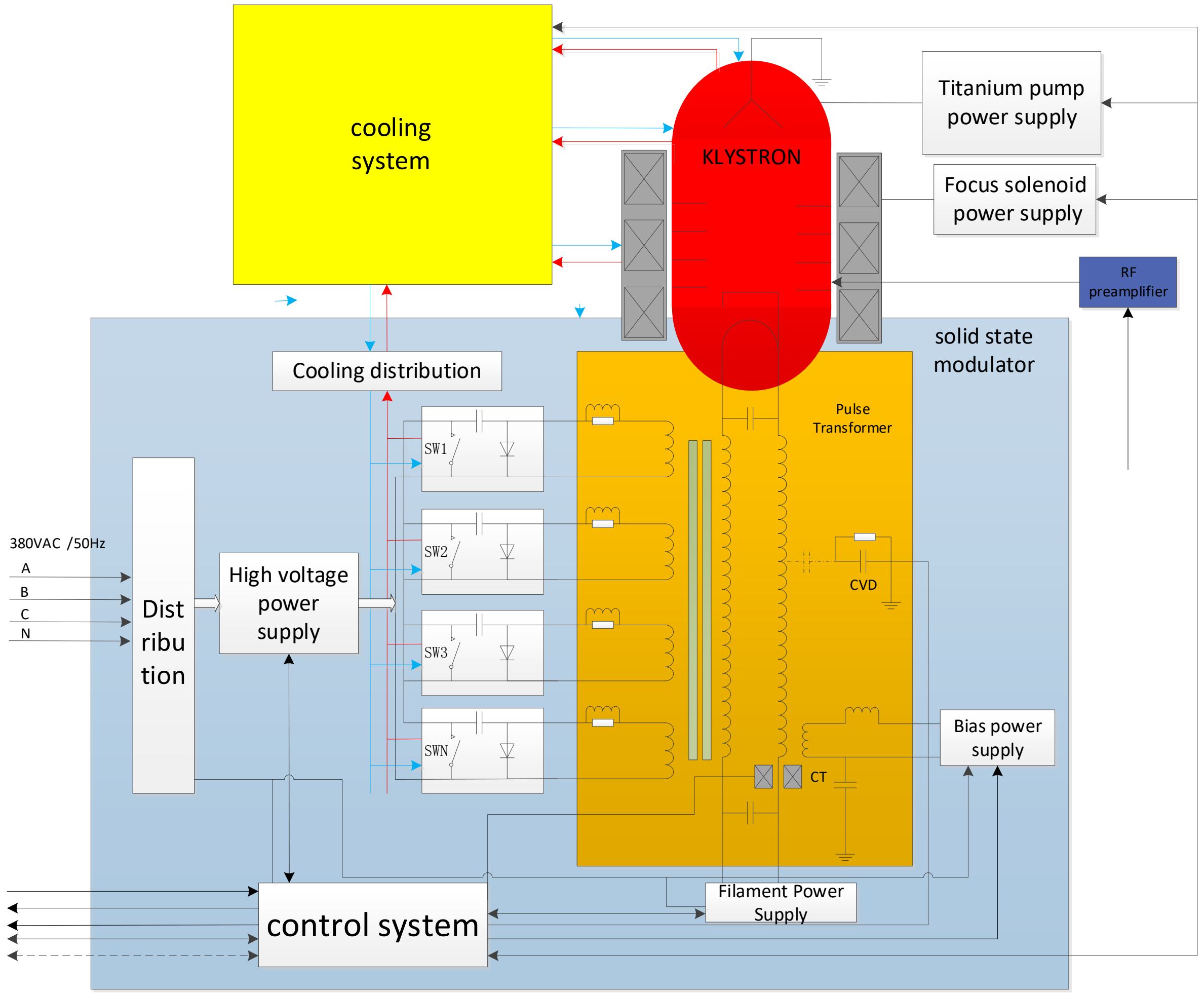

Figure 4.10-3: Block Diagram of a Single Linear RF Power Source System

Each 45-MW microwave power unit consists of the klystron, a matching high-voltage pulse modulator, and an auxiliary power supply. A power supply of about 1 kV DC is provided to the solid-state pulse modulator, which converts it into a pulsed voltage input to the primary winding of the pulse transformer. The primary-to-secondary turns ratio is tailored to deliver the required cathode pulse voltage. The auxiliary power supply includes units for the titanium pump, cathode heater, and focusing solenoid. The pulse modulator converts electrical energy into microwave power, which is then fed to the accelerating structures in the linacs.

An engineering layout diagram of a single 45-MW microwave power source unit is shown in Figure 4.10-4.

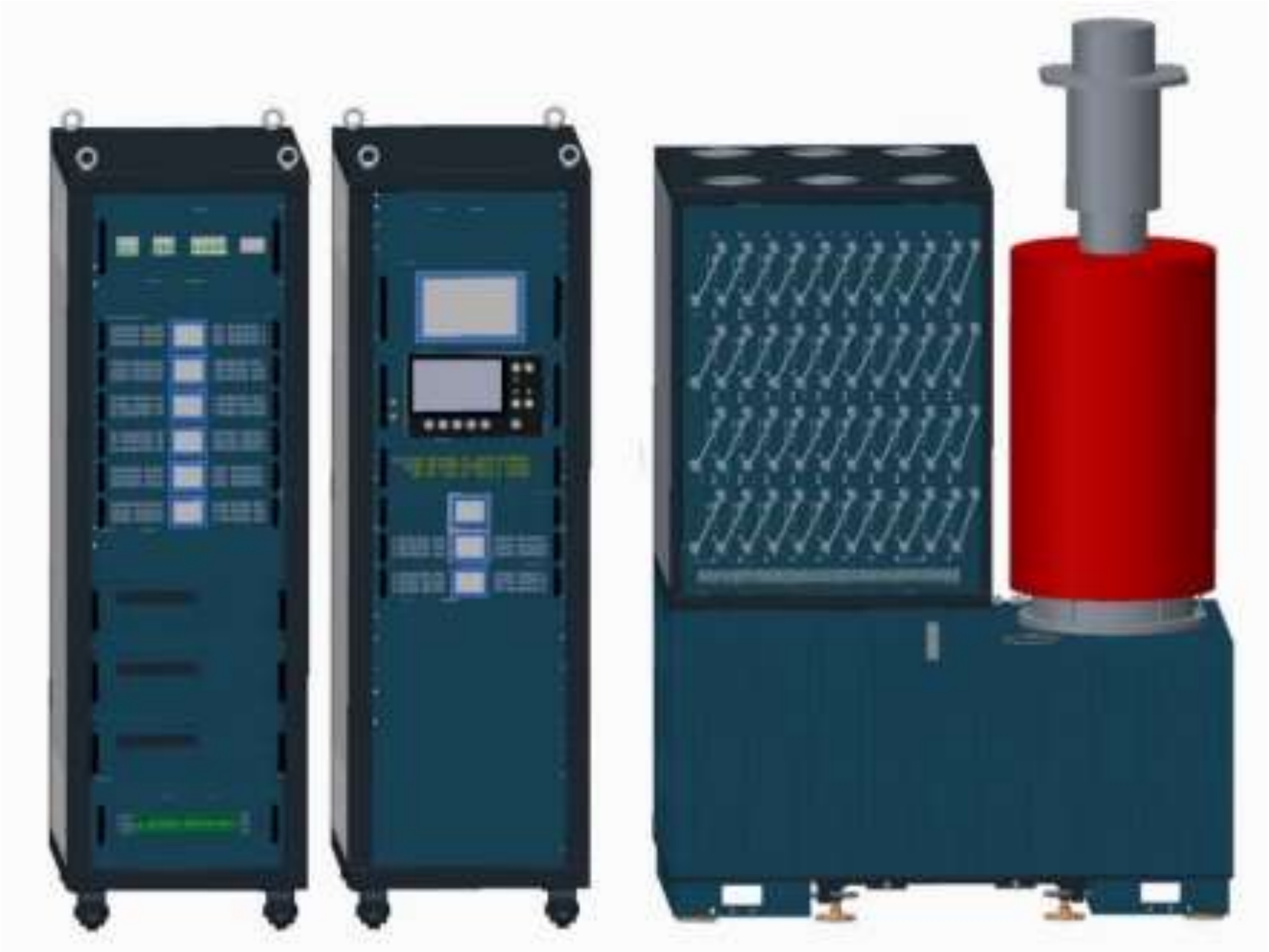

Figure 4.10-4: Engineering Layout of the Linac Microwave Power Source Unit

### 4.10.4 Feasibility Analysis

For pulse repetition rates up to 50 Hz, commercial klystrons such as Canon's E37302A are technically mature and widely deployed in accelerator facilities. Domestic manufacturers such as Hanguang Technology and CETC No.12 Institute have the capability to develop 2998 MHz klystrons, and their high-power designs are under testing at the Wuhan Light Source. To mitigate procurement risks, it is necessary to advance the maturity of these domestic klystrons through beam test platforms.

For higher repetition rates (up to 90 Hz), the limiting factor is the klystron's performance. According to discussions with Canon and Hanguang, high-repetition variants of the existing products are technically feasible, and further development can be pursued during STCF's R&D phase.

The pulsed modulator adopts an advanced fractional-turn transformer-type solid-state design, offering high integration, maintainability, and reliability. This approach is already applied at the Hefei Advanced Light Facility (HALF). The system's auxiliary power units (focusing solenoid, filament, biasing transformer, titanium pump supplies) use mature, digitally controlled technology, widely adopted at facilities such as HLS and HALF.

### 4.10.5 Summary

The linac microwave power source system is based on a technically advanced and feasible overall design, with high system maturity.

## 4.11 Positron Source System

### 4.11.1 Design Requirements and Specifications

The STCF positron source adopts the conventional method of bombarding a tungsten target with an electron beam to produce the required positron beam. As one of the core components of the STCF injector, the positron source must support both the off-axis injection and swap-out injection schemes in the collider rings, with priority given to off-axis injection during initial construction. The off-axis scheme uses an electron beam with an energy of 1.5 GeV, a single bunch charge of 10 nC, and a repetition rate of 30 Hz (1.5 GeV/10 nC/30 Hz) to bombard the target. The swap-out injection scheme requires a more demanding electron beam of 2.5 GeV/11.6 nC/90 Hz, leading to significantly higher average power on target and stricter demands on positron yield, target thermal resilience, and cooling performance. With the compatible injector design scheme, the initial phase utilizing off-axis injection achieved target beam parameters of 1.0 GeV/11.6 nC at 30 Hz. In the following upgrade to swap-out injection, the electron beam on the positron target is enhanced to 2.5 GeV/11.6 nC at 90 Hz.

To develop a high-intensity, high-quality positron source suitable for STCF, it is essential to master the design and manufacturing techniques of key components such as the positron conversion target, adiabatic matching device (AMD), and the capture accelerating structures. The main parameter specifications of the STCF positron source are listed in Table 4.11-1.

Table 4.11-1: Main Parameter Specifications of the STCF Positron Source

| Parameter | Off-axis Injection | Swap-out Injection |
|---|---|---|
| Positron beam for collider injection | 1.0 nC / 30 Hz | 8.5 nC / 30 Hz |
| Electron beam on target | 1.0 GeV/11.6 nC/30 Hz | 2.5 GeV/11.6 nC/90 Hz |
| Average power (W) | 348 | 2443 |
| Peak energy deposition density (PEDD, J/g) | 5.6 | 32.2 |
| AMD magnetic field (T) | 0.5–5 | 0.5–5 |
| Positron conversion efficiency (%) | 9.1 | 25 |
| Solenoidal magnetic field (T) | 0.5 | 0.5 |
| Aperture of accelerating structure (mm) | 30 | 30 |

### 4.11.2 Key Technologies and Technical Approaches

Given the high positron current demand of STCF, only the positron sources based on conventional electron beam bombarding onto a heavy-element target can meet the injection requirements. The positron source comprises three main parts: the conversion target system, matching system, and capture accelerator system.

#### *4.11.2.1 Positron Conversion Target System*

The primary function of the conversion target is to produce electron–positron pairs from a high-energy electron beam. Tungsten is typically used as the target material due to its high atomic number, high-temperature resistance, favorable performance, and low cost. Target structures include amorphous tungsten, single-crystal tungsten, and multilayer composite targets. STCF plans to adopt a conventional single-layer amorphous tungsten target for its balance between stability and maturity.

Since the target directly absorbs the electron beam energy, thermal management is critical. STCF designs two cooling schemes based on the injection modes: a conventional fixed tungsten target for the off-axis injection and a wobbling water-cooled target for the swap-out injection to improve heat dissipation.

#### *4.11.2.2 Positron Matching System*

Post-target positrons have large divergence angles and energy spreads. The matching system collects and transforms these positrons in phase space. Two common designs are the Quarter-Wavelength Transformer (QWT) and Adiabatic Matching Device (AMD). Since the AMD provides superior energy dispersion acceptance compared to the QWT, the STCF positron source adopts the AMD as its matching system, leveraging the pulsed high-strength magnetic field to facilitate the capture and phase-space manipulation of positrons.

#### *4.11.2.3 Capture and Acceleration Section*

As positrons pass through the AMD under a longitudinally varying magnetic field, they undergo Larmor rotation. Low-energy positrons spiral outward, and high-energy positrons remain near the axis, leading to a large transverse size and bunch length at the AMD exit. To improve capture efficiency, STCF uses large-aperture (30 mm) S-band accelerating structures in the pre-acceleration section. A 0.5-T solenoidal field suppresses beam divergence and accelerates the positrons to 200 MeV. A chicane structure is used to deflect the beam, and a collimator scrapes off positrons that do not meet the damping requirements. Finally, positrons are accelerated to 1 GeV via standard linear acceleration. More details on beam dynamics can be found in Chapter 3.

### 4.11.3 Design Scheme and System Composition

#### *4.11.3.1 Positron Target Design*

PEDD is a critical parameter to evaluate the thermal shock tolerance of a tungsten conversion target. SLAC research suggests that the upper PEDD limit for tungsten is 35 J/g [127, 128]; exceeding this causes serious damage. For the off-axis injection (1.0 GeV/11.6 nC/30 Hz), simulations (Figure 4.11-1) show a safe PEDD of 12.6 J/g. The target is fixed and surrounded by a water-cooled copper jacket (Figure 4.11-2).

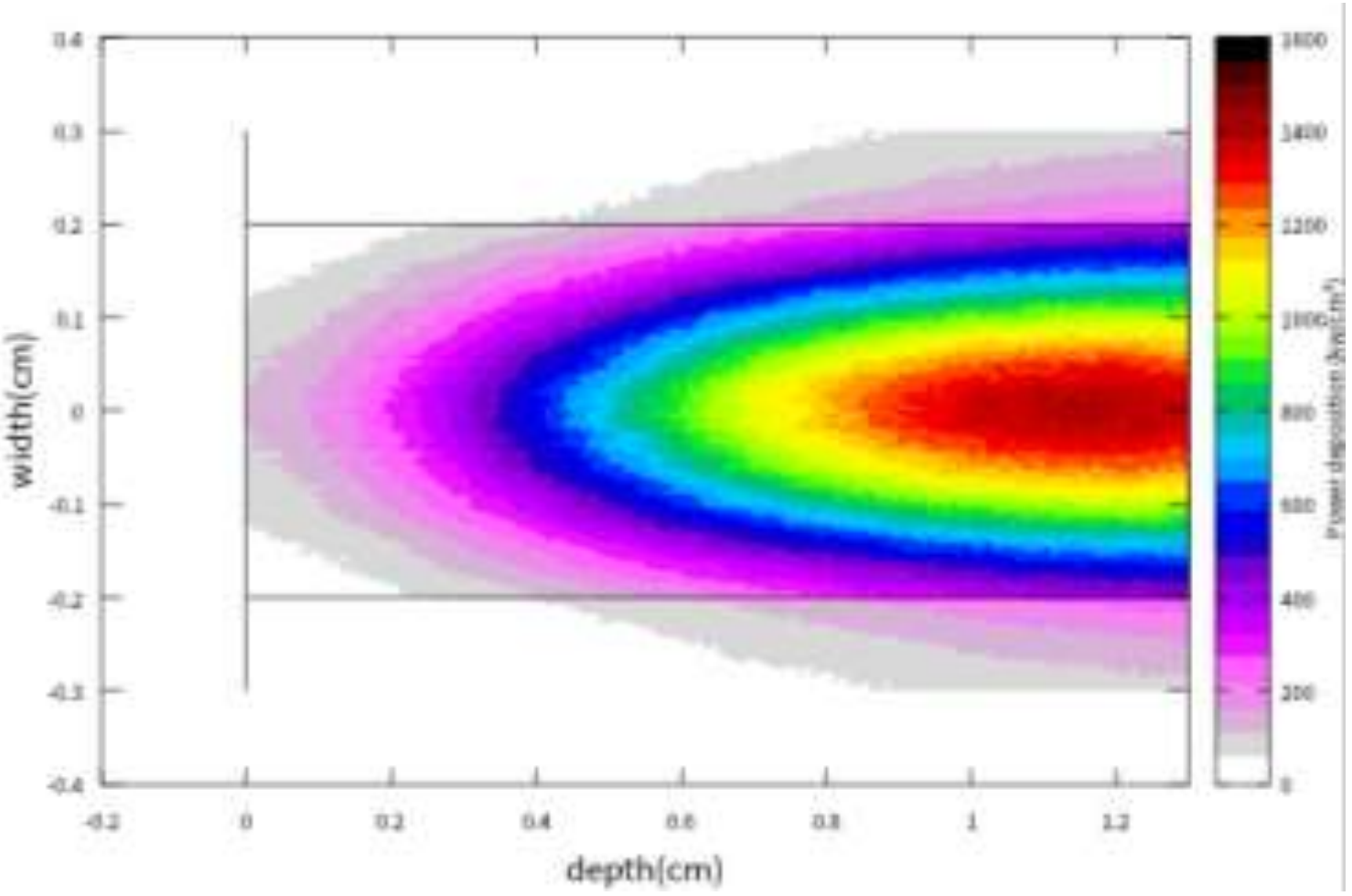

Figure 4.11-1: Power Deposition Simulation for the Off-axis Injection

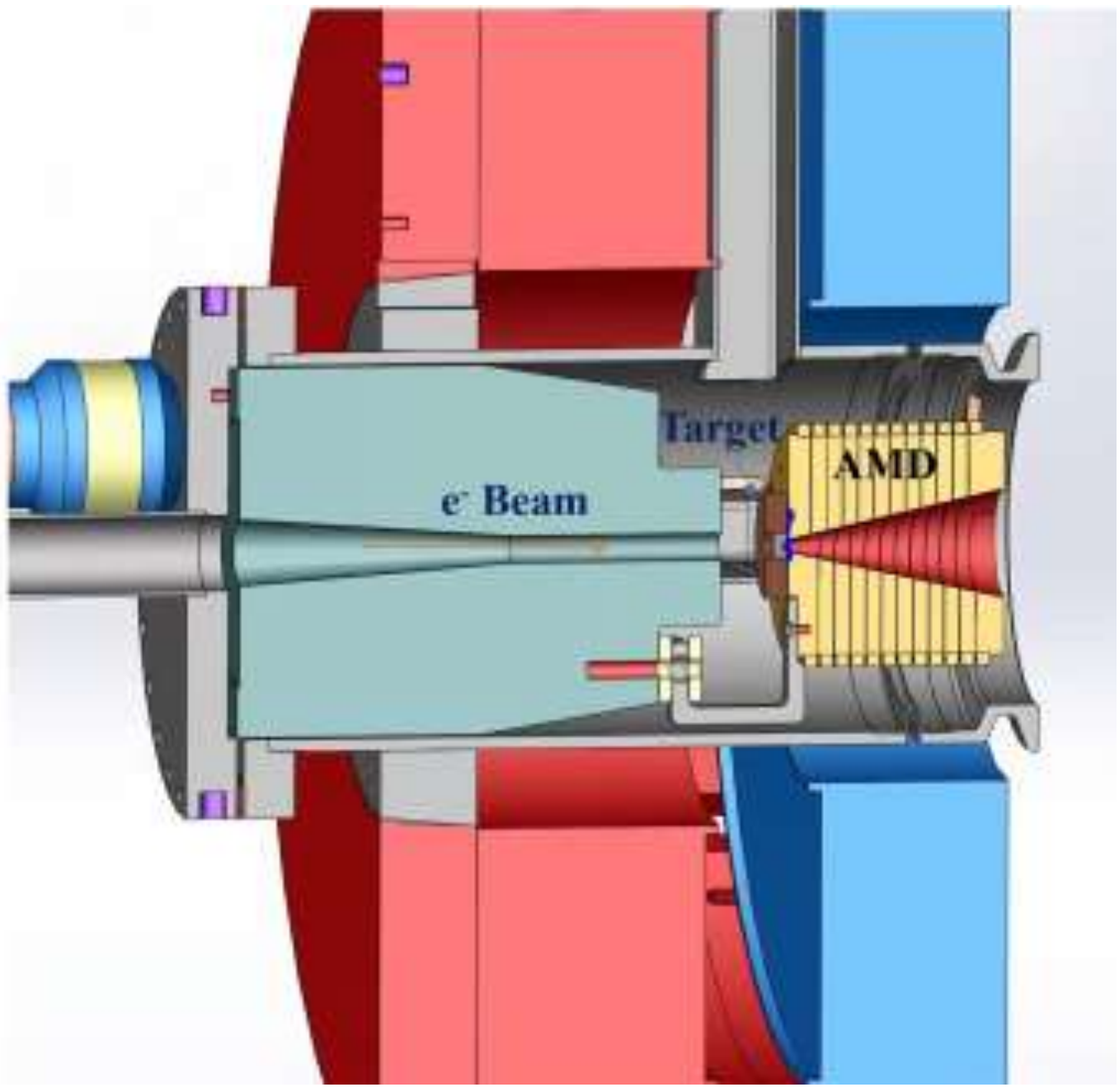

Figure 4.11-2: Positron Target Design for the Off-axis Injection

For the swap-out injection mode, an electron beam of 2.5 GeV/11.6 nC/90 Hz and 1 mm in diameter is used for positron production. Simulations (as shown in Figure 4.11-3) indicate a PEDD of 32.2 J/g in the tungsten target, close to the damage threshold. Thus, a new target design is required to ensure lifetime and reliability.

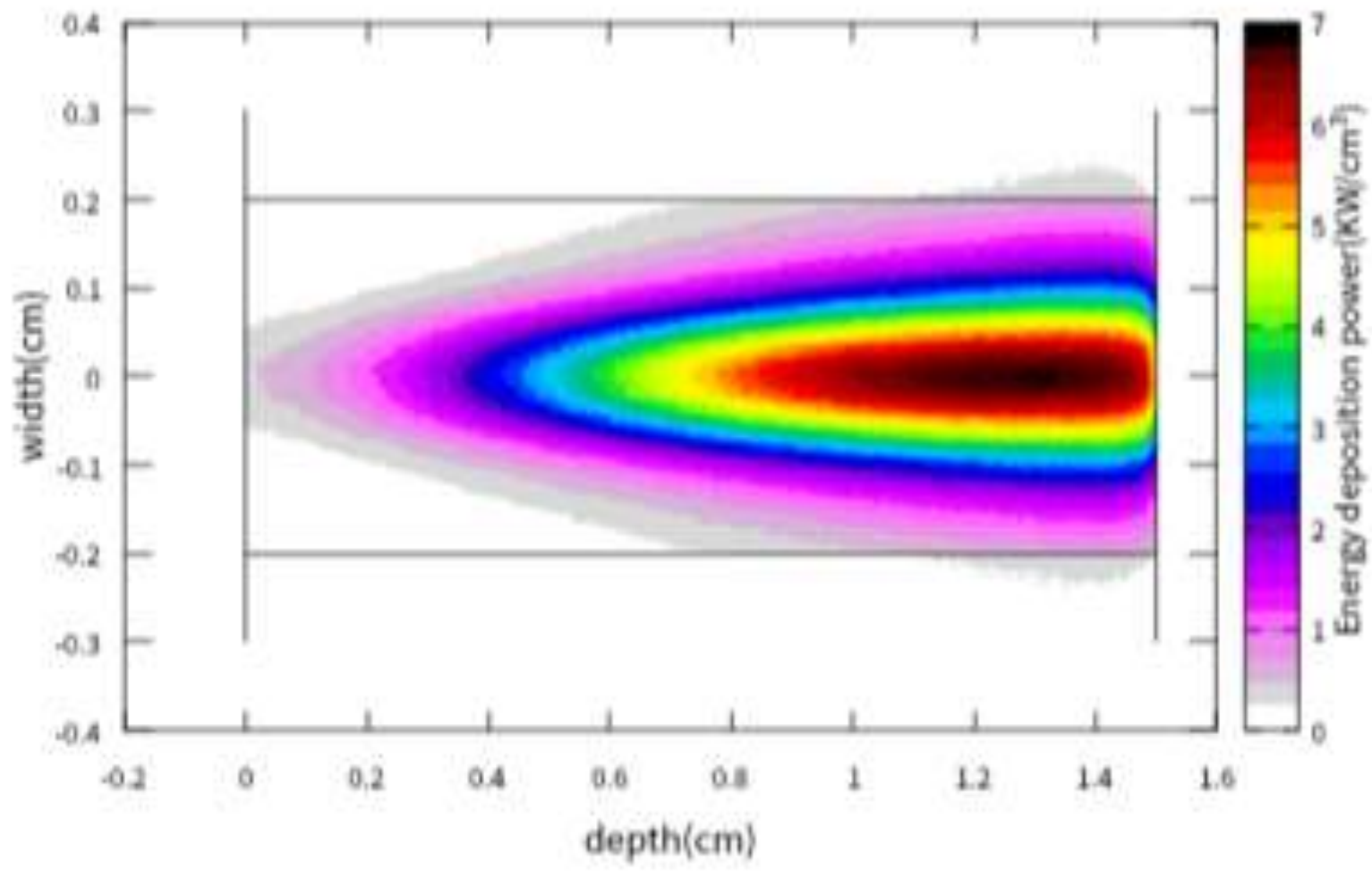

Figure 4.11-3: Power Deposition Simulation for Swap-out Injection

High-energy pulsed electron beams cause thermal shock and long-term heating effects. Shock can damage the target mechanically, while the high temperatures due to the heat deposition can change the tungsten lattice. Experimental results show that tungsten recrystallization begins at 800-900 °C, leading to decreased yield strength and positron yield [129]. ILC uses a rotating water-cooled target [130], but this design has vacuum sealing challenges, which is unsuitable for STCF.

To address thermal issues, STCF proposes a wobbling water-cooled target design (Figure 4.11-4). Simulations show that without motion, the target temperature peaks at 1515 °C, above the recrystallization point. With a wobbling motion, the temperature drops to 394 °C, preventing lattice damage and ensuring target longevity (Figure 4.11-5).

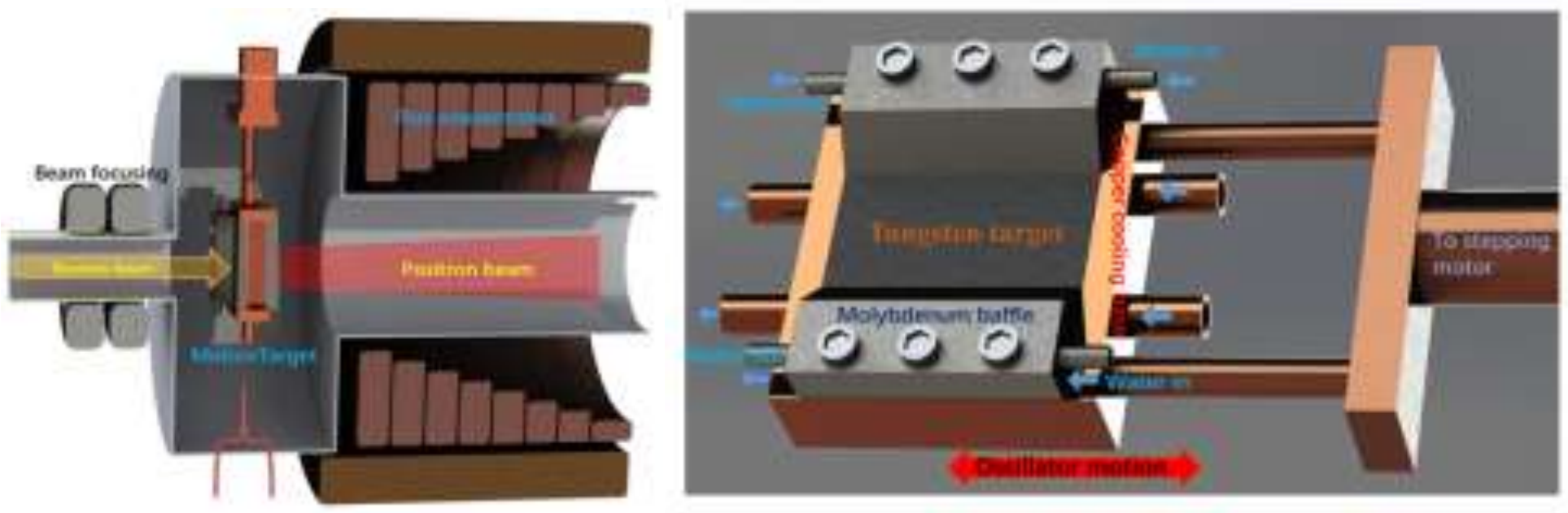

Figure 4.11-4: Wobbling Water-cooled Positron Target System

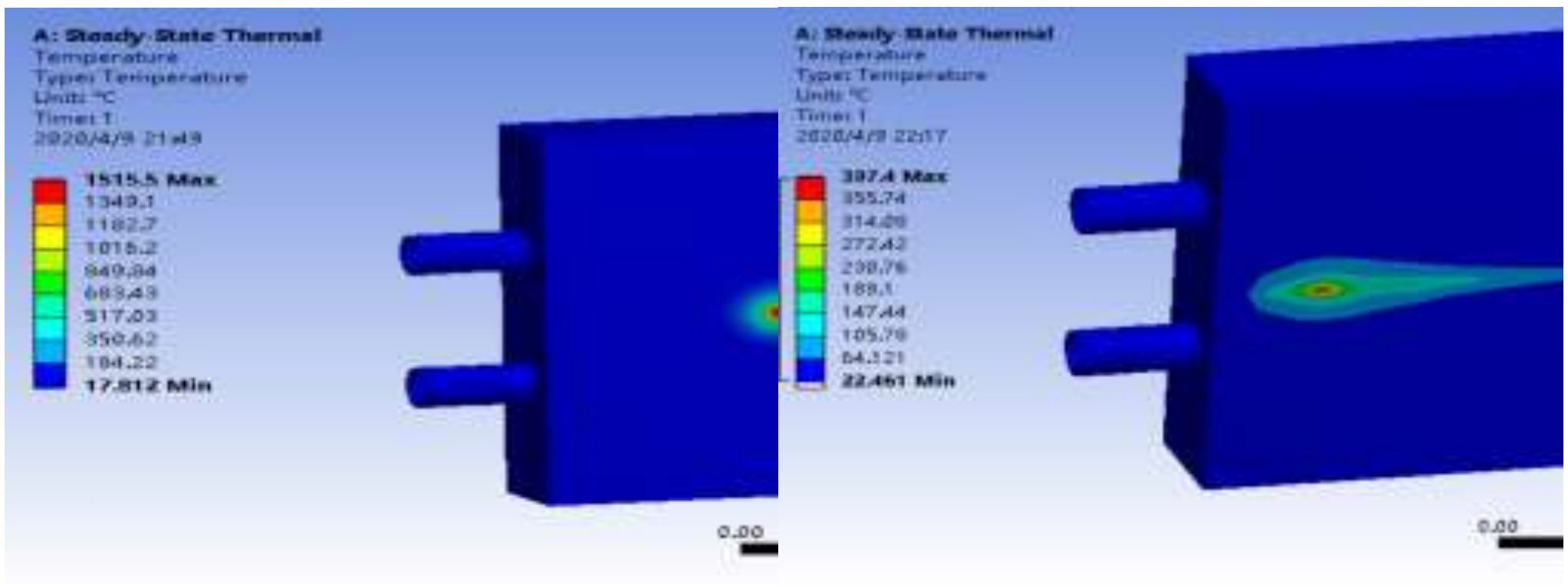

Figure 4.11-5: Thermal Simulation of Oscillating Target System

#### *4.11.3.2 AMD Design*

The positron beam produced by the electron bombardment of the conversion target features a wide energy spectrum and large angular divergence. To ensure effective matching with downstream accelerating structures, positrons must be captured and transformed into a beam with small divergence and larger transverse size. STCF will adopt an Adiabatic Matching Device (AMD) to perform this transformation and optimize positron collection, capture, and acceleration. The design parameters of the STCF AMD are shown in Table 4.11-2.

Table 4.11-2: STCF AMD Design Parameters

| Parameter | Value |
|---|---|
| Peak pulsed current | ⩾15 kA |
| Pulse width (FWHM) | 5 ± 0.5 µs |
| Pulse waveform | Half-sine wave |
| Repetition rate | 50 Hz |
| Long-term stability | 0.5% (8 h) |
| Charging peak voltage | 15 kV |
| Time jitter | ±20 ns |
| Equivalent load inductance | 0.8 µH |

The AMD consists of a radially centered conductor surrounded by a primary coil. The central conductor is formed by several thin disks connected together. When a time-varying current flows through the primary coil, magnetic flux is induced in the conical central conductor. Due to the skin effect, the induced current flows along the inner surface, generating a peak magnetic field at the center. The design target for the AMD peak field is above 5 T. Figure 4.11-6 shows simulation results under an excitation current of 15 kA and a pulse width of 5 µs, indicating a peak field exceeding 6 T, which satisfies the design requirements.

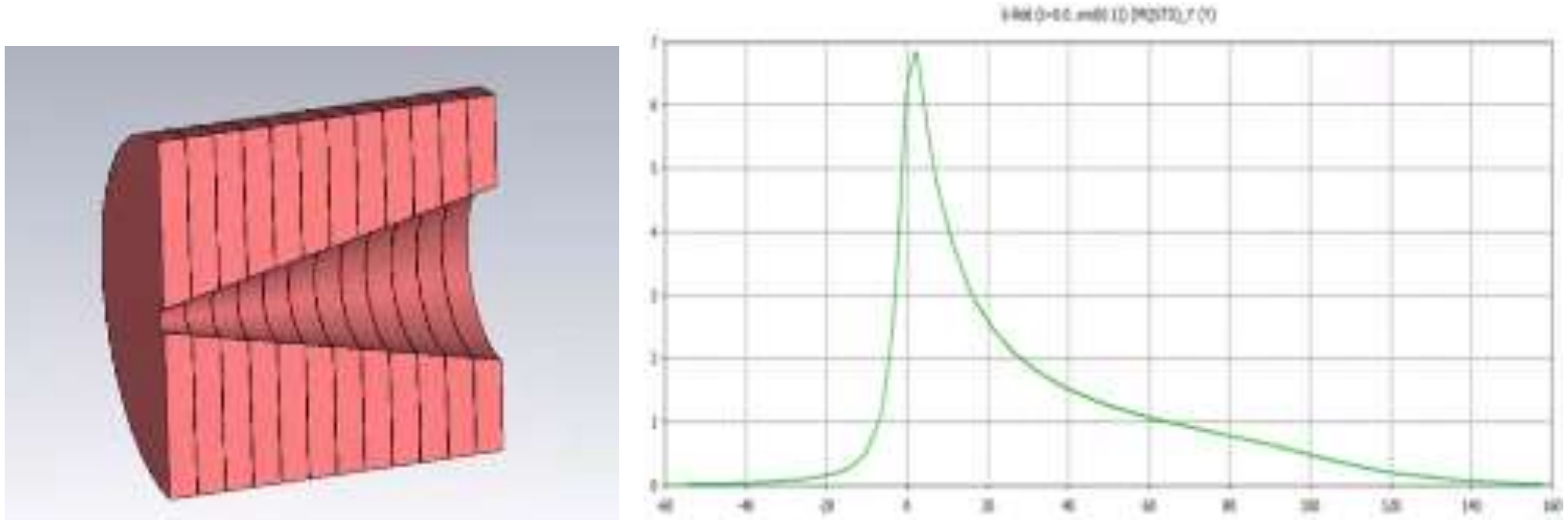

Figure 4.11-6: AMD Pulse Magnetic Field Simulation

#### *4.11.3.3 Positron Capture and Acceleration Design*

The key challenge of a positron accelerator lies in the capture and matching acceleration section, which differs from conventional electron accelerators. To enhance capture efficiency by increasing transverse acceptance, large-aperture accelerating structures are often used. For example, a 1298-MHz L-band cavity (with an aperture of 35 mm in diameter) provides about three times the acceptance of a 20-mm aperture S-band cavity. However, L-band structures have larger outer diameters, which significantly raises the cost of the solenoid magnet surrounding the accelerating structure.

Therefore, STCF adopts a compromised solution by using a 30-mm aperture S-band accelerating structure, optimizing the RF phase to enhance capture and acceleration efficiency while reducing cost. This large-aperture S-band accelerating section is positioned immediately after the positron target. A solenoidal field of 0.5 T surrounds the structure to suppress beam divergence and increase positron yield. Downstream, the positron beam passes through a chicane for charge separation, and a collimator removes off-spec particles. The positrons are then accelerated conventionally up to 1 GeV. Figure 4.11-7 shows the structure of the capture and acceleration section. For detailed lattice design, refer to Chapter 3 on injector beam dynamics.

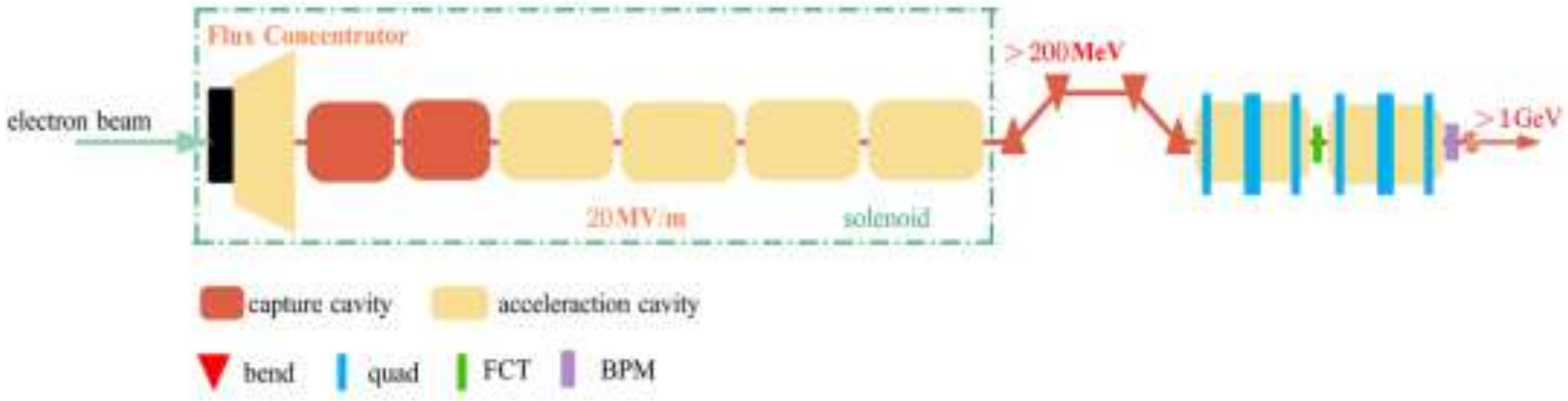

Figure 4.11-7: Positron Capture and Acceleration Section Layout

### 4.11.4 Feasibility Analysis

Positron generation via high-energy electron bombardment is a well-established method. In the off-axis injection scheme of STCF, the design for producing positron bunches of 1.0 nC with an electron beam of 1.0 GeV/11.6 nC/30 Hz does not present significant challenges. However, for the swap-out injection scheme, it is required to have an electron beam of 2.5 GeV/11.6 nC/90 Hz to bombard the tungsten target, and this imposes much higher demands on the positron source—posing a first-of-its-kind challenge in China.

Such high average power may create thermal management and lifetime issues for the positron target. These risks can be mitigated through optimized target designs, such as the implementation of movable targets. The STCF's key technology R&D project includes a positron-electron beam test platform for validating critical components and simulating high-power beam impacts. Preliminary experimental studies have been conducted, providing a strong technical foundation for future development.

In conclusion, despite certain technical challenges, the positron source design is technically feasible. STCF possesses the theoretical foundation, experimental experience, and engineering capabilities to ensure the positron source meets the collider's injection demands.

#### 4.11.5 Summary

The positron source is a crucial subsystem for achieving high luminosity in STCF. It utilizes conventional high-energy electron bombardment on a target to generate positrons. To meet the luminosity goals, the source must deliver a high-current positron beam, which places stringent demands on the target's yield efficiency, thermal resilience, and heat dissipation. This requires the development of a high-power, high-reliability positron target and an optimized positron collection, matching, and acceleration system. These elements together ensure the positron source meets the requirements of both injection schemes under consideration for STCF.

## 4.12 Control System

The accelerator control system is responsible for the comprehensive control over all equipment within the injector and collider rings, serving as the operational platform for both the commissioning and routine operation of STCF. It must fulfill the requirements of accelerator physicists and engineering staff for device monitoring, data acquisition, and analysis, while maintaining sufficient flexibility for future upgrades. Specifically, the control system shall:

- Initialize the accelerator and transition it into an operational state under safe interlock conditions.
- Enable the operators to adjust device parameters as required by accelerator physicists to achieve optimal operation.
- Utilize beam diagnostics to optimize beam performance and meet the physics requirements for beam commissioning.
- Safely bring the accelerator to shutdown status under interlock supervision.

Additionally, the system shall

- Support device status inspection and alarm functionalities;
- Provide a timing system to synchronize signals among subsystems;
- Maintain a central database for machine parameters and operational data to support online and offline physics analysis;
- Establish a beam tuning software development platform to assist in achieving the accelerator physics goals.

### 4.12.1 Design Requirements and Specifications

Based on the practical requirements of STCF and referencing the control system designs of BEPCII and SuperKEKB, the main design requirements and specifications of the control system are as follows:

- Control Network: Must support accelerator-wide data communication. Core switches in the backbone network must support port rates of 40 Gb/s, while access layer switches must support 10 Gb/s.
- Device Control: Must monitor and control all accelerator equipment. Parameter monitoring refresh rate ≥ 2 Hz; parameter setting response time < 200 ms.
- Timing System: Must ensure trigger signal synchronization. The delay and pulse width of each signal should be independently adjustable. Inter-signal jitter < 30 ps.
- Interlock System: Must ensure operational safety. Fast interlock intra-station signal response time < 10 μs, inter-station signal response time < 100 μs; slow interlock intra-station signal response time < 20 ms, inter-station signal response time < 100 ms.
- Beam Tuning Platform: In addition to the remote device control, the system must be open and extensible, with tools and interfaces for developing and debugging beam tuning software.

### 4.12.2 Key Technologies and Strategy

In recent years, large-scale accelerator control systems have widely adopted integrated solutions based on mature technologies and commercial products. This approach helps control development costs and timelines while ensuring system reliability and performance. Given the diversity of the devices and their wide geographic distribution of STCF, a distributed architecture will be adopted for the control system.

After evaluating major control frameworks used in large accelerator facilities, such as EPICS [131], TANGO [132], DOOCS [133], TINE [134], MADOCA [135], and commercial SCADA systems, EPICS stands out as one of the most mature and widely adopted solutions.

EPICS (Experimental Physics and Industrial Control System) is a software toolkit, based on the DCS (Distribute Control System) standard model. It is for developing and running distributed control software systems, originally co-developed in the early 1990s by LANL and ANL for large accelerator facilities. Since then, its development has become a global collaboration, with continuous contributions to core software, device drivers, record types, and application tools.

Designed around the open systems interconnection model, EPICS features modularity, scalability, and ease of maintenance and upgrade. It has been widely used in more than 100 major research infrastructures worldwide, including:

- Colliders: KEKB (Japan)
- Synchrotron light sources: Diamond (UK)
- Free electron lasers: LCLS I/II (USA)
- Spallation neutron sources: SNS (USA), ESS (Europe)

In China, EPICS forms the foundation of the control systems at BEPCII, HLS, SSRF, CSNS, SXFEL, and SHINE. Considering performance, development schedule, and resource availability, the STCF control system will adopt EPICS as its core framework.

### 4.12.3 Design Scheme and System Composition

Based on the EPICS system architecture, the STCF control system is organized into three layers:

- Operator Interface Layer (OPI);
- Input/Output Control Layer (IOC);
- Device Control Layer (DCL).

The overall architecture is shown in Figure 4.12-1.

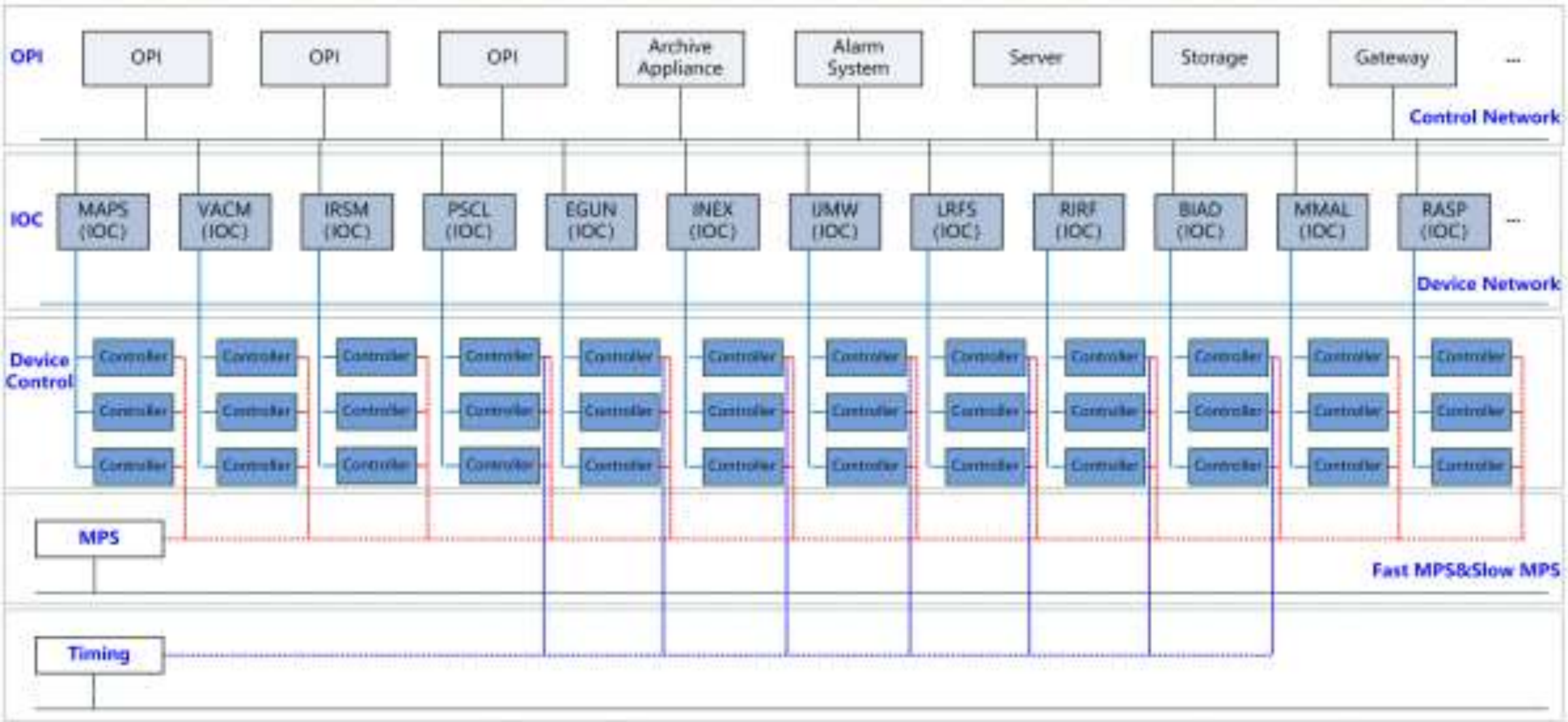

Figure 4.12-1: Overall architecture of the control system

The control system is composed of the following major components:

- Control platform

- Control network
- High-level application software
- Device control
- Interlock system
- Timing system.

***Control Platform***

The control platform provides a reliable and stable runtime environment for the control system. It includes central servers capable of long-term stable operation, storage systems, a high-reliability network system, OPI systems, and various support services such as account management, network file system, time synchronization, channel access, and version control to ensure the consistency, integrity, and security of control software and data.

***Control Network***

The control network serves as the nervous system of the entire control infrastructure, connecting all subsystems. Its performance and reliability are foundational to system effectiveness. It covers communication, network security, and operational management. A star topology is adopted, with the central server room at its core. Local control stations are connected to core switches via fiber optics. The core switch uses dual-redundancy. The network is divided into the backbone control network, local device networks, and dedicated data networks.

***High-Level Application Software***

These applications support both accelerator operation and beam tuning. Functions include system status monitoring and control, real-time and historical data access and analysis, event logging and alarm management, and accelerator physics tuning. Key components include graphical user interfaces, data archivers, alarm systems, and beam tuning toolkits.

***Device Control***

The system provides comprehensive control of all accelerator subsystems. Based on the controlled device types, it is divided into multiple subsystems: magnet power supply control, vacuum control, cryogenic control, electron source control, injection/extraction control, RF and microwave control, power system control, beam diagnostics, motion control, etc. All control software will be developed and run under EPICS. Control hardware includes protocol converter servers, network switches, IOC controllers, and other auxiliary servers.

***Interlock System***

The interlock system, also known as the Machine Protection System (MPS), is a crucial part of the control system that protects critical equipment from beam-related damage. It functions by aborting the beam or disabling power when abnormal conditions are detected. The system includes both a PLC-based slow interlock and an FPGA-based fast interlock system. A layered design is adopted, with a master station handling inter-subsystem logic and slave stations managing internal logic and protection.

***Timing System***

The timing system provides a series of precision timing and trigger signals to devices such as electron guns, kickers, modulators, microwave systems, RF cavities, and beam diagnostics to ensure synchronized operation. The White Rabbit technology will be adopted, comprising master nodes, WR switches, and slave nodes. The master node locks to the RF reference signal, while the WR switch distributes the synchronized time and frequency signals via optical fibers to downstream switches or end nodes. The end nodes generate trigger signals for connected devices.

### 4.12.4 Feasibility Analysis

Key components of the STCF accelerator control system include the control platform, device control, interlock system, and timing system. EPICS, a control framework widely adopted by the global accelerator community, is selected for development and operation. Most of the control solutions have already been successfully validated in Hefei Light Source (HLS) and Shanghai Synchrotron Radiation Facility (SSRF), significantly reducing overall system risk. Moreover, mature and stable Device Support/Driver Support is available for EPICS-based device integration.

The fast and slow interlock systems, as well as the White Rabbit timing system, have already been deployed and are operating stably in the SHINE injector. This indicates that we have accumulated solid technical experience across the major subsystems of the STCF control system, and no significant technical barriers or unforeseen risks are anticipated during its construction.

### 4.12.4 Summary

Large-scale accelerator control systems are complex engineering challenges due to:

- High device count and variety
- Interface complexity

Critical success factors include:

- Unified architecture design
- Standardized protocols

Through extensive studies of international systems and STCF-specific requirements analysis, we have:

- Established a distributed EPICS-based architecture
- Defined preliminary standards, protocols, and specifications

This foundation ensures efficient progression to detailed design and implementation.

## 4.13 Mechanics System

The STCF accelerator complex comprises two major parts: the injector and the collider rings. Within the collider rings, the MDI (Machine Detector Interface) section at the interaction region (IR) is structurally unique.

The collider consists of two intersecting electron and positron storage rings at the IR. The injector features two design schemes based on the off-axis injection and the swap-out injection in the collider rings. While their layouts differ, the main linac (0-2.5 GeV, alternating acceleration of electron and positron beams) is shared between both schemes. Other accelerator segments are also largely similar, and the mechanical support system of the linac is comparable to conventional linacs.

Each ring of the dual-ring collider contains four arc sections and four long straight sections. The primary devices are located within a radiation-shielded tunnel. The beamline elevation is 1.2 m, with the inner ring 3.5 m from the inner wall of the tunnel and the outer ring 1.5 m from the outer wall. The arc sections are composed of standard cells with essentially identical structures (Section 2.2).

The IR MDI mechanics system is highly specialized, characterized by a complex array of components, confined spatial constraints, and stringent alignment requirements. It is one of the most challenging aspects in the design of a collider [136]. The IR lies at the center of the detector and contains an ultra-thin IP chamber, superconducting magnets composed of multiple coils, and cryostats. It integrates systems related to cryogenics, magnets, vacuum, and detectors within a 7-meter-long region (with a diameter under 300 mm), where conventional mechanical and adjustment components cannot be installed or operated.

The STCF collider rings also require beam collimators, which adjust the physical aperture using movable absorbers to define beam acceptance, scattering or absorbing beam halo particles outside the acceptance. The collimators help localize beam losses, reduce radiation dose in critical areas, suppress experimental background in the IR, and protect sensitive instrumentation [137].

Beam dumps serve the critical function of absorbing high-energy discarded beams. As part of the engineering infrastructure, dumps must maximize the absorption of discarded beam energy and provide adequate radiation shielding to prevent adverse impacts on surrounding equipment, the environment, and personnel [138].

### 4.13.1 Design Requirements and Specifications

#### *4.13.1.1 Accelerator Mechanics Design Requirements*

The mechanical system will adopt a modular, stepwise integration approach. Major accelerator components will be grouped into independent support assemblies. These assemblies will be pre-installed offline and then integrated into units and finally into the full accelerator on site [136].

Design parameters vary depending on component requirements, but overall mechanical design must meet criteria related to adjuster selection, range, resolution, structural dynamic stability, cost-effectiveness, disassembly workflow, and alignment methodology. Table 4.13.1 summarizes the mechanical adjustment specifications.

Table 4.13.1: Mechanical adjustment specifications for the STCF accelerator components

| Parameter | Specification |
|---|---|
| Horizontal adjustment range (X/Y) | ±10 mm |
| Horizontal resolution (X/Y) | ⩽10 μm |
| Vertical adjustment range (Z) | ±8 mm |
| Vertical resolution (Z) | ⩽5 μm |

***Interaction Region Design Requirements***

The IR includes: the IP vacuum chamber inside the detector, bellows, remote vacuum connection (RVC) structures, beam position monitors, superconducting magnets and cryostats (IRSM), as well as surrounding room-temperature magnets, cryo-pipelines, and movable platforms.

Key component specifications:

1. **IP Chamber**
    - Material: Beryllium + Tantalum
    - Inner beryllium pipe diameter: 30 mm
    - IP chamber length: 1000 mm
    - Vacuum: better than $5\times10^{-8}$ Pa
2. **Movable Platform**
    - Foundation flatness: < 0.5 mm
    - Guide rail parallelism after installation: < 0.05 mm
    - Guide rail flatness after installation: < 0.05 mm
    - Alignment accuracy of platform: < 0.1 mm
    - Positioning accuracy of platform: < 0.2 mm
3. **RVC (Remote Vacuum Connection)**
    - Alignment accuracy between IRSM vacuum chamber and IR vacuum chamber bellows: < 0.1 mm

- Remote-controlled locking/unlocking capability
- Remote vacuum diagnostics and pumping capability
- Must meet vacuum requirements of the IR

#### *4.13.1.3 Beam Collimator Design Requirements*

Primary design specifications for the beam collimator include:

- Length (beam direction): 1000 mm
- Taper angle of scraper: < 0.15 rad
- Travel range of scraper: ≥ 24 mm (Din=60 mm); ≥ 23.5 mm (Din=67 mm)
- Motion positioning accuracy: < 50 μm
- Radiation tolerance: ≥ 1 MGy

#### *4.13.1.4 Beam Dump Design Requirements*

Main specifications for the beam dump system include:

- Positioning accuracy: ≤ 0.1 mm
- Environmental radiation dose: < 2.5 μSv/h
- Must meet vacuum specifications

### 4.13.2 Key Technologies and Technical Approaches

The primary challenges in accelerator mechanical support lie in meeting the stability and precision requirements imposed by beam physics and the accelerator components themselves. For most components not subject to extremely tight tolerances, conventional mechanical alignment and support designs are generally sufficient.

For the interaction region (IR), the mechanical system includes several critical technologies:

**Key Technology 1: IP Chamber Fabrication.** The central beryllium vacuum chamber, with a wall thickness of just 0.4-0.6 mm, presents major challenges in welding and assembly. An additional difficulty lies in ensuring uniform cooling performance when coolant is circulated through a cavity of 200 mm in length and only 1 mm in gap.

**Key Technology 2: Remote Vacuum Connector (RVC).** This device must offer reliable and convenient remote operation. Development challenges include the choice of sealing material for the vacuum interface, the precision of mechanical alignment, and the ability to maintain pressure and vacuum integrity.

**Key Technology 3: IP Chamber Integration with Detector.** The innermost layer of the detector (ITKW/ITKM) has a diameter of only 70 mm, whereas the IP chamber has a maximum radial size of 100 mm. This leaves extremely limited space for assembly. The current design requires the entire assembly to be integrated externally and installed as a unit. Ensuring accurate positioning and future maintainability is a significant design challenge.

**Key Technology of Beam Collimators.** The beam absorber (scraper) is subjected to multiple effects, including energy deposition, synchrotron radiation, Touschek scattering, and beam-induced impedance heating. The design must address these through simulation-driven material selection and structural optimization to reduce impedance while ensuring high reliability and long operational life.

### 4.13.3 Design Scheme and System Composition

#### *4.13.3.1 Accelerator Mechanical Design Scheme and System Composition*

**Injector Mechanics**

The injector system is designed under two injection schemes—off-axis injection and swap-out injection—corresponding to the collider ring layout, as shown in Fig. 1.2-1. The injector mechanical systems cover the linac segments for both electrons and positrons (including the shared main linac), the positron target station, the damping/accumulation ring, and associated beam transport lines.

The main component of the injector is the accelerating structure. Its support design largely follows standard practice established by domestic synchrotron light sources. Mechanical prototypes for STCF injector support structures will be further refined based on the finalized injector lattice.

**Collider Ring Mechanics**

The STCF collider adopts a dual-ring layout composed of an electron ring and a positron ring, as illustrated in Fig. 4.13-1. Both rings lie in the same horizontal plane and intersect at the interaction point (IP) and at the symmetric crossing point opposite the IP. Each ring is two-fold symmetric, and the arcs of the inner and outer rings have slightly different lengths due to the crossing geometry, resulting in a ~2 m spacing between rings.

Each ring consists of one interaction region, four large arc sections (each with a 60° bend), two short arc sections (each with a 30° bend), one crossing region, and several straight sections. The total ring circumference is 860.321 meters.

Taking the short arc section as a design example, Fig. 4.13-2 (left) shows the layout for the mechanical support system. The tunnel is 7.5 m wide, with each ring 2.5 m from the tunnel wall. The beamline height is 1.2 m, and the spacing between the two rings is 2 m. The tunnel height is 4.5 m, and ring-shaped lifting equipment is installed overhead.

In the local support structure, second-order (B2) magnets use separate supports, while fourth-order (B4) and sixth-order (B6) magnets are installed on a shared girder. Each magnet is equipped with three-directional mechanical adjusters, as shown in Fig. 4.13-2 (right).

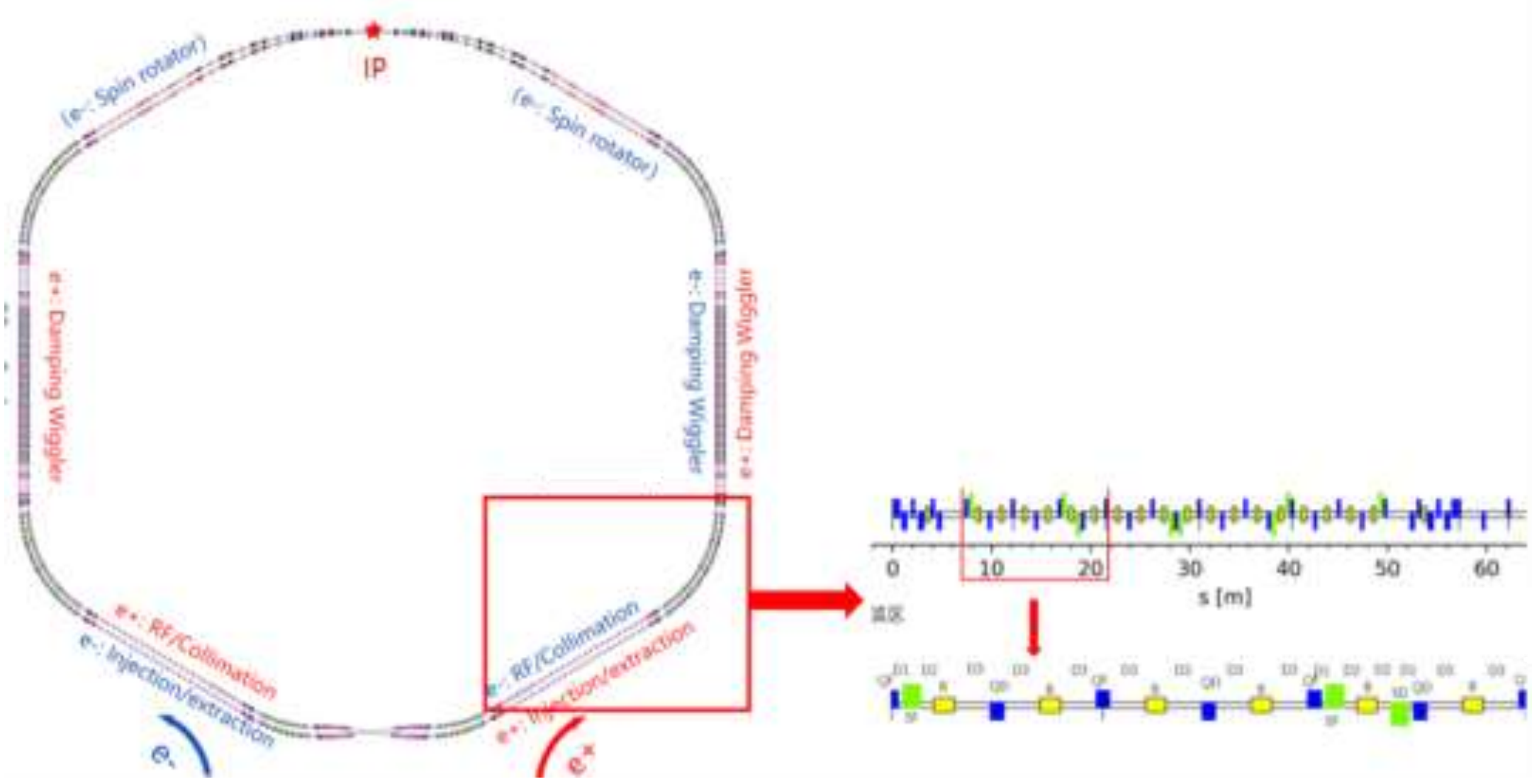

Fig. 4.13-1: Collider ring layout and short arc region beam physics design requirements

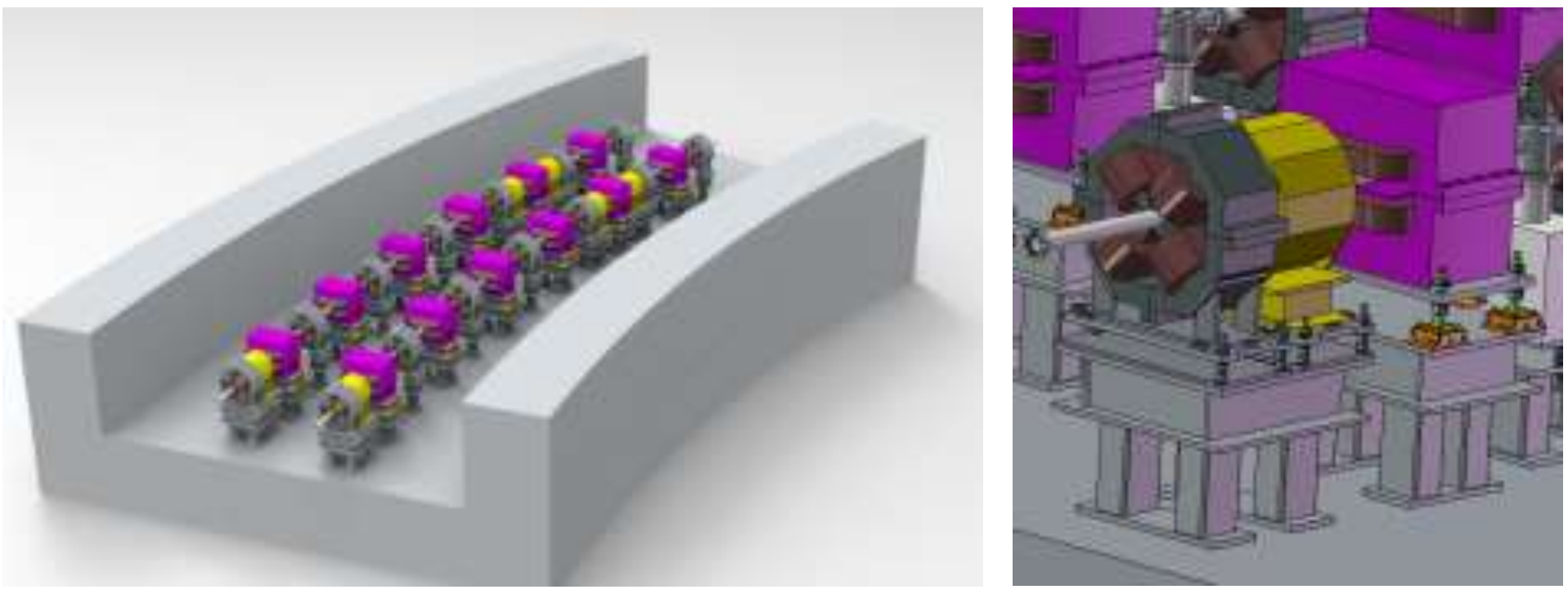

Fig. 4.13-2: Short arc standard unit layout (left) and magnet mechanical support system (right)

### *4.13.3.2 Structure and Design Scheme of the Interaction Region*

The composition of the interaction region (IR) is shown in Fig. 4.13-3. The STCF IR is defined as the ±8 m region centered on the interaction point (IP), and includes the beryllium IP vacuum chamber extending into the detector, bellows, remote vacuum connectors (RVCs), beam position monitors (BPMs), superconducting magnets and cryostats (IRSM), room-temperature magnets outside the cryostat, cryogenic piping, and support structures such as movable platforms.

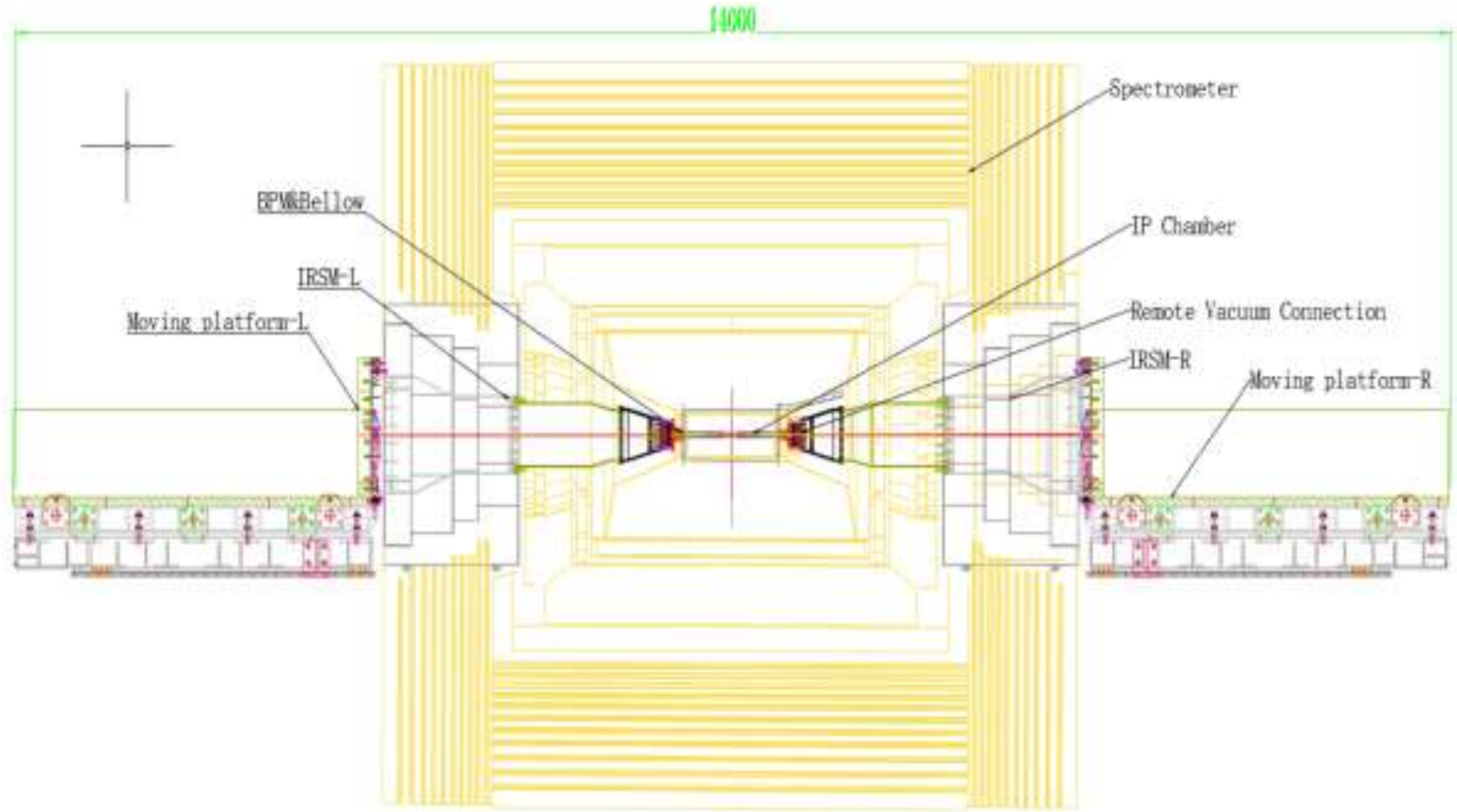

Fig. 4.13-3: Components of the interaction region

Figure 2.11-1 shows the crossing angle of the electron and positron beams in the IR, as well as the positions and lengths of the relevant beamline components. A beryllium vacuum chamber is placed at the collision point in the detector center, with bellows at both ends connecting to the upstream and downstream vacuum chambers. These bellows are designed to absorb thermal expansion of the beryllium chamber and provide flexible connections between vacuum sections. The outer vacuum chambers are connected to the superconducting magnet vacuum chambers via remote vacuum connectors (RVCs), which are critical mechanical-vacuum components of the IR. These connectors support automatic alignment and remote locking/unlocking of vacuum flanges, facilitating equipment installation and maintenance.

The IRSM magnets and upstream/downstream warm magnets are mounted on a movable platform. The cryogenic valve box and transfer lines that supply liquid helium to the IRSM magnets are also mounted on the platform's sides. Figure 4.13-4 shows the mechanical structure of the movable platform, based on the physical design.

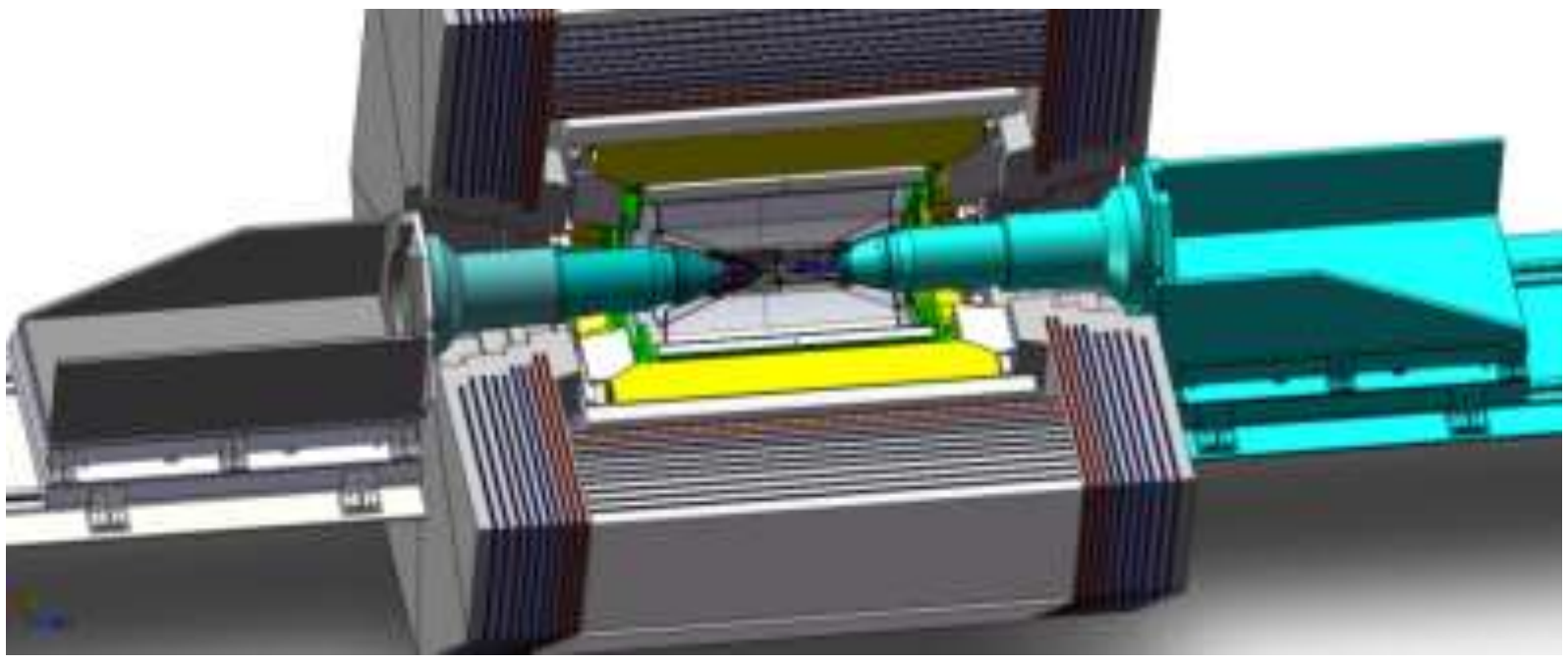

Fig. 4.13-4: Mechanical structure schematic of the interaction region

**IP Chamber Design Scheme**

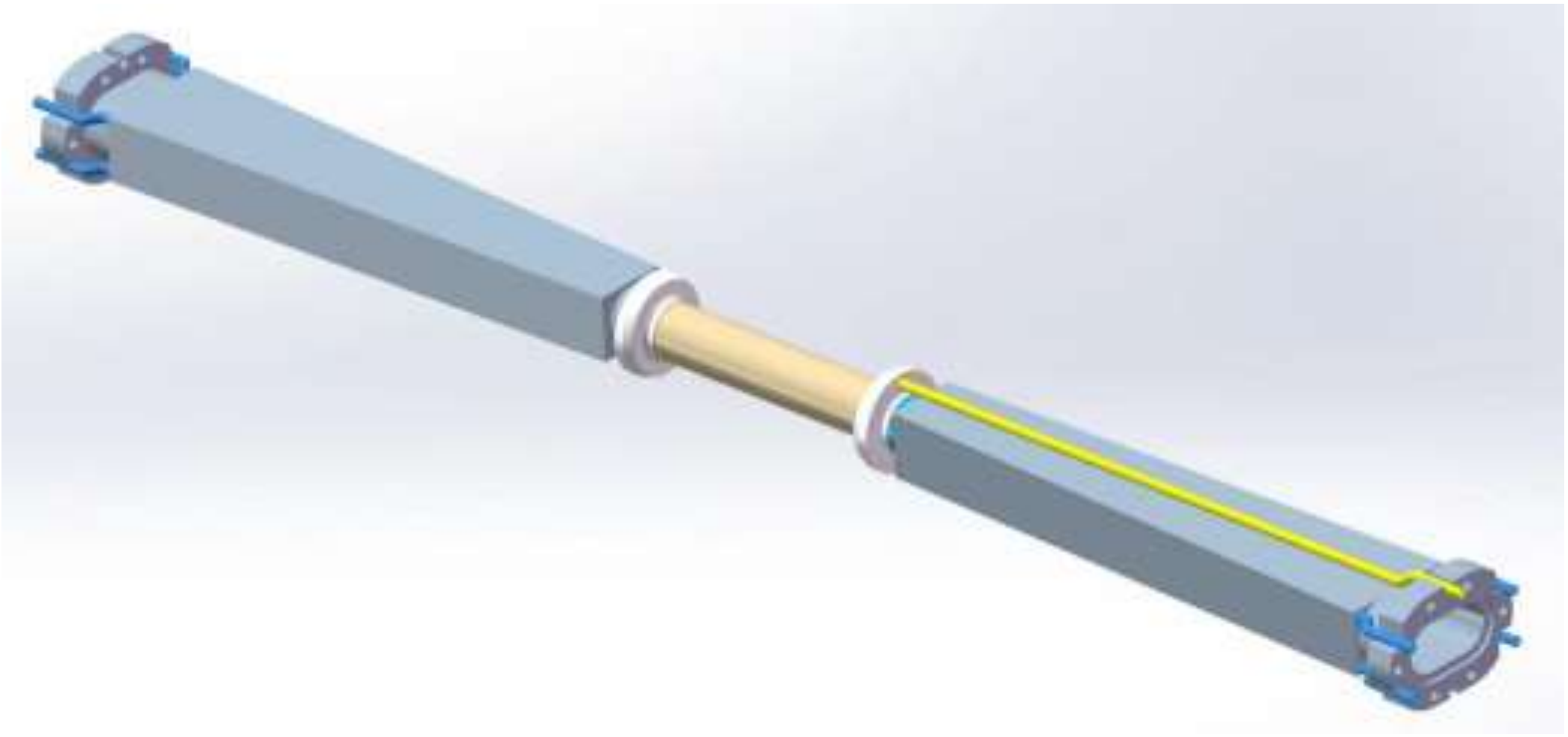

Fig. 4.13-5: IP beam pipe schematic

The IP beam pipe is designed symmetrically around the IP. To minimize material budget in the interaction region, beryllium is used due to its low density and high radiation transparency, which makes it ideal for high-energy physics applications [138]. According to the preliminary MDI design, the central beryllium beam pipe adopts a double-layer structure: the inner tube has an inner diameter of φ30 mm and a wall thickness of 0.6 mm, coated internally with 10 μm of gold. The outer tube has a wall thickness of 0.4 mm, with a 1.0 mm gap between layers through which electric spark oil or paraffin is circulated [136, 139, 140]. The beryllium section is 205 mm long, and the total beam pipe length is 1000 mm.

The beryllium section is flanked by tantalum pipes from circle diameters of φ30 mm expanding to a special ellipse with long diameter 60 mm and short diameter 30 mm at the flange ends (see Fig. 4.13-5). These tantalum sections are water-cooled, and their inner wall is coated with 10 μm of copper [5]. To avoid deformation due to long-span sagging, the beam pipe includes a support system with axial alignment capability and is fixed to the detector inner chamber for precise installation. The outer structure includes inlet/outlet piping for coolant circulation and integrated sensors for monitoring.

**Movable Platform Design Scheme**

To ensure the alignment accuracy of accelerator beamline components in the MDI region, the mechanical structure and motion precision of the movable platform (see Fig. 4.13-6) are critical. Due to space constraints, the IRSM on the platform is implemented as a cantilever structure. Finite element analysis assuming a uniform 2-ton load on the IRSM shows a deflection of 0.29 mm at the platform end, as illustrated in Fig. 4.13-7.

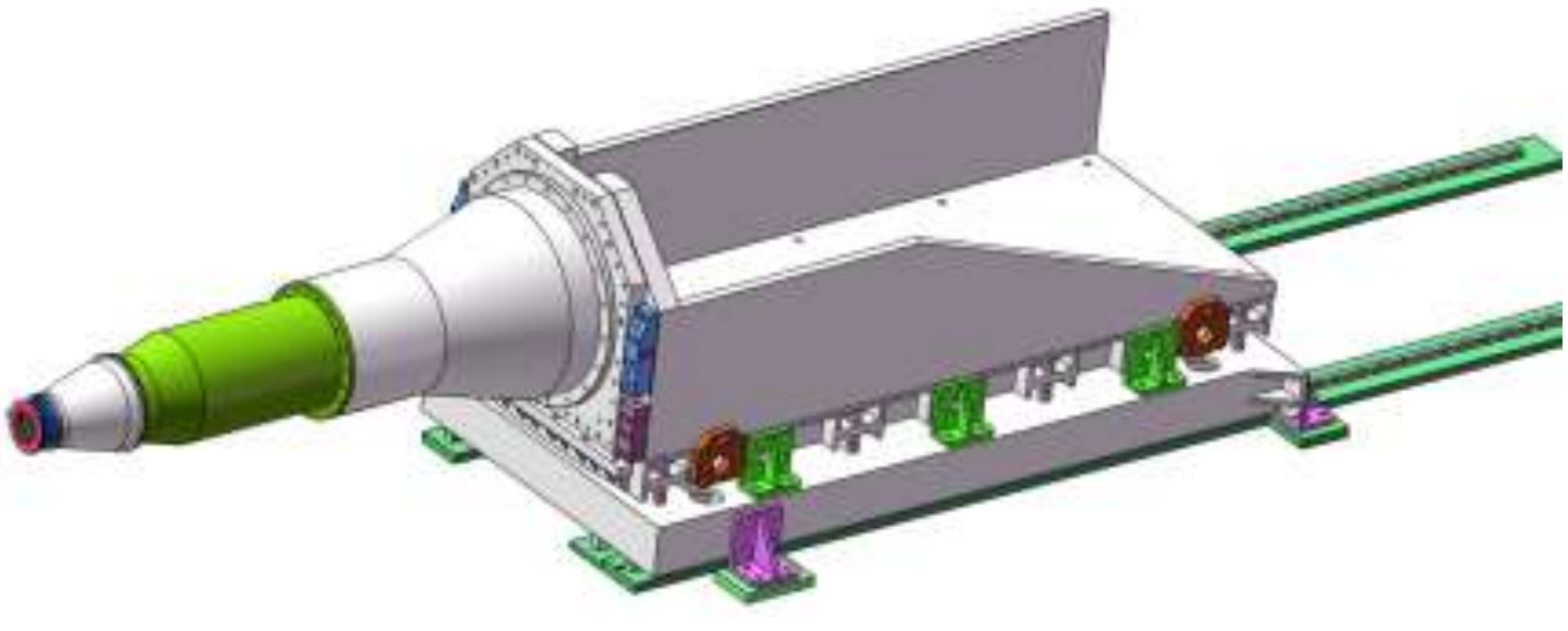

Fig. 4.13-6: Schematic of the movable platform design

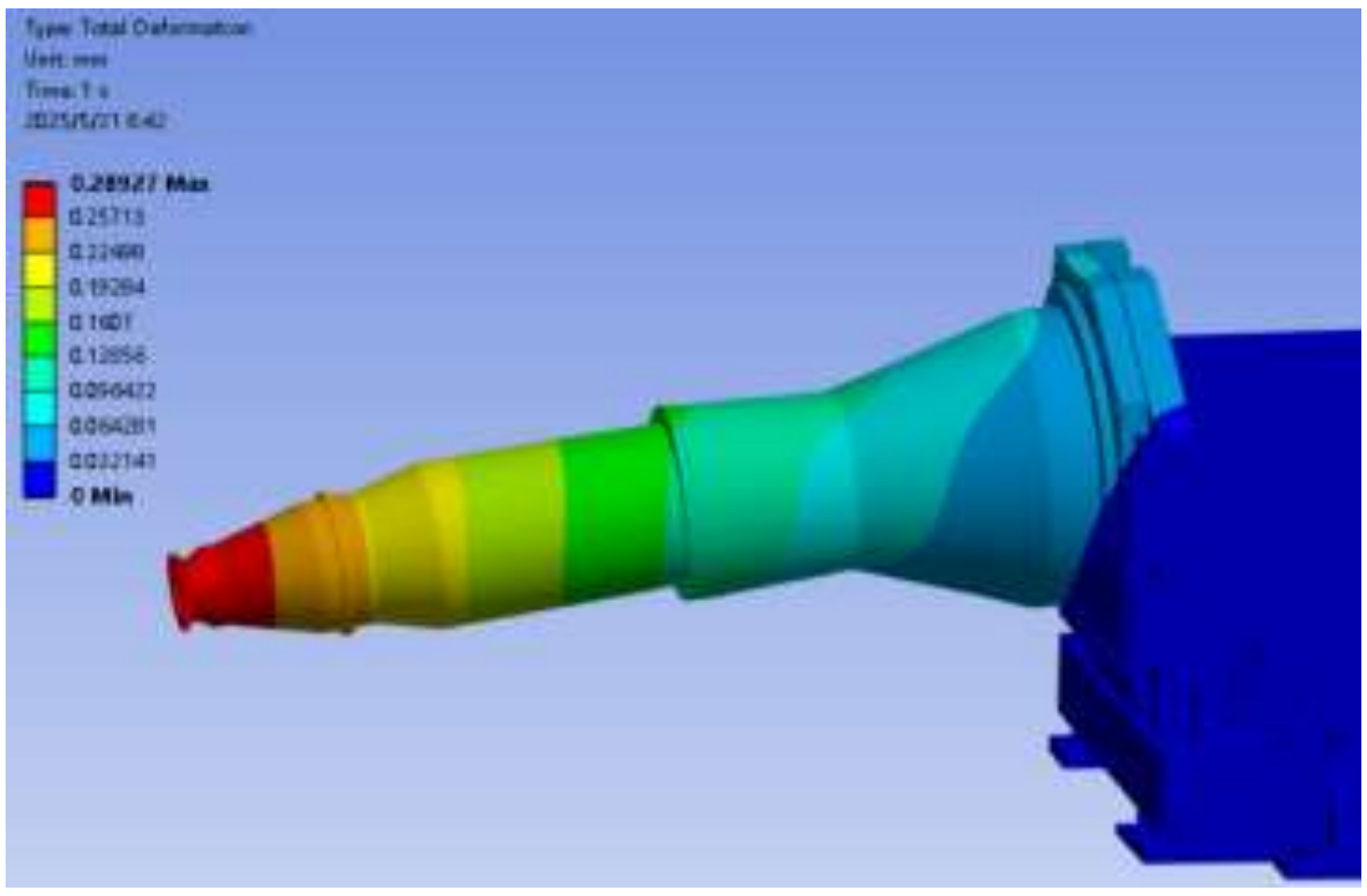

Fig. 4.13-7: Finite element simulation of front-end deformation

**RVC Structure Design Scheme**

Based on the RVC design used in SuperKEKB, a corresponding RVC concept has been proposed for STCF. When the IRSM is pushed into the detector region, initial flange alignment is achieved via two tapered pins. A remote handle rotates a gear disc, gradually pressing the vacuum flange faces together. Clamping force is provided by custom butterfly springs. A prototype of the STCF RVC structure is shown in Fig. 4.13-8.

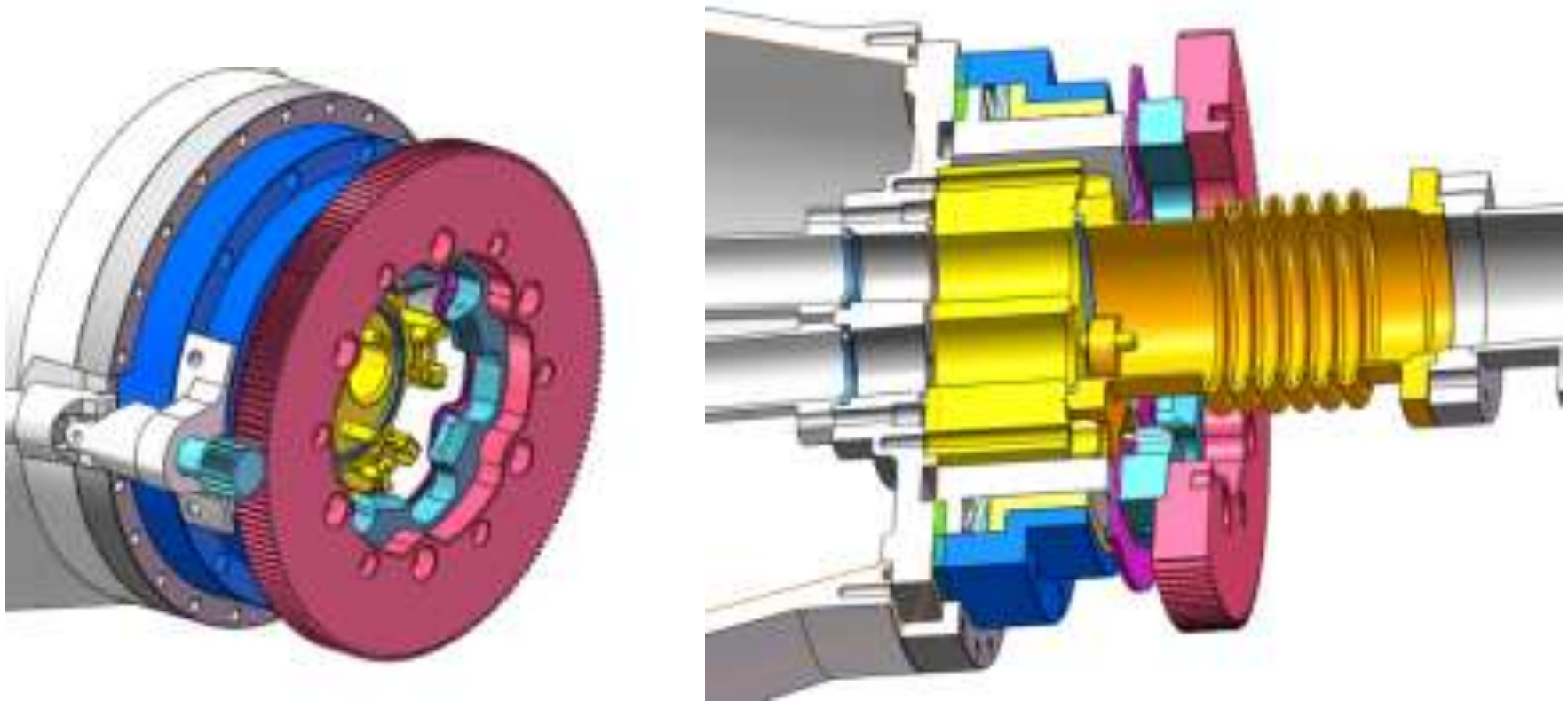

Fig. 4.13-8: Conceptual prototype of the RVC structure

### *4.13.3.3 Structure and Design Scheme of the Beam Collimators*

Based on design experience from major international colliders and synchrotron radiation sources, a STCF beam collimator mainly consists of the absorber, vacuum chamber, motion control system, support and adjustment mechanism, water cooling system, and radiation shielding structure, as shown in Fig. 4.13-9.

To cope with high thermal loads caused by direct beam impact, Touschek scattering, beam impedance, and synchrotron radiation, the absorber material is preliminarily selected as Glidcop-AL-15 (a dispersion-strengthened copper reinforced with nano-scale alumina). The vacuum chamber of the collimator has a length of 800 mm, with 100 mm long shielding bellows installed at both the upstream and downstream ends.

To reduce the impedance contribution, vertical internal surfaces of the vacuum chamber are capped with a plate to eliminate cavity-like structures. A tapered end section of 100 mm length is designed with optimized slanted surfaces to ensure a smooth longitudinal transition. The comb-shaped RF shielding fingers are used between the absorber and the vacuum chamber in the longitudinal gap along the beam, while spring-shaped RF shielding fingers are added across the vertical gaps.

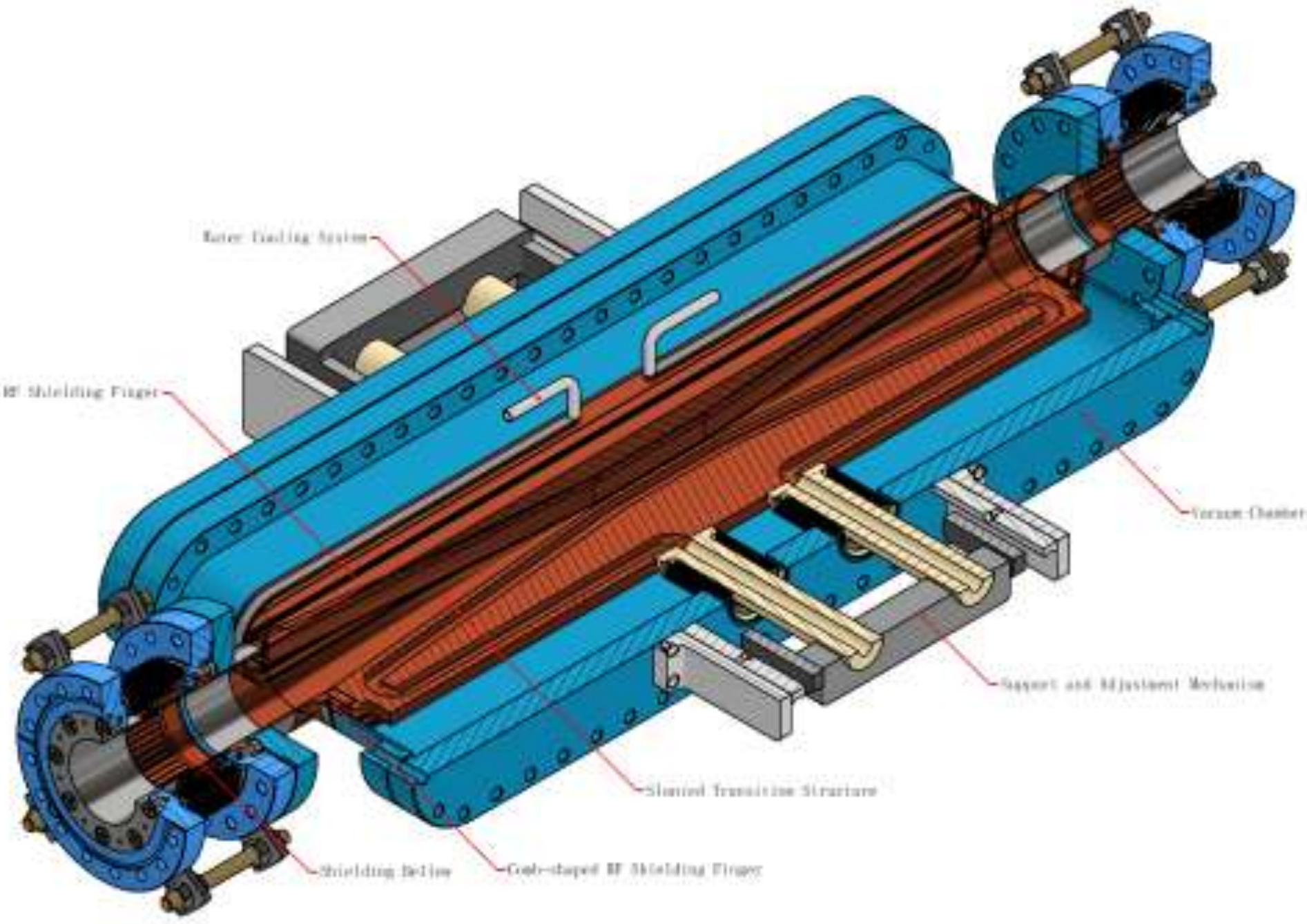

Fig. 4.13-9: Structural model of the collider ring beam collimators

### *4.13.3.4 Structure and Design Scheme of the Beam Dump Station*

The beam dump station mainly comprises an absorber, shielding structure, and support and adjustment mechanism, as shown in Fig. 4.13-10.

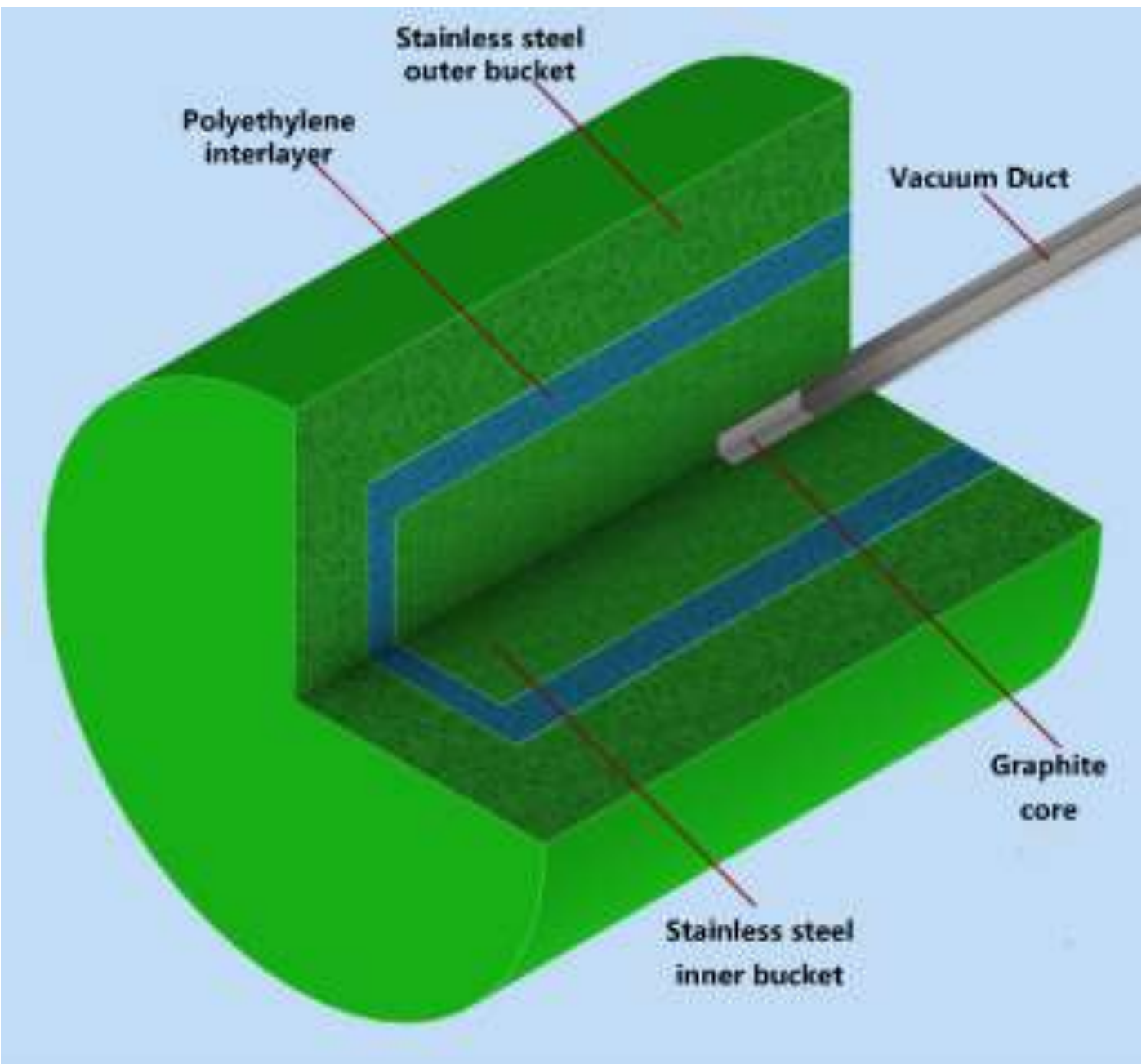

Fig. 4.13-10: Structural model of the beam dump station

### 4.13.4 Feasibility Analysis

The mechanical support schemes for accelerators have been widely adopted in major facilities worldwide and have been thoroughly validated for feasibility, thus providing a solid foundation for implementation.

The IP chamber is optimized based on experience from SuperKEKB, but further refinements in physics design, mechanical integration, fabrication, and interface with the detector are still required. It is recommended to conduct prototype development and mock installation tests. Only after repeated validation and meeting design specifications should the system be deployed for operational use.

The RVC (Remote Vacuum Connector) design must address both vacuum integrity and remote operability. Mechanical prototyping and pre-research are recommended, including the setup of a vacuum docking simulation platform. The system should be subjected to repeated testing to verify performance before final implementation.

The beam collimator adopts a relatively mature technical path. However, key processes such as material bonding, machining, and welding need to be further validated to ensure safety and reliability for large-scale engineering use.

The beam dump station design is already well-established. Thermal and radiation shielding analyses based on actual beam energy and power parameters have been performed to ensure that the design meets all application requirements.

### 4.13.5 Summary

The total beamline length of the STCF accelerator exceeds 2.5 km and includes a large number of beamline elements with varying requirements. The side-by-side layout of the double storage rings in the collider poses demanding challenges for mechanical system integration and tunnel space allocation.

The mechanical support requirements for the arc sections of the collider rings are similar to those in existing domestic and international accelerator facilities. With the recent development and operation of new accelerators in China, technical and manufacturing capabilities have matured sufficiently to meet these requirements.

However, the mechanical system of the interaction region poses a particular challenge for STCF. It involves numerous subsystems, complex interfaces, and high technical risks. A thorough review of successful and unsuccessful experiences from similar international facilities will be carried out. On this basis, in combination with the STCF overall design and China's current engineering capabilities, detailed scheme design and technical studies will be undertaken. Only after rigorous prototyping, testing, and verification will the system be implemented in the full-scale project.

## 4.14 Alignment System

### 4.14.1 Design Requirements and Specifications

The alignment task of the STCF accelerator is to ensure a relatively smooth trajectory along the magnetic centers of the components for the electron and positron beams under high-precision control. Therefore, alignment is generally implemented in two phases. During the construction period of STCF, a unified control network for the facility area was established through high-precision alignment instruments and optimized schemes. Based on this network, equipment is adjusted into its theoretical positions within a specified accuracy range, thereby providing an initial trajectory for accelerator operation. In the operation phase, online monitoring of equipment positions and trajectory smoothing is performed.

As shown in Table 4.14-1, the alignment precision requirements for the STCF collider rings are as follows: within and between supports, the relative horizontal alignment precision of components is 50 μm, and the vertical precision is 100 μm. Additionally, the attitude angle control precision for all magnet equipment must be better than 0.1 mrad. The absolute precision of the closed orbit circumference of the collider rings should be better than 2 mm.

**Table 4.14-1: Alignment Precision Requirements for the STCF Collider Rings**

| Parameter | Index |
|---|---|
| Relative horizontal precision within a support (mm) | 0.050 |
| Relative vertical precision within a support (mm) | 0.100 |
| Relative horizontal precision between supports (mm) | 0.050 |
| Relative vertical precision between supports (mm) | 0.100 |
| Magnet attitude angle precision (mrad) | 0.100 |
| Superconducting quadrupole magnets in the IP region (mm) | 0.020 |
| Ring closed orbit circumference precision (mm) | 2.000 |

Overall, the alignment accuracy requirements are highest and most demanding for the collider rings. The damping and accumulation rings require similar precision but are less challenging due to their smaller scale. The linacs and transport lines of the STCF injector has the lowest precision requirements, as listed in Table 4.14-2.

Table 4.14-2: Alignment Precision Requirements for the STCF Injector

| Parameter | Precision |
|---|---|
| Accelerator structure position precision (mm) | 0.1 |
| Accelerator structure attitude angle precision (mrad) | 0.1 |
| Corrector magnet position precision (mm) | 0.5 |
| Corrector magnet attitude angle precision (mrad) | 0.5 |

Moreover, the equipment alignment and monitoring in the constrained and compact interaction region (IR), as well as the smoothing and differential control of the dual-ring structure, are among the key alignment challenges.

### 4.14.2 Key Technologies and Roadmap Selection

Considering the high-precision alignment and monitoring in the constrained IR region and the specific features of the dual-ring structure in the STCF tunnel, the following core alignment technologies are identified:

#### *4.14.2.1 FSI-Based High-Precision Positioning Technology for Compact Regions*

For precise alignment and positioning of superconducting quadrupole magnets in the compact MDI region of the STCF, traditional methods, such as multi-laser systems requiring cooperative targets, are not practical. Frequency Scanning Interferometry (FSI) is employed instead. FSI uses a laser with a linearly varying frequency as the light source. Due to time delays from the measured distance, the reference and return beams generate a small, stable frequency difference upon overlap, producing an interference beat signal. This method offers a wide measurement range, high precision, strong anti-interference ability, and, most importantly, low return power requirements (suitable for non-cooperative diffuse reflection surfaces). FSI is being considered in the alignment systems of the FCC-ee [141, 142] and CEPC study projects.

#### *4.14.2.2 Multi-Sensor Fusion Monitoring of Deformation in Confined Spaces*

Distributed fiber optic sensor technologies, commonly used in tunnels and bridges, are adopted for long-term deformation monitoring of key equipment nodes in the MDI region. These technologies offer real-time data, high precision, interference resistance, and support for distributed sensing. Fiber Bragg Grating (FBG) sensors can achieve distance measurement precision of tens of microns, with each sensor segment covering centimeter-scale ranges. By placing multiple FBG sensors at critical locations, long-term monitoring of mechanical deformation becomes feasible. This approach has also been applied in FCC-ee [3] and CEPC, with the potential to be supplemented by CCD and other fusion sensor technologies.

#### *4.14.2.3 Trajectory Smoothing and Ring-to-Ring Differential Control for Dual-Ring Layouts*

Trajectory smoothing is essential for high-precision alignment and is a key guarantee for accelerator tuning. Given that the STCF collider adopts a dual-ring structure for electrons and positrons, with limited tunnel space and correlated alignment errors between the two rings, dedicated research is needed on alignment-based trajectory smoothing and differential control in such a configuration [144].

#### *4.14.2.4 Accurate Magnet Center Determination and High-Precision Transfer Techniques*

Precise extraction of the magnetic centers of magnets is critical for successful alignment and beam tuning. Depending on the magnet's special physical design, methods such as rotating coil, vibrating wire, magnetic targets, image-based, capacitive wire position sensors, coordinate measurement machines (CMMs), and precision photogrammetry are employed for reference point extraction and high-precision mapping [145].

#### *4.14.2.5 Accuracy and Efficiency Enhancement Under Strong Reference Constraints in Tunnel Alignment Networks*

Due to the STCF's large scale and the long-term need for alignment maintenance, it is necessary to improve precision, efficiency, and reliability under strong-reference-constrained tunnel alignment control networks. This is achieved by providing geometric constraints (position and attitude) to the laser tracker stations, thereby controlling cumulative measurement errors in the secondary network and resolving directional instability in intermediate sections [146-149].

### 4.14.3 Design Scheme and System Composition

The STCF alignment system is designed with seven key processes:

- Establishment and measurement of the first-level surface control network;
- Establishment and measurement of the second-level tunnel control network;
- Extraction and calibration of physical centers of equipment;
- Pre-alignment installation of magnet units on support structures;
- Initial alignment of equipment inside the tunnel;
- Beamline trajectory smoothing and dual-ring differential control alignment;
- Monitoring of ground settlement and deformation.

#### *4.14.3.1 Establishment and Measurement of First-Level Surface Control Network*

The first-level control network consists of permanent control points for the injector, collider rings, and surface installations. It is typically constructed by downward excavation and permanent rock pile foundations, with vertical projections aligned to the building roof. Measurements are performed using a combination of GNSS receivers and total stations. This network serves two purposes:

- To maintain global spatial relationships between the linacs, transport lines, damping ring, and collider rings;
- To provide absolute accuracy constraints so that relative positional requirements from the physics design are met across the facility.

#### *4.14.3.2 Establishment and Measurement of Second-Level Tunnel Control Network*

The second-level control network is deployed along the tunnel every 3-6 meters, with five reference points per segment: two on the floor, one each on the inner and outer walls, and one on the ceiling. The reference points are custom-machined and permanently bonded to the tunnel surfaces with epoxy. This network reestablishes a unified facility coordinate system and provides essential spatial positioning references for equipment installation.

#### *4.14.3.3 Extraction and Calibration of Physical Centers of STCF Equipment*

The physical center of STCF equipment typically refers to the magnetic or electric center. Since these centers are not directly observable, they must be transformed into mechanical centers accessible for alignment. This transformation introduces inevitable conversion error, so strict control of extraction accuracy is essential. For components with moderate accuracy requirements, the mechanical center is used as a substitute. Using standard edge method laser trackers, the mechanical center is measured and extracted with an accuracy up to 0.05 mm.

#### *4.14.3.4 Pre-Alignment Installation of Equipment on Support Structures*

Pre-alignment installation is performed in controlled laboratory environments using high-precision tools. Equipment is locally aligned on support modules before tunnel installation. This process improves local alignment precision and significantly reduces installation time and complexity during tunnel assembly.

#### *4.14.3.5 Initial Alignment of Equipment in the Tunnel*

Once accelerator equipment is coarsely positioned inside the tunnel, laser trackers use the tunnel control network to reestablish the facility-wide coordinate system. This enables accurate alignment of magnets and other components to their theoretical spatial positions.

#### *4.14.3.6 Beamline Trajectory Smoothing and Dual-Ring Differential Control Alignment*

Due to tunnel foundation settlement, environmental effects, and installation errors, deviations from theoretical trajectories may arise, resulting in trajectory irregularities and affecting machine tuning. A global smoothing algorithm is used post-installation to minimize local corrections while achieving globally optimized beamline smoothness. Given the unique dual-ring layout of STCF, additional studies and optimization are required for differential control and trajectory smoothing between the two rings to meet operational physics requirements.

#### *4.14.3.7 Monitoring of Ground Deformation*

Foundation movement can lead to cumulative displacement of equipment. Long-term monitoring is needed to track such deformation and enable correction during operational maintenance.

### 4.14.4 Feasibility Analysis

Unlike fourth-generation diffraction-limited synchrotron light sources that require sub-micron alignment precision over whole storage rings, at STCF, the exigent alignment requirements are mostly focused on the interaction region. Multiple fourth-generation sources, such as HEPS, have been successfully constructed and commissioned both in China and abroad. The technical solutions for STCF, by comparison, are generally mature and feasible.

Among the STCF subsystems, the collider rings present the highest alignment precision challenge, followed by the damping ring, linacs, and beam transport lines. The interaction region (IR) of the collider rings is particularly complex due to limited space, cryogenic conditions, and the need for precision in difficult-to-access locations. The dual-ring structure also adds complexity compared to single-ring light sources.

Targeted research has been conducted on five key technologies:

- FSI-based high-precision alignment in confined spaces;
- Multi-sensor deformation monitoring in elongated enclosed environments;
- Dual-ring trajectory smoothing and differential control;
- High-precision magnetic center extraction;
- High-efficiency accuracy enhancement under strong-reference tunnel control networks.

These studies indicate that the alignment requirements of STCF can be realized with a good confidence.

### 4.14.5 Summary

Through the systematic description of the STCF alignment system including the design, specifications, core technologies, system composition, and feasibility analysis, it can be concluded that the proposed technical approach is highly executable and capable of meeting the alignment requirements of the STCF accelerator.

## 4.15 Radiation Protection System

The STCF accelerator is a high-energy electron-positron collider composed of multiple linac sections, a damping/accumulation ring, several beam transport lines, and two collider rings. The maximum beam energy for both electrons and positrons is 3.5 GeV, with a center-of-mass energy range of 2-7 GeV and a maximum circulating current of 2 A in the collider rings. In addition to the synchrotron X-ray radiation emitted during normal operation, other radiation sources include bremsstrahlung generated by interactions between the beam and residual gas, and secondary particles produced when beam losses interact with accelerator components and shielding materials via nuclear reactions. These can result in exposure to personnel and the surrounding environment.

The primary goal of radiation shielding is to ensure that the dose outside of the shielding remains within design and regulatory limits. Following the ALARA principle (As Low As Reasonably Achievable), radiation exposure to people and the environment should be minimized. The radiation protection system is therefore an essential part of the STCF facility. It includes studies and design related to radiation fields during operation and maintenance, shielding calculations and layout, dose estimation, monitoring systems, personnel safety interlocks, personal and environmental dose monitoring, evaluation of gaseous/liquid/solid radioactive waste impacts, and general radiation safety management.

### 4.15.1 System Design Requirements and Criteria

#### *4.15.1.1 Dose Constraint Values and Limits*

According to the Basic Standards for Ionizing Radiation Protection and the Safety of Radiation Sources (GB 18871-2002) [150] and Regulations for Radiation Protection of Particle Accelerators (GB 5172-85), combined with the specific needs of the STCF project:

- Basic Dose Limits:
  - For occupational exposure, the average effective dose limit over five consecutive years is 20 mSv/year.
  - For members of the public: the annual effective dose limit for individuals in critical groups is 1 mSv.
- Dose Constraints (used for design purposes):
  - STCF radiation workers: 5 mSv/year

- Visiting scientists or users: 1 mSv/year, assuming up to 400 hours of work per year
- General public: 0.1 mSv/year

#### *4.15.1.2 Dose Rate Control Levels*

Under normal operating conditions, the ambient dose equivalent rate at 30 cm from the shielding surface of the beamline enclosure must be less than 2.5 μSv/h.

Under beam-loss accident scenarios, the maximum accumulated dose at the outer surface of the shielding must be less than 1 mSv per incident.

#### *4.15.1.3 Control Levels for Induced Radioactive Gases and Clearance Levels for Solid Waste*

Induced radioactive gas control levels are based on the exemption activity concentration limits in Appendix A, Table A1 of GB 18871-2002, assuming an air density of $1.225\times10^{-3}$ g/cm³. The control limits during tunnel operation are given in Bq/cm³:

| Isotope | H-3 | Be-7 | C-14 | O-15 | S-35 | Ar-37 | Cl-38 | Ar-41 |
|---|---|---|---|---|---|---|---|---|
| Limit | 1.23E+03 | 1.23E+00 | 1.23E+01 | 1.23E-01 | 1.23E+02 | 1.22E+03 | 1.22E-02 | 1.23E-01 |

For radioactive solid waste, the clearance levels are also based on GB 18871-2002, Appendix A1. Below are selected isotopes and their corresponding clearance criteria:

| Isotope | Activity Conc. (Bq/g) | Total Activity (Bq) | Isotope | Activity Conc. (Bq/g) | Total Activity (Bq) |
|---|---|---|---|---|---|
| H-3 | 1.0E+06 | 1.0E+09 | Mn-56 | 1.0E+01 | 1.0E+05 |
| Sc-46 | 1.0E+01 | 1.0E+00 | Co-56 | 1.0E+01 | 1.0E+05 |
| V-48 | 1.0E+01 | 1.0E+06 | Co-57 | 1.0E+02 | 1.0E+06 |
| Cr-51 | 1.0E+03 | 1.0E+05 | Co-58 | 1.0E+01 | 1.0E+06 |
| Mn-51 | 1.0E+01 | 1.0E+07 | Mo-99 | 1.0E+02 | 1.0E+06 |
| Mn-52 | 1.0E+01 | 1.0E+05 | Mn-52m | 1.0E+01 | 1.0E+05 |
| Mn-54 | 1.0E+01 | 1.0E+06 | Co-58m | 1.0E+04 | 1.0E+07 |
| Fe-55 | 1.0E+04 | 1.0E+06 | Tc-99m | 1.0E+02 | 1.0E+07 |

#### *4.15.1.4 Wastewater Discharge Standards*

According to GB 18871-2002, Section 8.6.2, radioactive wastewater may be discharged to regular municipal sewage systems with a flow rate at least 10 times higher than the discharge rate, after treatment and approval by regulatory authorities. Discharge limits are:

- Monthly total discharged activity: ≤ 10 ALImin
- Per-discharge activity: ≤ 1 ALImin

- After each discharge, flush the system with water at least 3× the discharge volume.

*Note: ALImin refers to the smaller value between the annual intake limits for ingestion or inhalation relevant to occupational exposure.*

### 4.15.2 Key Technologies and Technical Approaches

#### *4.15.2.1 Source Term Analysis*

As a high-energy electron–positron accelerator, the STCF involves a variety of beam–matter interactions, including electron/positron interactions with residual gas, collimators/targets, vacuum chamber walls, and beam–beam collisions. Electromagnetic cascades and nuclear reactions dominate the interaction processes, giving rise to ionizing radiation fields (instantaneous and mixed radiation fields) and the formation of radioactive nuclei inside target materials (induced radioactivity). The radiation sources include both prompt radiation and induced radioactivity.

Sources of prompt radiation include:

- Random beam losses from particles in the beam halo striking the vacuum chamber wall (modeled as uniform loss);
- Sudden loss of an entire bunch (100% loss of beam bunch);
- Bremsstrahlung generated by beam interaction with residual gas;
- Point losses from positron targets, positron linac sections, the electron and positron storage rings, damping/accumulation rings, and main linacs;
- Radiation from electron–positron collisions;
- Beam losses during machine tuning and at beam dumps (from bunch swap-out or beam abort scenarios during injection in the swap-out mode).
- The secondary particles generated in prompt radiation may carry energies up to the full incident beam energy and can also initiate activation processes.

Induced radioactivity arises when particles lost during beam operation interact with components, vacuum walls, targets, etc., creating prompt radiation that further induces nuclear reactions with surrounding structures, shielding, air, and water. This converts stable nuclei into radioactive isotopes, resulting in activation of materials. After beam shutdown, prompt radiation ceases, but induced radioactivity persists. The produced isotopes decay over time, emitting gamma and beta radiation, which can pose risks to personnel and the environment. Evaluation and mitigation strategies must be implemented.

Key activation products include:

- Activated structural materials (e.g., Cu, Fe, stainless steel, Al, Al alloys): $^{54}Mn$、$^{52m}Mn$、$^{56}Mn$、$^{58}V$、$^{51}Cr$
- Activated shielding concrete: $^{22}Na$、$^{24}Na$、$^{54}Mn$

- Activated air: $^{3}$H、$^{7}$Be、$^{11}$C、$^{13}$N、$^{15}$O、$^{41}$Ar
- Activated cooling water: $^{11}$C、$^{13}$N、$^{15}$O、$^{7}$Be、$^{3}$H
- Activated soil and groundwater: $^{7}$Be、$^{3}$H、$^{22}$Na

#### *4.15.2.2 Shielding Calculation Methods*

The radiation shielding design for STCF follows internationally recognized methodologies for shielding of electron accelerators, combining empirical formulas with Monte Carlo simulations. The final design is based on a combination of simulation results and proven engineering practices.

- Empirical formulas: Primarily based on Jenkins and Sakano's methods.
- Monte Carlo simulations: Performed using FLUKA.
- Skyshine (air-scattered radiation): Calculated using the semi-empirical Stapleton formula.
- Engineering references: Include experiences from BEPC [151-154], Beijing High Energy Photo Source [155], Hefei Advanced Light Facility (HALF), and similar domestic and international facilities.

#### *4.15.2.3 Radiation Monitoring System*

The STCF radiation monitoring subsystem will be a state-of-the-art, fully integrated system designed in compliance with the latest legal requirements and international standards. The design will incorporate:

- Results from preliminary hazard assessments;
- The latest technological developments;
- Unique beam time structure and radiation field characteristics.

The system will continuously measure environmental dose rates in both accelerator operational areas and surrounding interior/exterior zones. If dose levels within a controlled radiation zone exceed predefined thresholds, alarms will be triggered, including remote alerts to the control room. The system will also support:

- Remote monitoring,
- Long-term data storage,
- Offline data analysis.

A key challenge in this subsystem is ensuring detector responsiveness to STCF-specific conditions: ultra-high beam loss rates, high beam energies, complex secondary particle distributions, and time-structured radiation. It will be necessary to evaluate whether conventional detectors meet these conditions or whether new detectors must be developed.

#### *4.15.2.4 Personnel Safety Interlock Subsystem (PSIS)*

The STCF Personnel Safety Interlock Subsystem (PSIS), due to its unique application environment, must not only comply with low-voltage electrical system design standards but also adhere to radiation protection regulations and standards. Drawing from best practices and operational experience in other national and international accelerator facilities, the PSIS for STCF is designed with the following principles [153]:

- Hardware reliability;
- Optimal shutdown logic;
- Fail-safe operation;
- Redundancy;
- Defense-in-depth;
- Self-diagnosis and inspection;
- Human-centered design.

A critical component is the logical structure of the interlock system. The PSIS must be tightly integrated with the radiation monitoring system to ensure coordinated and safe machine operation.

### 4.15.3 Design Scheme and System Architecture

#### *4.15.3.1 STCF Accelerator Tunnel Shielding Design*

In the shielding design, FLUKA simulations were performed under two beam loss scenarios—uniform loss along the beam envelope and sudden full bunch loss. Neutron and photon energy spectra were calculated at various distances from copper and stainless-steel targets impacted by high-energy electron beams. Based on the shielding design objectives, radiation shielding formulas, and beam loss assumptions, radiation dose rates and primary shielding thicknesses for each accelerator section under various operating conditions were derived. Radiation dose calculations account for both public and occupational occupancy times and occupancy factors: public occupancy is assumed year-round with a factor of 1, while workers are assumed to be present up to 2000 hours annually. Depending on location, occupancy factors of 1 or 1/10 are applied to accessible areas outside the shielding walls.

To meet STCF radiation protection objectives, dose rates on the surface of shielding walls and ground surface above underground tunnels were calculated using the specified beam loss scenarios and shielding thicknesses. These calculations ensure that, even under assumed beam loss conditions, the shielding walls reduce dose rates in accessible regions to levels that meet STCF design goals.

Shielding strategies vary by tunnel location. The injector's accelerator sections are tentatively located 6.5 meters underground, and employ concrete combined with soil overburden for shielding. The collider ring tunnel is located at or near ground level, with concrete shielding as the primary method. For special areas, shielding may include heavy concrete combined with steel plates. Personnel access corridors adopt maze-type shielding and shielding doors, with shielding thickness determined through source term simulations.

#### *4.15.3.2 Personnel Safety Interlock Subsystem*

The Personnel Safety Interlock Subsystem (PSIS) is built upon the interlock control infrastructure and assisted by an access control monitoring system. Following principles of digitalization, networking, integration, and human-centered design, the system ensures personnel safety in radiation-controlled areas such as tunnels and prevents radiation accidents.

The PSIS is led by programmable logic controllers (PLCs), supplemented by access control systems and interlock keys, with the following core functions:

- Prevent personnel from entering zones where prompt radiation is being generated.
- Prevent radiation generation while personnel remain in controlled zones.
- Prevent uncontrolled radiation leakage from controlled areas.

A conceptual system diagram is shown in Figure 4.15-1 [153].

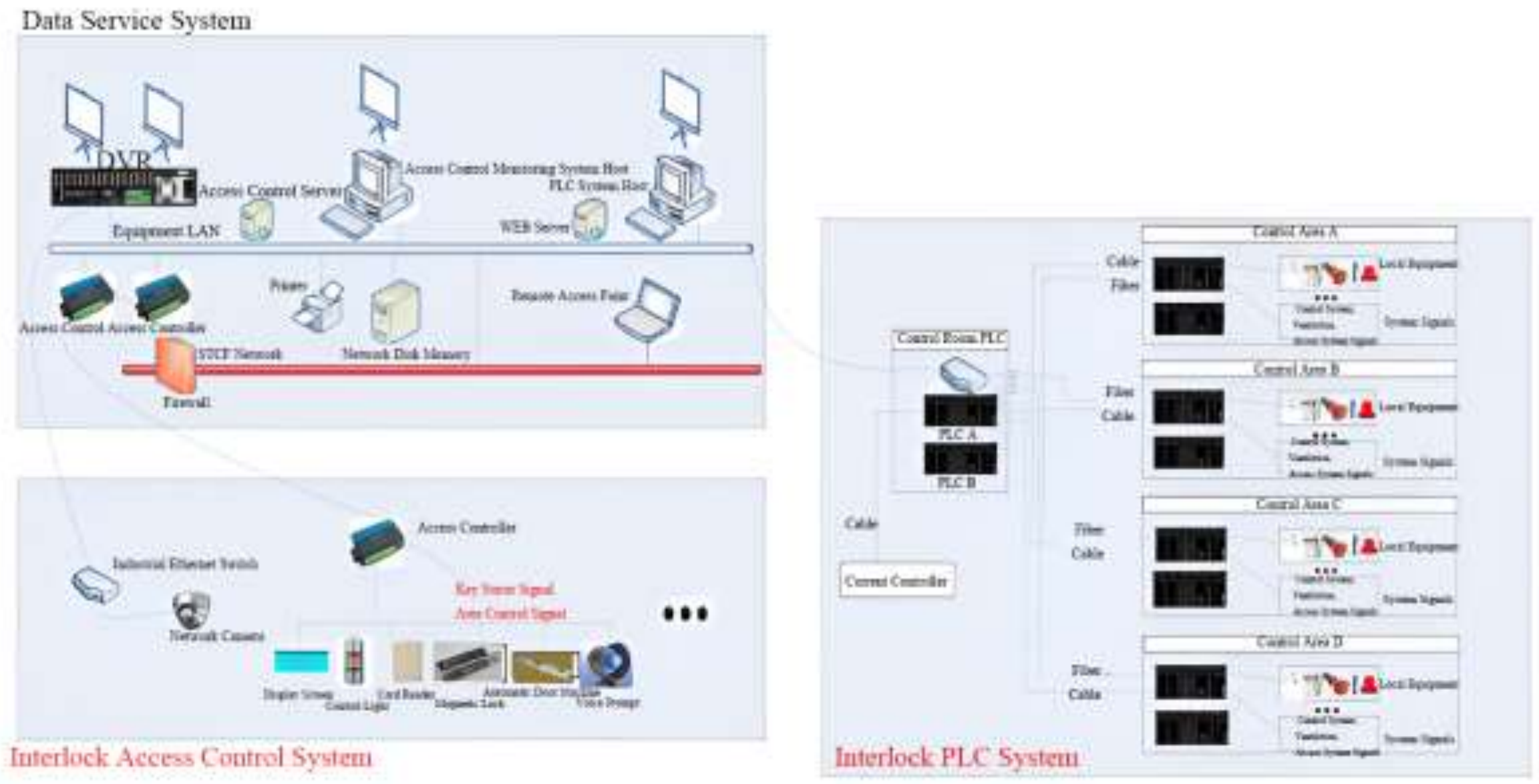

Figure 4.15-1: Basic Architecture of the Personnel Safety Interlock Subsystem

When the accelerator is operating in beam mode, interlocks are activated for the corresponding areas, prohibiting personnel access. After shutdown, access for maintenance is permitted. Each interlock zone includes a number of interlock doors and shielding doors, all equipped with status indicators, interlock control boxes, access control readers, emergency open buttons, and interlock key systems.

Inside each tunnel zone, emergency stop buttons are installed every 10 meters along walkways, with at least two stop points per zone. Search and clearance buttons and audio-visual alarms are installed at exits, remote corners, and blind spots. PSIS input and output signals are exchanged with the accelerator management system, ventilation control, high-power equipment systems, and on-site interlock hardware, including stop buttons, clearance buttons, access readers, and interlock key devices.

#### *4.15.3.3 Radiation Monitoring Subsystem*

The radiation monitoring system includes area monitoring, individual dose monitoring, and environmental monitoring.

- Area Monitoring: Combines fixed online monitors with mobile survey instruments.
- Environmental Monitoring: Uses fixed radiation monitors for continuous observation and periodic sampling and analysis of environmental media.
- Personnel Dosimetry: Primarily via passive dose badges, supplemented with personal alarm dosimeters. Personnel are monitored quarterly.

Routine area radiation monitoring ensures workplace safety and verifies protective measures to prevent accidental exposure of staff and the public.

As illustrated in Figure 4.15-2, the system consists of:

- Local detectors for real-time area/environment dose rate measurements;
- Data transmission and acquisition systems to send information to the central control and data storage center;
- Functions include threshold alarms, data archiving, information publication, and interfacing with interlock controls.

To ensure control and communication across varying conditions, four communication channels are supported: Ethernet, wireless, GPRS, and offline logging.

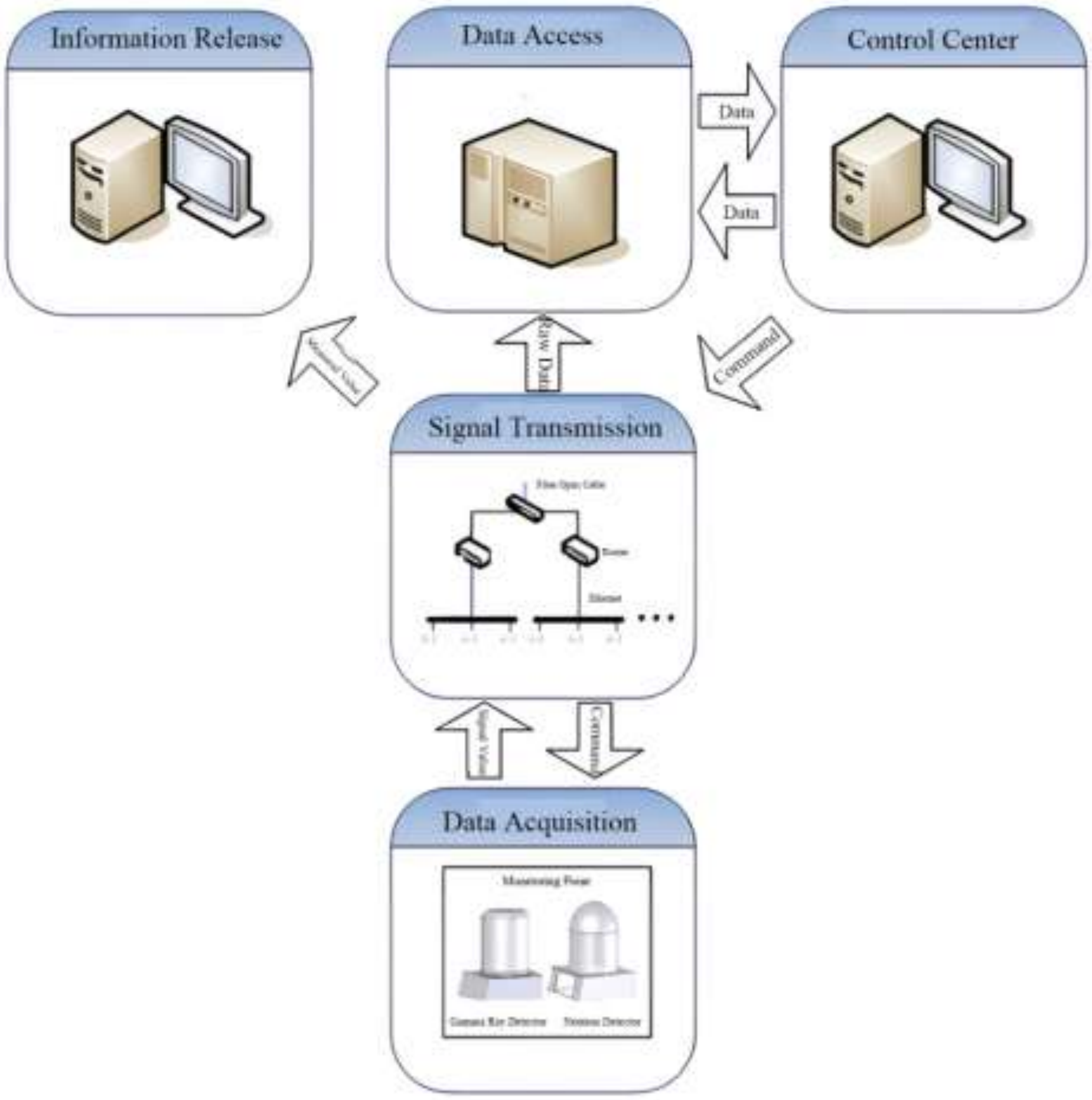

Figure 4.15-2: Radiation Monitoring Subsystem

### 4.15.4 Feasibility Analysis

The STCF project team brings together researchers from across the country with expertise in relevant fields. Domestically, there is already substantial experience in the design and operation of radiation protection systems for high-energy electron accelerators, which provides a solid foundation for the successful implementation of this project. The shielding calculation software required for this project is internationally recognized, mature, and open-source, with controlled access rights. Therefore, the proposed research topics, technical approach, and design scheme are feasible.

### 4.15.5 Summary

The radiation protection system for the STCF facility encompasses source term analysis, shielding calculations, the radiation monitoring subsystem, and the radiation safety interlock subsystem. The facility's complex structure, high beam loss rate, unique beam timing structure, and complex secondary particles pose significant challenges to radiation protection design. However, by drawing on the existing design experience of major domestic electron accelerator facilities, the project team is well-positioned to complete the radiation protection system for STCF. Continuous iteration and close integration with the physical design will be essential to ultimately deliver a radiation protection solution that meets all safety and regulatory requirements.

# 5 Key Technology R&D

## 5.1 Overview

As described in Chapter 1, the STCF accelerator will extensively adopt mature accelerator technologies. Over the past two decades, China has built numerous accelerator facilities at the international level, establishing a solid foundation in accelerator technology. At the same time, the capabilities of China's manufacturing industry have also improved significantly, enabling domestic mastery of world-class accelerator technologies and equipment fabrication techniques—an important advantage for the construction of the STCF accelerator.

However, STCF will also require a number of technologies that are not yet fully mastered by the project team or even within China. In some cases, the required technical specifications exceed the current international state-of-the-art, or the technologies themselves are not yet fully validated globally. Despite this, the adoption of these novel technologies is crucial to realizing the performance goals of a new-generation collider. It is a well-established tradition in the international accelerator and high-energy physics communities to drive technical advancement through sustained R&D and targeted pre-construction technology development. This approach is not only essential for delivering cutting-edge accelerator facilities but also contributes significantly to the advancement of accelerator science as a whole.

The STCF project will similarly pursue a strategy of strengthening national collaboration, forming strong partnerships, and organizing dedicated R&D campaigns to master key technologies during both the pre-research and construction phases. Where necessary, temporary substitution with mature technologies may be adopted during the construction phase, with an acceptance of modest trade-offs in device performance or advancement. These temporary solutions would be replaced with advanced technologies once they are fully developed and verified.

During the STCF conceptual design phase, the accelerator team conducted systematic studies of the physical design and coordinated closely between physical and technical system design. This process allowed the team to clearly identify the full set of technical requirements for building the STCF accelerator, including both well-established technologies and key technologies yet to be mastered. Through active international collaboration, the team also gained a comprehensive understanding of the design approaches and technical levels associated with third-generation electron-positron colliders worldwide. This provides a strong foundation for launching targeted R&D on key technologies and preparing engineering-oriented technical pre-research.

Given the varying degrees of technical maturity, difficulty, and criticality of different technologies required for the STCF accelerator, it is necessary to organize phased R&D programs. During early project development, the team received strong support from the Ministry of Science and Technology, the National Natural Science Foundation of China (NSFC), the Chinese Academy of Sciences, and the University of Science and Technology of China (USTC), enabling significant progress. However, the STCF accelerator is highly complex and requires a large-scale collaborative effort for comprehensive design and research, something that was not fully achieved during the initial stage. Thus, the R&D work conducted in the early phase was not fully systematic and only partially addressed certain key technologies.

Since 2023, with joint support from Anhui Province, the City of Hefei, and USTC through a provincial science and technology special initiative, the STCF team has been able to conduct focused R&D on critical technologies, resulting in major advances, especially in accelerator-related efforts. Section 5.2 will present specific progress and deployment of these accelerator-related technology R&D initiatives. Additional key technologies and engineering techniques still requiring validation—including follow-up work from the current R&D program—will be organized in the next project phase to ensure that STCF can commence construction as planned in 2027 or 2028. These efforts are detailed in Section 5.3.

## 5.2 Ongoing R&D on Key Technologies

With the support of the Anhui Provincial "STCF Key Technology R&D Project" and the National Key R&D Program project titled "Physics and Key Technology Research on High-Luminosity Electron-Positron Accelerators and Detectors in the GeV Energy Range", the STCF accelerator team is actively carrying out R&D on several critical accelerator technologies. These include: superconducting magnet technology for the interaction region, high-power RF technology for the collider rings, high-precision beam diagnostics and fast feedback systems, beam injection technology, MDI mechanical integration, high-charge S-band photocathode electron sources, large-aperture S-band accelerating structures, and high-power positron target technology.

For the injector-related technologies—such as the photocathode electron source, large-aperture accelerating structures, and positron target—the team plans to integrate prototype systems into a newly constructed test platform for electron and positron beam experiments. The major R&D efforts under way are outlined below.

### 5.2.1 Superconducting Magnet Technology for the Interaction Region

As a third-generation electron-positron collider, STCF adopts a large crossing-angle collision scheme and requires the final focusing superconducting quadrupoles to be placed as close as possible to the interaction point (IP). However, this design faces significant constraints, making the development of the IR superconducting magnets a major technical challenge. On both sides of the IP, a set of double-aperture combined superconducting magnets is placed deep within the detector, which must provide strong focusing to minimize the beam size at the collision point. These magnets must also incorporate orbit correction, higher-order multipole correction, and solenoidal field compensation functions.

The core of each magnet set consists of two double-aperture quadrupole magnets, and also includes compensation solenoids, correction coils, harmonic compensation coils, and anti-solenoid windings. These magnets are characterized by multiple coil types, high gradient, and high magnetic field precision. The current layout of the interaction region magnets is shown in Figure 2.11-1.

The complexity of this magnet system is further increased by strict spatial constraints, including limited separation between the apertures and limited outer dimensions. Internationally, there has been little precedent for such systems. The only operational example is the SuperKEKB collider in Japan, which uses an asymmetric-energy beam scheme and has different magnet requirements from STCF. Its performance has been suboptimal, and upgrades are under consideration. Within China, there has been no prior R&D experience on double-aperture combined superconducting IR magnets. However, the Institute of High Energy Physics (IHEP) developed a single-aperture combined superconducting magnet for BEPCII using the serpentine winding technique that was originally developed by BNL, and a single-aperture prototype for CEPC using the Cos2θ winding technique. These efforts provide useful references for STCF.

The STCF team has identified IR superconducting magnet development as a top-priority key technology. Experts with superconducting magnet R&D experience in China were consulted to evaluate four design technologies—CCT, Cos2θ, DCT, and serpentine winding. After comparative studies and receiving input from international experts, the CCT (canted cosine-theta) scheme was selected as the baseline for STCF IR magnet prototyping.

The prototyping is divided into two stages: the first develops the CCT-based design, manufacturing process, and a double-aperture quadrupole prototype; the second develops a full engineering prototype of the combined magnet system based on refined accelerator optics requirements. The design and prototype goals are listed in Table 5.2-1.

Table 5.2-1: Design and Prototype Targets for the IR Superconducting Magnet

| Item | Parameter | Design Target | Prototype Target |
|---|---|---|---|
| Magnet assembly | Double-aperture angle | 60 mrad | 60 mrad |
| Distance from IP to Q-center | | 1100 mm | 1100 mm |
| Quadrupole and correction coils | Field gradient | ⩾ 50 T/m | ⩾ 50 T/m |
| | High-order field error | ⩽ 0.2 ‰ | ⩽ 1.0 ‰ |
| Cross-talk | | ⩽ 30 Gauss | N/A |
| Magnetic length | | ⩽ 400 mm | ⩽ 400 mm |
| Coil inner diameter | | ~22 mm | ~22 mm |

| Item | Parameter | Design Target | Prototype Target |
|---|---|---|---|
| Coil outer diameter | | ⩽ 27 mm | ⩽ 27 mm |
| Anti-solenoid coil | Field strength | ⩾ 1 T | N/A |
| | Integrated field (within ±2 m of IP) | Suppress to ⩽ 1% of detector solenoid field | N/A |

The R&D process includes: (1) considering accelerator optics and spatial constraints to carry out 2D/3D magnetic field calculations using tools such as OPERA; (2) structural design of the full system including thermal, electrical, and mechanical properties, using 3D CAD and finite element analysis tools; (3) selection of coil materials and support structure designs suited to cryogenic operation; (4) machining and winding trials of CCT coil supports; (5) fabrication of a prototype superconducting magnet; (6) vertical cryogenic testing to verify operational current and magnetic field quality; (7) refinement of the magnetic design and process for future engineering models.

### 5.2.2 RF Technology R&D for the Collider Ring

STCF will operate with beams of a high beam current (about 2 A) and low-emittance, making the beam–cavity interaction effects such as beam loading and higher-order mode (HOM) instabilities critical concerns. These challenges impose stringent requirements on the RF system, particularly the development of accelerating cavities that require high-power (about 300 kW), low R/Q, and HOM deeply damped.

Although similar challenges have been addressed internationally, domestic R&D in this area lags significantly behind. Furthermore, advances in technology now require cavity designs that are more effective and economical than earlier generations. After reviewing RF cavities used in global storage rings of both e+e- colliders and synchrotron light sources where both room-temperature and superconducting types are being used, we identified the room-temperature TM020 cavity of 500 MHz that was developed jointly by Spring-8 and KEK as a promising candidate for STCF (see Section 4.3 for details).

While the NanoTerasu light source in Japan has recently deployed this cavity type, STCF's much higher beam current presents significantly greater demands on both the cavity and power coupler. This necessitates dedicated domestic development of the TM020 cavity and high-power coupler, including mechanical design, fabrication techniques, and tuning systems. Development of the low-level RF (LLRF) system is also essential, as it forms the basis for stable RF operation and beam feedback.

The ring RF technology development will proceed in stages. The current phase focuses on building a TM020 prototype cavity to study its mechanical and RF properties, along with high-power coupler development and tuning strategies. In the next phase, improvements based on this work will enable construction of an engineering prototype that meets all operational requirements for STCF.

### 5.2.3 Beam Instrumentation Technology R&D

Compared with fourth-generation synchrotron light sources under construction, STCF operates with much higher beam currents and achieves extremely small beam spots at the interaction point. Compared to similar colliders like SuperKEKB, STCF operates at lower energies, where collective effects such as intra-beam scattering are more pronounced. Under conditions of intense radiation background and strong collective effects, achieving accurate and timely beam parameter measurements—and providing precise, fast tuning capabilities—has become a key limiting factor for STCF performance. The overall performance targets for STCF's beam diagnostics system must significantly exceed the current levels of BEPCII and SuperKEKB to ensure stable beam operation under high-luminosity conditions. Specifically, this requires the development of advanced diagnostics for high-precision bunch-by-bunch spatial resolution, suppression of instabilities via bunch-by-bunch feedback, and interaction-point-based luminosity feedback. These efforts are currently being actively pursued.

#### *5.2.3.1 Bunch-by-Bunch Precision Diagnostics for the Collider Rings*

To reduce noise introduced by multiple front-end components, a low-noise signal conditioning front-end is under design, paired with a high-sampling-rate, high-bandwidth multi-channel oscilloscope for initial signal observation. The oscilloscope can also serve as a tool for random sampling and offline data analysis.

The team is developing a phase-resolved sampling technique. By sampling at two fixed-delay points near the signal peak (on either side of the zero crossing), the method effectively suppresses the impact of phase jitter on peak value determination. Combined with a modal-matching amplitude and phase extraction algorithm, this approach allows for 3D position and charge measurements using only 8 original sampling channels, achieving a higher signal-to-noise ratio and integration. An integrated bunch-by-bunch 3D position signal processor is under development, featuring an eight-channel data acquisition system. The prototype includes both an analog front-end electronics module (with 8-channel ADCs) and a digital processing module based on FPGA, and supports external clocking and triggering.

Based on this system, transient analysis of off-axis beam injection can be performed. After injection, differences in the initial conditions between the injected and stored bunches cause the injected bunches to undergo damped oscillations in both longitudinal and transverse dimensions around the stored beam's equilibrium position. Eventually, they merge fully. Bunch-by-bunch precision diagnostics enables detailed observation of this evolution.

#### *5.2.3.2 Collision Stability and Feedback Technologies*

This segment involves two critical areas: a fast bunch-by-bunch feedback technique to suppress collective beam instabilities, and a luminosity-oriented feedback technique at the interaction point.

The bunch-by-bunch feedback system includes RF signal processors, digital processors, broadband linear power amplifiers, and transverse and longitudinal kickers. Given STCF's high beam current (about 2 A), multiple kickers may be required in the same direction. Custom processors are being developed to support multiple independent-phase feedback outputs. The broadband amplifiers, a key part of the system, are currently dominated by foreign suppliers at high cost. The STCF team is pursuing domestic alternatives to reduce dependency and cost.

Beam-beam interactions and micro-vibrations affecting the final focus superconducting magnets, particularly in the vertical plane, can cause closed orbit distortions and collision angle changes at the IP, which cannot be fully compensated, leading to luminosity jitter and reduced average luminosity. Developing orbit feedback at the interaction point can help mitigate this. For example, a single-chamber dual-beam BPM with eight electrodes, currently under international study, can measure the relative displacement of the electron and positron beams. In cooperation with the detector's luminosity monitoring system, this information enables real-time orbit corrections to stabilize luminosity. These technologies are under exploratory study by the STCF beam instrumentation team during the pre-research phase.

### 5.2.4 MDI Technology R&D

The Machine-Detector Interface (MDI), located at the interaction region, is one of the most complex challenges in collider engineering design (see Sections 2.12 and 4.13). It must account for many interrelated technical domains. For example, the accelerator physics design specifies requirements for the superconducting magnet assembly near the interaction point, including position, aperture, and field configuration. At the same time, the detector imposes stringent constraints on the transverse boundaries of accelerator components. The close spacing between the dual-aperture superconducting quadrupoles and the limited volume available for coil supports and cryostats severely constrain mechanical integration. Moreover, the vacuum system must achieve an ultra-high vacuum without leaving space for dedicated pumps. This necessitates careful consideration of beam losses, beam-gas interactions, and radiation backgrounds caused by bremsstrahlung and synchrotron radiation, which also contribute to local heat loads.

Additionally, because this section is located inside the large spectrometer, mechanical support, maintenance procedures, and the interface with external vacuum sections (RVCs) are complicated. It is thus essential in the R&D stage to study the multiple constraints on the MDI, propose viable structural solutions, and develop prototype models. These activities must iteratively co-evolve with the accelerator and detector physics designs, which will continue to change in subsequent STCF design phases.

The prototyping effort begins with a 3D mechanical model, built using geometry data from the collider optics, superconducting magnet layout, vacuum system, and detector structure. This model supports integrated assembly planning, handling strategies, and co-design with other systems. A full-scale (1:1) mechanical integration prototype (one side) will then be fabricated to verify the mechanical design and installation process.

### 5.2.5 Collider Ring Injection Technology R&D

Due to STCF's small dynamic aperture and extremely short beam lifetime, each of the collider rings requires an injection and extraction system with exceptional performance. In particular, the proposed swap-out injection scheme imposes stringent timing requirements: the pulsed kicker magnet must respond within just a few nanoseconds and operate at high repetition rates, posing major technical challenges.

In the R&D phase, the team is not only supporting the ongoing injection system design (see Section 4.4) but is also developing and testing two kicker magnet prototypes. One is an ultra-fast pulsed kicker for the swap-out injection scheme; the other is a nonlinear pulsed kicker for the traditional off-axis injection scheme. Although similar devices have been developed and even operated both internationally and domestically, the technology is still evolving. STCF places higher demands on performance than current systems. The goal is to achieve full mastery of pulsed injection kicker technologies that meet STCF requirements.

### 5.2.6 Injector-Related Technology R&D and the $e^+/e^-$ Test Platform

#### *5.2.6.1 Key Injector Technologies*

To meet the ultra-high luminosity design goals of STCF, the collider must operate with high beam currents, small dynamic aperture, and short beam lifetimes. This requires top-up injection using high-charge, low-emittance electron and positron bunches at a moderate repetition rate. Minimizing the luminosity perturbation from injected beams is critical.

Whether using off-axis injection or swap-out injection, the injector must support high-repetition, high-charge beams with excellent quality. This places much stricter demands on beam quality and control precision than BEPCII. It is therefore necessary to conduct targeted R&D into key enabling technologies for the injector.

Current R&D efforts include the development of a high-charge photoinjector, an S-band RF system with energy compression, a high-power solid-state modulator, large-aperture S-band positron accelerating structures, a high-power positron target, an adiabatic matching device (AMD), and diagnostics for precision measurement of bunch charge and bunch length.

#### *5.2.6.2 Electron-Positron Beam Test Platform*

To support the injector-related key technologies described in the previous section, the STCF team has designed an integrated electron-positron beam test platform, also referred to as the beam test facility, where multiple R&D components can be tested in a unified experimental environment. The platform comprises a high-charge photoinjector, a 100 MeV electron linac, a positron production target and adiabatic matching device (AMD), a 100 MeV positron linac, along with comprehensive beam diagnostics and a beam dump system. The system parameters are summarized in Table 5.2-1.

This test platform is being co-developed with the klystron and accelerating structure test facility of the Hefei Advanced Light Facility (HALF), sharing infrastructure and space. The layout is shown in Fig. 5.2-2.

The test facility will cover the following six major technical tasks:

1) Develop a high-charge, low-emittance S-band photocathode electron gun;
2) Construct a 100-MeV electron linac to accelerate electrons from the photoinjector and deliver them either to the positron production target or to an upstream beam dump;
3) Build a positron target and an adiabatic matching device (AMD);
4) Construct a 100-MeV positron linac to capture and accelerate positrons using large-aperture accelerating structures;
5) Transport both electron and positron beams to a diagnostics beamline and a downstream beam dump;
6) Carry out precision beam diagnostics for both species to validate physical models of the STCF injector and verify beam quality, stability, and measurement techniques.

Table 5.2-1: Electron-Positron Beam Test Platform Design and Acceptance Parameters

| No. | Parameter Description | Design Target |
|---|---|---|
| 1 | Electron beam energy | 100 MeV |
| 2 | Electron bunch charge | 8 nC |
| 3 | RF frequency | 2998 MHz |
| 4 | Macro-pulse repetition rate | Max 50 Hz (1 Hz at high charge) |
| 5 | Energy stability | 0.1% rms |
| 6 | Bunch length (rms) | 4 ps |
| 7 | Beam spot at positron target | ~2 mm rms |
| 8 | Positron beam energy | 100 MeV |
| 9 | Positron yield ($e^+/e^-$) | 0.1% |

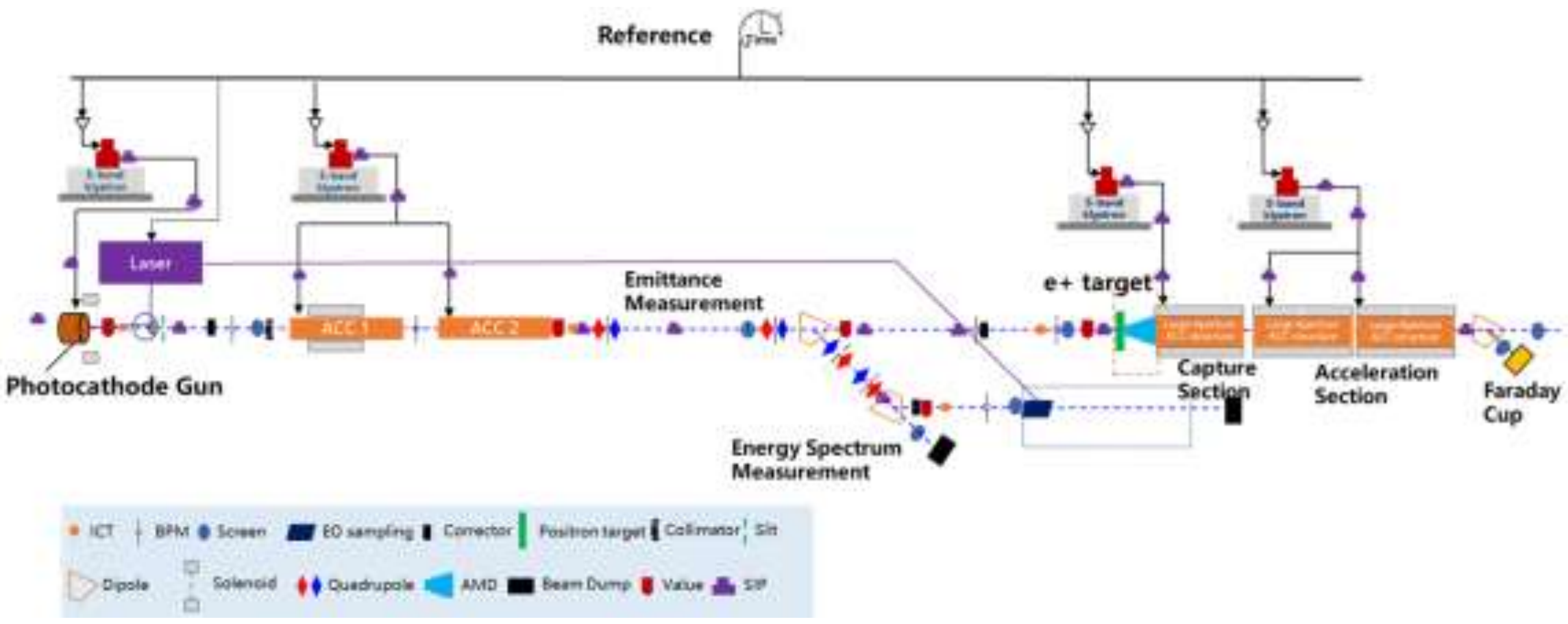

Figure. 5.2-2: Schematic Layout of the Electron-Positron Beam Test Platform

## 5.3 Future R&D on Key Accelerator Technologies

### 5.3.1 Continuation of Current Key Technology R&D

Some of the ongoing STCF accelerator technology development efforts involve significant technical challenges that necessitate a staged approach. The first stage is primarily exploratory and focuses on prototype-level validation, aiming to identify technical bottlenecks and develop feasible processes. This is often referred to as the "proof-of-principle" stage. The second stage builds upon this foundation to achieve full technical performance through system integration and engineering-level implementation, culminating in the development of qualified engineering prototypes.

One of the core areas requiring continued development is the superconducting magnet system in the interaction region. Building on the current prototype of a double-aperture superconducting quadrupole magnet, the next step involves developing a full engineering model. This model must incorporate mechanical and spatial constraints defined by the MDI region, satisfy updated physical parameters driven by machine optics, and integrate high-order harmonic correction coils and anti-solenoid compensation coils. Furthermore, the coil and compact cryostat structures must be optimized in tandem. The resulting engineering prototype will need to pass rigorous performance validation before it is deemed ready for deployment.

Another key area is the high-power RF system for the storage rings. The initial stage of research has focused on the structural design, machining process, and low-level RF (LLRF) controls of a TM020-type room-temperature cavity. However, to meet the STCF's high-current operational requirements, subsequent R&D must address the development of high-power input couplers and structural optimizations that improve power handling and electromagnetic performance. The final objective is to produce a TM020-type cavity that fully meets STCF specifications at the engineering prototype level.

### 5.3.2 Additional Accelerator Technology Development

Following the completion of the STCF conceptual design and external expert evaluations, it has become evident that several technologies—originally considered secondary or unconfirmed in demand—are in fact indispensable to STCF construction and must be addressed through targeted R&D. These technologies are not yet sufficiently mature to be directly used in STCF and must now be prioritized for development.

In many of these cases, the foundational knowledge or technical infrastructure already exists domestically. However, current performance levels fail to satisfy the demanding criteria required for STCF. As such, these technologies must now be elevated to the status of key technical areas, and efforts should be launched to develop engineering-level prototypes capable of meeting project requirements.

Additionally, there are technologies that, while currently available from mature international suppliers, have not yet been domestically developed. In anticipation of potential supply chain risks during STCF construction, it is essential to begin parallel domestic development of such systems. Even if the initial performance lags behind foreign products, these efforts will help ensure future self-reliance and mitigate strategic vulnerabilities in the engineering phase of STCF.

# 6 Requirements for Utility Infrastructure

Large-scale accelerator facilities impose stringent demands on utility systems, significantly exceeding those of civilian buildings or conventional research laboratories. These systems are a critical component of the engineering design and construction process. Based on the preliminary technical scheme of the STCF accelerator, the infrastructure requirements are largely defined and include the construction of injector tunnels, collider ring tunnels, experimental and testing halls, and various equipment rooms. Four to five equipment access points will be established at key locations such as the linac and damping ring area, beam crossing points of the collider rings, and the experimental detector hall.

The collider rings and injector place extensive requirements on public infrastructure, covering civil construction, power supply and energy management, HVAC systems, process cooling water, cryogenics, water supply and drainage, geodetic surveying and alignment, networking and computing infrastructure, transportation and logistics, and system installation and integration. Furthermore, as the machine devices will operate in radiation zones, infrastructure must not only support operational demands but also comply with radiation protection and safety requirements.

## 6.1 Electrical Power and System Grounding Requirements

The STCF accelerator's major power consumers include the collider rings, linac, damping ring, and beam transport lines. Large energy-consuming systems are the RF power sources, magnet power supplies, microwave sources for the linac, cooling water pumps, and air conditioning units. The estimated total power capacity is approximately 80 MVA.

A dedicated 110 kV substation with dual external transmission lines will be constructed within the STCF site. A 10 kV distribution network will deliver electricity to substations in each building, where it will be converted down and distributed to the accelerator's devices. To ensure power quality for sensitive systems, power conditioning devices and appropriate UPS or energy storage systems will be installed.

Dedicated grounding networks will be deployed in tunnel areas for systems such as the RF, power sources, and injection systems. A combined or M-type grounding system will be used, with a grounding resistance of less than 0.5 Ω to minimize interference from common-mode conduction. Copper busbars will be routed along tunnel walls with reserved grounding terminals. The grounding system will conform to the IEC62561 and GB/T33588 standards and satisfy electromagnetic compatibility requirements.

## 6.2 Water and Cooling Requirements

During accelerator operation, most of the electrical energy consumed by the machine devices will be converted into heat, which must be removed to maintain the devices within operational temperature limits and ensure stability. For example, magnets generate heat in their coils when excitation, necessitating direct internal water cooling. Any fluctuation in cooling water temperature can lead to magnetic field variations, adversely affecting beam stability. Therefore, the temperature stability of the cooling water must meet the precision requirements of magnets and other components.

Apart from water-cooled magnets, other devices requiring process cooling water include DC and pulsed power supplies, electronics racks, RF equipment, and cryogenic devices. The total thermal load from the STCF accelerator devices is approximately 30 MW. Most water-cooled loads operate at a typical pressure of 0.4 MPa, with a maximum pressure of 0.6 MPa, and require inlet water temperature to be maintained at 22±0.5 °C.

Both primary and secondary cooling water will be sourced from the municipal water supply, with a quantity of approximately 1500 $m^3$/h. Several distributed water stations around the accelerator will supply secondary cooling water to the machine devices. Heat removed by secondary water is transferred to primary water, which is cooled by air or used for heat recovery.

The campus water supply and drainage system must also meet the needs of both equipment and personnel. Wastewater from cooling towers and filtration systems needs retreatment, and a comprehensive drainage network will be implemented throughout the facility.

## 6.3 Compressed Air Requirements

The compressed air system at STCF is designed to supply clean, stable, and reliable compressed air to terminal devices requiring pneumatic actuation, including vacuum valves and pneumatic components in the injector and collider ring areas. The purified compressed air must meet the quality standards outlined in Table 6.3-1.

Table 6.3-1: Purified Compressed Air Quality Requirements

| Type | Oil Content ($mg/m^3$) | Contained Particle Size (μm) | Dew Point (°C) |
|---|---|---|---|
| Instrumentation-grade air | < 0.01 | < 0.1 | -40 |

The estimated compressed air consumption is approximately 500 $Nm^3$/h, with a working pressure of 0.6 MPa.

## 6.4 Ventilation and Air Conditioning Requirements

To maintain beam orbit stability, in addition to strict control of power supply stability and mechanical deformation of magnets with their supports, the tunnel environment, particularly inside the collider rings, is subject to stringent requirements. The ambient temperature in the collider ring tunnel must be maintained at 24 ± 0.1 °C, with relative humidity ≤ 50%.

A transverse ventilation scheme is adopted along the collider ring tunnel, with distributed ventilation sectors throughout the ring. The ductwork and air outlets are specially designed to ensure consistent humidity and temperature control during operation. To ensure continued airflow during duct failures, air is supplied bi-directionally from both ends using symmetrical system configurations.

Rapid air exchange devices will be installed to provide fresh air for personnel entering the tunnel and to remove smoke or hazardous gases in emergencies. Fresh air should be exchanged at a rate of 0.5 air changes per hour.

## 6.5 Cryogenic Requirements

STCF employs two superconducting magnet subsystems located in the interaction region (IR), each consisting of multiple magnet coils housed in a cryostat extending into the detector volume on either side of the interaction point. These magnets require cryogenic support at 4.5 K with a pressure of 1.5 bar. Each superconducting magnet has a thermal load of approximately 37.2 W and requires 2.0 g/s of cold helium gas for current lead cooling. The thermal shield operates at 77 K, and the magnet cooldown rate must be controlled below 10 K per hour.

To ensure the stable operation of both IR superconducting magnet systems, a dedicated cryogenic system will be constructed to supply low-temperature fluids at the required pressure and temperature while removing thermal loads. The system will be designed with a focus on reliability, safety, and maintainability. Cooling will be provided via subcooled liquid helium at a pressure between saturation and the critical point (2.3 bar), with two local valve boxes controlling helium flow into and return from the cryostats.

In addition to formal operation cooling, the cryogenic system must also support prototyping and performance testing. It will consist of both a helium and nitrogen subsystem. The helium system includes helium storage tanks, refrigerators, cryogenic transfer lines, valve boxes, and helium recovery and purification systems. The nitrogen system supplies liquid nitrogen for refrigerator precooling, purifier cooling, and thermal shielding cooling.

To support reliable cryogenic operations, the facility needs sufficient utilities—water cooling, electricity, compressed air, and ventilation. The helium compressor, turbine expanders, and other components require approximately 30 $m^3/h$ of cooling water and 800 kW of electrical power. Electrical power is supplied through three-phase five-line or single-phase three-line systems from the main utility distribution panel. Key electrical loads include helium compressors, heaters, mechanical pumps, turbomolecular pumps, cranes, fans, HVAC systems, and lighting.

Compressed air is required to operate all pneumatic valves in the cryogenic system, such as control valves, shut-off valves, and gate valves. A 15 $m^3$ air storage tank will be installed, and the cryogenic system will consume about 45 $Nm^3/h$ of instrumentation compressed air.

## 6.6 Geodetic Surveying

The STCF facility requires the establishment of a comprehensive geodetic infrastructure, including a dedicated geodetic reference network and precise gravity leveling. This infrastructure must serve during all the project phases—from the initial construction through the long-term operation—ensuring accurate spatial location and alignment of accelerator components.

## 6.7 Ground Settlement and Micro-vibration Requirements

Ground vibrations—originating from micro-seismic activity, surrounding environmental sources (e.g., machinery or human activity), and internal facility devices—can cause beam trajectory drifting and emittance growth. Site selection and foundational engineering must aim to minimize such effects. Taking into account beam dynamics, mechanical transmission, and feedback control, the STCF collider tunnel floor must meet a ground vibration criterion: the RMS displacement integrated over the 1–100 Hz frequency range within any 1-second interval must be less than 30 nm.

The injector tunnel floor must support loads of up to 100 $kN/m^2$, with flatness deviations of less than 4 mm per 25 m and less than 10 mm over the full length, including transfer lines. The collider tunnel floor must support up to 200 $kN/m^2$, with flatness criteria of less than 4 mm per 25 m and less than 10 mm around the full circumference.

During operation, uneven ground settlement can significantly affect beam trajectories. The allowable differential settlement rate is set at less than 10 μm per 10 m per year for the collider tunnel, and less than 20 μm per 10 m per year for the injector tunnel. Real-time, long-term ground movement monitoring technologies will be implemented to guarantee these criteria continuously.

## 6.8 Communications, Computing, and Data Services

STCF accelerator control and operations require robust infrastructure for data transmission, telephony, and wireless communication—all of which will be supported by a dedicated fiber-optic network. A central data center will be established to support operations, diagnostics, and analysis, with appropriate capacity for data storage, computing resources, and secure network access.

## 6.9 Digital Modeling

To enable a fully integrated visualization and planning management for the STCF project, 3D digital models will be developed for all key areas, including accelerator tunnels, detector halls, and device rooms. These models will be integrated with engineering designs to support layout optimization, ensure compliance with the machine requirements, and reserve sufficient space for future upgrades.

### 6.10 Transport and Logistics

The transportation, storage, and installation of STCF's technical components will be planned and executed based on detailed specifications for device types, quantity, size, weight, testing locations, operational positioning, storage conditions, and sensitivity to vibration or shock. Custom-made transport and installation tools will be designed in accordance with the spatial constraints of the tunnels and the overall engineering layout.

Logistics planning will also address the movement of personnel, equipment, and materials within tunnel sections and equipment halls. Detailed analysis of workflow bottlenecks—such as the number of magnets that can be installed per day—will inform the development of an efficient, safe, and coordinated logistics strategy that meets installation timelines and overall project milestones.

## References

[1] K. Oide et al., Design of beam optics for the Future Circular Collider $e+e-$-collider rings, Phys. Rev. Accel. Beams 19 (2016)

[2] H. Peng, Y. Zheng and X. Zhou, Physics 49, 513 (2020).

[3] P. Raimondi, M. Zobov, D. Shatilov, in: Proceedings of EPAC08, 2008, p. 2620.

[4] A. Bogomyagkov et al., Touschek lifetime and luminosity optimization for Russian Super Charm Tau factory, Journal of Instrumentation, 19 P02018, (2024.)

[5] A. Bogomyagkov, Chromaticity correction of the interaction region. IAS Program on High Energy Physics, Hong Kong18-21 January (2016).

[6] R. Brinkmann, Optimization of a Final Focus System for Large Momentum Bandwidth, DESY M-90-14, November 1990

[7] K.L. Brown, and R.V. Servranckx, First- and second-order charged particle optics. AIP Conference Proceedings, 127, 62–138. (1985). https://doi.org/10.1063/1.35177

[8] T. Liu, PAMKIT. https://pypi.org/project/PAMKIT/

[9] SAD home page. (n.d.). Retrieved October 31, 2022, from https://acc-physics.kek.jp/SAD

[10] Y. Ohnishi, T. Abe, K. Akai, et al., SuperKEKB operation using crab waist collision scheme, Eur. Phys. J. Plus (2021) 136:1023.

[11] P. Raimondi, Local chromatic correction Arc & Final Focus, FCC Physics Workshop, Boston, March 26th, 2024.

[12] The CEPC Study Group, CEPC Technical Design Report - Accelerator, Radiation Detection Technology and Methods, 8(1), (2024) 1-1105

[13] K. Brown, R. Servranckx, First- and second-order charged particle optics, SLAC-PUB-3381, 1984.

[14] M. Donald, R. Helm, J. Irwin, H. Moshammer, et al., Localized Chromaticity Correction of Low-Beta Insertions in Storage Rings, SLAC-PUB-6197, April 1997.

[15] A. Bogomyagkov, E. Levichev and P. Piminov, Final focus designs for crab waist colliders, PHYS. REV. ACCEL. BEAMS 19, 121005 (2016).

[16] S. Glukhov, E. Levichev, P. Piminov, and D. Shatilov. Dynamic aperture studies in e+e− factories with crab waist. IR'07 proceedings, 2007.

[17] K. Ohmi, and H. Koiso. Dynamic aperture limit caused by IR nonlinearities in extremely low-beta B factories. Proceedings of IPAC'10, Kyoto, Japan, 2010.

[18] K. Hirosawa, et al., Advanced damper system with a flexible and fine-tunable filter for longitudinal coupled-bunch instabilities caused by the accelerating mode in SuperKEKB, Nuclear Inst. and Methods in Physics Research, A 953, 163007 (2020).

[19] H. Ego, et al., Compact HOM-damping structure of a beam-accelerating TM020 mode rf cavity, Nuclear Inst. and Methods in Physics Research, A 1064, 169418 (2024).

[20] P. Baudrenghien and T. Mastoridis, Fundamental cavity impedance and longitudinal coupled-bunch instabilities at the High Luminosity Large Hadron Collider, Phys. Rev. Accel. Beams 20, 011004 (2017).

[21] T. He, and Z. Bai, Graphics-processing-unit-accelerated simulation for longitudinal beam dynamics of arbitrary bunch trains in electron storage rings, Phys. Rev. Accel. Beams 24, 104401 (2021).

[22] J. M. Byrd, et al., Transient beam loading effects in harmonic rf systems for light sources, Phys. Rev. ST Accel. Beams 5, 092001 (2002).

[23] D. Zhou, K. Ohmi, Y. Funakoshi, Y. Ohnishi, Y. Zhang, Simulations and experimental results of beam-beam effects in SuperKEKB, Phys. Rev. Accel. and Beams 26, 071001 (2023).

[24] K. Ohmi, M. Tawada, Y. Cai, S. Kamada, K. Oide, and J. Qiang, Luminosity limit due to the beam-beam interactions with or without crossing angle, Phys. Rev. ST Accel. Beams 7, 104401 (2004).

[25] K. Ohmi, N. Kuroo, K. Oide, D. Zhou, and F. Zimmermann, Coherent Beam-Beam Instability in Colli- sions with a Large Crossing Angle, Phys. Rev. Lett. 119, 134801 (2017).

[26] M. Zobov, D. Alesini, M. Biagini, C. Biscari, A. Bocci, R. Boni, M. Boscolo, F. Bossi, B. Buonomo, A. Clozza et al., Test of "Crab-Waist" Collisions at the DAΦNE Φ Factory, Phys. Rev. Lett. 104, 174801 (2010).

[27] A. Chao, M. Tigner, H. Weise, and F. Zimmermann (Eds.), Handbook of accelerator physics and engineering, World scientific (2023).

[28] R. Nagaoka and K. Bane, Collective effects in a diffraction-limited storage ring, J. Synchrotron Rad. 21, (2014) 937-960.

[29] A. Blednykh, et al., Microwave instability threshold from coherent wiggler radiation impedance in storage rings, Phys. Rev. Accel. Beams 26, 051002 (2023).

[30] K. Ng, Physics of intensity Dependent Beam instabilities [M]. Singapore: World Scientific, 2006.

[31] Y. Funakoshi, T. Abe, K. Akai , et al., The SuperKEKB has broken the world record of the Luminosity, Proceedings of IPAC 2022, (2022) 1-5

[32] R. Cimino and T. Demma, Electron cloud in accelerators, Int. J. Mod. Phys. A 1430023 (2014).

[33] KEK B-factory Design Report, KEK Report 95-7, (1995).

[34] H. Robert, Status of the APS-U Project, IPAC2021, DOI: 10.18429/JACoW-IPAC2021-MOXA02.

[35] M. Aiba, B. Goddard, K. Oide, et al. Top-up injection schemes for future circular lepton collider, Nuclear Instruments & Methods in Physics Research A 880, (2018) 98-106.

[36] M. Aiba, Review of Top-up Injection Schemes for Electron Storage Rings, Proc. of IPAC2018, Vancouver, BC, Canada, (2018) 1745-1750.

[37] SuperKEKB Design Report, (2020).

https://www-linac.kek.jp/linac-com/report/skb-tdr/

[38] T.K. Charles, B. Holzer, R. Tomas, et al. Alignment & stability challenges for FCC-ee. EPJ Techniques and Instrumentation, 10(1): 8, (2023)

[39] R. Assmann, P. Raimondi, G. Roy et al., Emittance optimization with dispersion free steering at LEP. Physical Review Special Topics-Accelerators and Beams, 3, 121001 (2000).

[40] J. Safranek, Experimental determination of storage ring optics using orbit response measurements. Nucl. Instrum. Methods Phys. Res. A 388, (1997) 27-36.

[41] D.E. Khechen, Y. Funakoshi, D. Jehanno et al. First beam loss measurements in the SuperKEKB positron ring using the fast luminosity monitor diamond sensors. Physical Review Accelerators and Beams, 22, 062801 (2019).

[42] T. Ishibashi, S. Terui, Y. Suetsugu. Low impedance movable collimators for SuperKEKB[J]. Proceedings of the IPAC2017, (2017) 14-19.

[43] A. Bogomyagkov, E. Levichev, D. Shatilov. Beam-beam effects investigation and parameters optimization for a circular e+ e− collider at very high energies. Physical Review Special Topics-Accelerators and Beams, 17, 041004 (2014).

[44] Y. Funakoshi, K. Ohmi, Y. Ohnishi, K. Kanazawa, Y. Suetsugu, H. Nakayama and H. Nakano, Small-Beta Collimation at SuperKEKB to Stop Beam-Gas Scattered Particles and to Avoid Transverse Mode Coupling Instability, Conf. Proc. C 1205201, 1104 (2012)

[45] Li Ma, Chuang Zhang (editors), Design and development of accelerator for the major upgrading project of Beijing Electron Positron Collider, Shanghai Scientific and Technical Publishers, 2014. (in Chinese)

[46] PEP-II: An Asymmetric B Factory: Conceptual Design Report June 1993, SLAC Report, SLAC-418, LBL-PUB-5379, CALT-68-1869, (1993)

[47] S. Terui, Y. Suetsugu, T. Ishibashi et al. Development of a hybrid collimator bonding tantalum and carbon-fiber-composite for SuperKEKB. Nuclear Instruments and Methods in Physics Research Section A 1059, 168971 (2024)

[48] S. Terui T. Ishibashi, T. Abe et al. Low-Z collimator for SuperKEKB. Nuclear Instruments and Methods in Physics Research Section A 1047, 167857 (2023)

[49] S. Terui, Y. Funakoshi, T. Ishibashi et al. Collimator challenges at SuperKEKB and their countermeasures using nonlinear collimator. Physical Review Accelerators and Beams, 27, 081001 (2024)

[50] Liu Zuping. Introduction to the Physics of Synchrotron Radiation Sources. University of Science and Technology of China Press, (2009) 121-186. (in Chinese)

[51] S.Y. Lee. Accelerator Physics (Fourth Edition). World Scientific Publishing Company, (2019) 432-500.

[52] A. Wolski. Damping Ring Design and Physics Issues [EB/OL], (2007). https://pcwww.liv.ac.uk/~awolski/Teaching/USPAS/Houston/DampingRings-Lecture4.pdf,

[53] Zhang Xinmin. Superconducting Cryogenic Technology and Particle Accelerators. (1994: (in Chinese)

[54] He Yongzhou. The research on magnetic properties of permanent magnet and magnetic field of prototype for Cryogenic Permanent Magnet Undulator, PhD Thesis, University of Chinese Academy of Sciences (Shanghai Institute of Applied Physics), 2015. (in Chinese)

[55] J.T. Tanabe. Iron dominated electromagnets: Design, fabrication, assembly and measurements. World Scientific, (2005)

[56] Miao Zhang. Study on the Effect of Undulator Dynamic Field Integral and the Shimming Method; PhD Thesis, University of Chinese Academy of Sciences (Shanghai Institute of Applied Physics), (2016) (in Chinese)

[57] Zhang Qing-Lei. Study on the Effects of Insertion Devices at SSRF; PhD Thesis, University of Chinese Academy of Sciences (Shanghai Institute of Applied Physics), (2015). (in Chinese)

[58] B.W. Montague, Elementary spinor algebra for polarized beams in storage rings, Particle Accelerators, Vol.11, (1981) 219-231

[59] A. Latina, N. Solyak, D. Schulte, A spin rotator for the compact linear collider, Proceedings of IPAC10, Kyoto, Japan, (2010) 4608-4610

[60] Y. Wang, M. Borland, Pelegant: A parallel accelerator simulation code for electron generation and tracking. AIP Conf. Proc. 877, (2006) 241-247

[61] FERMI@Elettra Conceptual Design Report, Sincrotrone Trieste, January 2007
[62] A. Abada, et al., FCC-ee: The Lepton Collider, The European Physical Journal Special Topics, 228 (2019) 261-623.
[63] J.P. Delahaye, J.P. Potier, Reverse bending magnets in a combined-function lattice for the CLIC damping ring, 1989 IEEE Particle Accelerator Conference, Vol. 1613, (1989) 1611-1613.
[64] M. Kikuchi, Reverse-bend FODO lattice applied to damping ring for SuperKEKB, Nuclear Instruments and Methods in Physics Research Section A 556, (2006) 13-19.
[65] H. Xu et al. Development of an L-band photocathode RF gun at Tsinghua University. Nuclear Instruments and Methods in Physics Research Section A 985, 164675 (2021).
[66] M. Krasilnikov, Z. Aboulbanine, G. Adhikari, et al. RF Performance of a Next-Generation L-Band RF Gun at PITZ, Proc. of LINAC2022, Liverpool, UK (2022) 699-702
[67] D. Gu, X. Li, Z. Wang, M. Zhang, Physics Design and Beam Dynamics Optimization of the SHINE Accelerator, Proc. of FLS2023, Luzern, Switzerland (2023) 174-176
[68] M. Kikuchi, Reverse-bend FODO lattice applied to damping ring for SuperKEKB, Nuclear Instruments and Methods in Physics Research Section A 556, (2006) 13-19.
[69] YANG Penghui, Study on the Magnetic Focusing Structure and Related Dynamics of Diffraction Limit Storage Ring, PhD Thesis, University of Science and Technology of China, (2021) (in Chinese)
[70] J. Tanabe, Iron Dominated Electromagnets Design, Fabrication, Assembly, and Measurements, USA: World Scientific Publishing Company，2005
[71] ZHAO Jijiu and YIN Zhaosheng, Particle Accelerator Technology, Beijing: Higher Education Press, 2006 (in Chinese)
[72] ESRF-EBS project, EBS storage ring technical report (ESRF, 2018), Chapter 4: Power Supplies and Electrical Engineering
[73] R. Bartolini et al., SEE-LS: A 4th Generation Synchrotron Light Source for Science and Technology, CERN Yellow Report: CERN-2020-001, Chapter 11: Power supplies
[74] Advanced Photon Source Upgrade Project Preliminary Design Report - Chapter 4: Accelerator Upgrade, ANL Report: APSU-2.01-RPT-002 (2017)
[75] SLS 2.0 Storage Ring Technical Design Report - Chapter 2.3: Power supplies, PSI Bericht Nr. 21-02, November 2021
[76] T. Oki, S. Nakamura, T. Adachi, High-stability Magnet Power Supplies for SuperKEKB, Proc of IPAC2017, (2017) 3391-3393.
[77] H. Ego, et al., Compact HOM-damping structure of a beam-accelerating TM020 mode rf cavity, Nuclear Inst. and Methods in Physics Research, A 1064, 169418 (2024).
[78] T. Inagaki, H. Tanaka, et al., High-power tests of the compactly HOM-damped TM020 cavities for a next generation light source, Proceedings of IPAC2023, Venice, Italy, (2023) 2635-2638
[79] Z. Xiong, et al., The design of LLRF system for STCF storage ring, Modern Physics Letters A 2440001 (2024).

[80] C. Pappas, S. De Santis, J.E. Galvin et al., FAST KICKER SYSTEMS FOR ALS-U, Proc. of IPAC2014, Dresden, Germany, (2014) 564-566

[81] J.H. Chen, H. Shi, L. Wang L et al., Strip-line kicker and fast pulser R&D for the HEPS on-axis injection system, Nuclear instruments & methods in physics research. Section A 920 (2019)1-6.

[82] K. Fan, I. Sakai, Y. Arakaki, Modeling of eddy current effects in an opposite-field septum, Nuclear Instruments and Methods in Physics Research Section A 597, (2008) 142-148.

[83] S. Lee, J-H. Han, Septum magnet design for compact storage ring. Nuclear Instruments and Methods in Physics Research Section A 1023, 165972 (2022)

[84] C. Mitsuda, H. Takaki, R. Takai et al., Suppression of eddy-current effects in beam injection using a pulsed sextupole magnet with a new ceramic chamber. Physical Review Accelerators and Beams, 25, 112401 (2022)

[85] Wang Lei, Luo Qing, Wang Lin, Li Weimin, Design and research of HLS pulsed octupole magnet injection scheme. Nuclear Techniques, 38(6), (2015) 24-29. (in Chinese)

[86] T. Miyajima, Y. Kobayashi, S. Nagahashi, Development of a pulsed octupole magnet system for studying the dynamics of transverse beam instabilities in electron storage rings. Nuclear Instruments and Methods in Physics Research Section A 581, (2007) 589-600.

[87] Y. Suetsugu, K. Shibata, H. Hisamatsu et al., Development of copper beam ducts with antechambers for advanced high-current particle storage rings. Vacuum, 84(5), (2009) 694-698.

[88] K. Kanazawa et al., Experience at the KEK B-Factory vacuum system, Prog. Theor. Exp. Phys., (2013) 03A005.

[89] An Asymmetric B Factory Based on PEP: Conceptual Design Report (Vacuum Part), SLAC, SLAC-0372, Feb, 1991.

[90] J. Dorfan, A. Hutton, M.S. Zisman et al., PEP-II Asymmetric B-Factory: R&D Results, LBL-32098, ESG-204, LLNL-UCRL-JC-110287, SLAC PUB-5785, ABC-75, (1992)

[91] D. Teytelman, J. Fox et al., Set-up of PEP-II Longitudinal Feedback Systems for Even/Odd Bunch Spacings, AIP Conference Proceedings, (2002) 474-482

[92] Y. Suetsugu, K. Kanazawa, K. Shibata et al., Design and construction of the SuperKEKB vacuum system, J. Vac. Sci. Technol., A 30, 031602 (2012)

[93] J. Wang, D. Wang, A. Wang, et al., Design of the Button-Type Beam Position Monitor for the Electron Storage Ring of Hefei Advanced Light Facility, *2024 9th International Conference on Electronic Technology and Information Science (ICETIS)*, Hangzhou, China, (2024) 1-5, doi: 10.1109/ICETIS61828.2024.10593791.

[94] K. Yoshihara, T. Abe, M. Aversano, A. Gale, et al., Development and implementation of advanced beam diagnostic and abort systems in SuperKEKB, Nuclear Instruments and Methods in Physics Research Section A 1072, 170117(2025)

[95] Jin-Liu Su, Yu-Dong Liu, Sai-Ke, et al., Longitudinal impedance measurements and simulations of a three-metal-strip kicker, Nuclear Science and Techniques, 34, (2023)115-126

[96] LIU Tao, WANG Qian, LUO Qing, Design of bunch length and charge monitor based on cavity resonator for injector of Super Tau-Charm Facility, NUCLEAR TECHNIQUES, 10, 100204 (2024) (in Chinese)

[97] Xingyi Xu, Yongbin Leng, Yimei Zhou, et al., Bunch-by-bunch three-dimensional position and charge measurement in a storage ring, Physical Review Accelerators and Beams, 24, 032802 (2021)

[98] Y. X. Han, L. W. Lai, Y. M. Zhou, et al., High performance generic beam diagnostic signal processor for SHINE, Proc. of *IBIC 2024*, Beijing, China, (2024) 385-388

[99] M. Arinaga, J. W. Flanagan, H. Fukuma, et al., Beam instrumentation for the SuperKEKB rings, Proc. of *IBIC2012*, Tsukuba, Japan, (2012) 6-10

[100] M. Tobiyama, E. Kikutani, J. W. Flanagan and S. Hiramatsu, Bunch by bunch feedback systems for the KEKB rings, Proc. of the 2001 Particle Accelerator Conference (Cat. No.01CH37268), Chicago, IL, USA, (2001) 1246-1248

[101] J. Liu, L. Zhao, L. Zhan, S. Liu and Q. An, Bunch-by-Bunch Beam Transverse Feedback Electronics Designed for SSRF, IEEE Transactions on Nuclear Science, Vol. 64, No. 6, (2017) 1395-1400

[102] T. Nakamura, Transverse and Longitudinal Bunch-by-Bunch Feedback for Storage Rings. Proc. of IPAC2018, Vancouver, BC, Canada, (2018) 1198-1203

[103] M. Tobiyama et al., Bunch By Bunch Feedback System For Super KEKB Rings. Proc. of the 13th Annual Meeting of Particle Accelerator Society of Japan Chiba, Japan, (2016) 144-148

[104] Y. Funakoshi, H. Fukuma, T. Kawamoto, et al., Interaction Point Orbit Feedback System at SuperKEKB, Proc. of IPAC2015, Richmond, USA, (2015) 921-923

[105] Y. Funakoshi, H. Fukuma, T. Kawamoto, et al., Recent Progress Of Dithering System at SuperKEKB, Proc. of IPAC2017, Copenhagen, Denmark, (2017)1827-1829

[106] K. Iida, N. Nakamura, H. Sakai, et al., Measurement of an electron-beam size with a beam profile monitor using Fresnel zone plates, Nuclear Instruments and Methods in Physics Research Section A 506 (1-2), (2003) 41-49

[107] Y. Leng, G. Huang, Z. Zhang, et al., The beam-based calibration of an X-ray pinhole camera at SSRF, Chinese physics C, 36(1), (2012) 80-83

[108] N. Samadi, X. Shi, L. Dallin, et al., Source size measurement options for low-emittance light sources, Physical Review Accelerators and Beams, 23(2), 024801, (2020)

[109] W. J. Corbett, W. X. Cheng, A. S. Fisher, and X. Huang, Bunch Length and Impedance Measurements at SPEAR3, Proc. of EPAC'08, Genoa, Italy, (2008) 1595-1597.

[110] P. Emma, T. Raubenhemier, Systematic approach to damping ring design, Physical Review special topics-Accelerators and Beams, 4(2), 021001 (2001)

[111] N. Ohuchi et al., A SuperKEKB beam final focus superconducting magnet system, Nuclear Inst. and Methods in Physics Research, A 1021, 165930 (2022).

[112] C. Shen, Y. Zhu, and F. Chen, Design and Optimization of the Superconducting Quadrupole Magnet Q1a in CEPC Interaction Region, IEEE Trans. Appl. Supercond., Vol. 32, No. 6, (2022) 1-4

[113] M. Koratzinos et al., The FCC-ee interaction region magnet design, Proc. of IPAC2016, Busan, Korea, (2016) 3824-3827

[114] V. Shkaruba et al., Superconducting CCT quadrupole test at Budker INP, Vol. 39, No. 40, 2440010 (2024)

[115] Z. Zong et al., Cryogenic systems of SuperKEKB final focusing superconducting magnets, Nucl. Instruments Methods Phys. Res. Sect., A1058, 168855 (2024).

[116] N. Ohuchi et al., Design and Construction of the Magnet Cryostats for the SuperKEKB Interaction Region, IEEE Trans. Appl. Supercond., Vol. 28, No. 3, (2018) 1-4

[117] B. Liu, M. Gu, C. Zhang, Y. Chi, *New electron gun system for BEPCII*. Proc. of PAC 2005. Knoxville, Tennessee, (2005) 1-3

[118] Hanxun Xu, Jiaru Shi, Yingchao Du, et al., *Development of an L-band photocathode RF gun at Tsinghua University.* Nuclear Instruments and Methods in Physics Research Section A 958, 164675 (2021)

[119] M.L. Stutzman, P.A. Adderley, M.A. Mamun, M. Poelker, *Vacuum characterization and improvement for the Jefferson Lab polarized electron source*. Proc. of IPAC2015, Richmond, VA, USA, (2015) 3540-3543

[120] C. Serpico, et al., High gradient, high reliability, and low wake-field accelerating structures for the FERMI FEL, Rev. Sci. Instrum. 88, 073303 (2017)

[121] H.W. Pommerenke, et al., RF Design of Traveling-Wave Accelerating Structures for the FCC-ee Pre-injector Complex, Proc. of Linac2022, Liverpool, (2022) 707-710.

[122] J. Branlard, G. Ayvazyan, V. Ayvazyan, et al. The European XFEL LLRF system. Proc. of IPAC 2012, (2012) 55-57.

[123] X. Li, H. Sun, W. Long et al., Design and performance of LLRF system for CSNS/RCS, Chinese Phys. C, 39, 027002 (2015)

[124] P. Orel, S. Zorzut, P. Lemut et al., Next Generation CW Reference Clock Transfer System With Femtosecond Stability, Proc. of PAC2013, Pasadena, (2013) 1358-1360

[125] H. Ma, J. Rose, Upgrade and Operation Experience of Solid-State Switching Klystron Modulator in NSLS-II Linac, Proc. of NAPAC2019, Lansing, MI, USA, (2019) 519-521

[126] C. D. Beard, J. Alex, H.-H. Braun, et al., RF System Performance in SwissFEL, Proc. of LINAC, Liverpool, UK, (2022) 679-684

[127] L. Zang, T. Kamitani, SuperKEKB positron source target protection scheme, Proc. of IPAC2013, Shanghai, China (2013) 315-317

[128] V. Bharadwaj, Y. Batygin and J. Sheppard et al., Analysis of beam-induced damage to the SLC positron production target, SLAC-PUB-9438, (2002)

[129] A. Zhang, L. Xu, J. Sun et al., Nuclear Instruments and Methods in Physics Research Section A 1039, 167107 (2022)

[130] H. Nagoshi, M. Kuribayashi, M. Kuriki et al., A design of an electron driven positron source for the international linear collider, Nuclear Instruments and Methods in Physics Research Section A 953, 163134 (2020)

[131] EPICS：http://www.aps.anl.gov/epics/

[132] TANGO：http://www.tango-controls.org/

[133] DOOCS：http://tesla.desy.de/doocs/doocs.html

[134] TINE：http://tine.desy.de/

[135] MADOCA:http://www.spring8.or.jp/en/about_us/manage_structure/jasri/control_system/madoca/

[136] K. Kanazawa, SuperKEKB mechanical assembly at IR, Workshop on the mechanical optimization of the FCC-ee MDI, Geneva, Switzerland, (2018): https://indico.cern.ch/event/694811/

[137] S. Terui, T. Ishibashi, T. Abe, et al., Low-Z collimator for SuperKEKB. Nuclear Inst. and Methods in Physics Research, A 1047, 167857 (2023)

[138] T. Ishibashi, S. Terui, Y. Suetsugu, K. Watanabe, and M. Shirai, Movable collimator system for SuperKEKB, Physical Review Accelerators and Beams 23, 053501 (2020)

[139] Li Xunfeng, Research on the Beam Pipe and Its Thermal Control System of the Beijing Electron-Positron Collider, PhD Thesis, University of Science and Technology Beijing, 2008.

[140] Zheng Lifang, Li Xunfeng, Wang, Li, Beam Pipe System of the Beijing Electron-Positron Collider [Monograph], Beijing: Metallurgical Industry Press, 2024.8. (in Chinese)

[141] V. Rude, M. Duquenne, L. Mans et al., Validation of the crab-cavities internal monitoring strategy, Proc. of 14th International Workshop on Accelerator Alignment, Grenoble, France, (2016):

https://www.slac.stanford.edu/econf/C1610034/papers/642.pdf

[142] M. Sosin, F. Micolon, V. Rude et al., Robust Optical Instrumentation for Accelerator Alignment Using Frequency Scanning Interferometry, Proc. of IPAC2021, Campinas, Brazil, (2021) 2204-2206

[143] Léonard Watrelot, Sosin M, Stéphane Durand. Frequency scanning interferometry based deformation monitoring system for the alignment of the FCC-ee machine detector interface, Meas. Sci. Technol. 34, 075006 (2023)

[144] Liu Zhonghe, Stability analysis of tunnel control network and beam orbit smoothing of particle accelerator, PhD Thesis, PLA Strategic Support Force Information Engineering University, (2020) (in Chinese)

[145] Yang Yanbing, Research on the magnetic center measurement methods for accelerator magnetic components, PhD Thesis, University of Chinese Academy of Sciences, (2022) (in Chinese)

[146] Tao Luo, et al, Error Analysis and Application of Laser Tracker's Bundle Adjustment in the Tunnel Alignment Measurement of Particle Accelerator, Geomatics and Information Science of Wuhan University, 48(6), (2023) 919-925. (in Chinese)

[147] Tao Luo, Zhijun Qi, Wei Wang, et al, Improved Adjustment Method Applied in Transformation of Laser Tracker, Chinese Journal of Lasers, 51(2), 0204001 (2024) (in Chinese)

[148] Zhaoyi Wang, Tao Luo, Wei Wang, et al., Fixing positions and orientations of laser trackers during bundle adjustment in multi-station measurements, Measurement Science and Technology, 32(3), 035017 (2021)

[149] Zhijun Qi, Wei Wang, Wenxian Zeng et al., Two Efficient Recursive Total Least Squares Solutions Based on the Grouping Strategy, IEEE Transactions on Instrumentation and Measurement, Vol.73, 100910 (2024)

[150] National Standard of the People's Republic of China, Basic standards for protection against ionizing radiation and for the safety of radiation sources, GB 18871 - 2002 (in Chinese)

[151] M. Yan, Q. Wu, Y. Ding et al., Study on Radiation Field in the Injection Area of Beijing Electron-Positron Collider (BEPC II), Radiation Protection, 37 (1), (2017) 12-17 (in Chinese)

[152] M. Yan, X. Gong, Q. Zhang et al., Measurement of Prompt Radiation Field in the Linear Accelerator Tunnel of BEPC II, Nuclear Electronics & Detection Technology, 36 (8), (2016) 880-884 (in Chinese)

[153] T. Li, Z. Chen, Q. Zhang et al., Personal Safety Interlock Access Control System for the Major Upgrade Project of Beijing Electron-Positron Collider (BEPC II), Proc. of the National Symposium on Radioactive Effluent and Environmental Monitoring and Assessment, (2002) 544-549 (in Chinese)

[154] J. Li, Y. Tang, B. Shao et al., Environmental Radiation During Beam Tuning and Operation of Beijing Electron-Positron Collider, Radiation Protection, 10(1), (1990) 1-10 (in Chinese)

[155] Q. Liu, Study on Key Problems of Radiation Protection for the Accelerator at the High Energy Photon Source, PhD Thesis, University of Chinese Academy of Sciences. (2022) (in Chinese)